\documentclass[structabstract]{aa}  

\usepackage{graphicx}
\usepackage{float}
\usepackage[varg]{txfonts}
\usepackage{amsmath}
\usepackage{longtable}
\usepackage{longtable,lscape}
\usepackage{lscape}
\usepackage{epsfig}

\newcommand\be{\begin{equation}}
\newcommand\en{\end{equation}}

\usepackage{txfonts}
\begin{document} 

\title{The Herschel/PACS view of the Cep\,OB2 region: \\Global protoplanetary disk evolution and clumpy star formation\thanks{Based on observations obtained with the
{\it Herschel} Space Telescope within Open Time proposal "Disk dispersal in Cep\,OB2", OT1\_asicilia\_1. {\it Herschel} is an ESA space
observatory with science instruments provided by European-led PI consortia and with important participation from NASA.}}

\author{Aurora Sicilia-Aguilar\inst{1,2}, Veronica Roccatagliata\inst{3},  Konstantin Getman\inst{4}, 
Pablo Rivi\`{e}re-Marichalar\inst{5,2}, Tilman Birnstiel\inst{6}, Bruno Mer\'{i}n\inst{7}, 
Min Fang\inst{2}, Thomas Henning\inst{8},  Carlos Eiroa\inst{2}, Thayne Currie\inst{9}}

\institute{\inst{1}SUPA, School of Physics and Astronomy, University of St Andrews, North Haugh, St Andrews KY16 9SS, UK\\
	\email{asa5@st-andrews.ac.uk}\\
	\inst{2}Departamento de F\'{\i}sica Te\'{o}rica, Facultad de Ciencias, Universidad Aut\'{o}noma de Madrid, 28049 Cantoblanco, Madrid, Spain \\
	\inst{3}Universit\"ats-Sternwarte M\"unchen, Ludwig-Maximilians-Universit\"at,Scheinerstr.~1, 81679 M\"unchen, Germany \\
	\inst{4}Department of Astronomy \& Astrophysics, 525 Davey Laboratory, Pennsylvania State University, University Park PA 16802\\
	\inst{5}Kapteyn Astronomical Institute, P.O. Box 800, 9700 AV Groningen, The Netherlands\\
	\inst{6}Harvard-Smithsonian Center for Astrophysics, 60 Garden Street, Cambridge MA 02138, USA\\
	\inst{7}Herschel Science Centre, ESAC-ESA. PO Box 78. E-28691 Villanueva de la Ca\~{n}ada, Madrid, Spain \\
	\inst{8}Max-Planck-Institut f\"{u}r Astronomie, K\"{o}nigstuhl 17, 69117 Heidelberg, Germany\\
	\inst{9}Department of Astronomy \& Astrophysics, University of Toronto, Canada\\}

   \date{Submitted July 24 2014, accepted October 8 2014}

\abstract
{The Cep\,OB2 region, with its two intermediate-aged clusters Tr\,37 and NGC\,7160, is a 
paradigm of sequential star formation and an ideal site for studies of protoplanetary disk evolution.}
{We use Herschel data to study 
the protoplanetary disks and the star formation history of the region.}
{Herschel/PACS observations at 70 and 160\,$\mu$m probe 
the disk properties (mass, dust sizes, structure) and evolutionary state
of a large number of young stars. Far-IR data also trace the
remnant cloud material and small-scale cloud structure. }
{We detect 95 protoplanetary disks at 70\,$\mu$m, 41 at
160\,$\mu$m, and obtain upper limits for more than 130 objects. 
The detection fraction at 70\,$\mu$m depends on the spectral type (88\% for K4 or earlier stars,
17\% for M3 or later stars) and 
on the disk type ($\sim$50\% for full and pre-transitional disks, $\sim$35\% for
transitional disks,  no low-excess/depleted disks detected). Non-accreting
disks are not detected, suggesting significantly lower masses. Accreting transition
and pre-transition disks have systematically higher 70$\mu$m excesses than full disks,
suggestive of more massive, flared and/or thicker disks. Herschel data also reveal several mini-clusters in Tr\,37, 
small, compact structures containing a few young stars surrounded by nebulosity. }
{Far-IR data are an excellent probe of the evolution of
disks that are too faint for submillimetre observations.
We find a strong link between far-IR emission and accretion, and between the inner and outer
disk structure. 
Herschel confirms the dichotomy between accreting and 
non-accreting transition disks. Accretion is a powerful measure of global disk evolution: 
Substantial mass depletion and global evolution need to occur to shut down accretion 
in a protoplanetary disk, even if the disk has inner holes. Disks likely follow different
evolutionary paths: Low disk masses do not imply opening inner holes, and having inner holes does
not require low disk masses.
The mini-clusters reveal multi-episodic star formation in Tr\,37.  
The long survival of mini-clusters suggest that they formed
from the fragmentation of the same core. Their various morphologies favour different
formation/triggering mechanisms acting within the same cluster.
The beads-on-a-string structure in one mini-cluster is consistent with gravitational fragmentation/focusing
acting on very small scales (solar-mass stars in $\sim$0.5 pc filaments). 
Multi-episodic star formation could also produce evolutionary variations between disks in the same region.
Finally, Herschel also unveils what could be the first heavy mass loss episode of the O6.5 star
HD\,206267 in Tr\,37.}


\authorrunning{Sicilia-Aguilar et al.}

\titlerunning{Herschel observations of Cep\,OB2}

\maketitle


\section{Introduction \label{intro}}

The Cep\,OB2 region is a classical example of sequential or triggered star formation (Patel et al. 1998).
Located at 870 pc distance (Contreras et al. 2002), it contains the young clusters Tr\,37 and NGC\,7160
(Platais et al. 1998). With mean ages $\sim$4 Myr for Tr\,37 
and $\sim$12 Myr for NGC\,7160 (Sicilia-Aguilar et al. 2005), the two regions are an ideal place for
the study of protoplanetary disk evolution at  critical ages
for disk dispersal (Sicilia-Aguilar et al. 2006; Hern\'{a}ndez et al. 2007). 
Spitzer IRAC and MIPS surveys have been used to study the
disk population among the cluster members (Sicilia-Aguilar et al. 2006a; Mercer et al. 2009 [M09]; 
Morales-Calder\'{o}n et al. 2009 [MC09]; Getman et al. 2012 [G12]; Sicilia-Aguilar et al. 2013b [SA13]), 
estimating a disk fraction $\sim$48$\pm$5\% for solar-type stars (spectral types late G/early M) in Tr\,37,
while only 2 disks are found among the more than 60 NGC\,7160 low-mass members.
Spitzer photometry also revealed significant disk evolution, with the mean disk excesses being
lower than in Taurus (Sicilia-Aguilar et al. 2006a). Spitzer/IRS and IRAM millimetre data allowed to 
impose stronger constraints on the evolution of the disks, estimating the total disk mass in 
small dust, and revealing grain growth and inside-out evolution (Sicilia-Aguilar et al. 2007, 2011). 

Nevertheless, observations in the near- and mid-IR have a limited power to reveal the structure of
the disk as a whole. For evolved disks in the distant Cep\,OB2 region, 
single-dish submillimetre data are often unfeasible due to low fluxes, cloud contamination, and beam
dilution. 
Constraining the evolutionary status of a disk is a complex problem, given the high
degree of degeneracy between nearly all important parameters involved in disk evolution
(grain size, disk flaring, settling, presence of inner holes and gaps; e.g. D'Alessio et al. 2004, 2005;
Dullemond \& Dominik 2004; Whitney et al. 2013). Far-IR data
is thus a bridge between the inner and outer disk properties, and a key to measure 
global disk evolution, including global small dust depletion, settling, and
changes in the grain size distribution. 
Herschel/PACS data allow us to obtain details about the
global mass and structure of faint and very evolved disks, which typically have 
too low millimetre/submillimetre fluxes to be detected.

If Cep\,OB2 is a paradigm of sequential star formation (Patel et al. 1995), 
Tr\,37 is an example of multi-episodic star formation.
IC\,1396\,A is clearly younger than 
the main Tr\,37 cluster (Sicilia-Aguilar et al. 2005, 2006a; Sicilia-Aguilar et al. 2014, from
now on Paper I). In addition, previous Spitzer and optical
studies suggested the presence of smaller structures containing $<$10 spatially-related,
stars (Barentsen et al. 2001 [B11]; G12; 
SA13), which are probably younger than the main cluster population 
attending to their disk features and high accretion rates. 
These grouplets or mini-clusters are a reduced version of
the subclusters found in massive star-forming regions (MYStiX survey; 
Getman et al. 2014; Kuhn et al. 2014), 
although the latter are denser, presumably more numerous, and larger in size than the mini-clusters. 
Mini-clusters extend the idea of clumpy star formation in molecular clouds
down to less massive, less dense regions. Triggering (Getman et al. 2014) or sequential
formation of stars in a non-uniform cloud may offer an answer to such structures. Herschel data
allows us to track cooler (compared to Spitzer) and less dense (compared to millimetre observations)
cloud structures, revealing the details of the star formation history of the region.

In Paper I, we discussed the young IC\,1396\,A globule in Tr\,37,
and associated embedded protostars.
Here we present the results of our Herschel/PACS survey of disks and cloud structure
in the Tr\,37 and NGC\,7160 clusters. 
Observations and data reduction are discussed in Section \ref{data}. In Section \ref{analysis}
we analyse the disk properties and evolutionary status as seen by Herschel, and the cluster
structure. In Section \ref{discussion} we discuss the evolutionary signs observed in the disks in
the context of the dispersal of the gaseous and dusty disk and its potential outcomes,
and the possibility of multi-episodic star formation in Tr\,37. Finally, our results
are summarised in Section \ref{conclu}.

\section{Observations and data reduction \label{data}}

\subsection{Herschel/PACS observations and map-making \label{obs}}

The Cep\,OB2 clusters, Tr\,37 (centered at 21:38:09, +57:26:48 J2000)
 and NGC\,7160 (centered at 21:53:40, 62:36:10 J2000), were observed with the 
ESA Herschel Space Observatory (Pilbratt et al. 2010) using 
the Photodetector Array Camera and Spectrometer (PACS; Poglitsch et al. 2010),
as part of the OT program ``Disk dispersal in Cep\,OB2" (PI A. Sicilia-Aguilar).
We obtained a total of 23h observing time in parallel mode at 70 and 160\,$\mu$m.
Table \ref{obs-table} lists AORs, pointings, sensitivity, and map areas.
Observations comprise 6 scan maps and 19 mini-maps.
Large scan maps were designed to cover most of the cluster members
and the extended emission, while mini-maps were selected to trace fainter
sources with deeper integration.
There is some overlap at the edges between the 
maps that increase the sensitivity in localized areas. 
For the large maps, we took two AORs per pointing, corresponding to scan and cross-scan
(at 45 and 135 degrees with respect to the array). For mini-maps, we observed
two AORs per pointing with scan and cross-scan at 70 and 110 degrees with respect to the array.
Observations took place between November 2012 and January 2013.

\begin{figure*}
\centering
\begin{tabular}{c}
\includegraphics[width=1.0\linewidth]{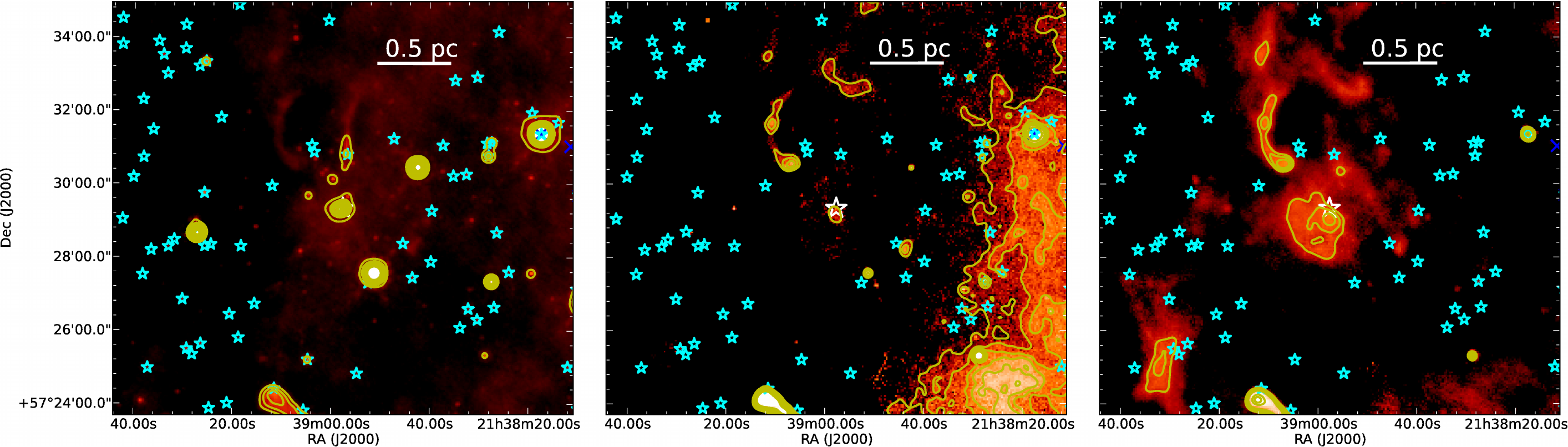}\\
\includegraphics[width=1.0\linewidth]{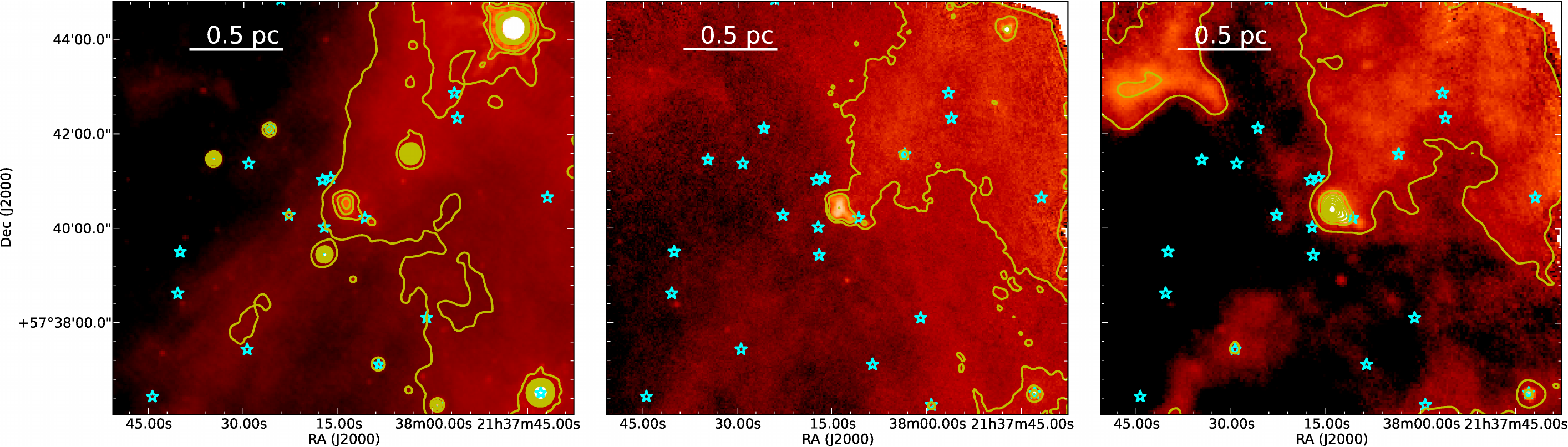}\\
\end{tabular}
\caption{Images at 24, 70, and 160\,$\mu$m (from left to right) of the
center (up) and the northern part (down) of Tr\,37. The O6.5 star HD\,206267 is marked with 
a large white star in the upper diagrams. 
Other cluster members (with and without disks) are marked with small cyan stars.  The
extended object at 21:38:14, +57:40:27 is likely an unrelated planetary
nebula (based on cluster age considerations).\label{centernorth-fig}}
\end{figure*}

The data were reduced using 
{\sc Hipe}\footnote{{\sc Hipe} is a joint development by the Herschel Science Ground Segment 
Consortium, consisting of ESA, the NASA Herschel Science Center, and the HIFI, 
PACS and SPIRE consortia.} environment, version 9.0 (Ott et al. 2010) 
and the Unimap\footnote{See http://w3.uniroma1.it/unimap/ for further details} 
software (Piazzo et al. 2012). We use the PACS data calibration of 2013 February 14.
Unimap is recommended for PACS observations of
sources with complex background (Paladini et al. 
2013\footnote{http://herschel.esac.esa.int/twiki/pub/Public/PacsCalibrationWeb/
pacs\_mapmaking\_report\_ex\_sum\_v3.pdf}).
The strongest extended emission in the region is associated the IC\,1396\,A globule
(Paper I). The sources presented in this work are point-like (disks) or 
small ($<$15") extended structures. For
intermediate-brightness sources, the differences in the photometry resulting from
{\sc Hipe} and Unimap maps are negligible, but very faint sources are washed out if iterative masking techniques
are applied, or artificially enhanced if independent masking is provided within {\sc Hipe}.
The map construction starts from Level 0 products, flags bad and saturated pixels,
converts ADUs to Volts, and computes the coordinates with aberration correction. 
Unimap processes the standard Level 1 products using a Generalised Least Squares approach (GLS)
to estimate the noise spectrum.
It also fixes cosmic rays, glitches, and drifts.
The initial GLS map is subsequently corrected from the distortion introduced by GLS, finally resulting in
post-processed, weighted GLS (WGLS) maps.
The final maps are projected with pixel sizes 2" for the 70~$\mu$m image, and 3" for 
the 160~$\mu$m image using the UniHIPE software, also available from Unimap.
Overlapping AORs of the large scan maps were combined together. Combining large maps and
mini-maps does not significantly improve the S/N of sources far from the centre, so we treated both kinds of maps separately.
Figure \ref{centernorth-fig} show some parts of Tr\,37
observed with Herschel/PACS.

\begin{table*}
\caption{List of Herschel/PACS AORs and fields.} 
\label{obs-table}
\begin{tabular}{l c c c c c c c c}
\hline\hline
Field & AOR Numbers & RA   & DEC & Duration &  Date  & Type & Size & 1$\sigma_{70\mu m}$/1$\sigma_{160\mu m}$ \\
     &             &    (J2000)     &    (J2000)      &  (s) &    &  &  & (mJy)  \\
\hline
01-580 & 1342261383/4 & 21:53:37.07 & +62:28:47.0 & 2$\times$895 & 2013-01-18 & M & 6.5'$\times$2.5'  &  1.1/3.0 \\
01-1152 & 1342261385/6 & 21:53:19.83 & +62:34:00.5 & 2$\times$445 & 2013-01-18 & M & 6.5'$\times$2.5'  & 1.8/4.1  \\
DG-481 & 1342261387/8 & 21:52:21.13 & +62:45:03.5 & 2$\times$445 & 2013-01-18 & M & 6.5'$\times$2.5'  & 1.8/4.1  \\
North & 1342256224/5 & 21:38:20.15 & +57:41:01.6 & 2$\times$2244 & 2012-11-29 & S & 7.0'$\times$7.0'  & 2.5/5.6  \\
West & 1342256961/2 & 21:35:18.90 & +57:32:15.0 & 2$\times$4563 & 2012-12-11 & S & 10.5'$\times$10.5' & 2.3/5.1 \\
72-875 & 1342256963/4 & 21:35:49.75 & +57:24:04.1 & 2$\times$895 & 2012-12-11 & M & 6.5'$\times$2.5'  &  1.1/3.0 \\
54-1781 & 1342258024/5 & 21:38:16.13 & +57:19:35.8 & 2$\times$445 & 2013-01-01 & M & 6.5'$\times$2.5'  & 1.8/4.1  \\
23-798 & 1342259517/8 & 21:41:28.65 & +57:36:43.3 & 2$\times$445 & 2013-01-11 & M & 6.5'$\times$2.5'  & 1.8/4.1  \\
IC1396A & 1342259791/2 & 21:37:07.24 & +57:29:41.9 & 2$\times$3101 & 2013-01-16 & S & 8.0'$\times$8.0'  & 2.4/5.4  \\
Cluster & 1342259793/4 & 21:38:18.77 & +57:31:36.8 & 2$\times$10082 & 2013-01-16 & S & 12.5'$\times$12.5'  & 1.6/3.6  \\
East & 1342259795/6 & 21:39:58.97 & +57:32:36.1 & 2$\times$8156 & 2013-01-17 & S & 13.5'$\times$13.5'  & 2.0/4.3  \\
93-720 & 1342261393/4 & 21:40:10.00 & +58:00:03.7 & 2$\times$445 & 2013-01-18 & M & 6.5'$\times$2.5'  &  1.8/4.1 \\
21393104 & 1342261395/6 & 21:39:31.05 & +57:47:14.0 & 2$\times$445 & 2013-01-18 & M & 6.5'$\times$2.5'  & 1.8/4.1  \\
91-506 & 1342261397/8 & 21:38:58.07 & +57:43:34.4 & 2$\times$445 & 2013-01-18 & M & 6.5'$\times$2.5'  & 1.8/4.1  \\
92-393 & 1342261399/400 & 21:39:44.08 & +57:42:16.0 & 2$\times$895 & 2013-01-18 & M & 6.5'$\times$2.5'  & 1.1/3.0   \\
KUN-196 & 1342261401/2 & 21:40:15.09 & +57:40:51.3 & 2$\times$445 & 2013-01-18  & M & 6.5'$\times$2.5'  & 1.8/4.1  \\
24-1796 & 1342261403/4 & 21:40:11.83 & +57:40:12.2 & 2$\times$445 & 2013-01-18  & M & 6.5'$\times$2.5'  & 1.8/4.1  \\
13-1250 & 1342261431/2 & 21:39:12.14 & +57:36:16.5 & 2$\times$445 & 2013-01-18  & M & 6.5'$\times$2.5'  & 1.8/4.1  \\
21-998 & 1342261433/4 & 21:39:34.80 & +57:23:27.8 & 2$\times$445 & 2013-01-18  & M & 6.5'$\times$2.5'  & 1.8/4.1  \\
21-33 & 1342261435/6 & 21:39:35.62 & +57:18:22.1 & 2$\times$445 & 2013-01-18  & M & 6.5'$\times$2.5'  & 1.8/4.1  \\
21362507 & 1342261853/4 & 21:36:25.08 & +57:27:50.3 & 2$\times$445 & 2013-01-23  & M & 6.5'$\times$2.5'  & 1.8/4.1  \\
11-1209 & 1342261855/6 & 21:36:58.51 & +57:23:25.8 & 2$\times$445 & 2013-01-23  & M & 6.5'$\times$2.5'  & 1.8/4.1  \\
12-1091 & 1342261857/8 & 21:37:57.62 & +57:22:47.7 & 2$\times$445 & 2013-01-23  & M & 6.5'$\times$2.5'  & 1.8/4.1  \\
54-1547 & 1342261859/60 & 21:38:44.46 & +57:18:09.1 & 2$\times$445 & 2013-01-23  & M & 6.5'$\times$2.5'  &  1.8/4.1 \\
21391145 & 1342261861/2 & 21:39:07.50 & +57:23:52.6 & 2$\times$541 & 2013-01-23  & S & 4.5'$\times$3.0'  & 3.0/6.9  \\
\hline
\end{tabular}
\tablefoot{List of the observed fields and information about the pointings. The first three pointings
belong to the cluster NGC\,7160, while the rest belong to Tr\,37.
The size of the fields correspond approximately to the places with high
coverage. The 1$\sigma$ value corresponds 
to the average sensitivity estimated by Hspot at the center 
of the map. The local sensitivity decreases near extended structures or at the edges of the maps, 
and increases in the combined map in regions covered by more than one map. The map type is marked as scan-map (S) or mini-map (M). }
\end{table*}

\subsection{Point source photometry \label{photometry}}

We performed point source aperture photometry for all the known young stars with
disks in Cep\,OB2 located within our Herschel/PACS fields.  A list of spectroscopically confirmed cluster
members was drawn from Sicilia-Aguilar et al. 2004, 2005, 2006b, and SA13.
We also analysed other candidate members based on
Spitzer IR excesses, H$\alpha$ emission, and X-ray emission (M09; MC09; B11; G12). 
The IRAF\footnote{IRAF is distributed by the National Optical Astronomy 
Observatories, which are operated by the Association of Universities for Research
in Astronomy, Inc., under cooperative agreement with the National Science Foundation.} 
package $apphot$ was used to automatically calculate the aperture photometry on
the source position (considering 2MASS coordinates), to subtract the sky, and
to derive the corresponding errors.
Since the background is highly variable throughout most of the fields, we 
used small apertures (5" for 70\,$\mu$m, 8" for 160\,$\mu$m), together with the
aperture corrections (1.764 for 70\,$\mu$m, 1.919 for 160\,$\mu$m) recommended by 
Herschel\footnote{http://herschel.esac.esa.int/twiki/pub/Public/PacsCalibrationWeb/ pacs\_bolo\_fluxcal\_report\_v1.pdf}.
Sky annuli were placed between 12-18" for the 70\,$\mu$m maps, and 18-27" for the 160\,$\mu$m maps.
The aperture corrections are derived considering larger sky annuli, but we adopted a compromise
between sky location and the strongly variable background
found in these regions (see also Sicilia-Aguilar et al. 2013a).
The errors are calculated taking into account 
the average sky rms and the correlated noise due to undersampling, estimated as 
(3.2/$pixsize$)$^{0.68}$/0.95, where $pixsize$ is the pixel size in the projected
maps (A. Mora private communication). We also include a general error of 10\% to account for
flux calibration, aperture, and mapping effects on the final flux. This error is
consistent with the typical variations observed in the photometry of 
intermediate-brightness sources derived from maps
constructed with different tools (Unimap vs {\sc Hipe} standard 
scripts). Colour corrections  for objects with
relatively flat $\nu$F$_\nu$ distributions are very small compared to 
the flux errors (see PACS calibration
manual\footnote{http://herschel.esac.esa.int/twiki/pub/Public/PacsCalibrationWeb/ cc\_report\_v1.pdf};
Howard et al. 2013), so the photometry is presented uncorrected "as is".
 
Automated algorithms for finding point-sources often fail in far-IR images 
with extended structures.
Therefore, in addition to checking
source positioning, centering, and S/N (detections must be $\geq$3$\sigma$ over the
background), we did a careful visual inspection to control that
the sources were real and point-like. 
We do not expect the disks to have
extended emission (the PACS PSF at 70\,$\mu$m is 5.6", equivalent to $\sim$5000 AU
at 870\,pc), so extended sources are regarded as
``cloud contamination" and thus non-detections. 
Special attention was paid to sources showing coordinate mismatches
$>$2" that could be a sign of misidentification with red background sources.
We only consider as real Herschel detections those belonging to objects with IR excesses 
(from Spitzer and/or WISE data) in their spectral energy
distributions (SED). Sources with apparent PACS counterparts but no IR excess
are probably spurious because the Herschel detection limits are several orders of 
magnitude over the photospheric emission of late-type stars or associated cold/debris disks. 
For completeness, we explored objects with marginal excess at 8 and/or 24$\mu$m only, 
but none of them is detected with Herschel.
Some of the H$\alpha$ candidates in B11 ($\sim$15\%) do not show any disk emission
at Spitzer wavelengths, and are thus not considered.

Source detectability varies greatly across the maps, depending on the presence of 
cloud/background emission. For non-detected sources in clean areas, we can estimate significant
upper limits. For sources associated with bright regions such as IC\,1396\,A, the
upper limits are of the order of Jy and thus do not offer any further constraint to the SEDs.
Sources at the edge of the maps have also uncertain upper limits and are not considered. For cases
with low, smooth background, we estimated a 3$\sigma$ upper limit to the flux considering the
sky rms at the position of the source. For a few
objects, the automated photometry classified them as detections, but visual inspection of the maps 
revealed that although there is some emission consistent with the source, the local background
structures (e.g. nebular patches, nearby bright objects) 
could affect the detectability. We consider these as ``marginal detections",
since the source appears to be detected and its flux could be of the same order than the
``marginal" detection, but there is a high risk of contamination.

The photometry for all objects and their SEDs are presented in Appendix \ref{photometry-app}.
From our spectroscopically identified sources, (483 in Tr\,37 and 145 in NGC\,7160, many of which are out
of the Herschel fields), we detect 77 at 70\,$\mu$m (only one in NGC\,7160),
36 of them also at 160\,$\mu$m. All detections are listed in 
Table \ref{photometry-table} in Appendix \ref{photometry-app}. 
The only object detected in NGC\,7160 is the only surviving accreting star
in this cluster. From the 270 targets from M09, MC09, B11, and G12 (many of which are outside the Herschel fields), 
18 are detected at
70\,$\mu$m and 5 of them are also detected at 160\,$\mu$m. All these objects are listed in Table \ref{other-table}.
Two further objects, 21373786+5728467 and 21373885+5732494, have potential far-IR detection, but since
the rest of data does not show disk-like excesses, we classify them as most likely background-contaminated.
We also obtain significant upper limits for 2 objects with disks
in NGC\,7160 and 117  more in Tr\,37, among spectroscopically-identified members (Table \ref{uplims-table}). 
A further 25 objects without spectroscopic confirmation have significant upper limits (Table \ref{otheruplims-table}).
Their properties (including membership confirmation, spectral type, and presence of accretion) 
are less well-established than for spectroscopically confirmed members, but the
strong selection criteria offered by X-ray, IR excess, and H$\alpha$ photometry ensure a very reduced
potential contamination by non-members.

The maps also reveal several tens of point sources in the area that have not been classified
as YSO by previous surveys. Due to the thorough search for YSO in the center of the clusters
(optical, X-ray, H$\alpha$, and Spitzer/WISE searches from various
teams), most of the new point
sources probably correspond to extragalactic objects, so we leave them out of the current work.

\subsection{Ancillary data \label{ancillary}}

There is a large collection
of ancillary data available for most sources, including optical photometry 
and spectroscopy (Contreras et al. 2002; Sicilia-Aguilar et al. 2004, 2005; SA13; B11), 
2MASS near-IR photometry (Cutri et al. 2003), Spitzer IRAC/MIPS fluxes (Sicilia-Aguilar
et al. 2006a; SA13; M09; MC09; G12). 
Objects with optical spectroscopy and photometry have also spectral
type, extinction estimates, and
strong constraints on the presence of accretion. For a subset of objects with 
accurate spectral types and extinction, individual, instantaneous
accretion rates are also available (Sicilia-Aguilar et al. 2010). Finally, for a few objects we also have
Spitzer/IRS spectra and 1.3mm continuum emission measurements and/or upper limits 
(Sicilia-Aguilar et al. 2007, 2011). Many objects have been detected with WISE, but 
we give priority to Spitzer data, since WISE photometry is strongly affected by source
blending and cloud emission in such distant
and relatively crowded fields. All the Spitzer photometry for the sources
presented here was derived self-consistently following the procedures in
(SA13).

All the available data\footnote{Ancillary data available upon request.} 
is included to trace the SEDs of the sources
(Section \ref{seds}). The
data are non-simultaneous. Optical variability is
common in young stars (Herbst et al. 2000; Brice\~{n}o et al. 2001; Sicilia-Aguilar et al. 2005),
but most of the CepOB2 objects do not seem to be strongly variable in the IR.
Only in a few cases the SEDs betray potential IR variability, such as 21-998, 21394850+572049544,
and 21395813+5728335. For very variable sources, the SED classification may be uncertain.

\begin{figure*}
\centering
\begin{tabular}{cccc}
\includegraphics[width=0.24\linewidth]{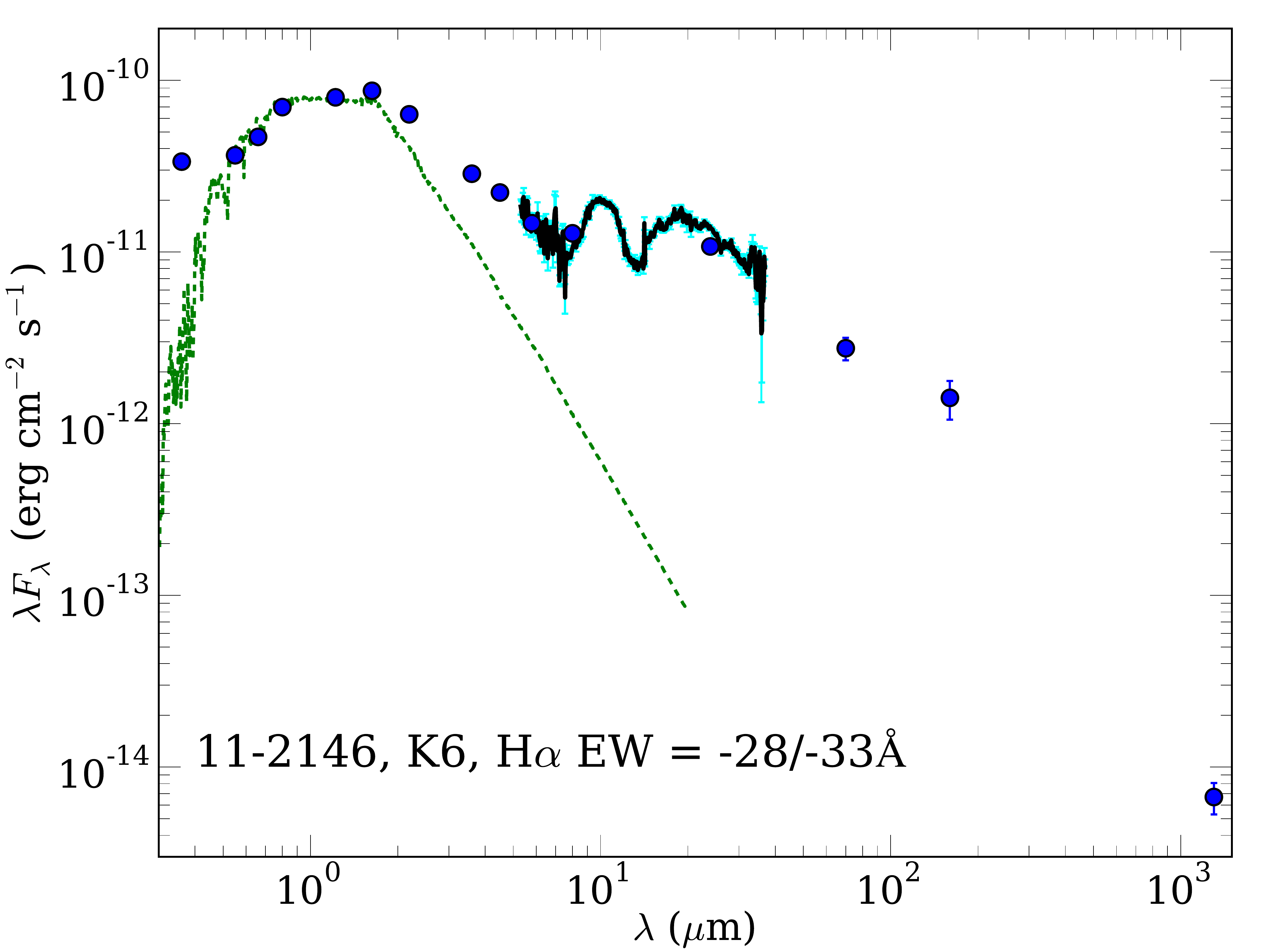} &
\includegraphics[width=0.24\linewidth]{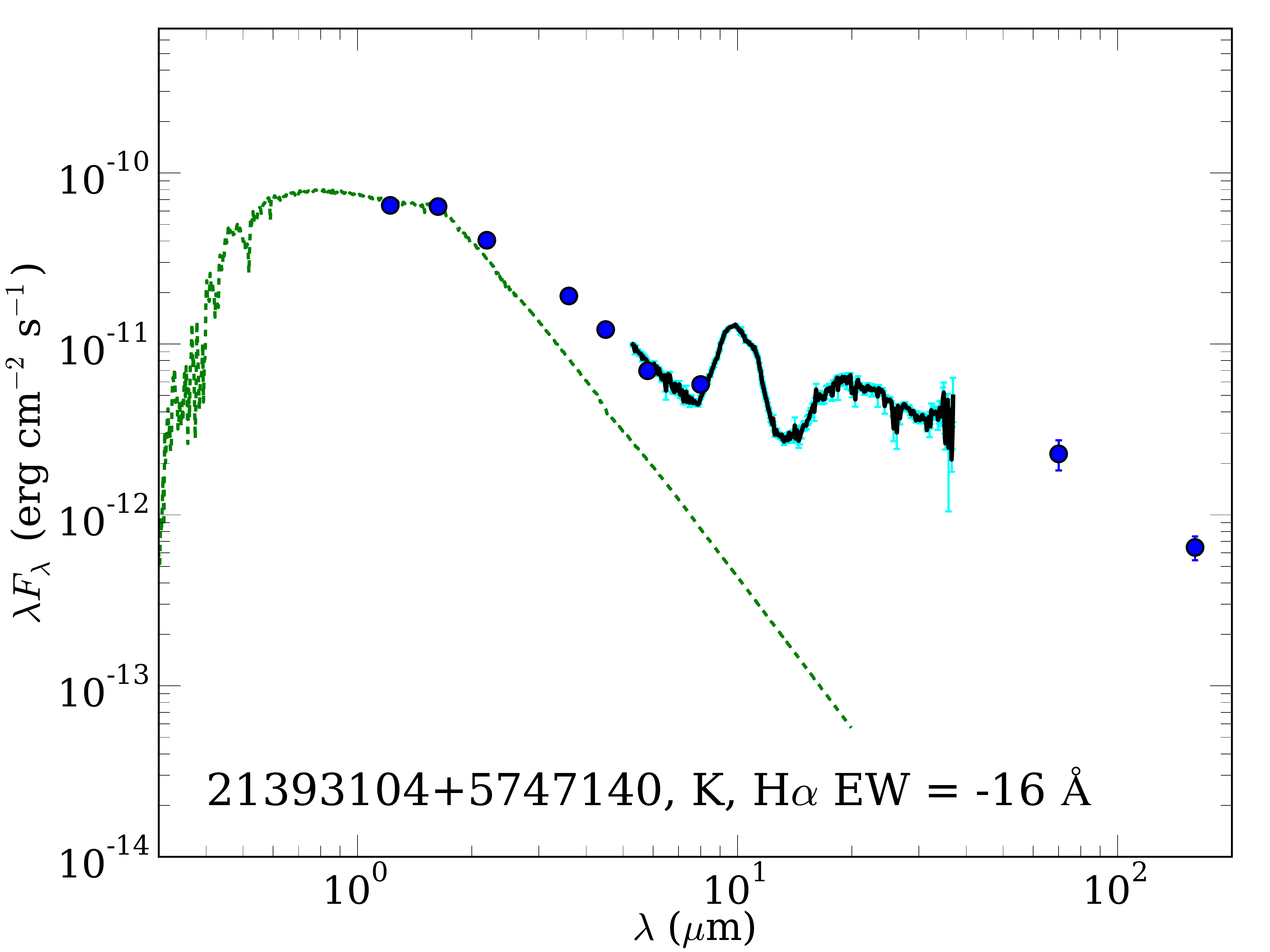} &
\includegraphics[width=0.24\linewidth]{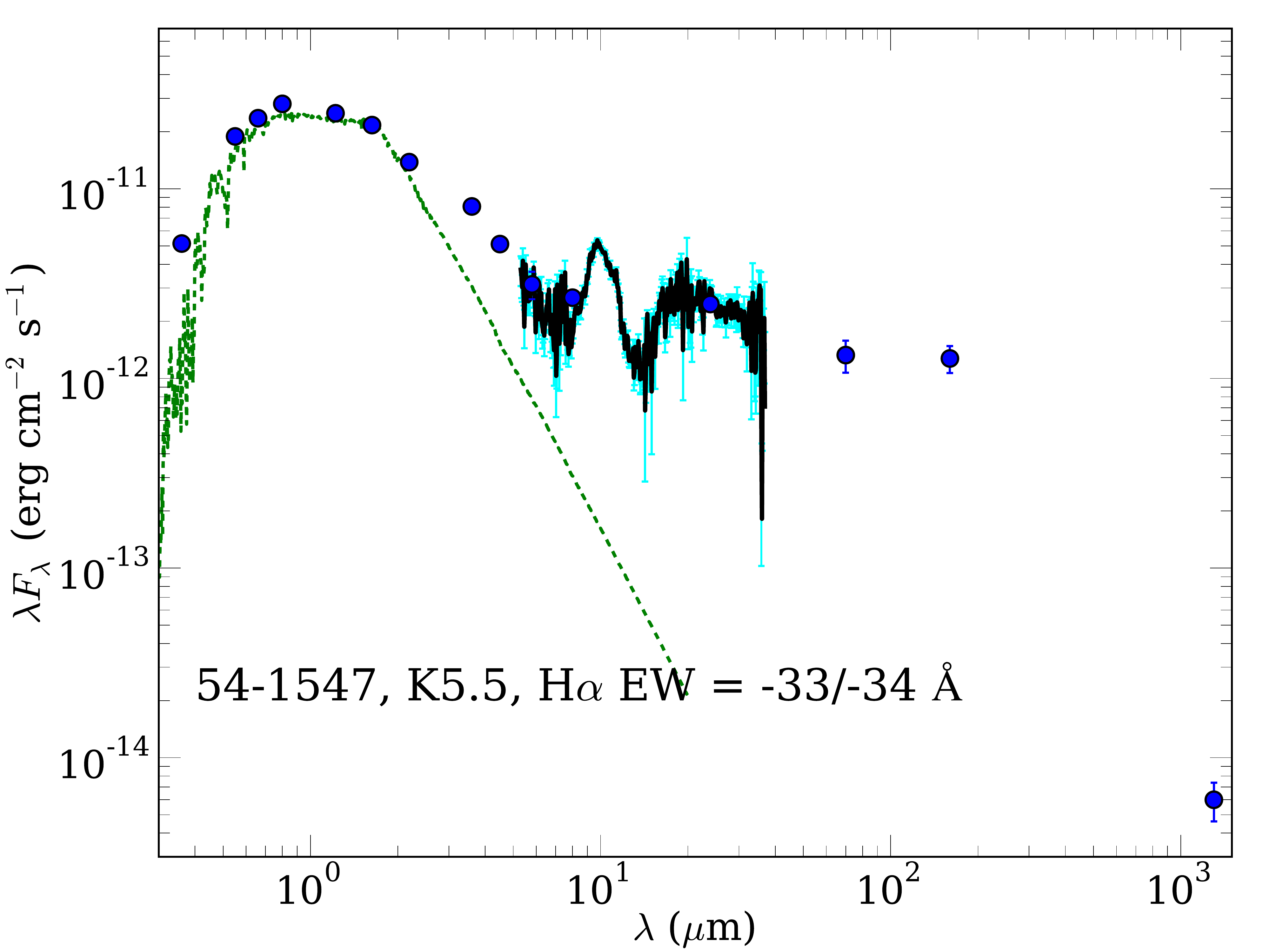} &
\includegraphics[width=0.24\linewidth]{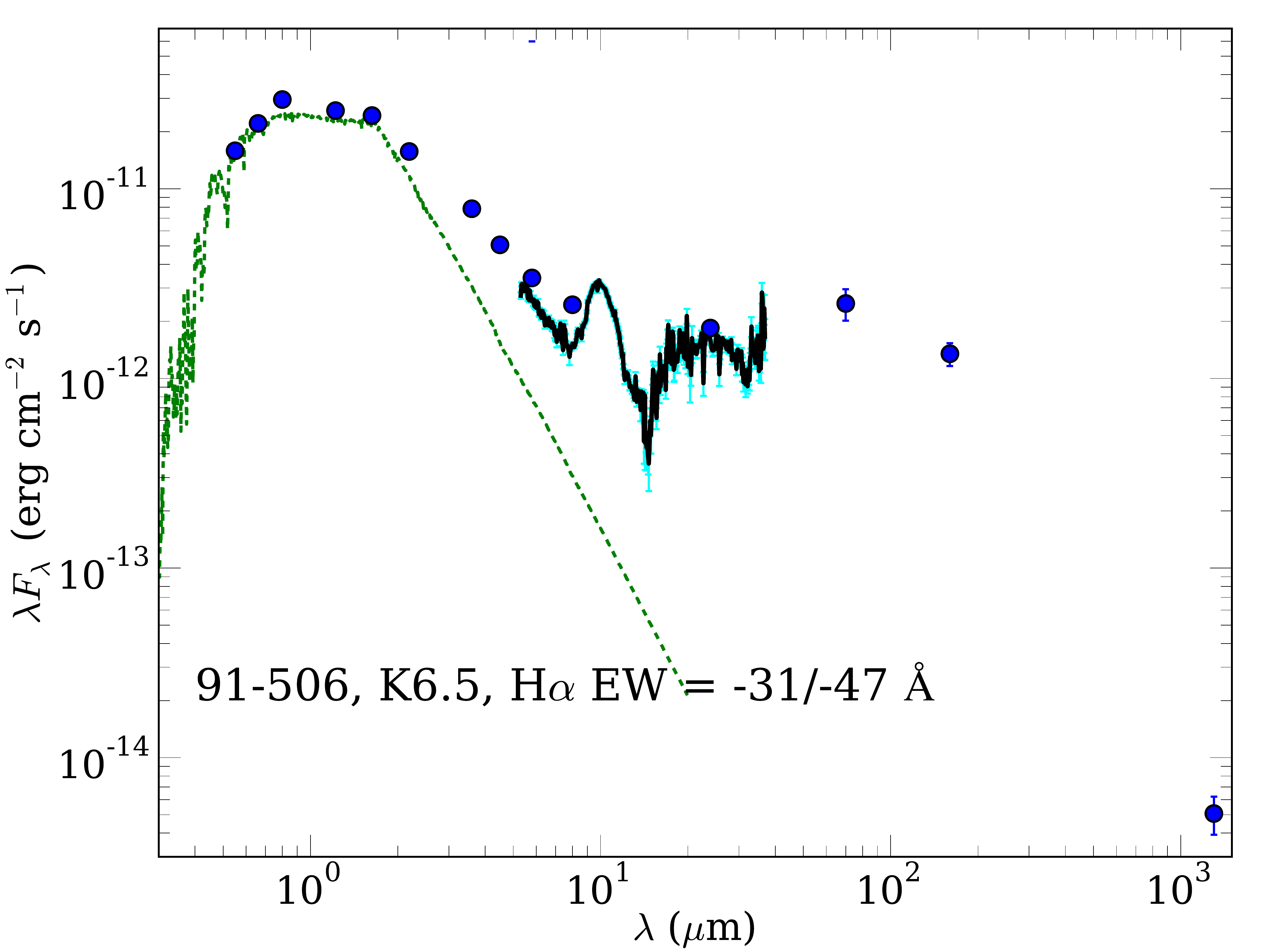} \\
\includegraphics[width=0.24\linewidth]{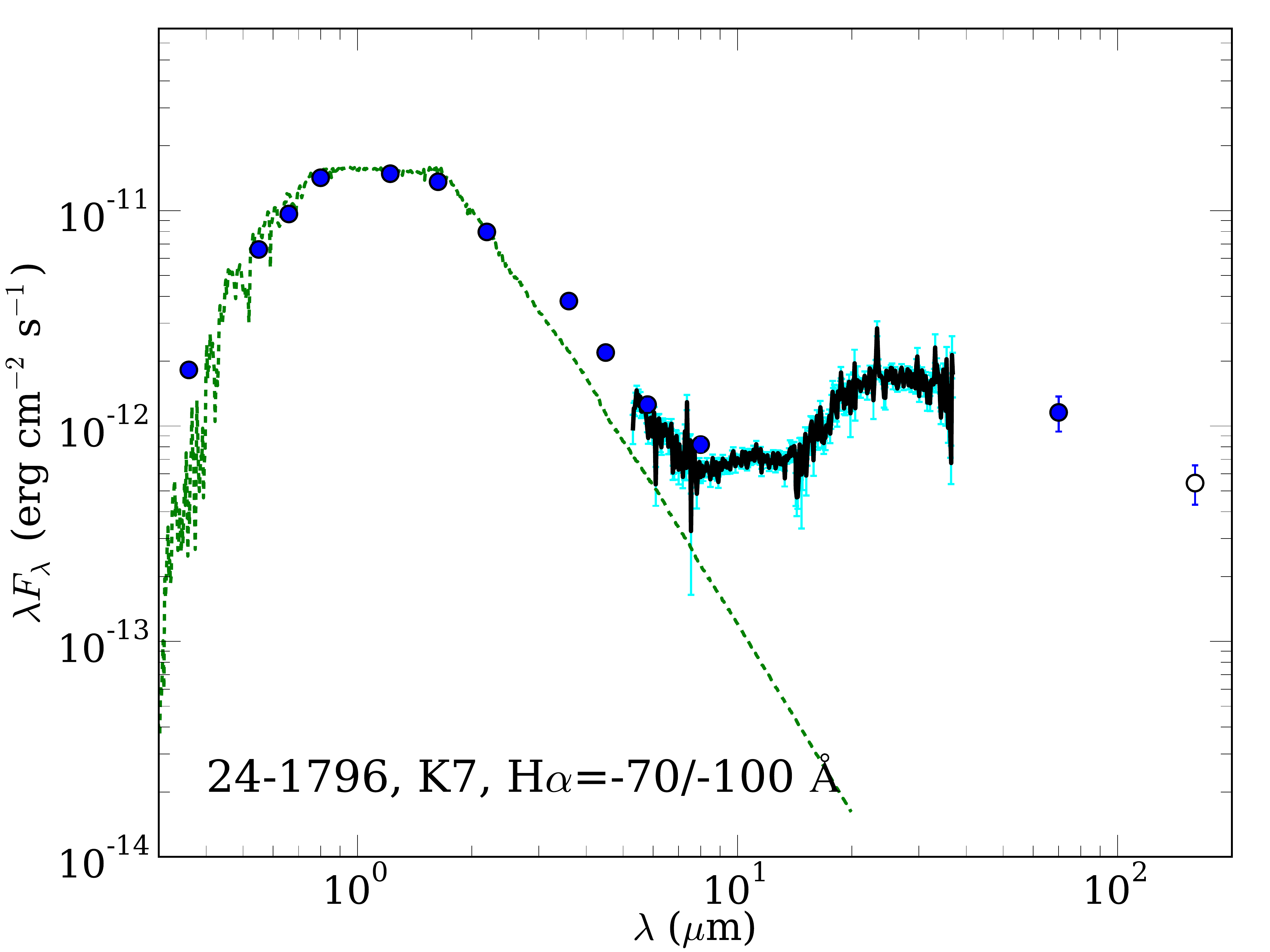} &
\includegraphics[width=0.24\linewidth]{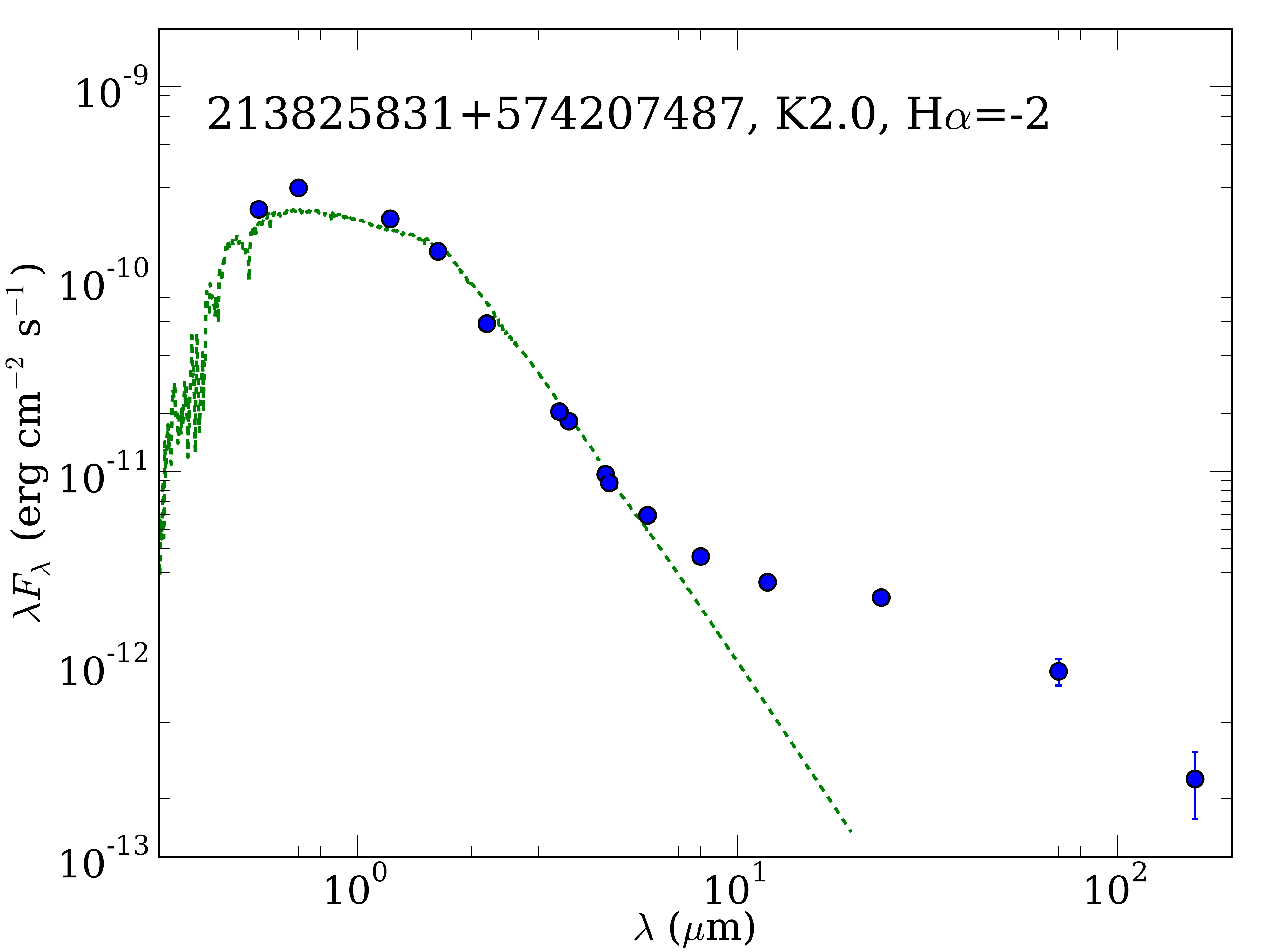} &
\includegraphics[width=0.24\linewidth]{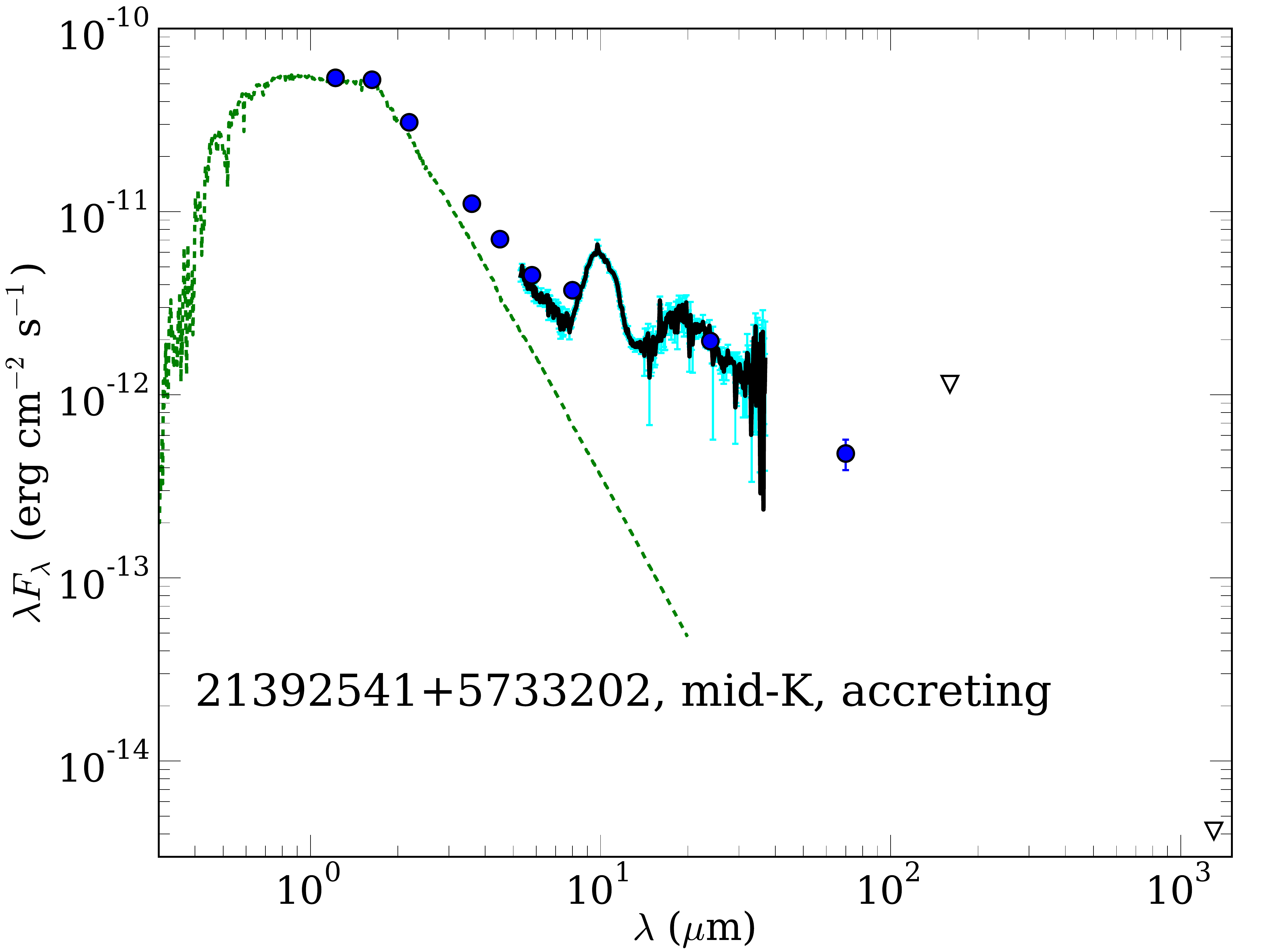} &
\includegraphics[width=0.24\linewidth]{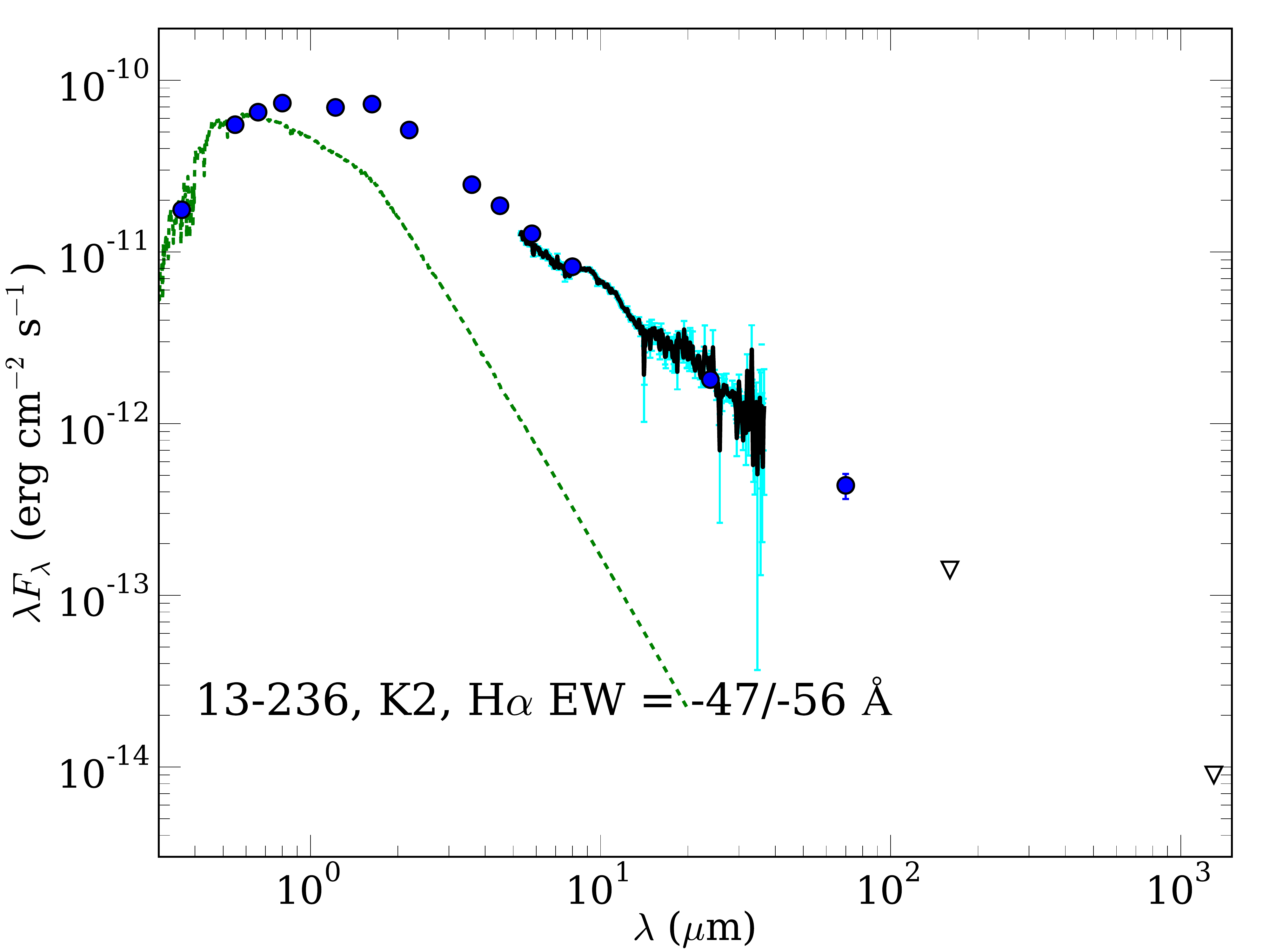} \\
\end{tabular}
\caption{Examples of SEDs of objects detected with Herschel, including optical, Spitzer (IRAC/MIPS photometry and
IRS spectra), WISE, and IRAM 1.3mm data. Errorbars are often smaller than the symbols (full circles). Upper limits
are marked as inverted open triangles, marginal detections are open circles. In the upper row, 11-2146 and 21393104+5747140 are full disks, 
54-1547 and 91-506 are pre-transitional disks. In the lower row, 24-1796 and 213825831+574207487 are
transitional disks, and 21392541+5733202 and 13-236 are disks with signs of strong dust processing/low mass in
small grains.  Spectral types and H$\alpha$ equivalent width in emission
are also listed. A MARCS model photosphere (Gustafsson et al. 2008) is shown
as a dotted line for comparison. \label{sedssample-fig}}
\end{figure*}

\section{Analysis \label{analysis}}

\subsection{SEDs and disk classification \label{seds}}

The disks were classified following the scheme in SA13
according to their Spitzer fluxes and after a careful inspection of their SEDs. 
Table \ref{SEDclass-table} lists the criteria for SED classification and the
possible interpretations in terms of disk structure and evolutionary status. 
For low-mass stars (spectral types G-M), our classification
distinguishes between full disks (with excesses at all wavelengths),
transitional disks with inside-out evolution (TD; with nearly photospheric colours, [3.6]-[4.5]$<$0.2 mag, 
and large excesses at longer wavelengths), ``kink" or pre-transitional disks (PTD; with moderate-to-low 
near-IR excesses, [3.6]-[4.5]$\sim$0.2-0.4 mag, and strong silicate features and/or a sharp 
change in slope between 8-12 and 24$\mu$m), and depleted/low-excess disks (with near-IR excesses 
in the range of PTD but a steep slope at all wavelengths, resulting in low 24$\mu$m excesses;
Currie et al. 2009). The classification is not unique, as some of the depleted/low-excess
disks also satisfy the criteria for TD.  
For the high-mass stars, the classification was done by visual inspection of the SEDs and  comparison
to stellar templates (MARCS models, Gustafsson et al. 2008). 
Class I sources in IC\,1396\,A were presented in Paper I.
For a few objects where part of
the data is missing or inconsistent (usually, due to nearby bright objects, cloud
emission, or IR variability), we adopt an ``uncertain" SED type and exclude them from the discussion.

\begin{table*}
\caption{SED classification.} 
\label{SEDclass-table}
\begin{tabular}{l c l}
\hline\hline
Disk Class           		& Criteria & Interpretation \\
\hline
Transition disk (TD)  		& [3.6]-[4.5]$<$0.2 mag         & Inner hole, radial variations of dust properties \\			
				& 				& and/or settling, inside-out evolution.\\
\\
Kink/Pre-transition disk (PTD) 	& [3.6]-[4.5]$\sim$0.2-0.4 mag, & Holes/gaps with small grains, \\
			 	& strong silicate,  		& inside-out evolution,\\
			 	& and sharp slope change 		& dust filtering by planet/companion.\\
\\
Low excess/Depleted disk (D) 	& [3.6]-[4.5]$\sim$0.2-0.4 mag,     & Small dust depletion, strong settling, \\
				& and similar slope up to 24$\mu$m  & global disk evolution.\\
\\
Full disks (F) 			&  None of the above		& Consistent with homogeneous, continuous, flared CTTS disk. \\
\hline
\end{tabular}
\tablefoot{The SED classification used in the paper, together with the various physical interpretations of
the SED morphology in each disk class.}
\end{table*}

The Herschel far-IR fluxes are in very good agreement with the Spitzer photometry and IRS spectra.
Figure \ref{sedssample-fig} shows examples of disks with different SEDs types.
We find consistently lower Herschel fluxes for objects with low mid-IR excesses, showing that
24$\mu$m data offer a good prediction of the global disk structure. This is reflected in
the detection rates of objects: while more than half (46 out of 89) 
of the disks classified
as "full" are detected (11 out of 17 in case of PTD),
only 1/3 of the TD (19 out of 54) and none of the settled/depleted/low-excess disks (out of 19) is detected. 
The detection rate is
strongly dependent on the spectral type: The only star for which we reach the photospheric
levels at 70\,$\mu$m is the O6.5 star HD\,206267. We detect some debris disk candidates
around A and B stars, but none around late-type
stars. For low-mass stars with Spitzer-confirmed
disks and spectral types K4 or earlier, the detection rate is close to 90\% (15 out of 17), but it
drops to $\sim$50\% for late-K stars (26 out of 49) and to $<$30\% for M-type
stars (23 out of 55 for early-M stars, 4 out of 24 for stars with spectral types M3 or later).
Therefore, predictions for stars with spectral types K and earlier are representative for
the whole class, while conclusions on the lowest-mass population are subject to
higher uncertainty and concern only the most massive and flared disks.

The 160\,$\mu$m fluxes are sometimes higher than expected from the SED shape 
and the 70\,$\mu$m data, especially for objects with non-negligible background.
There is extended 160\,$\mu$m emission in some parts
of the cluster (see Figure \ref{centernorth-fig}), and some grouplets of stars are also
associated with small nebular structures (see Section \ref{cloud}). Therefore, even in well-detected
sources, extended nebulosity may contaminate the beam at long wavelengths. 
Extended nebular material could also
affect some of the IRAM/1.3mm detections.

\subsection{The Cep~OB2 disks as seen by Herschel: disk structure and its Far-IR imprint \label{diskproperties}}

\begin{table}
\caption{Common parameters in the RADMC disk models.} 
\label{commonpars-table}
\begin{tabular}{l l}
\hline\hline
Parameter & Value  \\
\hline
Stellar mass 		& M$_*$=1\,M$_\odot$   \\ 
Effective temperature 	& T$_{eff}$=4275 K  \\
Stellar radius		& R$_*$=1.7\,R$_\odot$  \\
Disk outer radius	& R$_{out}$=200 AU \\
Gas-to-dust ratio	& R$_{gas/dust}$=100 \\
\hline
\end{tabular}
\tablefoot{The parameters are representative for a young
K6 star. The outer disk radius is largely unconstrained 
with the available data. Deviations from the standard
gas to dust ratio could also happen in evolved disks.}
\end{table}

To examine the far-IR properties of disks in a more quantitative way, we define the
spectral indices $\alpha$ as the SED slope between two wavelengths $\lambda_1$ and $\lambda_2$ (Lada et al. 1987):
\begin{equation}
	\alpha(\lambda_1-\lambda_2) = \frac{log(\lambda_1 F_{\lambda_1})-log(\lambda_2 F_{\lambda_2})}{log(\lambda_1)-log(\lambda_2)} \label{alphas}
\end{equation}
where F$_{\lambda_x}$ is the flux at wavelength $\lambda_x$.
The Spitzer SED spectral indices (or equivalent colour indices) were used to classify
the disks according to the mentioned types (Sicilia-Aguilar et al. 2006a, 2011; SA13).
Now the Herschel data allow us to explore in an independent way the global disk properties of
objects with inner disks in different evolutionary states.

\begin{figure*}
\centering
\includegraphics[width=1.0\linewidth]{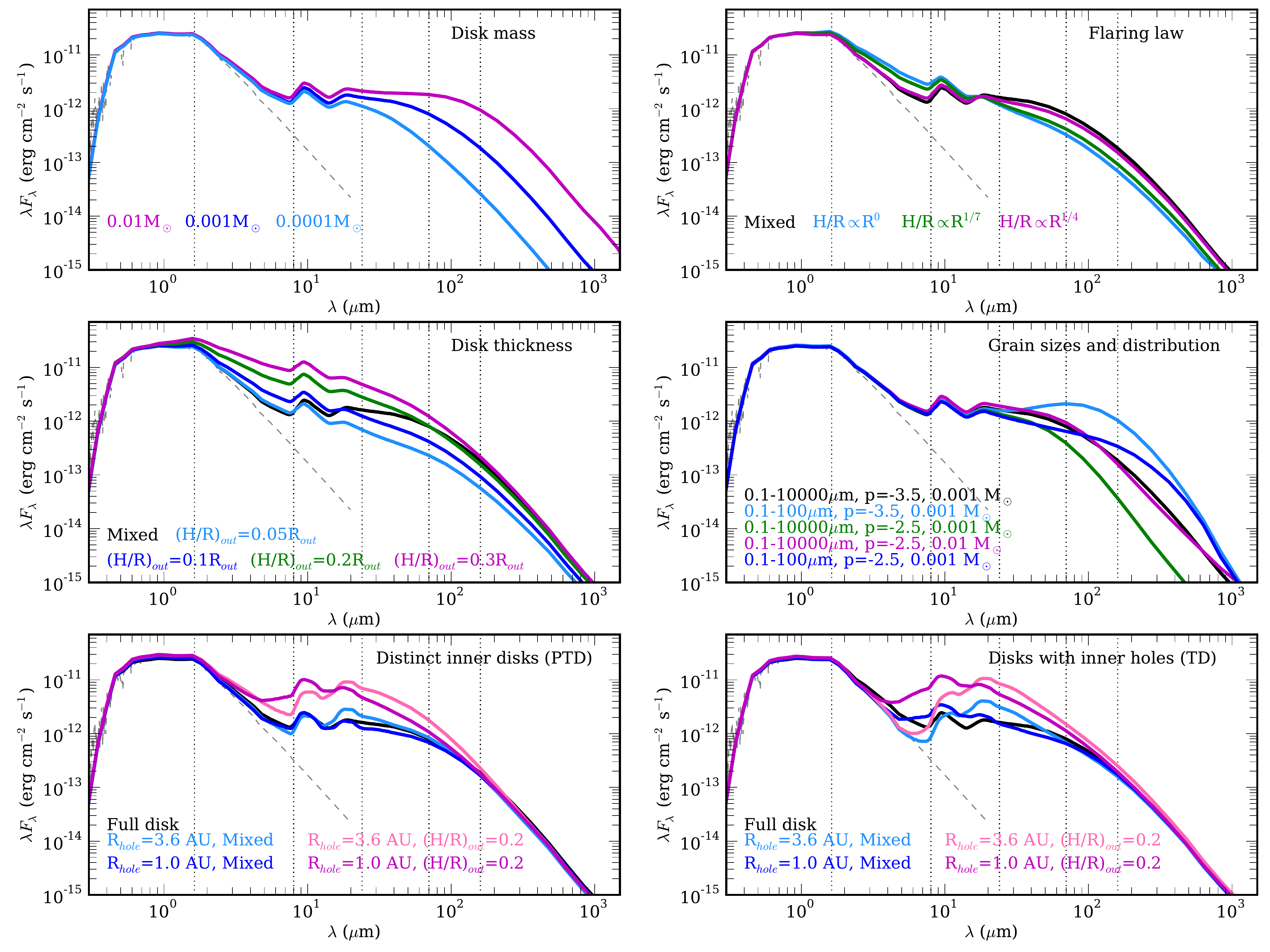}
\caption{Comparison of the SEDs of some of the RADMC models in Table \ref{model-table}.
Unless otherwise specified, the models assume well mixed gas and dust,
a disk mass of 0.001 M$_\odot$ and a standard grain distribution with grain 
sizes 0.1-10000\,$\mu$m and power law exponent p=-3.5. For examples changing the
flaring, thickness, and some of the TD/PTD cases, we show both well-mixed (``Mixed") 
and decoupled dust/gas distributions (given by various H/R relations and outer disk
thickness (H/R)$_{out}$).\label{models-fig}}
\end{figure*}

\begin{figure*}
\centering
\includegraphics[width=1.0\linewidth]{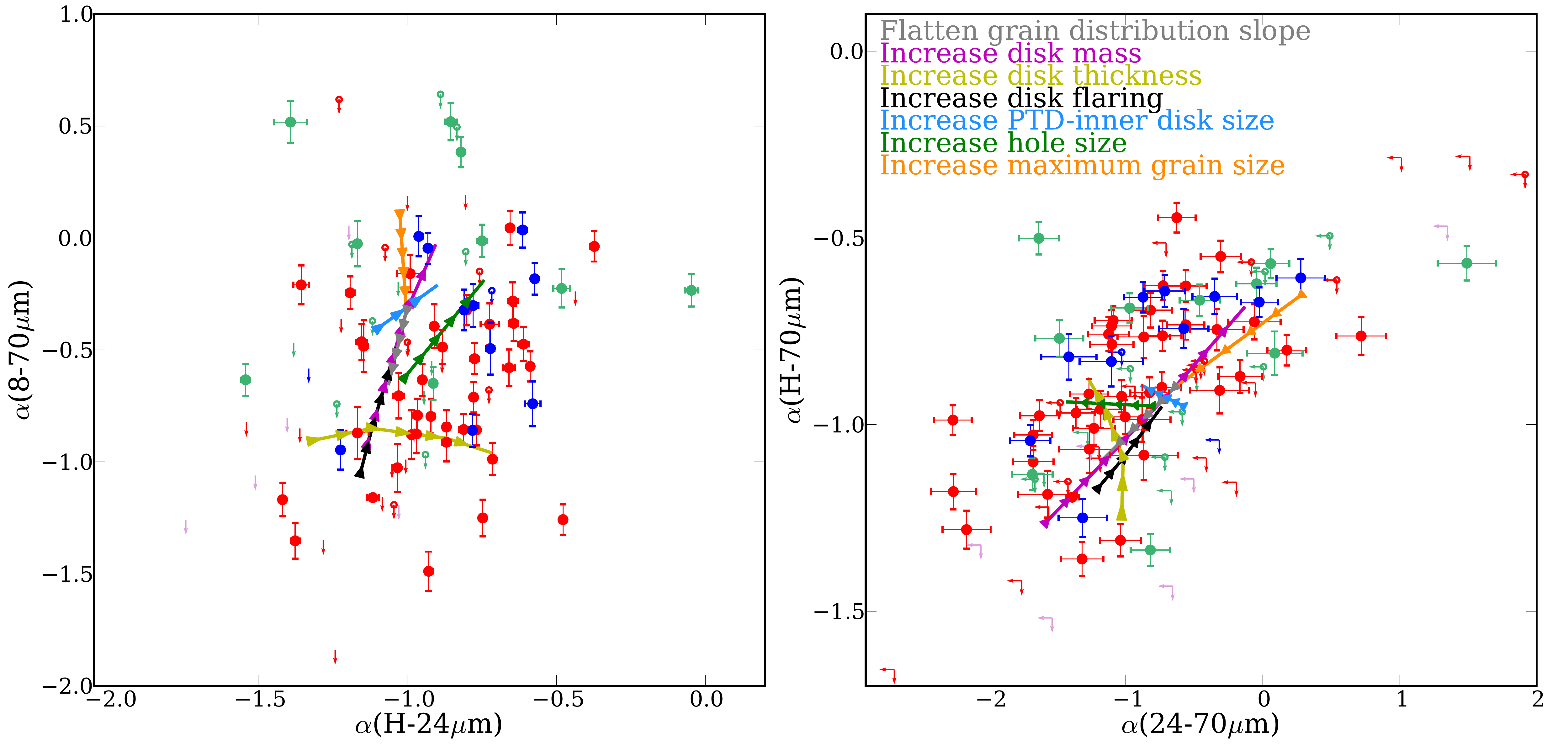}
\caption{A plot with the mid- and far-IR SED spectral indices for stars with different disk types,
classified according to their Spitzer colours. Note the degeneracy that arises between
different disk structures when only simple diagrams are considered.
Full disks are marked in
red, PTD are marked in blue, TD are marked in green, and low excess/depleted disks
are marked in pink. Circles represent detections
 and left/downwards pointing arrows are upper limits. Marginal detections are marked as
upper limits with an additional open circle. We only include upper limits for stars with spectral types
K7 or earlier, given that the data are very incomplete for M-type stars.
The SED spectral indices for the models in Figure \ref{models-fig} are shown as continuous lines,
with arrows marking the direction of the change when various disk properties are modified (see legend). 
\label{alphaalpha-fig}}
\end{figure*}

To understand the effect of the disk structure on the far-IR fluxes, 
we created a set of template models with fixed stellar properties and various disk parameters
(see Table \ref{commonpars-table}).
For a stellar model, we chose a star with T$_{eff}$=4275 K, M$_*$=1 M$_\odot$, and R$_*$=1.7R$_\odot$,
which corresponds approximately to a solar-type star at an age $\sim$3-4 Myr (Siess et al. 2000), similar to 
the typical late-K stars in Cep\,OB2. We then constructed various simple models using the radiative-transfer
code RADMC (Dullemond \& Dominik 2004), and varying the disk mass (assuming a gas-to-dust ratio 100), 
disk thickness, flaring law, grain sizes, and
power-law exponent of the grain size distribution. We also modelled ``kink"/PTD and TD by introducing
radial variations of the inner disk rim (adding inner holes), grain properties, and disk structure. 
The outer disk
radius was kept constant and equal to 200 AU (note that in the absence of millimetre data, 
there is a strong degeneracy between disk radius and
disk mass). The RADMC models assume that dust grains of all sizes are well-mixed and coupled to the
gas, although size-independent dust settling can be simulated by modifying the scale height 
and assuming that the dust is decoupled from the gas in hydrostatic equilibrium.
The details of the individual models are described in Appendix \ref{models-app}. The models are not
constructed to fit any object in particular nor to cover all possible disk structures, 
but to give a broad view of the disk structures observed in Cep\,OB2. Figure \ref{models-fig}
shows the global effects on the SED shape as the disk structure and properties are varied
in a controlled way.

\begin{figure}[h!]
\centering
\includegraphics[width=1.0\linewidth]{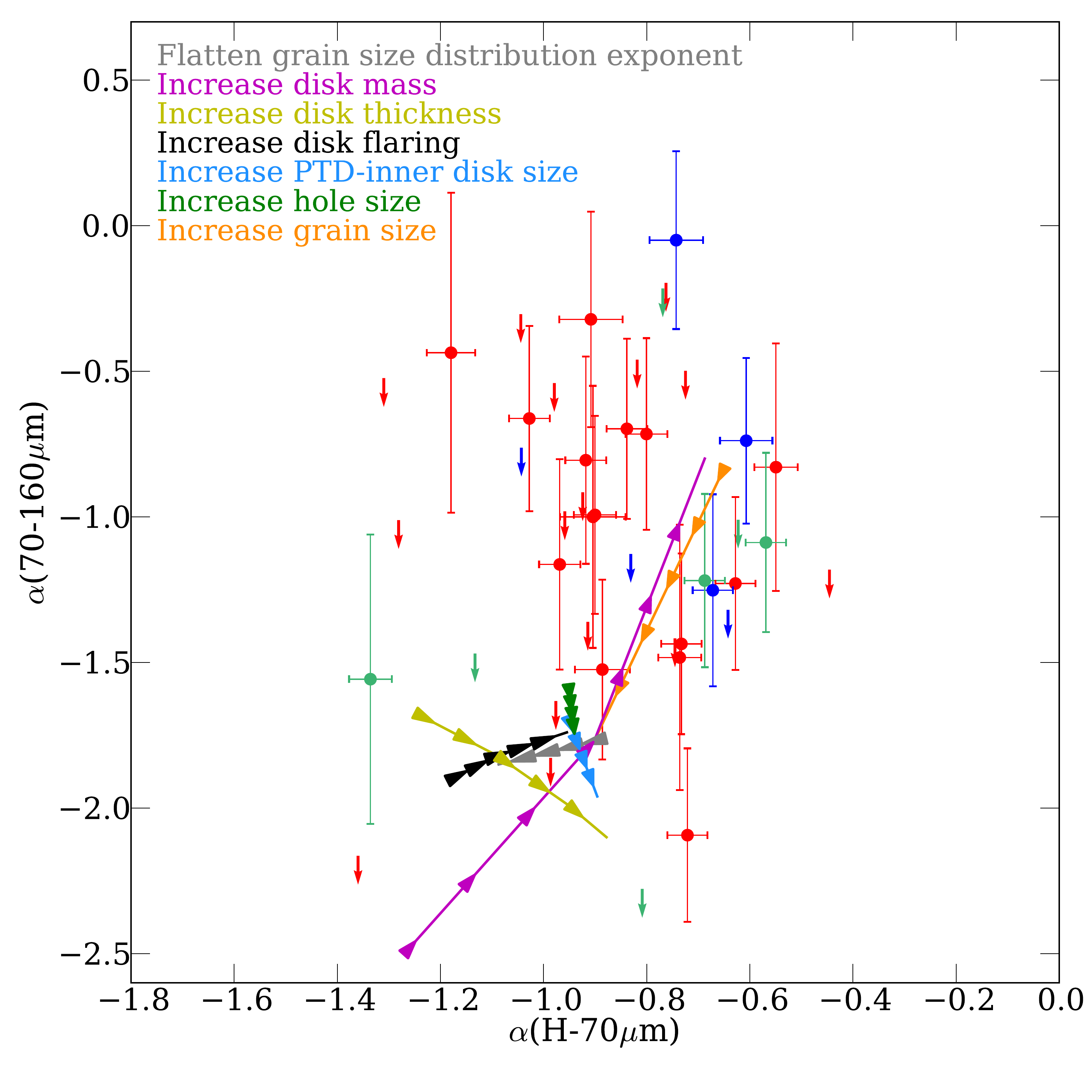}
\caption{Observed SED spectral indices $\alpha$ (including 160$\mu$m) and predicted trends from the models in Figure \ref{models-fig}. 
Detections are marked as dots with errorbars (red for full disks, blue for PTD, green for TD), and
upper limits are shown as arrows. We only consider disks detected at 70\,$\mu$m and with no evidence
of nebular/extended emission at 160\,$\mu$m. The thick lines show the variation of the spectral indices when
the disk properties are changed.
 \label{alpha160-fig}}
\end{figure}

By examining the model SEDs and spectral indices, we can explore
the observable effects of various structural and physical
changes in the disks, and compare this to the Cep\,OB2 observations. Figure \ref{alphaalpha-fig} shows
how the SED spectral index changes when various disks parameters are varied: disk mass
(for full disks with well-mixed gas and dust), disk flaring law (for disks
with constant mass, ranging from non-flared to very flared), disk thickness
(for disks with constant mass and assuming a standard flaring law with
H/R$\propto$R$^{1/7}$), presence of distinct inner disks of increasing
sizes (as in PTD, for disks with constant mass and well-mixed dust and gas),
and presence of inner holes of increasing sizes (as in TD, for disks with constant mass 
and well-mixed dust and gas). The data on the Cep\,OB2 stars is also displayed. 

Figure \ref{alphaalpha-fig} demonstrates
the high degree of degeneracy between physical properties of disks when only colour-colour
diagrams are used. Disk mass and flaring have a similar
effect on the $\alpha$ indices that measure the global far-IR excess, 
while other parameters like disk thickness and presence of
inner holes/gaps (what we call ``transitionality") dominate the effect on the 
mixed mid- and far-IR spectral indices. For
most of our stars H band emission is photospheric (Sicilia-Aguilar et al. 2005), so $\alpha$(H-70\,$\mu$m) is 
sensitive to the total far-IR excess, being thus a good indicator of the global disk mass and disk flaring.
Inside-out evolution and ``transitionality" can be well traced by $\alpha$(8-70\,$\mu$m), although this
index is in part affected 
by the strength of the silicate feature. TD in Cep\,OB2 show a range of silicate features (with some of them having no or
very weak silicate emission, and others having strong features), while
PTD have stronger silicate features than normal disks (and also stronger features than our models
here; Sicilia-Aguilar et al. 2007, 2011). This can make the observed difference between
full and TD/PTD in the $\alpha$(8-70\,$\mu$m) index stronger than predicted by the models. 
The behaviour of $\alpha$(24-70\,$\mu$m) is more complex, being affected by settling, disk mass,
and to some extent, ``transitionality".

The grain sizes and grain size distribution leave detectable imprints in the SED slope
between 70 and 160\,$\mu$m (Figure \ref{models-fig}). 
Small-grain dominated distributions increase the 
far-IR flux and the 160\,$\mu$m emission, making $\lambda$F$_\lambda$ comparable at 70\,$\mu$m
and 160\,$\mu$m. Figure \ref{alpha160-fig} explores the $\alpha$(70-160) index. Although 
we expect to detect  at 160\,$\mu$m only the
most pristine, brighter disks, we find that the SEDs tend to fall on the side of significant
grain growth/dust processing. For some objects, such as 54-1547 and 21-2113,
the SED shape is consistent with a small-grain dominated disk, but since both
cases are associated with some nebular background emission, we
cannot exclude contamination. The very tight upper limits imposed in a few targets
reveal that some of the disks suffer from very strong grain processing and/or 
are very depleted of small dust grains at this stage of evolution. 
Therefore, the large
majority of disks in Tr\,37 are consistent with substantial populations of 
grains with sizes well over 100\,$\mu$m.

\subsection{Cloud structure and temperature \label{cloud}}

\begin{figure*}
\begin{tabular}{c}
\includegraphics[width=0.98\linewidth]{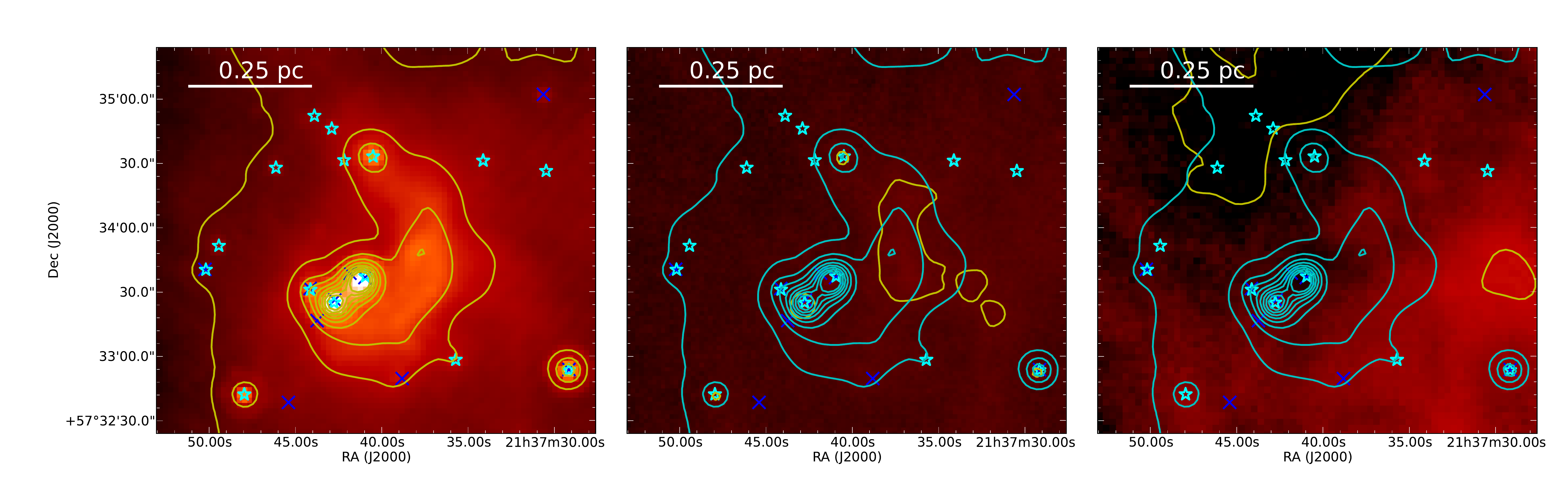} \\
\includegraphics[width=0.98\linewidth]{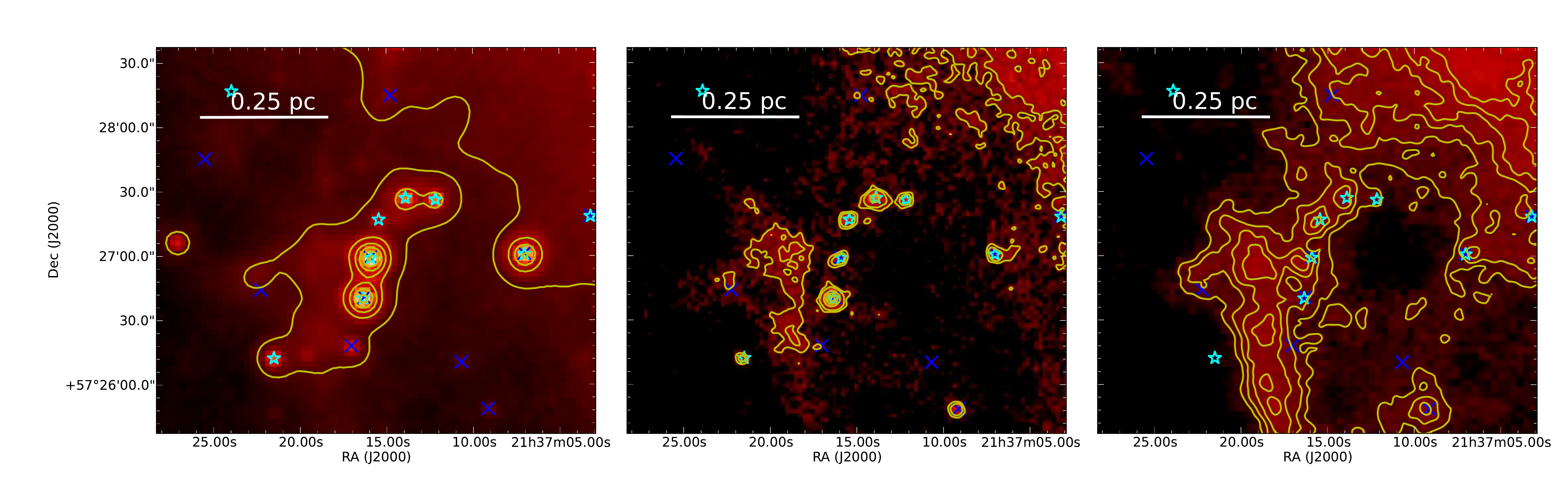} \\
\includegraphics[width=0.98\linewidth]{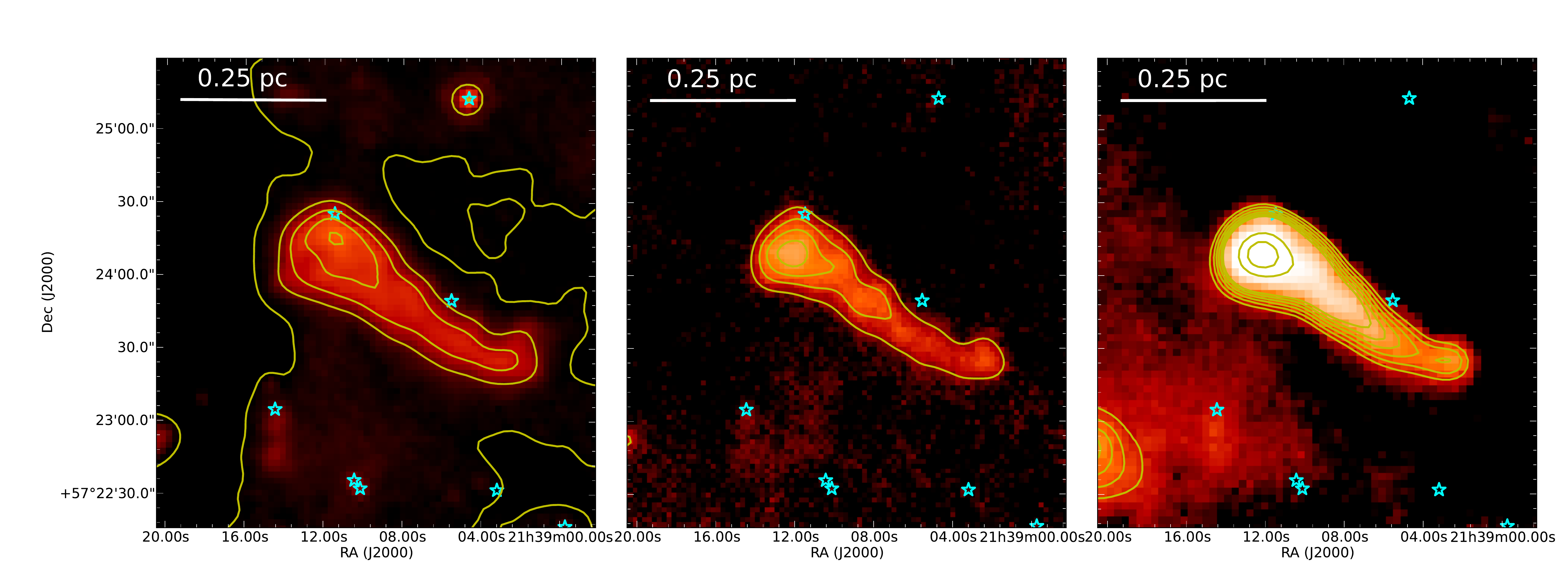} \\
\end{tabular}
\caption{Images at 24, 70, and 160\,$\mu$m (from left to right) of the mini-clusters
associated with CCDM J2137+5734 (top), 11-2031 (middle), and 213911452+572425205 (bottom).
Cluster members (with and without disks) are marked as small cyan stars.
Yellow contours at variable levels for each wavelength are shown to mark the structure of the mini-cluster. For
the cluster associated with CCDM J2137+5734, whose nebular structure is not detected at
Herschel wavelengths, we also overplot the 24\,$\mu$m contours in cyan in the PACS maps 
to mark the location of the warm dust. \label{miniclusters-fig}}
\end{figure*}

Several areas in Tr\,37 show extended far-IR emission. The 12 Myr-old
cluster NGC\,7160, in contrast, is very clean and without any signs of extended
emission at Spitzer and Herschel wavelengths, suggestive of complete cloud removal
at later evolutionary stages. Besides the 
large IC\,1396\,A globule (Paper I), we observe substantial patches of
nebulosity towards the west and north of Tr\,37. They probably correspond
to material at the edge of the bubble-shaped HII region surrounding 
HD\,206267 (Patel et al. 1998; B11).
The nebular emission is
often very inhomogeneous, containing significant small-scale structure.
We also find
small ($\sim$0.25 pc) and bright nebulous patches, containing several 
stars with strong mid- and far-IR emission. We call them ``mini-clusters" (SA13) 
and will be described in detail in Section \ref{miniclusters}. 
Figure \ref{miniclusters-fig} shows the nebular structures associated with
the B3+B5 stars CCDM J2137+5734 (Dommanget et al. 1983; a.k.a. HD\,206081; G12), 11-2031, and the clump
containing the emission line star 213911452+572425205 (SA13).

The observations at 70 and 160\,$\mu$m can be used to trace the temperature (Paper I)
and column density of the cloud structures. 
Temperatures and densities are only significant in regions dominated by cloud emission, since
clean areas are noise-dominated. After resampling both maps at the same resolution
(3"/pixel) and cancelling out fluxes at or below the local noise level (which changes from map to
map), the approximate pixel-by-pixel temperature is calculated from the
flux ratio by assimilating the emission to a modified black body function:
\begin{equation}
	\frac{F_{\nu,70}}{F_{\nu,160}} = \frac{B_{\nu,70}(T) (1-e^{-\tau_{\nu,70}})}{B_{\nu,160}(T) (1-e^{-\tau_{\nu,160}})}. \label{eq1}
\end{equation}

Following Roccatagliata et al. (2013), 
we assume that the emission is optically thin at Herschel/PACS wavelengths,
and consider that the optical depth $\tau$ has a power-law dependency with
the frequency ($\beta$=1.9) and is proportional to the mass absorption coefficient $k_\nu$ and the
column density $\Sigma$, arriving to:

\begin{equation}
	F_{\nu} = \Omega B_{\nu}(T) \tau_{\nu} =  \Omega B_{\nu}(T) k_{\nu} \Sigma. \label{eq2}
\end{equation}

Here, $\Omega$ is the solid angle subtended by the emitting region (3"$\times$3" pixel).
The Hydrogen column density  (N$_H$) is estimated considering the gas to dust ratio (R$_{gas/dust}$=100),
the mass of the Hydrogen atom (m$_H$) and the mean molecular weight ($\mu$ = 2.8):

\begin{equation}
	N_H= \frac{2 \Sigma R_{gas/dust}}{m_H \mu}. \label{eq3}
\end{equation}

We can thus
calculate the temperature and column density structure of the cloud by fitting the fluxes
at 70 and 160\,$\mu$m on a pixel-by-pixel base. Since we use two datapoints to fit two parameters,
we cannot estimate the uncertainties, but the variations of temperature and column density 
between neighbouring pixels give a good idea of the uncertainty levels. 
This method tends to produce slightly higher temperatures than more complex
SED-fitting techniques (Preibisch et al. 2013; Roccatagliata et al. 2013) 
since the short-wavelength emission is dominated by the hottest parts of the
cloud on the line-of-sight. It 
is also affected by the way of treating the dust opacity (Juvela et al. 2013).
The cloud is very
clumpy and inhomogeneous, so the PACS emission may come from different layers of material at various
locations and with different properties.
For compact and distinct structures such as mini-clusters, the overdensities observed
are probably a more reliable measure of the real column density and temperature of the structure,
so they can thus provide
information on the heating mechanisms in the region.

\begin{figure*}
\centering
\begin{tabular}{cc}
\includegraphics[width=0.46\linewidth]{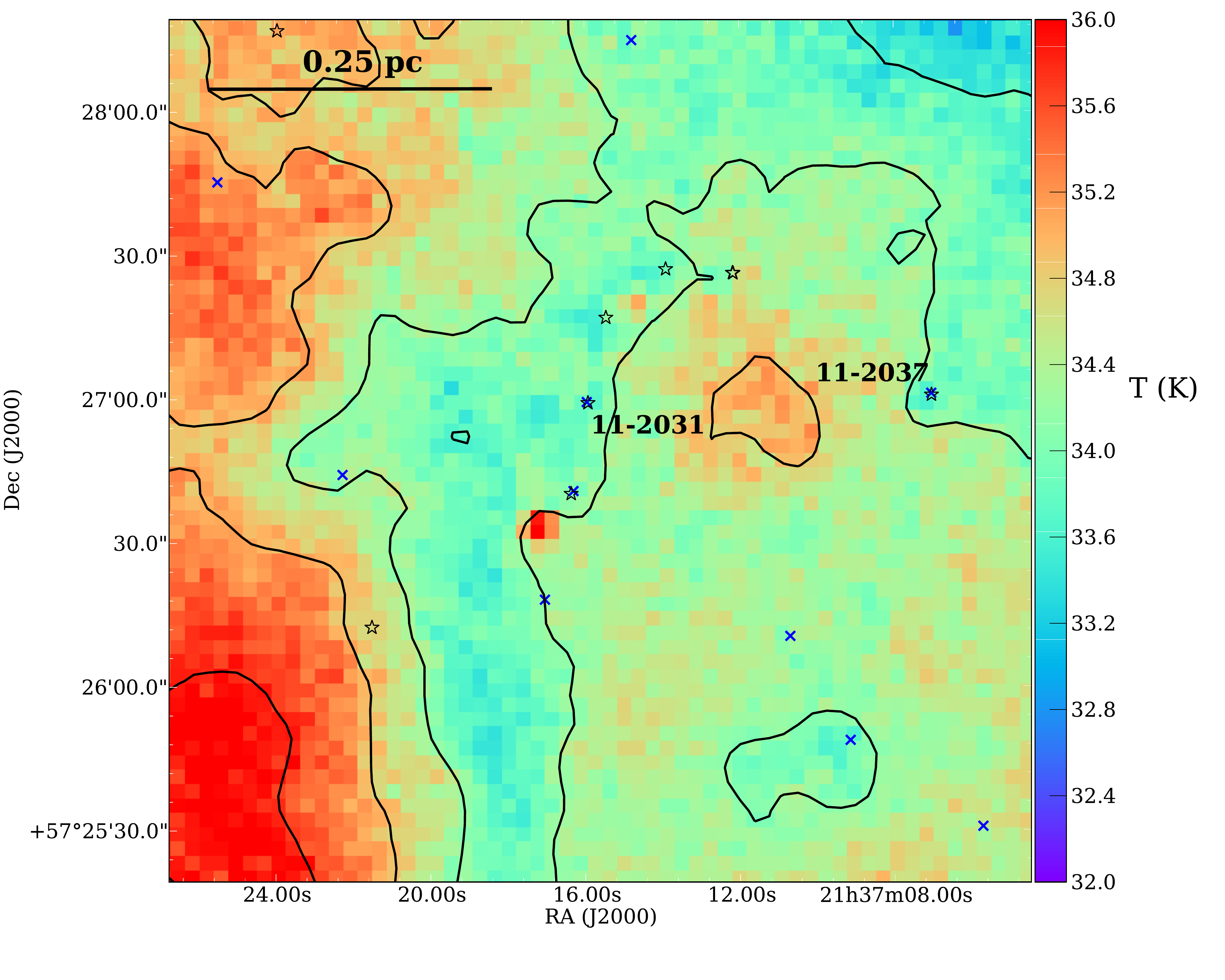} &
\includegraphics[width=0.46\linewidth]{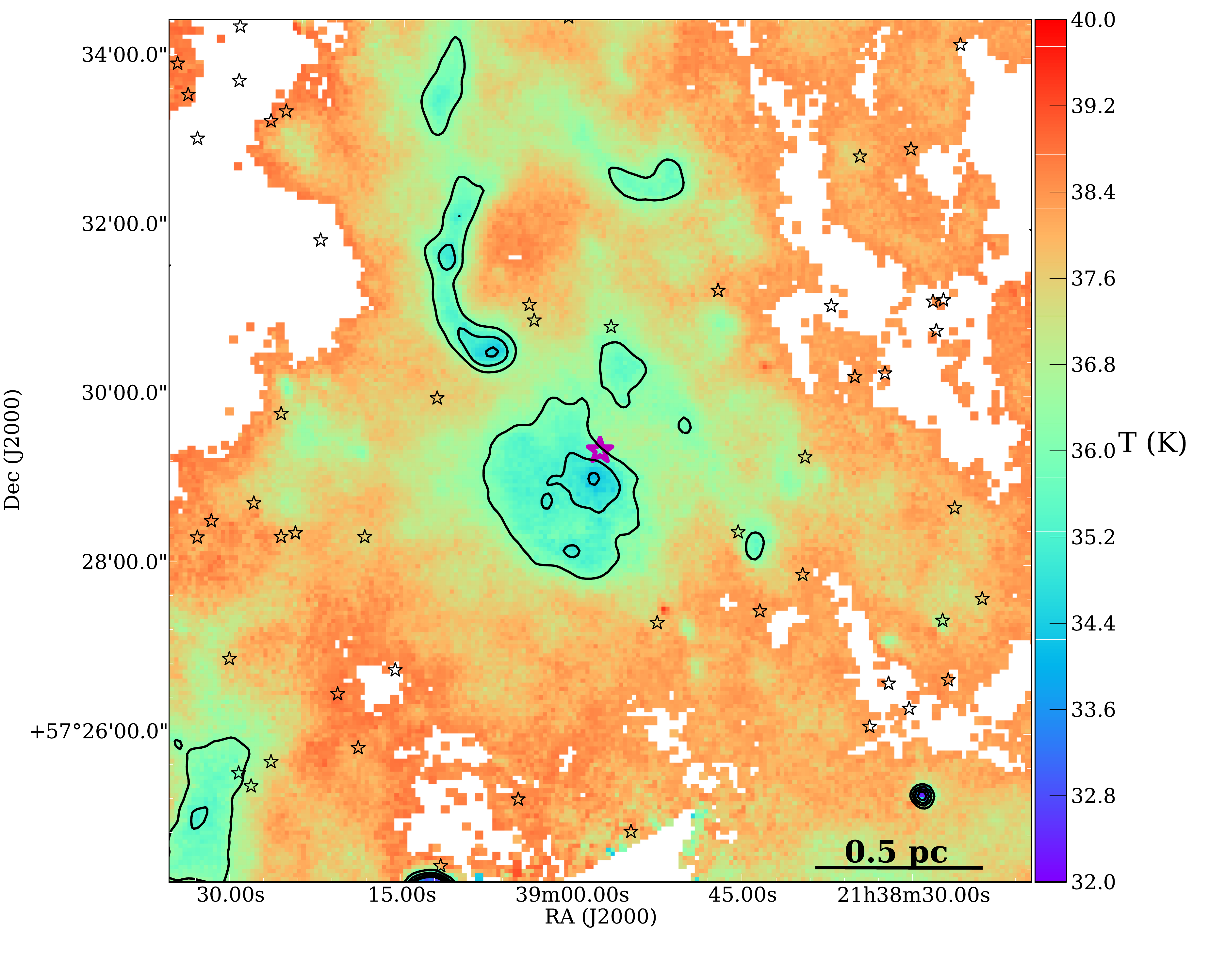} \\
\includegraphics[width=0.46\linewidth]{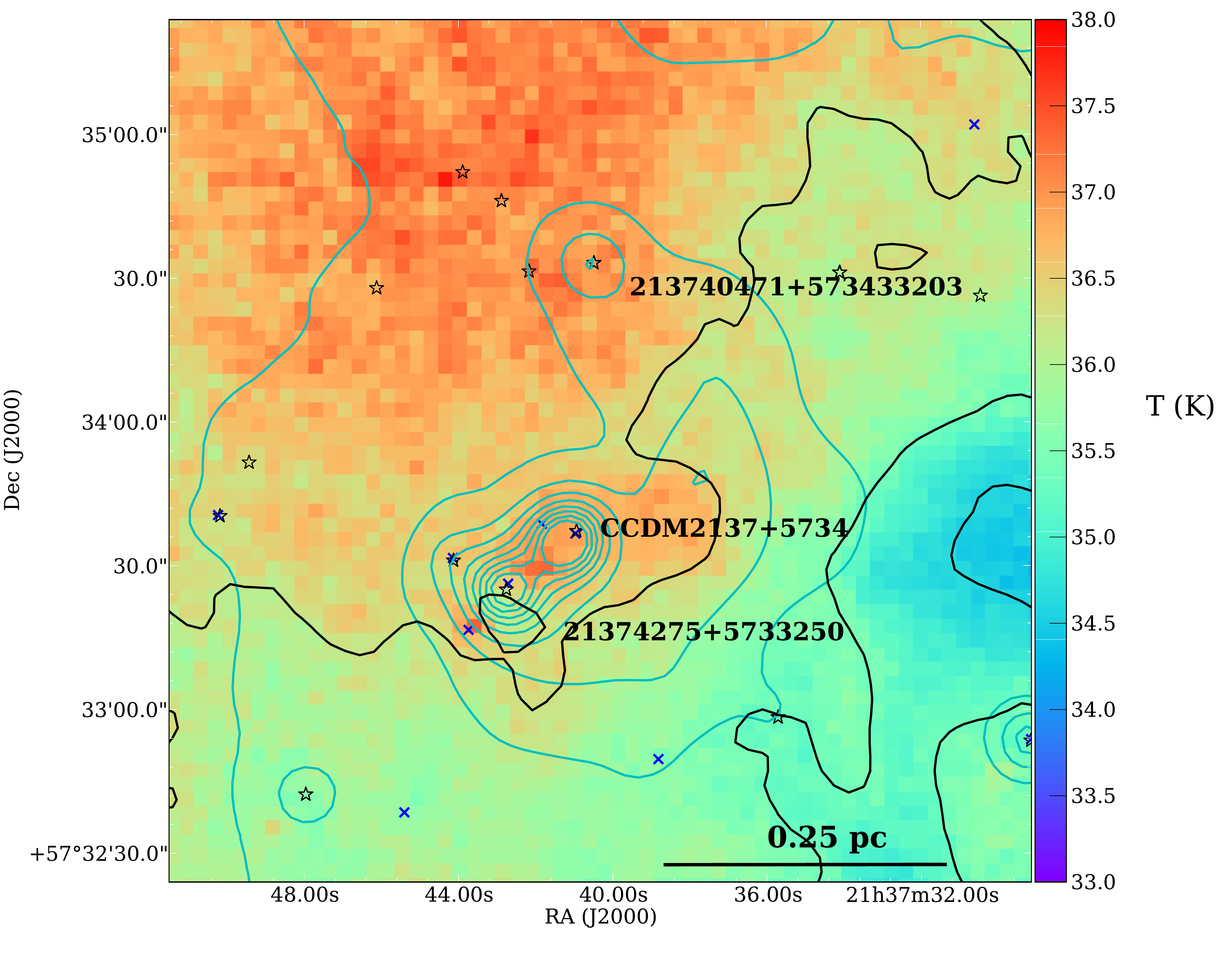} &
\includegraphics[width=0.46\linewidth]{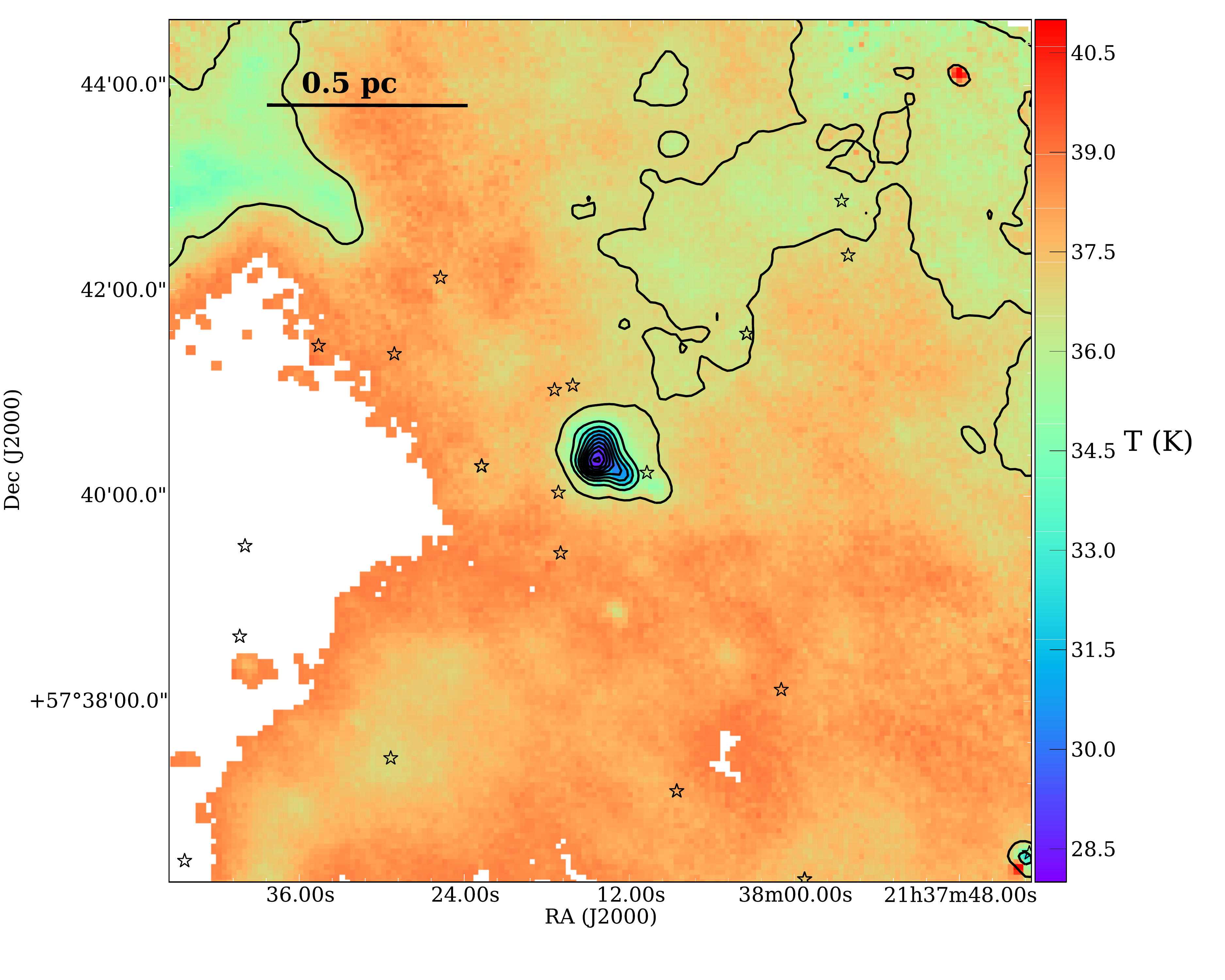}  \\
\includegraphics[width=0.46\linewidth]{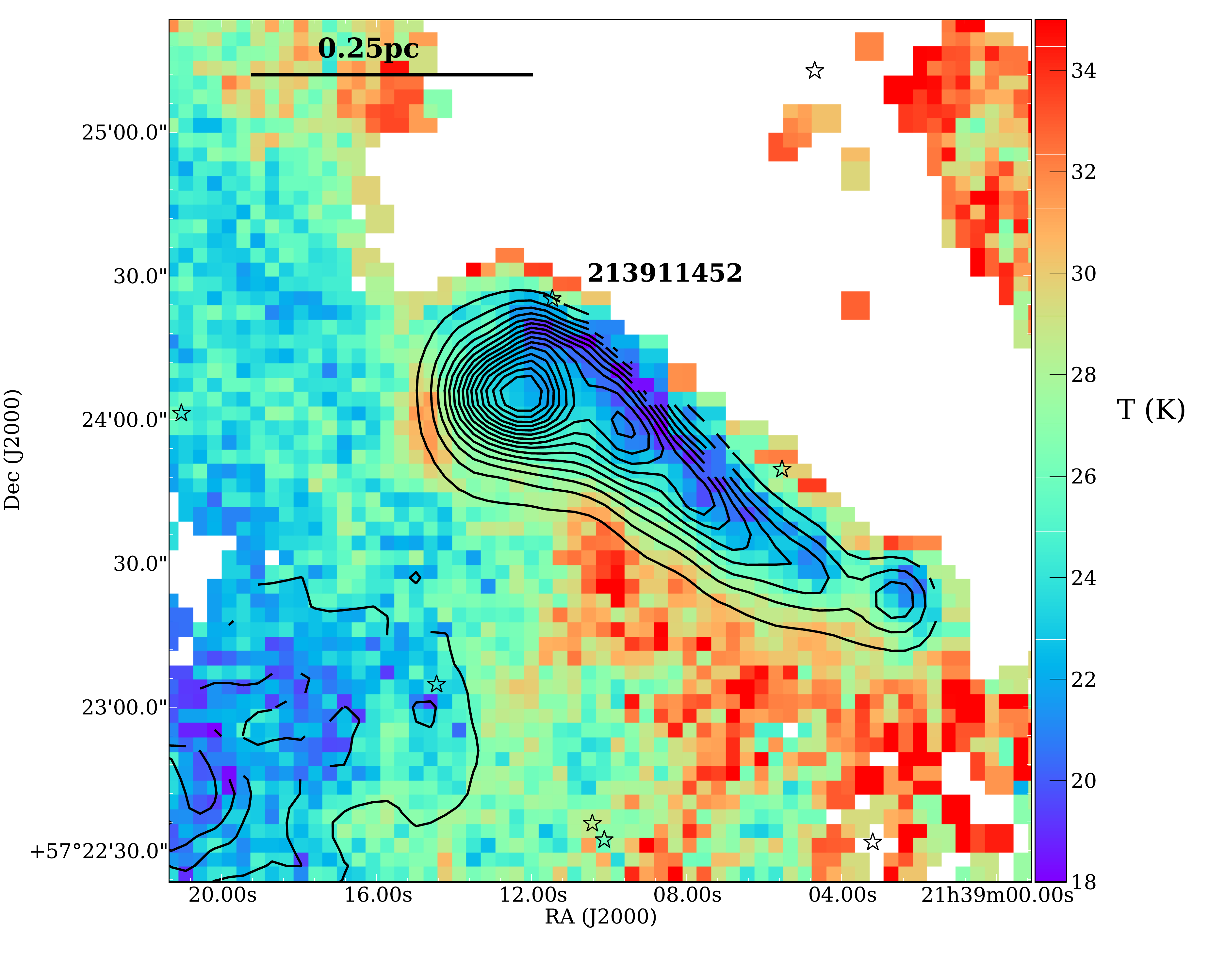} &
\includegraphics[width=0.46\linewidth]{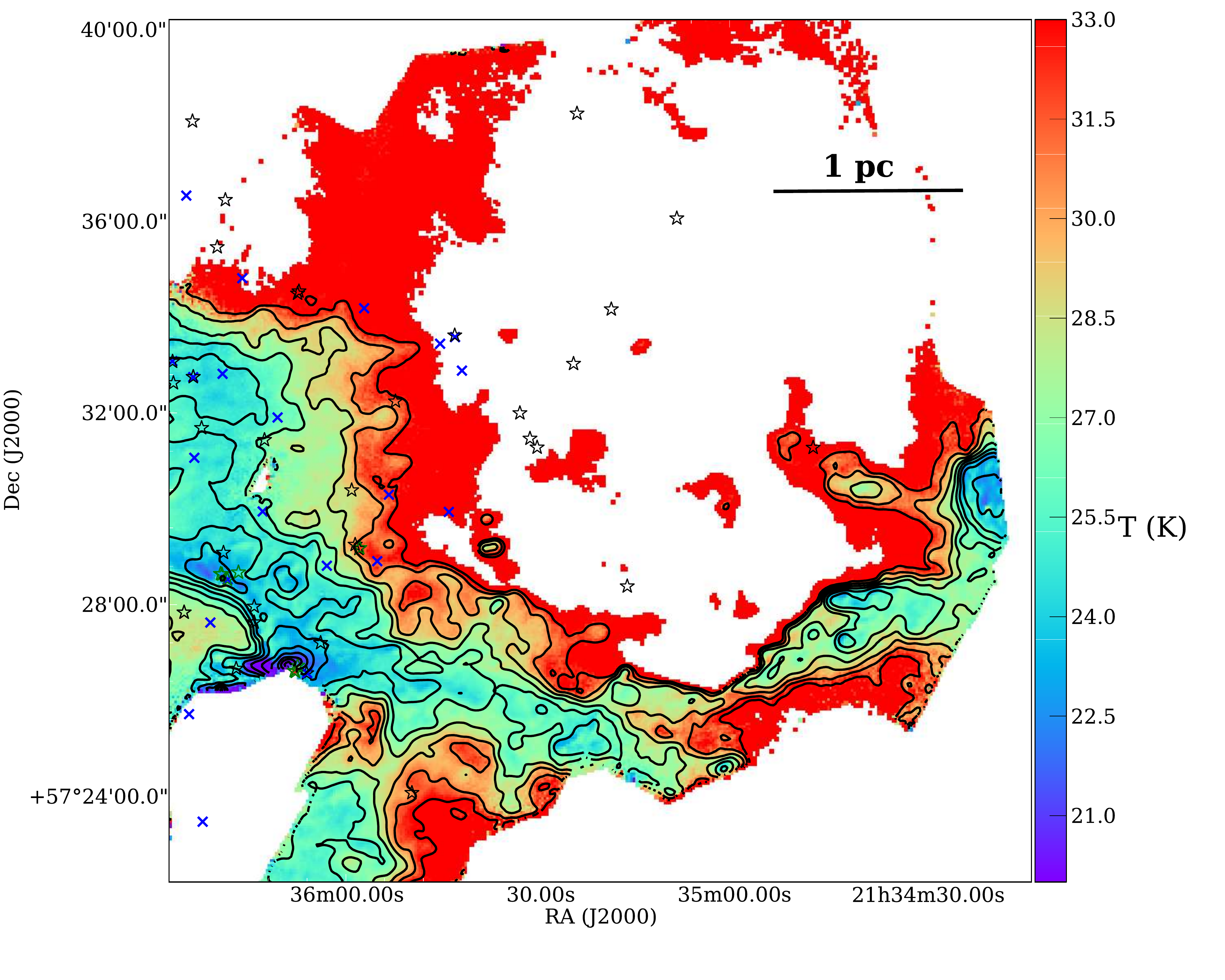}  \\
\end{tabular}
\caption{Approximate temperature maps derived from the Herschel/PACS data
for the mini-clusters (left column, top to bottom, mini-clusters associated with: 11-2031,
CCDM J2137+5734, and 213911452+572425205) 
and the center, north, and west of Tr\,37 (right column, from top to bottom).
The N$_H$ column density contours are plotted in black. For the
CCDM J2137+5734 mini-cluster, whose structure is not detected by Herschel,
we also plot the MIPS 24\,$\mu$m contours (cyan). The temperature and column density
scales are different for each region to show the fine structure. A temperature
scale is attached to each plot. The N$_H$ contours have been adjusted to each region, and correspond to
linear spacing between 2e+20-3e+20 cm$^{-2}$ (for the 11-2031 and CCDM J2137+5734 mini-clusters,
the Tr\,37 center map), 1e+20-1.5e+21 cm$^{-2}$ (for the 213911452 mini-cluster and the Tr\,37 north map),
and a log scale between 1e+20-1e+23 cm$^{-2}$ (for the Tr\,37 west map). For 
the 11-2031 mini-cluster, N$_H$ increases towards the centre of the filament (centre of the image). For 
CCDM J2137+5734, N$_H$ increases towards the south-west, as the mini-cluster itself does not present
any detectable overdensity with respect to the local background. 
Confirmed cluster members (with and without disks) are marked with stars, X-ray sources consistent with YSO are
marked withs crosses. The O6.5 star HD\,206267 is marked as a large magenta star in the
cluster center. Noise-dominated regions are excluded from the plots. Note that the temperature/column density
derivation is only valid in regions that are optically thin, and thus not relevant in very dense parts or
in pixels dominated by stellar emission. \label{temperature-fig}}
\end{figure*}

\begin{figure}
\centering
\includegraphics[width=1.0\linewidth]{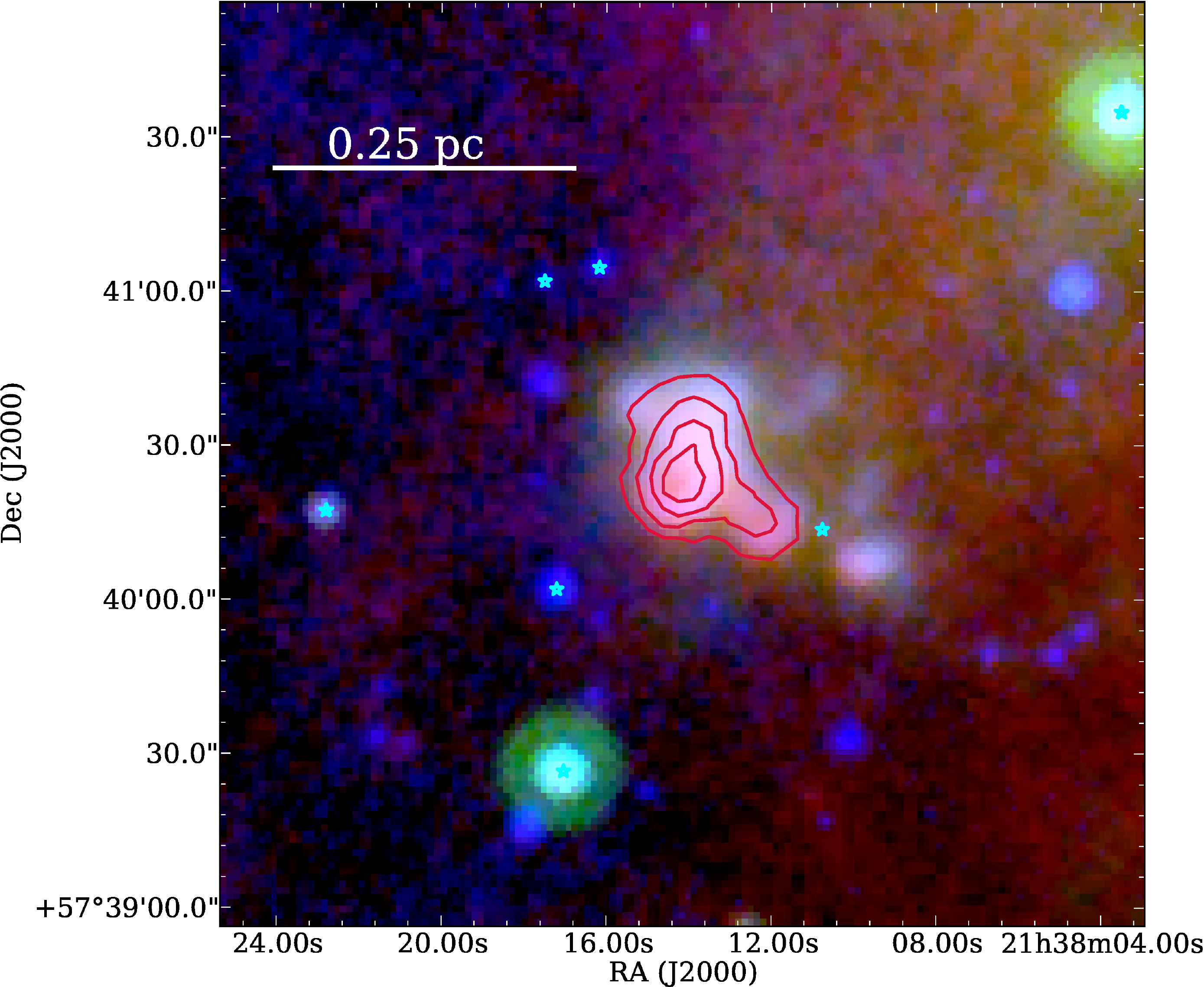}
\caption{Bright outflow-like object, consistent with a planetary nebula, located to the north of Tr\,37,
as observed at 8\,$\mu$m (blue) 24\,$\mu$m (green), and 70\,$\mu$m (red), with 160\,$\mu$m shown as crimson contours. 
\label{nuki-fig}}
\end{figure}

Figure \ref{temperature-fig} shows
the temperature and column density structure of the three mini-clusters, together with the
extended nebular regions around the center, west, and north of Tr 37. The temperature 
is dominated by the massive stars. The O6.5
star HD\,206267 is the main source of heating in the cluster, 
a similar situation to what has been found in more massive star-forming regions (Roccatagliata et al. 2013). 
Nevertheless, we also observe heating at smaller scales
due to the local stellar population. The mini-cluster
associated with the binary B star CCDM J2137+5734 is clearly hotter than the other grouplets, 
and despite being bright at 24\,$\mu$m, it has no significant nebular  
emission at 70 or 160\,$\mu$m. The mini-cluster associated with 11-2031 is 
in contrast better detected at Herschel wavelengths, while the denser clump near the star
213911452+572425205 has a remarkable cold center.
The northern part of Tr\,37 also contains a bright, extended object with an apparent ring-shaped structure
and a bipolar outflow (Figure \ref{nuki-fig}). The structure of the object in the IR and the faint emission in the
optical blue bands we observe in our available photometry are consistent with a candidate planetary 
nebula (E. Villaver, A. D\'{i}az, private communication), 
probably unrelated to Tr\,37 considering the age of the cluster.

Herschel also reveals some nebulosity around the
massive star HD\,206267 (a multiple trapezium-like system; Peter et al. 2012). 
The structure is brighter at 160\,$\mu$m and asymmetric with 
respect to HD\,206267. It does not show evidence of the action of the stellar 
winds or ionization from the massive star, although the stellar emission merges smoothly onto
the extended structure at 160\,$\mu$m. The star is known to have strong photoevaporating
winds that are sweeping the surrounding H II region and structures (Sharpless 1959; Osterbrock 1989)
and nearby protoplanetary disks (Balog et al. 2006). Elias et al. (2008) also observed
a scattering envelope and a well-developed wind around the object. Although chance projection
cannot be completely ruled out, it is likely that the observed extended structure
corresponds to material ejected by HD\,206267.
The star is relatively young and has not started yet its phase as a blue variable or
WR star, but there are some indications that massive stars may shed larger amounts of mass via
strong winds at an early stage. Similar bubble or ring-like structures have been observed in more
evolved massive stars with Herschel (Groenewegen et al. 2011; Vamvatira-Nakou et al. 2011, 2013; 
Cox et al. 2012). 
There is thus the possibility that Herschel is revealing us one of the first heavy mass loss 
episodes in HD\,206267. 

\subsection{Remarkable objects in the Cep\,OB2 region \label{individual}}

The 70\,$\mu$m flux predictions from Spitzer SED models (Sicilia-Aguilar et al. 2011; SA13),
are in good agreement with Herschel detections and upper limits (see Appendix \ref{individualobjects-app}
for details). Stringent upper limits also confirm 
the status of disks classified as depleted/settled based on Spitzer data.
The agreement between the Spitzer SED model
predictions and Herschel observations suggests that our interpretation of disks in terms of evolved objects
with large grains and relatively low small-dust masses is correct.
Herschel/PACS also reveals interesting features in some individual 
objects, described in the following paragraphs.

\subsubsection{GM\,Cep: A rare intermediate-mass star in the making \label{gmcep}}

The G8 star GM\,Cep is the brightest disk and the most conspicuous point source observed with Herschel/PACS
in Tr\,37. GM\,Cep has a high and 
variable accretion rate ($\sim$10$^{-7} - 10^{-6}$ M$_\odot$/yr)), variable extinction, and 
the strongest 1.3mm flux in the region (Sicilia-Aguilar et al. 2008, 2011; Semkov \& Peneva 2012). 
From its spectral type and high luminosity, it is probably the precursor of an A or B star, although
it is very different at optical and IR wavelengths from other HAeBe stars, 
such as MVA-426  in
Tr\,37. The Herschel/PACS data confirm that the disk around GM\,Cep is
more massive than the disk of the B7 star MVA-426 and that the
disks around late-type stars. GM\,Cep is located towards the cluster center, in a region where
most of the surrounding stars have mean ages around 4 Myr.
If GM\,Cep is
a A/B star in the making, it would either have an anomalous age among the surrounding stars
or a very long-lived disk, being a key object to study the
evolutionary paths of intermediate-mass stars.
 There is a significant difference
between the 70\,$\mu$m excess measured with Spitzer/MIPS and the Herschel results, but given the
MIPS non-linearities (Paladini et al. 2013) and the agreement between non-simultaneous
IRAS, IRAC, MIPS, and IRS data, it cannot be attributed to variability.

\subsubsection{Low far-IR flux in a strongly accreting star: 11-2037 \label{11-2037}}

The K4.5 star 11-2037 has a remarkable low 70\,$\mu$m flux for its Spitzer fluxes and 
its IRAM/1.3mm detection (Sicilia-Aguilar
et al. 2011). This is an example where our previous disk models fail to predict the
observed Herschel fluxes, which are lower than expected. The object is part of a 
mini-cluster (see Section \ref{miniclusters}), so nebular emission
could have led to contamination of the millimetre flux (and maybe of the 160\,$\mu$m flux
as well), but the sharp turn-down at 70\,$\mu$m is real: although the source is one of the
brighter sources compared to neighbouring disks at Spitzer wavelengths, it is clearly fainter 
at 70\,$\mu$m. The sharp decline in 70\,$\mu$m emission could
be due to a lack of dust, to strong settling/flattening at longer wavelengths, or to radially
variable structures such as wide gaps in the disk at larger distances.
Further data (in particular, confirmation of the millimetre flux) is needed to test this hypothesis.
If the dust mass is in fact lower than in other systems, it would offer a
very strong contrast with the expected gas mass from accretion, 
since the object also has strong H$\alpha$ emission and a high accretion rate ($\sim$10$^{-8}$M$_\odot$/yr;
Sicilia-Aguilar et al. 2005).

\subsubsection{Young in the gas, evolved in the dust: a survivor in NGC\,7160 \label{01-580}}

\begin{figure}
\centering
\includegraphics[width=0.99\linewidth]{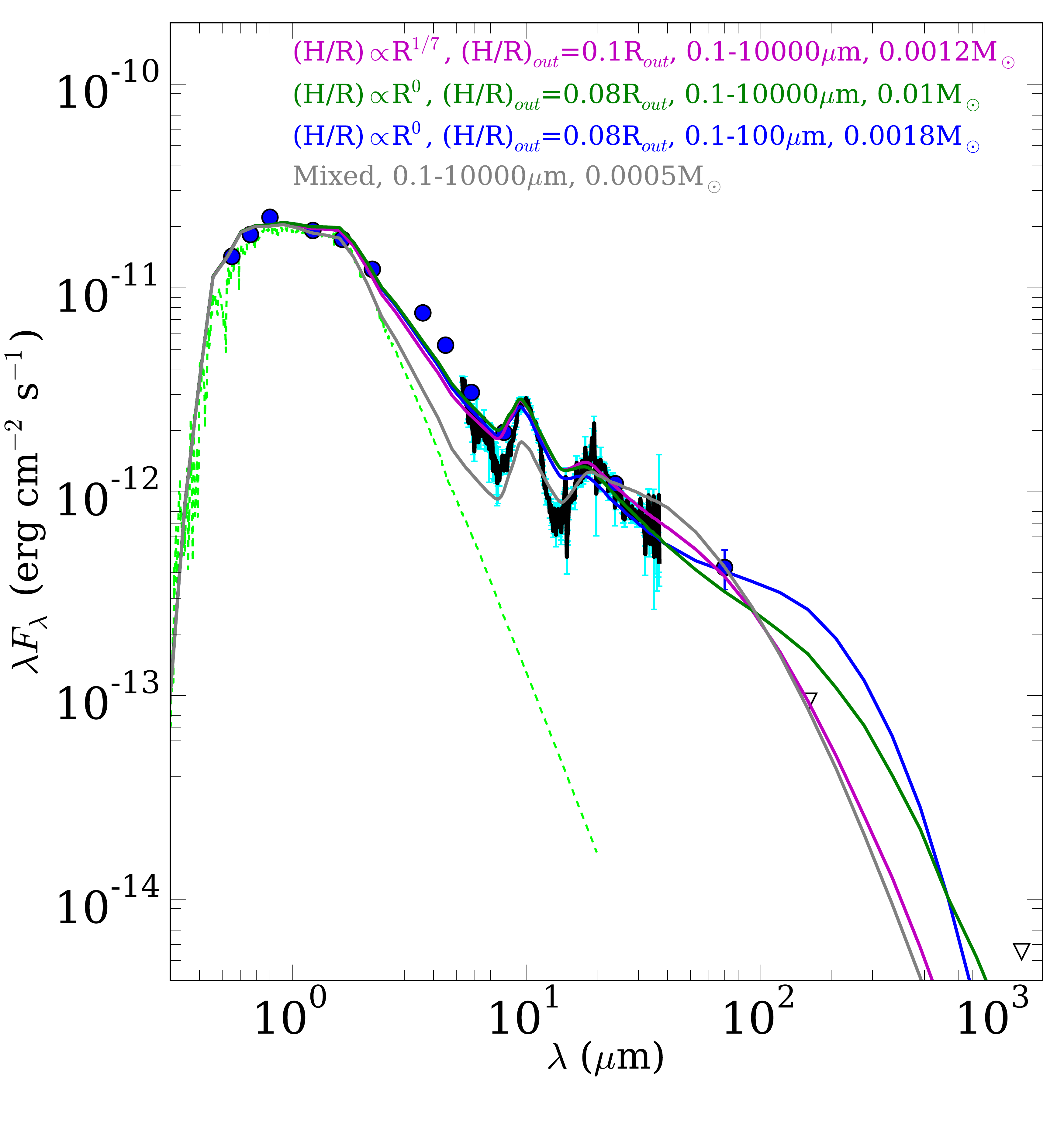}
\caption{Comparison of radiative transfer models for the disk of 01-580, the
only accreting star in NGC\,7160.  Blue dots mark the data (errors
are often smaller than the dots). Inverted triangles are upper limits.  The dashed
line is a MARCS photospheric model (Gustafsson et al. 2008).
The colour lines show models with different vertical structures (flaring laws H/R, 
outer disk thickness), grain size distributions, and disk masses. The model labelled as "Mixed" includes
well-mixed dust and gas. Even if the disk is assumed to be very settled, substantial
mass depletion is required to fit the 160$\mu$m upper limit.\label{01-580-fig}}
\end{figure}

The only object detected at 70\,$\mu$m in NGC\,7160 is the K4.5 star 01-580. It is also 
the only disk with confirmed accretion in this 12 Myr-old cluster, and one of the
strongest accretors among low-mass stars in the whole Cep\,OB2 region
(\.{M}=4$^{+3}_{-2}$ $\times$ 10$^{-8}$ M$_\odot$/yr; Sicilia-Aguilar et al. 2010). Remarkably, its
mid-IR flux is relatively modest compared to similar objects in Tr\,37, which had been previously
interpreted as a lack of small dust in the disk (Sicilia-Aguilar et al. 2011). 
The object is clearly detected at 70\,$\mu$m, but
not at 160\,$\mu$m. 

We run several models to reproduce its mid- and
far-IR SED (Figure \ref{01-580-fig}). Low far-IR fluxes can result from dust settling,
depletion of small-sized grains, and/or more dramatic shaping of the outer disk, such as
very large gaps or external disk truncation. Given the high accretion rate, it is unlikely
that the disk has very low gas mass, but the amount of mass in small dust could be low.
Models assuming well-mixed gas and dust allow only a very small amount of 
dust mass, especially if the maximum grain size is $<$100\,$\mu$m. Flattening the disk allows
us to increase the dust mass, but only as long as we still include substantial grain growth
(beyond $\sim$100\,$\mu$m) to reduce the 160\,$\mu$m excess.
The best-fitting models to reproduce the observed mid- and far-IR SED are those with either settled disks 
with relatively low mass and strong grain
growth, or flared disks with a reduced dust mass and also strong grain growth. 

Although the disk falls within our full disk
classification, its SED also betrays radial evolution,
including a very strong near-IR excess (not well reproduced in our models
and suggestive of a very thick inner disk or puffed inner wall)
and a very strong silicate feature (similar to PTD, and thus suggestive of a
non-uniform distribution of small grains).
The object has thus signs of both inside-out 
evolution and more generalised dust 
evolution (strong grain growth, setting, and/or depletion). The disk would also be a good candidate
for anomalous gas to dust ratio (which would allow to have a large gas mass despite having
a reduced dust content), or dust filtration to the innermost part of the disk, as has been
predicted (Hughes \& Armitage 2012) and suggested 
(Hughes et al. 2008) for other evolved objects. 

These features are not shared with strong accretors in Tr\,37, which are typically small-grain rich. 
Old-accreting TTS are also often found to be relatively primordial, compared to what
we observe in 01-580 (e.g. Ingleby et al. 2014). This suggests that at the old age of NGC\,7160, even if the accretion behaviour
is similar to that of younger disks, dust evolution has become unavoidable, 
as expected from models (Testi et al. 2014).
Tr\,37 also contains a few objects with steep mid- and far-IR SEDs (e.g. 11-1209 and 13-236 in
Figure \ref{oldmodels-fig}). 
Nevertheless, these objects have two significant differences compared to 01-580: They have very weak
silicate features (negligible for 11-1209) and lower accretion rates 
($\sim$5-6$\times$10$^{-9}$M$_\odot$/yr). The best-fitting models for these objects require settled
disks at all radial distances, which is not the case for 01-580. This makes 01-580 stand up
as a remarkably active, old protoplanetary disk, and not simply as the last survivor 
among average protoplanetary disks.

\section{Discussion \label{discussion}}

\subsection{Disk evolution in the light of Herschel data \label{disksevolution}}

\subsubsection{The dust and gas connection}

Exploring together the far-IR properties and accretion
helps to understand the connection between gas and dust in protoplanetary
disks. For most of the stars, we have constraints on accretion
(from high-resolution H$\alpha$ observations; Sicilia-Aguilar et al. 2006b) and
accretion measurements and upper limits (from U band; Sicilia-Aguilar et al. 2005, 2010). 
Based on Sicilia-Aguilar et al. (2010), we also consider an 
upper limit of 10$^{-11}$ M$_\odot$/yr for objects with
narrow H$\alpha$ detected in high-resolution spectroscopy (Sicilia-Aguilar et al. 2006b).
H$\alpha$ strengths and profiles for such objects are indistinguishable from 
those of diskless TTS, and thus consistent with no accretion,
although U band measurements are limited to accretion rates over $\sim$10$^{-11}$M$_\odot$/yr
for late K and M stars \footnote{As noted in Sicilia-Aguilar et al.(2010), the gap between non-accreting
objects and our accretion measurements, with nearly no objects with rates below
a few times 10$^{-10}$M$_\odot$/yr, is real and not a result of observational 
limitations. Accreting objects with upper limits to their U-band accretion rates 
have usually uncertain
spectral types, anomalous extinction values and/or larger photometry errors. Their
H$\alpha$ widths, profiles, and EW are fully consistent with typical CTTS, and not
indicative of particularly low accretion rates.} (Sicilia-Aguilar et al. 2010). Accretion rates derived from other
methods (e.g. H$\alpha$ photometry or spectroscopy) are not considered since they
show more scatter than U-band based measurements.

\begin{figure*}
\includegraphics[width=1.0\linewidth]{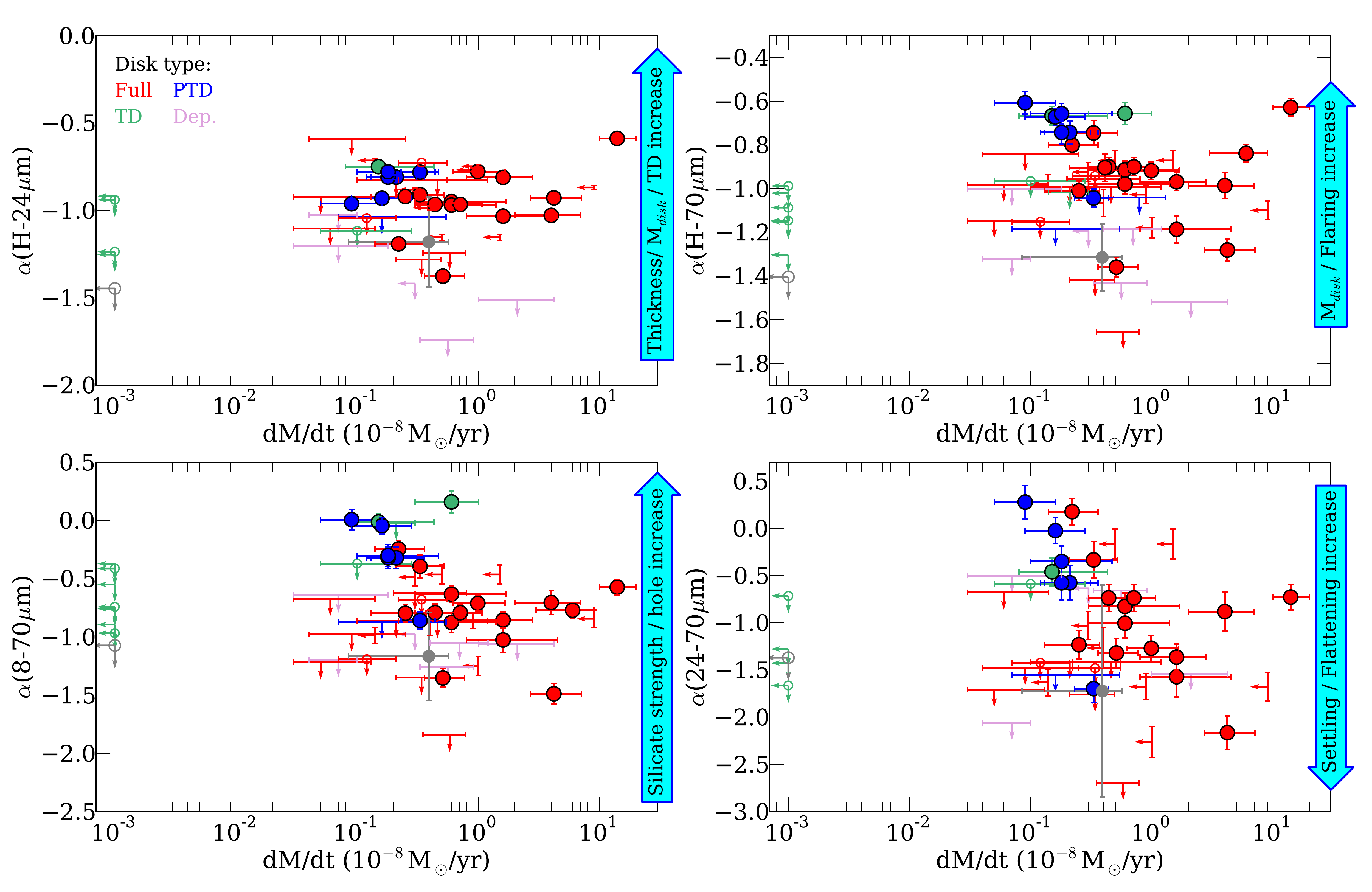}
\caption{The $\alpha$ spectral indices involving far-IR colours vs the accretion rates for stars with different disk
types, classified according to their Spitzer colours (see text). Full disks are marked in
red, PTD are marked in blue, TD are marked in green, and low excess/depleted disks
are marked in pink. Circles represent detections (in both photometry and accretion rate),
and upper limits in the accretion rate or spectral indices are marked by arrows. 
Marginal detections are marked with arrows but we also show an open circle at the
marginal value. The solid grey dot corresponds to the median value obtained by stacking the
images of the non-detected objects. Its errorbars represent the dispersion in the $\alpha$
index and the quartiles in the accretion rate. The open grey dot is the median value (marginal detection) obtained
by stacking the non-accreting disks. On the right, blue arrows 
show how certain changes in the disk properties would affect the observed spectral
indices, as explored in Section \ref{diskproperties}. \label{alphamdot-fig}}
\end{figure*}

None of the non-accreting disks around low-mass stars 
is clearly detected with Herschel. The K2.5 star 213655283+5725516683 is well-detected and has
a large hole and maybe no accretion, although a high-resolution spectrum would be needed to confirm 
this. Further 3 non-accreting TD are consistent with marginal detections (73-758, 13-350, 21384350.5727270),
although only 13-350 among them is clearly non-accreting. 
Disks without signs of accretion are rare and belong to the class of TD with inner holes
(of which approximately half are consistent with no accretion) or to the low-excess/dust-depleted/settled
SEDs (Sicilia-Aguilar et al. 2010; SA13). All full and PTD disks in Tr\,37 have accretion signatures, even though
the accretion rates range between few times 10$^{-10}$M$_\odot$/yr and $\sim$10$^{-7}$M$_\odot$/yr.
Since the detectability of the disks around similar-type stars depends mostly on the disk mass 
and the flaring/thickness, this is a strong sign that disks with no accretion
are either less massive and/or significantly settled compared to full disks.
Different grain properties, such as very strong and global grain
growth, could also reduce the far-IR emission in non-accreting disks, although
most of them have evidence of small grains in their inner disks as shown by
Spitzer IRS spectra (Sicilia-Aguilar et al. 2007, 2011).

Figure \ref{alphamdot-fig} displays the accretion rate \.{M} against different spectral indices
for objects with various SED types. Interestingly, full disks appear clearly separated from
transitional and pre-transitional ones.
A double-sided Kolmogorov-Smirnov (KS) test reveals a probability of 0.002 (0.2\%) that the
$\alpha$(H-70) of the PTD and TD are drawn from the same distribution than 
full disks, which is highly significant since the number of PTD and TD is small. In the
case of $\alpha$(8-70), the probability drops to 0.01\%, although this could be 
due to inner holes or gaps decreasing the continuum levels at 8$\mu$m:
$\alpha$(8-70\,$\mu$m) increases when the disk becomes more ``transitional". 
In addition, flattening/settling
of a disk for a given mass reduces the difference between the 24 and 70\,$\mu$m flux,
which can be observed in $\alpha$(24-70\,$\mu$m), although this index shows a stronger degeneracy.

Accreting TD and PTD in Tr\,37 have moderate but not significantly different accretion
rates, compared to full disks, since very strong accretors are rare in Tr\,37 and generally
correspond to stars with earlier spectral types (Sicilia-Aguilar et al. 2010). The difference becomes significant 
once a proxy for disk mass/flaring such as the SED slope between the near- and far-IR is included.
For the same accretion rate, PTD and TD tend to have higher $\alpha$(H-70\,$\mu$m)
spectral index than full disks. Imposing the condition of
having similar accretion rates, the KS test produces probabilities of observing the same 
differences from samples drawn for the same distribution of 6\% (for $\alpha$(H-24), not conclusive),
0.06\% (for $\alpha$(H-70), highly significant), 0.08\% (for $\alpha$(8-70), highly significant),
and 0.8\% (for $\alpha$(24-70), also significant). If we instead run the
KS test on the spectral indices derived from our models, we
find no significant differences between PTD/TD disks and full disks in any
of the mentioned spectral indices. 
The presence of a hole or a radial variation in disk properties \textit{per se} is not enough to
produce the observed difference if the remaining disk properties (disk mass, flaring, disk thickness, 
global dust grain properties) are kept the same for full and PTD/TD disks, even though vertically extended 
structures in the disks can alter the irradiation at different distances,
causing a redistribution of flux towards longer wavelengths (e.g. Varni\`{e}re et al. 2006).
If disks with and without gaps are assumed to have well-mixed dust and gas in hydrostatic equilibrium, the difference
between those with holes and without them is negligible for the same disk mass (see Figure \ref{models-fig}),
as long as the holes are small. 
Therefore, we find that disks identified with Spitzer as candidates to have inner holes and/or gaps
do differ in their global radial and/or vertical structure with respect to objects 
classified as full disks. What happens in the inner disk is thus not completely disconnected of
deeper changes in the global disk structure: Observing systematically higher 70\,$\mu$m fluxes for TD 
and PTD is an indication of them having more massive and/or more flared/thicker disks,
compared to the full disks.  

TD and PTD are good candidates for active giant
planet formation. Massive planets, capable of clearing the disk, 
can change the disk vertical structure, especially in the inner rim close to the
planets (e.g. Kley 1999). Changing the global disk
flaring would require more dramatic 
mechanisms such as radially-variable settling. Planet-induced gaps can also
produce radial variations in the dust properties, due to dust filtering
(Rice et al. 2006; Zhu et al. 2012).
Accretion rates in the presence of giant planets are expected to be lower than
what would correspond from viscous evolution for the total disk mass (Varni\`{e}re et al. 2006; 
Lubow \& D'Angelo 2006), which is in agreement with previous observations of TD (Najita et al. 2007). 
Nevertheless, this depends on the ability of the planet to block the gas flow (and not only dust) 
through the gap, which is uncertain (M\"{u}ller \& Kley 2013; Zhu \& Stone 2014)
and inconsistent with some of the accreting binary systems
known to date (e.g. Fang et al. 2014), including cases such as 82-272 in Tr\,37, which is
an accreting spectroscopic binary (Sicilia-Aguilar et al. 2006b).

If the main cause of the difference in the far-IR slopes were the diversity of disk masses
(as it happens in viscous accretion models; Hartmann et al. 1998), we
would expect to observe a trend between the $\alpha$(H-70\,$\mu$m)
indices of the full disks and their accretion rates. This is not observed
at significant levels. Variations in the disk vertical structure (flaring/thickness)
are common in disks (Sicilia-Aguilar et al. 2011) and
can affect $\alpha$(H-70\,$\mu$m), washing out potential correlations between the 70\,$\mu$m fluxes
and the accretion rate. 
There is also the possibility that the various physical conditions (or grain properties) throughout
the disk affect the viscosity and thus the global viscous transport (Isella et al. 2009).
In addition, the accretion rate baseline for detected objects is small, although the 
lack of detections of non-accreting disks
points to a correlation between the 70\,$\mu$m flux and the mass of the disk. 

Since many of the objects with low
accretion are undetected at 70$\mu$m, we followed 
Andrews \& Williams (2007) and stacked the non-detection images
of the 16 disks with measured accretion rates (excluding objects near IC\,1396\,A that have
complex background). This results in a 4.6$\sigma$ detection at 70$\mu$m (3.7$\pm$0.8 mJy).
We then estimate the median indices
to compare with the median accretion rate (3.9$\times$10$^{-9}$M$_\odot/yr$,
quartiles 5.1$\times$10$^{-9}$M$_\odot/yr$-8.5$\times$10$^{-10}$M$_\odot/yr$). The median
indices are consitent with steeper slopes for objects with lower accretion rates, although they
do not make the accretion-spectral index correlation significant. They also show that
non-detected disks accreting at low rates have small, but not tiny (comparable to
other detected disks), mean dust masses. If we repeat the exercise for the 8 confirmed 
(via high-resolution spectroscopy) non-accreting disks, including marginal detections, 
we obtain a marginal 3$\sigma$ detection of 2mJy.
The resulting spectral indices are consistent with strong disk evolution, although
half of them are not detected at 24$\mu$m either. This 
provides a further evidence that the mean disk mass for non-accreting disks is
significantly lower than for the accreting sources.

\subsubsection{Global disk evolution and the disk dispersal mechanisms}

Herschel far-IR observations confirm the existence of various distinct evolutionary
paths in protoplanetary disks, likely related to the diversity of physical processes
acting on disk removal and on the initial conditions. Having low dust masses or strongly 
settled disks as observed in dust-depleted objects does not trigger the immediate
opening of a hole. Conversely, opening a hole or gap does not require a previous depletion of 
disk mass, at least in the case of accreting TD/PTD. In addition, shutting down the accretion
of a disk seems only possible in case the disk mass has already decreased sufficiently.
This probably requires dramatic changes in the whole disk, which would explain why
the only non-accreting disks in Tr\,37 are either TD or dust-depleted.

Different physical
processes appear to be behind accreting and non-accreting disks,
as previously suggested (Sicilia-Aguilar et al. 2007, 2010, 2011). 
Herschel data
confirm the dichotomy in the transition disk classification. 
Non-accreting TD, despite having similar Spitzer colours than accreting 
TD (Figure \ref{alphamdot-fig}; Sicilia-Aguilar et al. 2011; SA13), are clearly separated by the
spectral indices involving 70\,$\mu$m, and their lower mean 70$\mu$m excesses 
are consistent with lower disk masses and/or lower flaring. 
Photoevaporation is a good candidate to explain the
lack of near-IR excesses together with the lack of accretion signatures in non-accreting TD.
Photoevaporation can open inner holes and shut down the accretion flow in a
short timescale, as long as viscous transport does not 
replenish the inner disk at a faster rate
(Clarke et al. 2001; Alexander et al. 2006; Gorti et al. 2009). 
Once the mass of the disk and the viscous transport rate have decreased below a certain limit,
photoevaporation would 
take over shutting down accretion and eventually removing the rest of the disk.
This theoretical picture is in agreement with the lack of disks accreting at very
low rates (Sicilia-Aguilar et al. 2010), although the lack of correlation
between the accretion rate and the far-IR indices suggests that the problem is
more complex.

None of the objects classified as low excess/dust-depleted disks is 
detected with Herschel, which is also a strong sign that these objects are subject to 
substantial mass depletion or deep structural differences. The upper limits
in Figure \ref{alphaalpha-fig} and \ref{alphamdot-fig} are consistent with depleted disks
occupying a different parameter space than full disks. Strong
settling at large radii or generalised small-dust depletion by grain growth
could explain these observations.
Unfortunately, the lack of detections among this class does not let us explore in
detail the causes of the depletion.

\subsection{Multi-episodic star formation in Tr\,37: Mini-clusters \label{miniclusters}}

Previous Spitzer studies revealed
diffuse nebulosity around some grouplets of stars (G12; SA13). 
Herschel data offer a new view of the Tr\,37 grouplets or mini-clusters (Figure \ref{miniclusters-fig}).
PACS data reveal the presence of cloud structures in the most remarkable 
mini-cluster, containing the star 11-2031.
The dense structure next to the emission-line star 213911452+572425205
(SA13) is also detected with Herschel. 
Not all mini-clusters are detected at the same wavelengths, revealing
differences in density and temperature. The following paragraphs discuss the morphology
and properties of the Tr\,37 mini-clusters and their potential origin. 

\subsubsection{Irregularly-shaped mini-clusters: The CCDM J2137+5734 grouplet}

\begin{table}
\caption{Stars associated with the CCDM J2137+5734 mini-cluster.} 
\label{minicluster2-table}
\begin{tabular}{l c c c}
\hline\hline
Name 		& Sp. Type & Disk type & A$_V$ (mag) \\
\hline
CCDM J2137+5734e & B3 & Db & 1.7 \\ 
CCDM J2137+5734w & B5 & Db & 1.5 \\
21374275+5733250 & F9 & F &  1.6 \\
21374376+5533169 & K/M & N & 1.6 \\
213744131+573331130 & K7/M1 & F & 1.9 \\
213742167+573431486 & M2.0 & TD & 1.4 \\
213740471+573433203 & M2.5 & TD & 2.0 \\
\hline
\end{tabular}
\tablefoot{Members of the CCDM J2137+5734 mini-cluster,
an irregular-shaped stellar grouplet. Spectral types, extinctions,
and ages from Contreras et al. (2002) and Sicilia-Aguilar et al. (2005, 2013). 
There are no individual age estimates for these objects. Disk classes according to
Table \ref{SEDclass-table}, Db stands for debris disk candidate}.
\end{table}

The most conspicuous Spitzer
mini-cluster, associated with the massive binary star CCDM J2137+5734 and
with remarkable 24\,$\mu$m extended emission (B11, G12, SA13),
does not show extended emission at PACS wavelengths. It contains at
least 5 confirmed members, several of them
detected at 70\,$\mu$m. There is a large variety of SED types
among its members (Table \ref{minicluster2-table}), including 
the diskless X-ray candidate 21374376+5733169 (G12),
and TD such as  213740471+573433203 at the tip of the cloudlet (with
strong 24\,$\mu$m emission) and 213742167+573431486. The IR excess of the second
could be affected by nebular contamination, even though it shows
strong H$\alpha$ emission from accretion.

This structure has the highest temperature and likely the lowest density among all the mini-clusters.
Herschel-based estimates of temperature and column density are not substantially
different from the surrounding background (local background N$_H \sim$1$\times$10$^{20}$cm$^{-2}$).
Although heating in Tr\,37 
is dominated by the central O6.5 star, the two B stars CCDM J2137+5734A/B may
affect the temperature and dispersal rate of the cloudlet at shorter distances, leading to
a thinner, warmer structure compared to other mini-clusters only
populated by low-mass stars. This is comparable to the findings of Roccatagliata et al. (2013),
although at a much smaller scale. 

\subsubsection{Filamentary mini-clusters: The 11-2031 grouplet}

\begin{table}
\caption{Stars associated with the 11-2037/11-2031 mini-cluster.} 
\label{minicluster-table}
\begin{tabular}{l c c c}
\hline\hline
Name 		& Sp. Type & Age (Myr) & A$_V$ (mag) \\
\hline
11-2037 	& K4.5    & 2.5 & 1.6 \\
11-2131 	& K6.5    & 3.0 & 2.3 \\
21371389+5727270$^*$ & late K & -- & 3.0 \\
21371545+5727170$^*$ & K/M  & -- & 2.3 \\
11-2031 	& K2.0    & 4.5 & 1.7 \\
213716349+57264020 & K7.0 & -- & 6.0:\\
21372152+5726123$^*$ & K/M & -- & 1.7 \\
\hline
\end{tabular}
\tablefoot{Members of the 11-2037/11-2031 mini-cluster, a stellar grouplet with
filamentary structure and associated nebulosity. Spectral types, extinctions,
and ages from Sicilia-Aguilar et al. (2005, 2013), except for objects marked with $^*$, 
for which they are derived in the present work from SED fitting. All SEDs are consistent with
"full disks".}
\end{table}

The 11-2031 mini-cluster has the shape of an elongated filament with 
6 confirmed young stars, plus a potential further member with 24\,$\mu$m emission. 
The star 21370909+5725485
is located at a similar distance than 11-2037, but since it is not 
surrounded by the same patch of nebulosity, it may not belong to the
group. All the mini-cluster members have similar spectral types
and masses around 1M$_\odot$ (considering the
evolutionary tracks of Siess et al. 2000 for a 3-4 Myr age), and they all have full disks
(Table \ref{minicluster-table}). 
There is a tendency for younger ages towards the
west of Tr\,37 (Sicilia-Aguilar et al. 2005), but
the mini-cluster members are not significantly younger than the main cluster,
considering the individual uncertainties.
Radial velocities available for two of the stars (11-2037 cz=-19.9$\pm$0.6 km/s;
11-2031 cz=-17.2$\pm$1.2 km/s) are consistent with the average cluster value (-15.2$\pm$3.6 km/s;
Sicilia-Aguilar et al. 2006b). But the SED types for the mini-cluster members
are clearly different from the bulk of Tr\,37 members: They all have full disks.
Since the disk fraction in Tr\,37 is $\sim$48\%, and nearly
1/3 of stars have disks classified as TD, PTD, or low-excess/mass depleted, 
the probability of finding 7 full disks among near neighbours by chance is below 0.1\%.

Herschel data reveals column density enhancements over the background 
in the range $\sim$2-7$\times$10$^{19}$cm$^{-2}$ 
(background levels are $\sim$2$\times$10$^{20}$cm$^{-2}$ at this location).
Integrating over the observed structure, we estimate a total mass of $\sim$0.05M$_\odot$ 
in nebular material. This would be similar to the mass of a few protoplanetary
disks around low-mass stars, albeit distributed over a very
large area ($\sim$0.036 pc$^2$). Since the filament contains 6 (maybe 7)
low-mass stars with disks, the mass in the remnant cloud is of the order
of the total mass in the protoplanetary disks. The 
cloud is optically thin and relatively warm, unlikely to be infalling onto the
objects, even though 
the filamentary nebulosity probably contains remnant material from the initial core.
The extinction toward the mini-cluster members is not significantly higher than the
cluster average (A$_V$=1.67$\pm$0.45; Sicilia-Aguilar et al. 2005), 
consistent with excesses around 0.1-0.2 mag expected from
the estimated column density. 
Only 213716349+572640200 is significantly more extincted than
the rest, which could be due to disk orientation or to the presence of remnant envelope
material (which could explain the brightness of the object at 160\,$\mu$m).

The stars are separated by less than 0.1pc projected distance, 
with 11-2037 at $\sim$0.2pc from the rest.
21372152+5726123 is also located at 0.2pc of 213716349+57264020, but if the
24\,$\mu$m excess source at 21:37:18.9 +57:26:25.7 could be confirmed  
as a young star, the separation would be also around 0.1pc.
This is very similar to the beads-on-a-string picture
predicted for gravitational fragmentation of a thin sheet or filament (Hartmann 2002; 
Ward-Thompson et al. 2010; Nielbock et al. 2012). Here, the filament
is much smaller in size than the usual structures, and the masses of the individual 
``beads" are also low ($\sim$1M$_\odot$). In the
simple model for an infinite cylinder of Heitsch et al. (2009), the critical
mass for a scale of 0.1pc is similar to the masses of the stars in the mini-cluster,
although this would imply an unrealistic star formation efficiency. By itself, such a low
mass filament would be not form stars unless shielded
from external radiation by further cloud material. One possibility to form 
such a small structure would be gravitational focusing in the collapse of an
asymmetric sheet (Burkert \& Hartmann 2004), which tends to concentrate
material towards the edges of the collapsing finite structure. In this case, 
it would be a sign that gravitational fragmentation operates in similar ways in
spatial scales ranging from fractions of a pc up to several hundred pc.

\subsubsection{Other cloud structures in Tr\,37}

The third mini-cluster contains the M1 star 213911452 +572425205 (which has
a full disk) and the probably diskless K6.5 star 213905519+572349596, being
the smallest one and thus the less representative. Unlike in the other
mini-clusters, both stars are at the rim of the extended nebular structure. 213911452+572425205
has a massive, young disk, with strong accretion and remarkable emission lines
(SA13). The globule, clearly detected at Spitzer and Herschel wavelengths, 
shows a remarkable dense and cold center.
Our column density estimates are in the range $\sim$1.0-17$\times$10$^{20}$cm$^{-2}$,
giving an approximate mass $\sim$0.15M$_\odot$ integrating over the globule surface. This
is comparable to the mass of a very low-mass star, but it is unlikely to
be cold and dense enough to undergo gravitational collapse.
The globule has a very steep density gradient, in contrast with the other two mini-clusters,
which merge smoothly with the cluster background.
There is no evidence of shock lines that suggest interaction between
the young stars and the nebula, so chance projection is also a possibility. If so, 
this globule could be a starless structure, maybe a very low-mass but dense clump left over of
the removal of the original cloud. 

Further small patches of nebulosity are found throughout Tr\,37, 
especially at 160\,$\mu$m. Several structures are seen 
towards the west of Tr\,37, behind the IC\,1396\,A ionization front (Figure \ref{temperature-fig}). 
They have rounded shapes and sizes ranging
from several tens of arcsec to 1 arcmin. Some smaller, rounded structures are also
visible through the cluster, but since they are typically
faint at Herschel wavelengths, it is not possible to determine accurate
temperatures or column densities. A few individual sources
seem to be associated with the nebular structures (e.g. 12-1091).
Nevertheless, given the abundance of small nebulosity patches not associated with
stellar sources, there is a high risk that they result from chance projection.

\subsubsection{The star-formation history of Tr\,37}

The diversity in stellar content shows
that the differences in the mini-clusters properties do not correspond to time evolution
within the same class of objects, but to essentially different grouplets,
which may have formed from clouds with different properties and/or various triggering
mechanisms, suggestive of a complex star-formation history in Tr\,37.
Paper I revealed multi-episodic star
formation triggered by the O6.5 star HD\,206267 in IC\,1396\,A (Paper I). Herschel/PACS observations
show an arc-shaped structure about 0.5pc in length, which contains at least 
one intermediate-mass Class 0 object most likely triggered by radiation-driven
implosion (RDI). Formation and collapse of a dense structure
that was initially inside a larger cloud is thus possible. The
similar-mass, similar-separation observed in the stars in the 11-2031 mini-cluster 
are more consistent with gravitational fragmentation, but
the mini-cluster associated with the binary B3+B5 star CCDM J2137+5734
shows a very different morphology and stellar content. Besides the intermediate-mass
binary, it contains a further relatively massive source (21374275+5733250, 
spectral type F9) at $\sim$16" separation, 
and some low-mass stars. It could thus have resulted from a similar structure as observed 
at the tip of IC\,1396\,A, which is consistent with an intermediate-mass protostar or a 
small stellar grouplet containing intermediate-mass protostars.

The differences in the mini-clusters associated with 11-2031 and with CCDM J2137+5734
suggest that star formation
in Tr\,37 has progressed in a multi-episodic way, similar to what we observe
in the IC\,1396\,A globule. The mini-clusters are 
very compact structures ($\sim$0.25 pc in size). 
Having clearly distinct grouplets of stars in a relatively old cluster
such as Tr\,37 adds further evidence to the hypothesis that the objects in a mini-cluster
originated within the same parental molecular core, as it has been suggested for
subclusters (Getman et al. 2014). Since the typical
velocity dispersions in star-forming clouds are of the order of 1-2 km/s 
(consistent with our kinematic study of Tr\,37; Sicilia-Aguilar et al. 2006b),
objects could move about 1-2 pc in 1 Myr. This is incompatible with maintaining
the mini-cluster structure over timescales similar to the ages of their members:
Even if part of the original members of the grouplet may have dispersed, the long survival 
of mini-clusters requires that their
members formed together from a comoving filament or core with little
velocity dispersion. Strong
dynamical interactions between the grouplet members would also need to be
prevented, as they would result in the dispersal of the mini-cluster members
on short timescales (Bate 2012).
The stellar and nebular content of the mini-clusters also shows that 
clumpy and multi-episodic star formation may proceed
in various ways at small scales within the same cloud. 
Although at present we do not have velocity information for the mini-clusters,
this could change our understanding of the star formation scenario in Tr\,37
and similar regions,
suggesting that cloud fragmentation may proceed in an irregular
way, with various small structures fragmenting and collapsing independently 
in time to give birth to small grouplets of stars.

\section{Summary and conclusions \label{conclu}}

Herschel appears as a powerful tool to understand protoplanetary disks as a whole,
revealing a strong connection between the innermost disk and outer/global disk evolution.
The results of our Herschel survey on protoplanetary disks in the Cep~OB2 clusters Tr\,37 and NGC\,7160 
are summarised below: 

\begin{itemize}
\item We detected 95 disks at 70\,$\mu$m and 41 at 160\,$\mu$m, adding significant upper limits
to more than 130 disks. The detection fraction is strongly dependent on the Spitzer
SED type and the stellar spectral type. More than 50\% of full disks and 
PTD are detected, while we obtain zero detections among 
low-excess/depleted disks. Nearly 90\% of the disks 
around stars K4 or earlier are detected, compared to less than 1/3 of the disks around M-type stars.
Therefore, our results for K stars are representative for the whole disk 
class, while for M-type objects we only explore the most massive and flared disks.

\item We find a large variety of disks among solar-type stars in Tr\,37.
Far-IR data breaks the degeneracy of disk structures inferred from 24-30$\mu$m data,
revealing details such as global settling/flattening, disk masses,
and changes in the small-dust grain distribution. 
Herschel confirms structural differences behind the
Spitzer-based classification in terms of full disks, disks with inside-out evolution (TD, PTD),
and low-excess/dust-depleted disks. 
Low-excess/depleted objects are confirmed (via stringent upper limits) to be consistent with low small-dust
content and/or global mass depletion.

\item The 70 and 160\,$\mu$m observations of disks in Tr\,37 are 
consistent with strong dust grain growth, with maximum grain sizes over
100\,$\mu$m. Even the most IR-luminous, massive and flared disks (expected to 
contain more pristine, less evolved grains than
disks with low excesses) are consistent with substantial grain growth,
suggesting that it is a rapid and universal process.

\item None of the non-accreting disks is clearly detected with Herschel.
Mean far-IR excesses for non-accreting disks 
are significantly lower than for acreting disks, which
is consistent with non-accreting disks being more evolved, less massive and/or more settled disks.

\item Herschel confirms the dichotomy of accreting vs
non-accreting TD as objects resulting from different physical processes. 
Stopping accretion in a disk requires substantial changes to its global structure. All non-accreting disks
are either TD or low-excess/depleted disks. 
Having a relatively massive and flared disk without 
accretion seems excedingly rare, so even objects with inner disk evolution (TD, PTD) 
are accreting if their disks are sufficiently massive. This is a strong evidence that 
accretion is mainly governed by the total mass content of
the disk, as expected from viscous evolutionary models. It also suggests that photoevaporation is
not efficient in opening a gap unless the disk mass has already decreased considerably.

\item We observe significantly higher $\alpha$(H-70\,$\mu$m) indices
for PTD/TD, compared to full disks with similar accretion rates.
Inner holes/gaps \textit{per se} do not affect much the
70\,$\mu$m flux, compared to other parameters such as disk mass and thickness/flaring.
One possibility could be that accreting TD and PTD disks are systematically more massive
than full disks with similar accretion rates,
in agreement with disk clearing by massive (planetary?) companions (Najita et al. 2007). 
Nevertheless, this 
would also predict a correlation between the accretion rate and the $\alpha$(H-70\,$\mu$m) index for full disks,
which is not observed. 
Other explanation could be the presence of higher vertical structures (walls, puffed-up rims)
in PTD/TD shifting of the emission peak of the disk to longer wavelengths.
This suggests that inside-out evolution and changes in the innermost disk are not disconnected from
the outer- or global disk evolution. 

\item At old ages, significant dust processing/settling seems unavoidable, even if the object is still 
gas rich and strongly
accreting. The only accreting disk in NGC\,7160, 01-580, is a surprising case combining a very low far-IR
flux with a remarkably high accretion rate. Studying such objects may hold the key to understanding 
disk removal and survival.
\end{itemize}

Herschel/PACS also reveal strong evidence for multi-episodic star formation in Tr\,37. 
Our Herschel observations allow us to conclude that:

\begin{itemize}
\item The extended cloud emission in Tr\,37 contains several
small structures or mini-clusters, compact ($\sim$0.25 pc) groups of young stars still surrounded by nebulosity. 
The physical properties of the mini-clusters (approximate dust temperature and column density)
depend on the type of stars that they contain (e.g. intermediate-mass vs only solar-mass
stars). 

\item Mini-clusters show that clumpy or multi-episodic star formation can occur
even at relatively small scales of fractions of pc. Clumpy and multi-episodic star 
formation may also contribute to the variations in age and
evolutionary stage observed between disks on relatively small spatial scales.

\item  The survival of the very compact mini-clusters after ages of 1-2 Myr indicates that
their members were formed within the same cloudlet or core and without much internal disruption
or interaction: Otherwise, the typical velocity dispersion of stars in clusters (1-2 km/s) 
would have already disrupted them by the ages of Tr~37.

\item The mini-cluster associated with the binary B star CCDM J2137+5734 has the lowest density and
highest temperature among the mini-clusters, and it could have originated in a structure similar
to the IC\,1396\,A-PACS-1 protostar at the tip of the IC\,1396\,A globule.

\item The mini-cluster associated with the low-mass star 11-2031  
bears strong resemblance with a filament with equidistant, similar-mass
stars (beads-on-a-string). With the total length of the filament being $\sim$0.5 pc, and the distance
between the $\sim$1M$_\odot$ stars being less than 0.1 pc, it could be an example of gravitational fragmentation and/or
gravitational focusing in a collapsing cloud holding down to very small scales.

\item Herschel also reveals a bright structure to the north of Tr\,37,
morphologically suggestive of a planetary nebula and most likely (given the cluster age) 
not associated with Tr\,37.

\item Finally, Herschel observations reveal what
could be one of the first episodes of heavy mass loss in the O6.5 star HD\,206267, a sign of 
unmistakable evolution in the heart of Tr\,37.
\end{itemize}

\vskip 0.5truecm
Acknowledgments: We thank the anonymous referee and the editor, M. Walmsley, for their comments that helped to
improve and clarify this paper. 
We thank Bruno Altieri from the Herschel Helpdesk for his valuable help with the 
data reduction, and Lorenzo Piazzo for making available the Unimap code. We also thank 
Sofia Sayzhenkova from the computing support at the Departamento de F\'{i}sica
Te\'{o}rica, and Y. Ascas\'{i}bar, E. Villaver, and A. D\'{i}az for their help identifying
the field planetary nebula found in Tr\,37.
A.S.A. acknowledges support by the Spanish MICINN/MINECO "Ram\'{o}n y Cajal" 
program, grant number RYC-2010-06164. A.S.A. and M.F. acknowledge support by
the action ``Proyectos de Investigaci\'{o}n fundamental no orientada", grant number
AYA2012-35008. C.E. is partly supported by Spanish MICINN/MINECO
grant AYA2011-26202. V.R. is supported by the DLR grant number 50 OR 1109
and by the {\it Bayerischen Gleichstellungsf\"orderung} (BGF). T.B. acknowledges support from 
NASA Origins of Solar Systems grant NNX12AJ04G.
This research has made use of the SIMBAD database,
operated at CDS, Strasbourg, France.
We also made use of Astropy, a 
community-developed core Python package for Astronomy (Astropy Collaboration, 2013)
and APLpy, an open-source plotting package for Python hosted at http://aplpy.github.com.

\clearpage
\onecolumn

\Online

\begin{appendix} 

\section{SEDs and photometry tables for all the YSO observed by Herschel\label{photometry-app}}

\begin{figure*}
\centering
\begin{tabular}{cccc}
\includegraphics[width=0.24\linewidth]{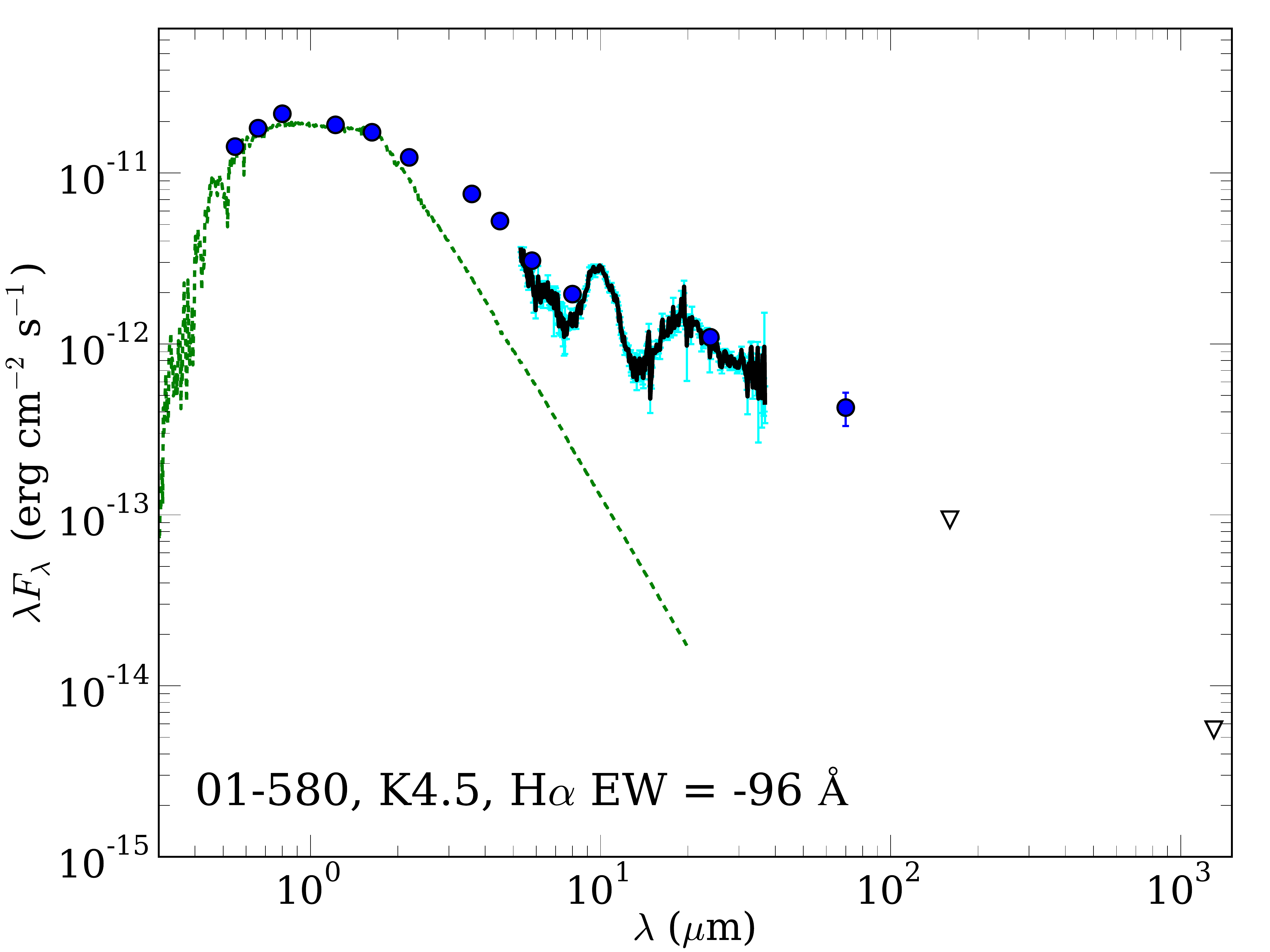} \\
\includegraphics[width=0.24\linewidth]{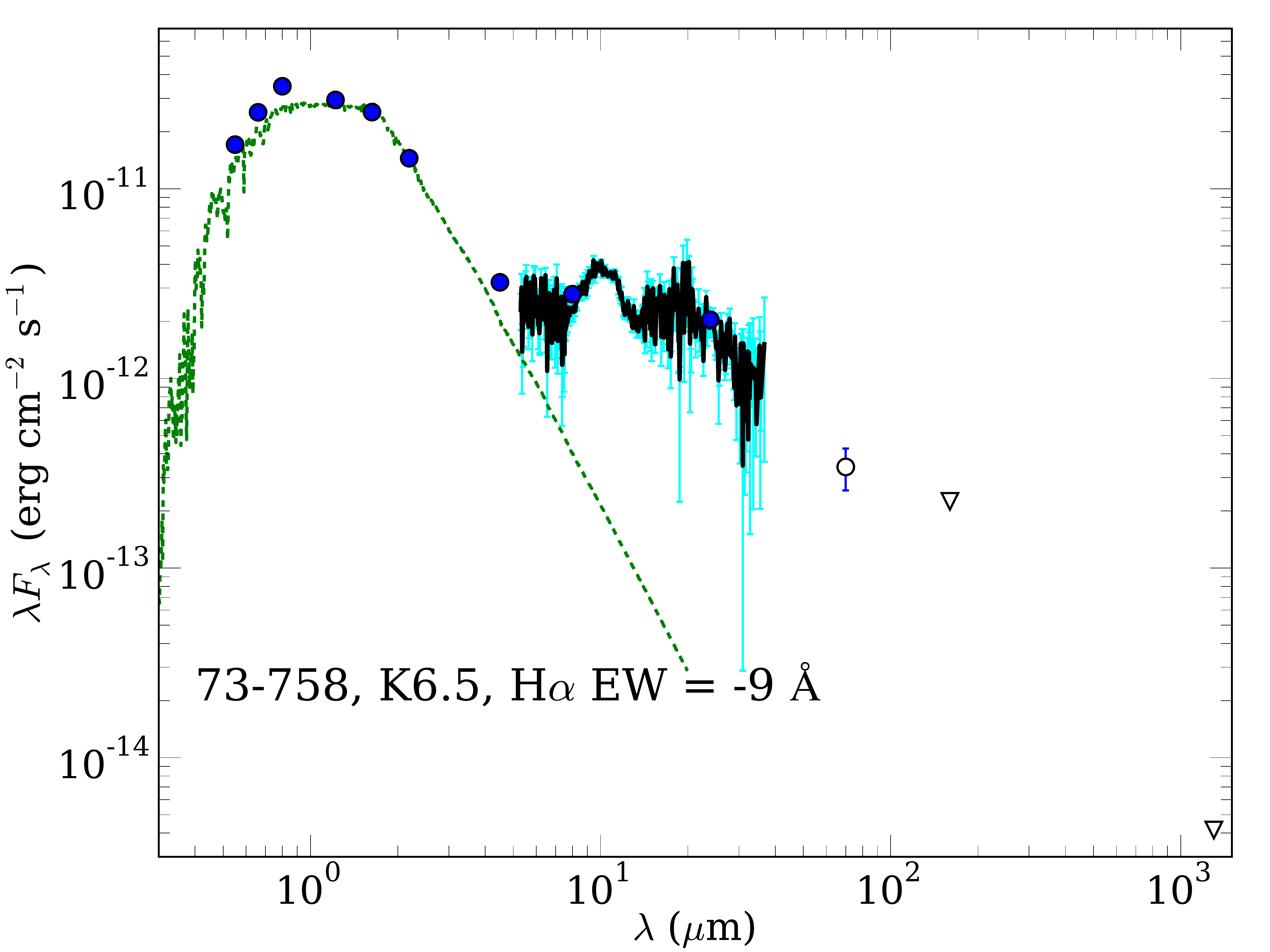} &
\includegraphics[width=0.24\linewidth]{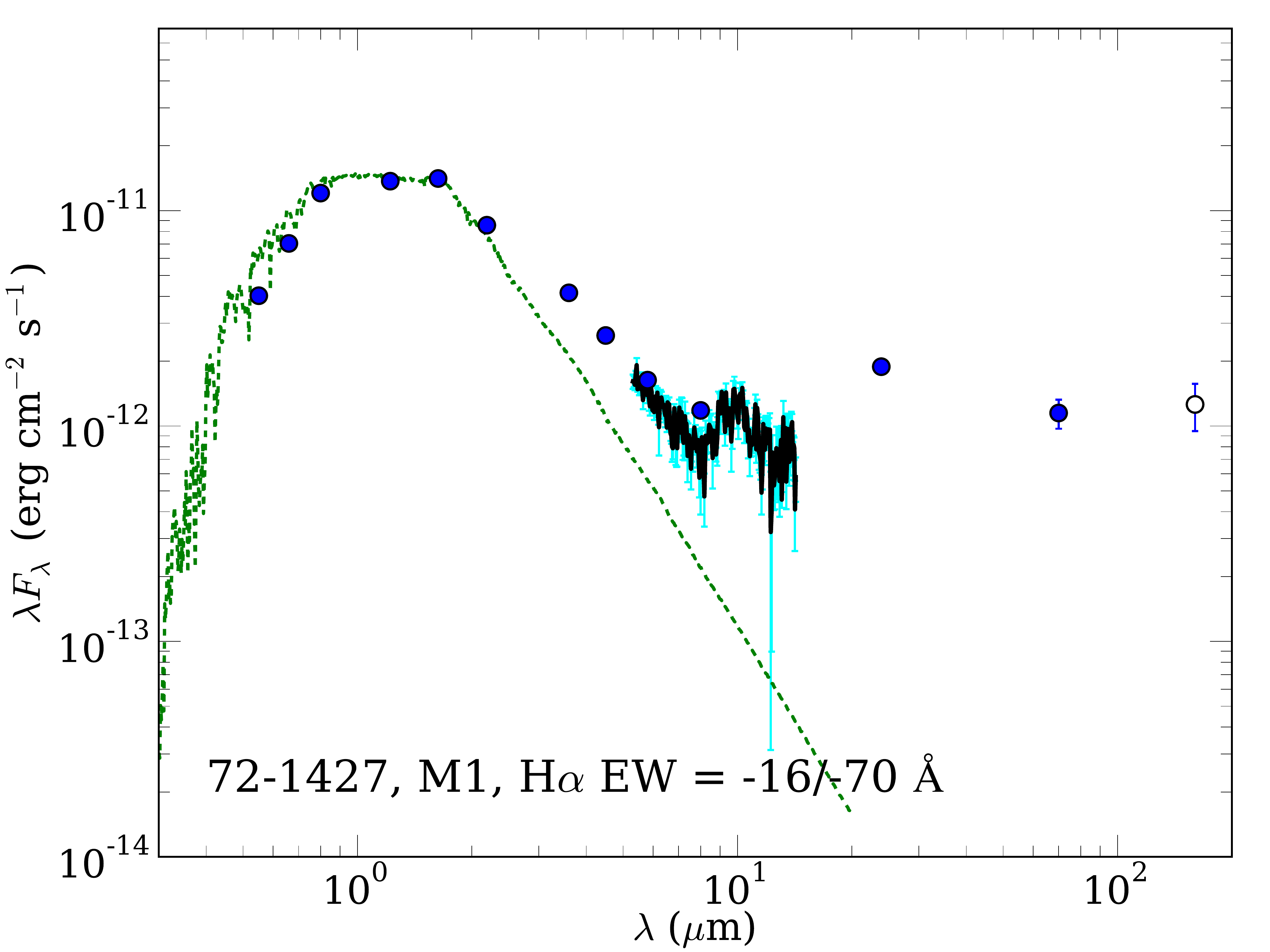} &
\includegraphics[width=0.24\linewidth]{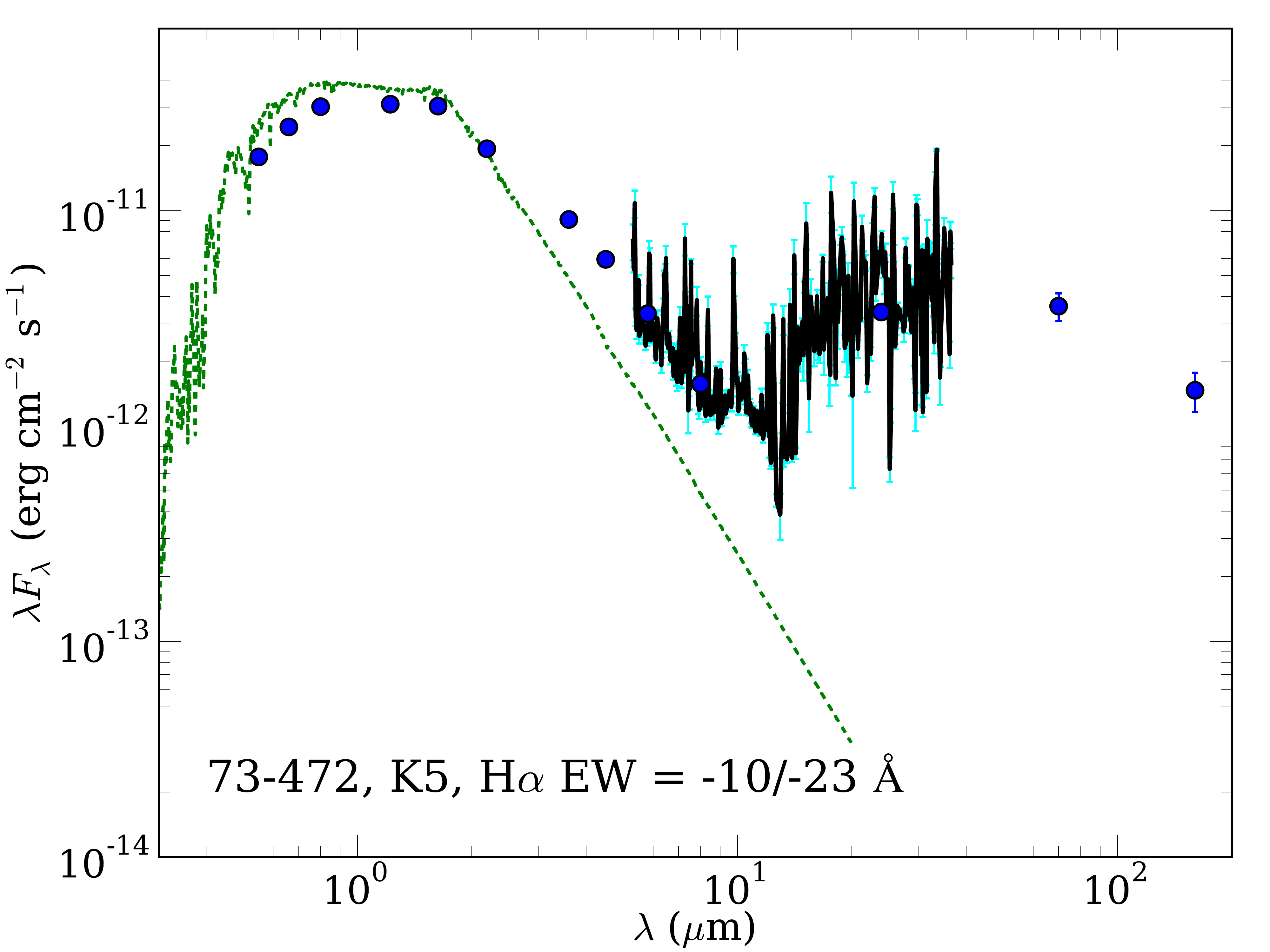} &
\includegraphics[width=0.24\linewidth]{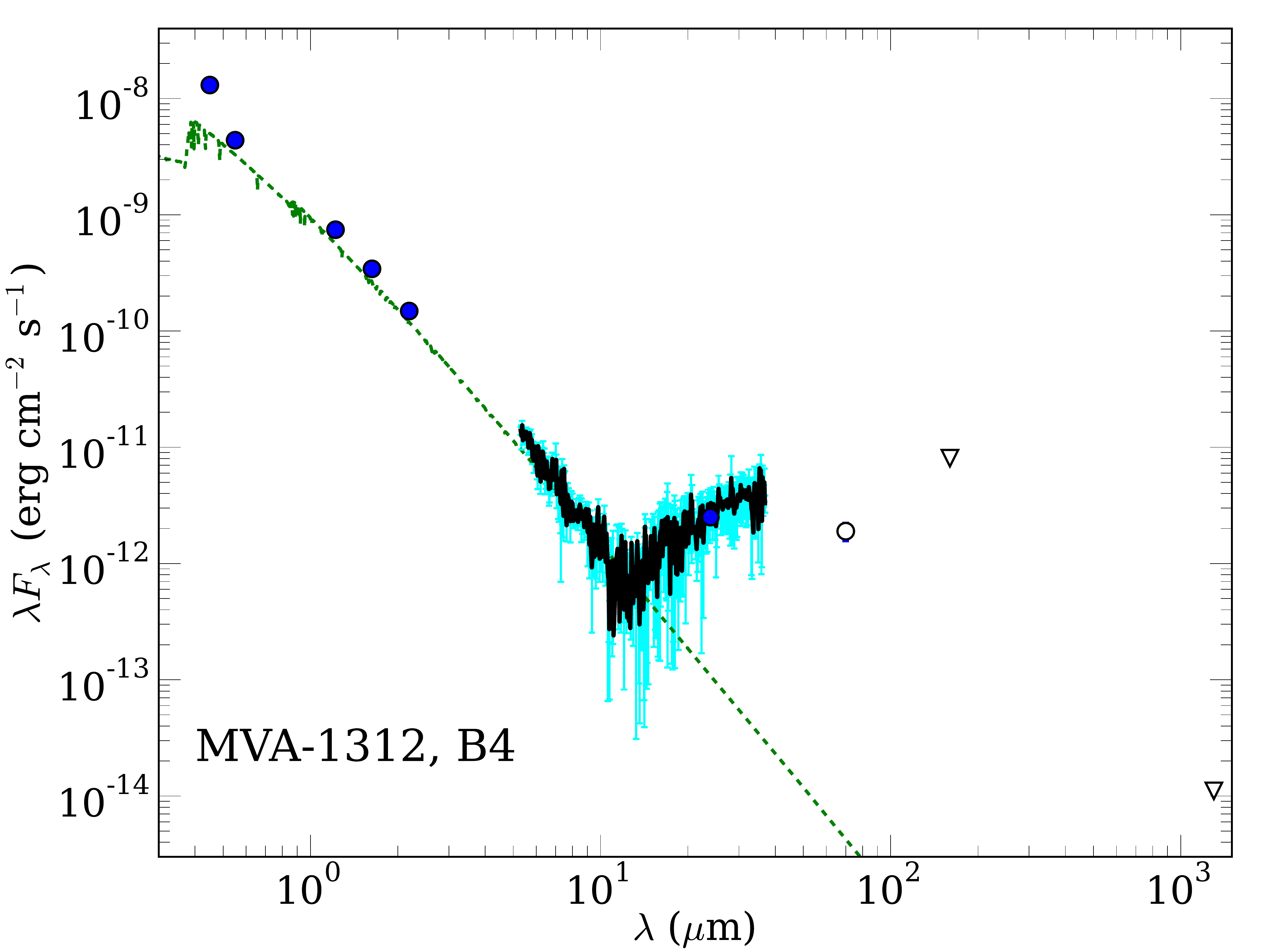} \\
\includegraphics[width=0.24\linewidth]{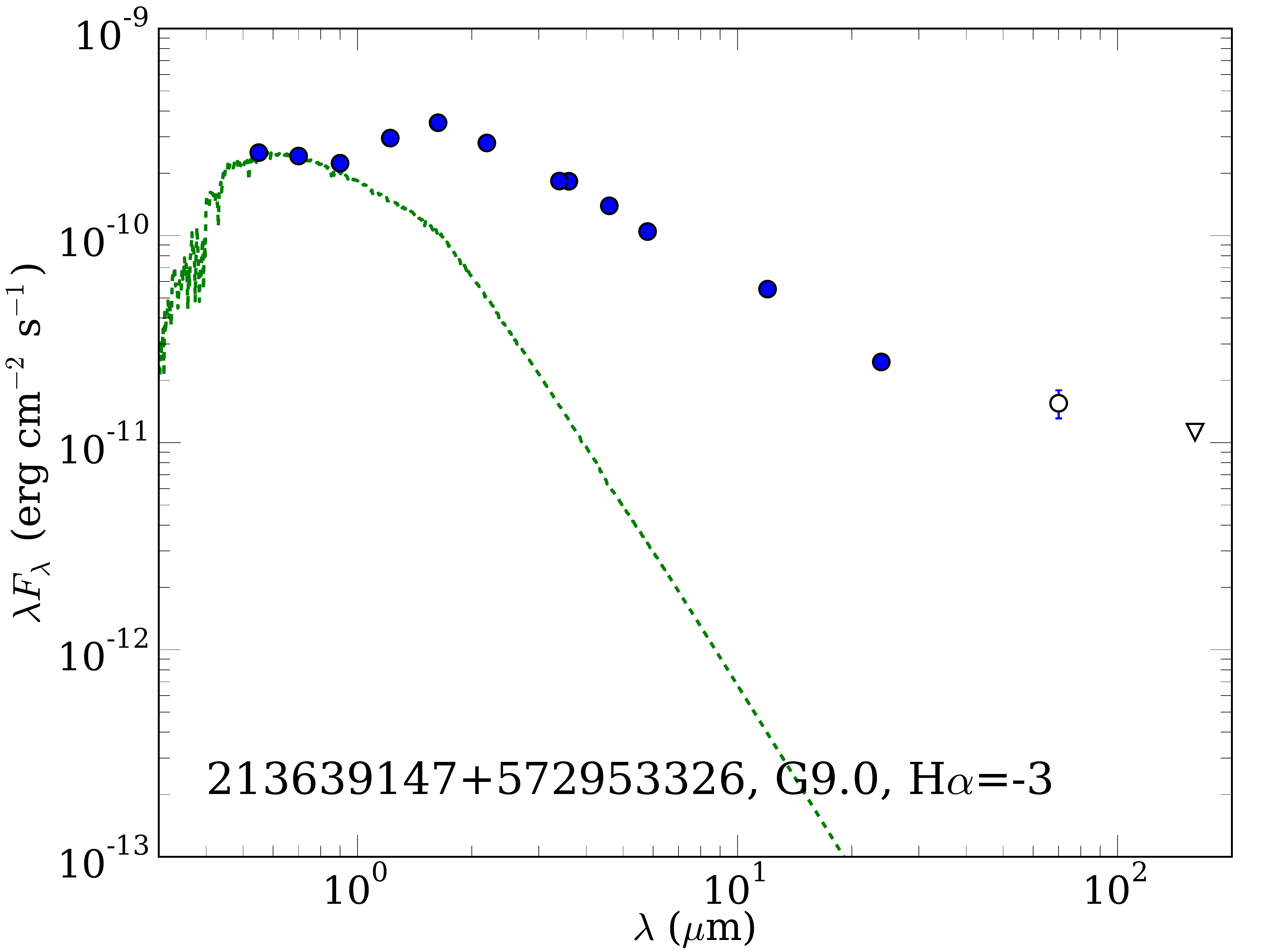} &
\includegraphics[width=0.24\linewidth]{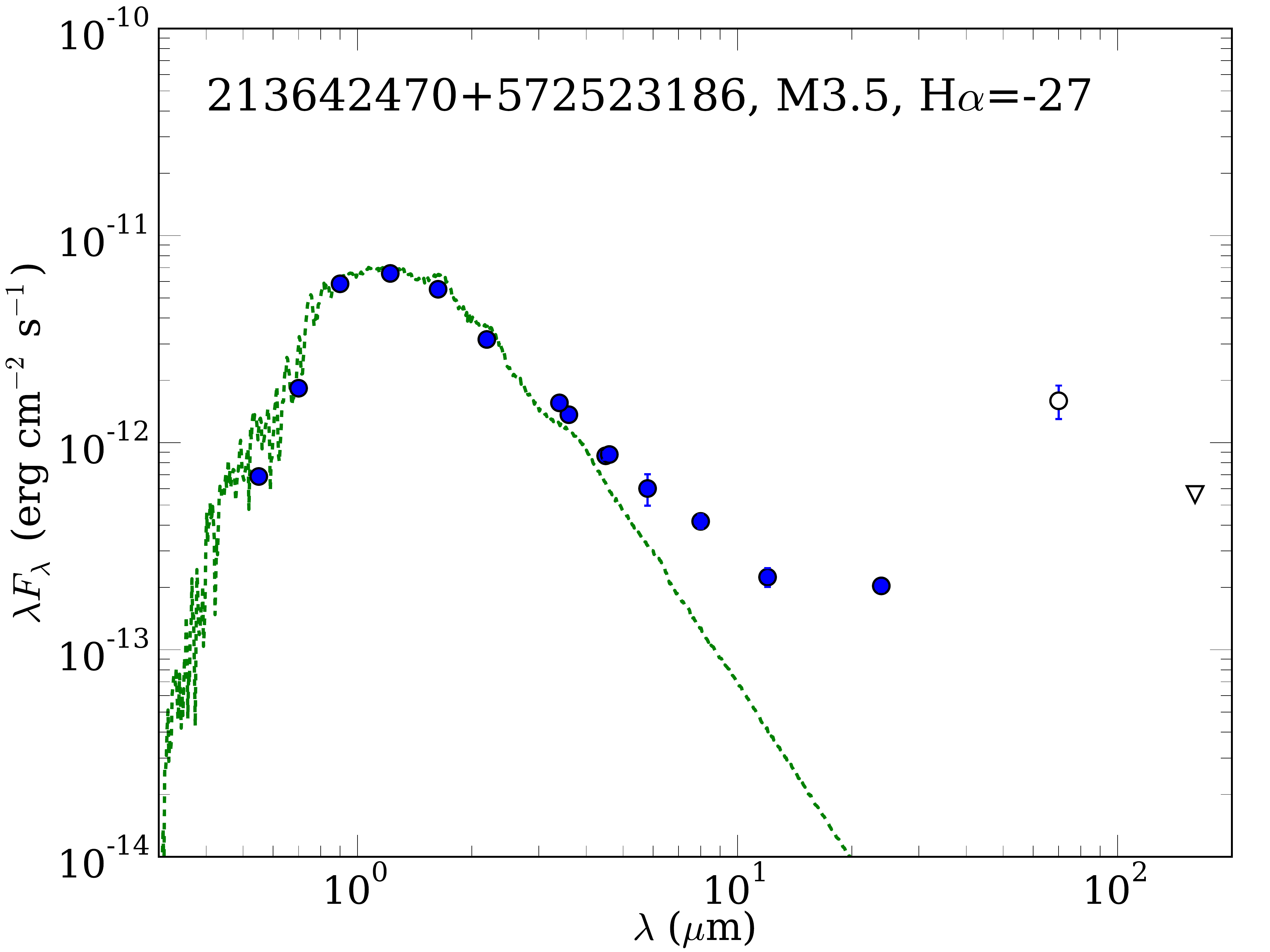} &
\includegraphics[width=0.24\linewidth]{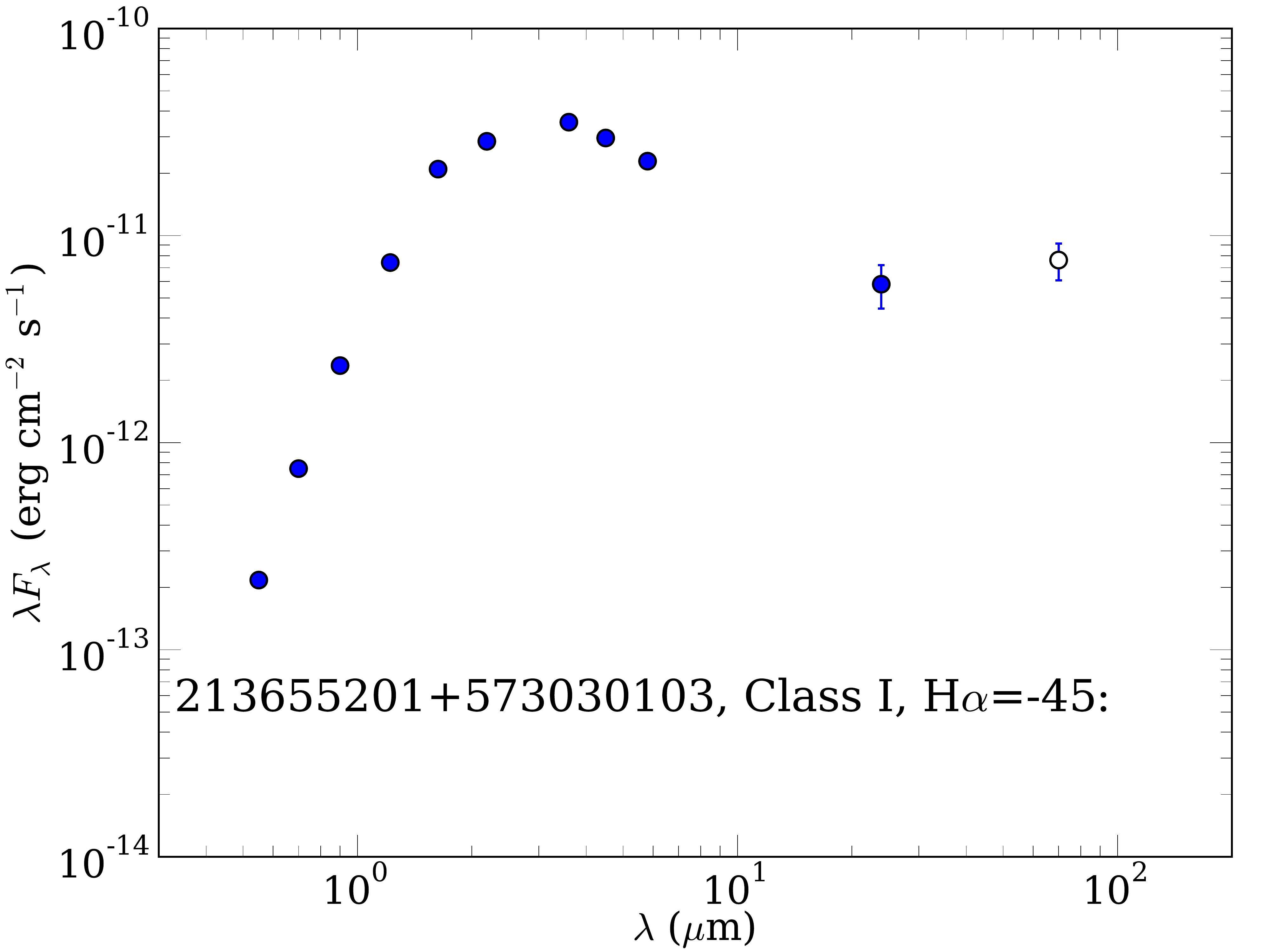} &
\includegraphics[width=0.24\linewidth]{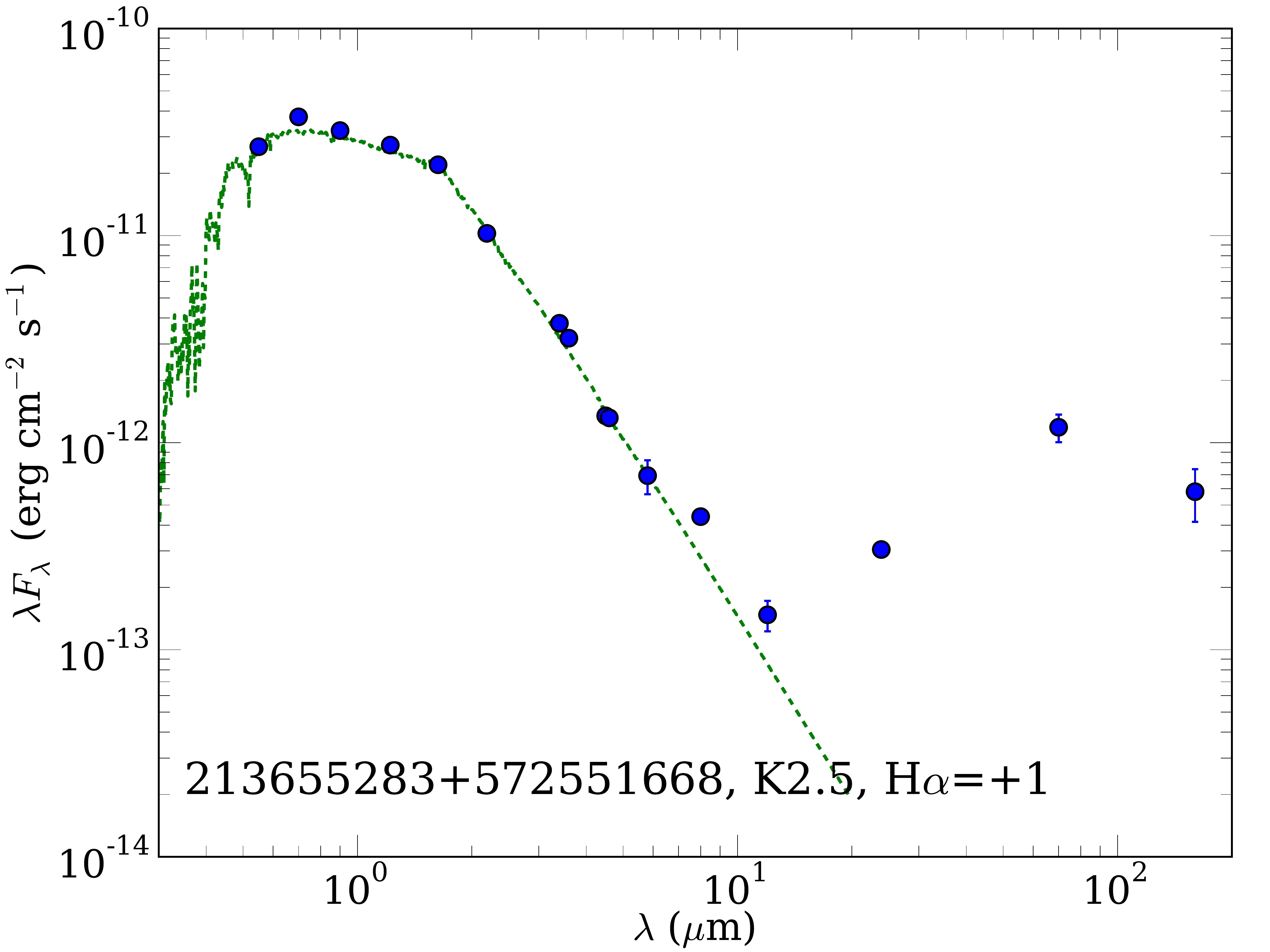} \\
\includegraphics[width=0.24\linewidth]{11_2146.pdf} &
\includegraphics[width=0.24\linewidth]{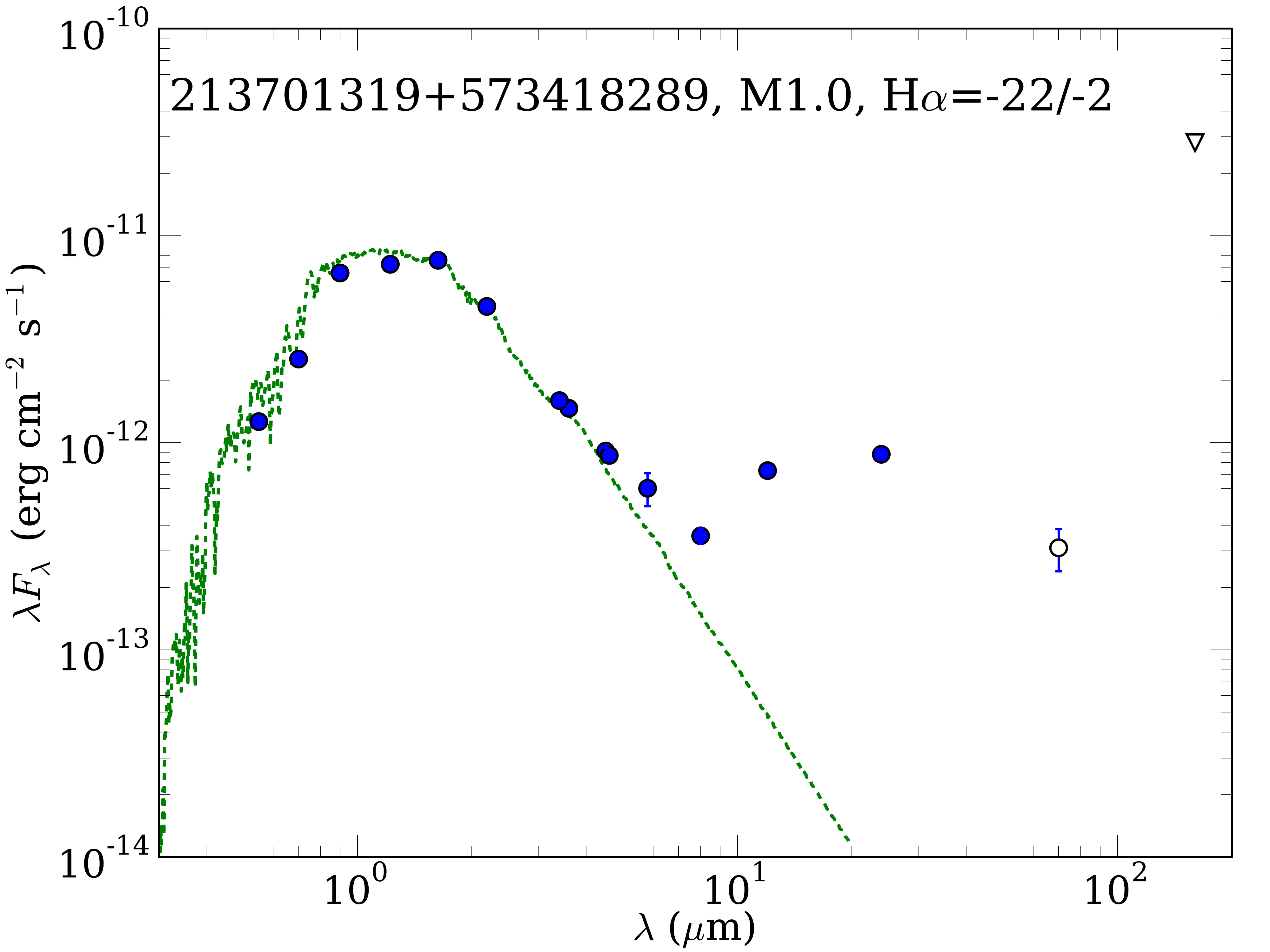} &
\includegraphics[width=0.24\linewidth]{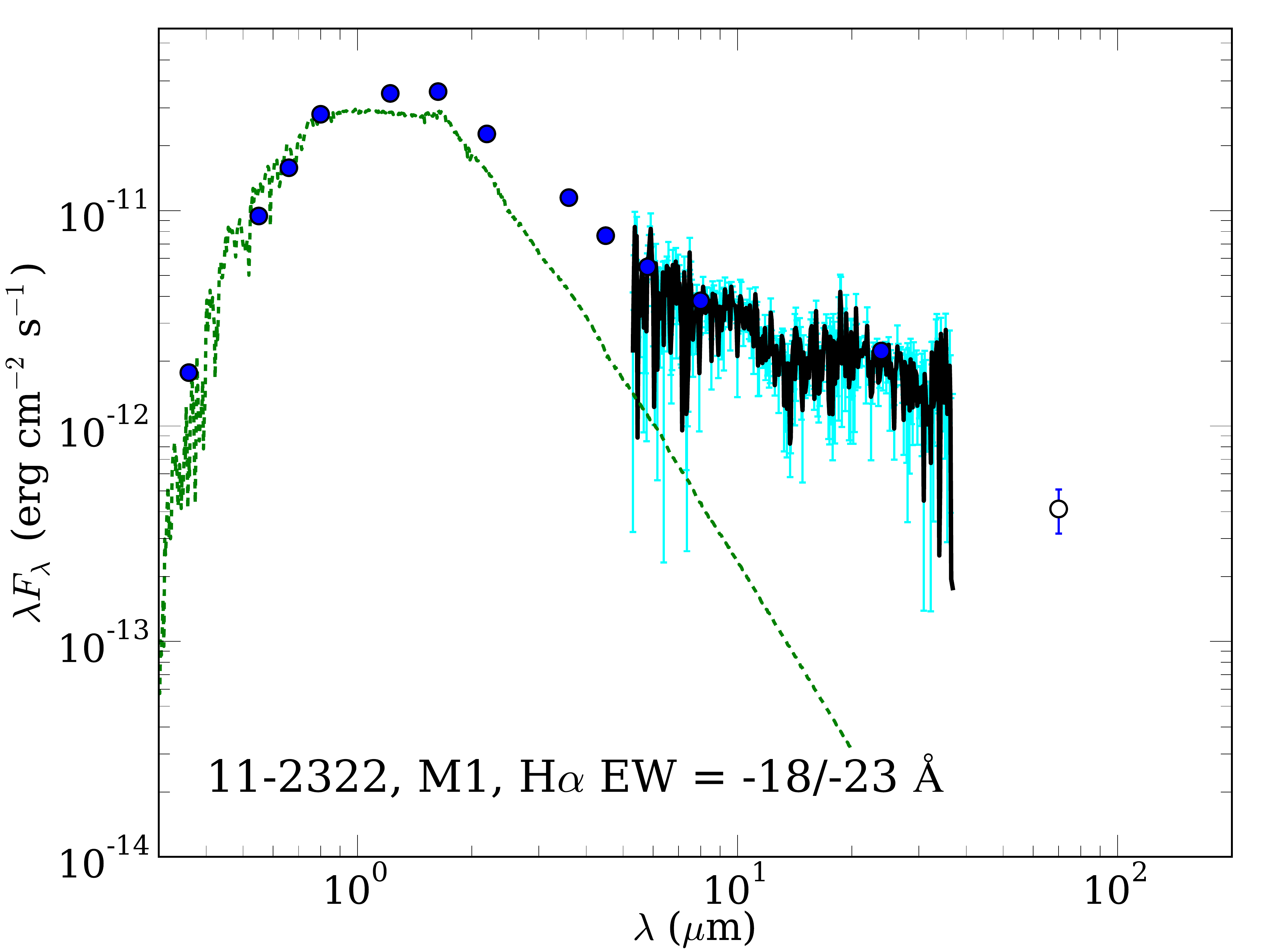} &
\includegraphics[width=0.24\linewidth]{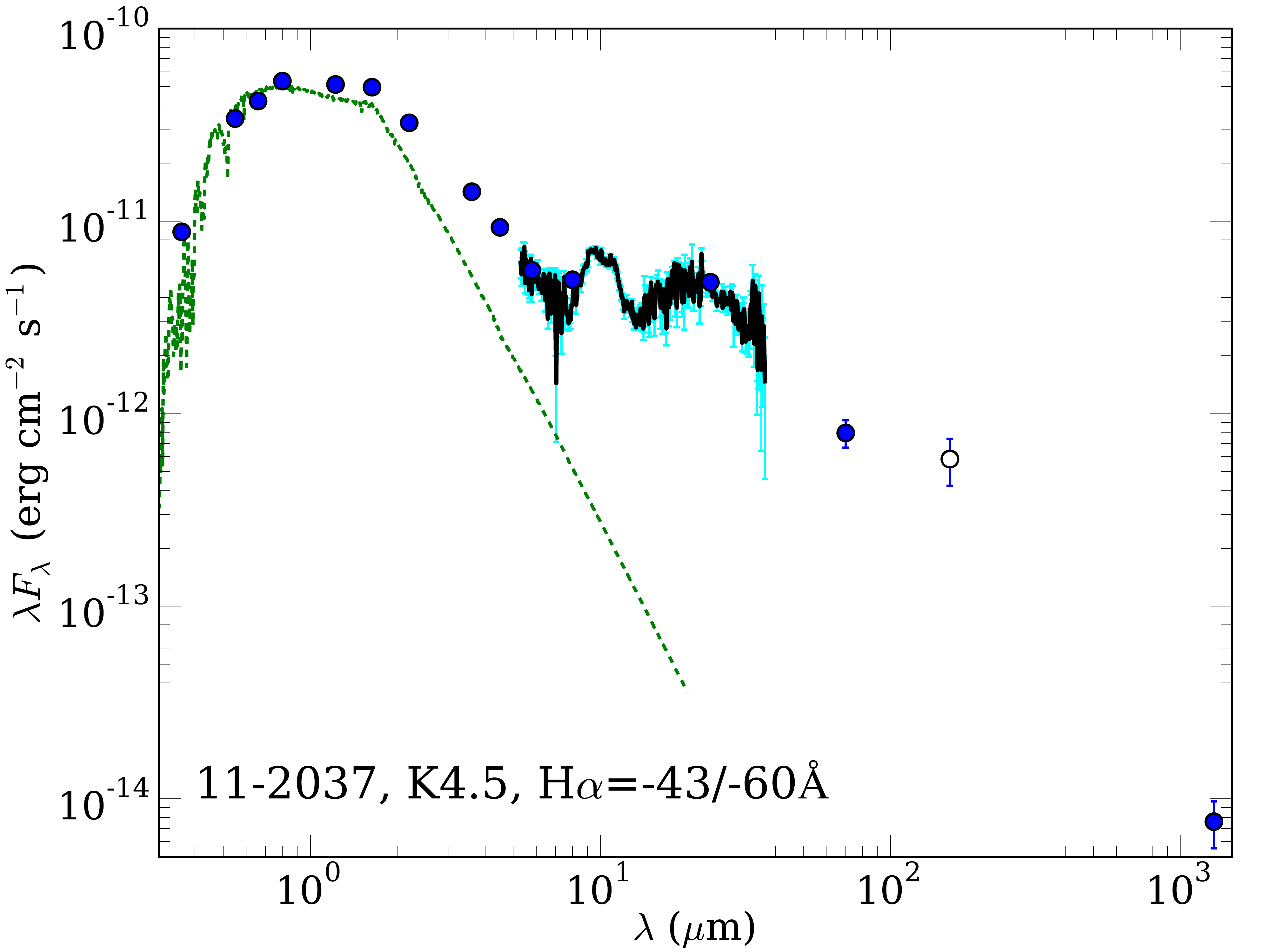} \\
\includegraphics[width=0.24\linewidth]{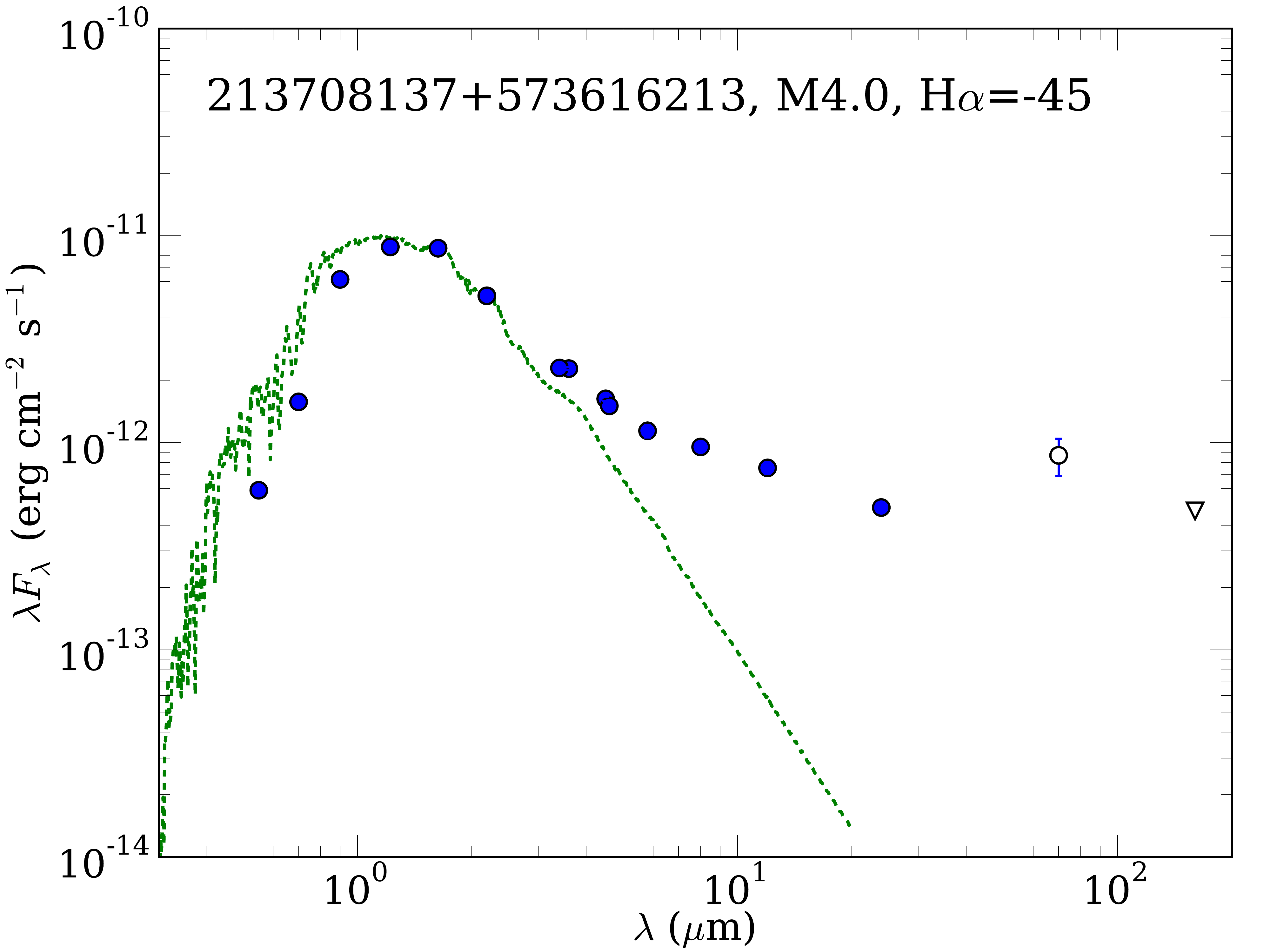} &
\includegraphics[width=0.24\linewidth]{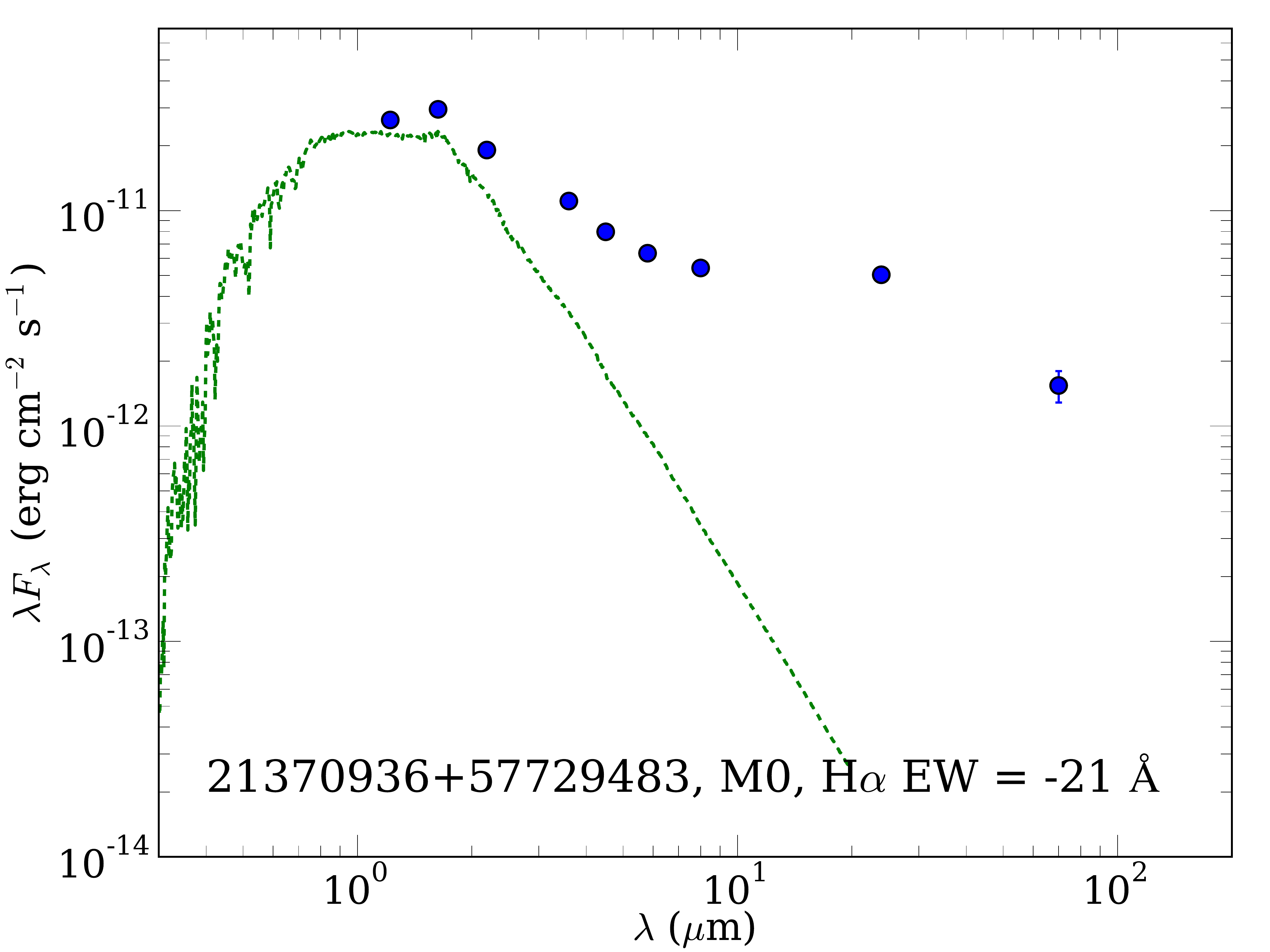} &
\includegraphics[width=0.24\linewidth]{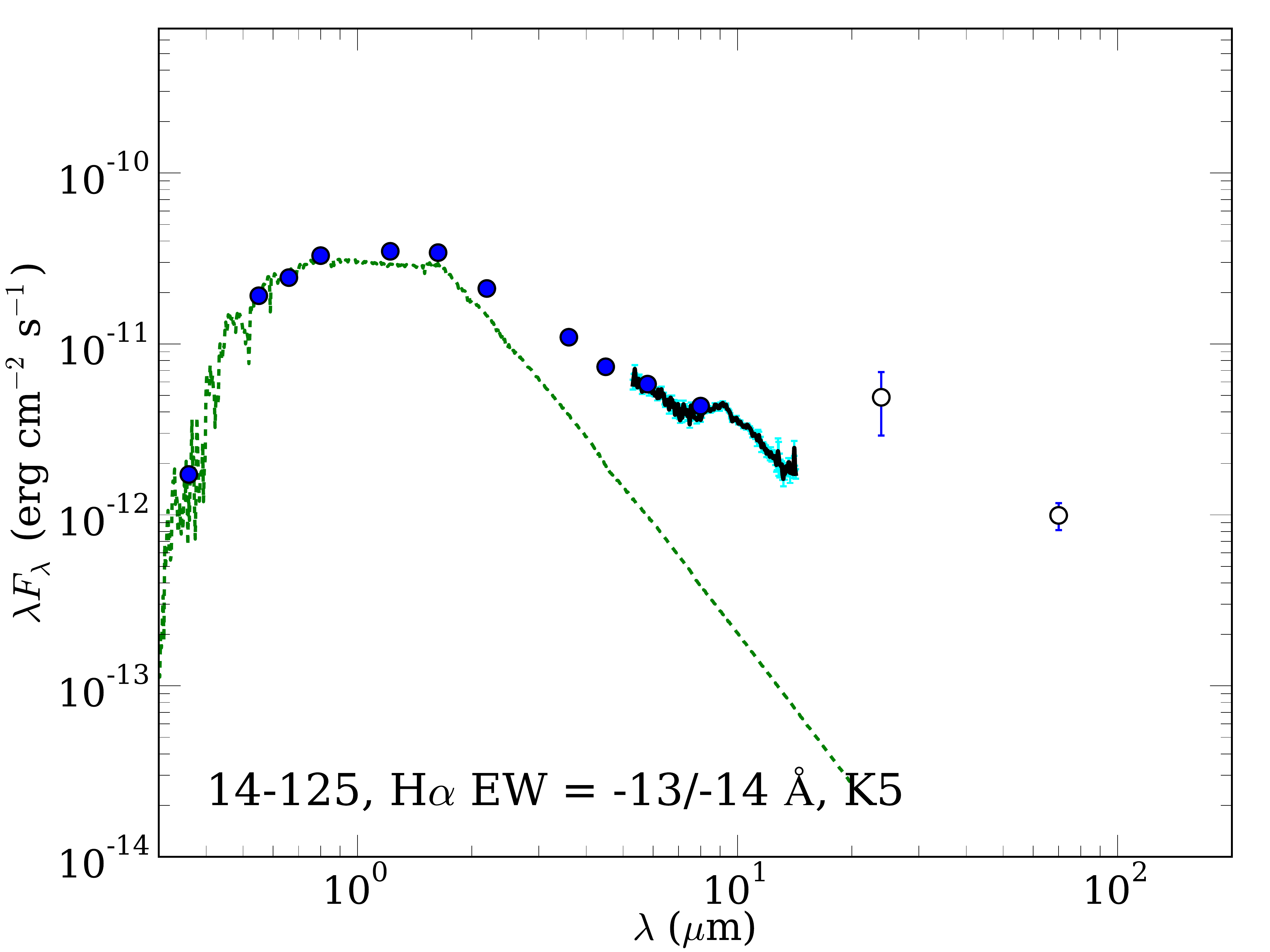} &
\includegraphics[width=0.24\linewidth]{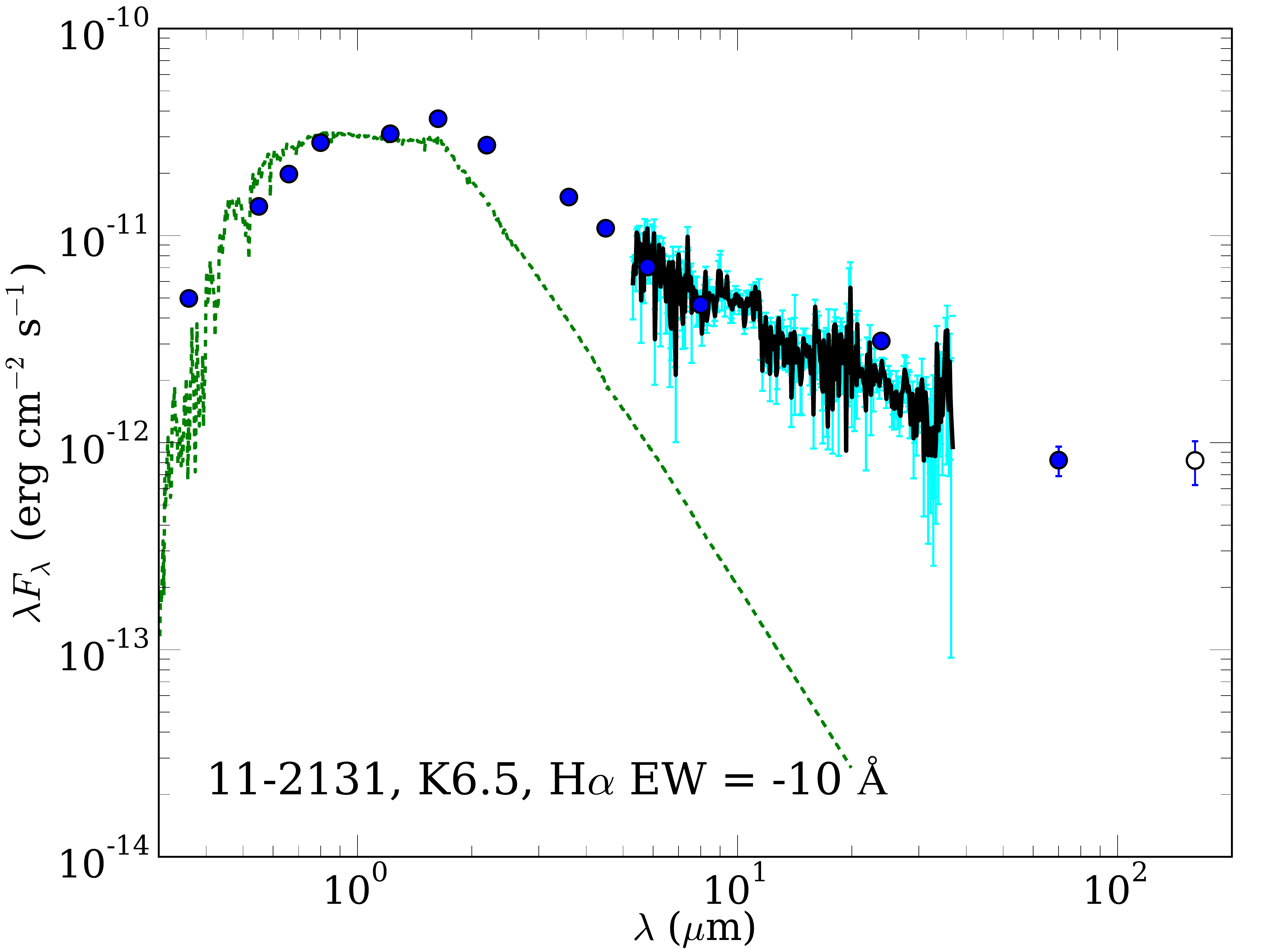} \\
\includegraphics[width=0.24\linewidth]{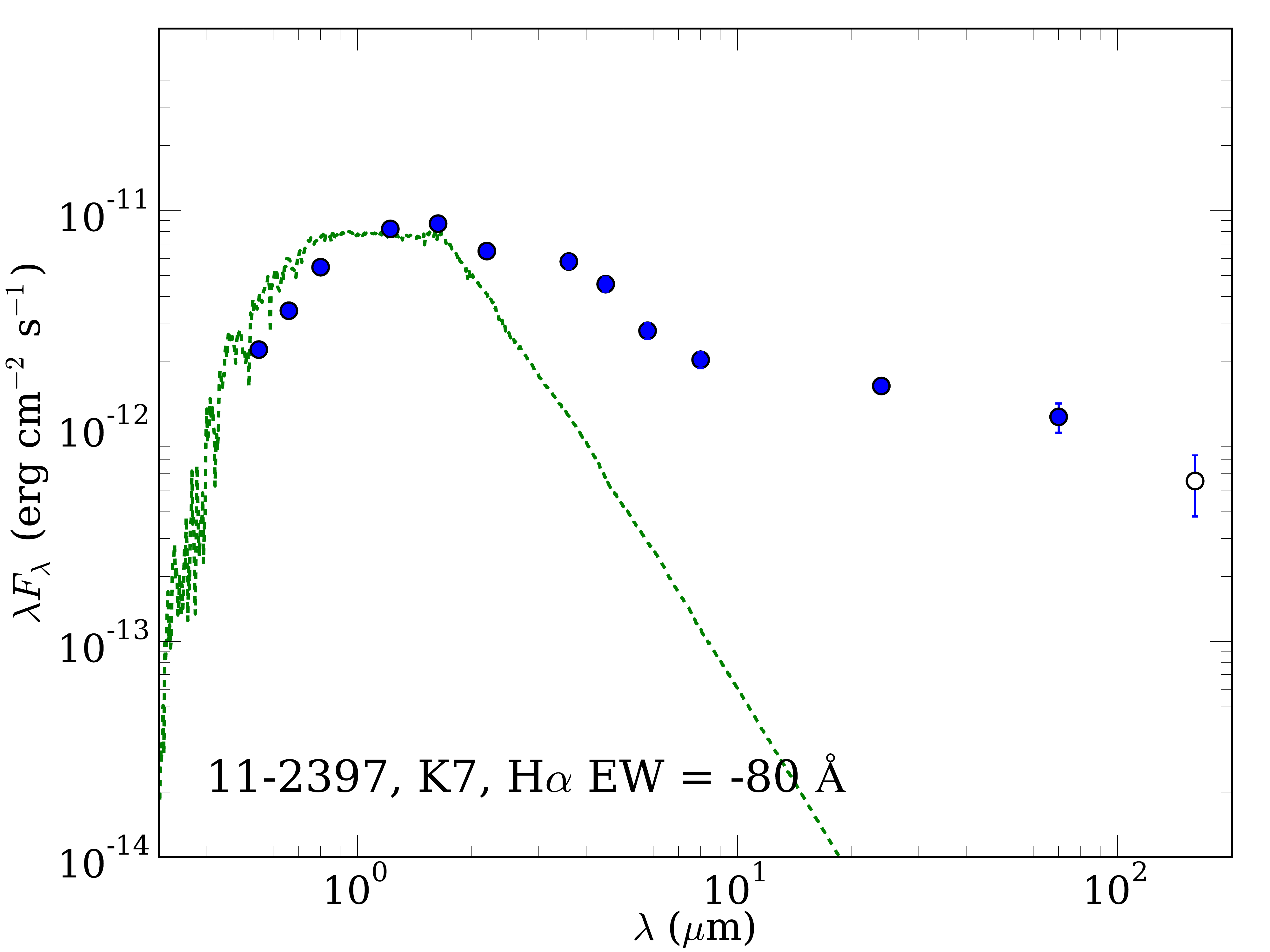} &
\includegraphics[width=0.24\linewidth]{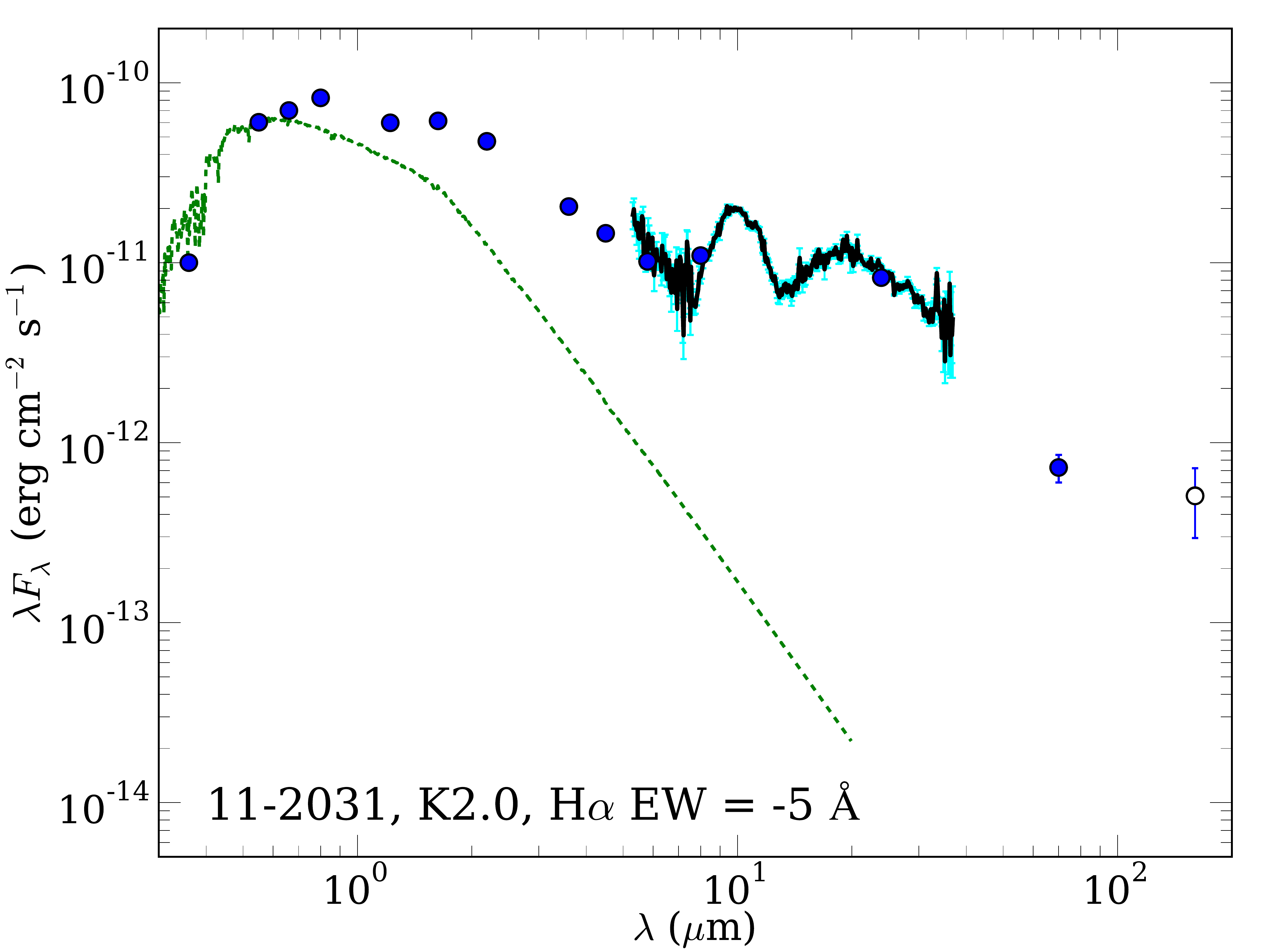} &
\includegraphics[width=0.24\linewidth]{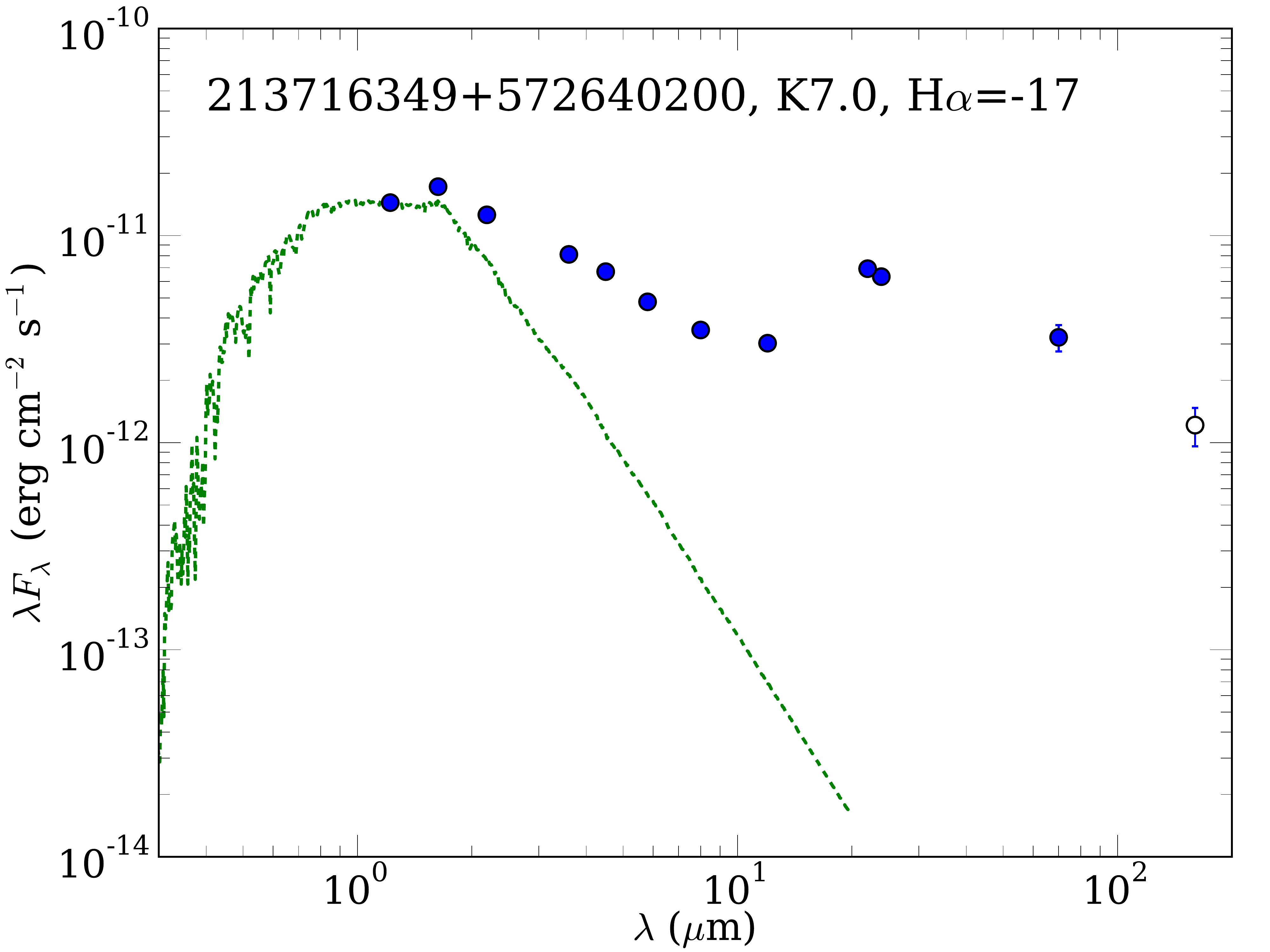} &
\includegraphics[width=0.24\linewidth]{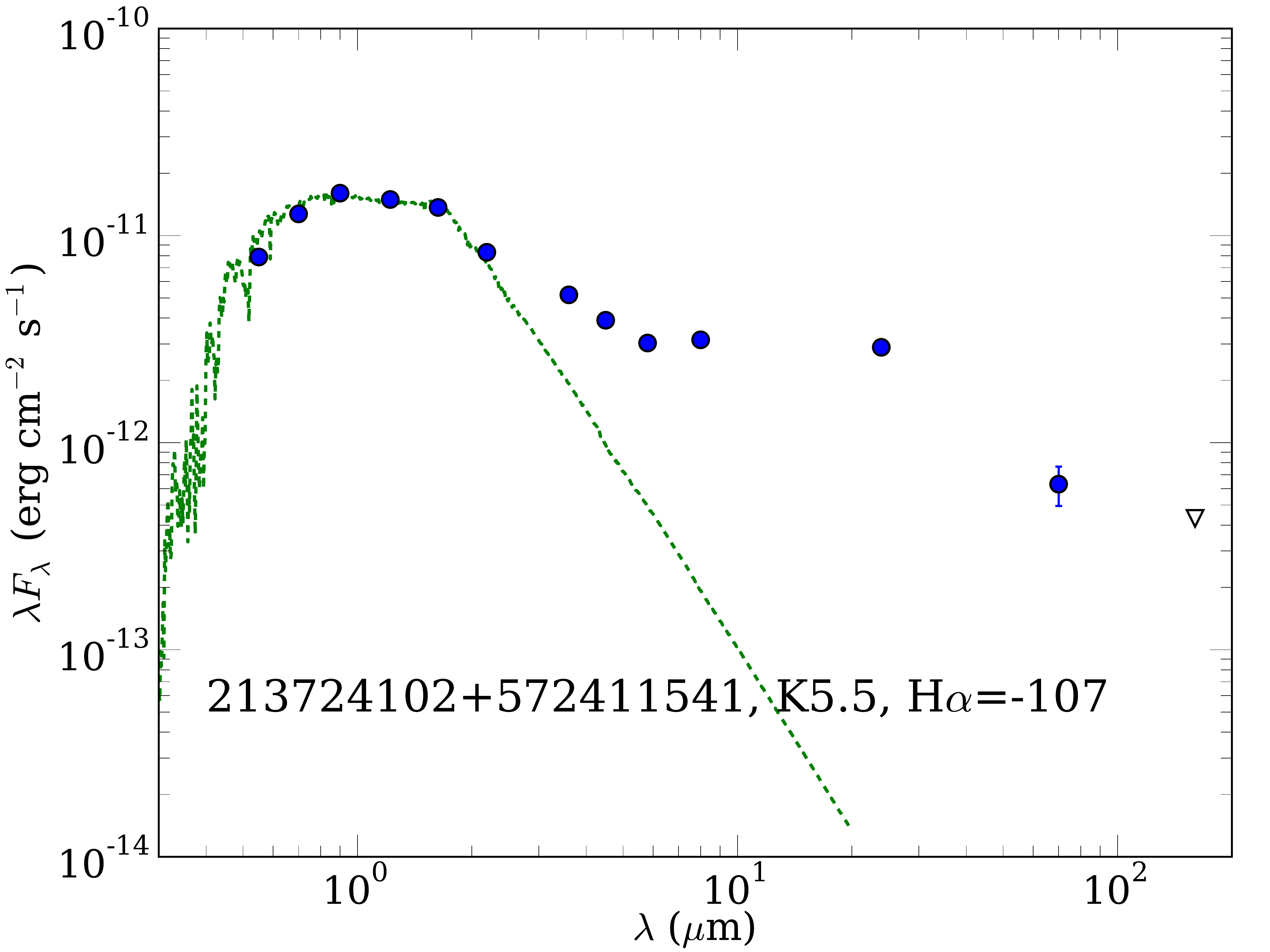} \\
\includegraphics[width=0.24\linewidth]{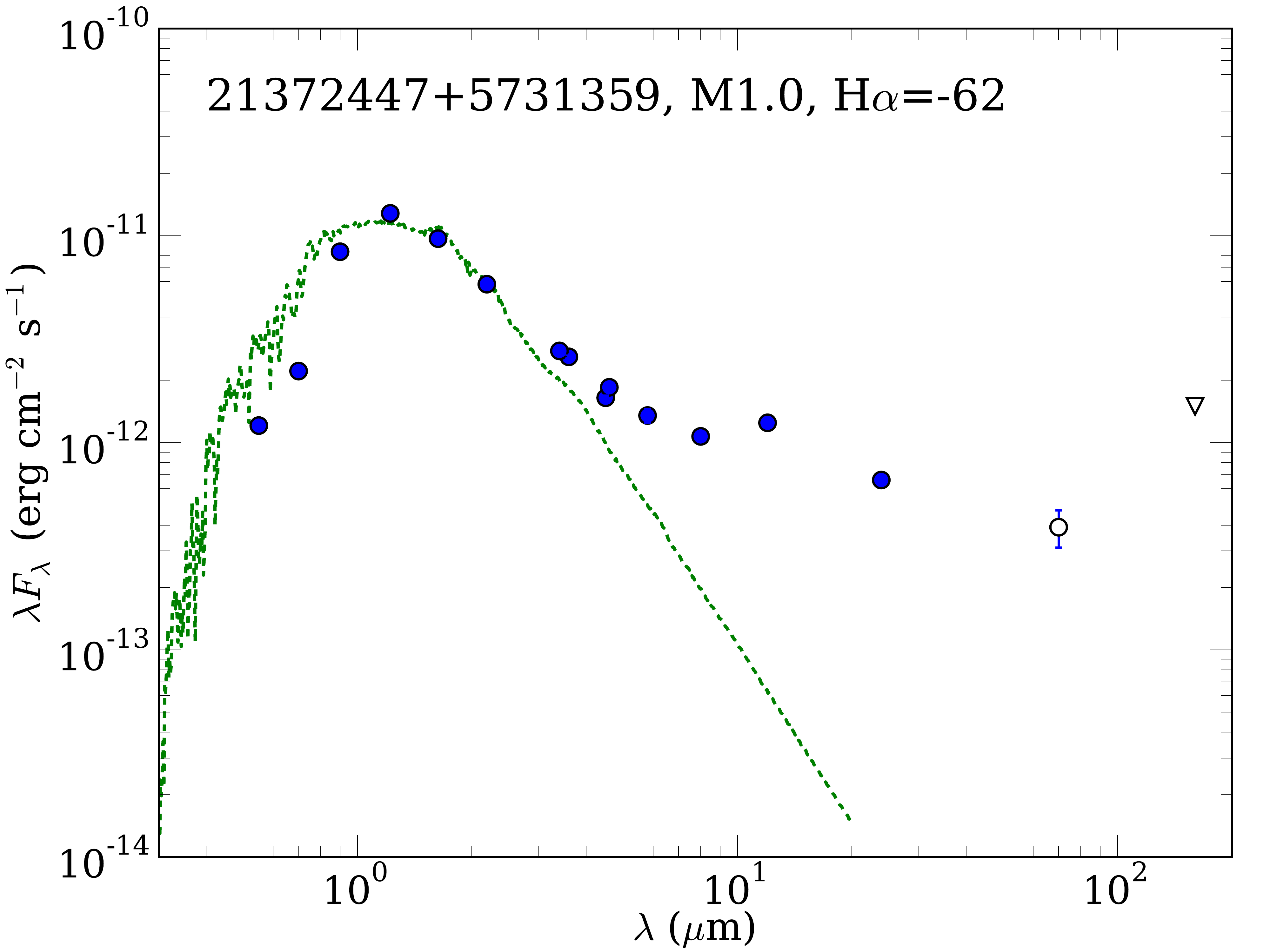} &
\includegraphics[width=0.24\linewidth]{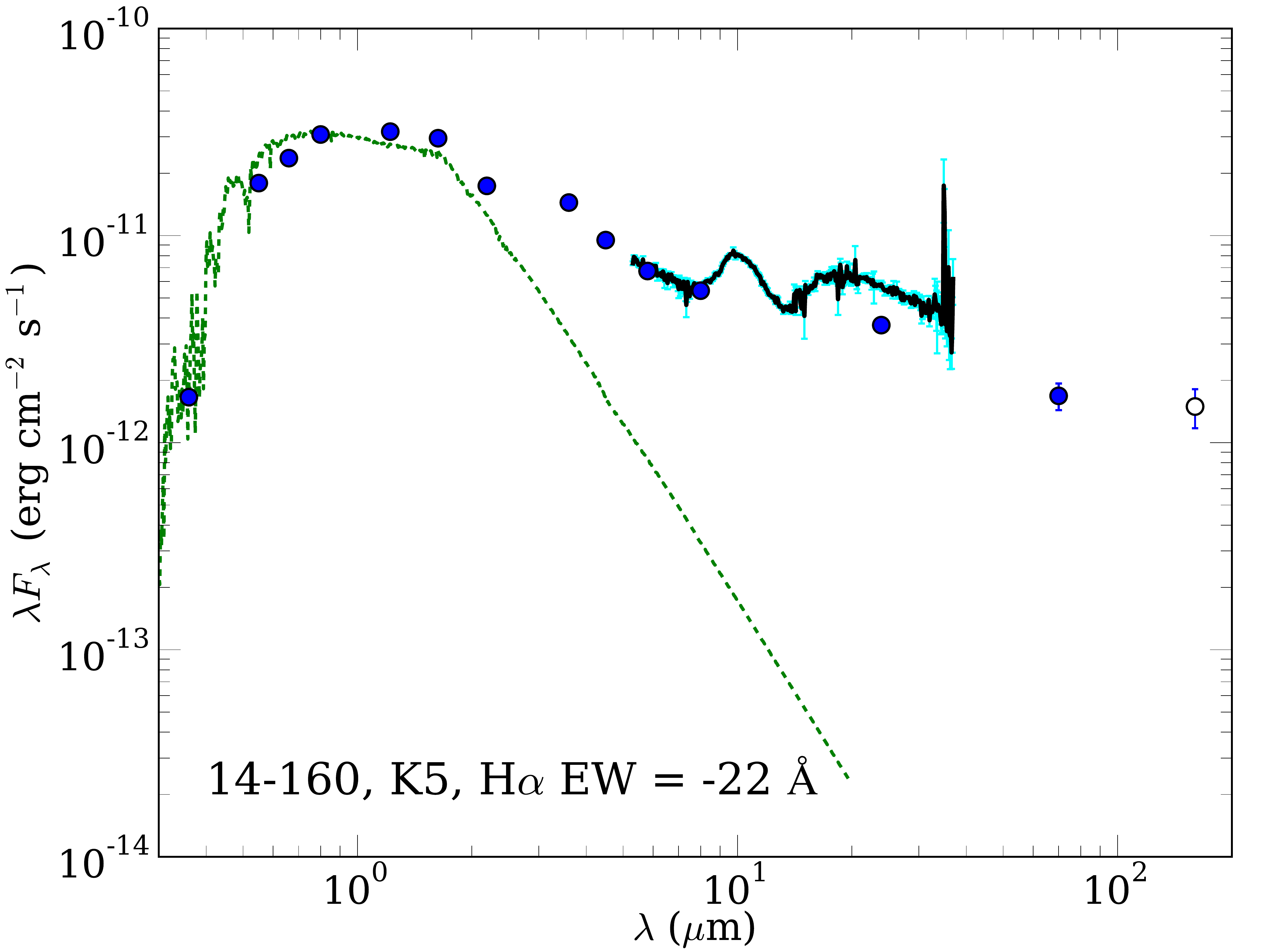} &
\includegraphics[width=0.24\linewidth]{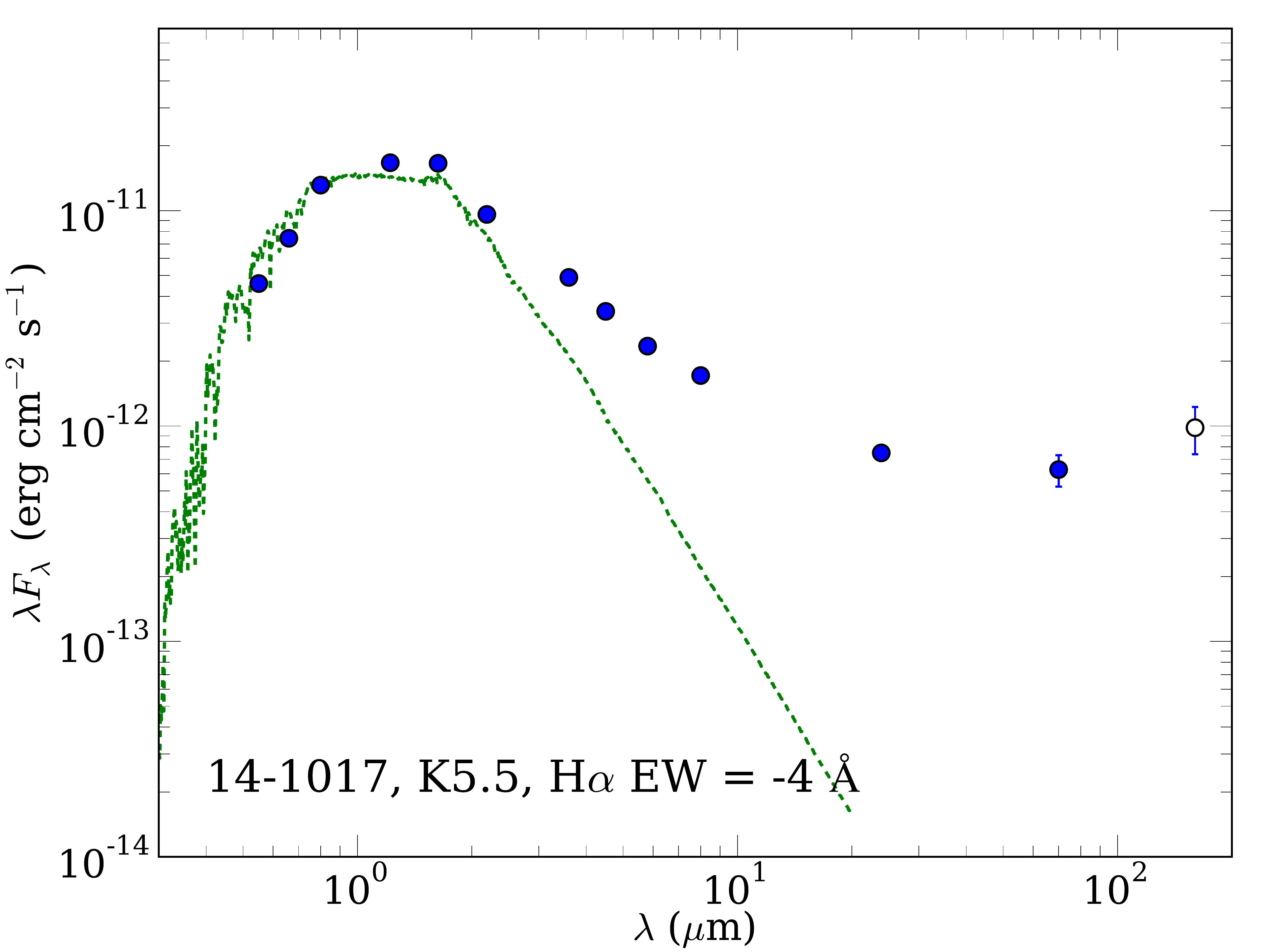} &
\includegraphics[width=0.24\linewidth]{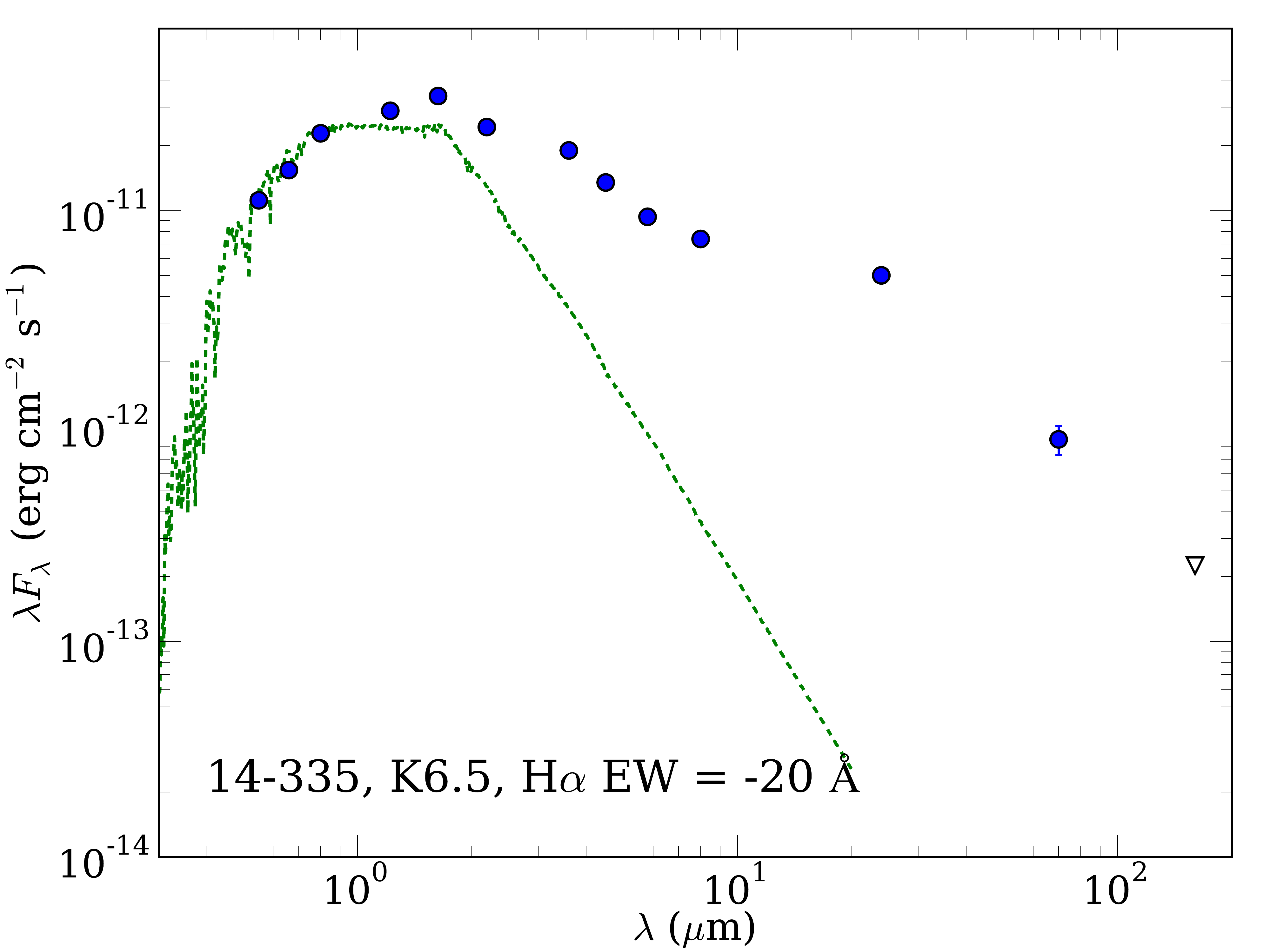} \\
\end{tabular}
\caption{SEDs of the objects detected with Herschel, including available optical, Spitzer (IRAC/MIPS photometry and
IRS spectra), WISE, and IRAM 1.3mm data. The first row contains the only NGC\,7160 member detected with
Herschel (01-580), the rest belong to Tr\,37.
Filled symbols mark detections at different wavelengths. Errorbars are shown in blue
for both photometry points (often smaller than the symbols) and the IRS spectra. Upper limits
are marked as inverted open triangles. Marginal detections (close to 3$\sigma$ or affected by nebulosity)
are marked as open circles. The information about spectral types and H$\alpha$ emission
from the literature is also listed. A photospheric MARCS model is displayed as a dotted line for comparison. \label{seds1-fig}}
\end{figure*}

\begin{figure*}
\centering
\begin{tabular}{cccc}
\includegraphics[width=0.24\linewidth]{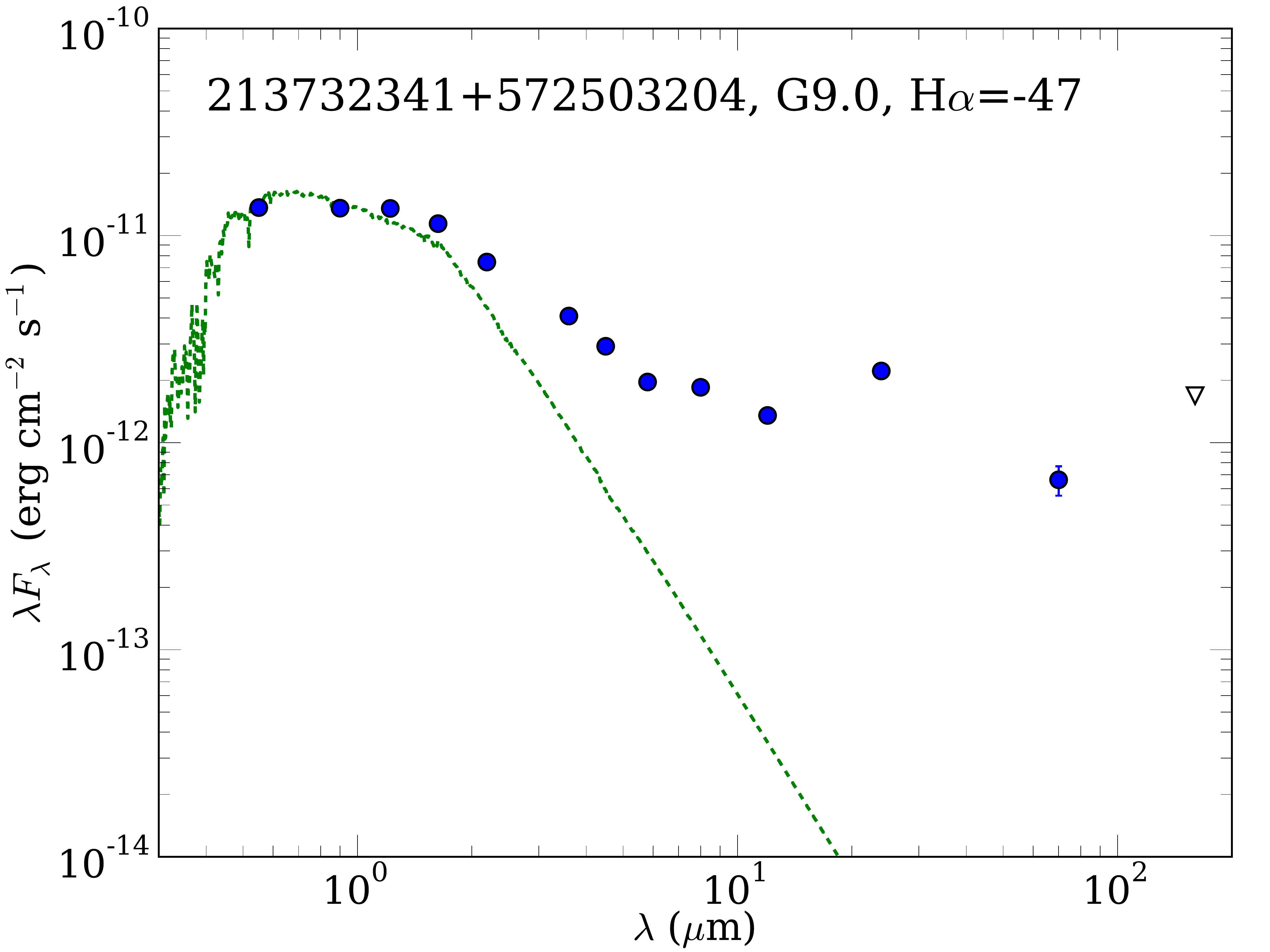} &
\includegraphics[width=0.24\linewidth]{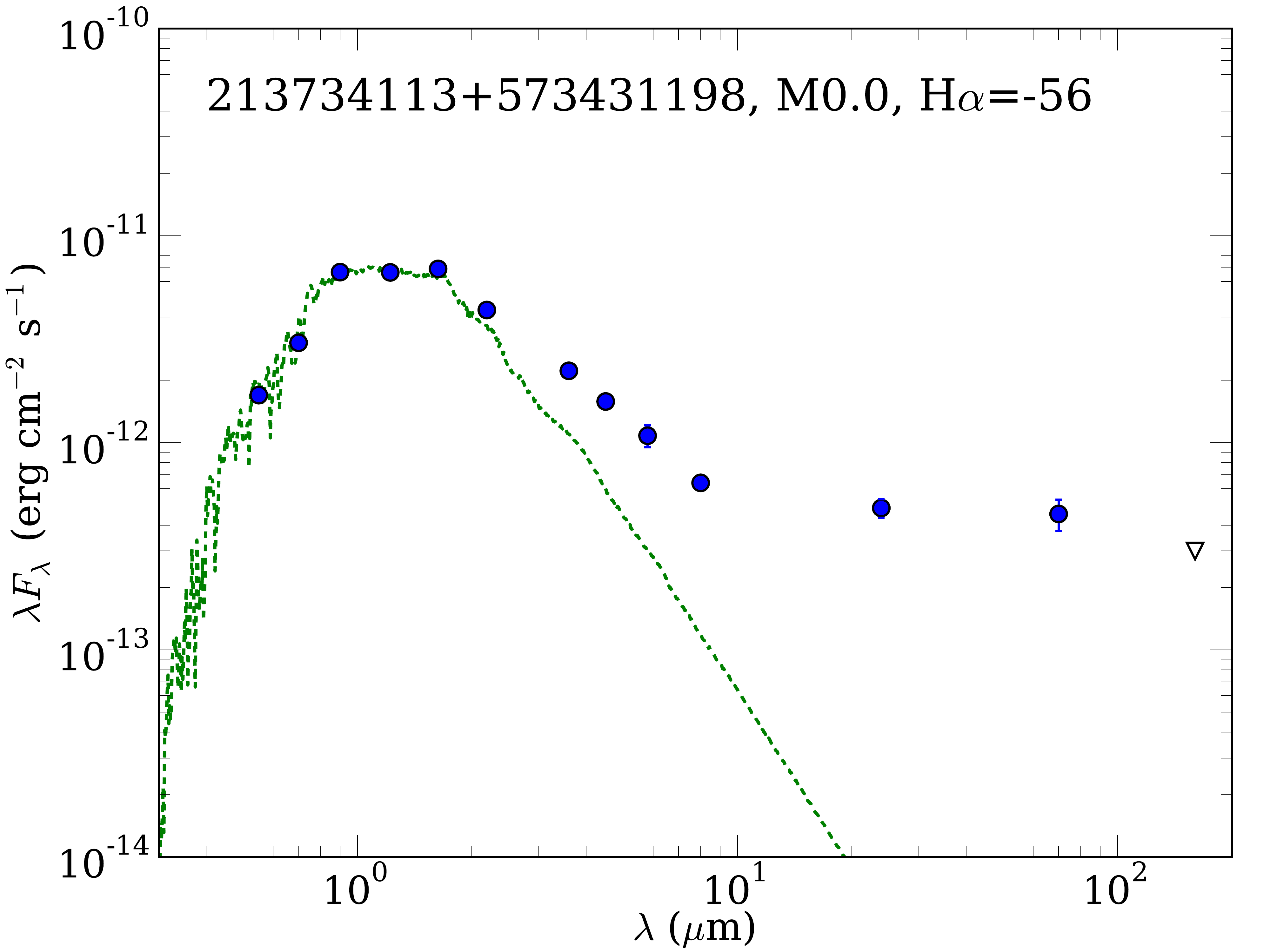} &
\includegraphics[width=0.24\linewidth]{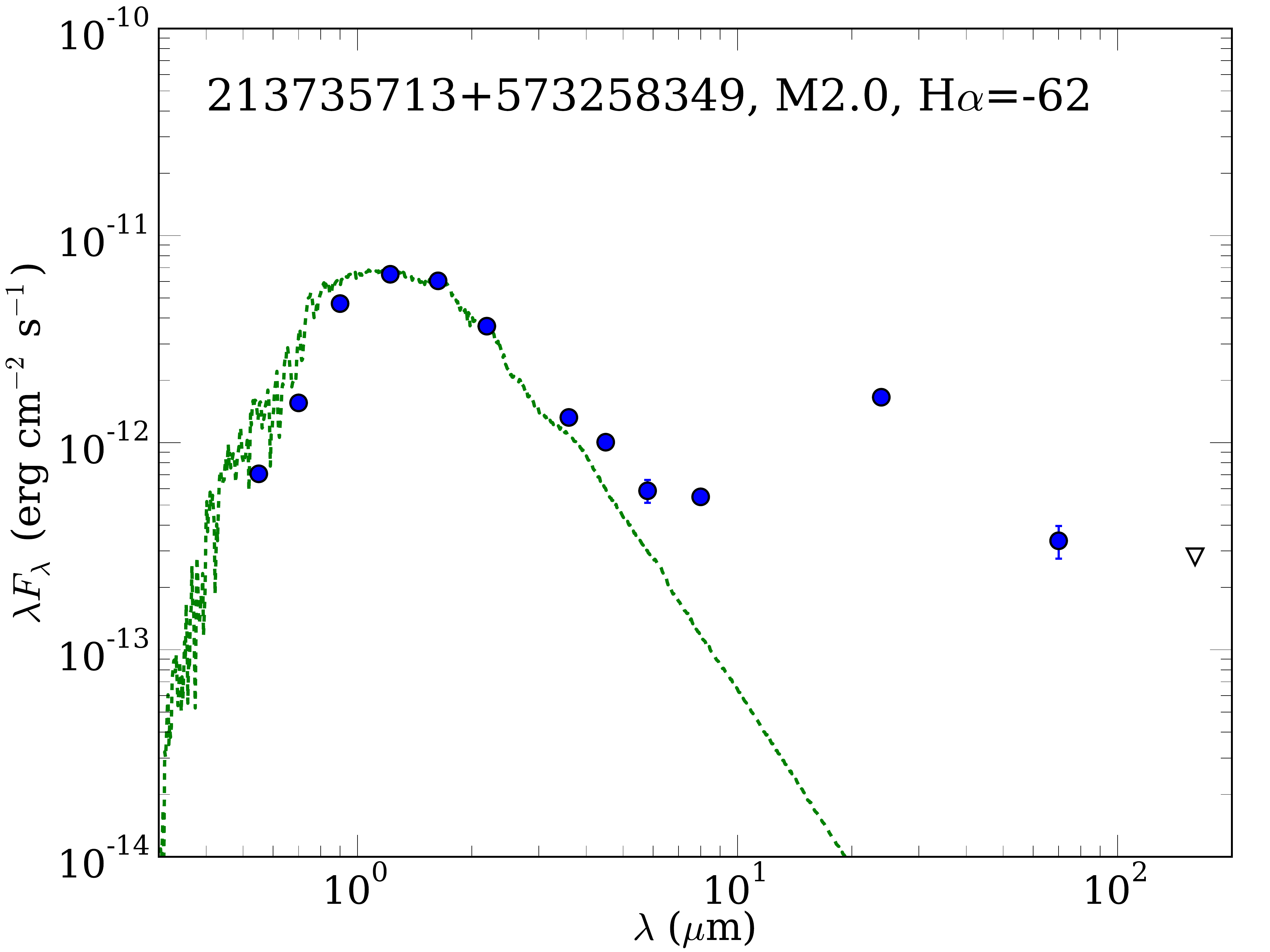} &
\includegraphics[width=0.24\linewidth]{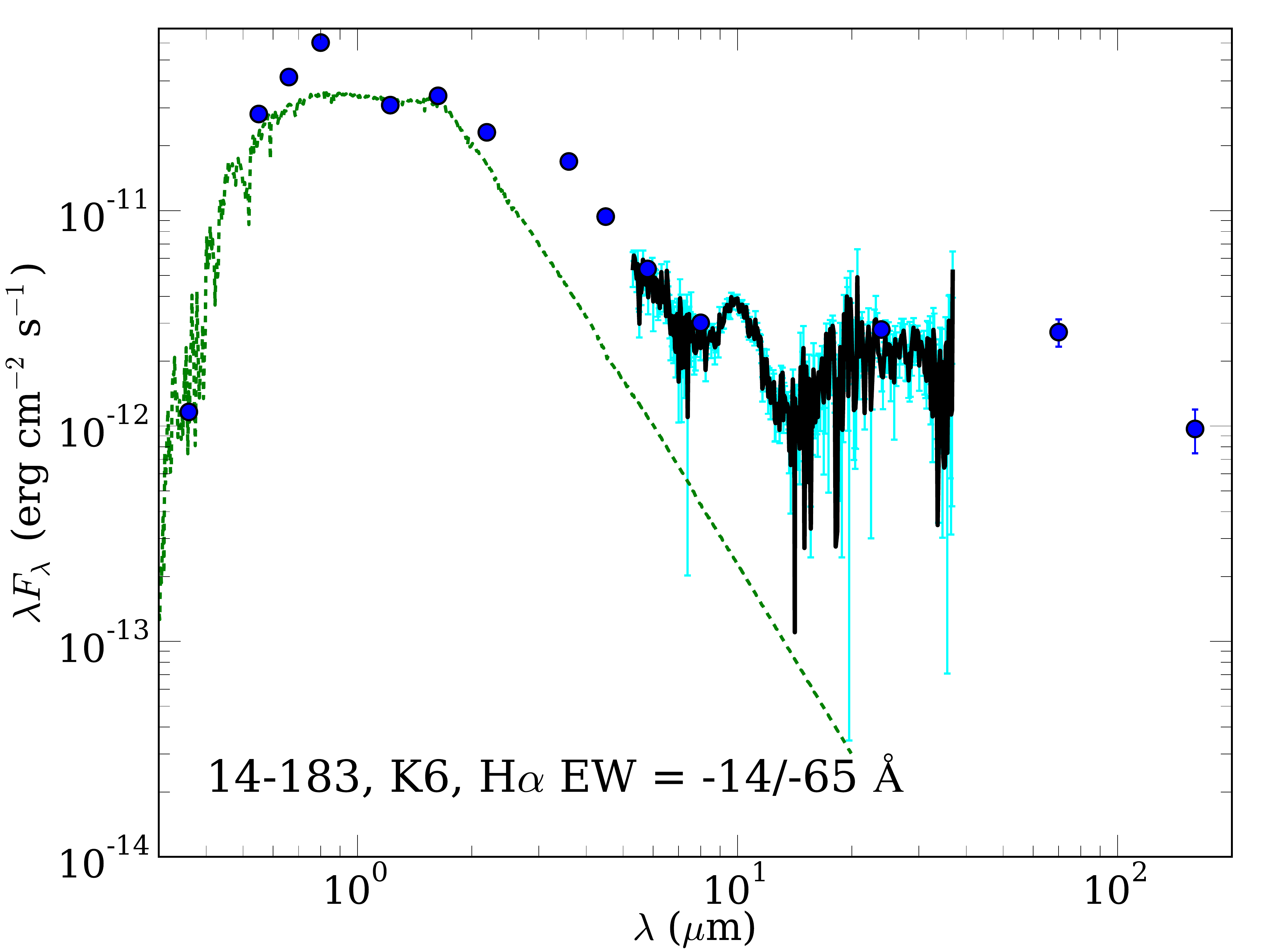} \\
\includegraphics[width=0.24\linewidth]{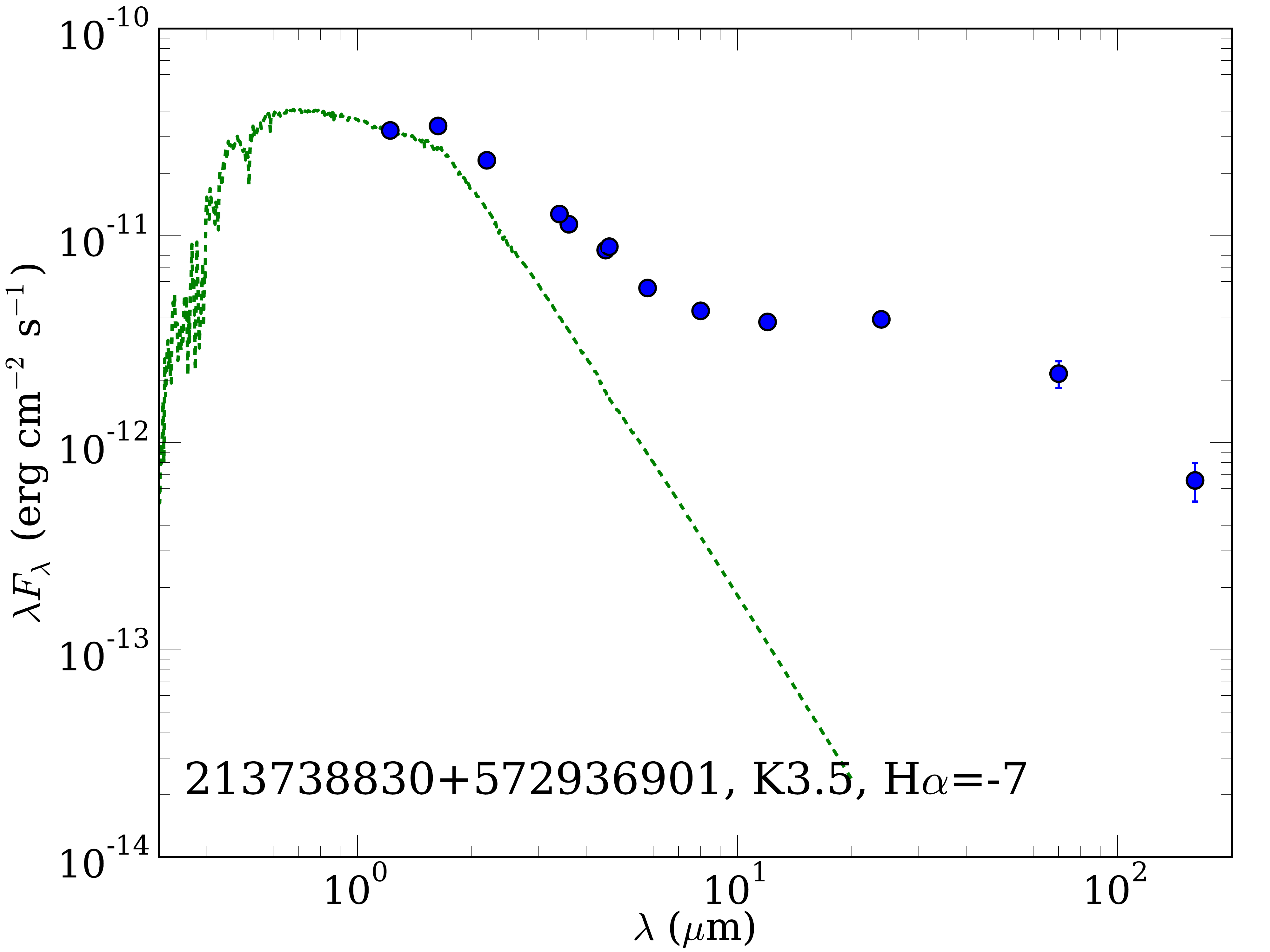} &
\includegraphics[width=0.24\linewidth]{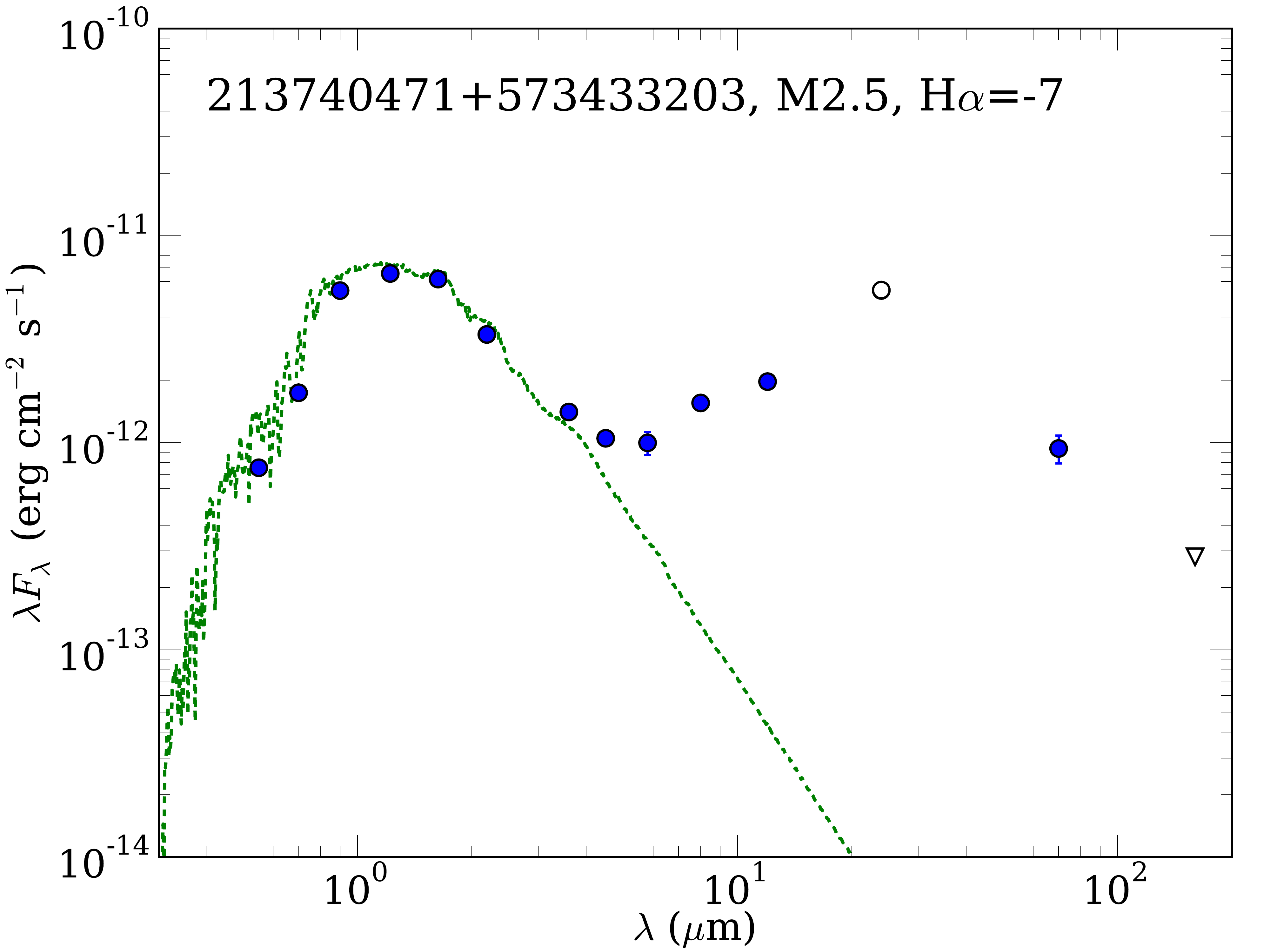} &
\includegraphics[width=0.24\linewidth]{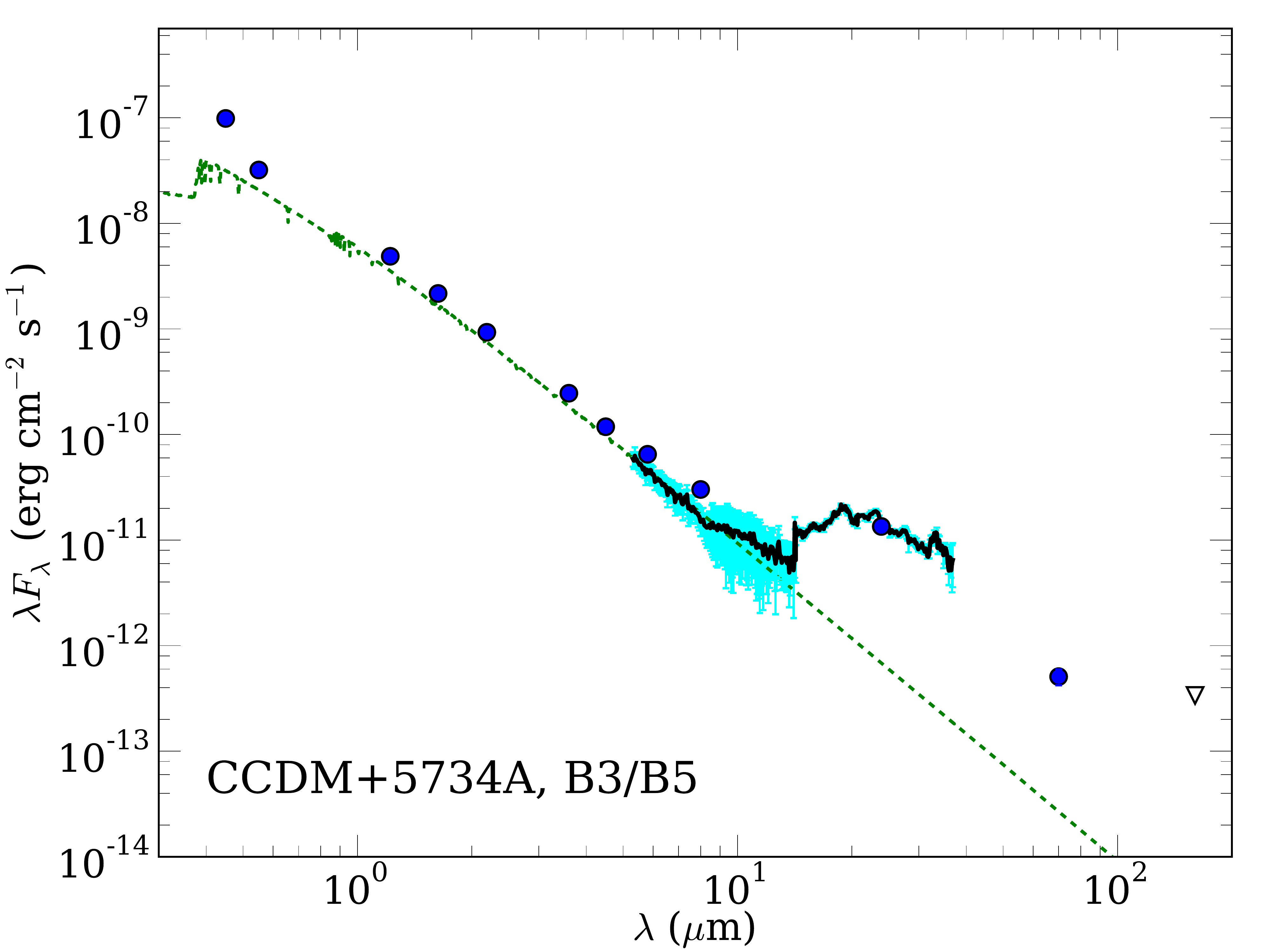} &
\includegraphics[width=0.24\linewidth]{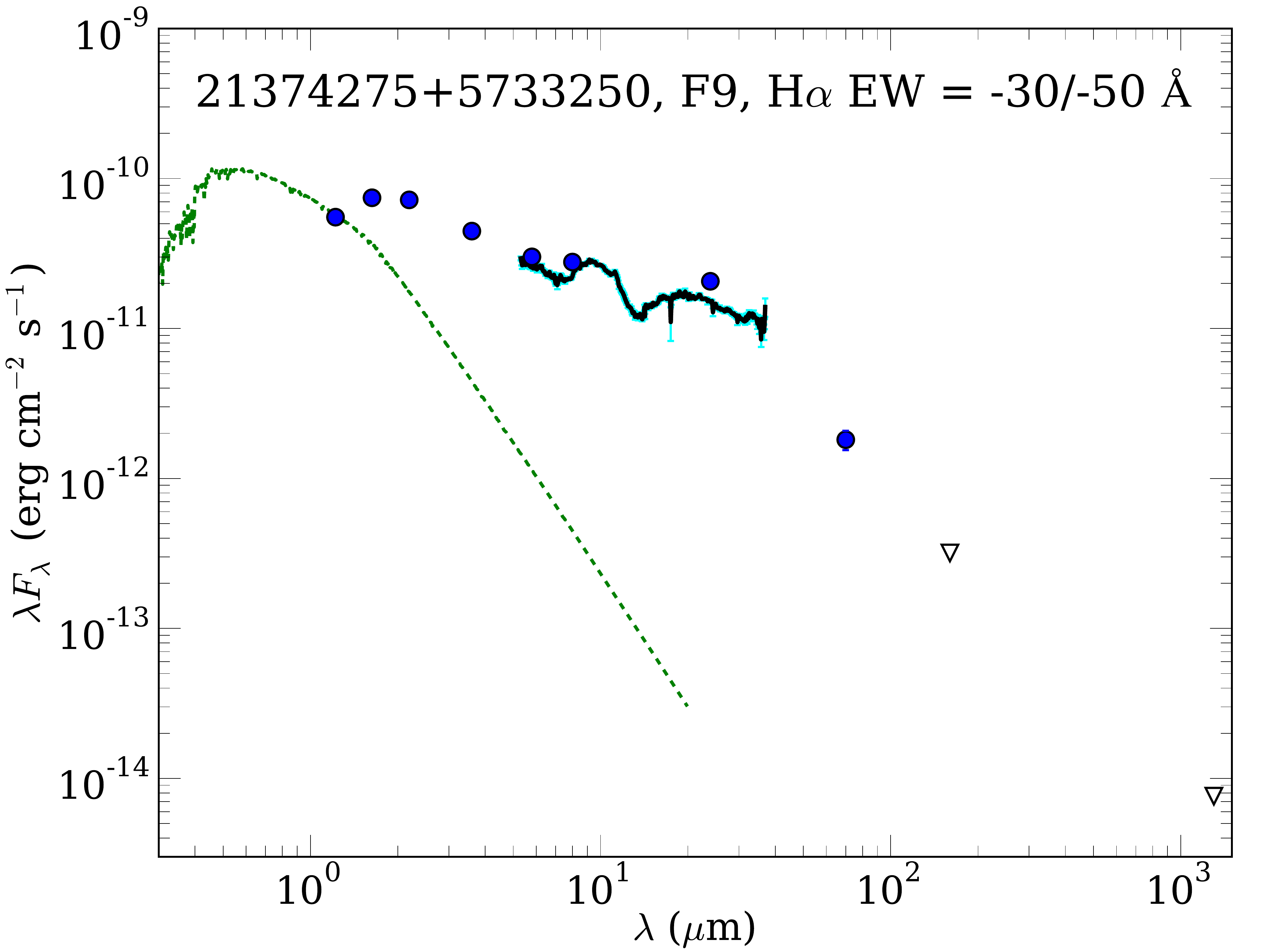} \\
\includegraphics[width=0.24\linewidth]{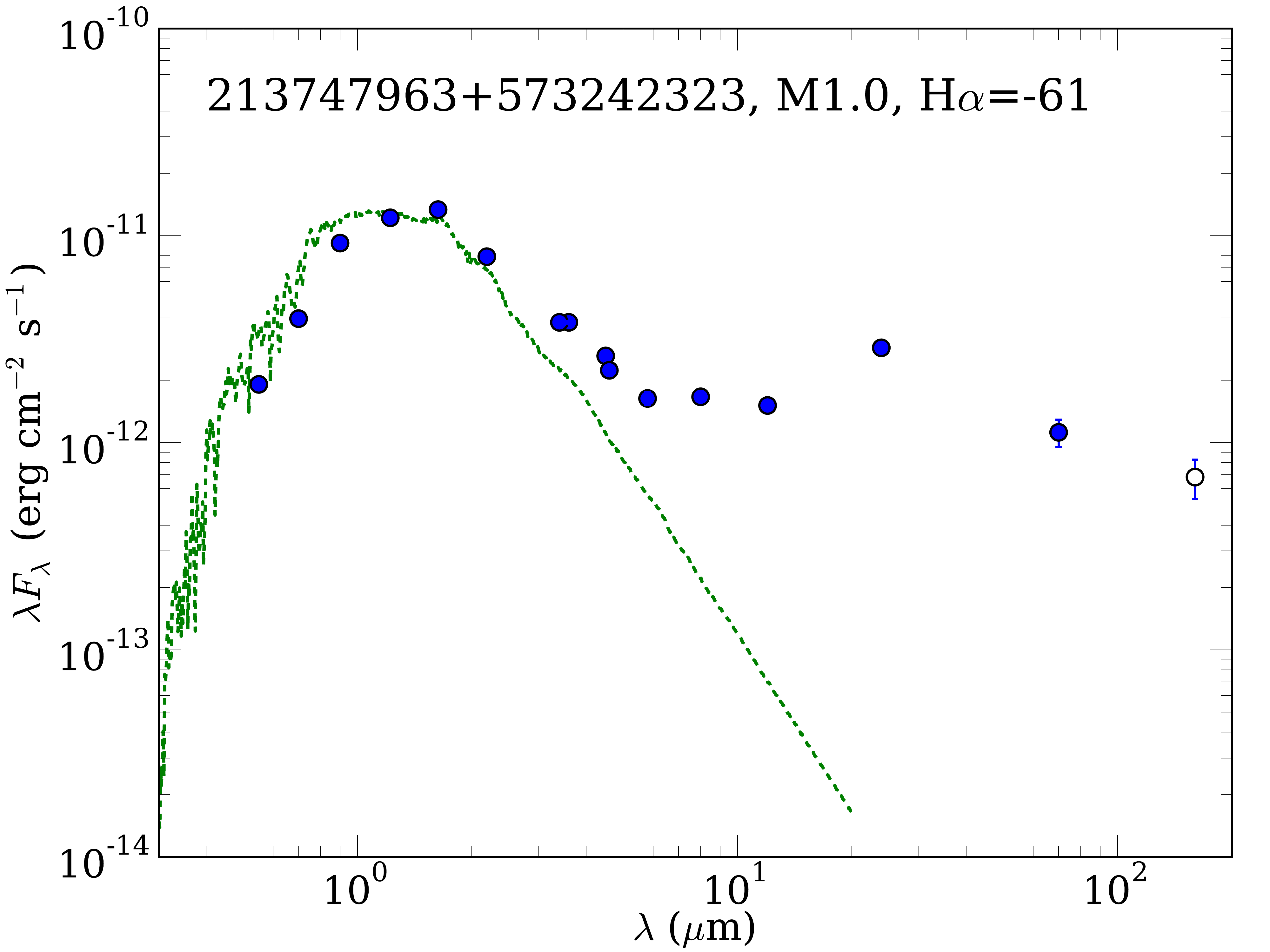} &
\includegraphics[width=0.24\linewidth]{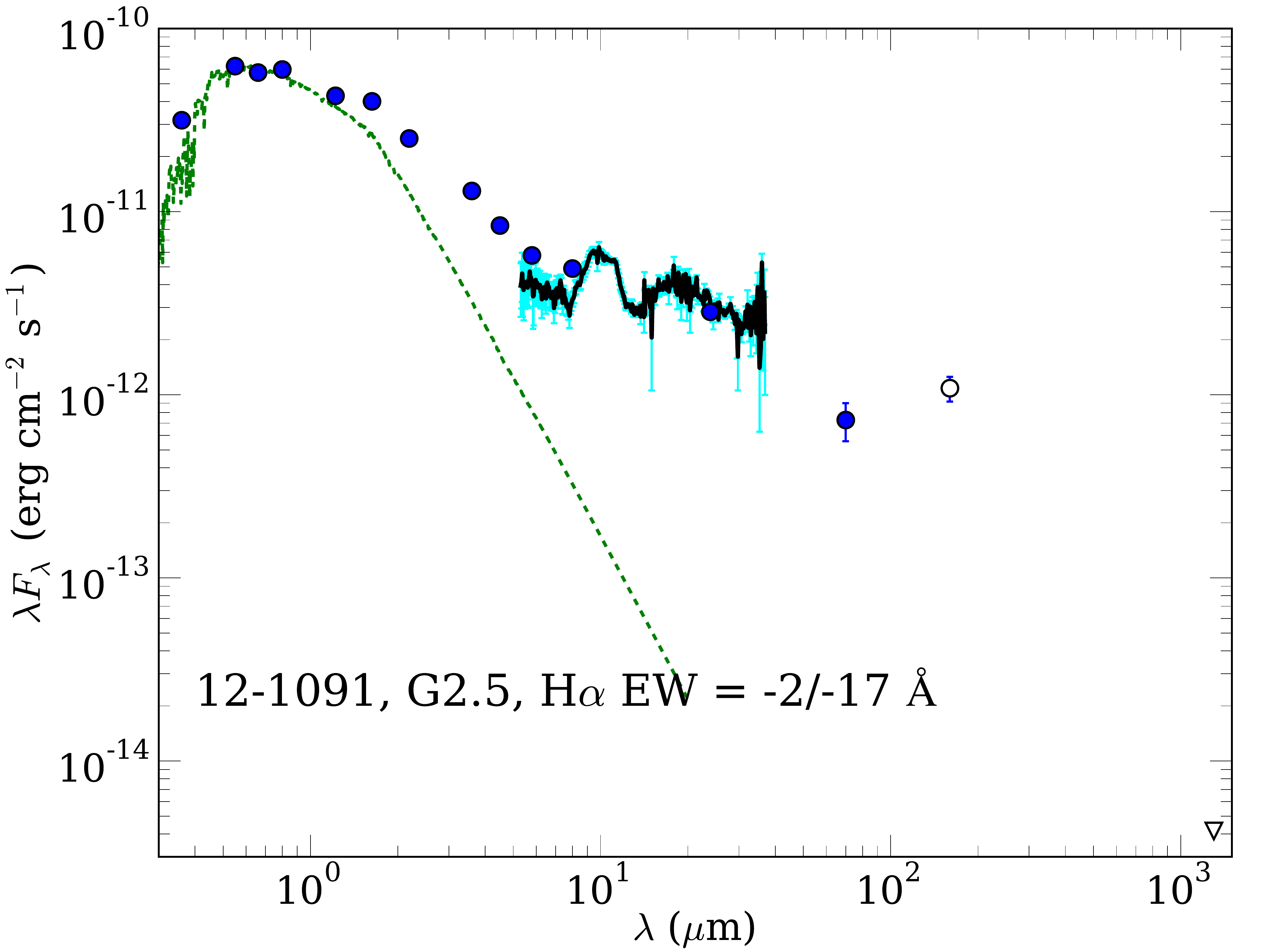} &
\includegraphics[width=0.24\linewidth]{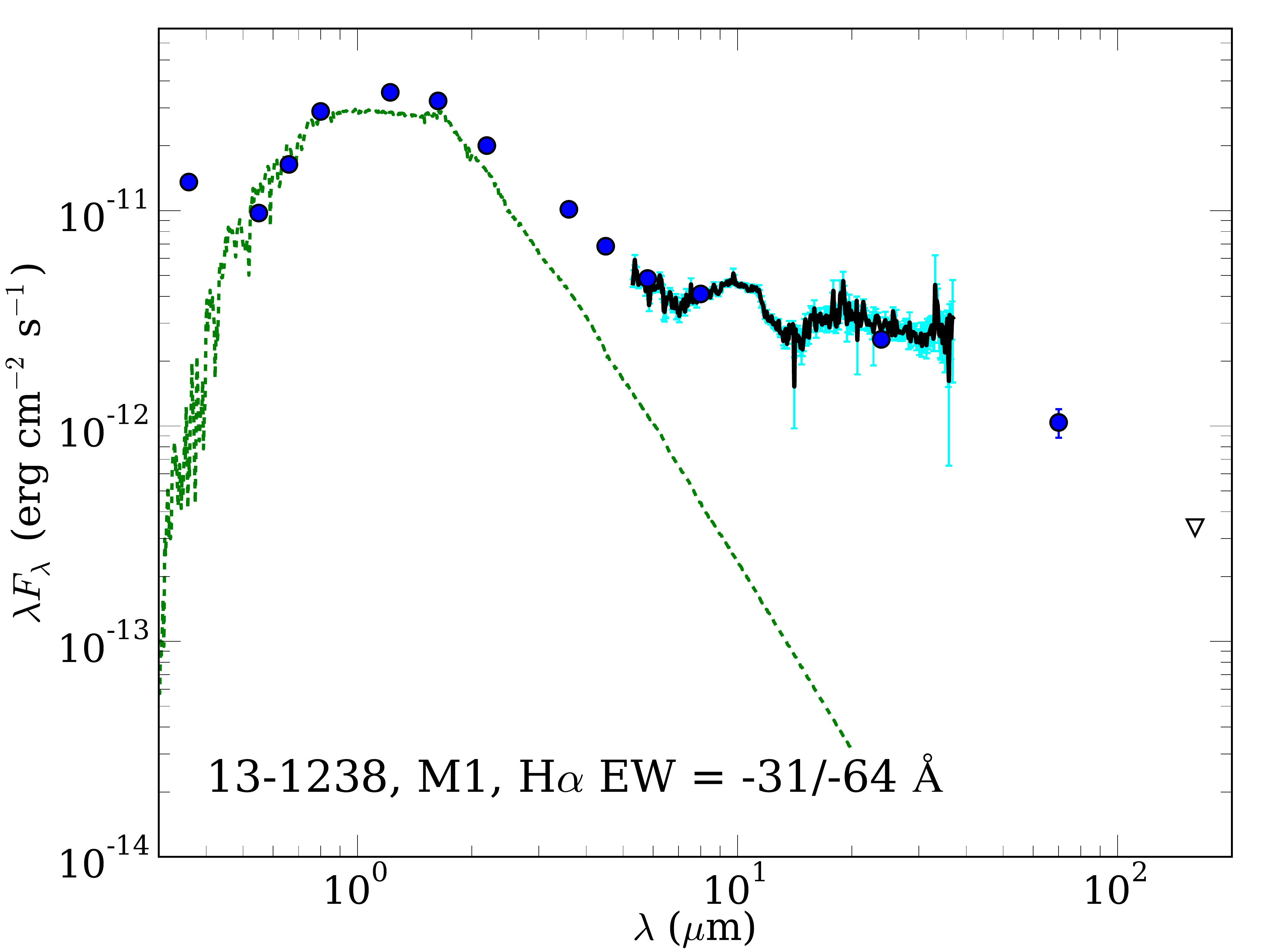} &
\includegraphics[width=0.24\linewidth]{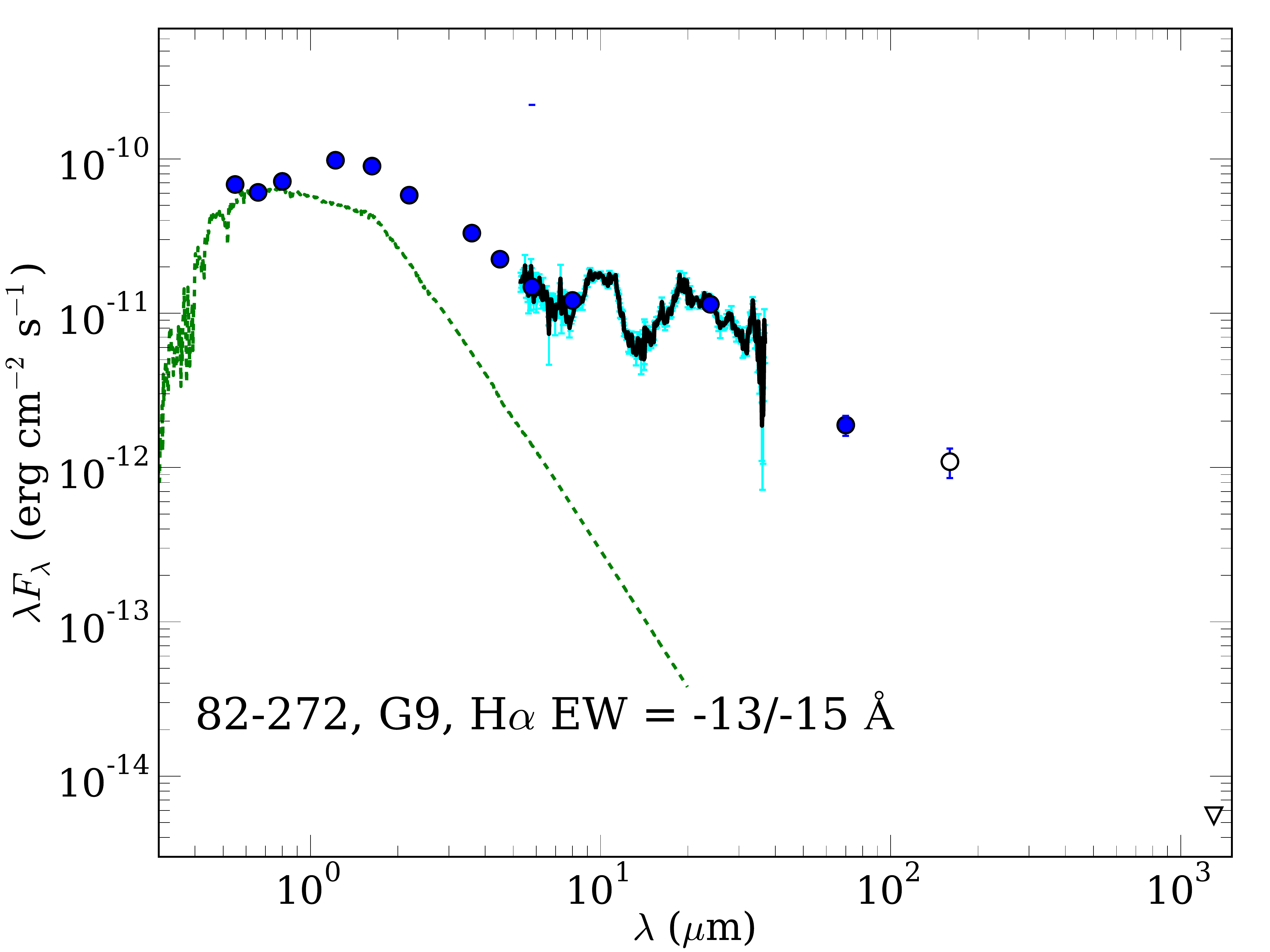} \\
\includegraphics[width=0.24\linewidth]{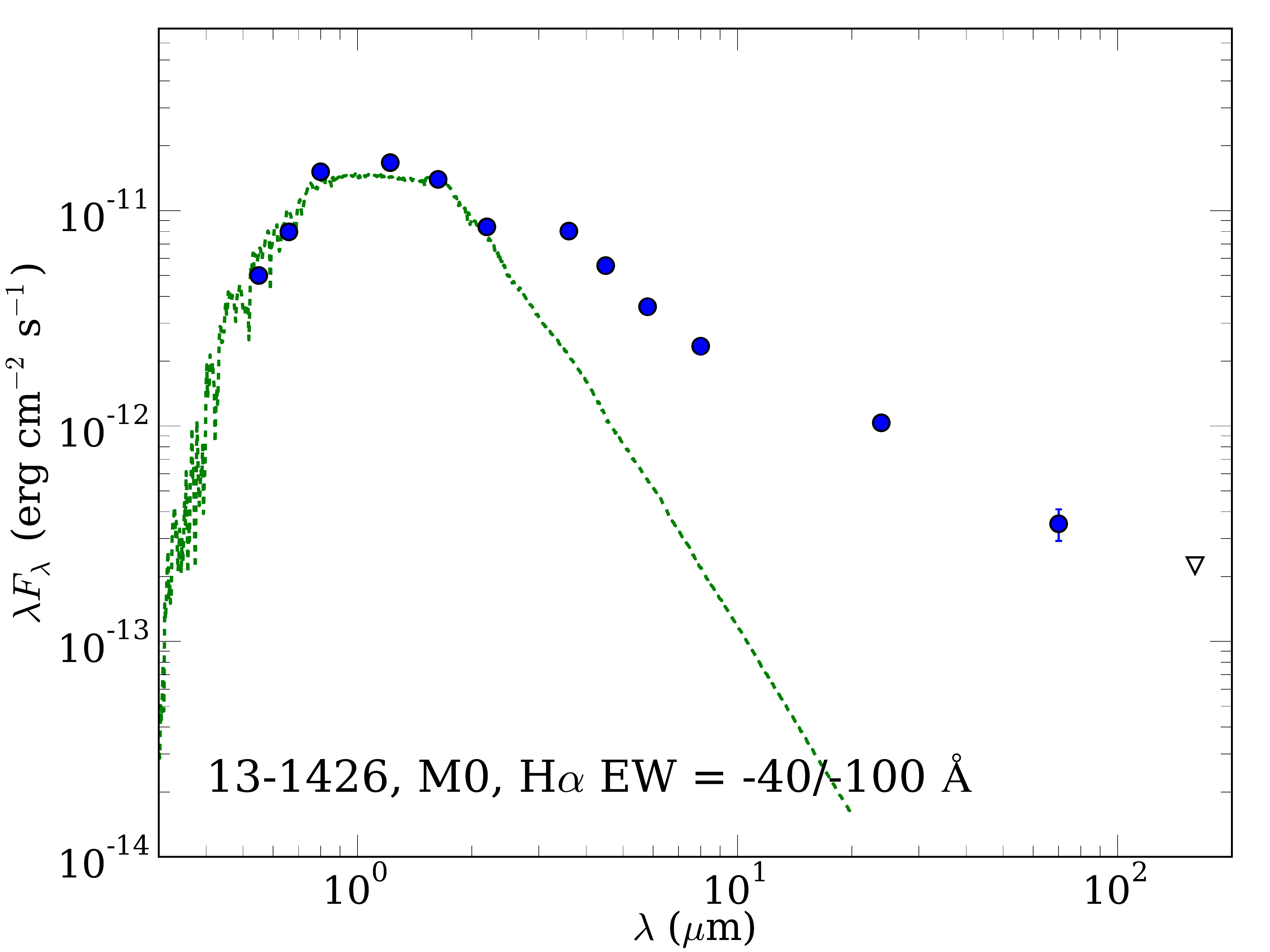} &
\includegraphics[width=0.24\linewidth]{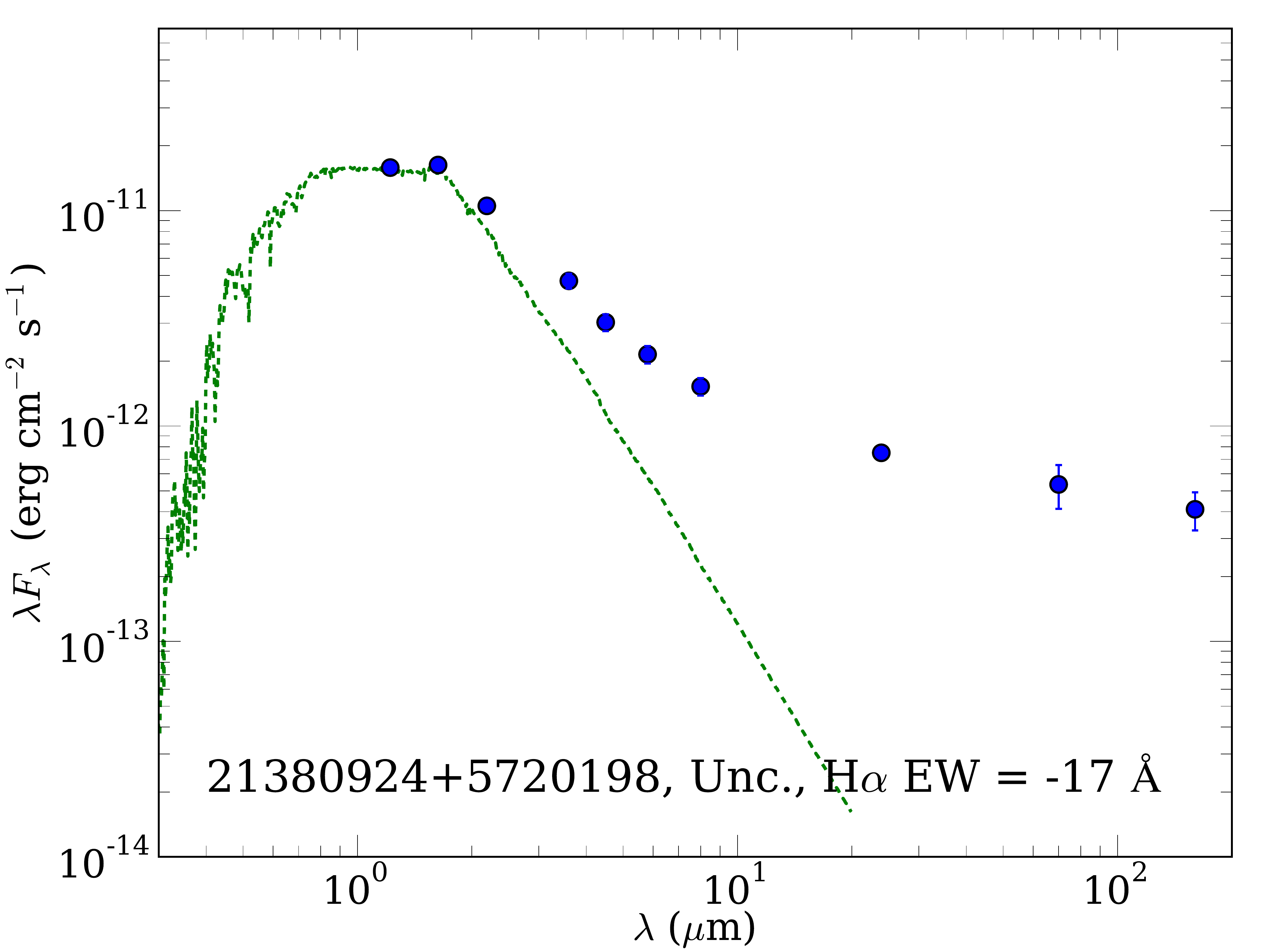} &
\includegraphics[width=0.24\linewidth]{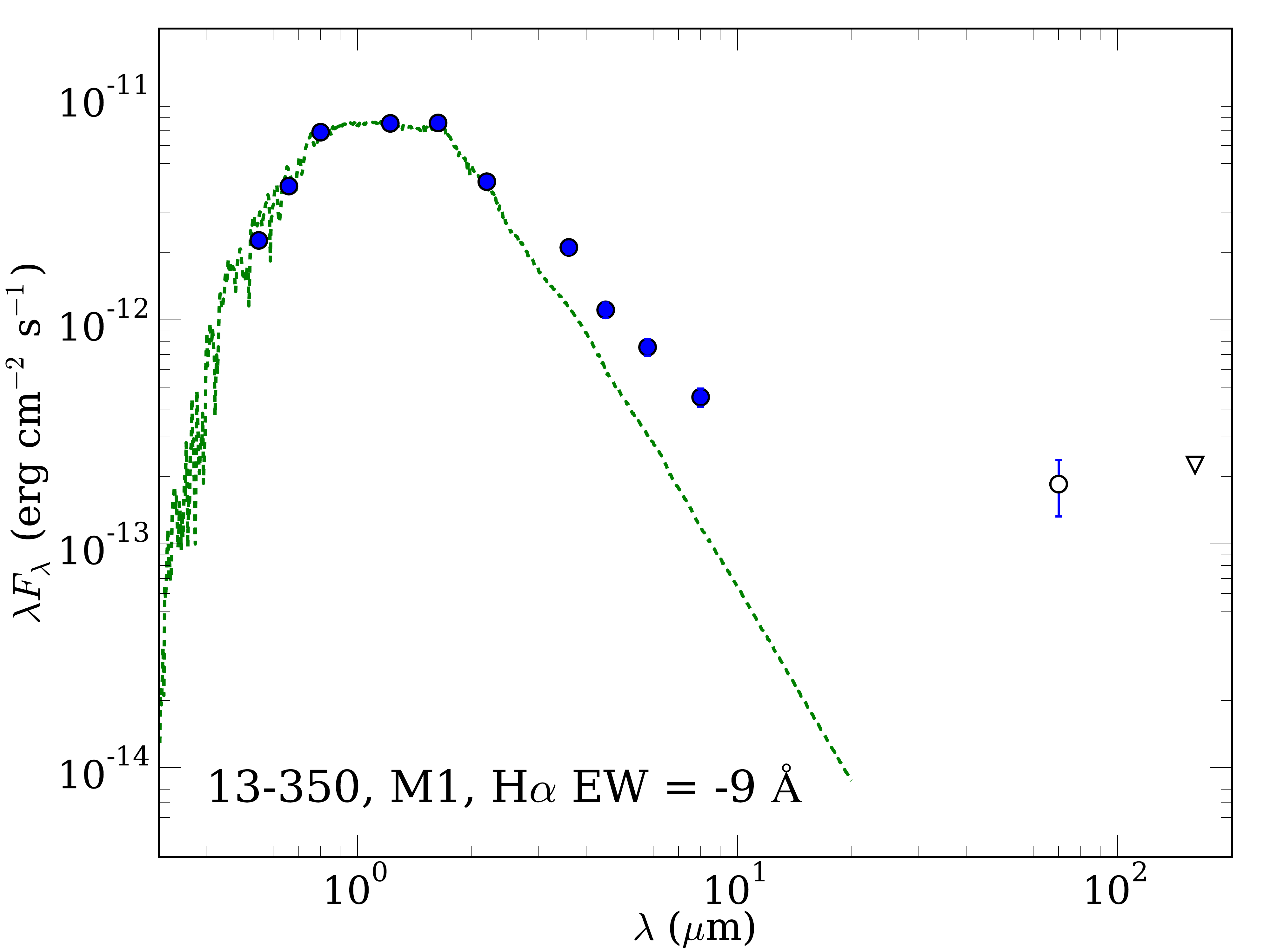} &
\includegraphics[width=0.24\linewidth]{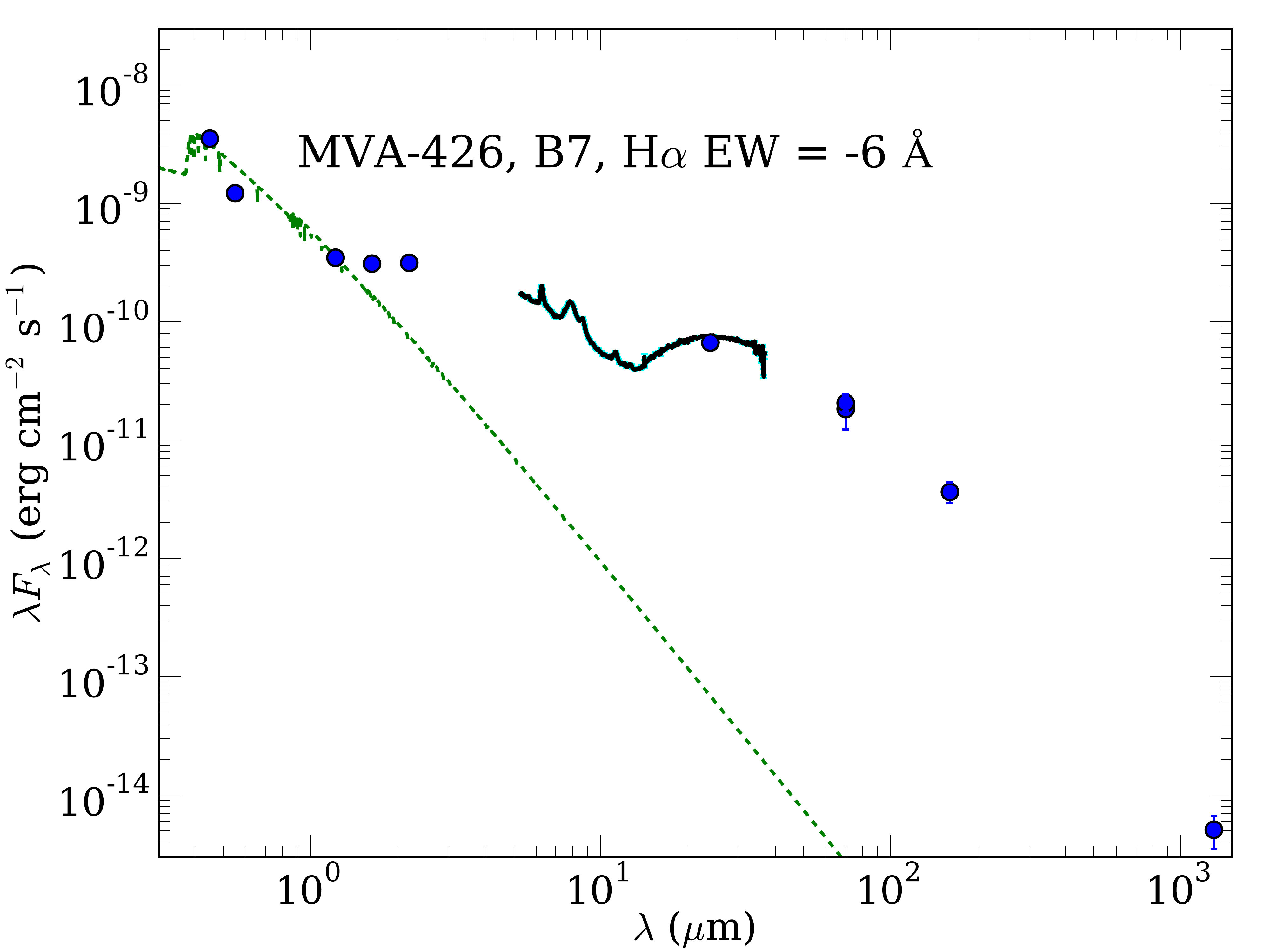} \\
\includegraphics[width=0.24\linewidth]{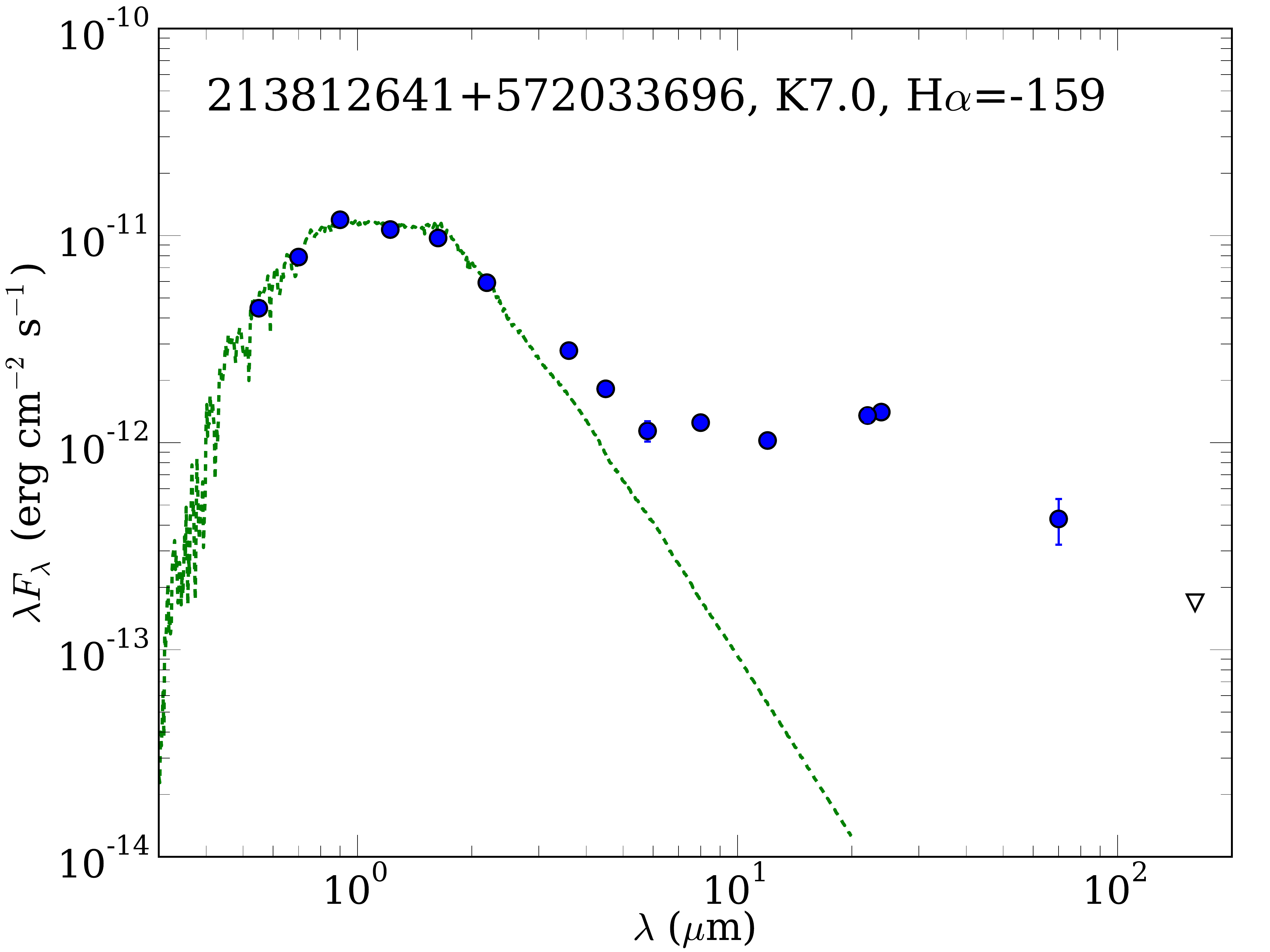} &
\includegraphics[width=0.24\linewidth]{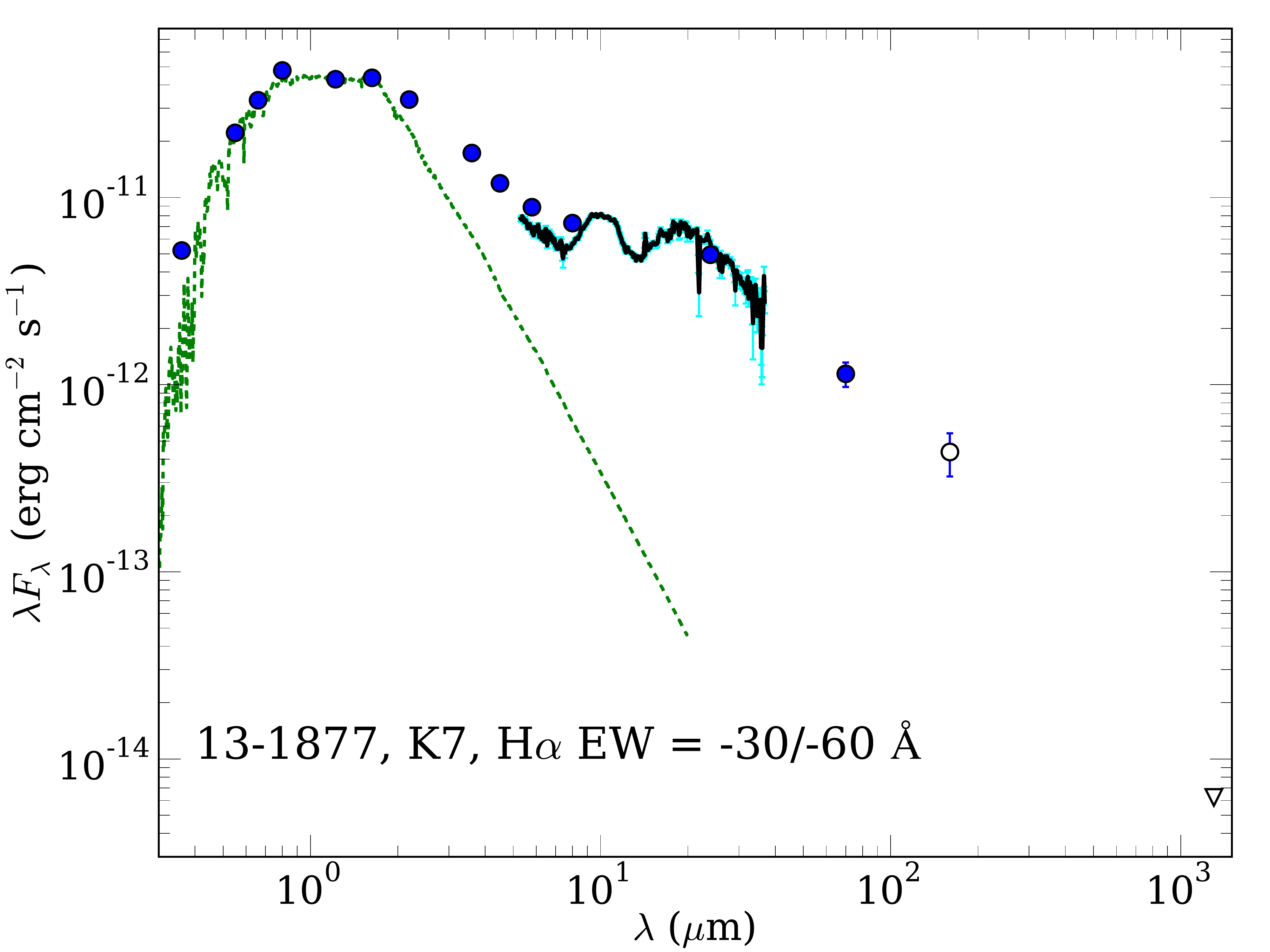} &
\includegraphics[width=0.24\linewidth]{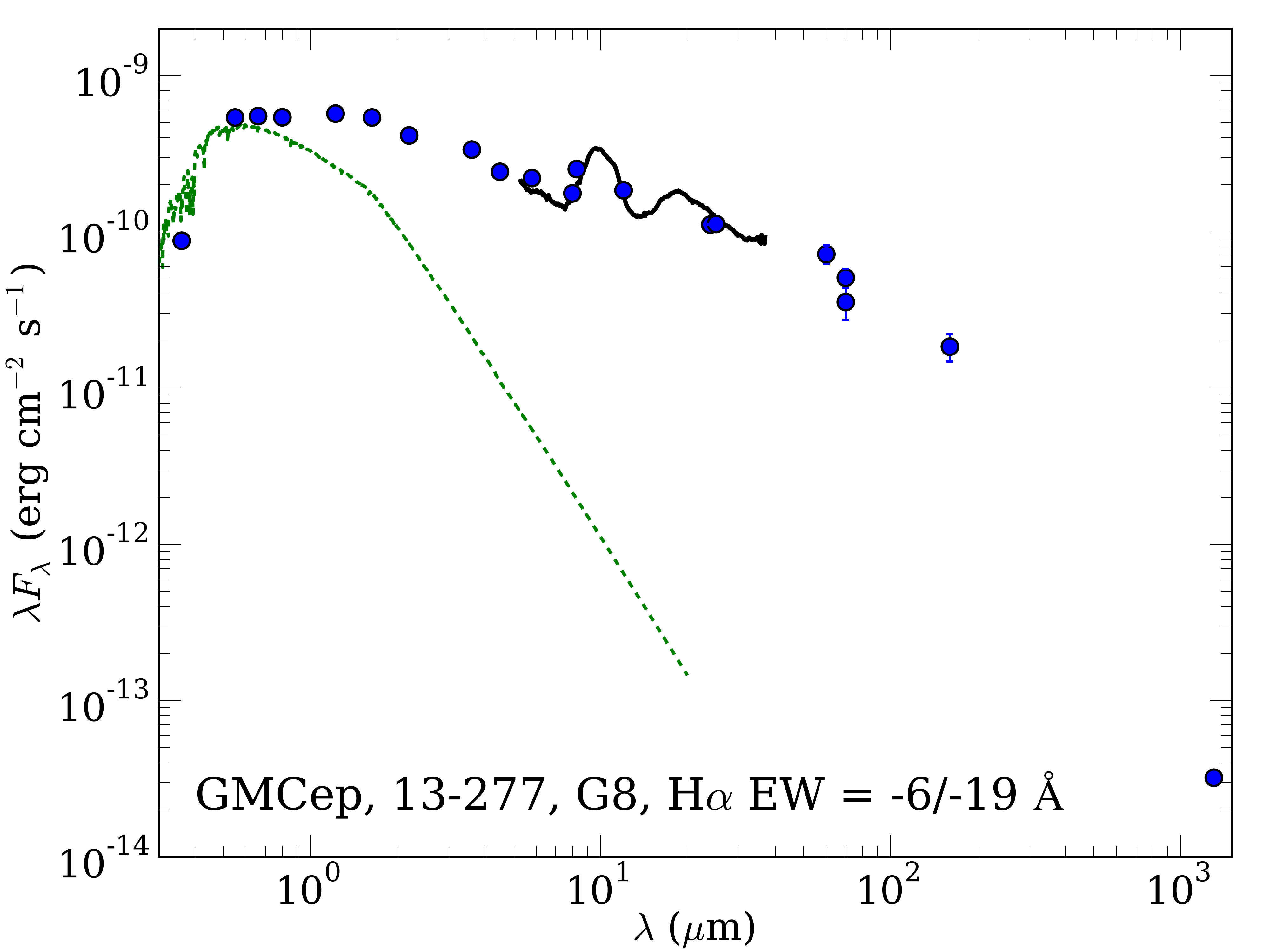} &
\includegraphics[width=0.24\linewidth]{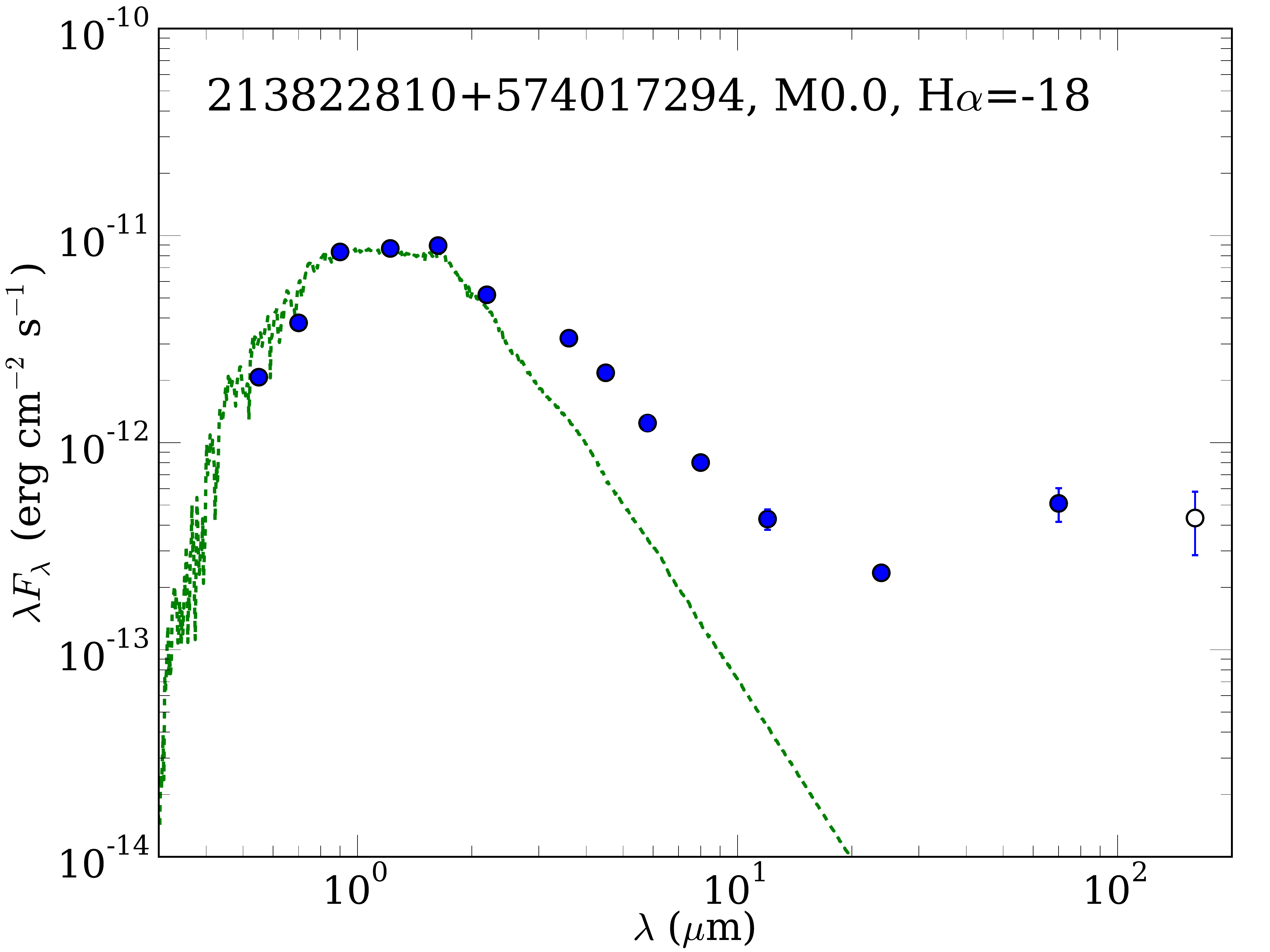} \\
\includegraphics[width=0.24\linewidth]{213825831+574207487.pdf} &
\includegraphics[width=0.24\linewidth]{13_236.pdf} &
\includegraphics[width=0.24\linewidth]{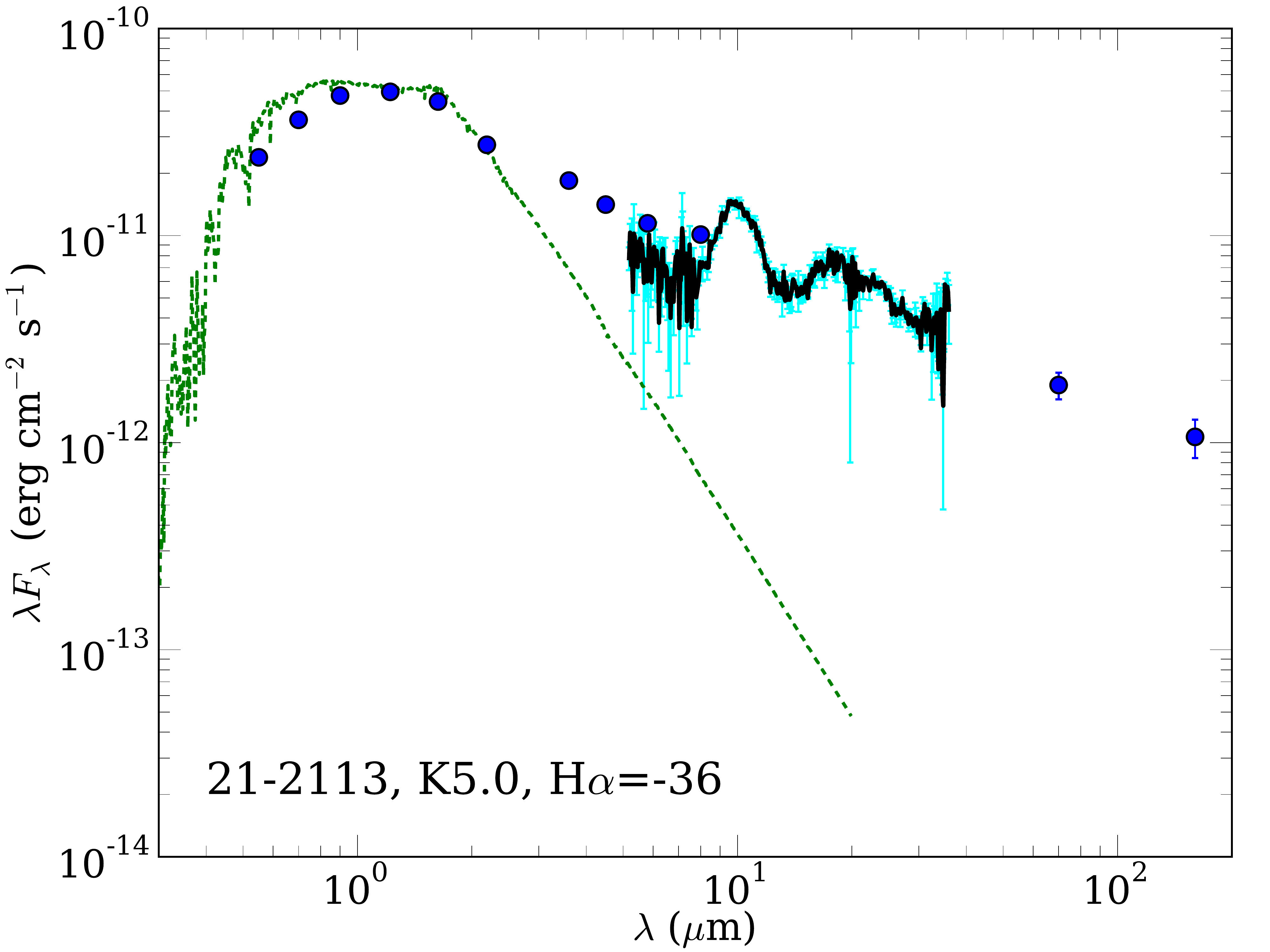} &
\includegraphics[width=0.24\linewidth]{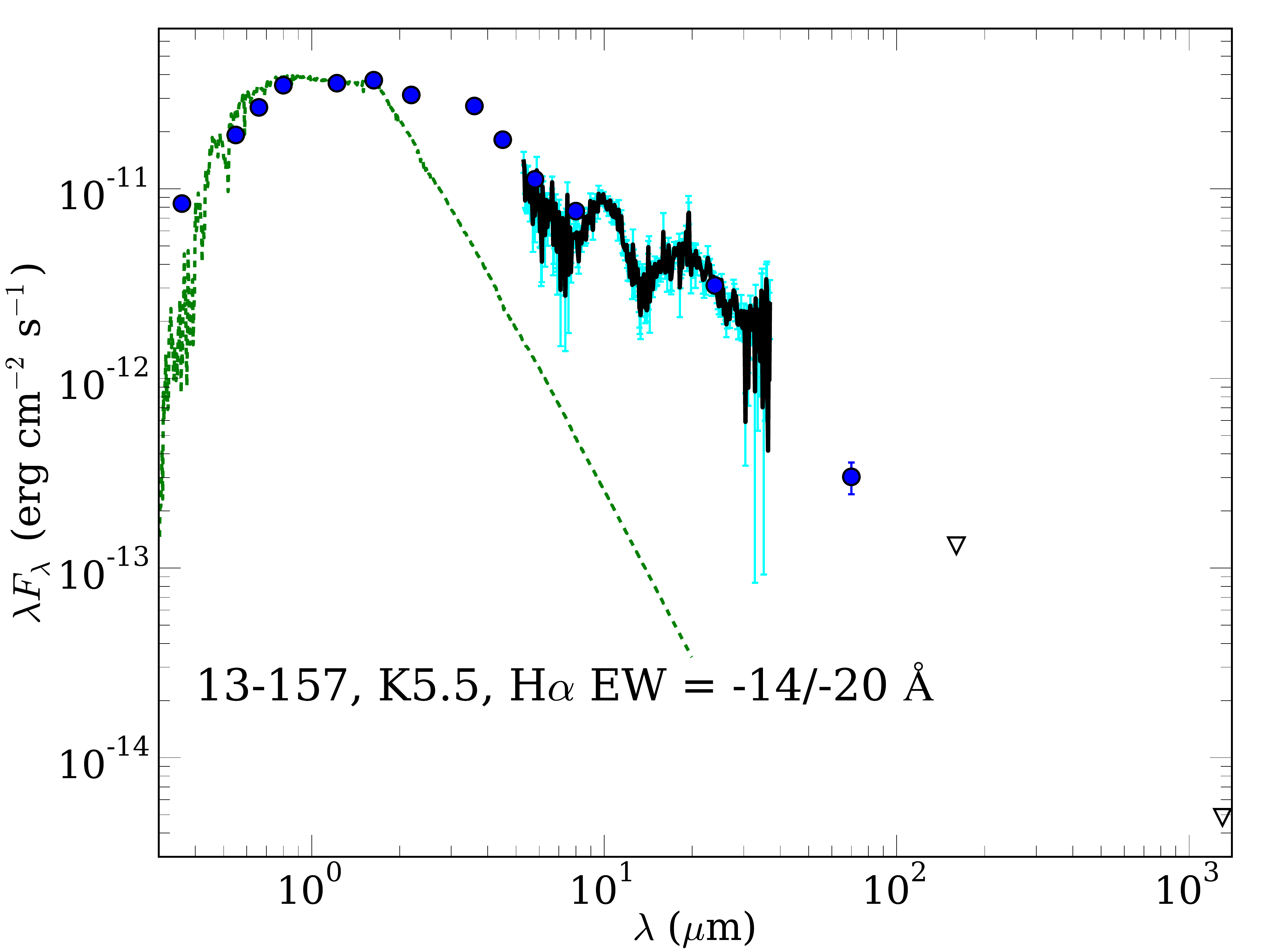} \\
\includegraphics[width=0.24\linewidth]{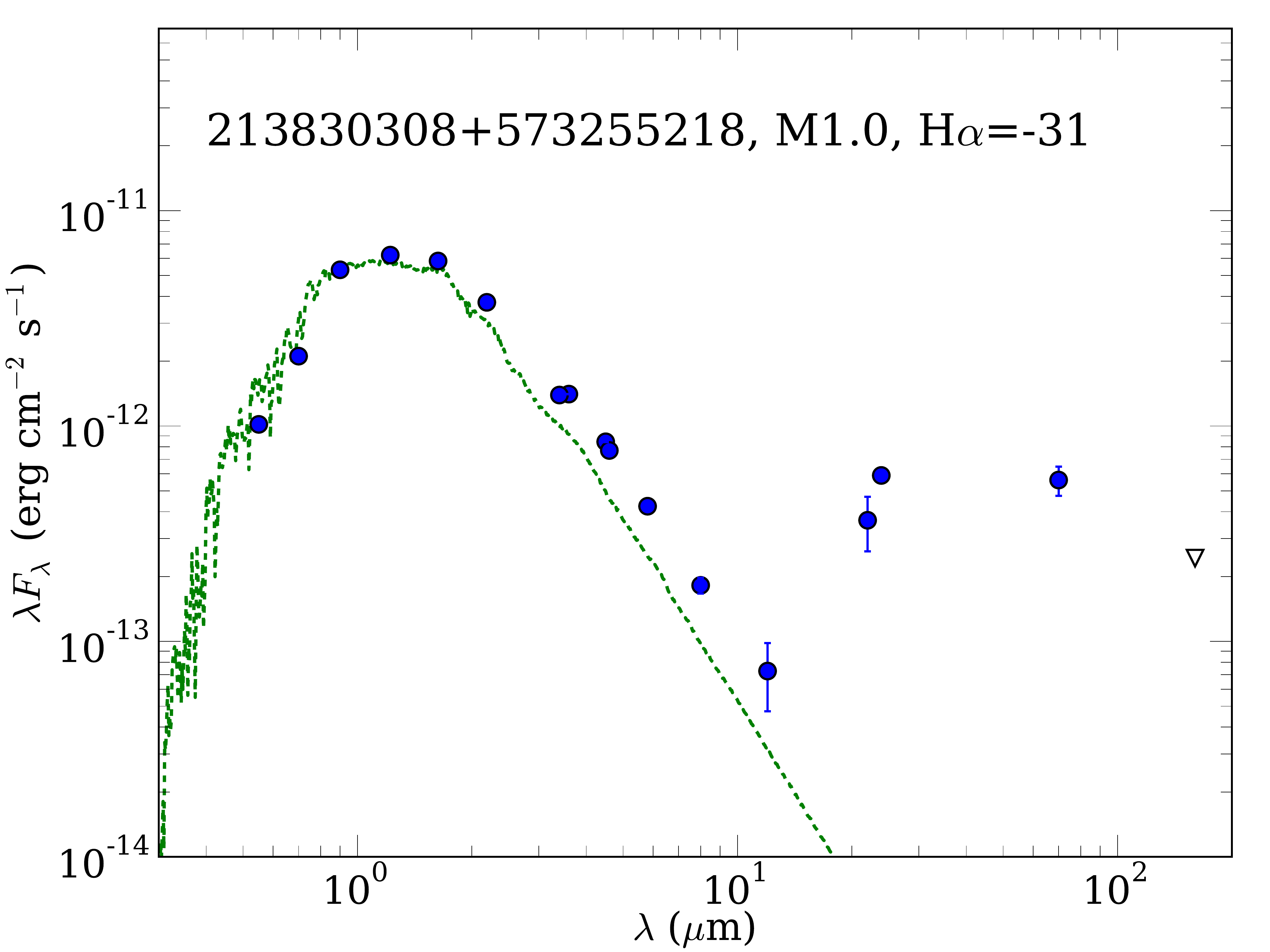} &
\includegraphics[width=0.24\linewidth]{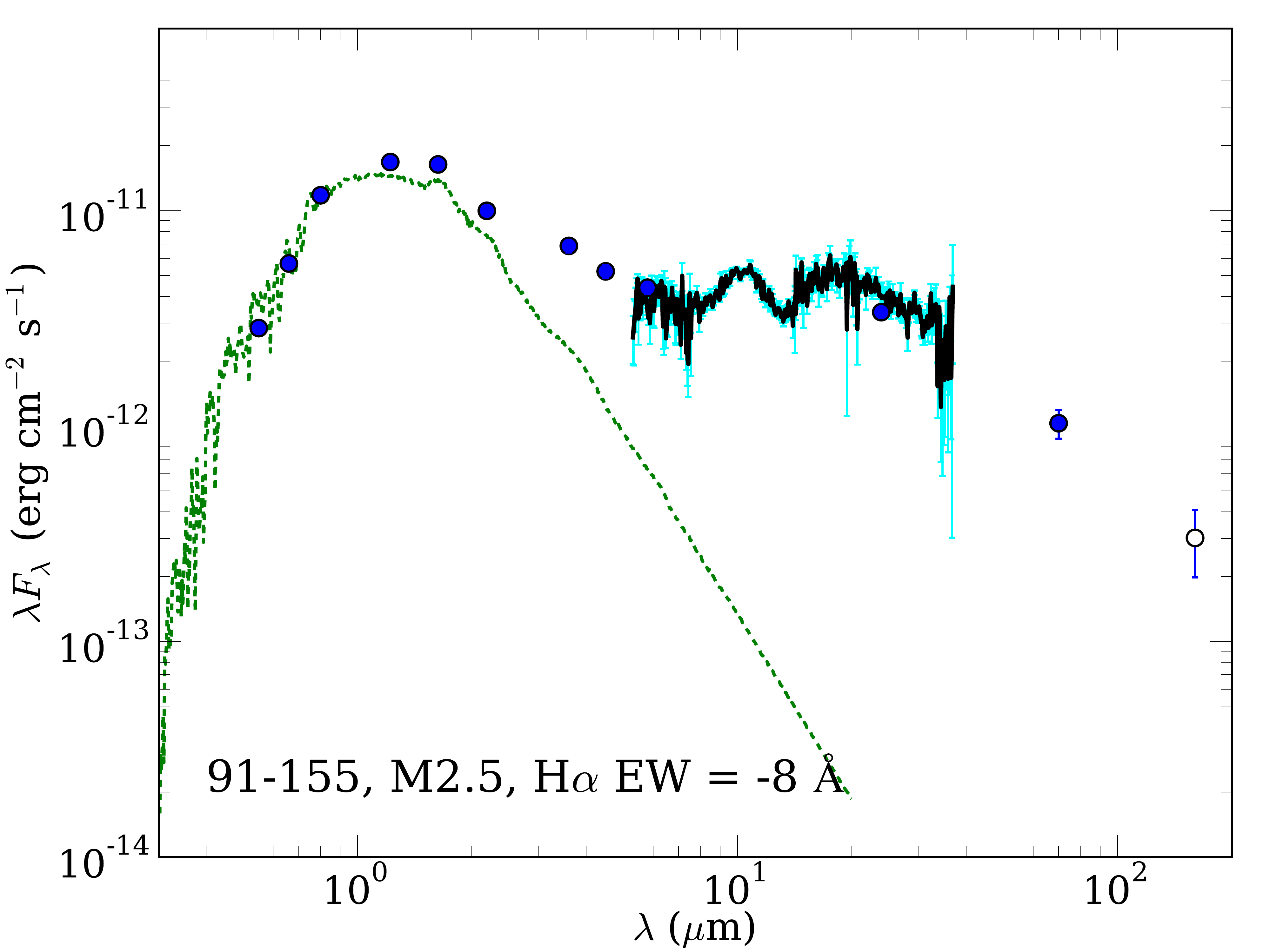} &
\includegraphics[width=0.24\linewidth]{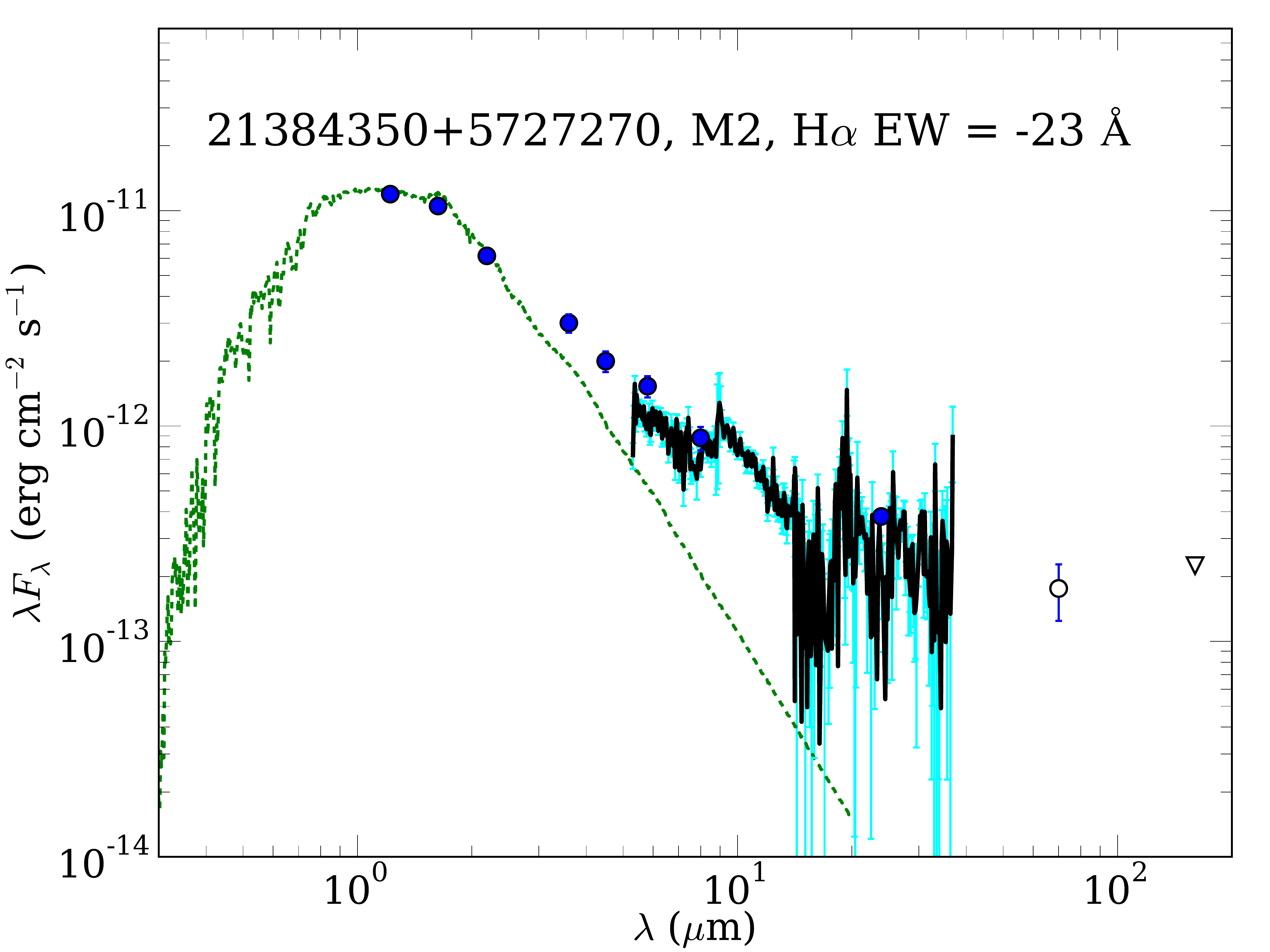} &
\includegraphics[width=0.24\linewidth]{54_1547.pdf} \\
\end{tabular}
\caption{SEDs of the objects detected with Herschel (continuation). All objects belong to Tr\,37. 
Symbols as in Figure \ref{seds1-fig}. 
\label{seds2-fig}}
\end{figure*}

\begin{figure*}
\centering
\begin{tabular}{cccc}
\includegraphics[width=0.24\linewidth]{91_506.pdf} &
\includegraphics[width=0.24\linewidth]{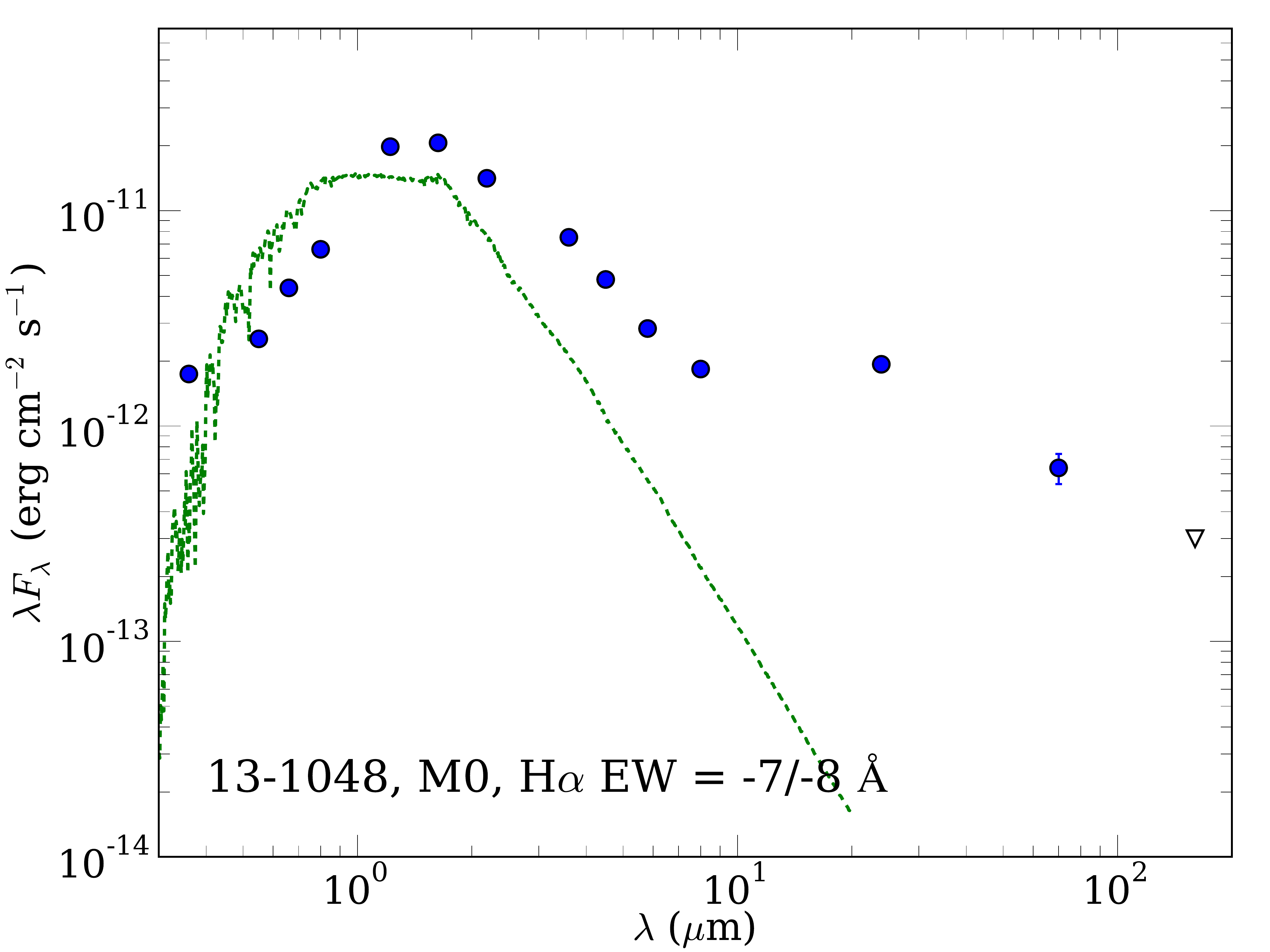} &
\includegraphics[width=0.24\linewidth]{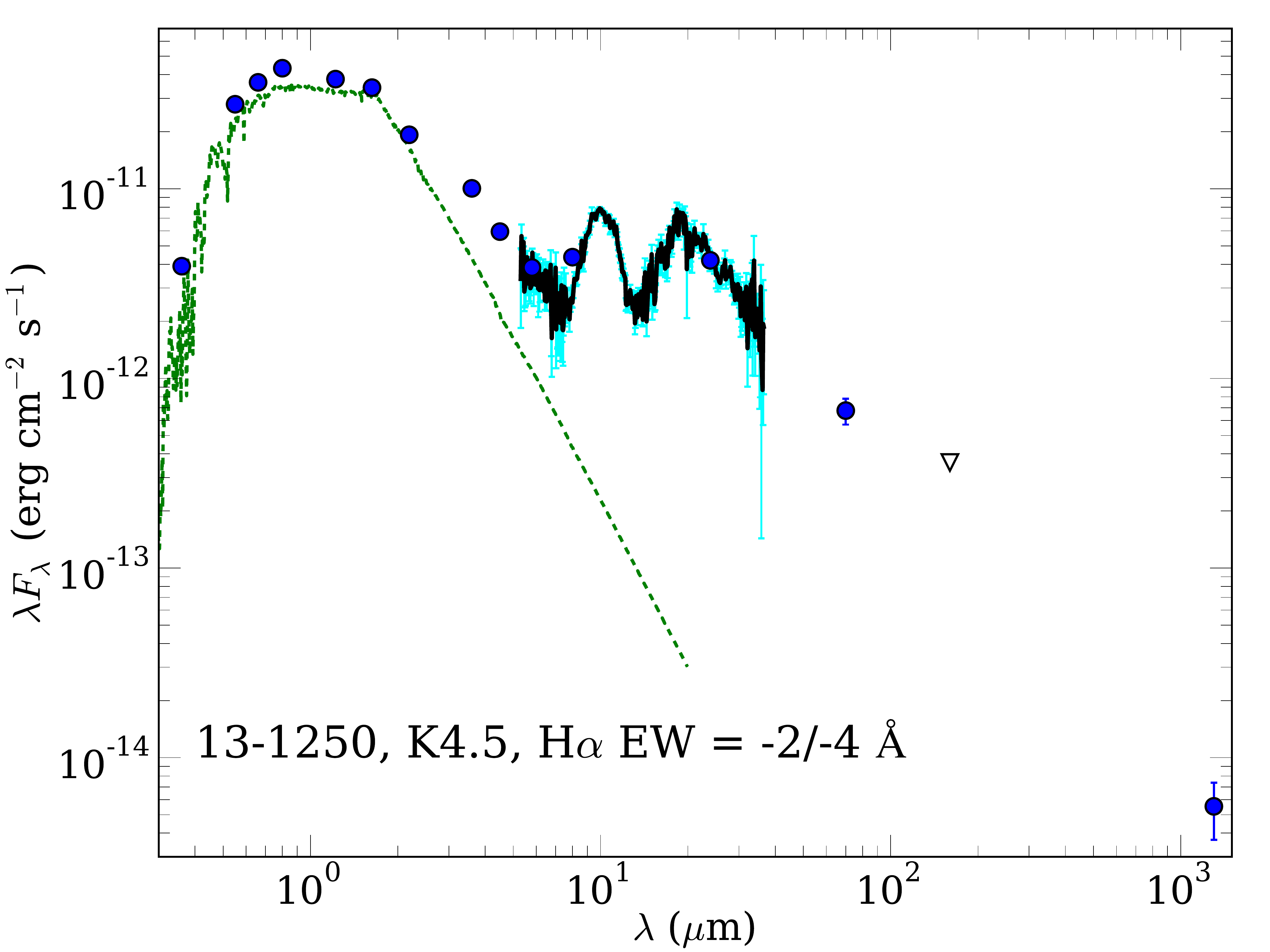} &
\includegraphics[width=0.24\linewidth]{21392541+5733202.pdf} \\
\includegraphics[width=0.24\linewidth]{21393104+5747140.pdf} &
\includegraphics[width=0.24\linewidth]{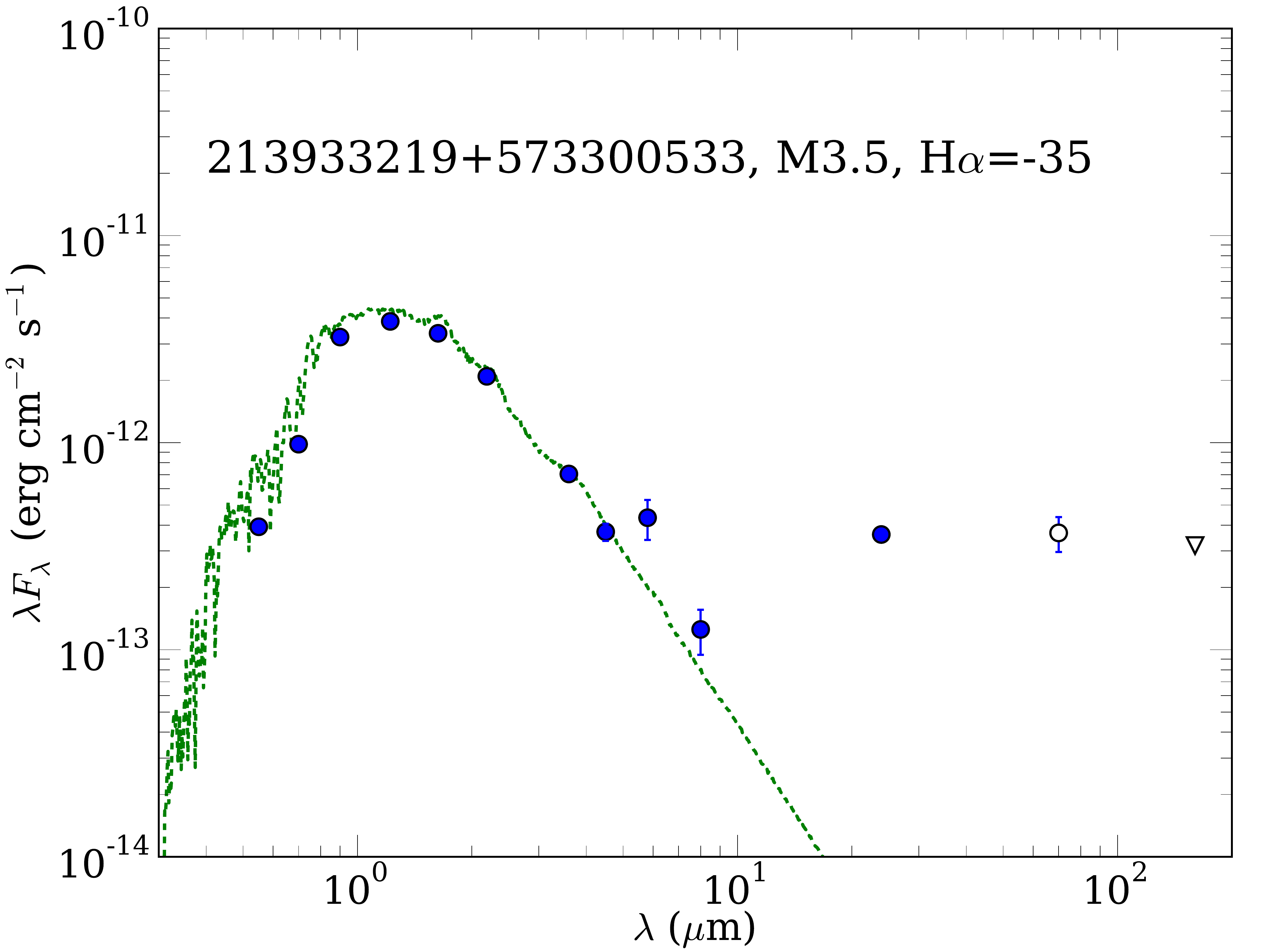} &
\includegraphics[width=0.24\linewidth]{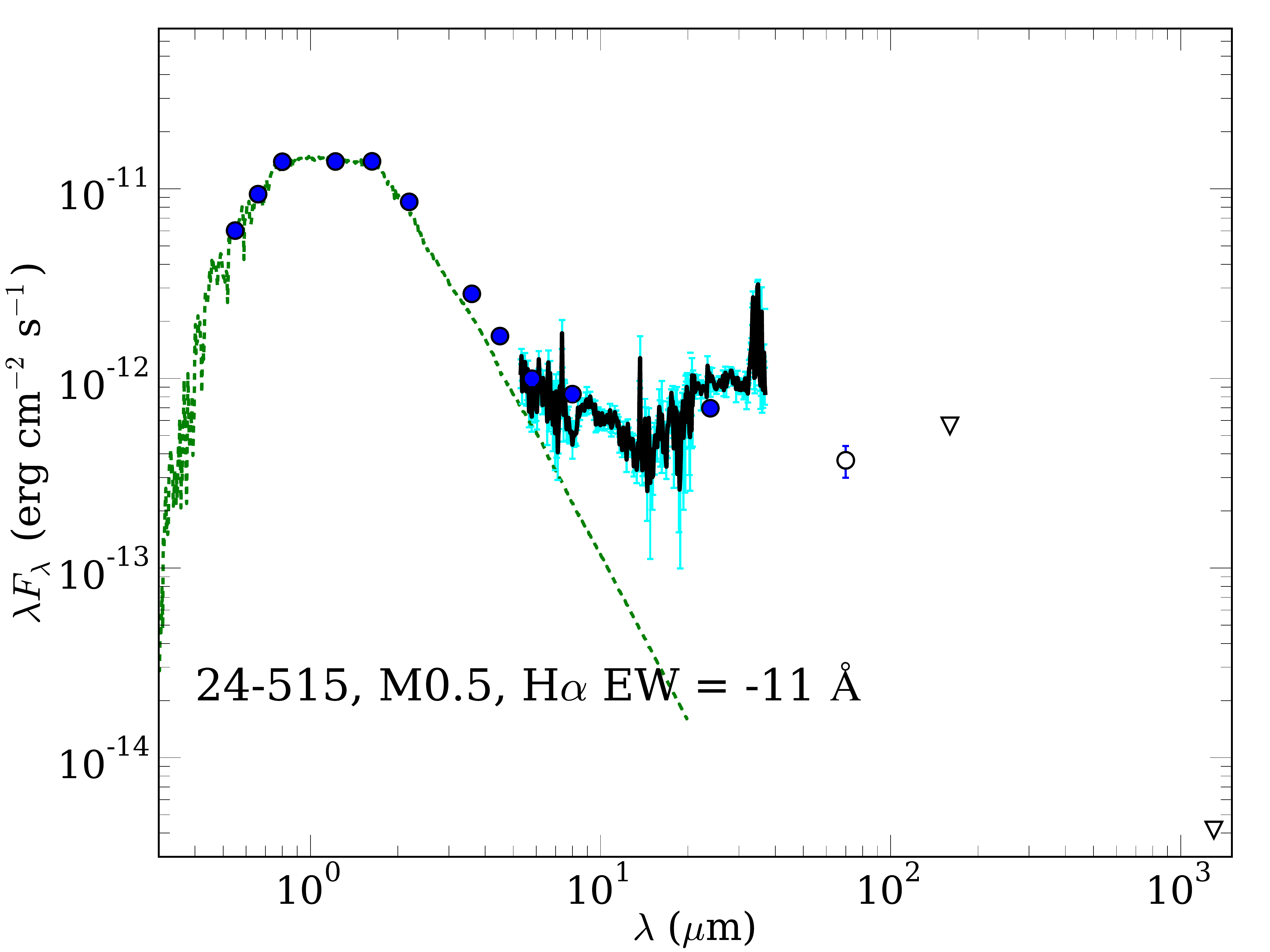} &
\includegraphics[width=0.24\linewidth]{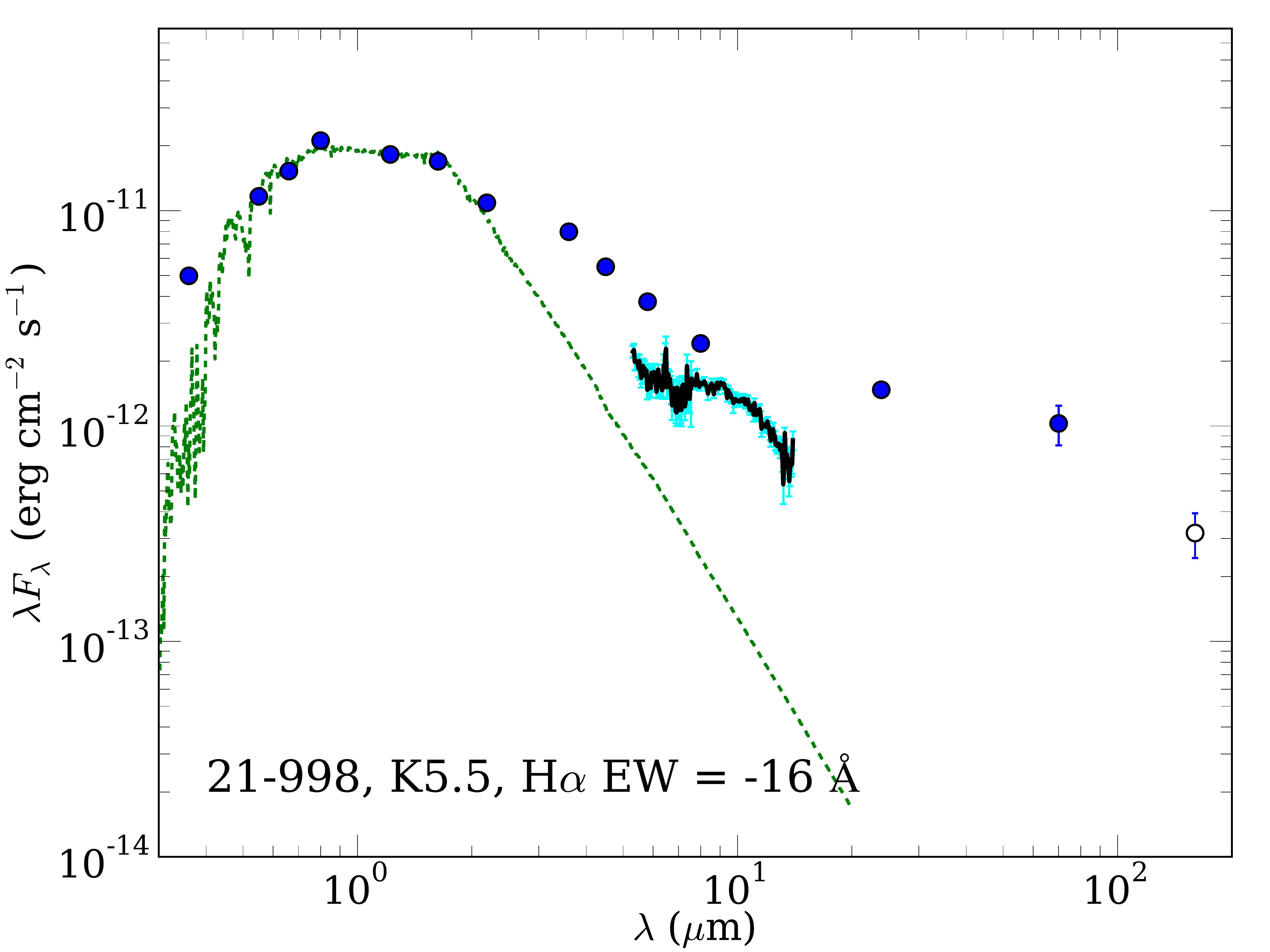} \\
\includegraphics[width=0.24\linewidth]{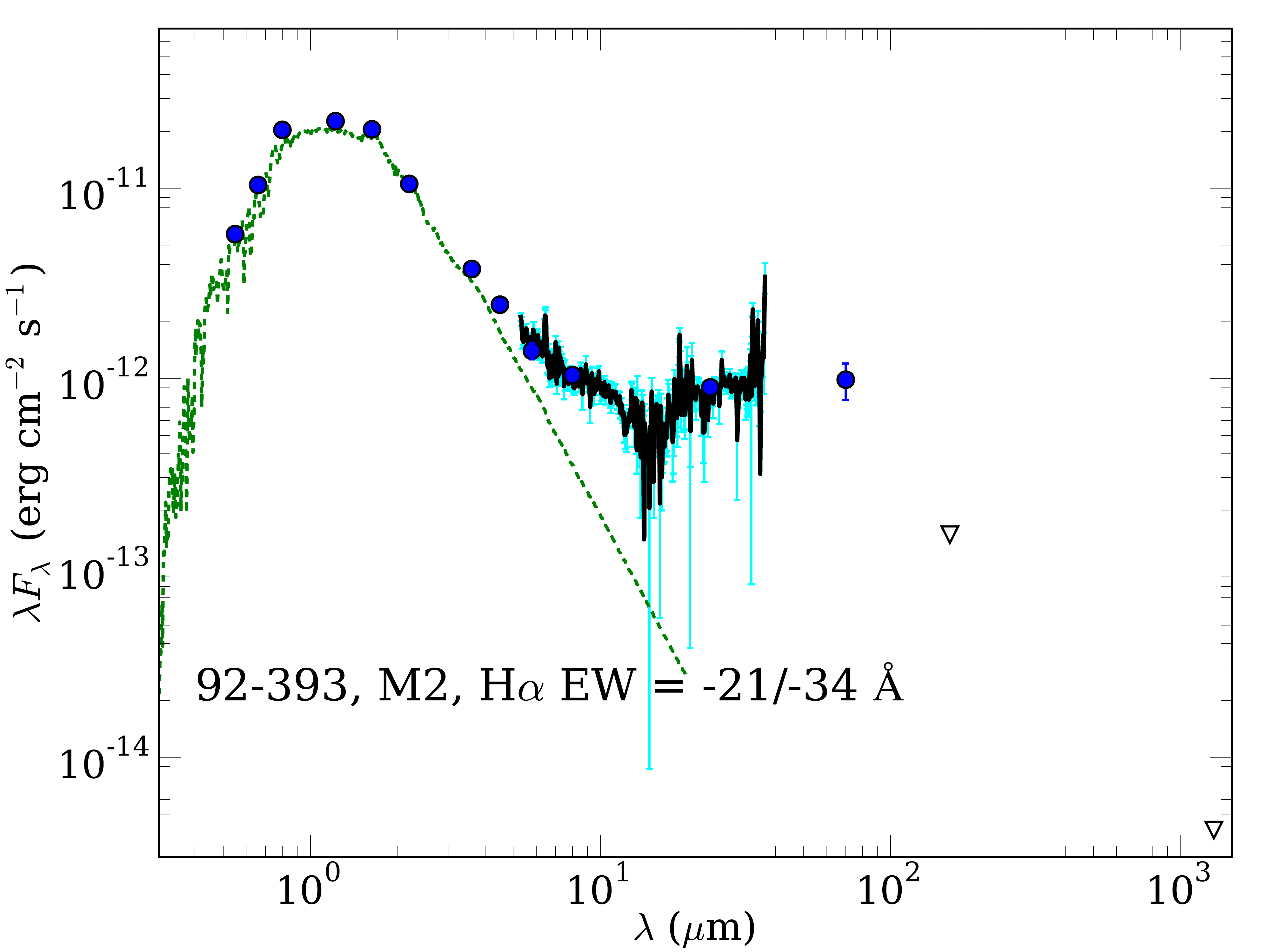} &
\includegraphics[width=0.24\linewidth]{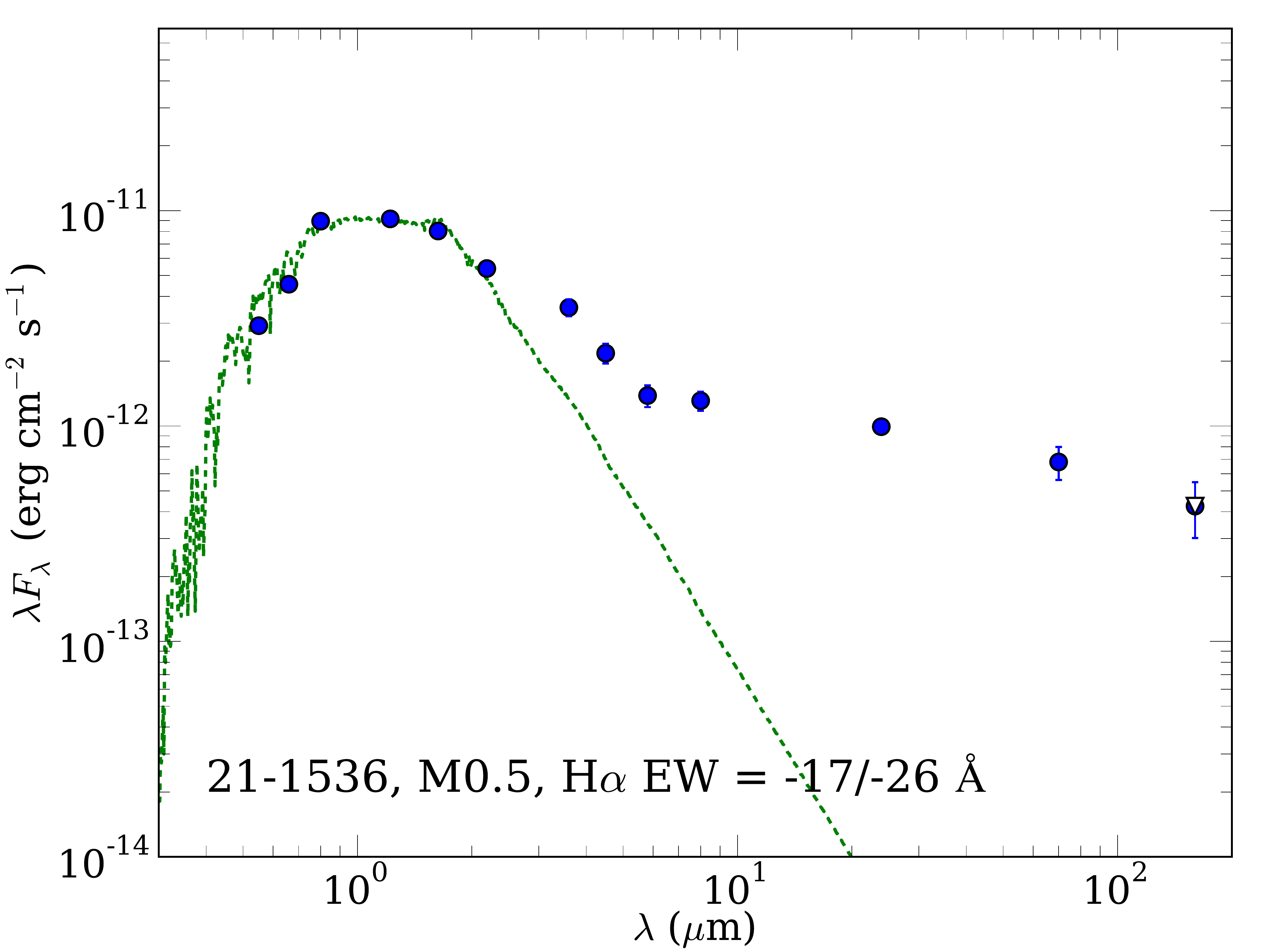} &
\includegraphics[width=0.24\linewidth]{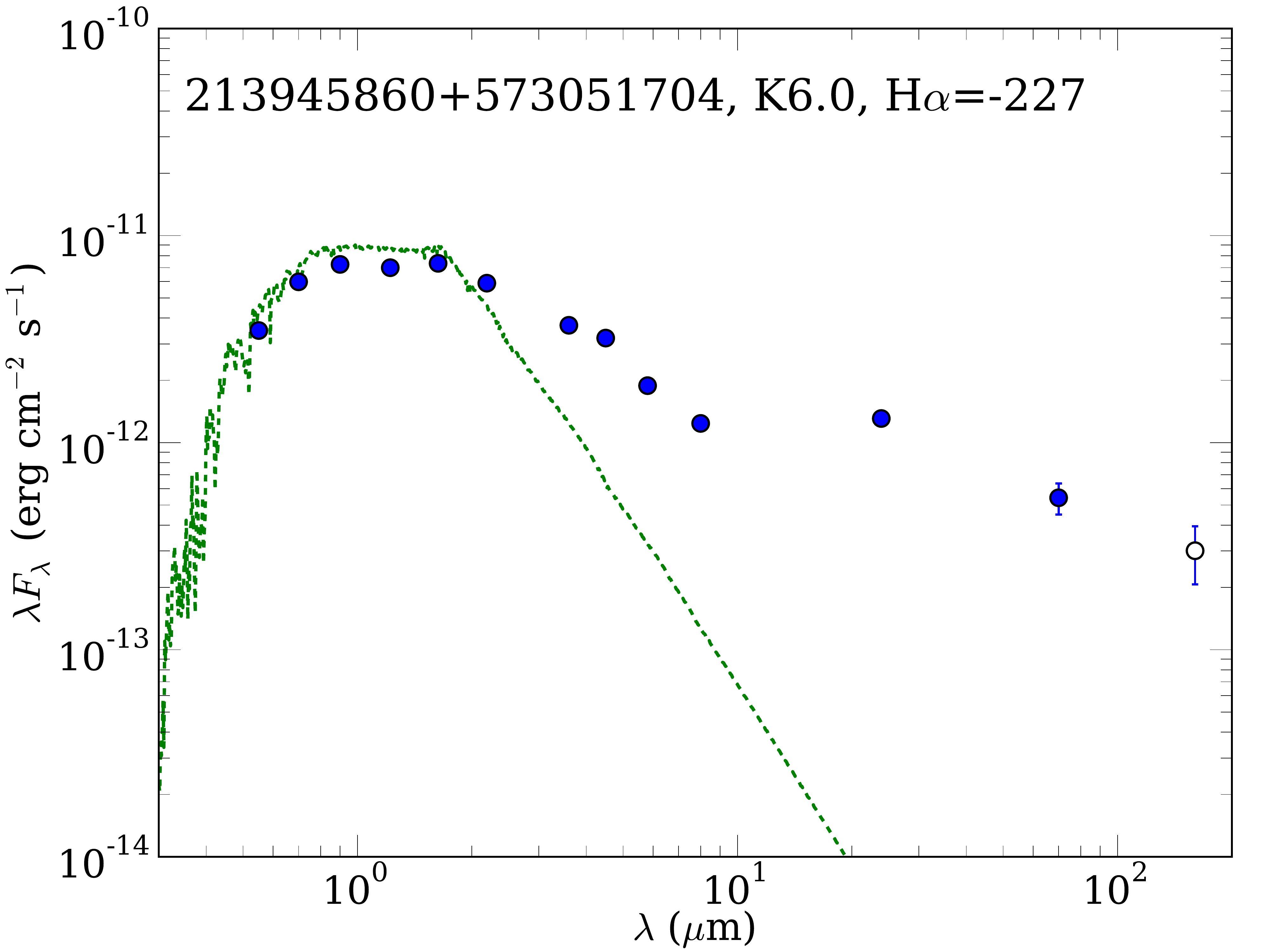} &
\includegraphics[width=0.24\linewidth]{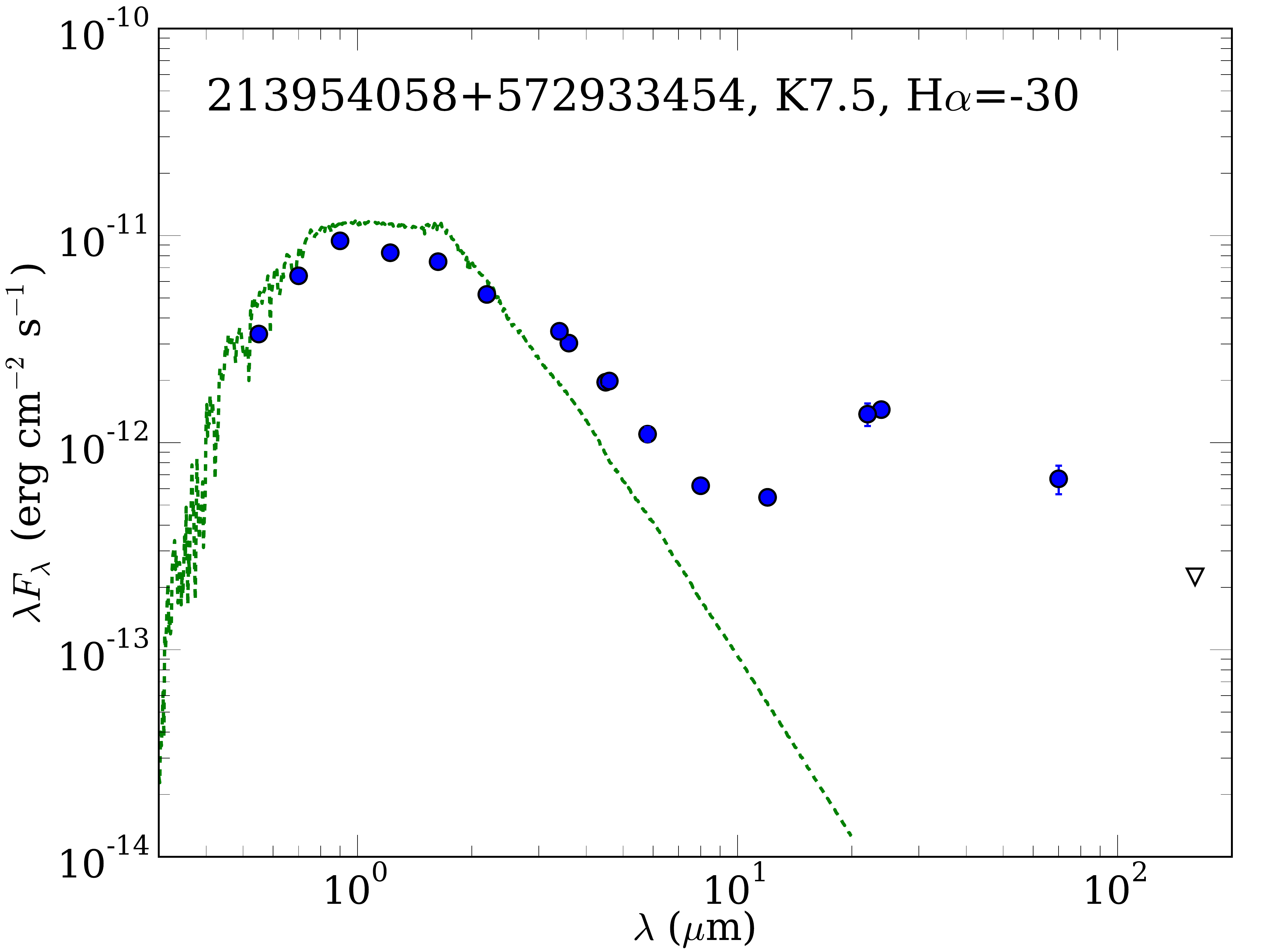} \\
\includegraphics[width=0.24\linewidth]{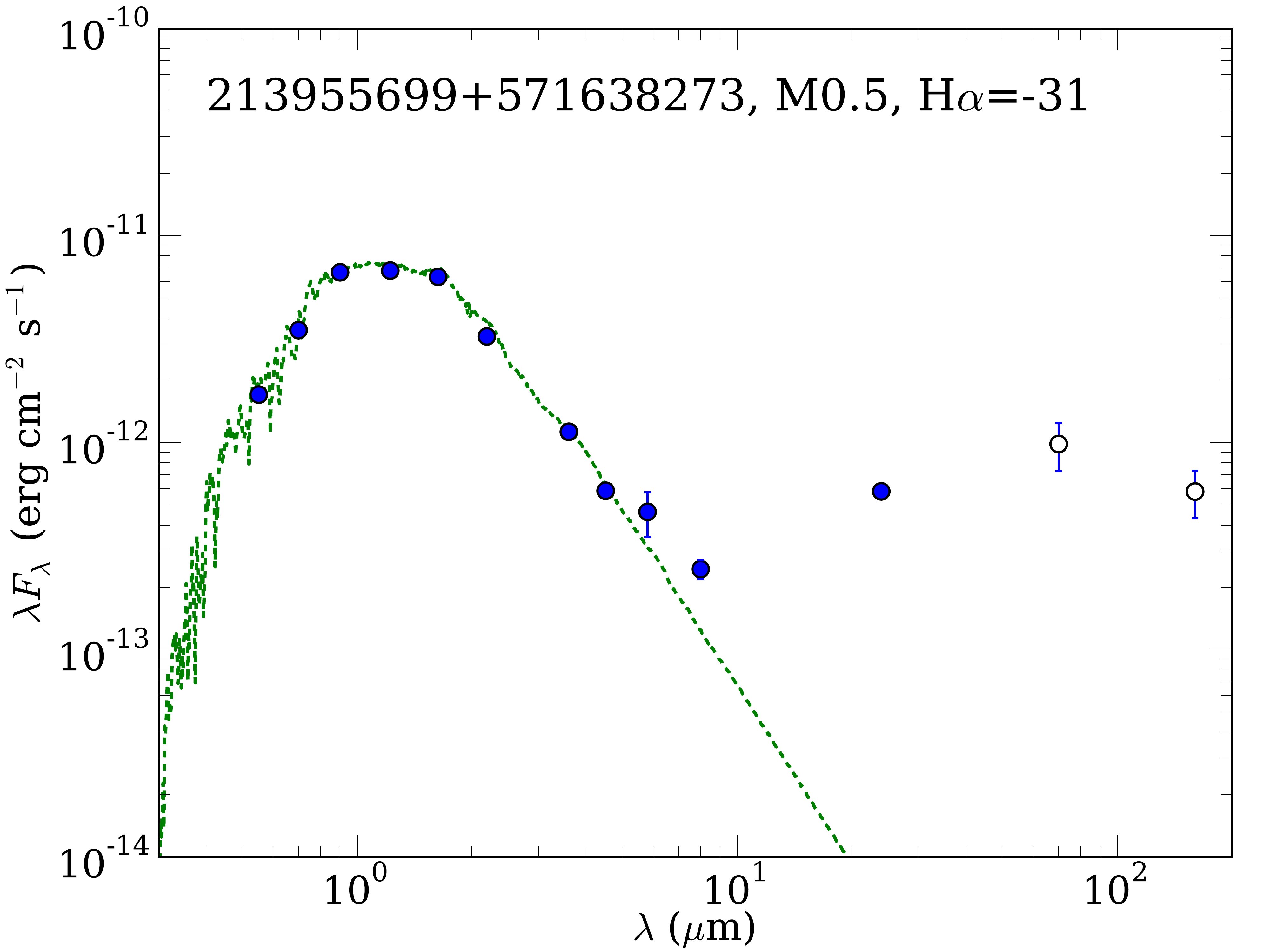} &
\includegraphics[width=0.24\linewidth]{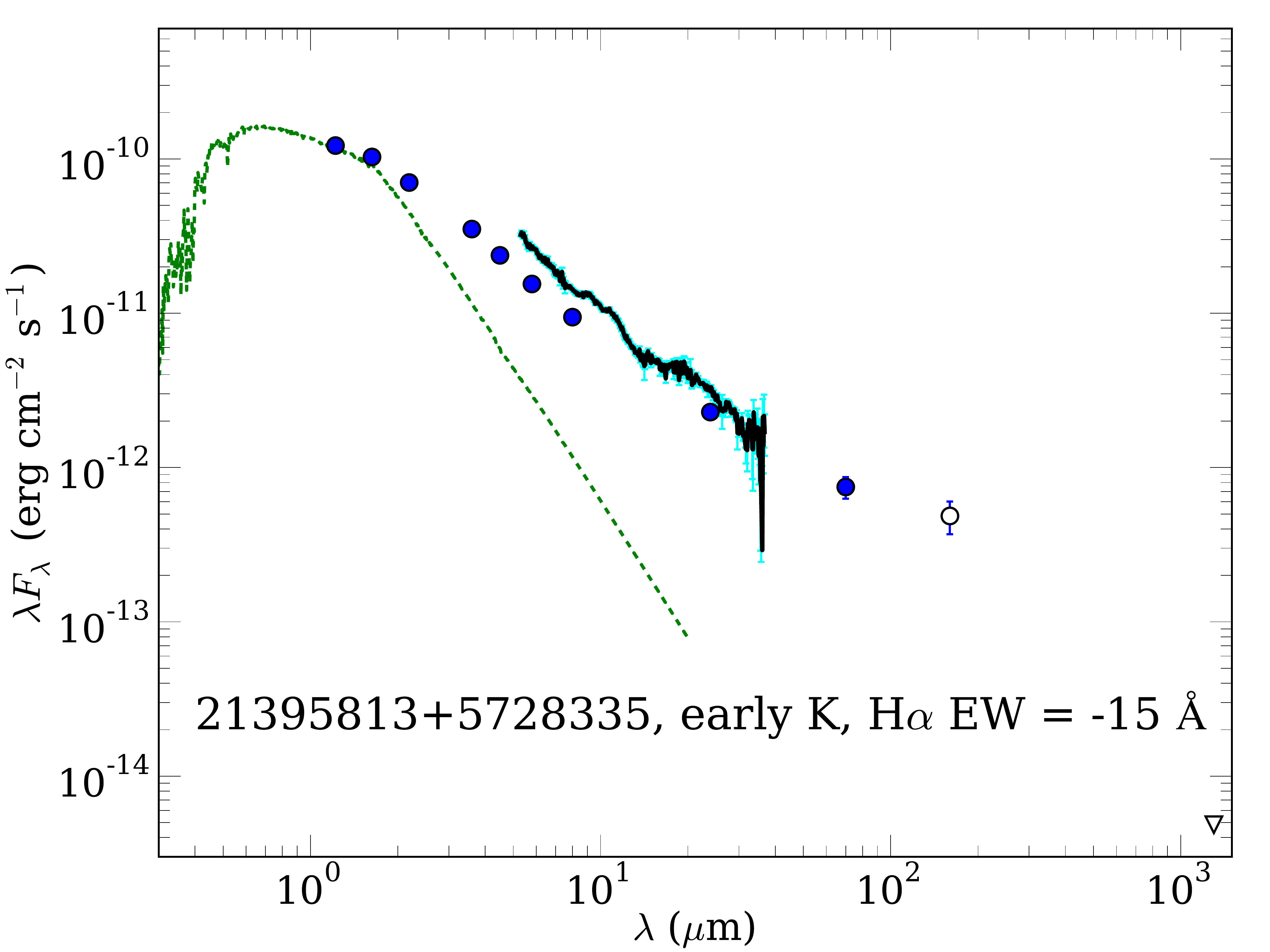} &
\includegraphics[width=0.24\linewidth]{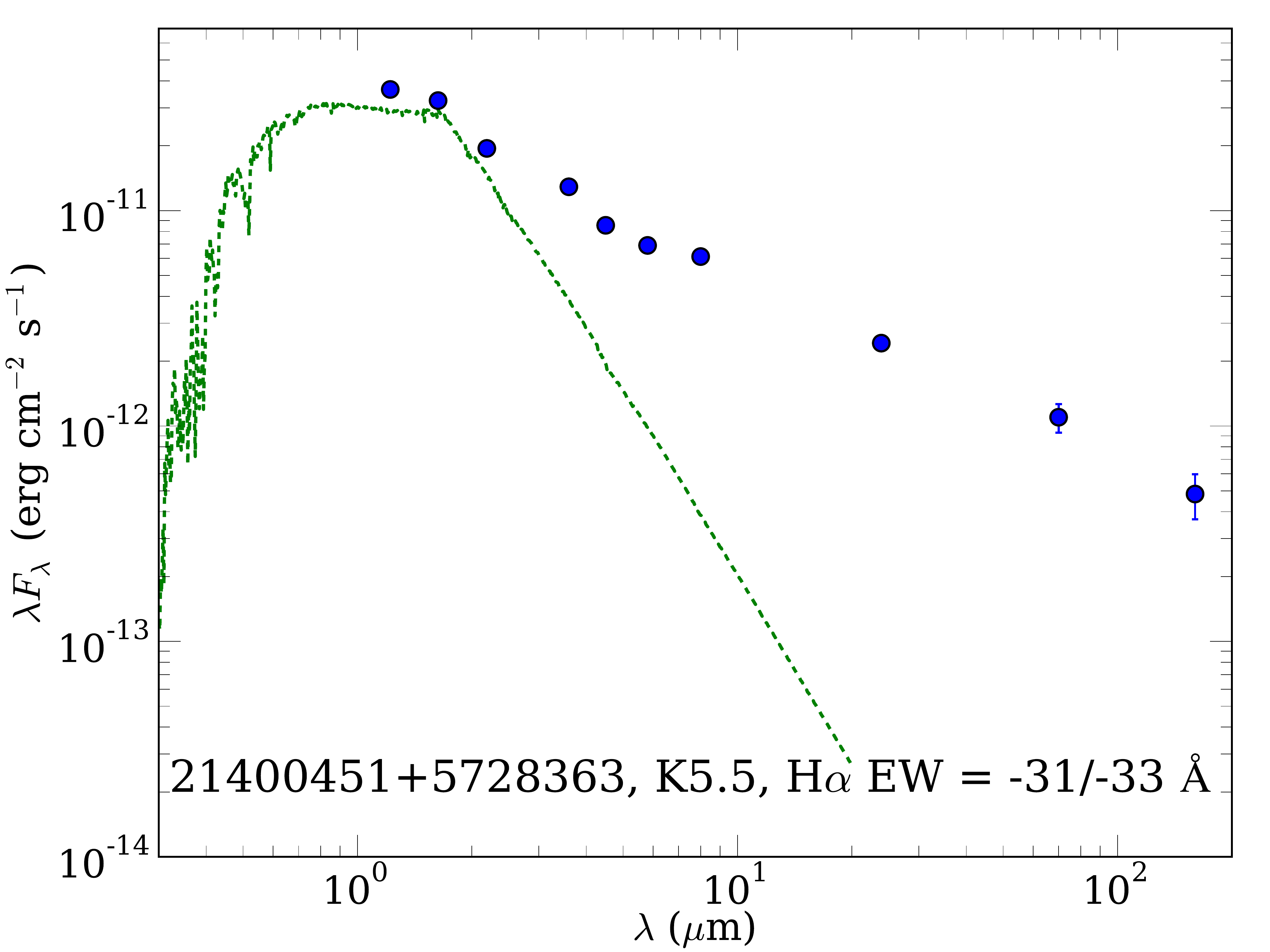} &
\includegraphics[width=0.24\linewidth]{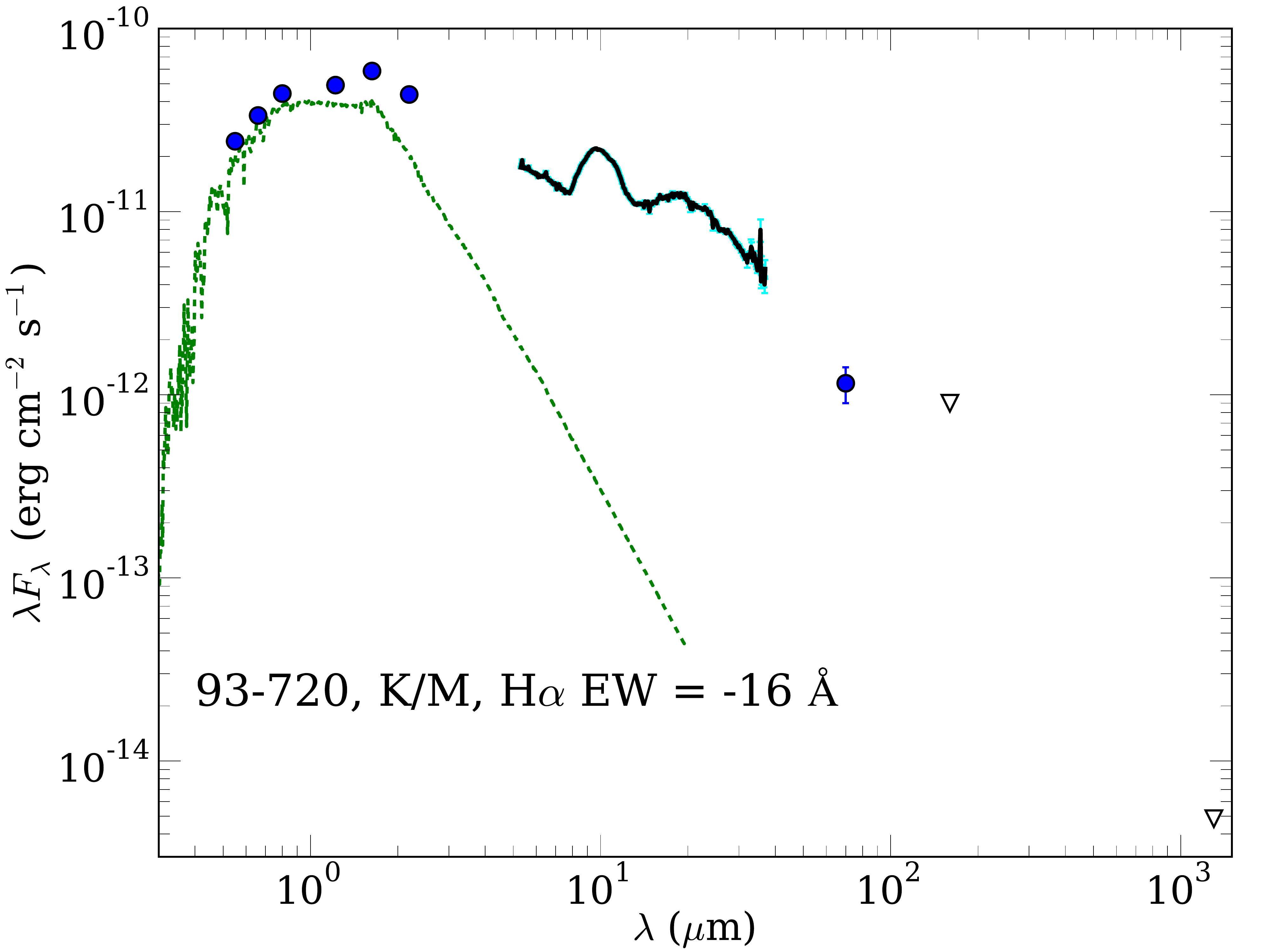} \\
\includegraphics[width=0.24\linewidth]{24_1796.pdf} &
\includegraphics[width=0.24\linewidth]{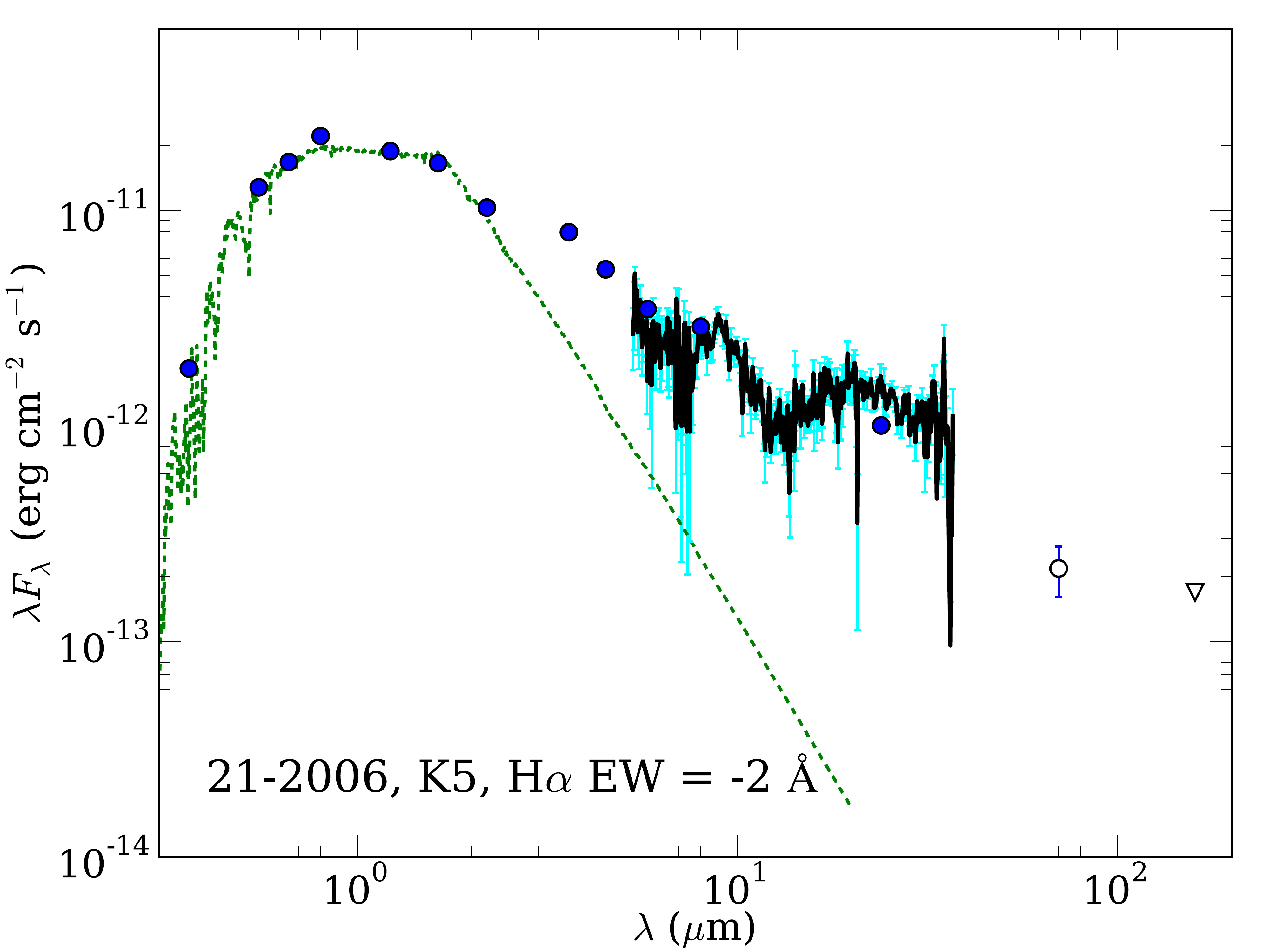} &
\includegraphics[width=0.24\linewidth]{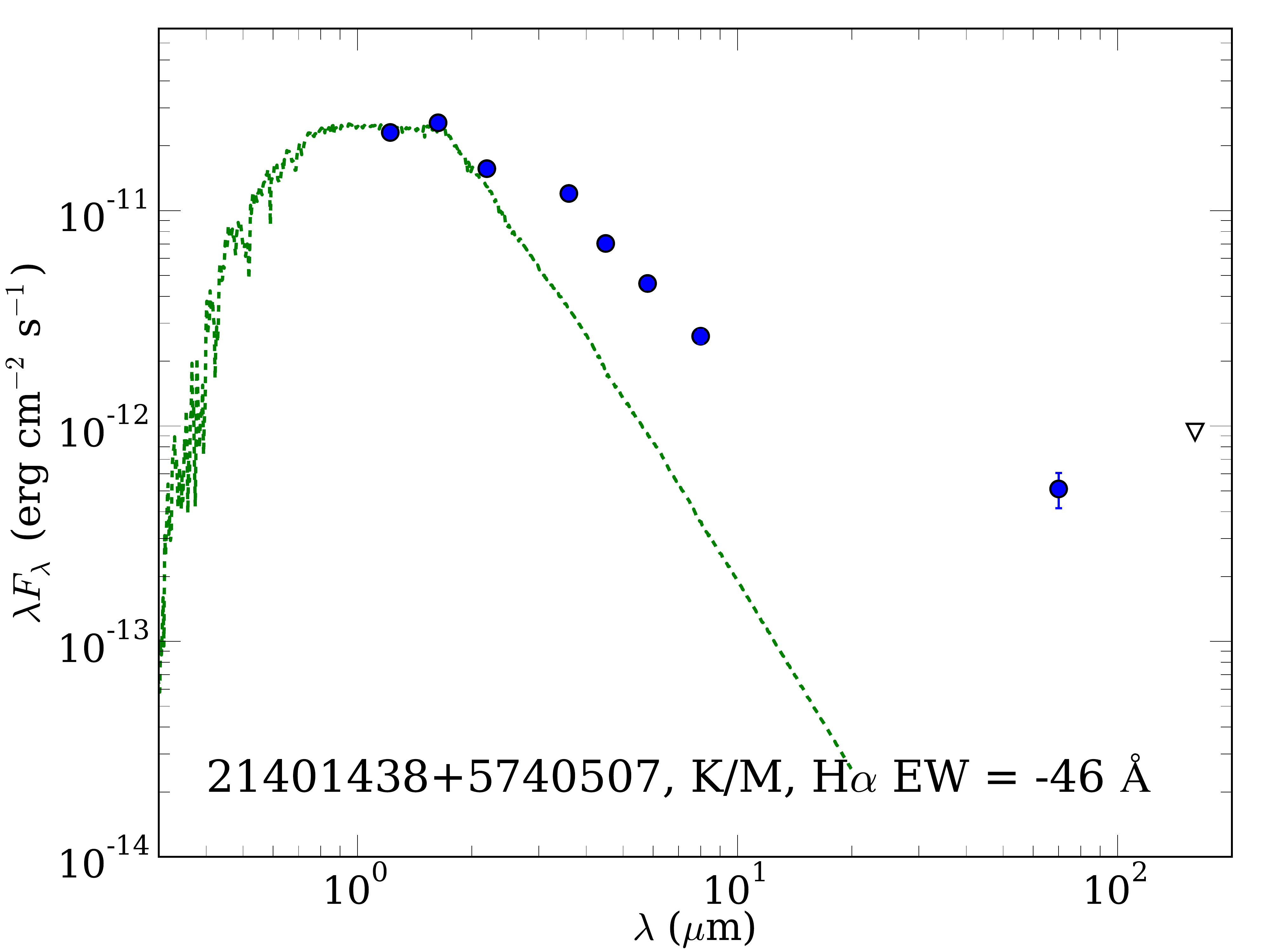} &
\includegraphics[width=0.24\linewidth]{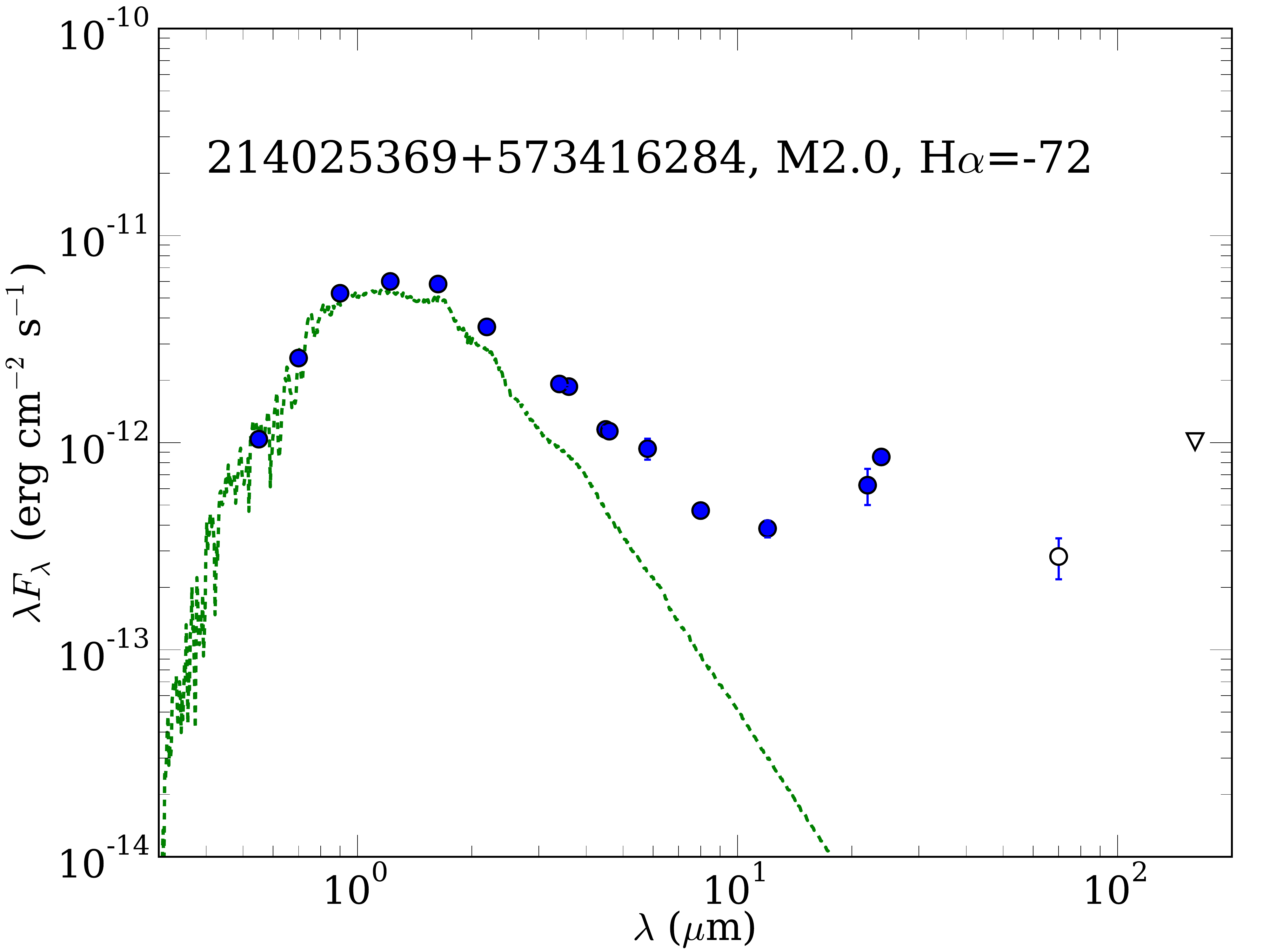} \\
\includegraphics[width=0.24\linewidth]{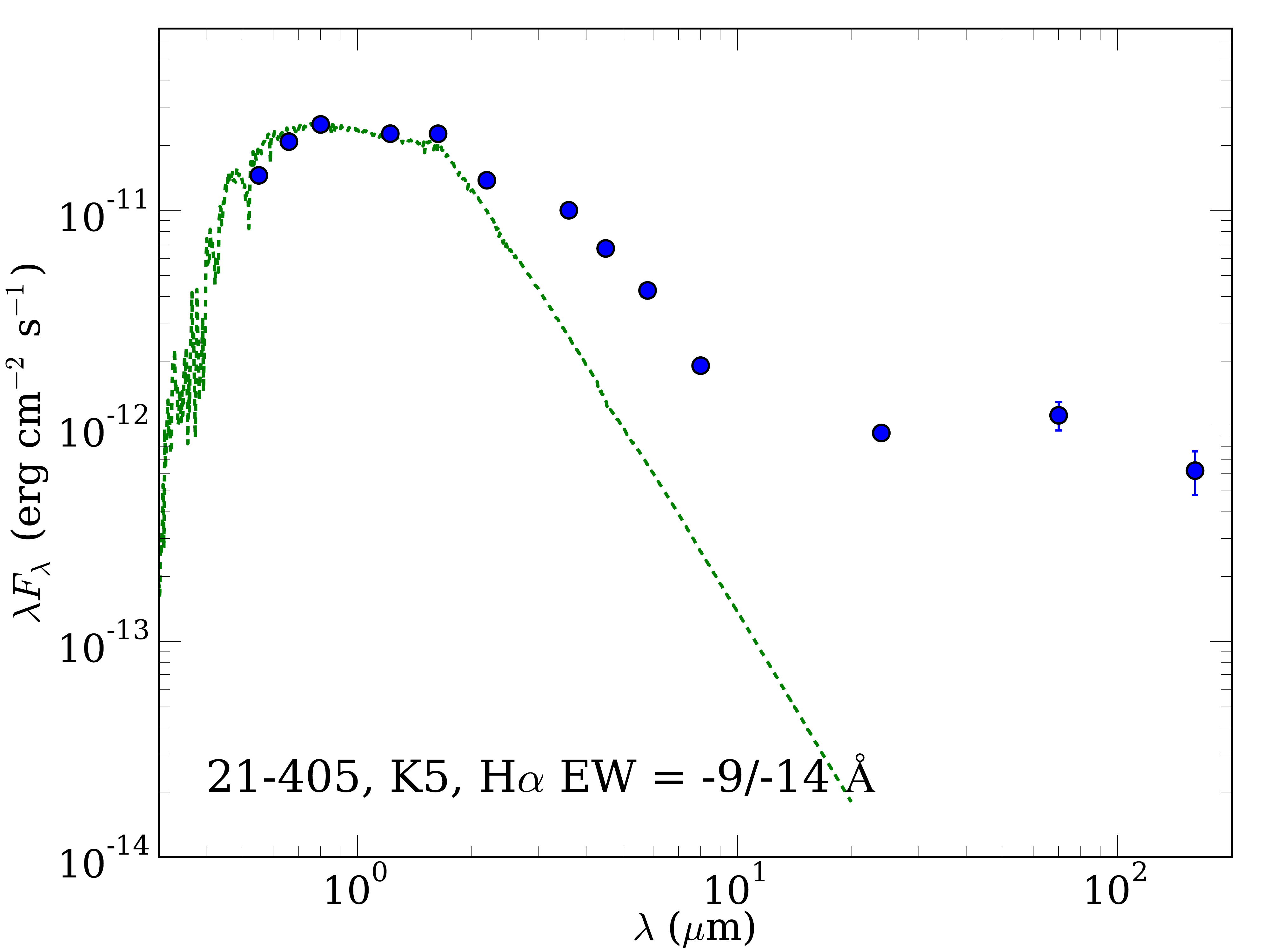} &
\includegraphics[width=0.24\linewidth]{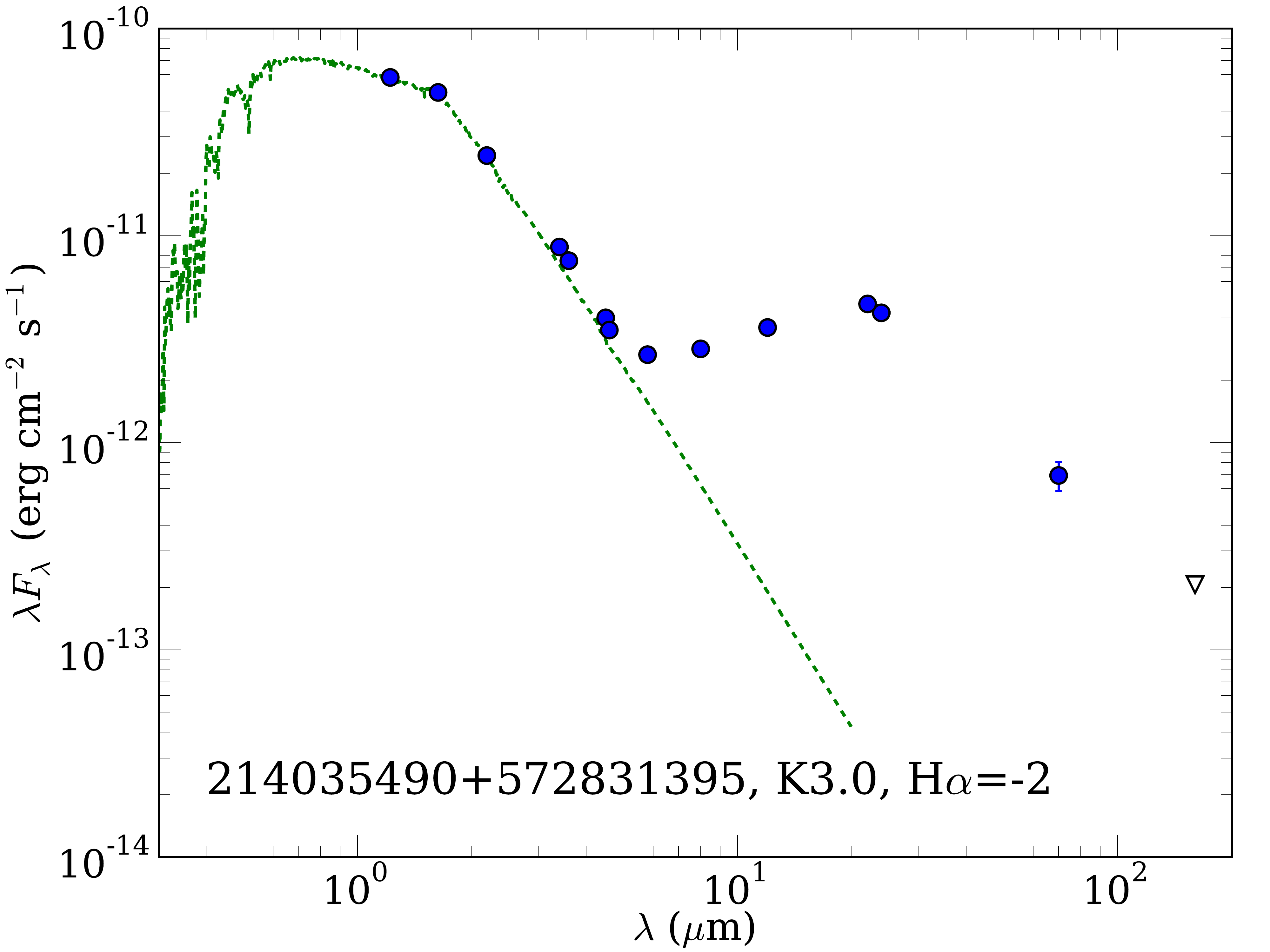} &
\includegraphics[width=0.24\linewidth]{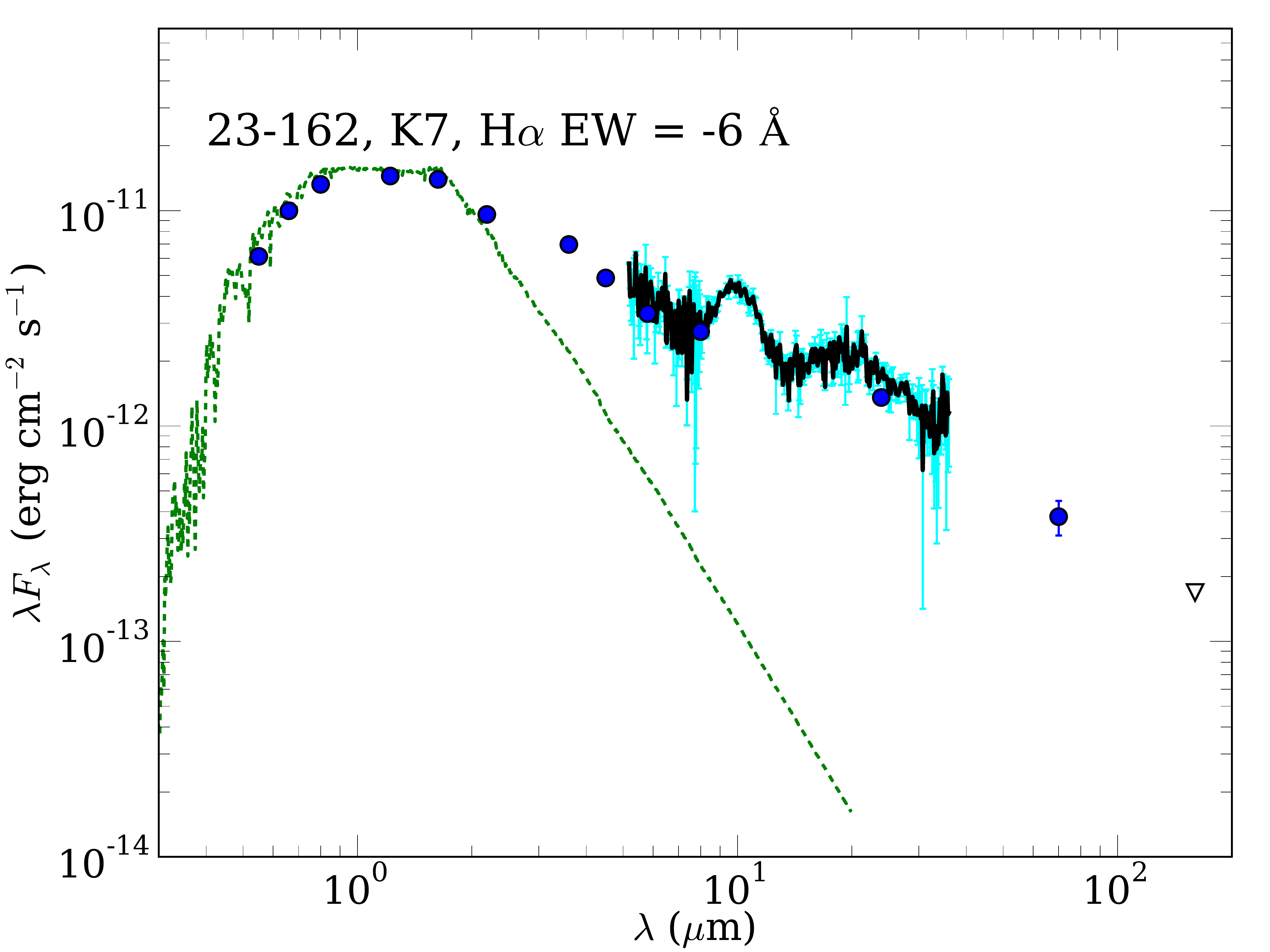} &
\includegraphics[width=0.24\linewidth]{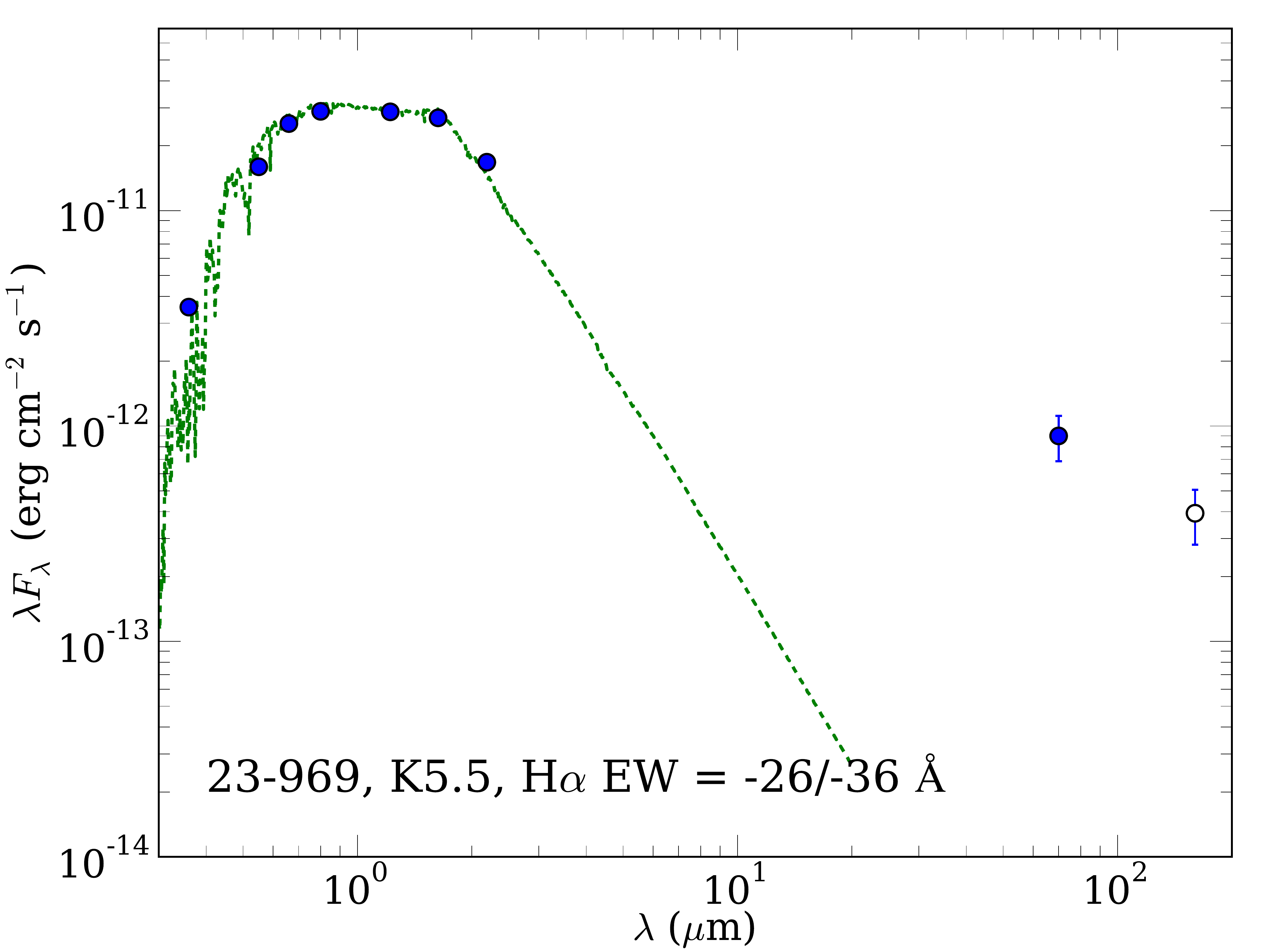} \\
\end{tabular}
\caption{SEDs of the objects detected with Herschel (continuation).  All objects belong to Tr\,37. 
Symbols as in Figure \ref{seds1-fig}.  
\label{seds3-fig}}
\end{figure*}

\begin{figure*}
\centering
\begin{tabular}{cccc}
\includegraphics[width=0.24\linewidth]{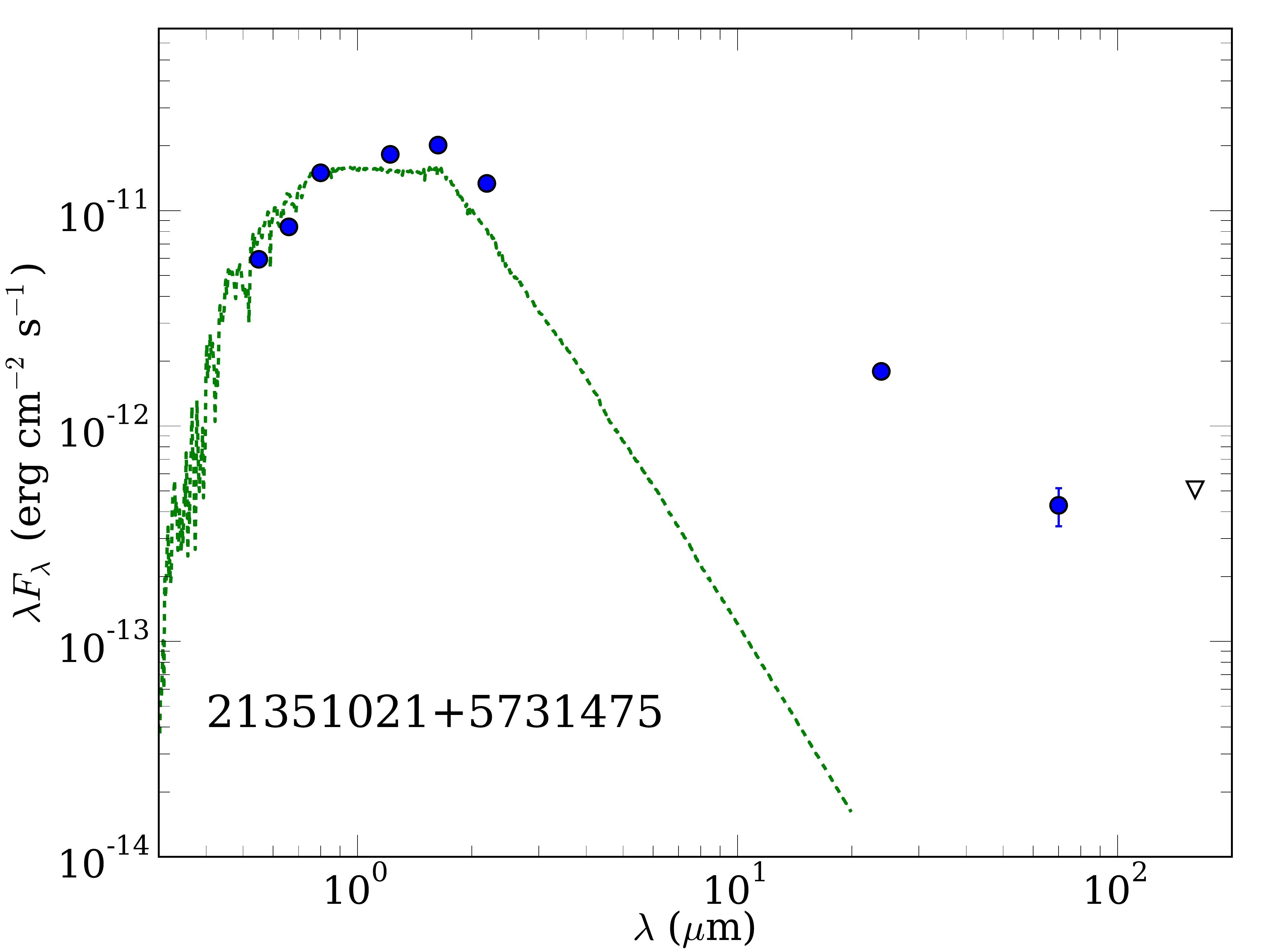} &
\includegraphics[width=0.24\linewidth]{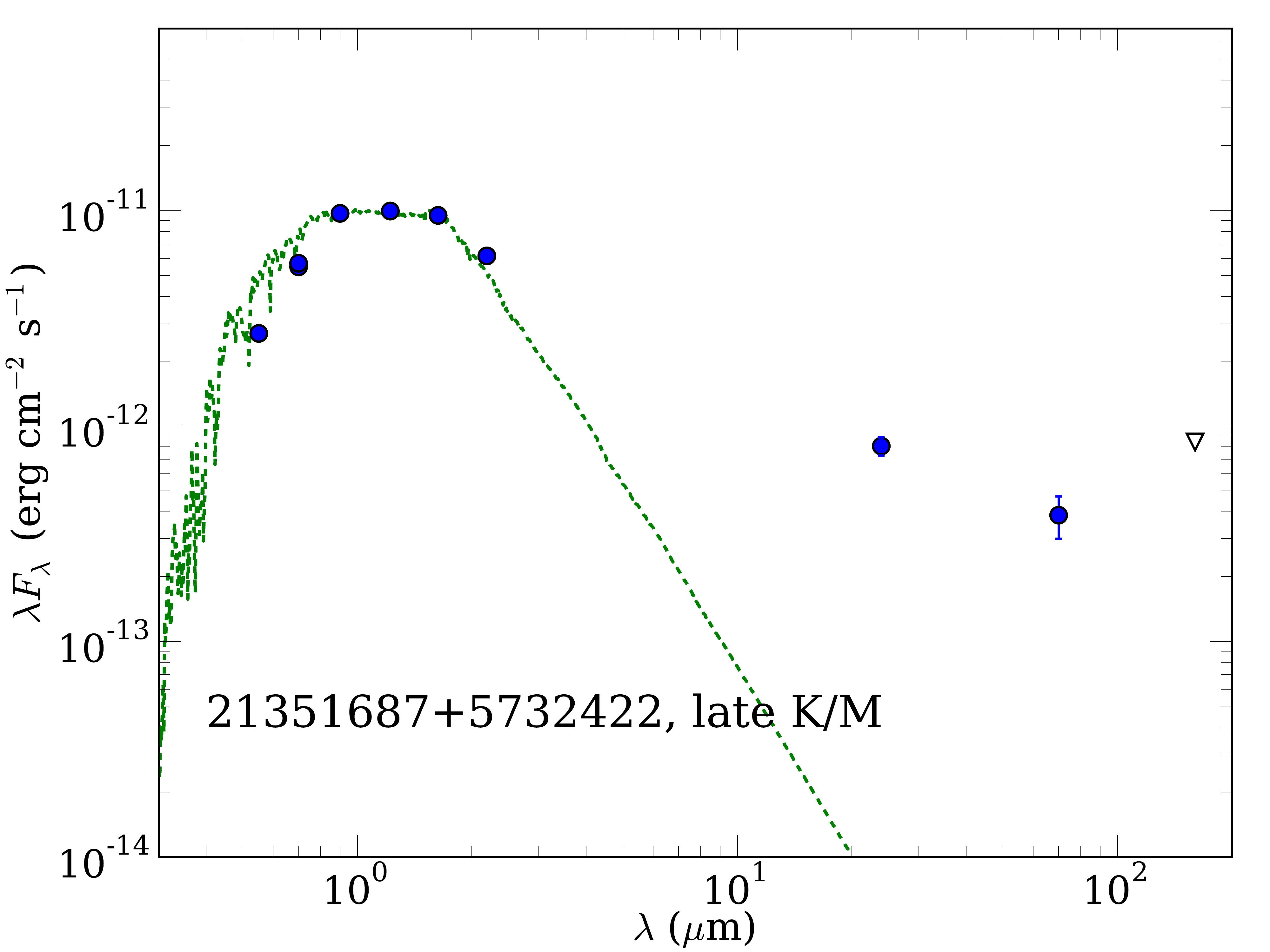} &
\includegraphics[width=0.24\linewidth]{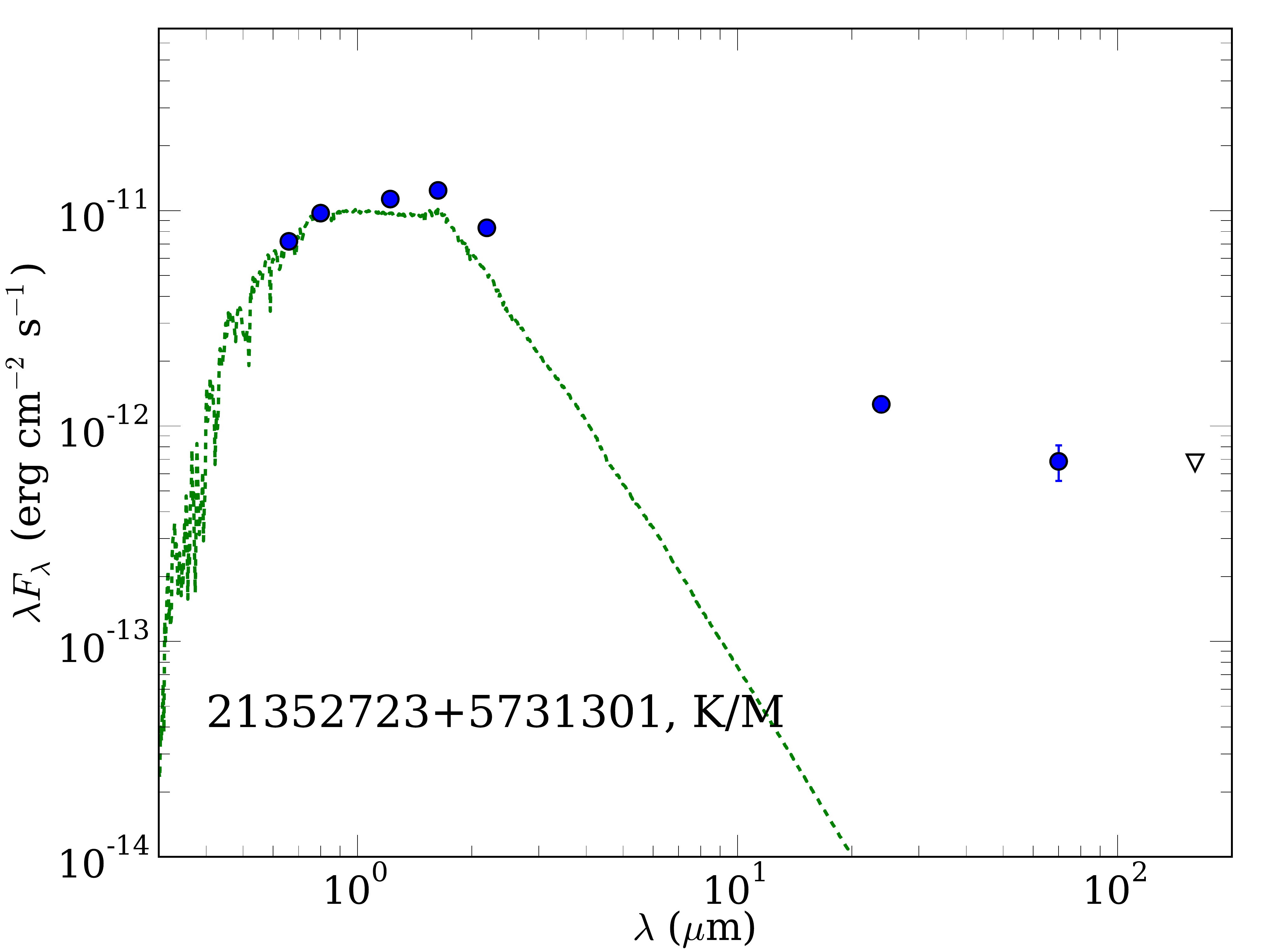} &
\includegraphics[width=0.24\linewidth]{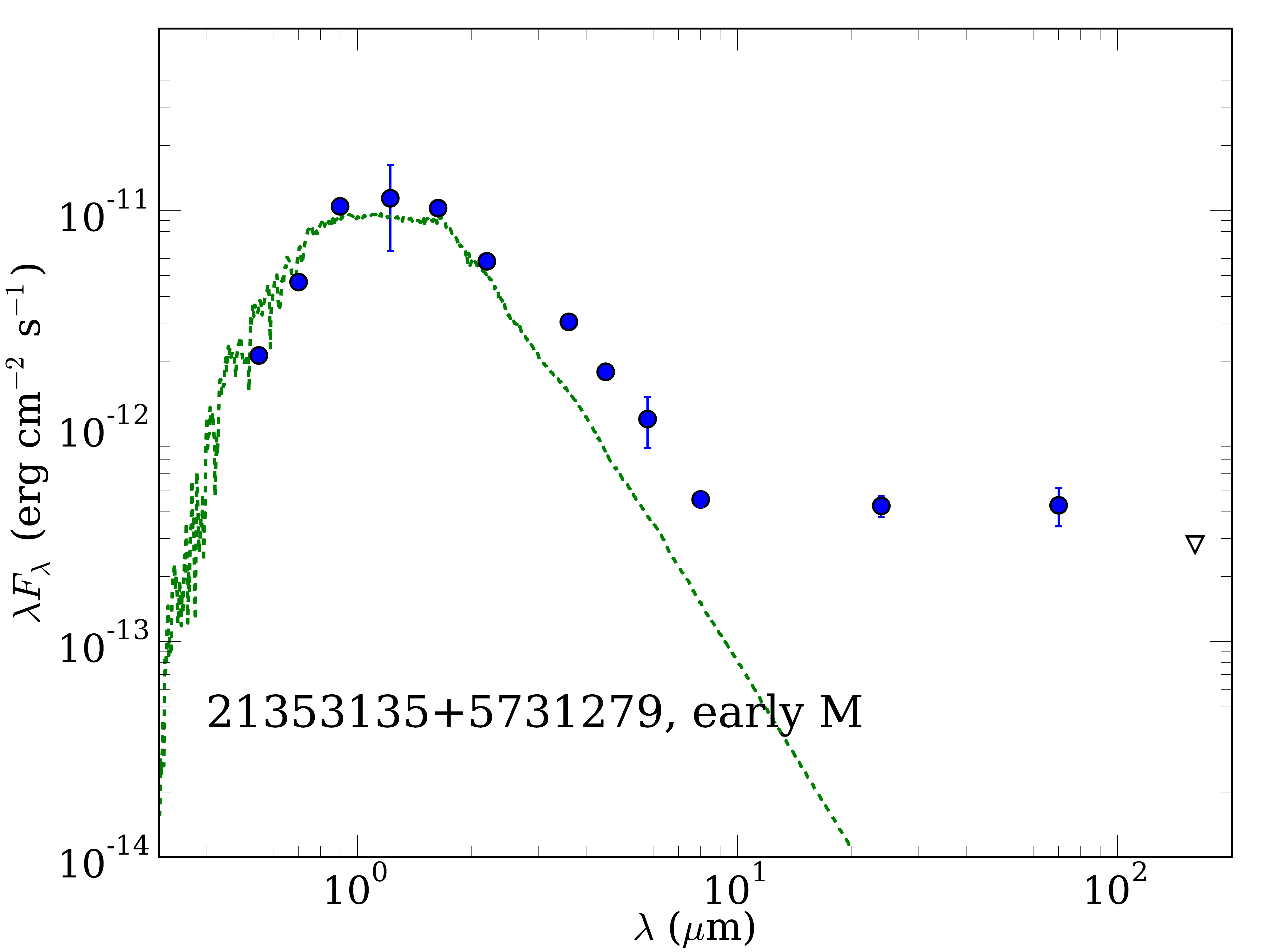} \\
\includegraphics[width=0.24\linewidth]{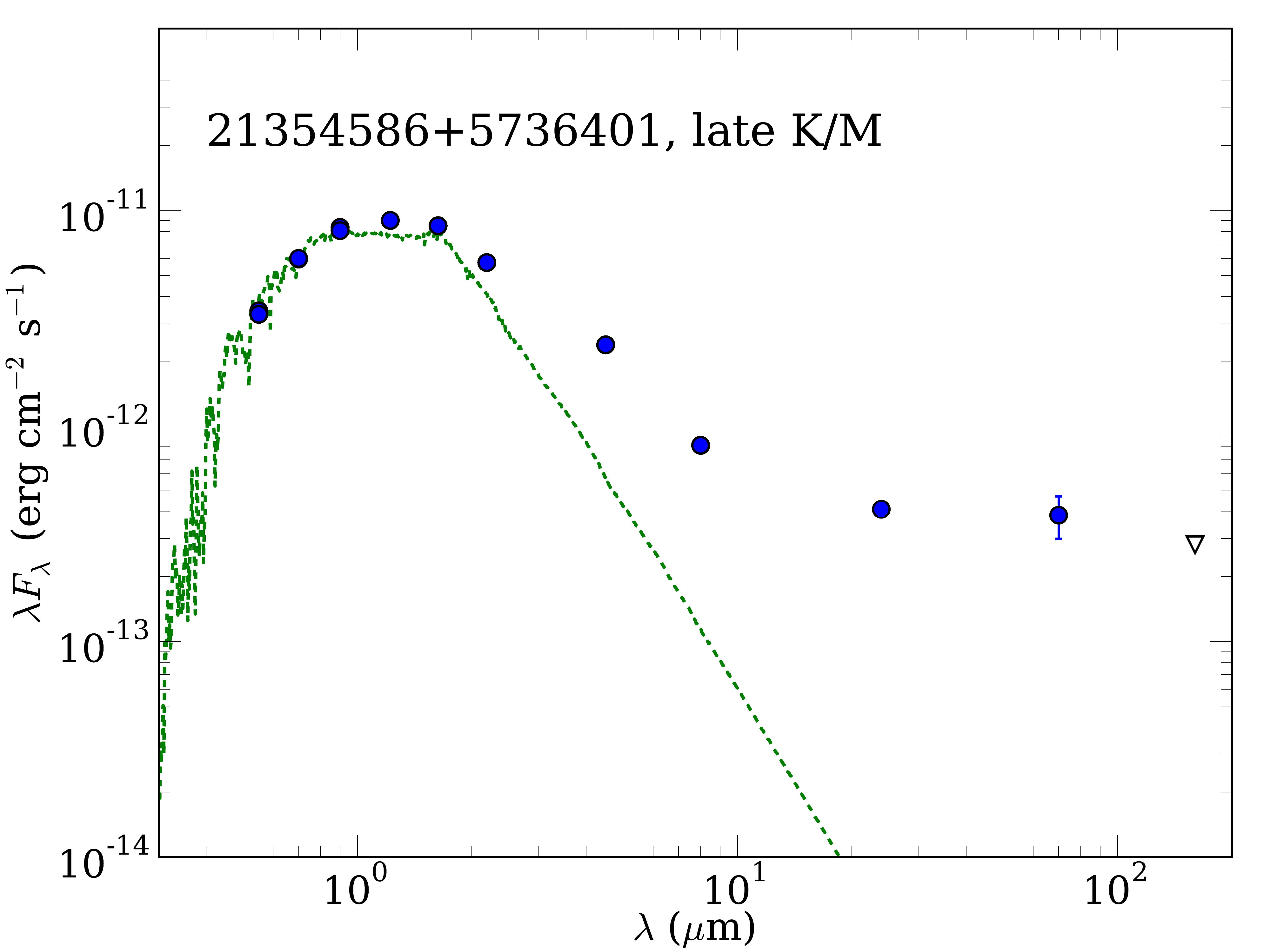} &
\includegraphics[width=0.24\linewidth]{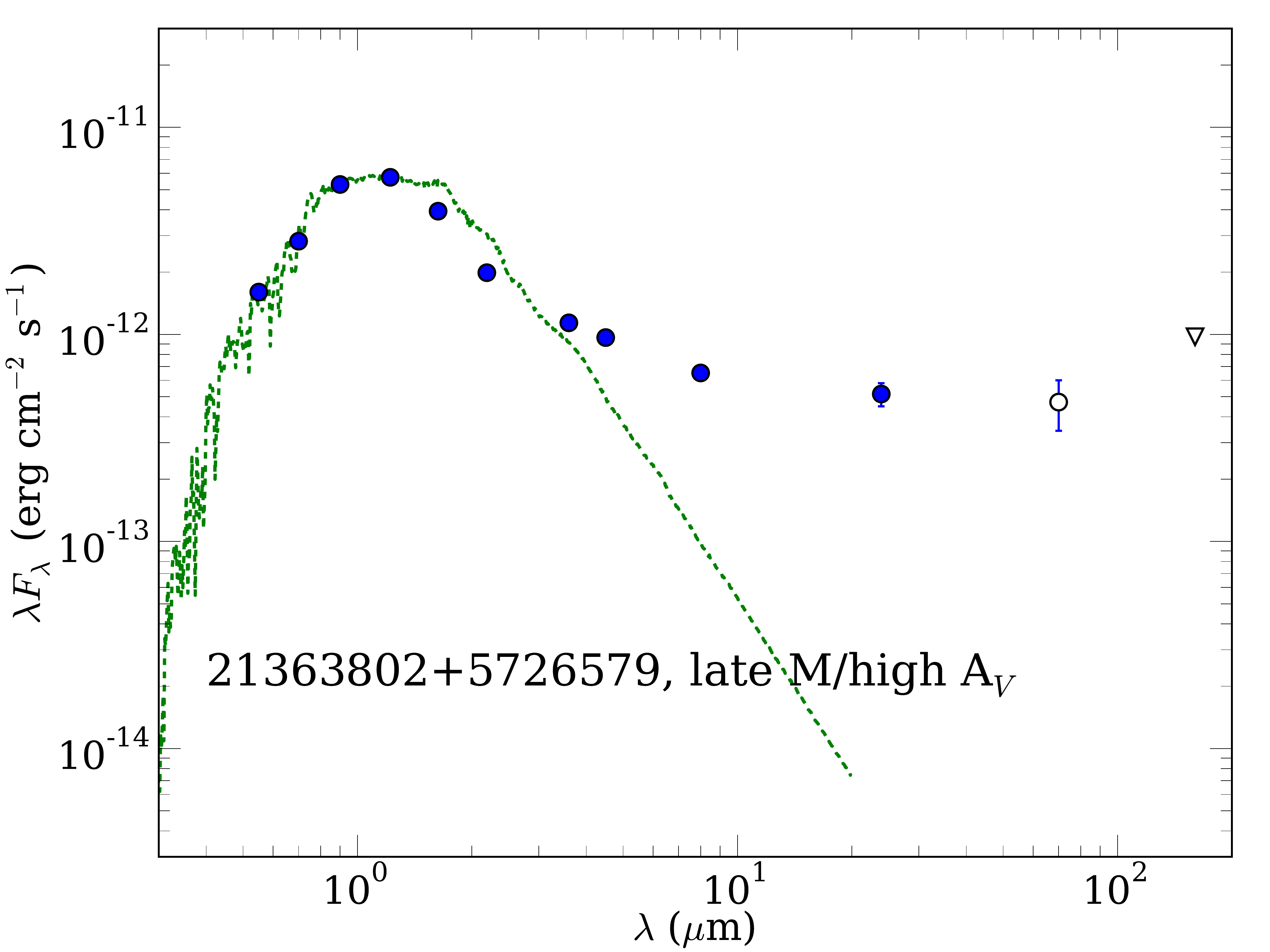} &
\includegraphics[width=0.24\linewidth]{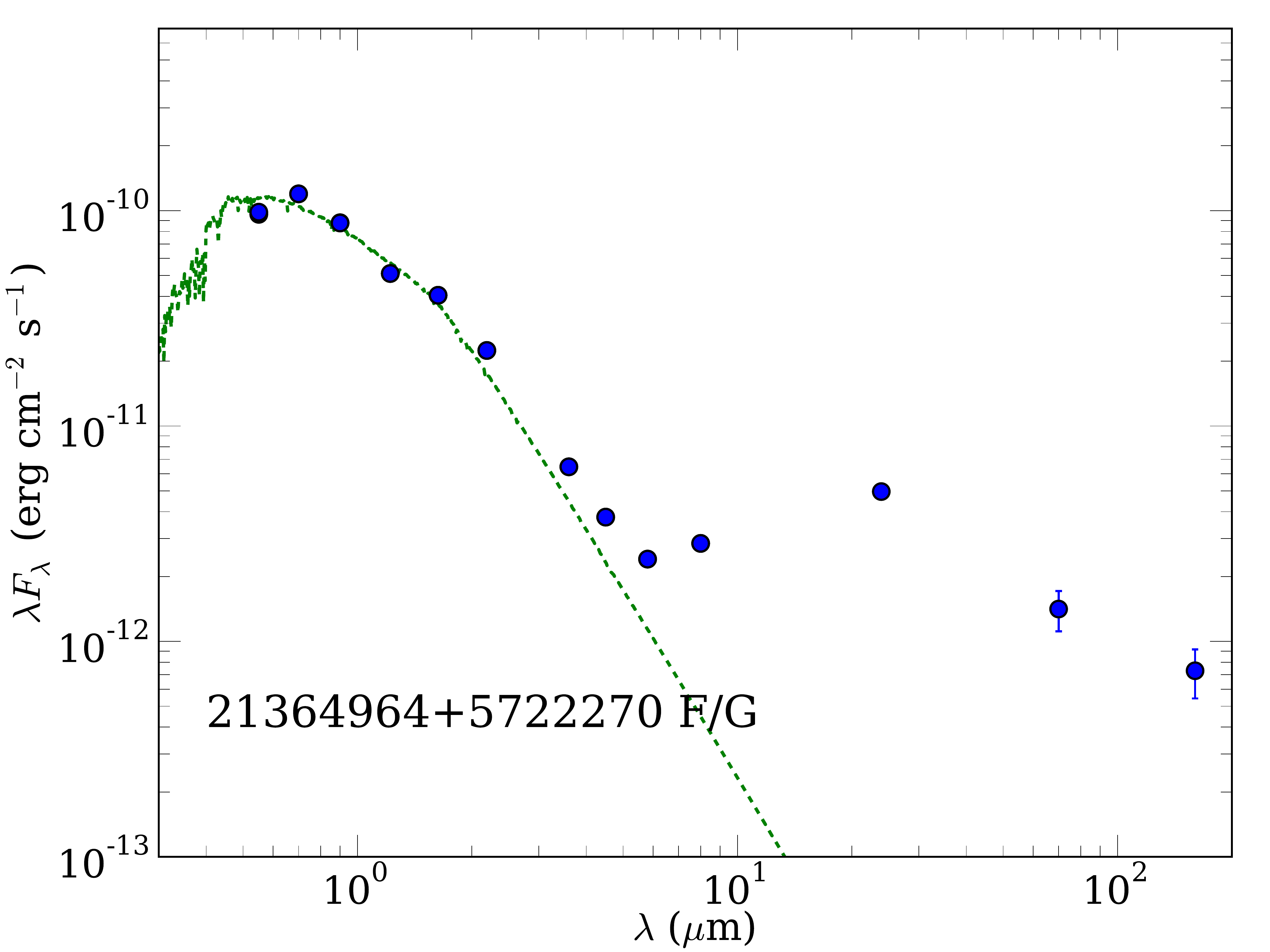} &
\includegraphics[width=0.24\linewidth]{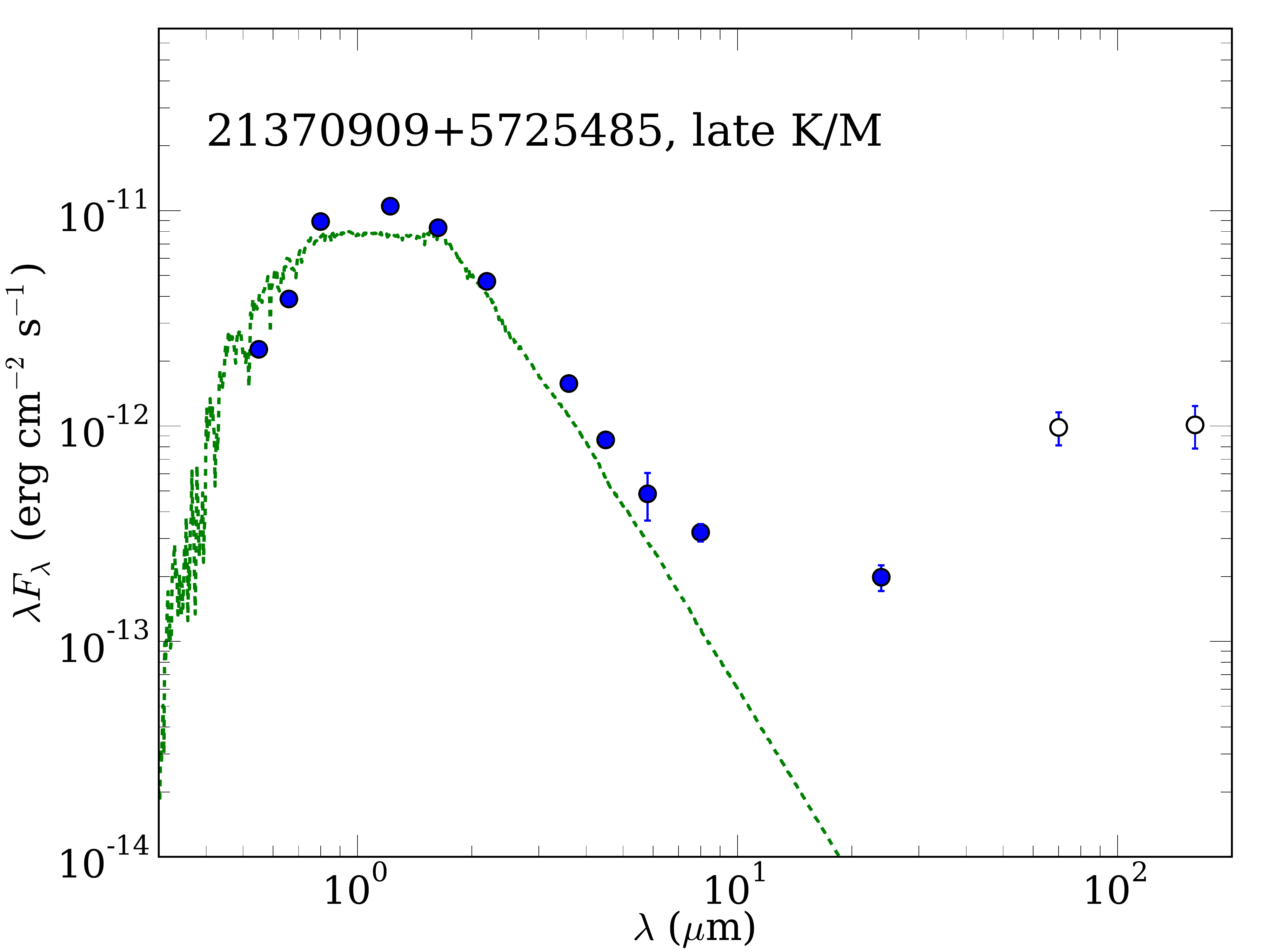} \\
\includegraphics[width=0.24\linewidth]{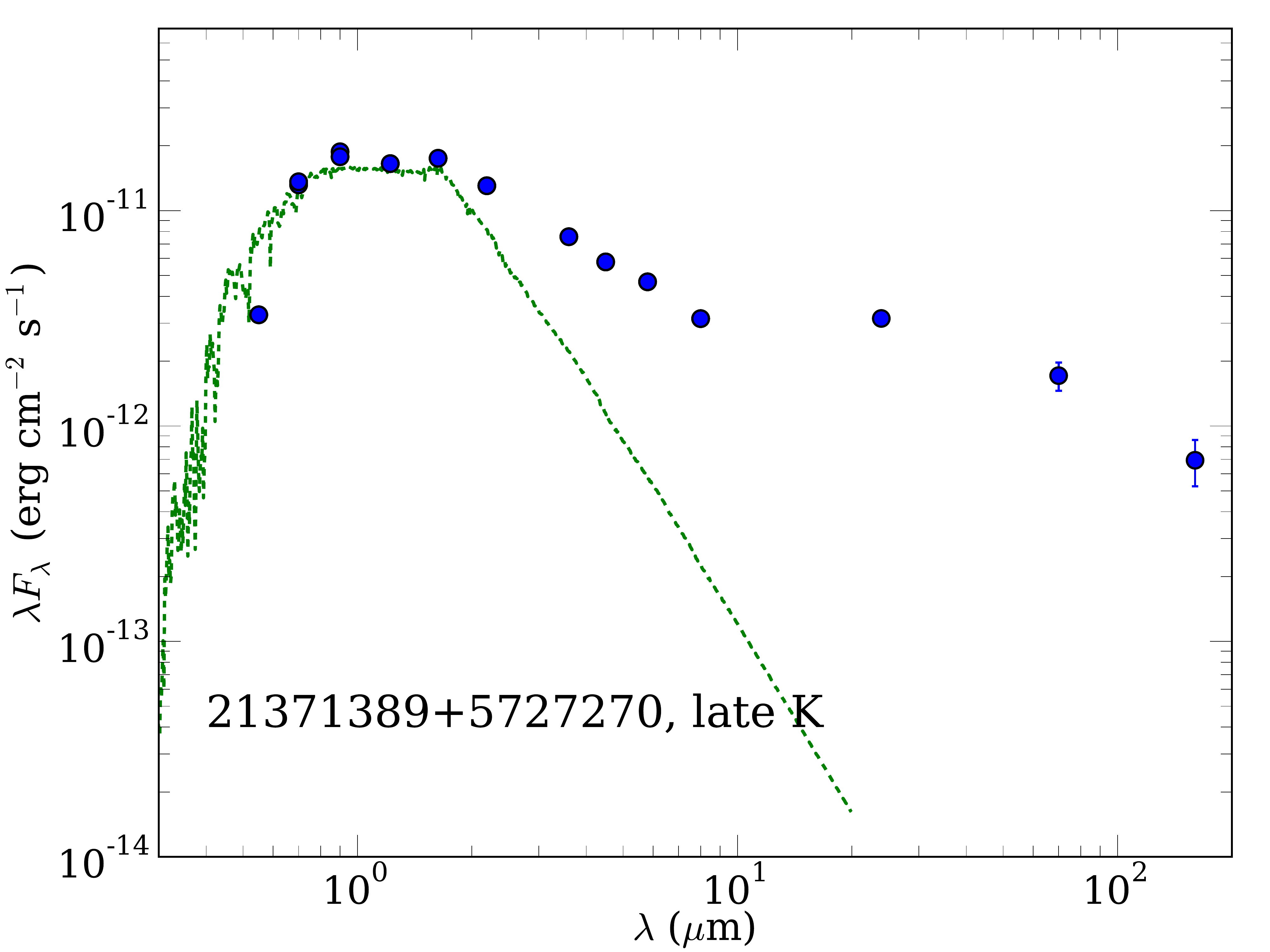} &
\includegraphics[width=0.24\linewidth]{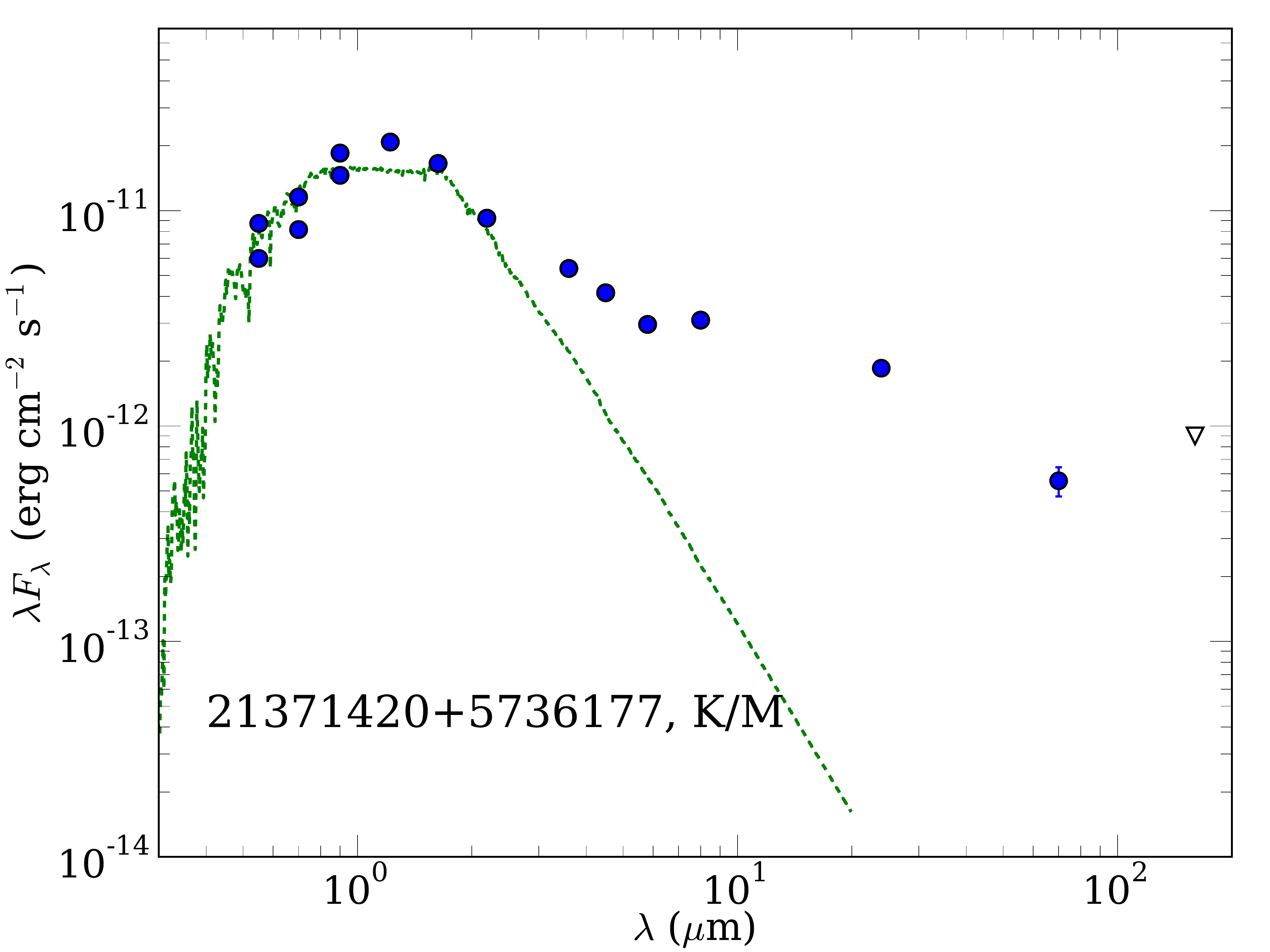} &
\includegraphics[width=0.24\linewidth]{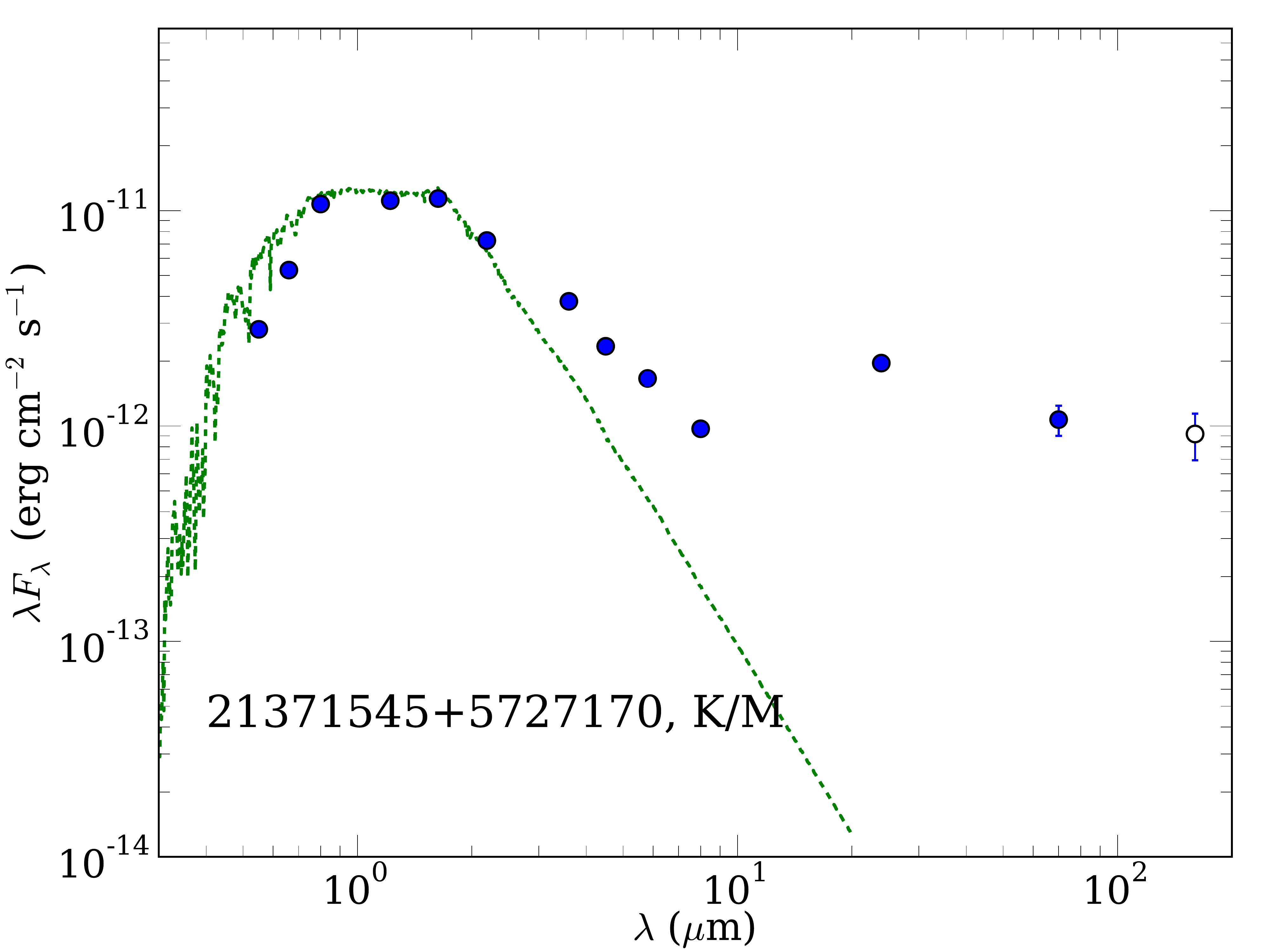} &
\includegraphics[width=0.24\linewidth]{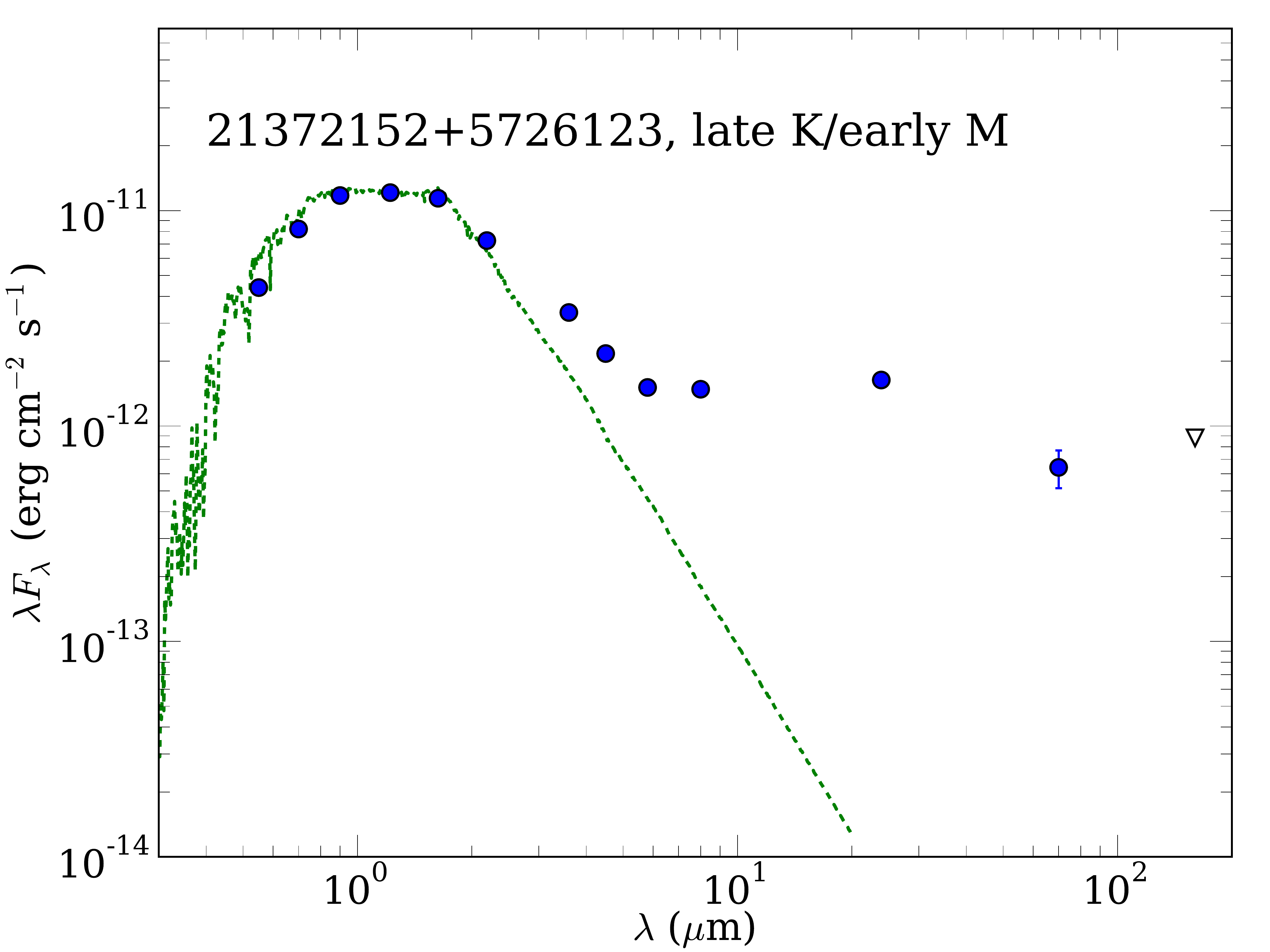} \\
\includegraphics[width=0.24\linewidth]{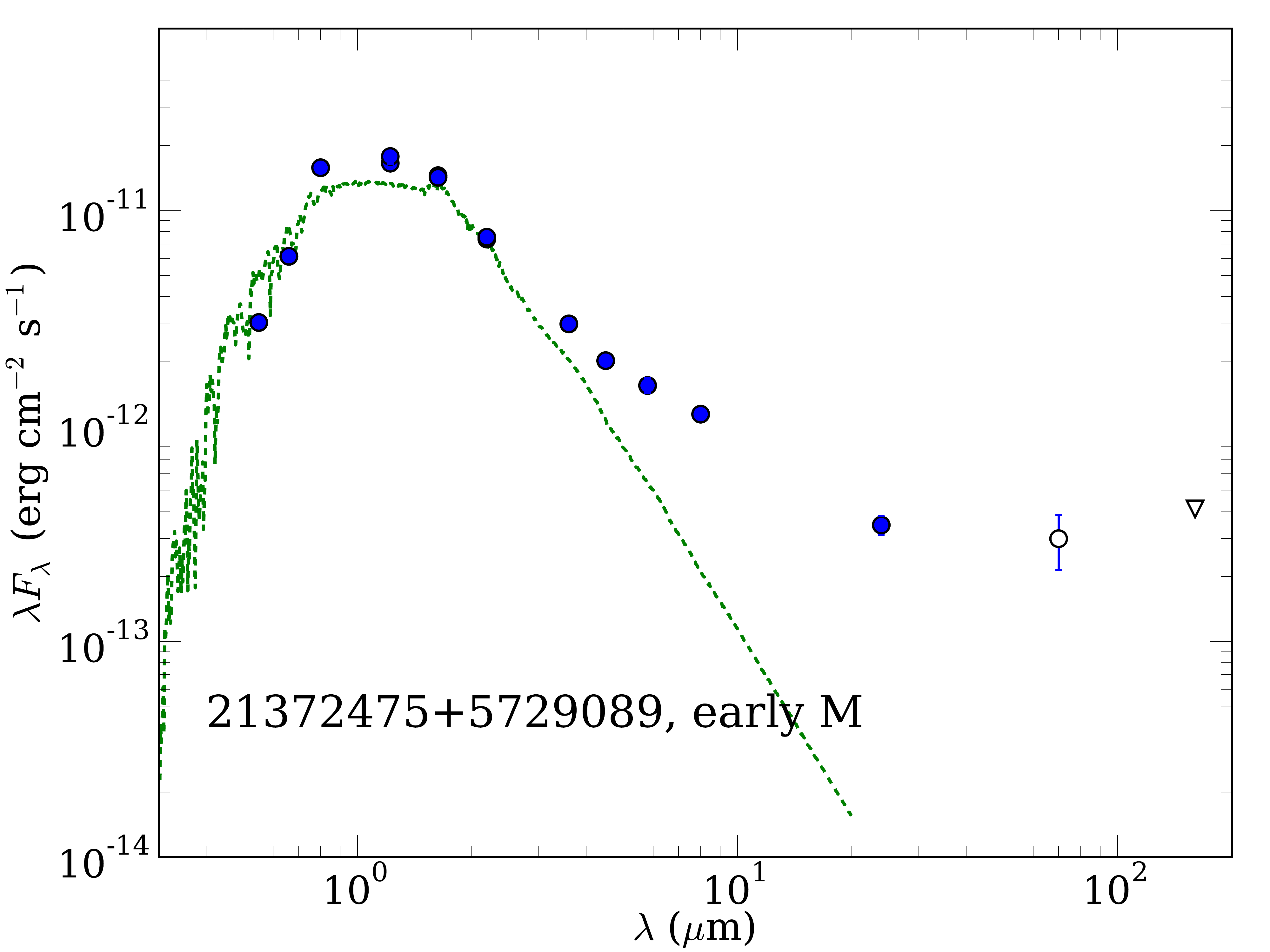} &
\includegraphics[width=0.24\linewidth]{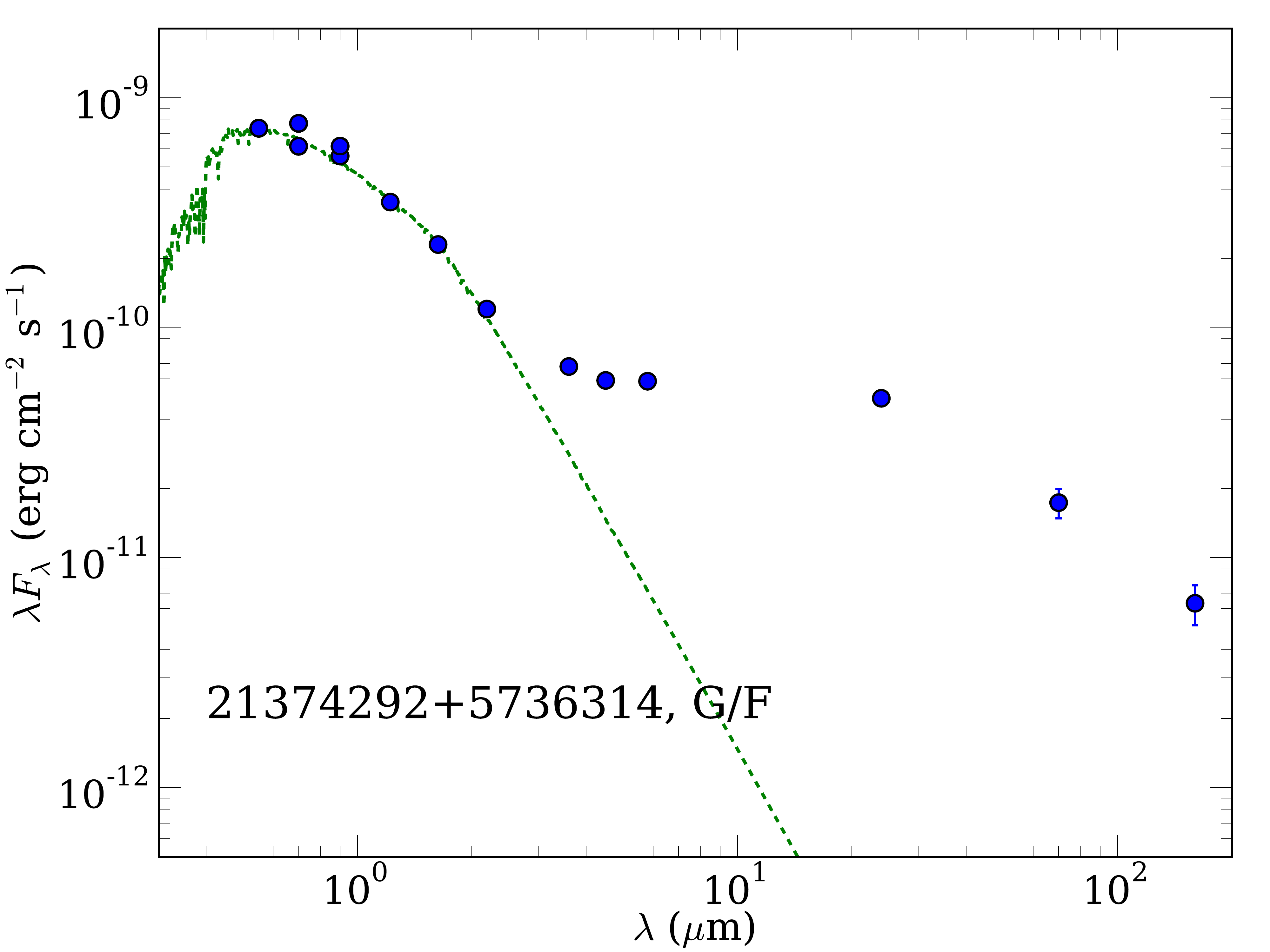} &
\includegraphics[width=0.24\linewidth]{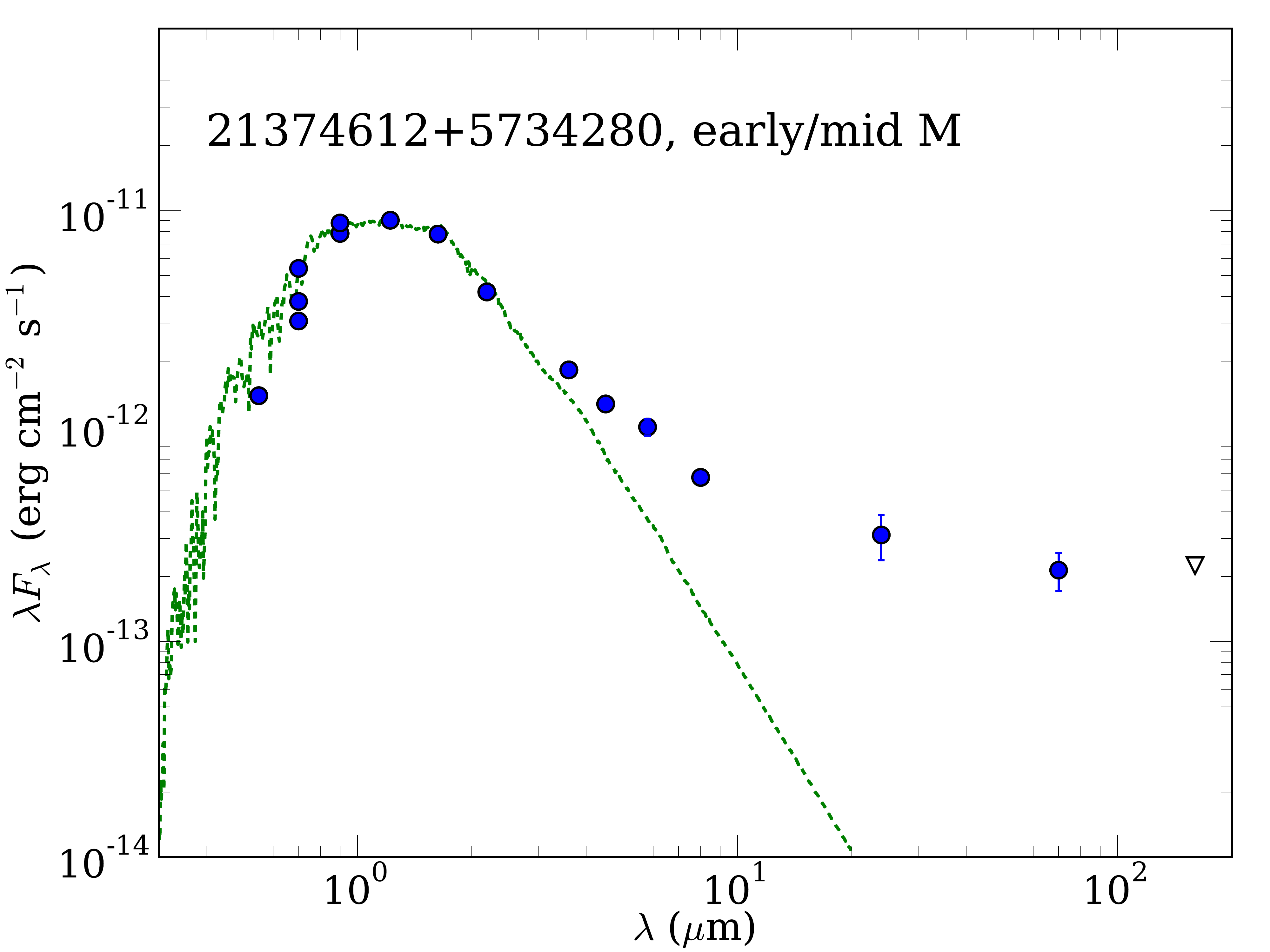} &
\includegraphics[width=0.24\linewidth]{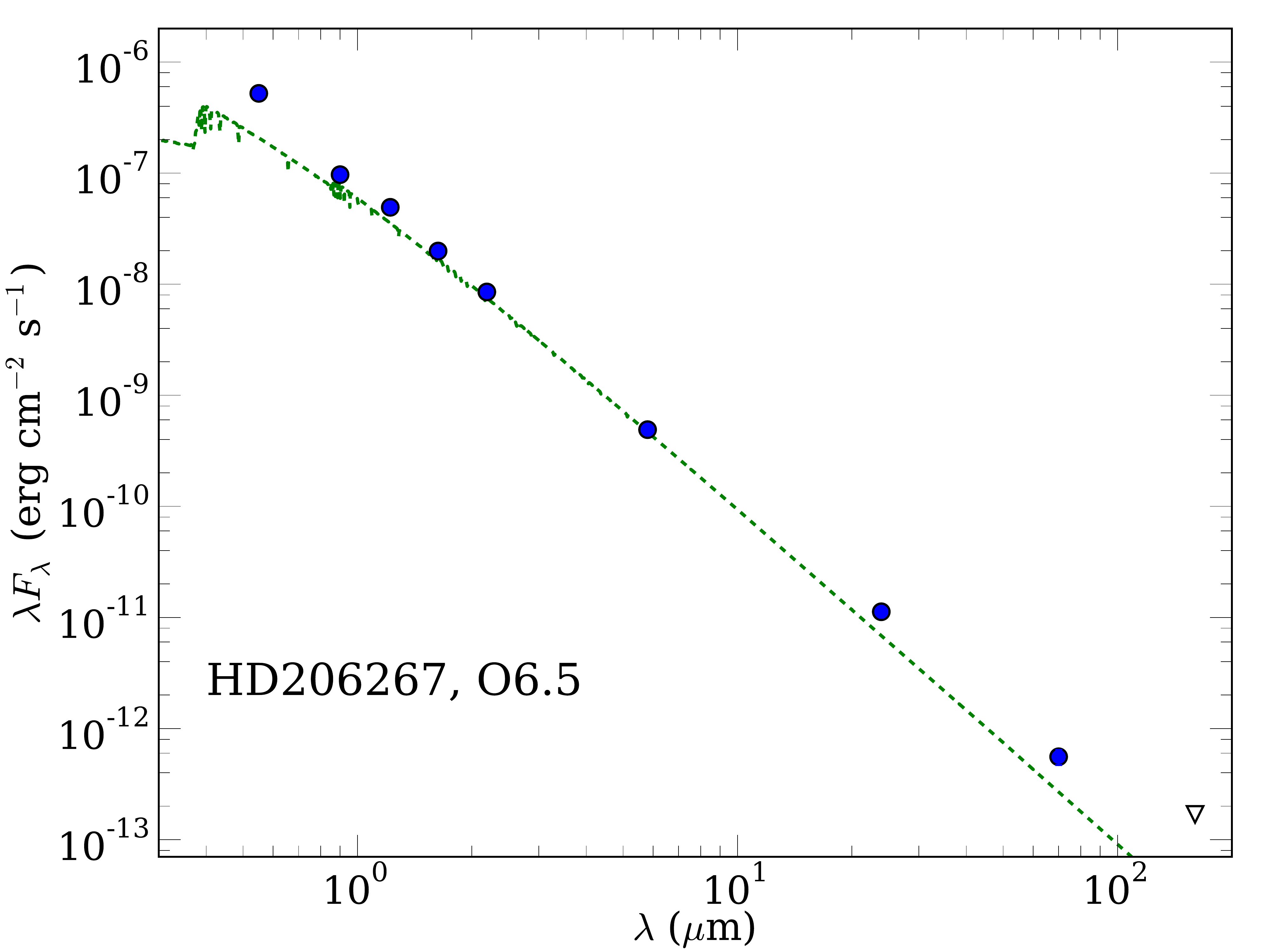} \\
\includegraphics[width=0.24\linewidth]{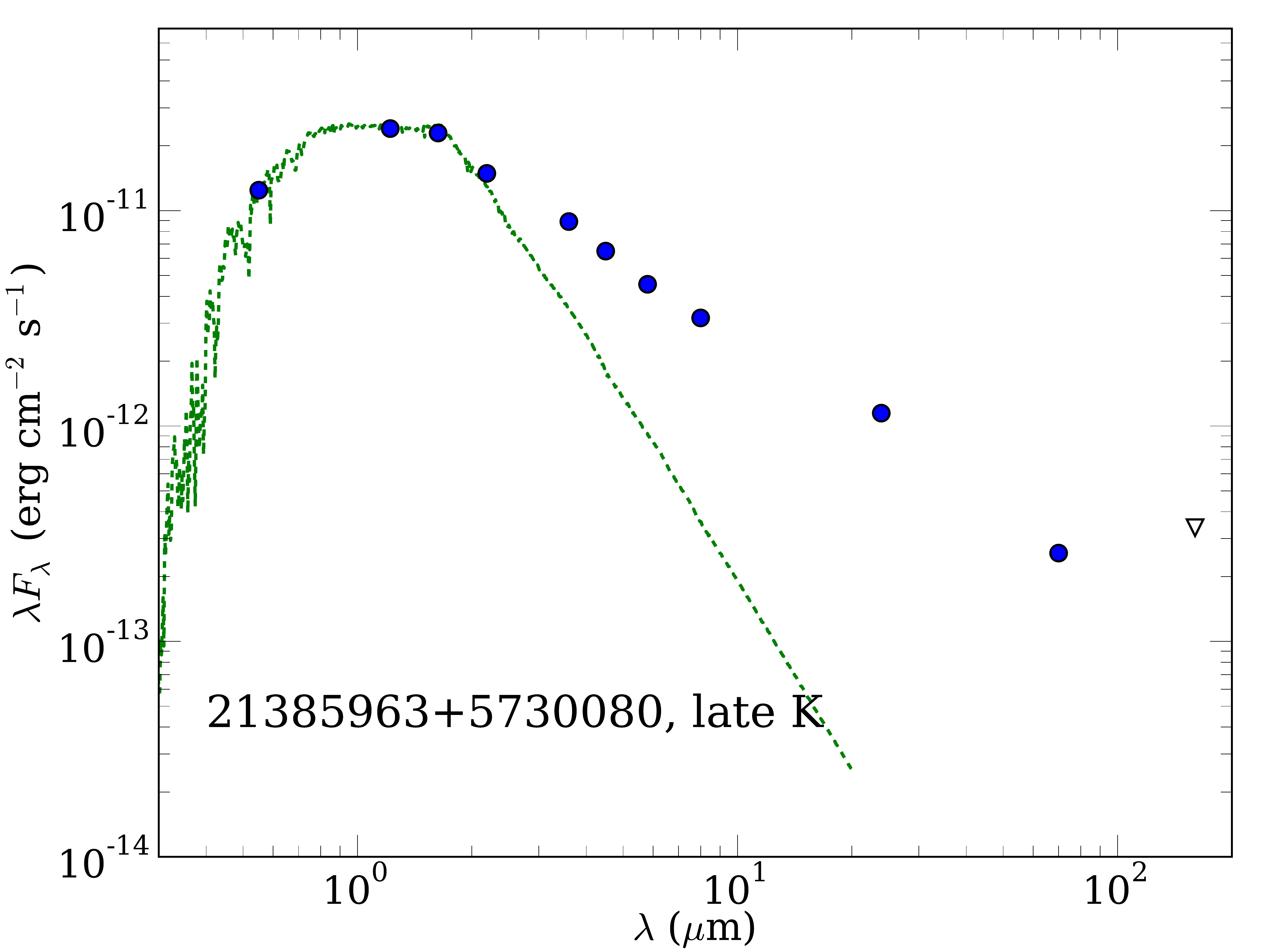} &
\includegraphics[width=0.24\linewidth]{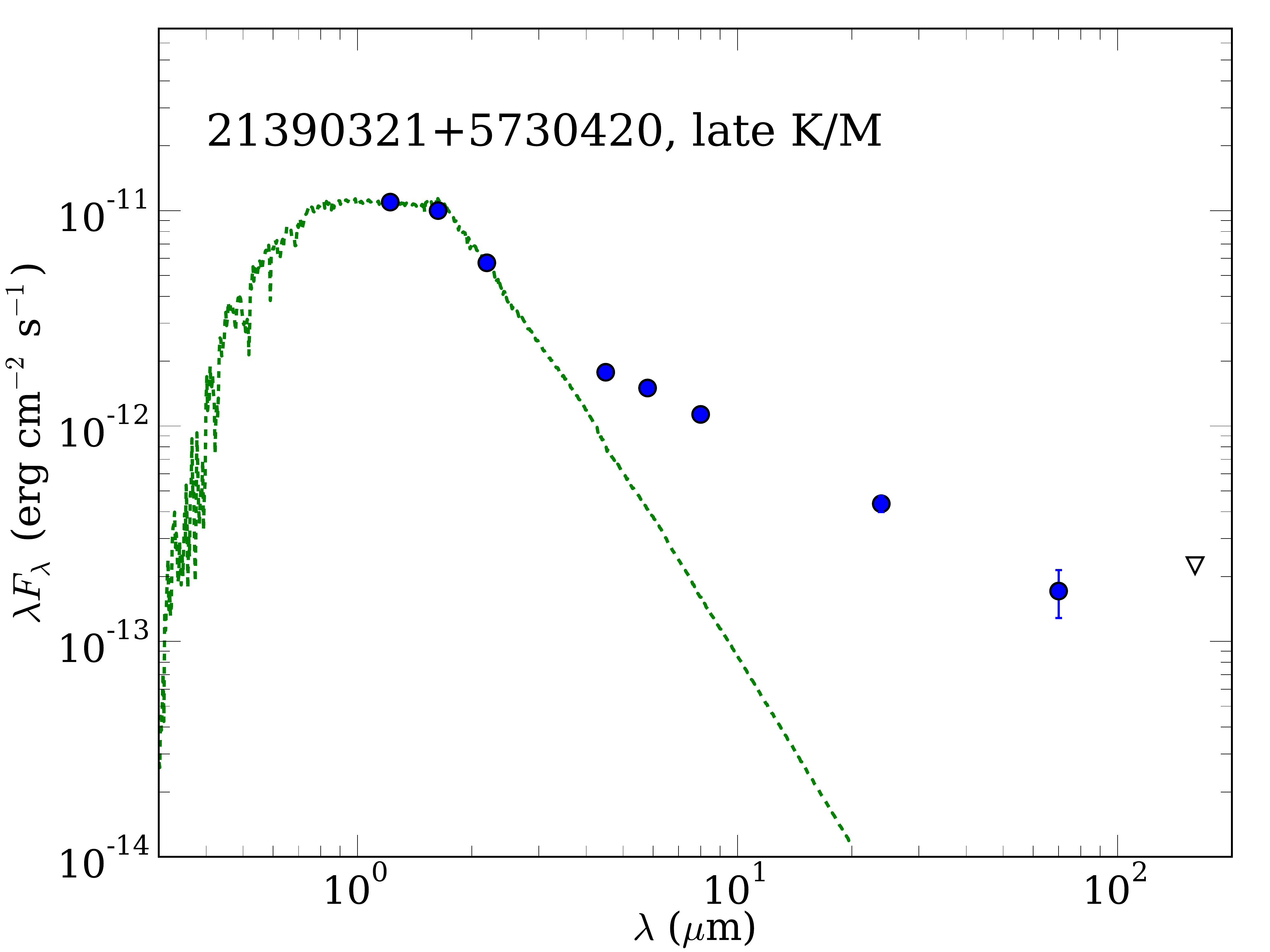} \\
\end{tabular}
\caption{SEDs of objects without optical confirmation (see Table \ref{other-table})
detected with Herschel. Available optical, Spitzer (IRAC/MIPS photometry and
IRS spectra), and
WISE, data are shown as filled dots. Errorbars are shown (although they are often smaller than the symbols). 
Upper limits are marked as inverted open triangles.  Marginal detections (close to 3$\sigma$ or affected by nebulosity)
are marked as open circles. A photospheric MARCS model is displayed for
comparison. Approximate spectral types (except for HD\,206267
are estimated from the comparison with the MARCS model, assuming
the typical cluster extinctions. \label{otherseds-fig}}
\end{figure*}

\begin{figure*}
\centering
\begin{tabular}{cccc}
\includegraphics[width=0.24\linewidth]{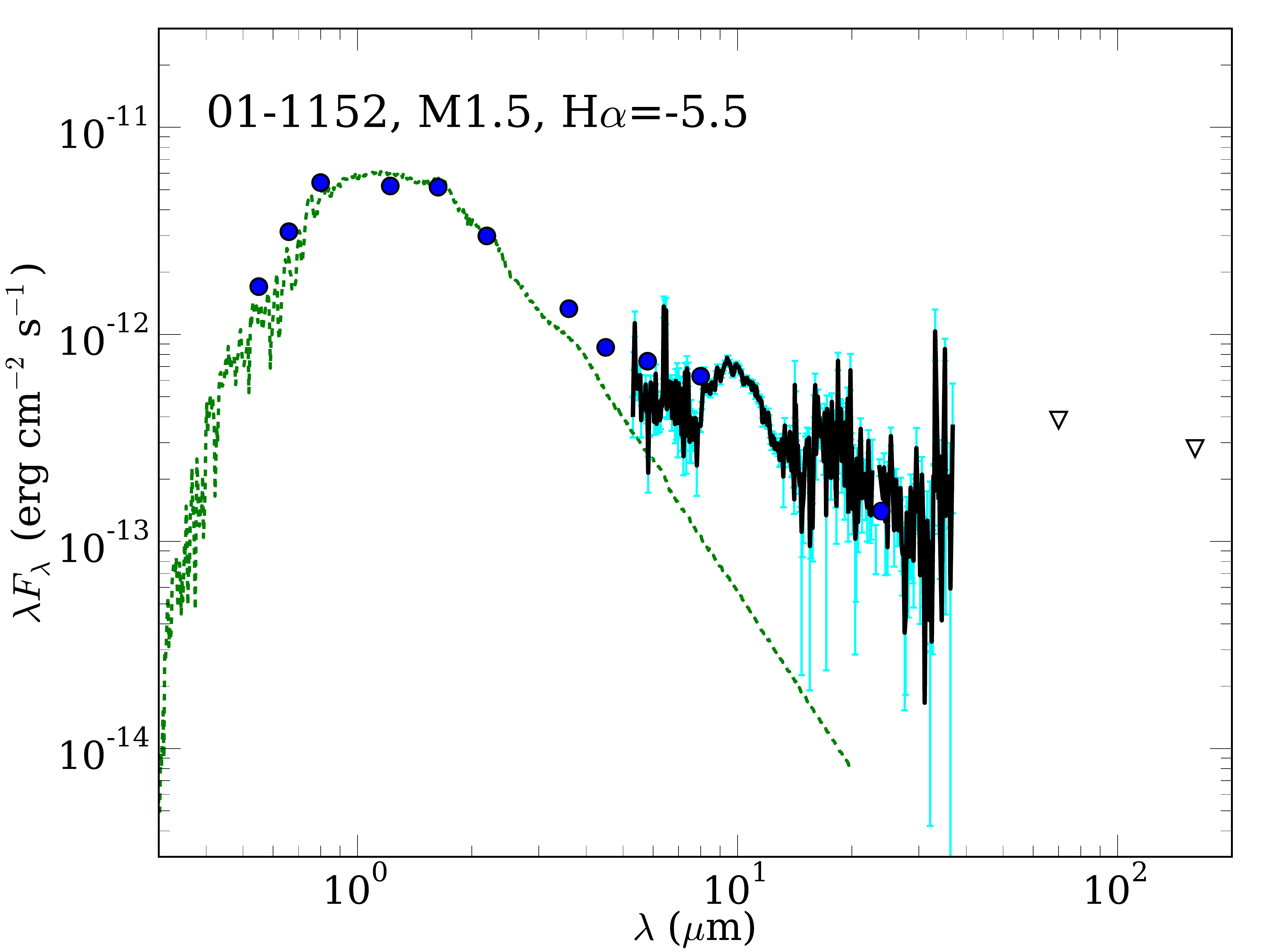} &
\includegraphics[width=0.24\linewidth]{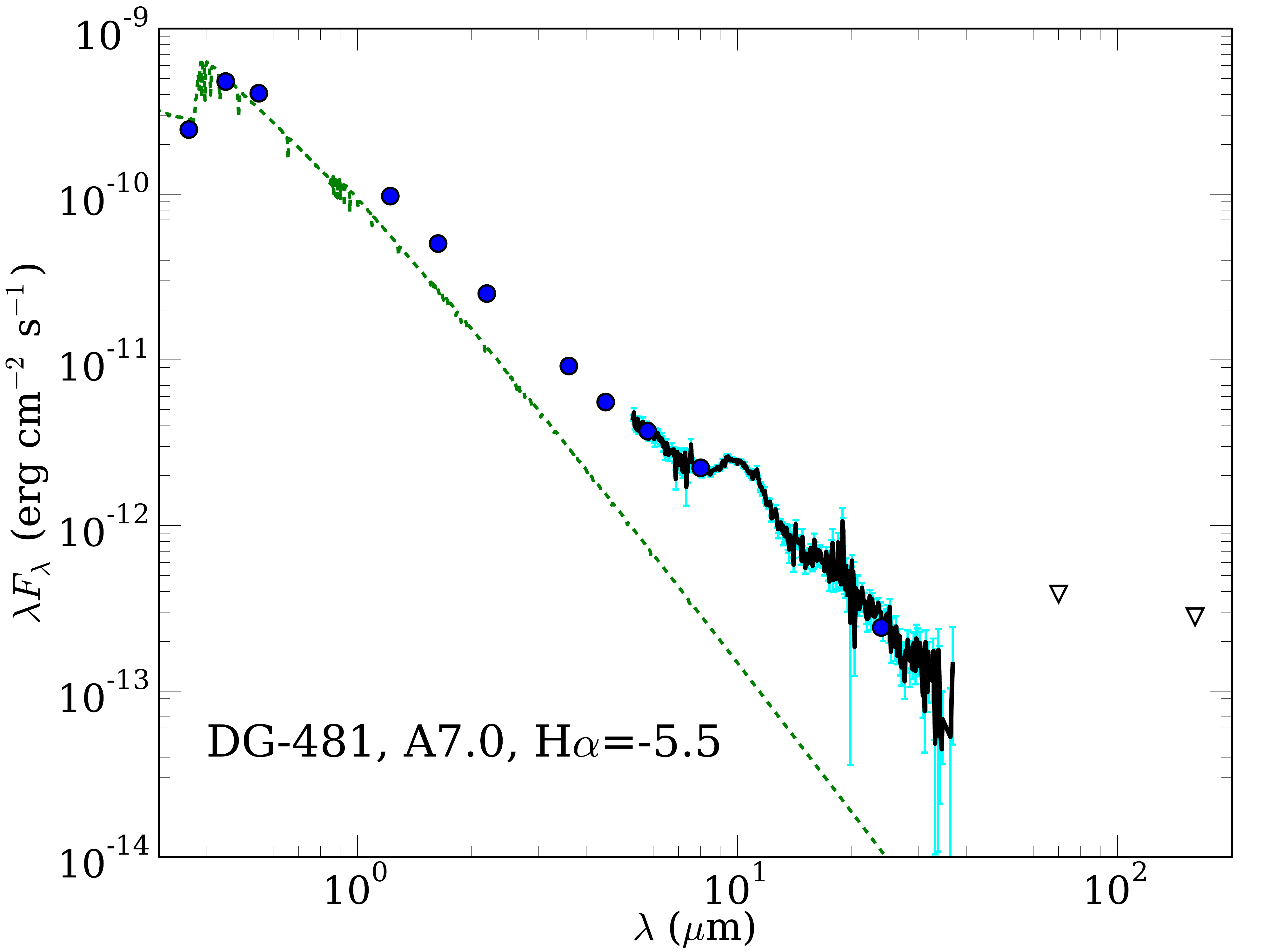} \\
\includegraphics[width=0.24\linewidth]{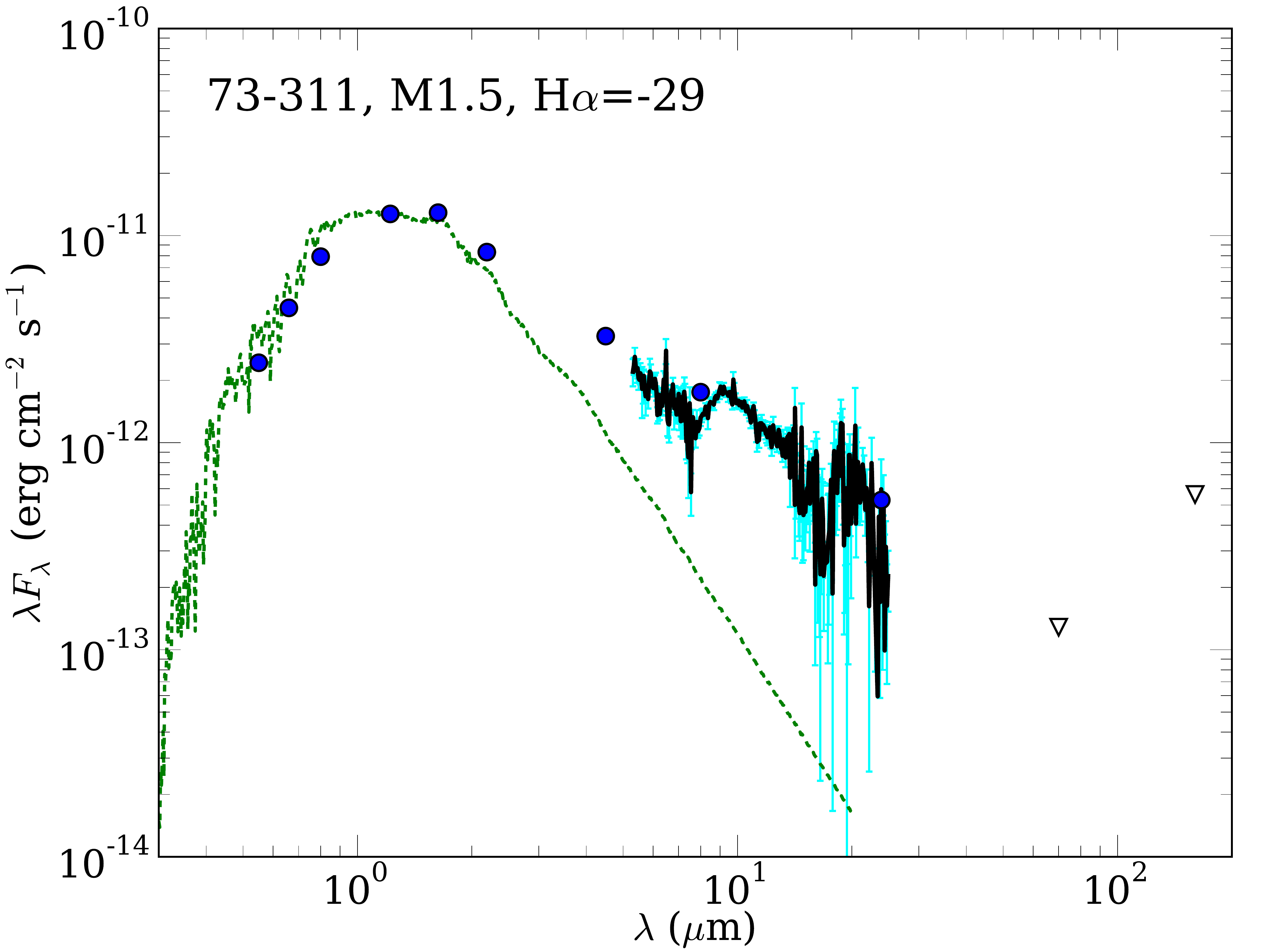} &
\includegraphics[width=0.24\linewidth]{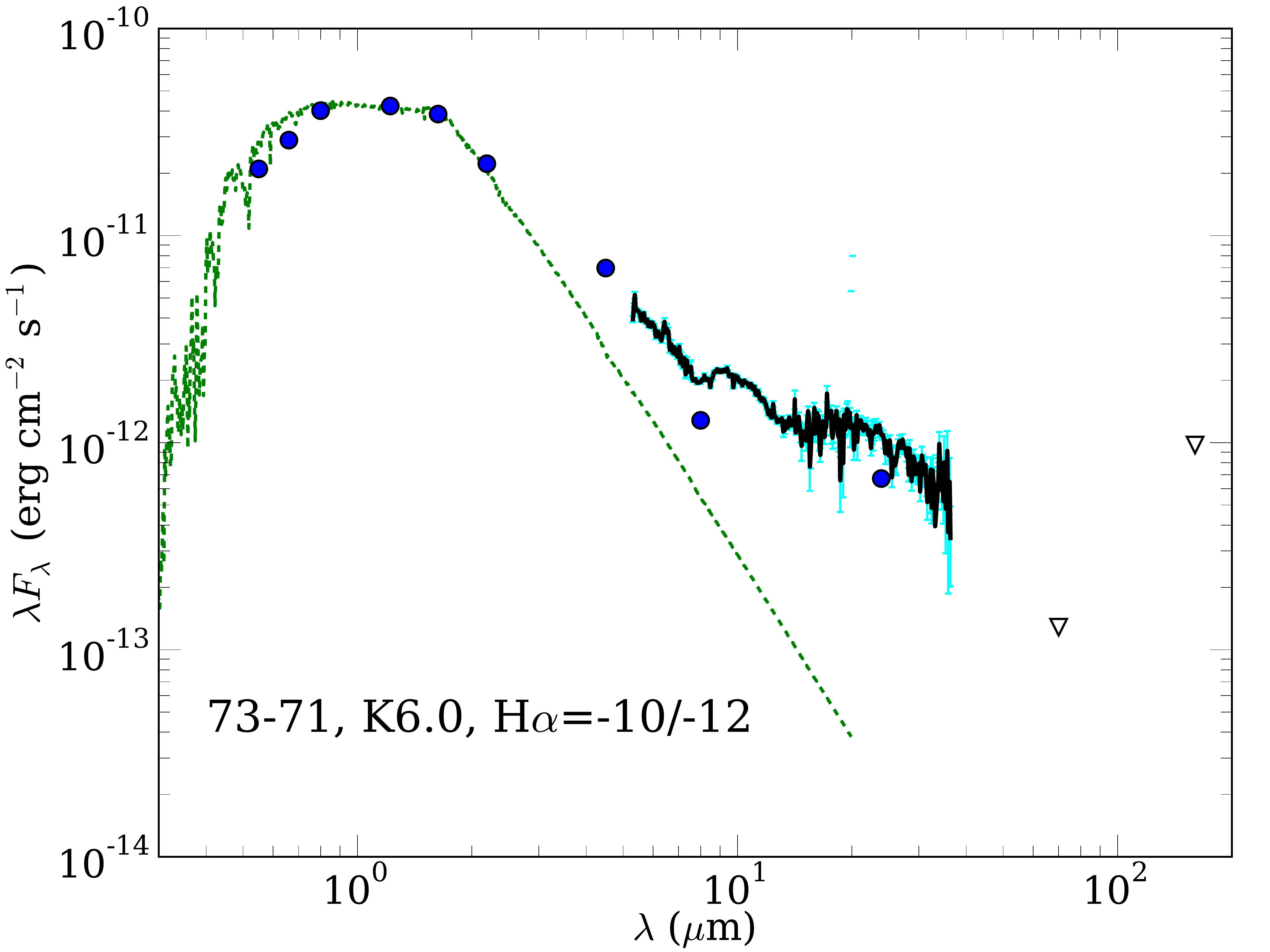} &
\includegraphics[width=0.24\linewidth]{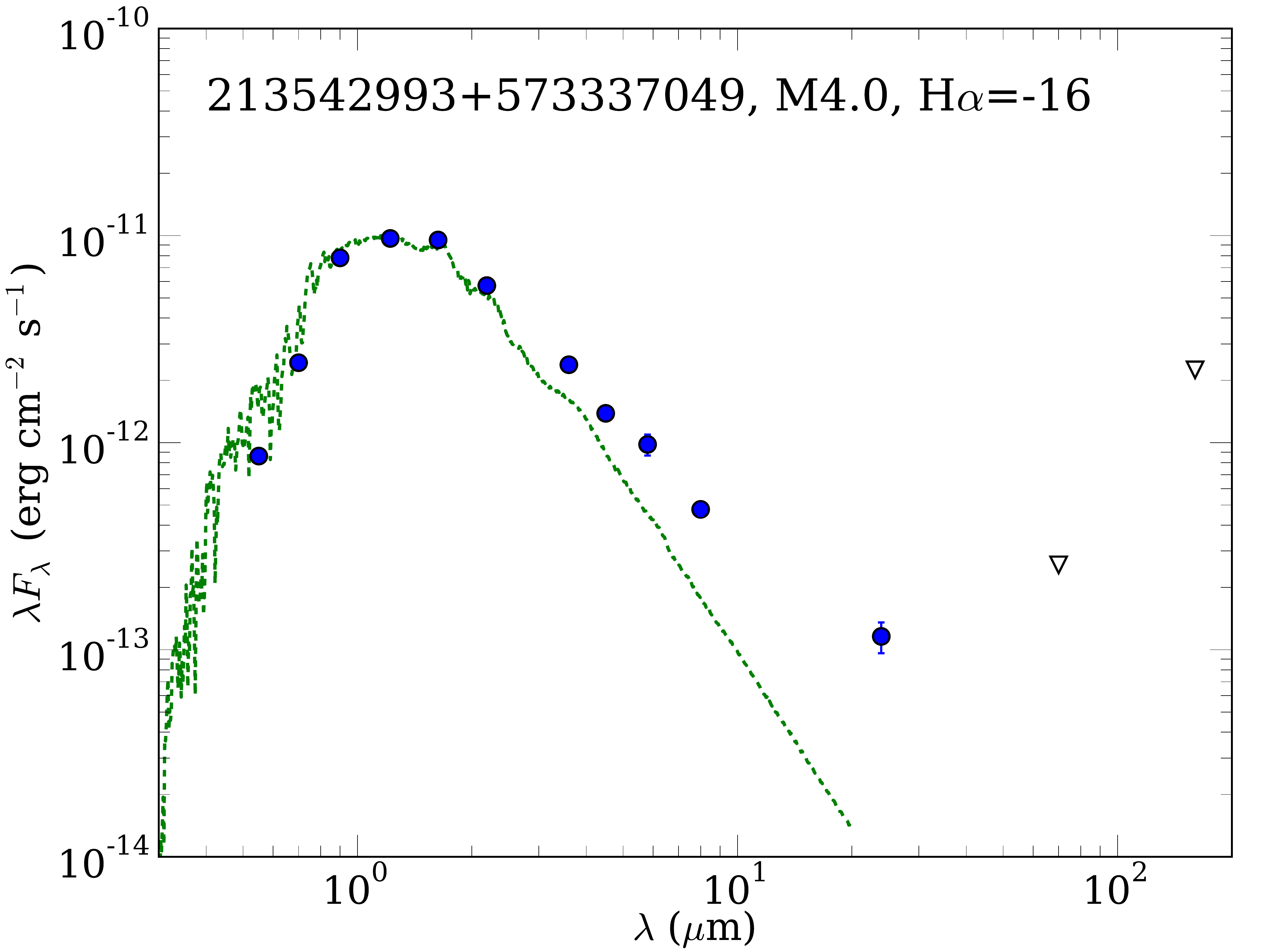} &
\includegraphics[width=0.24\linewidth]{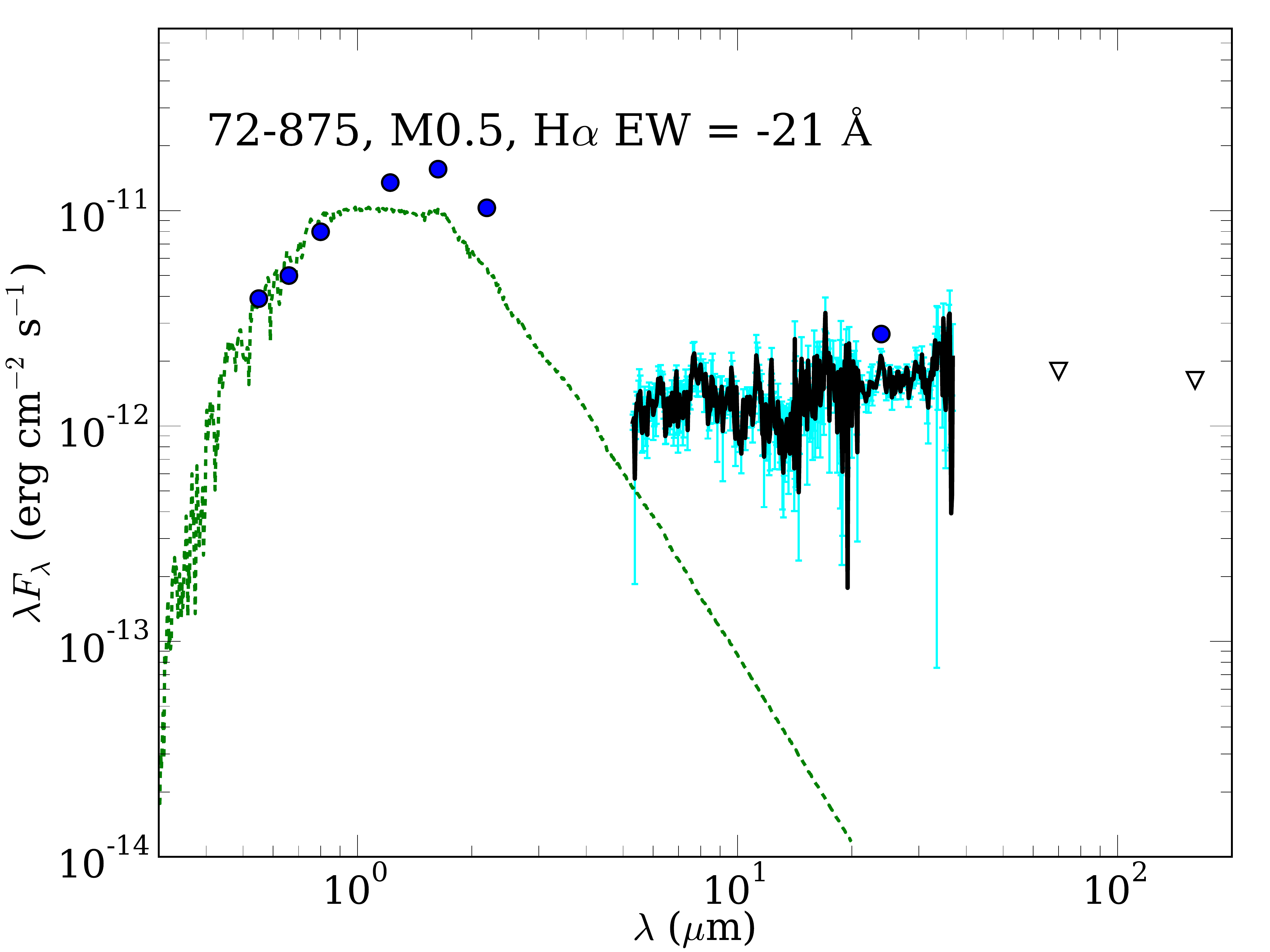} \\
\includegraphics[width=0.24\linewidth]{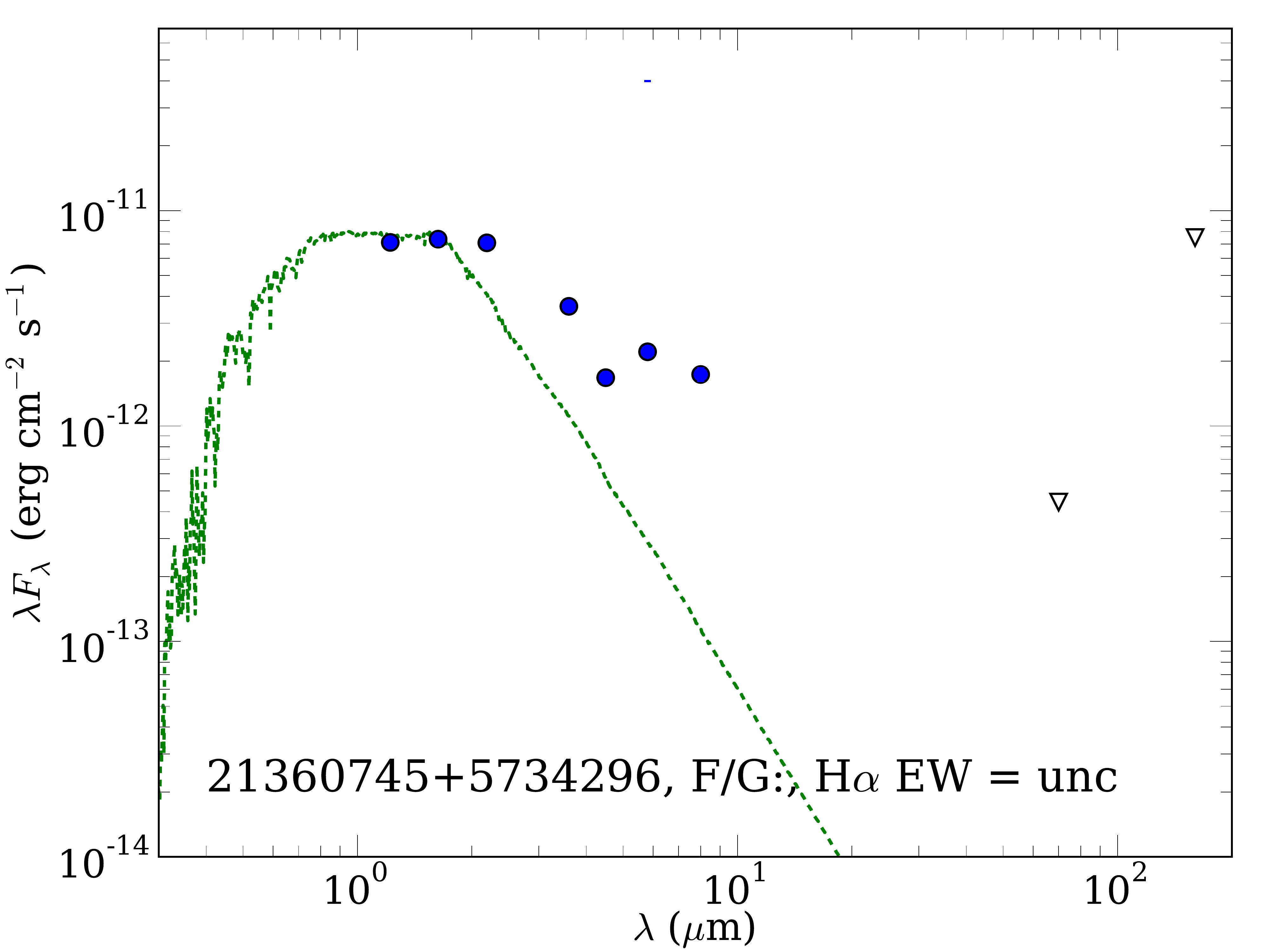} &
\includegraphics[width=0.24\linewidth]{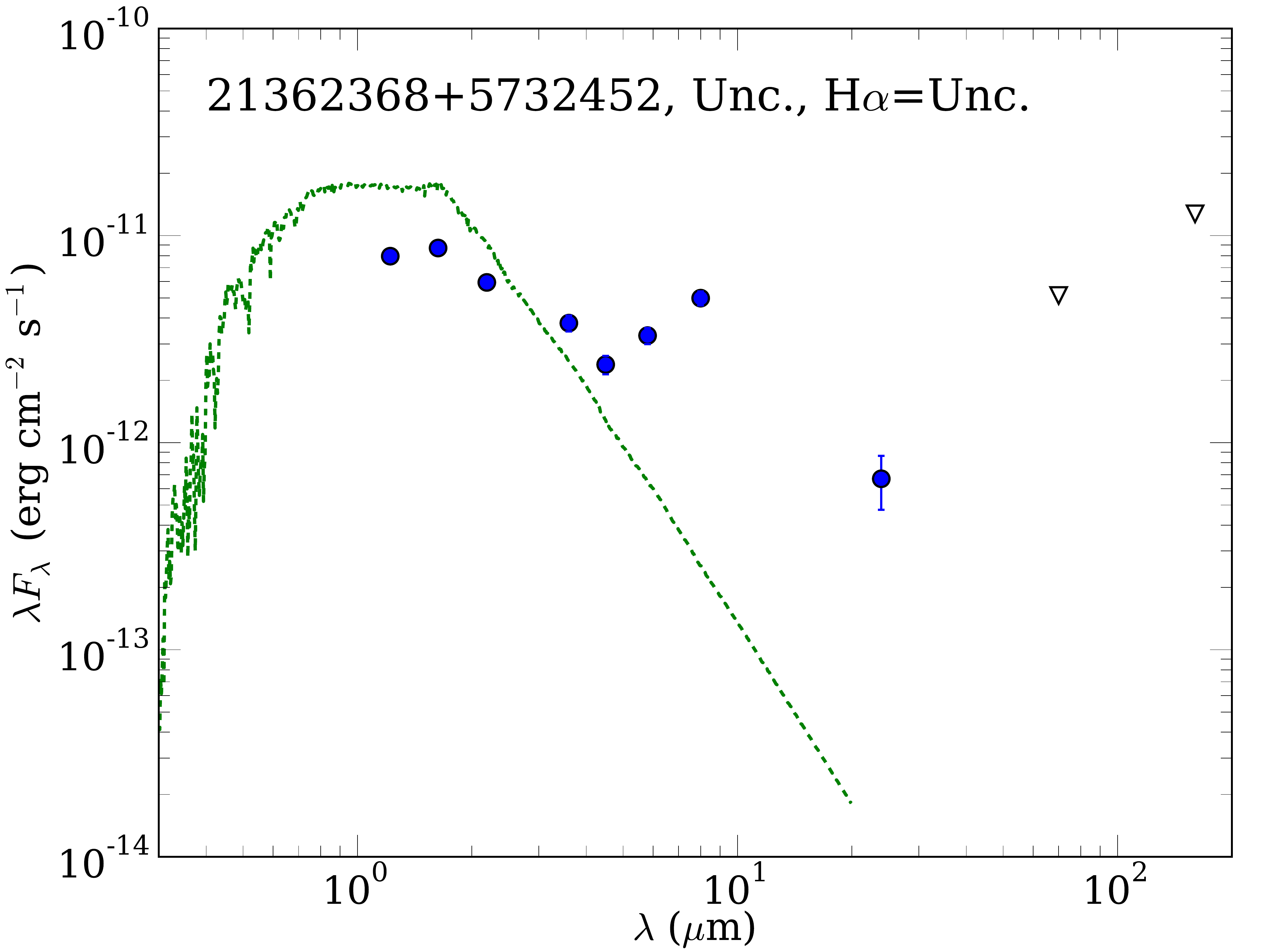} &
\includegraphics[width=0.24\linewidth]{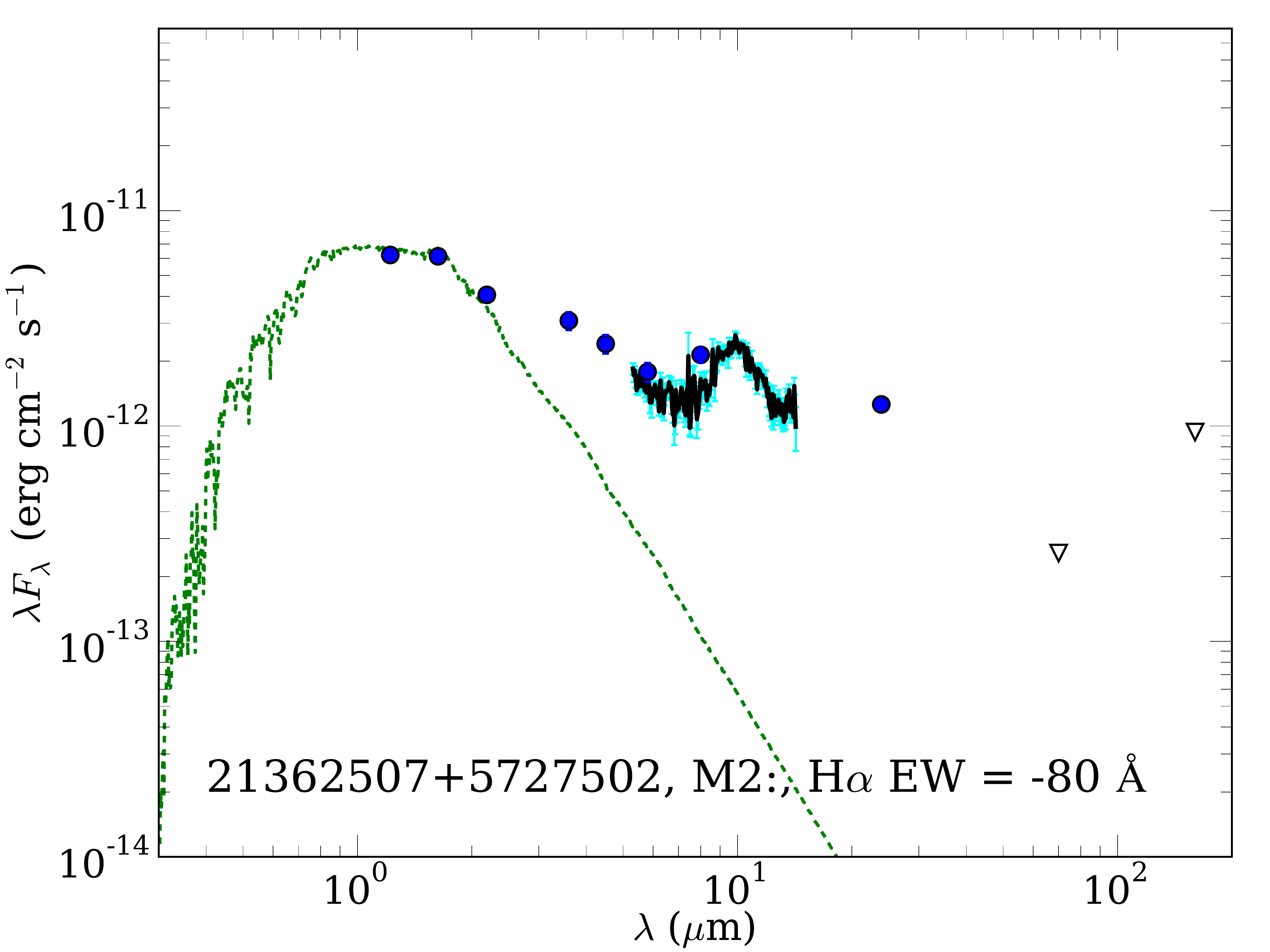} &
\includegraphics[width=0.24\linewidth]{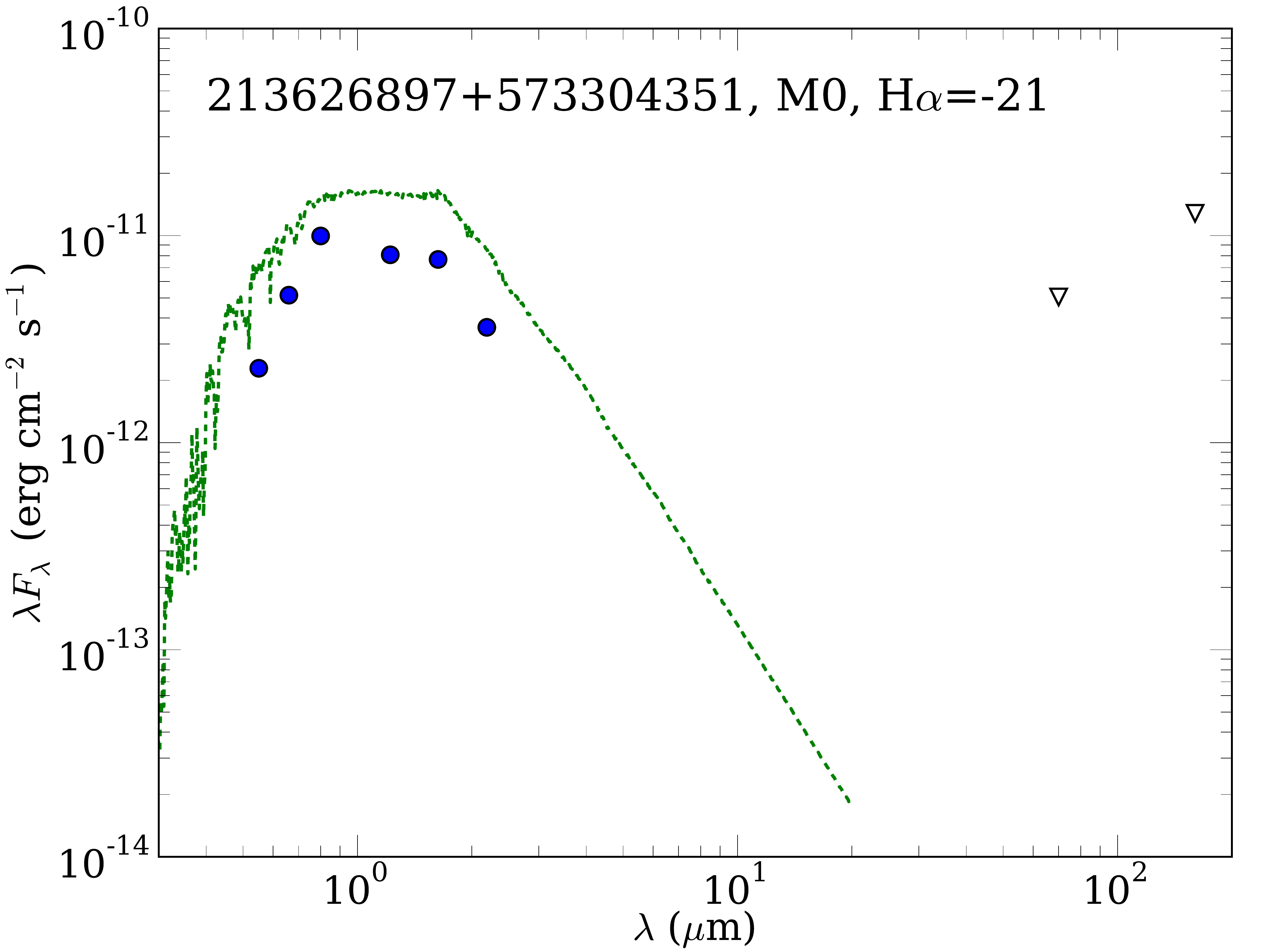} \\
\includegraphics[width=0.24\linewidth]{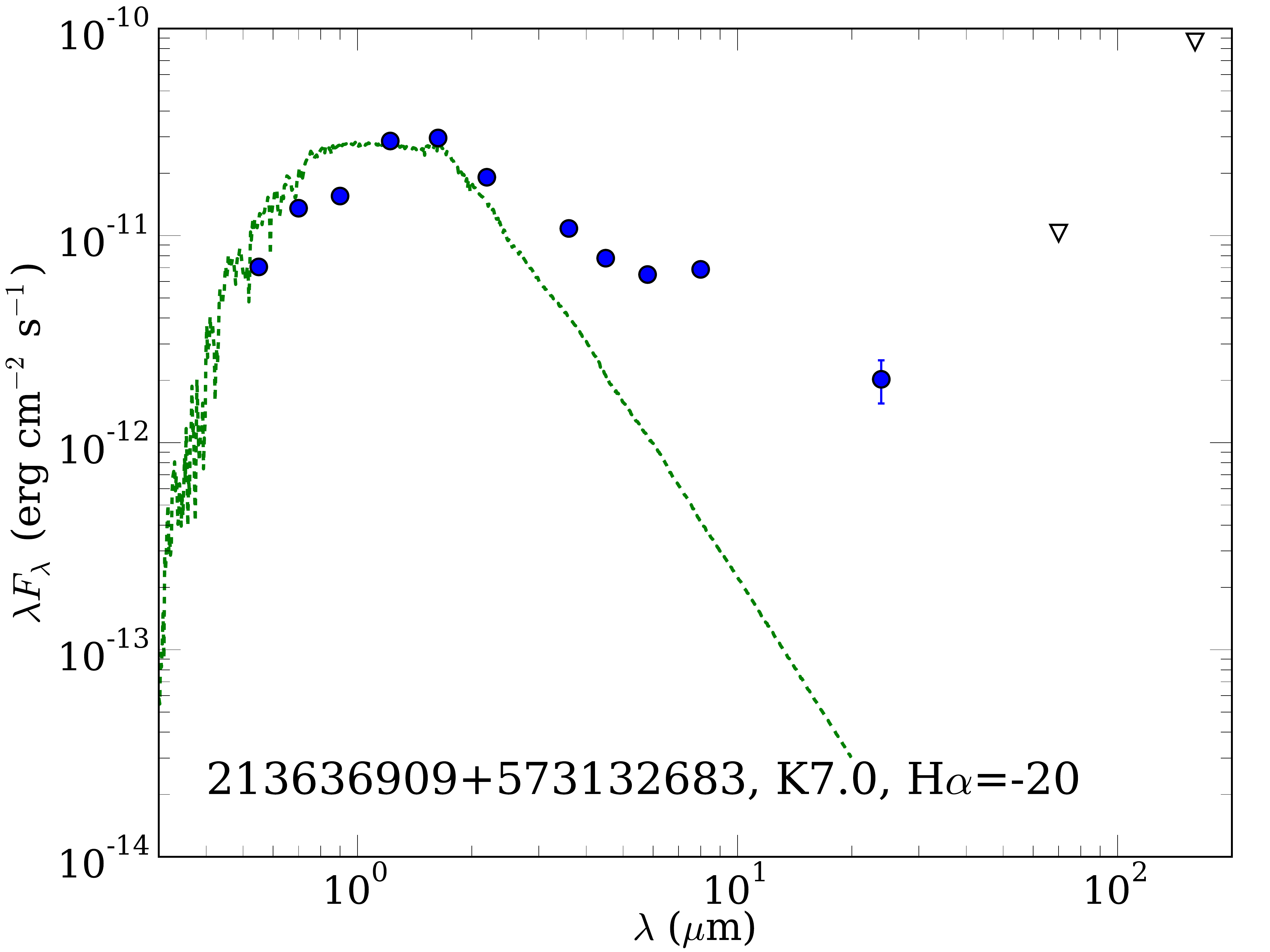} &
\includegraphics[width=0.24\linewidth]{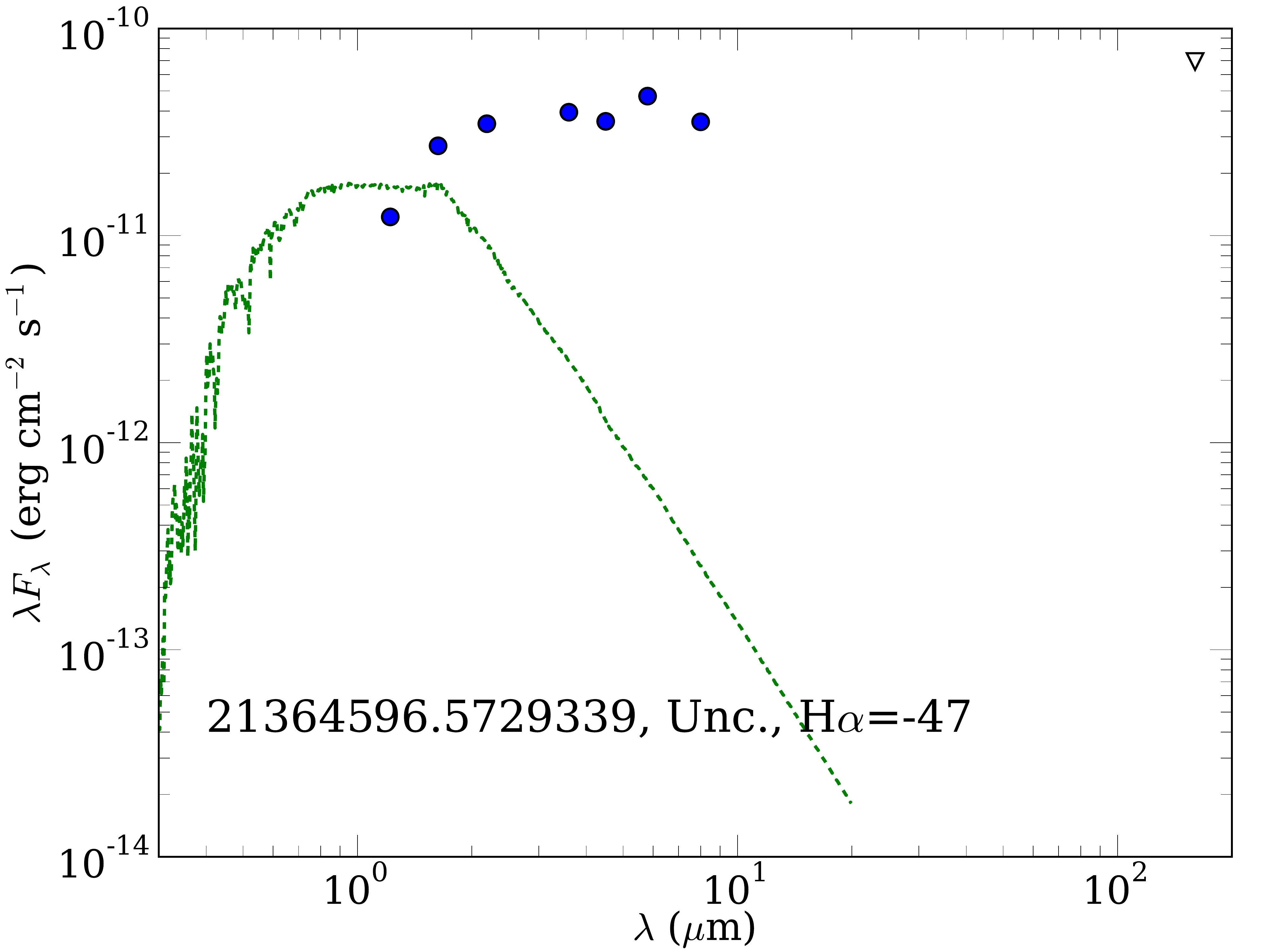} &
\includegraphics[width=0.24\linewidth]{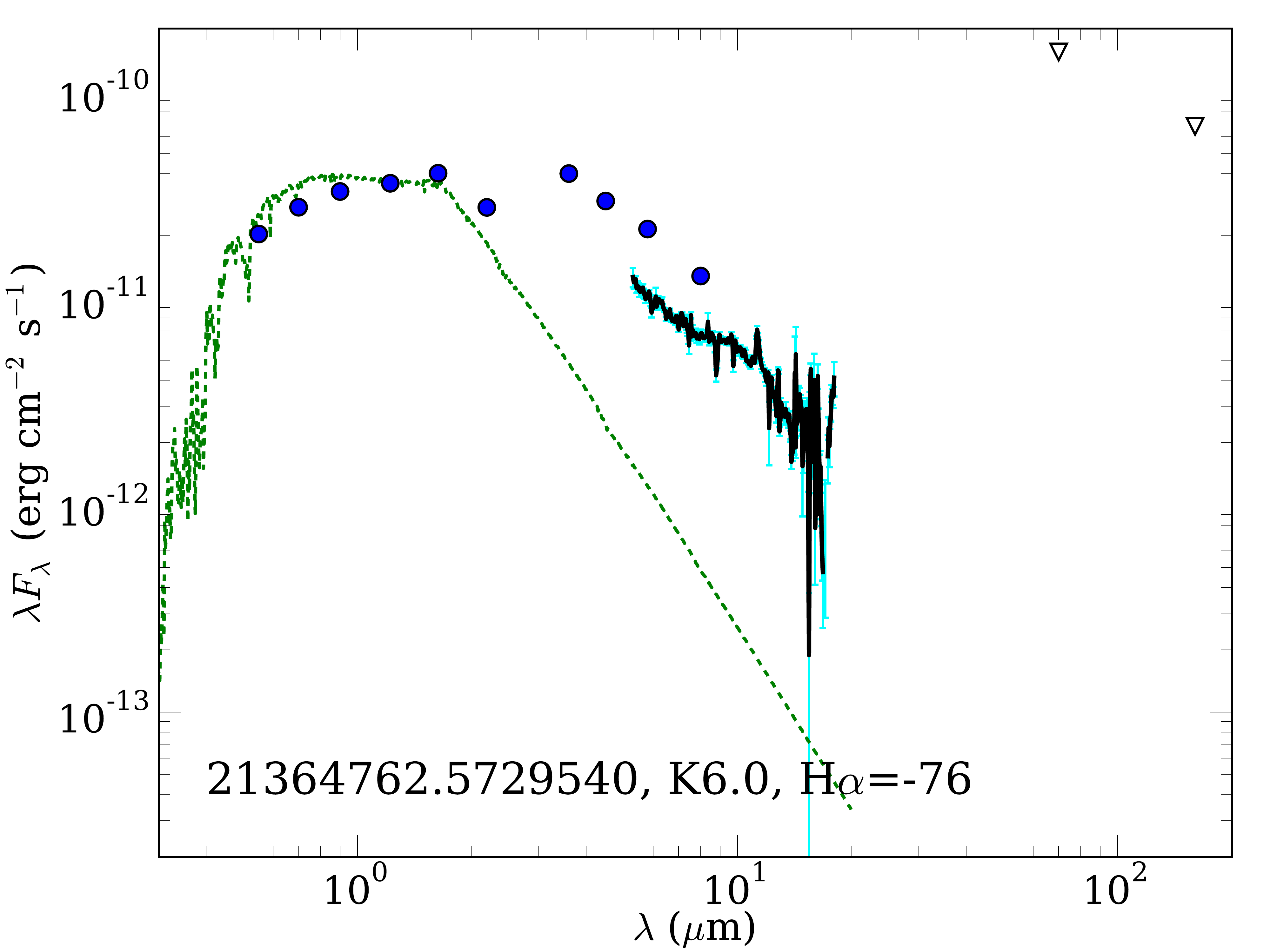} &
\includegraphics[width=0.24\linewidth]{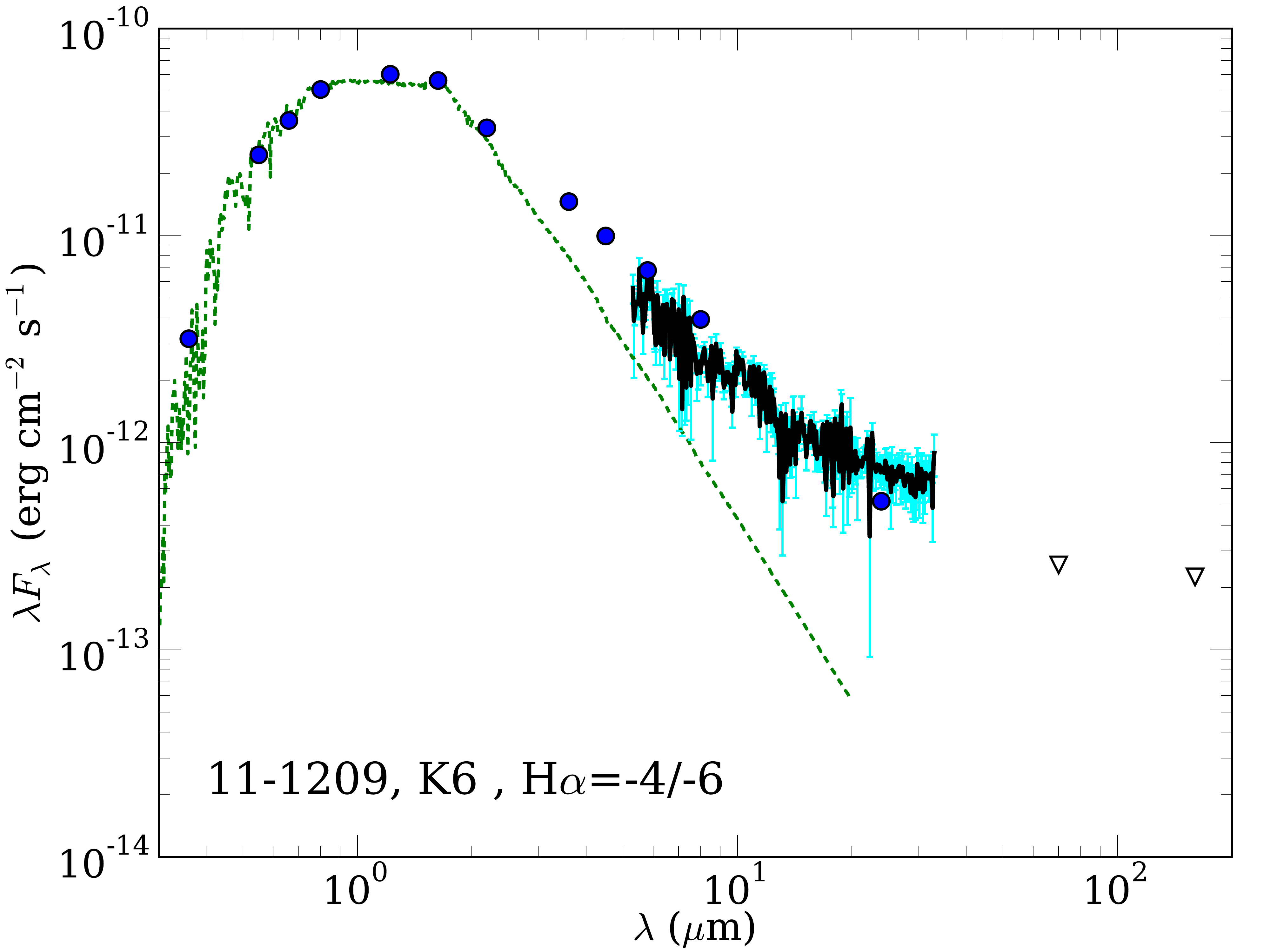} \\
\includegraphics[width=0.24\linewidth]{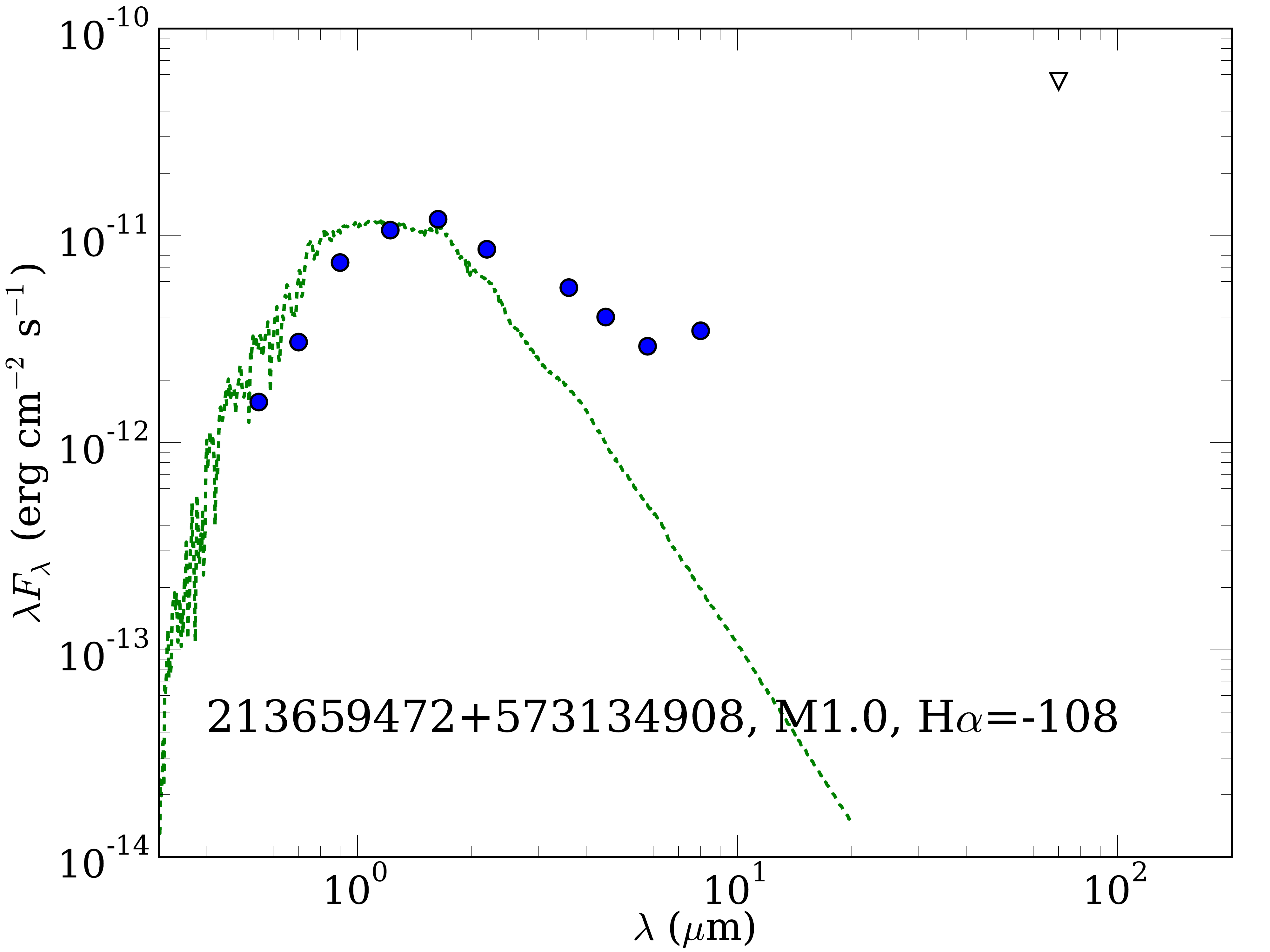} &
\includegraphics[width=0.24\linewidth]{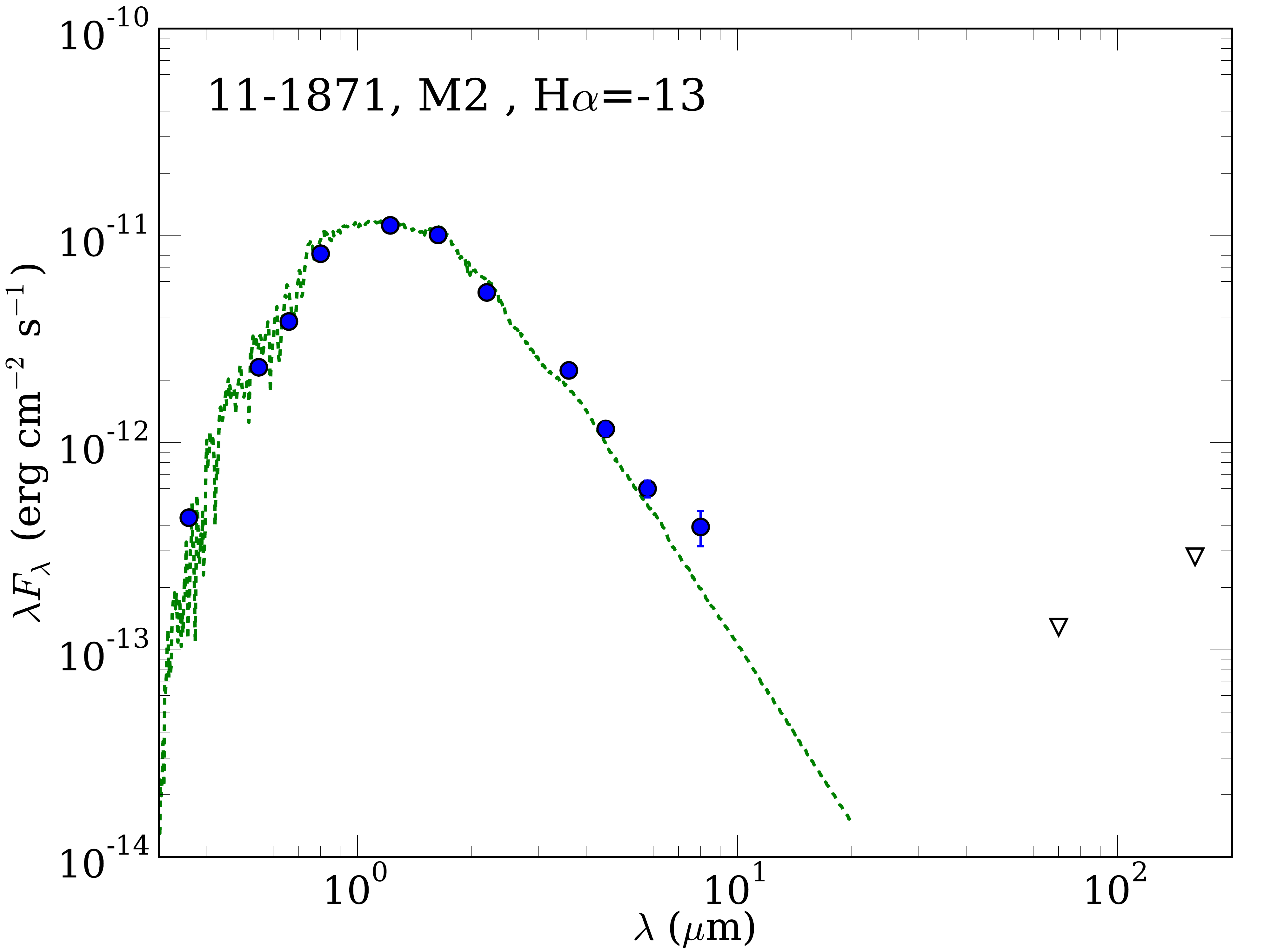} &
\includegraphics[width=0.24\linewidth]{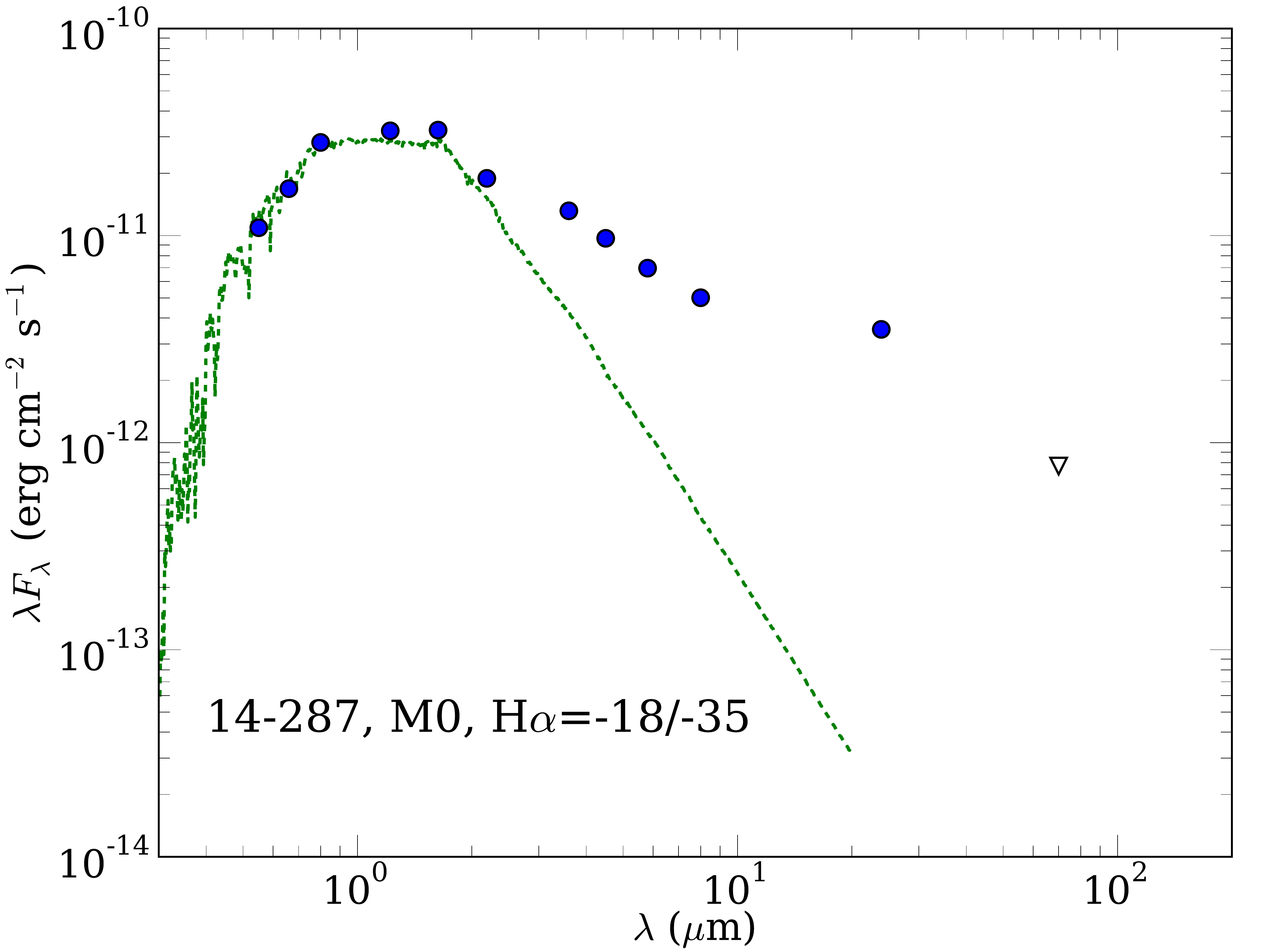} &
\includegraphics[width=0.24\linewidth]{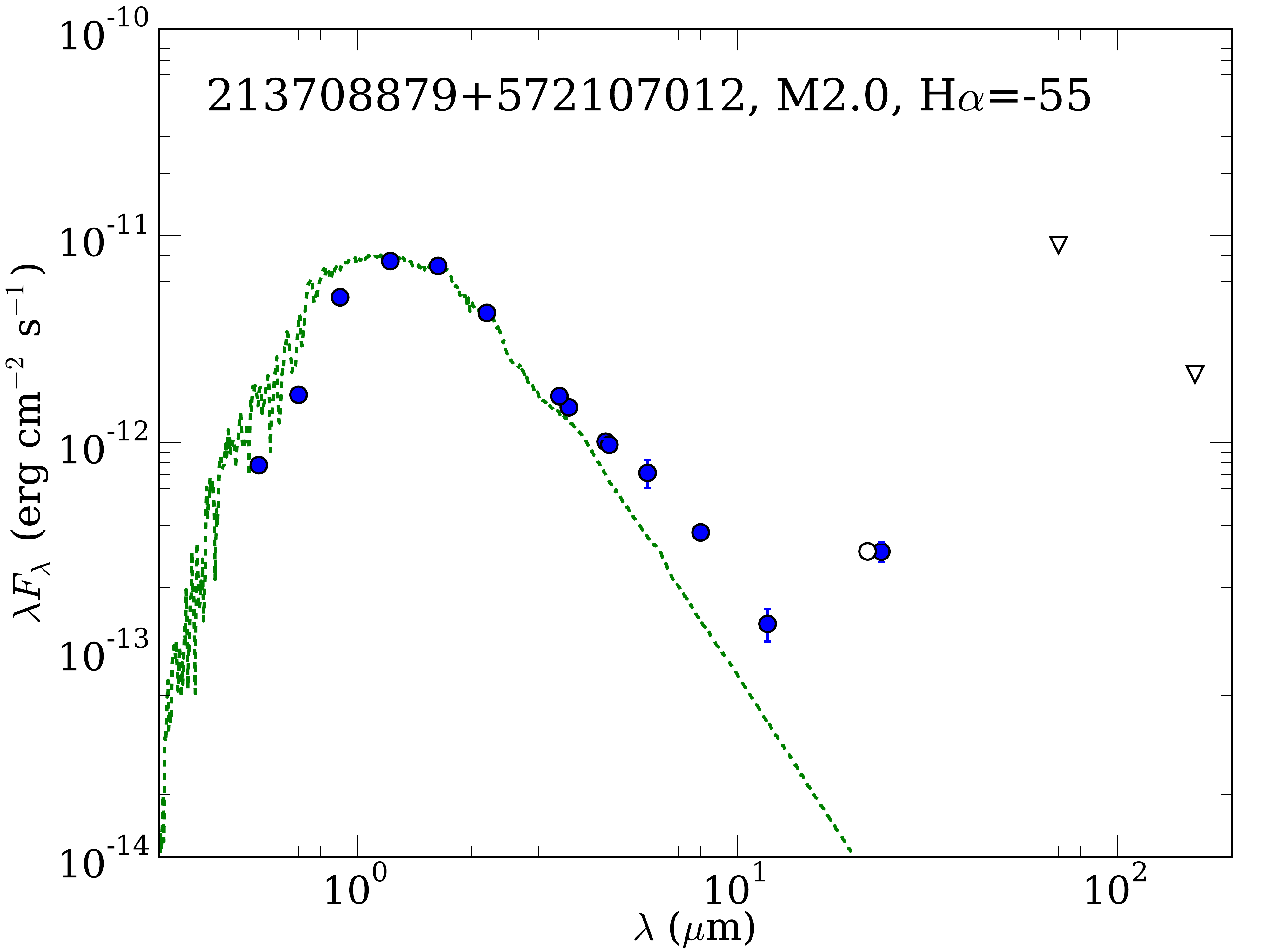} \\
\includegraphics[width=0.24\linewidth]{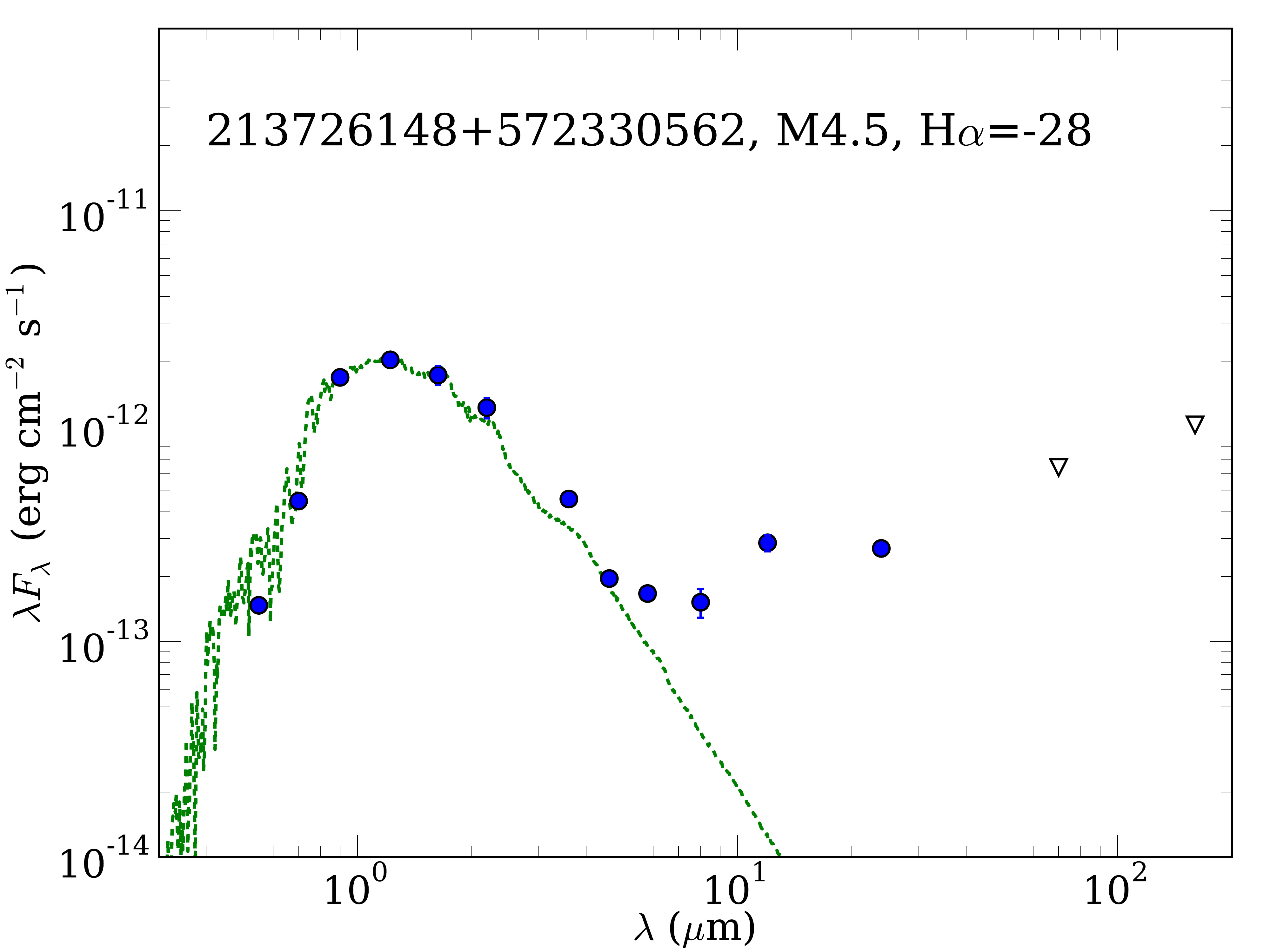} &
\includegraphics[width=0.24\linewidth]{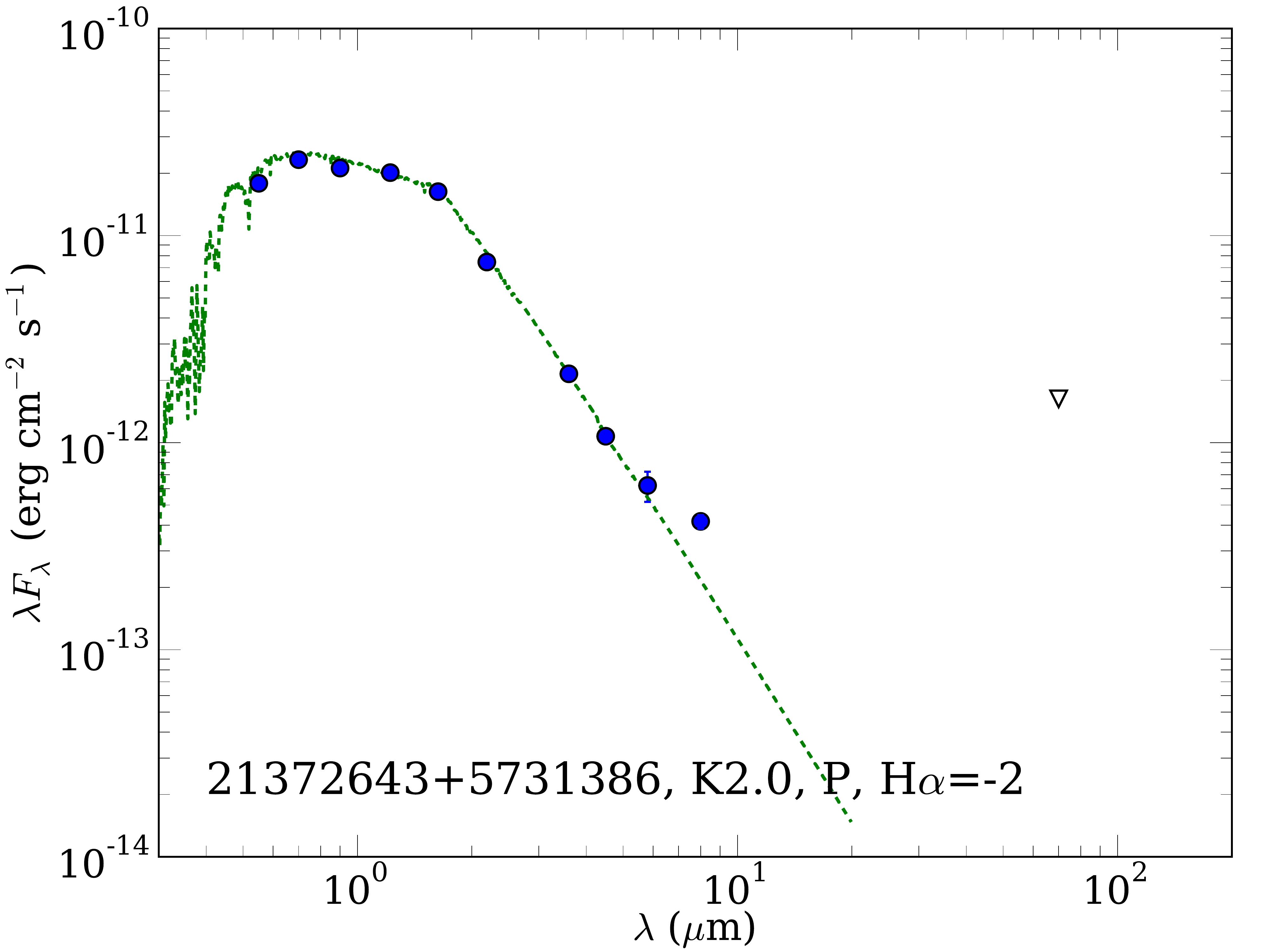} &
\includegraphics[width=0.24\linewidth]{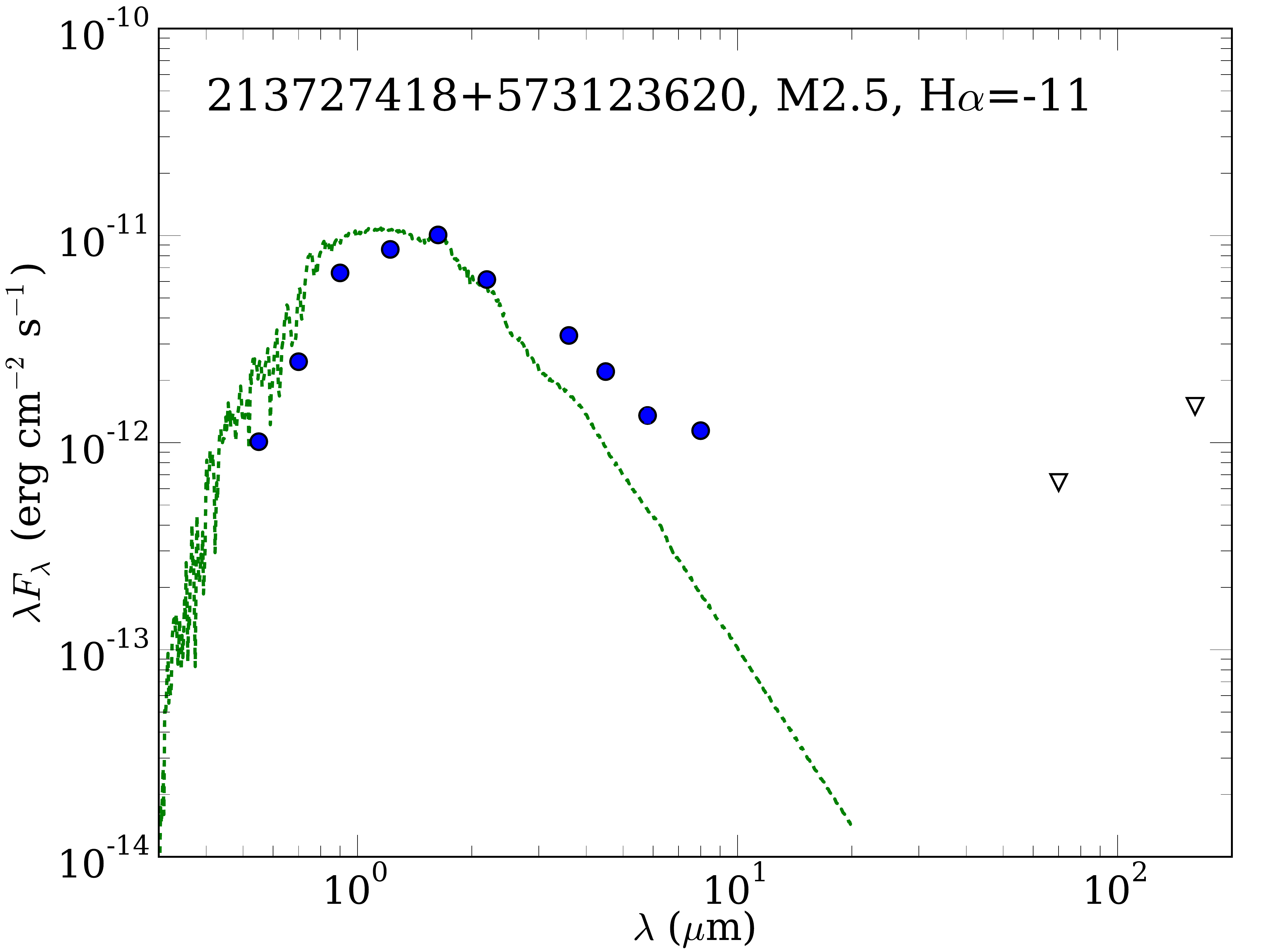} &
\includegraphics[width=0.24\linewidth]{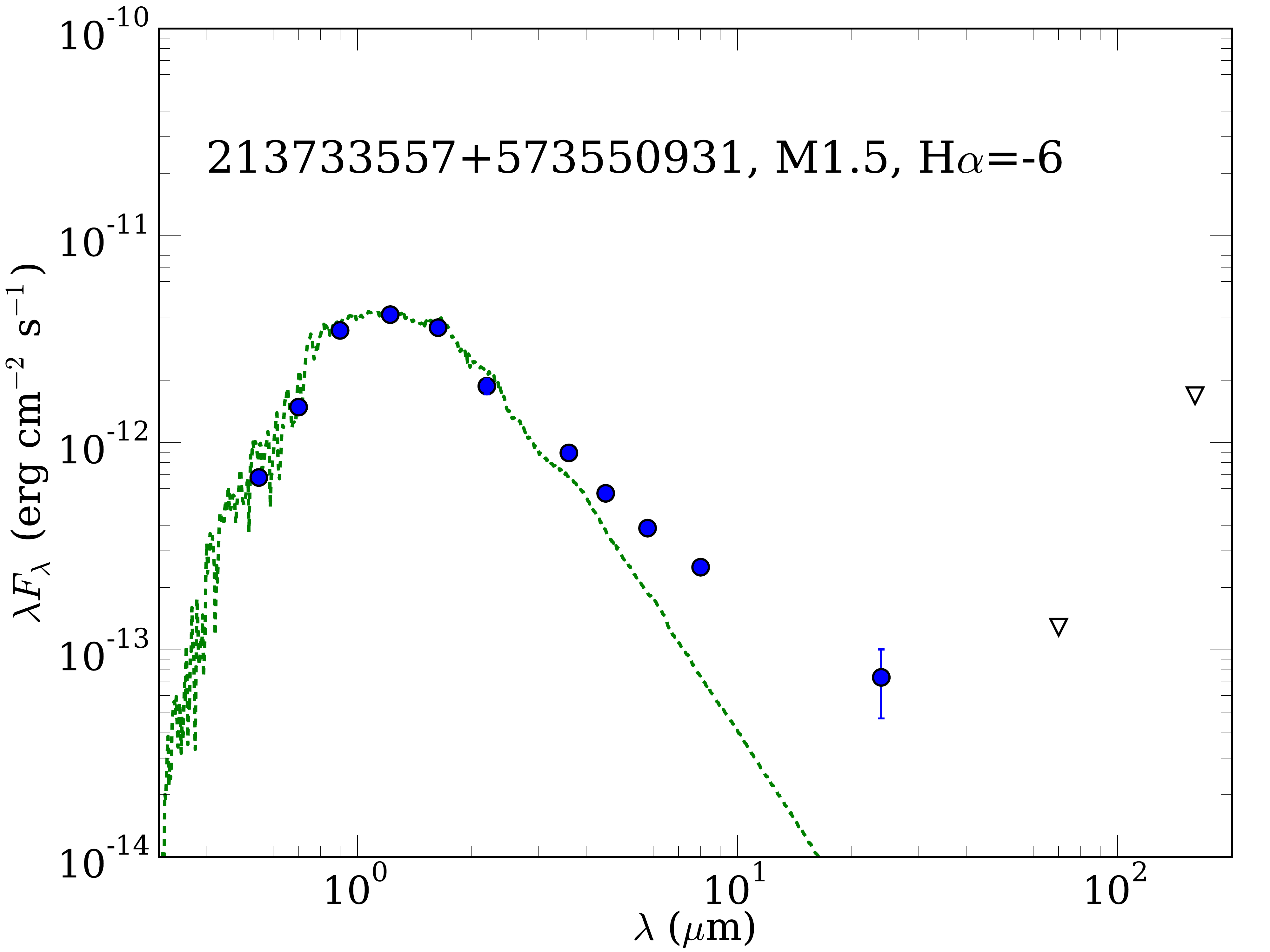} \\
\includegraphics[width=0.24\linewidth]{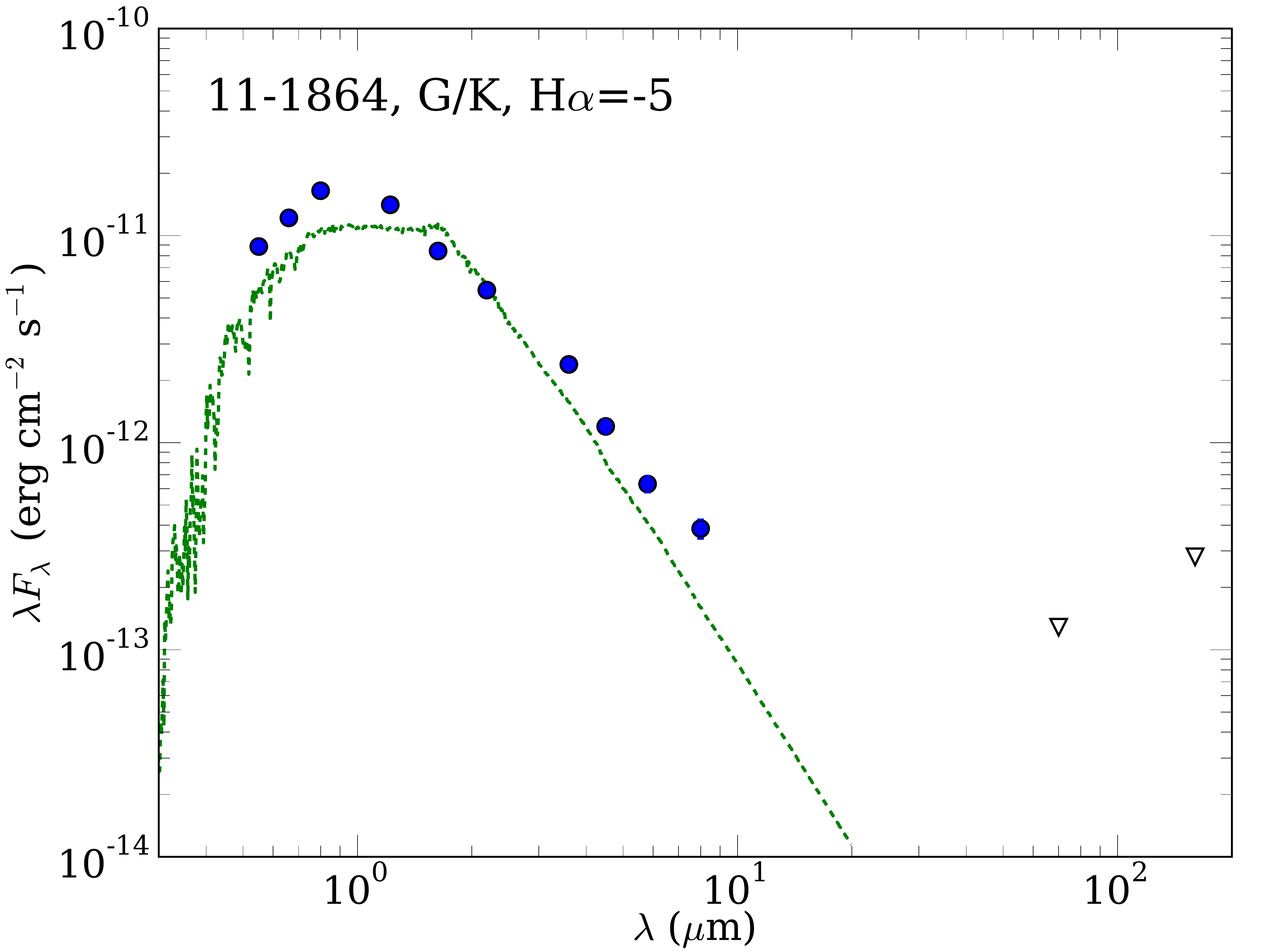} &
\includegraphics[width=0.24\linewidth]{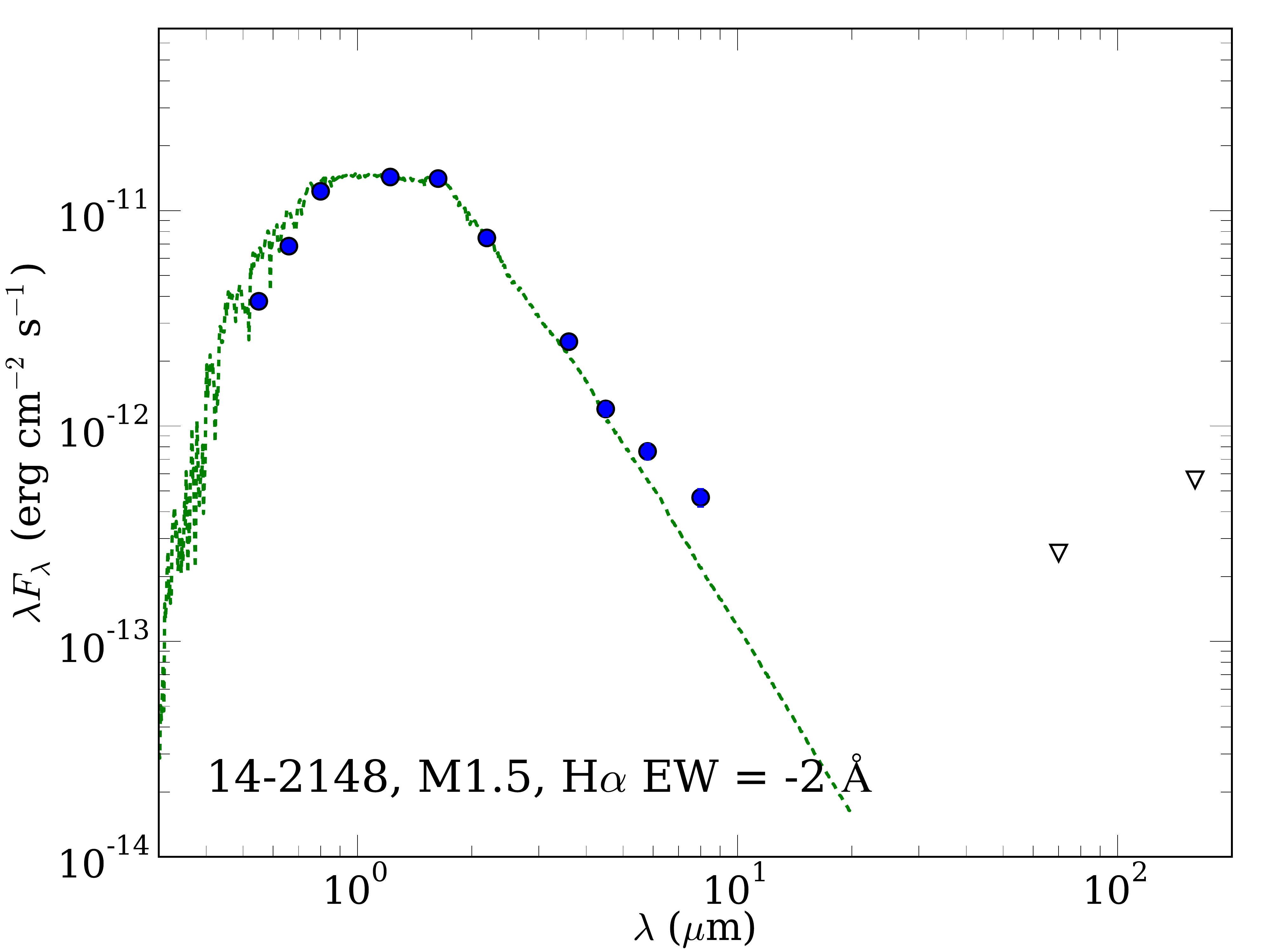} &
\includegraphics[width=0.24\linewidth]{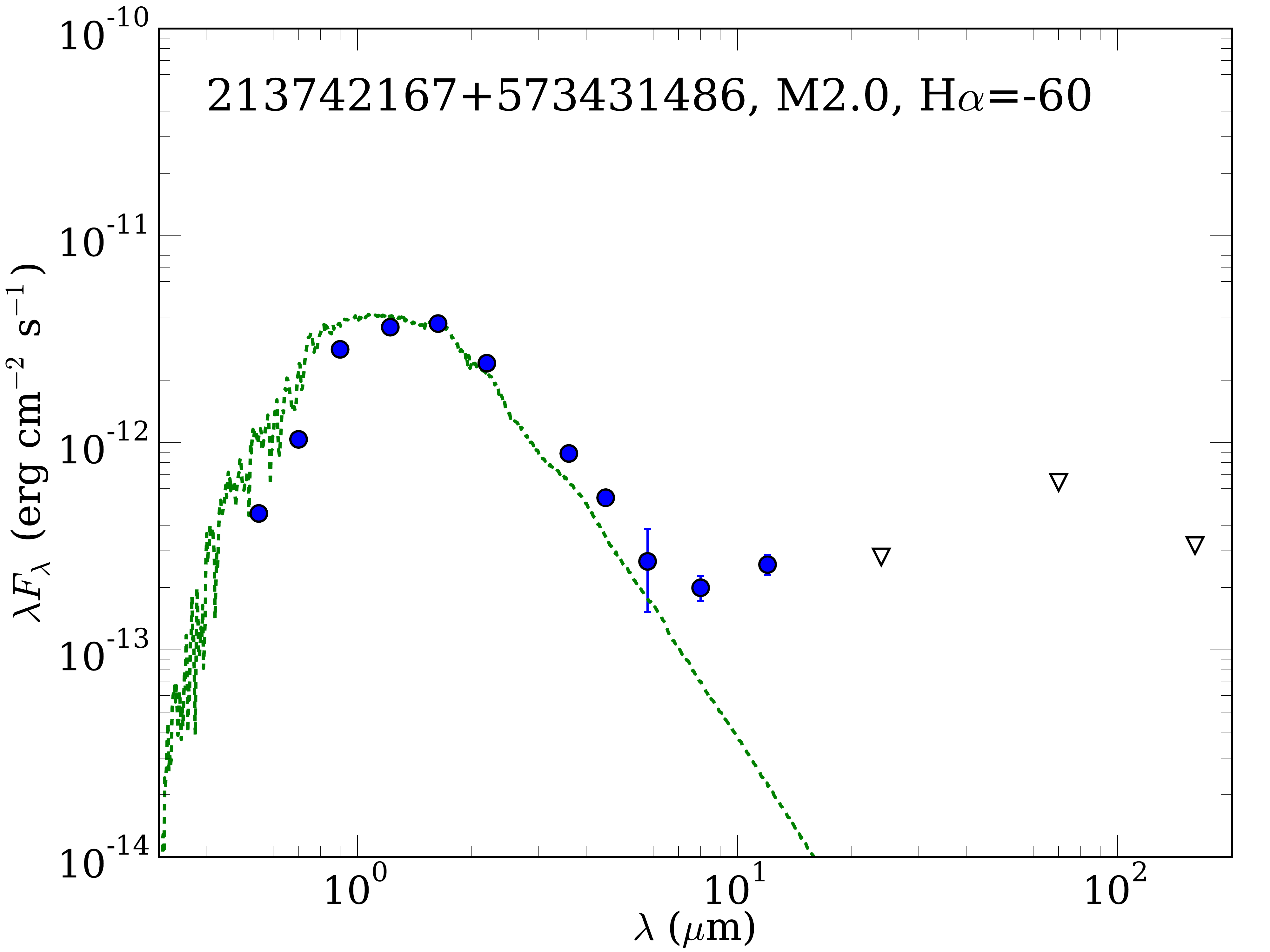} &
\includegraphics[width=0.24\linewidth]{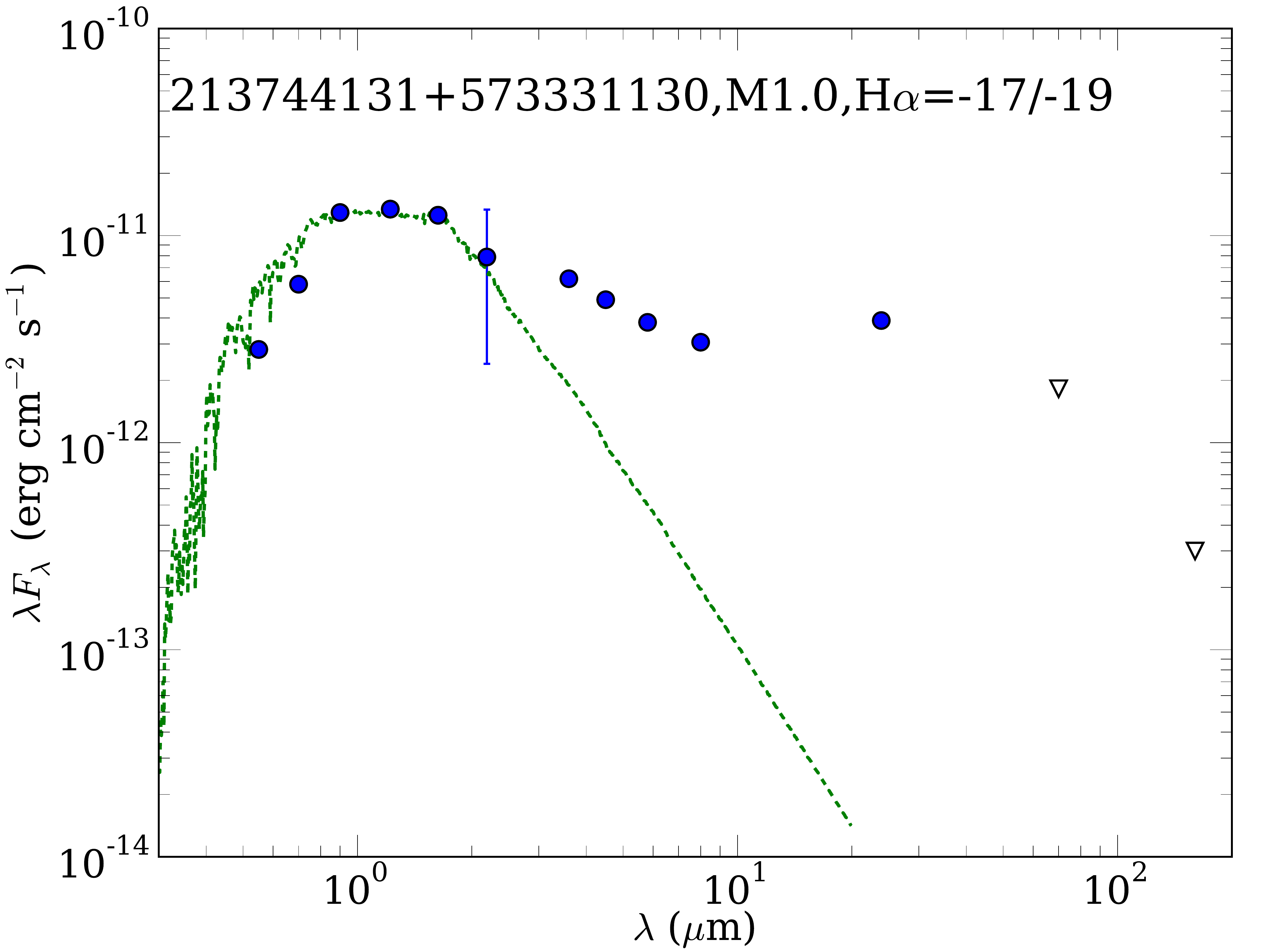} \\
\end{tabular}
\caption{SEDs of the objects with upper limits only. The first two objects are in the NGC\,7160 cluster,
the rest are in Tr\,37. Only objects with confirmed disks 
(Spitzer excesses) are displayed. 
Filled symbols mark detections at different wavelengths. Errorbars are shown in blue
for both photometry points (often smaller than the symbols) and the IRS spectra. Upper limits
are marked as inverted open triangles. Marginal detections (close to 3$\sigma$ or affected by nebulosity)
are marked as open circles. The information about spectral types and H$\alpha$ emission
from the literature is also listed. A photospheric MARCS model is
displayed for comparison.\label{uplims1-fig}}
\end{figure*}

\begin{figure*}
\centering
\begin{tabular}{cccc}
\includegraphics[width=0.24\linewidth]{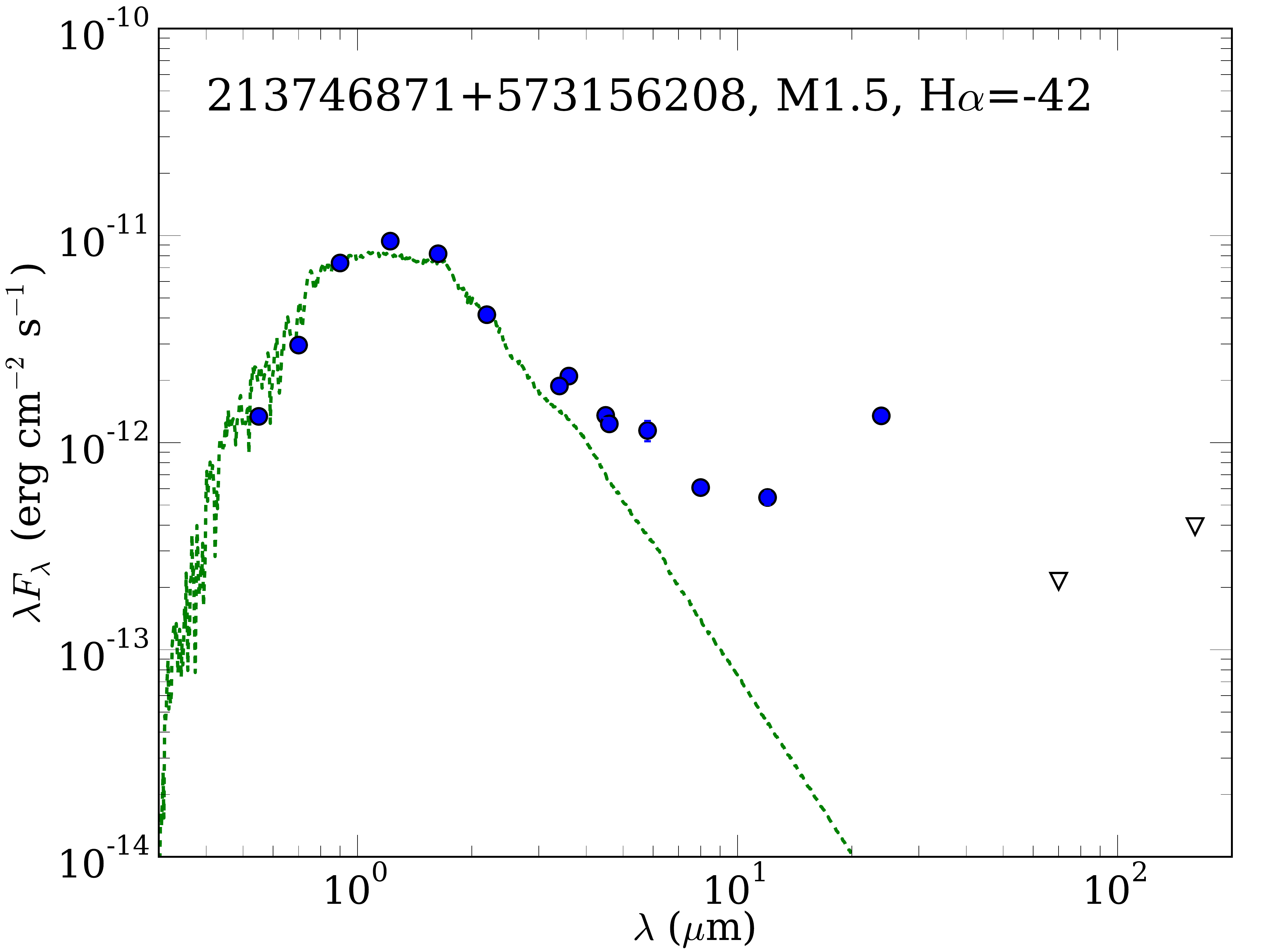} &
\includegraphics[width=0.24\linewidth]{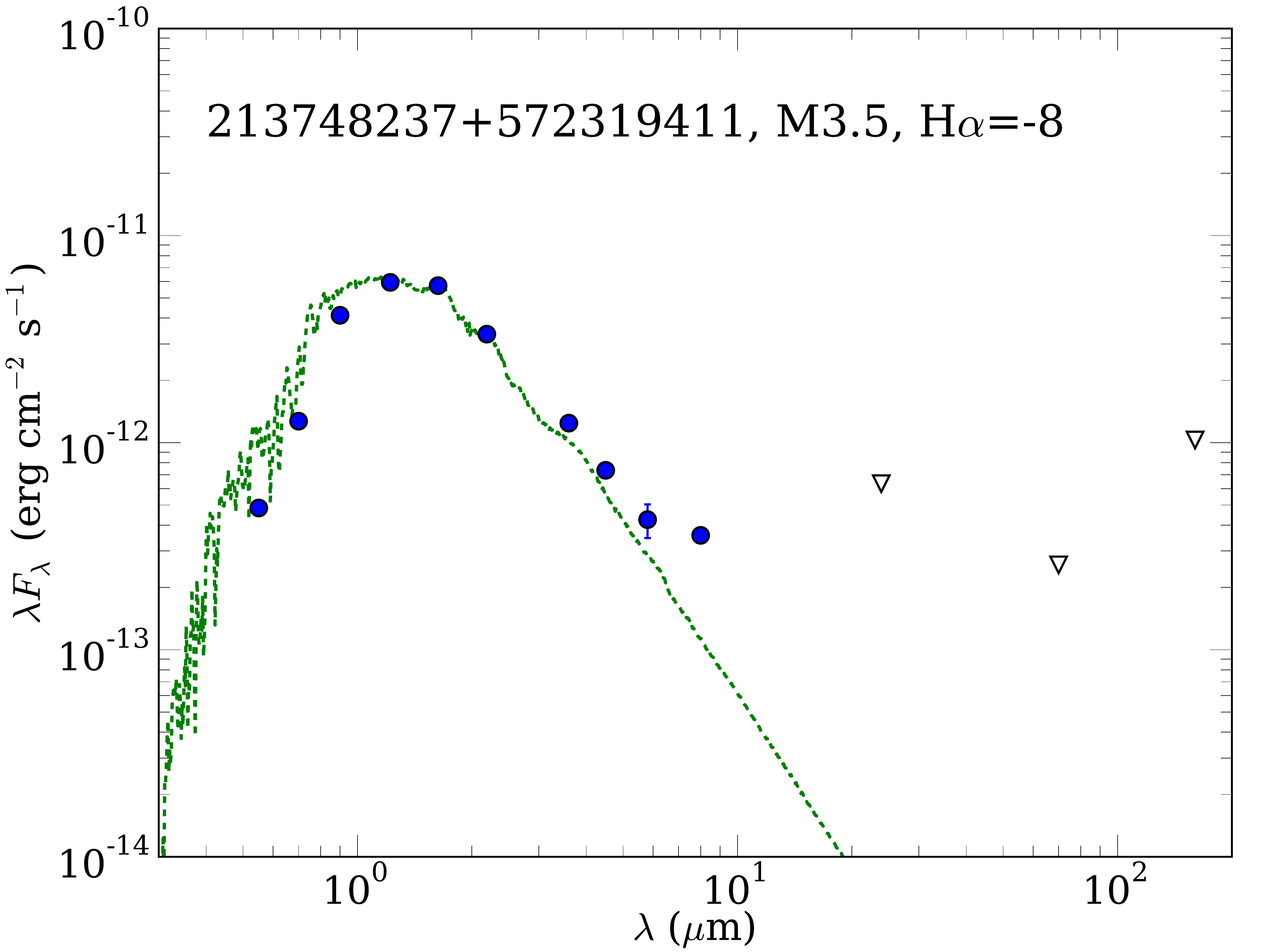} &
\includegraphics[width=0.24\linewidth]{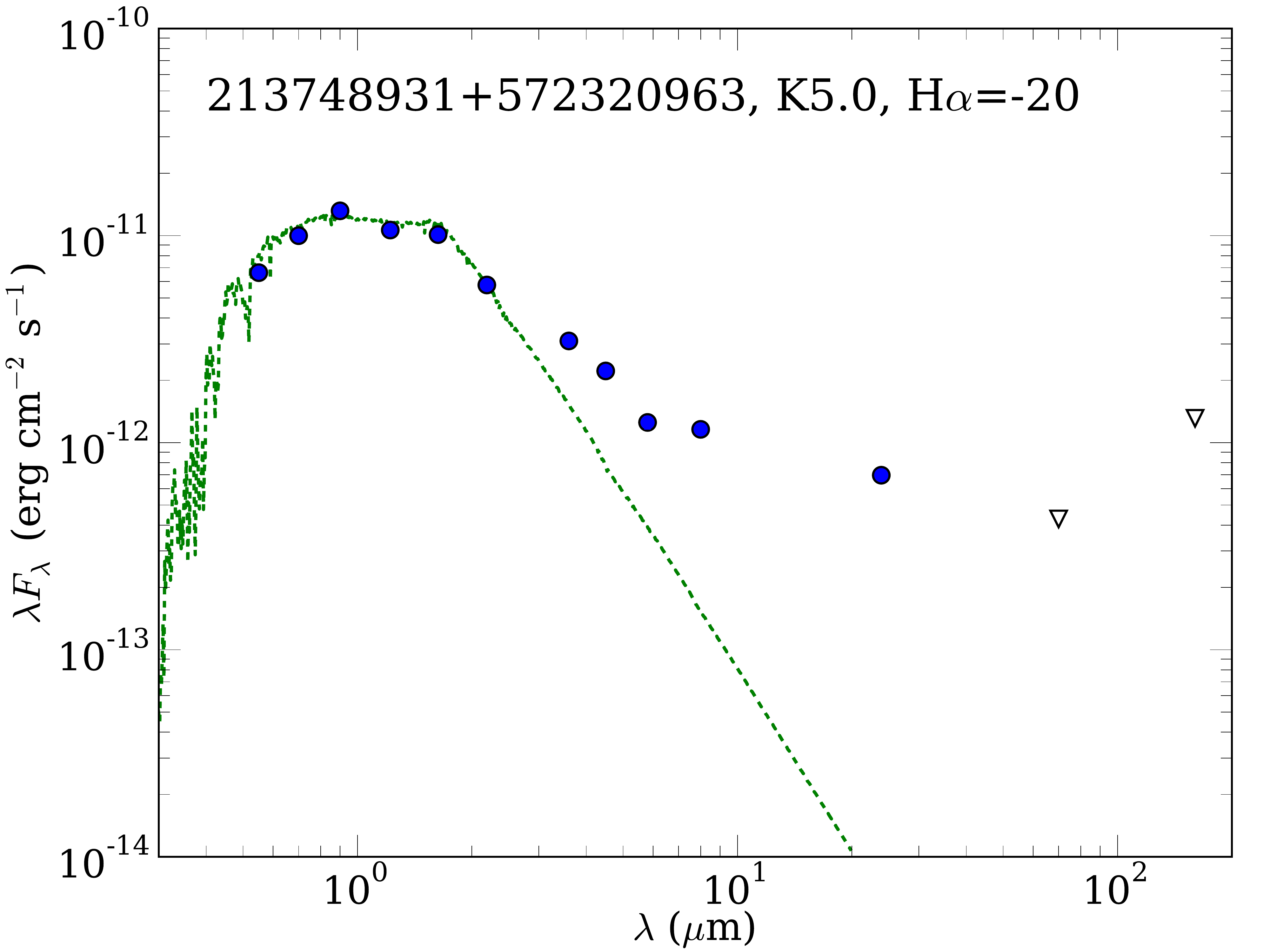} &
\includegraphics[width=0.24\linewidth]{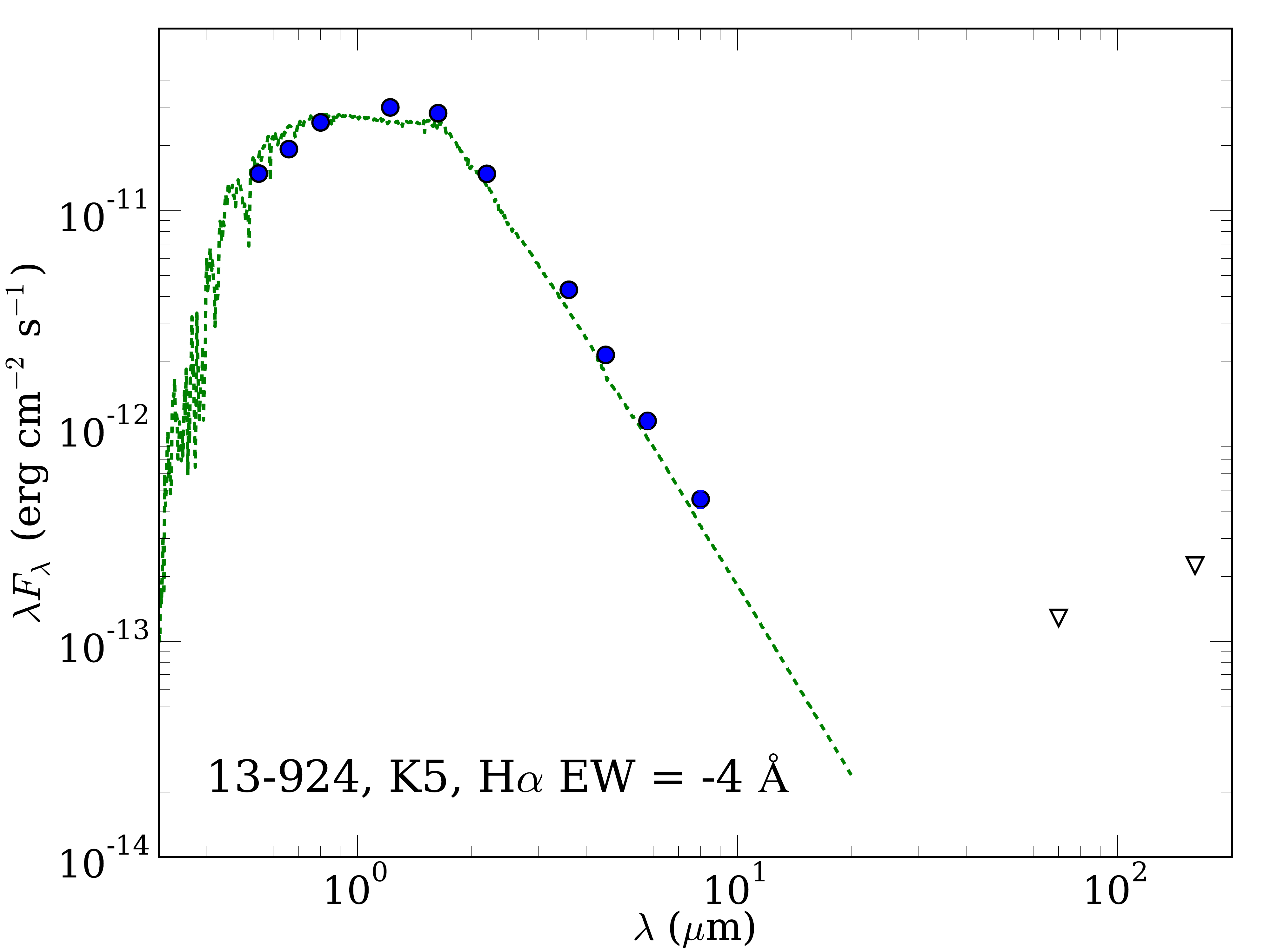} \\
\includegraphics[width=0.24\linewidth]{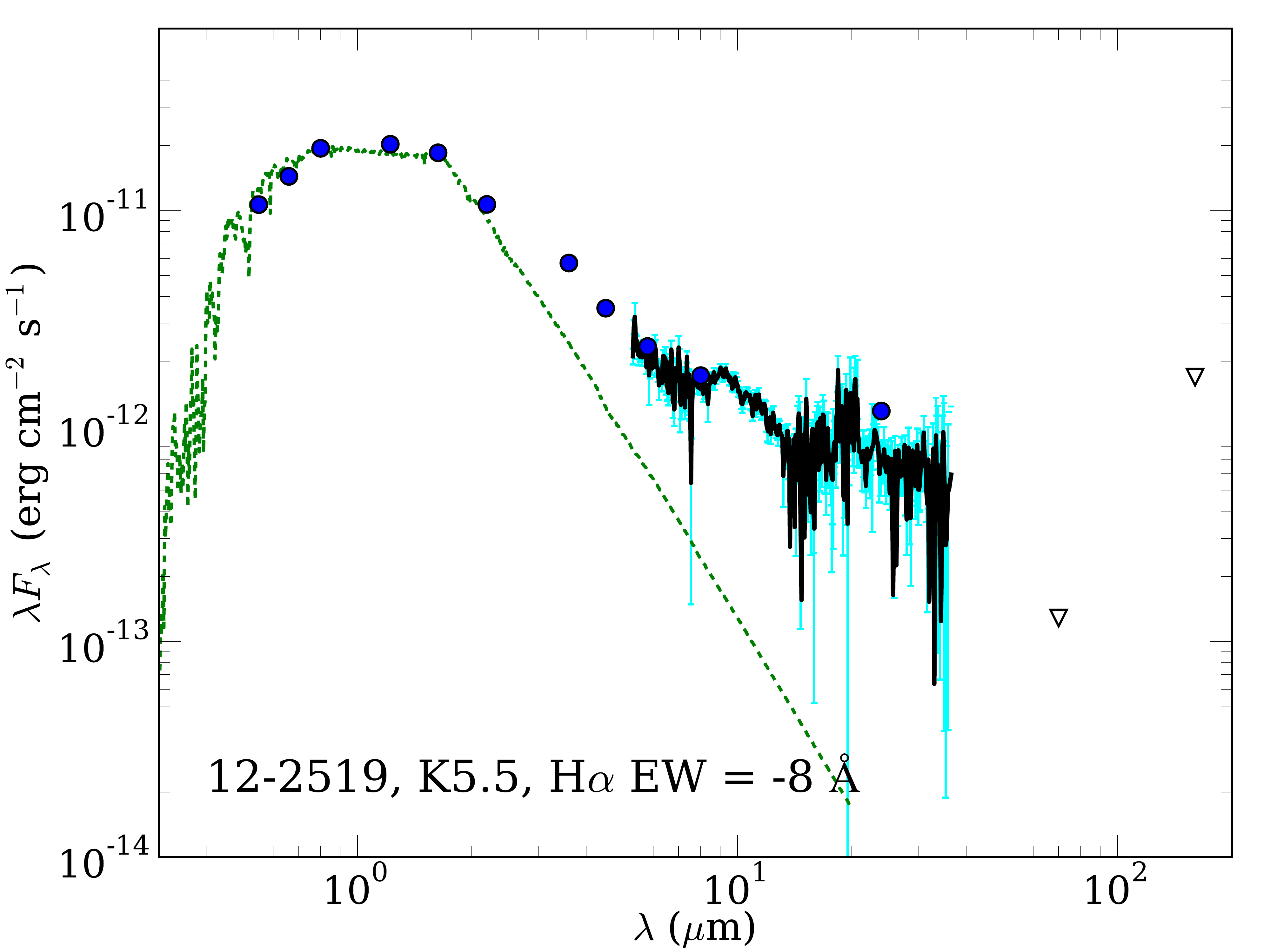} &
\includegraphics[width=0.24\linewidth]{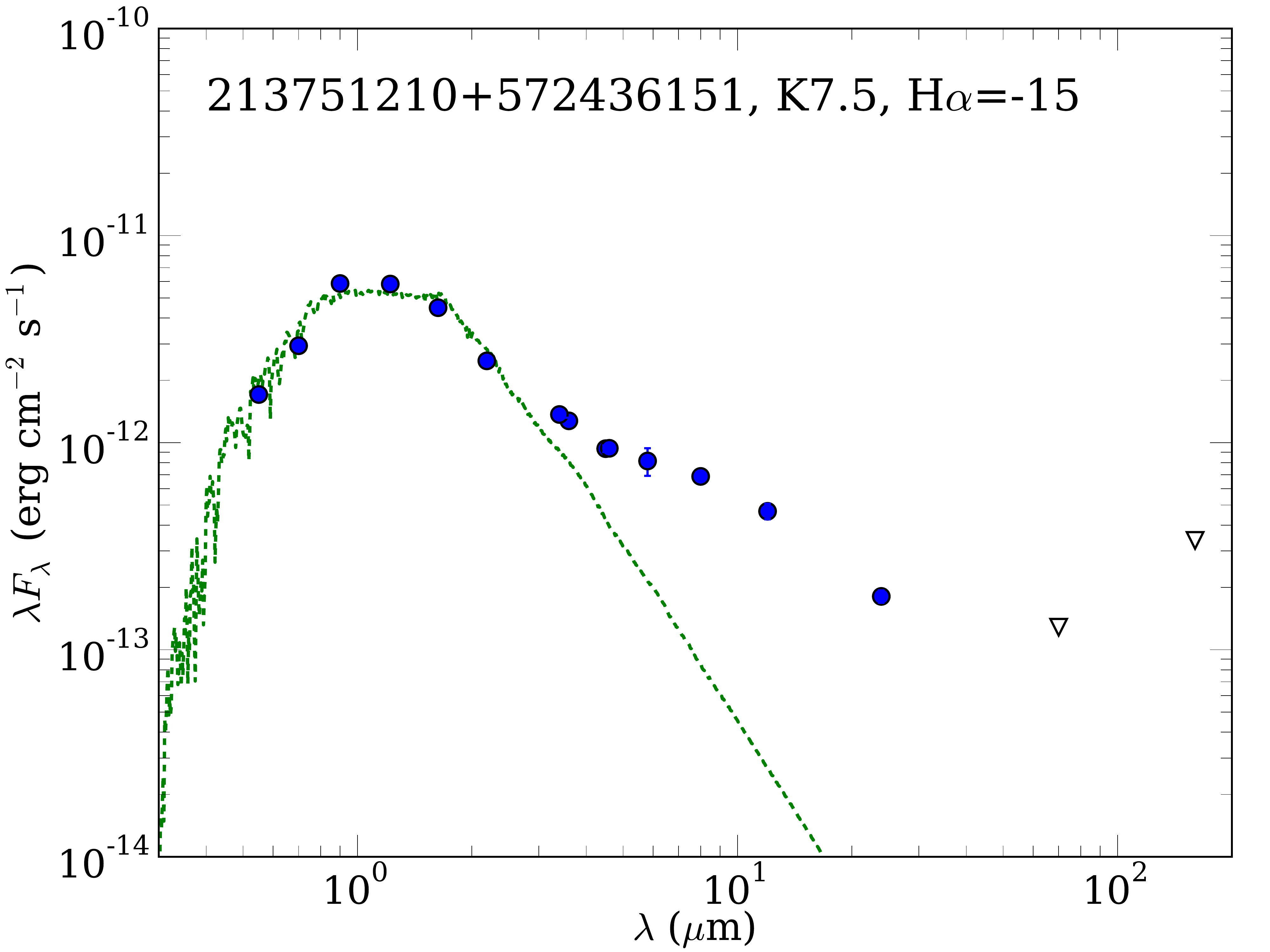} &
\includegraphics[width=0.24\linewidth]{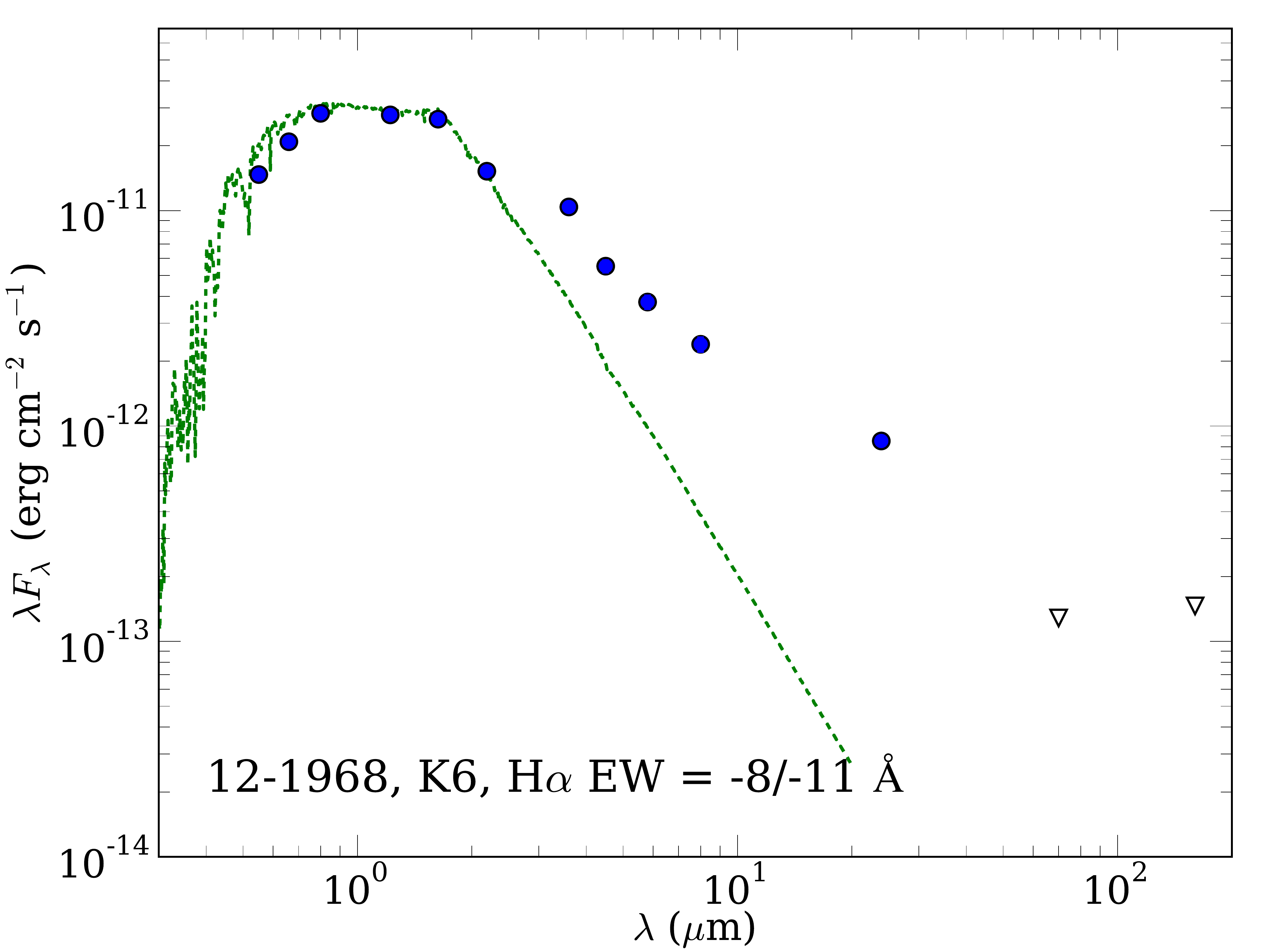} &
\includegraphics[width=0.24\linewidth]{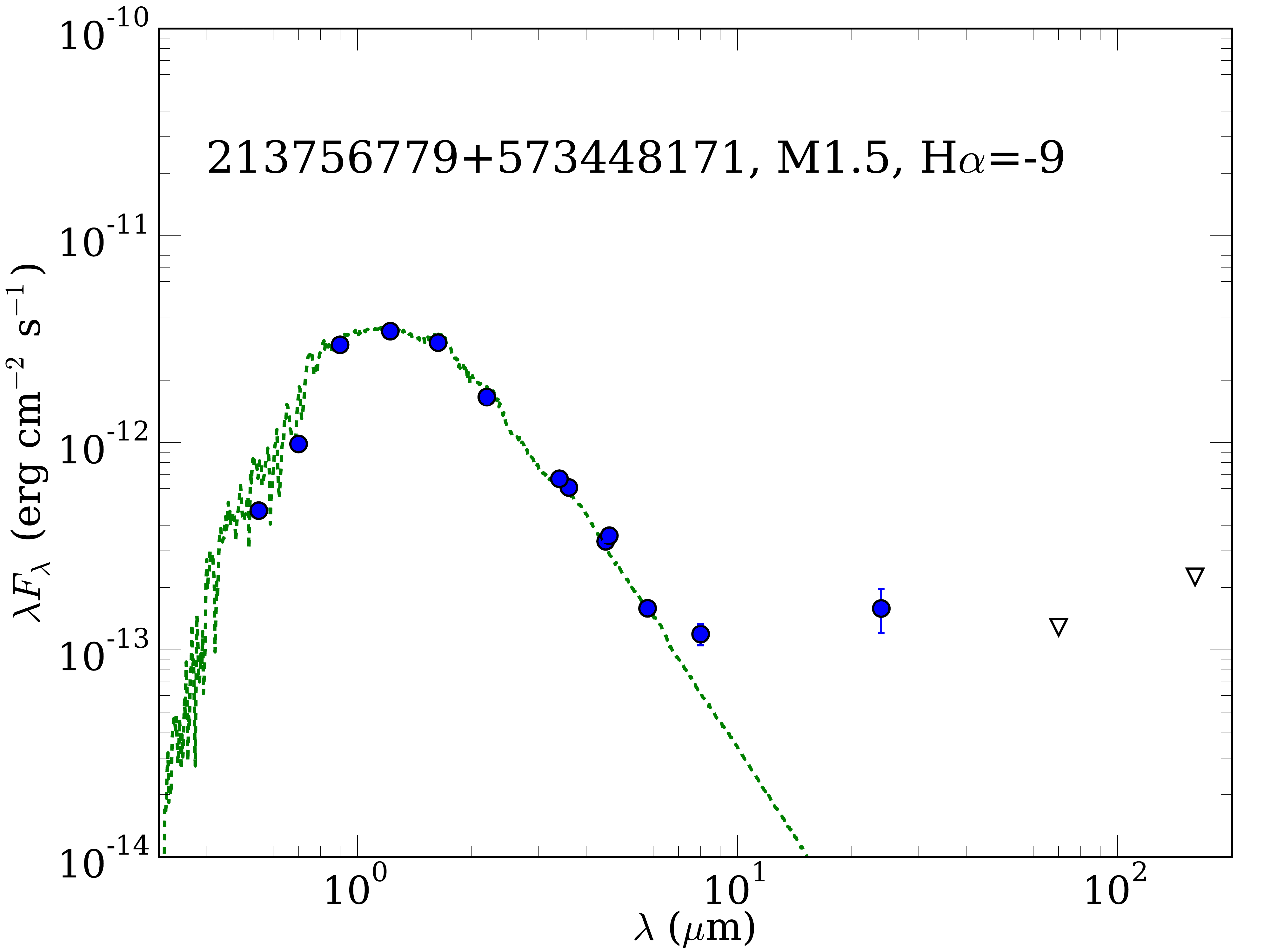} \\
\includegraphics[width=0.24\linewidth]{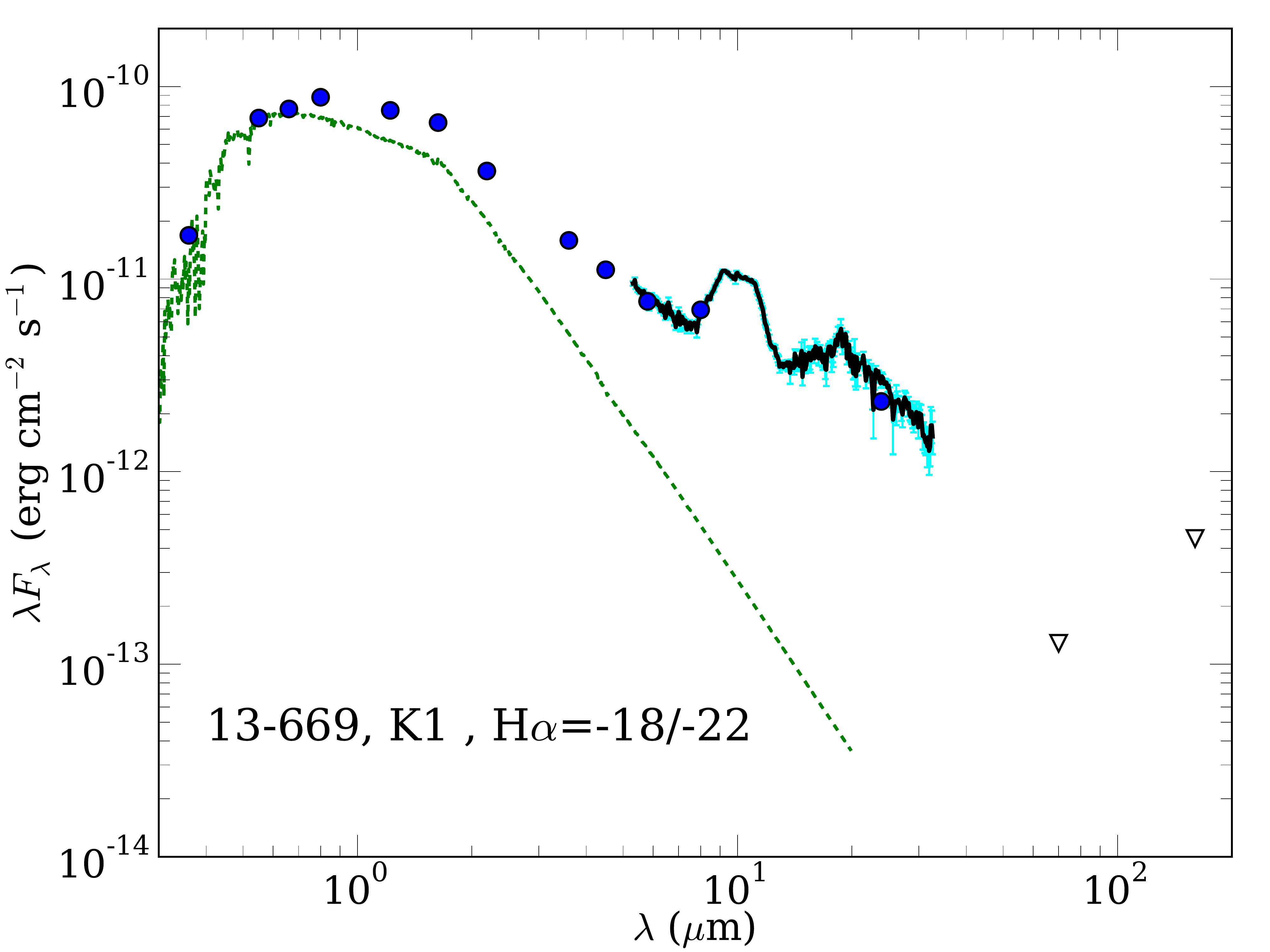} &
\includegraphics[width=0.24\linewidth]{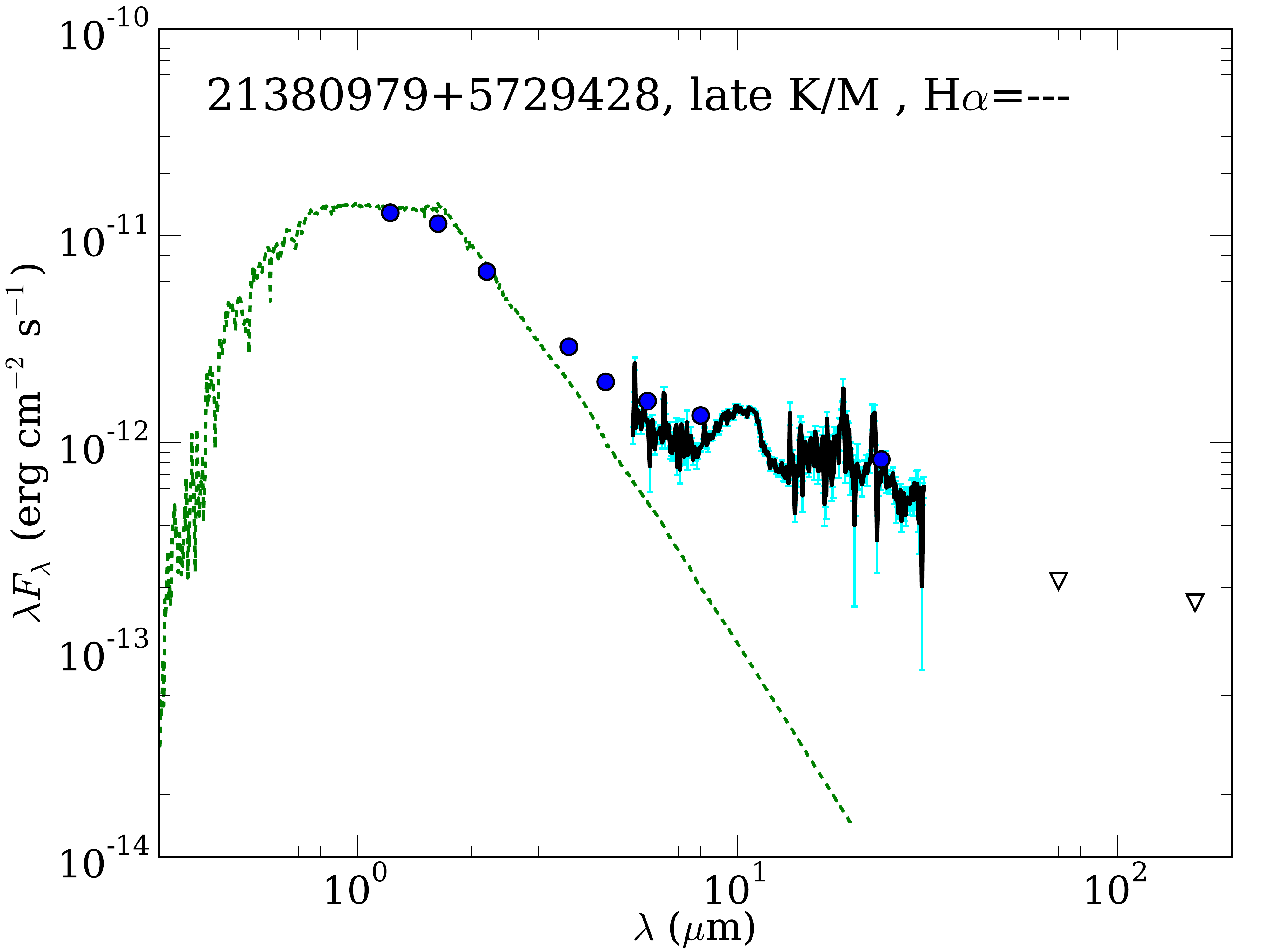} &
\includegraphics[width=0.24\linewidth]{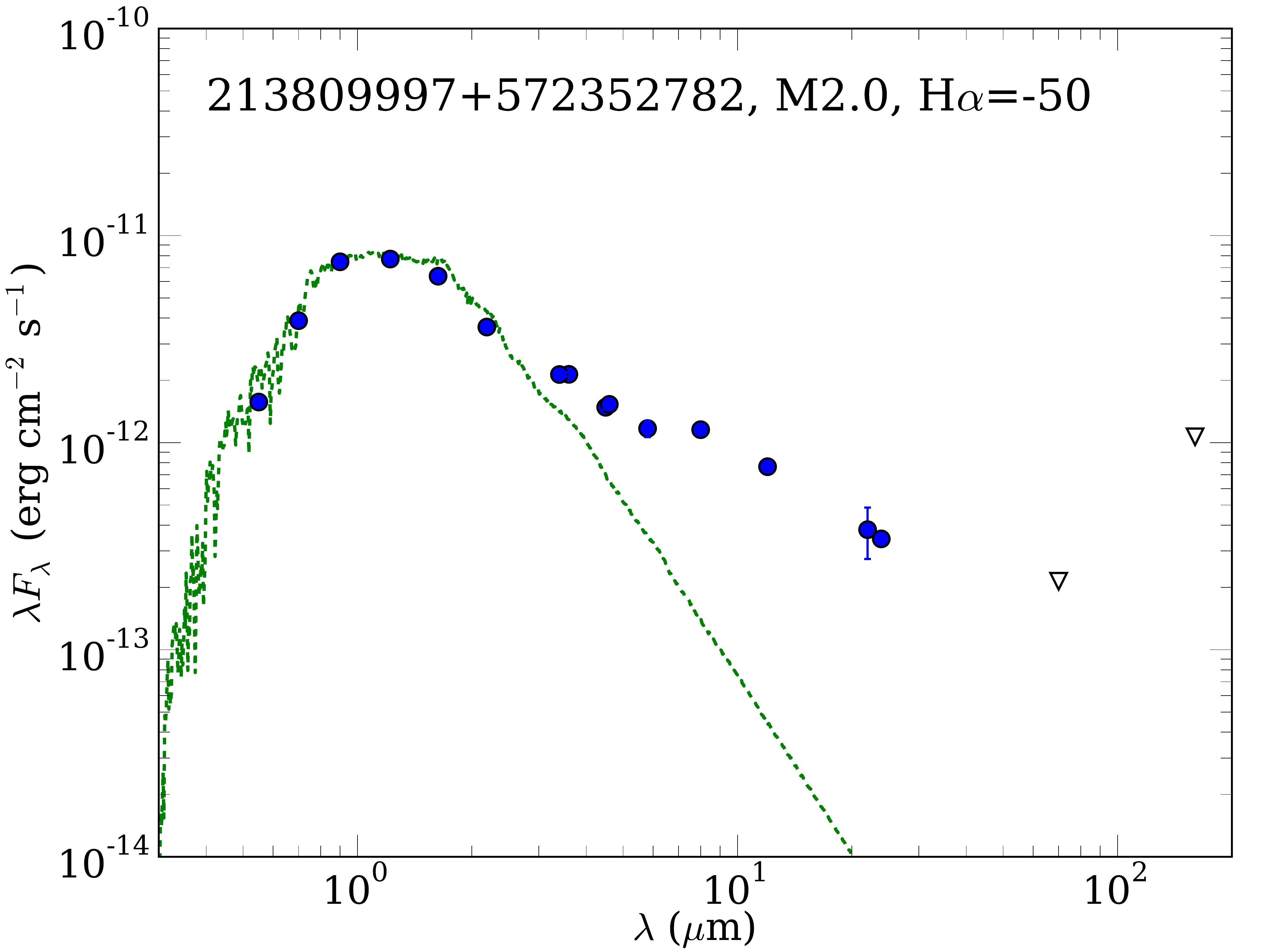} &
\includegraphics[width=0.24\linewidth]{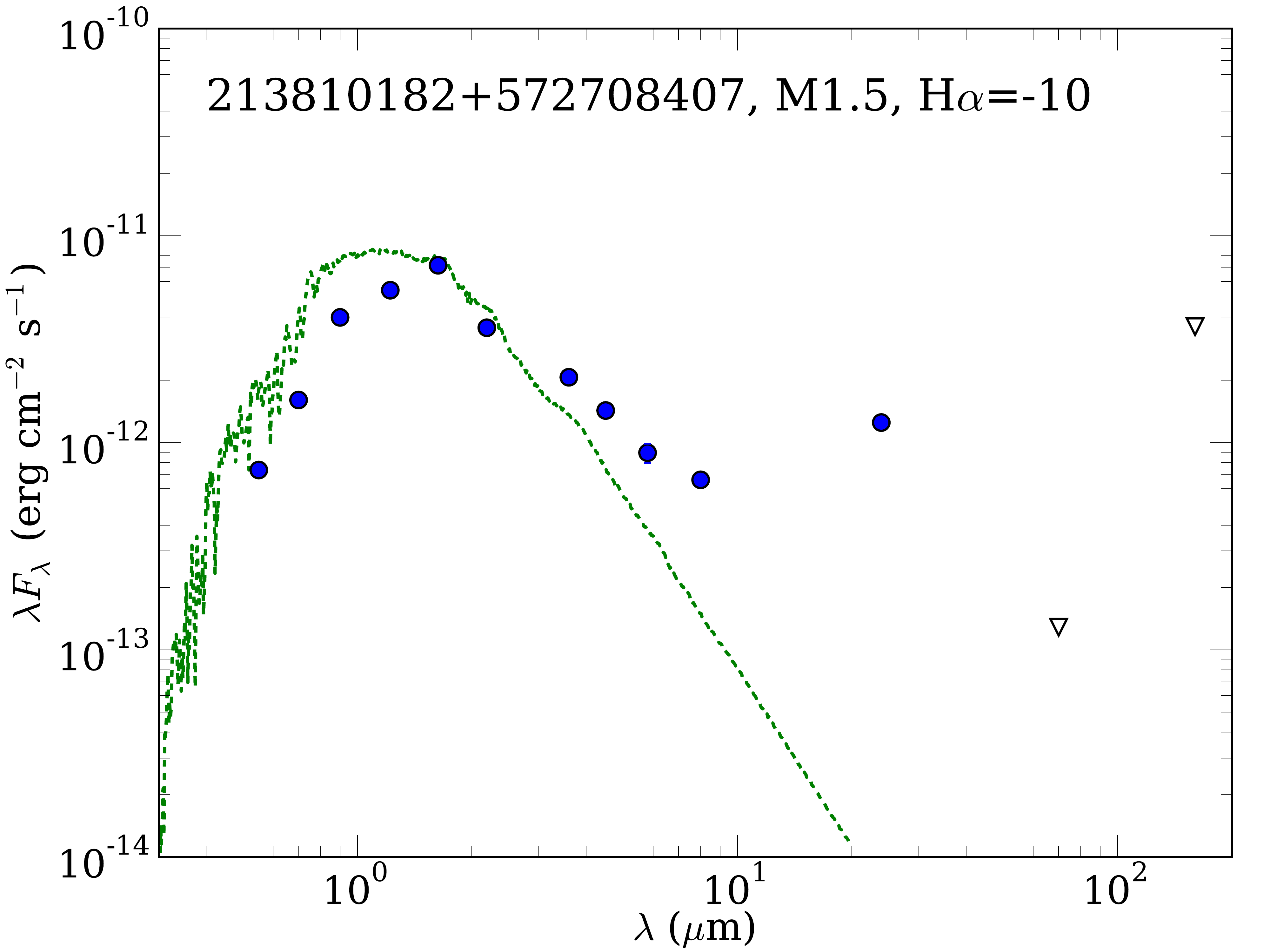} \\
\includegraphics[width=0.24\linewidth]{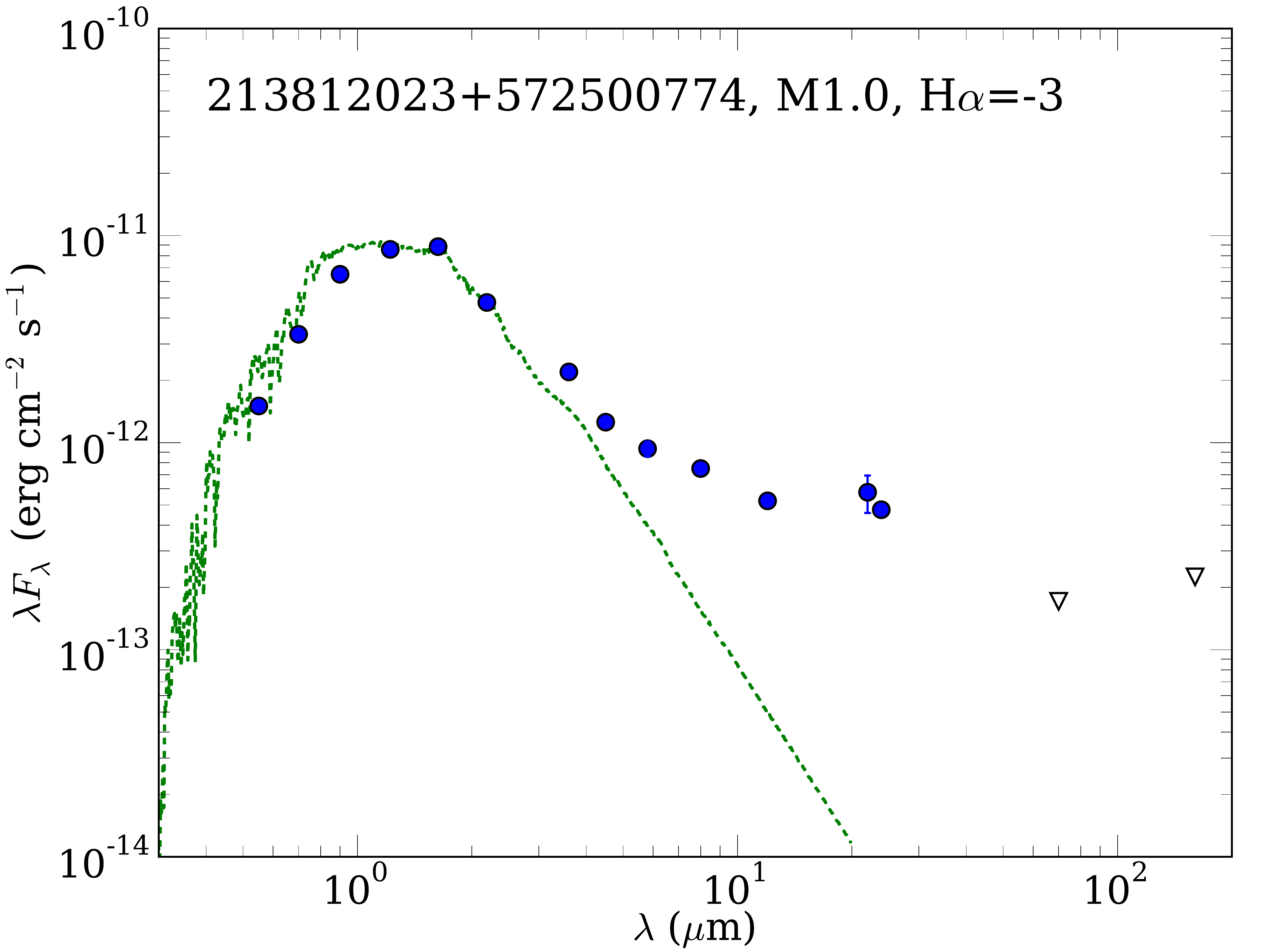} &
\includegraphics[width=0.24\linewidth]{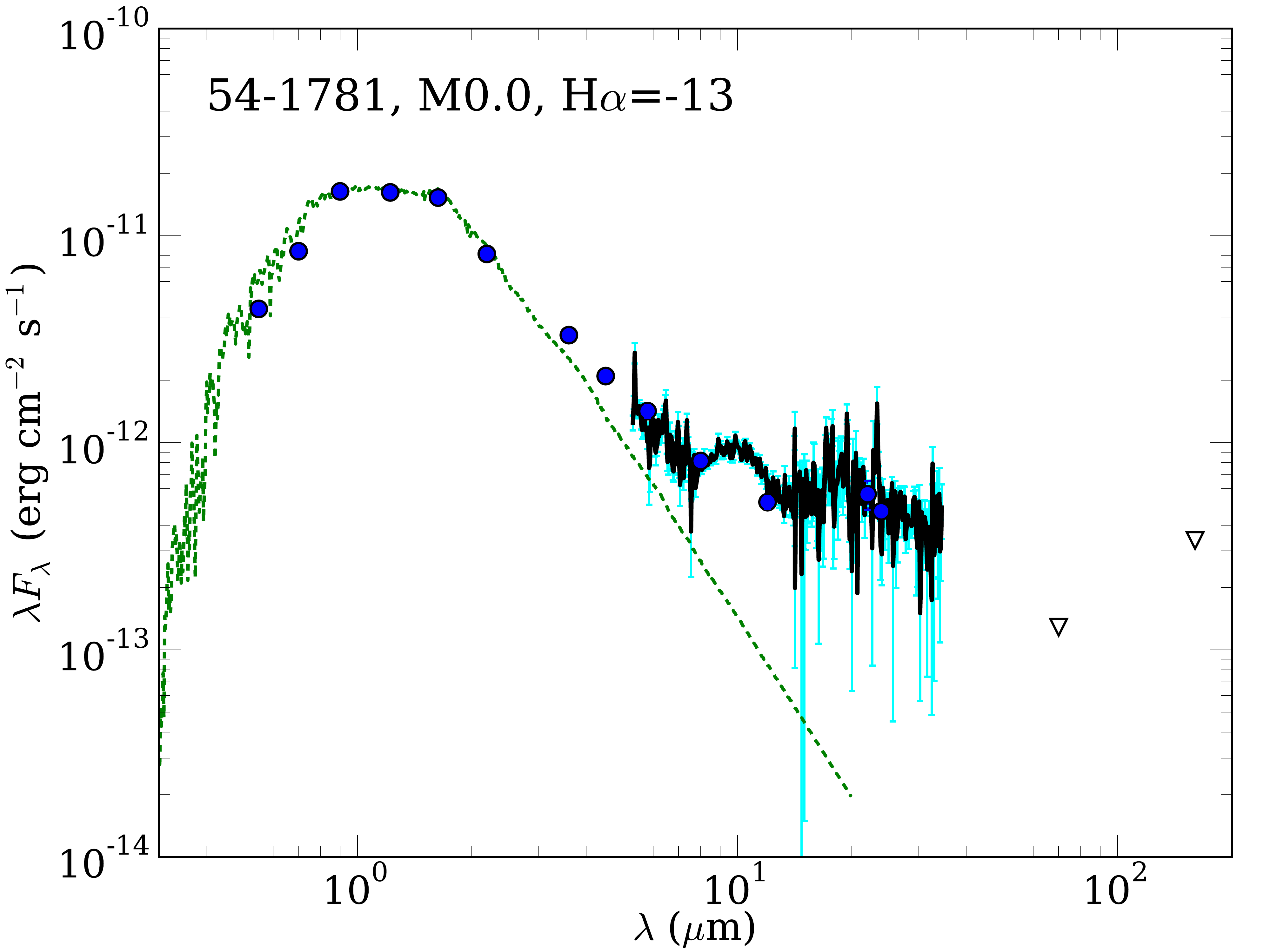} &
\includegraphics[width=0.24\linewidth]{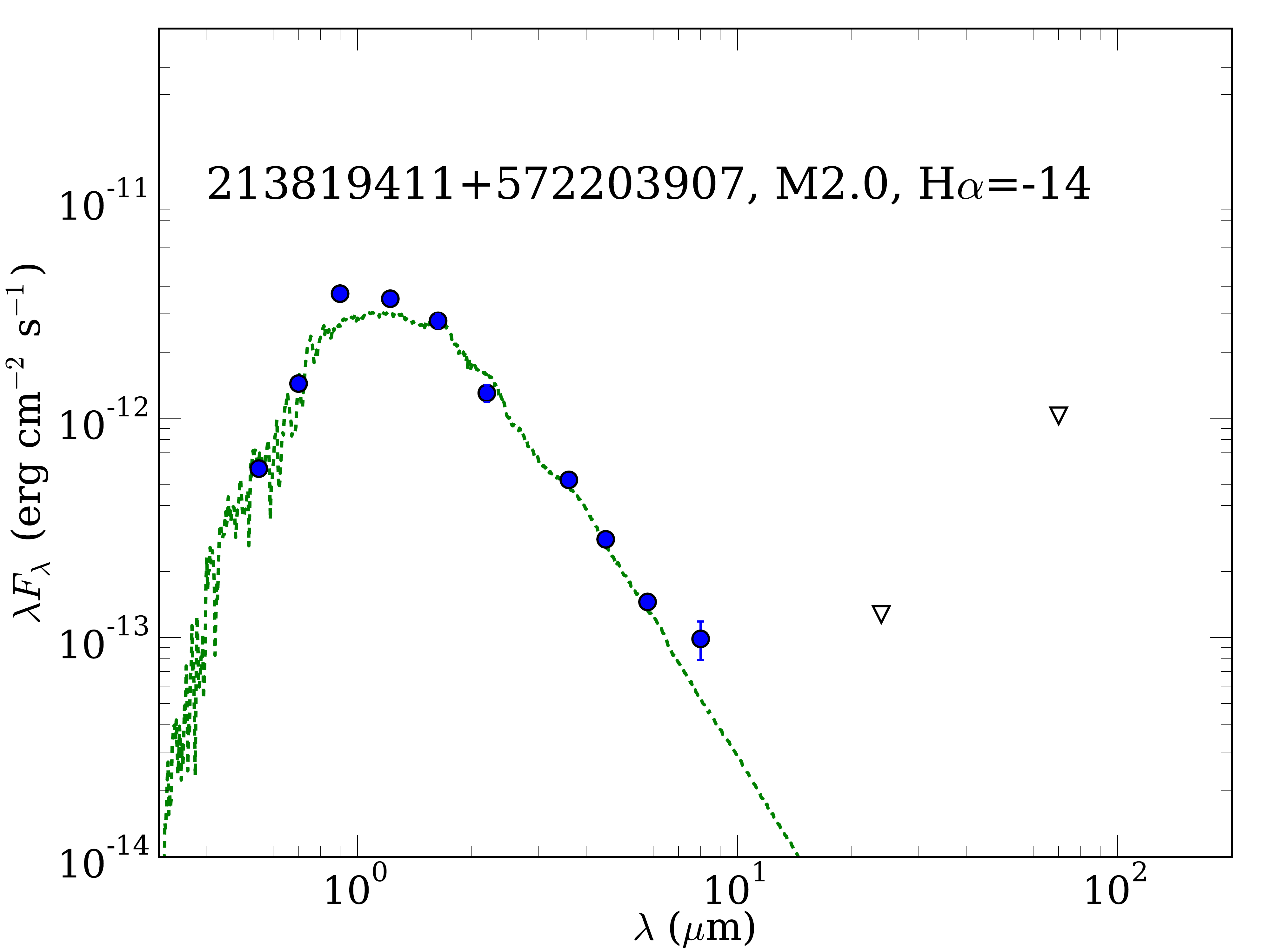} &
\includegraphics[width=0.24\linewidth]{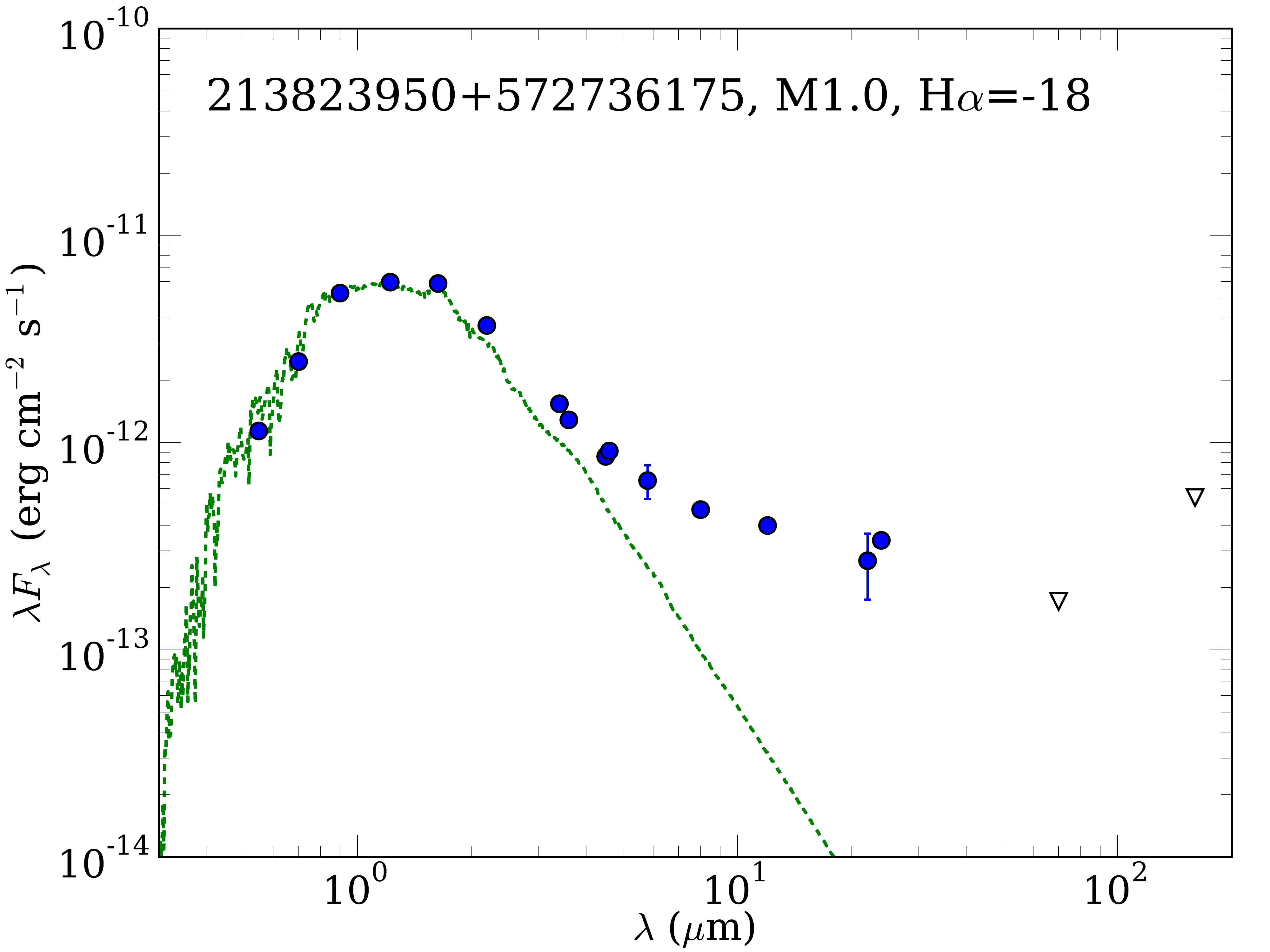} \\
\includegraphics[width=0.24\linewidth]{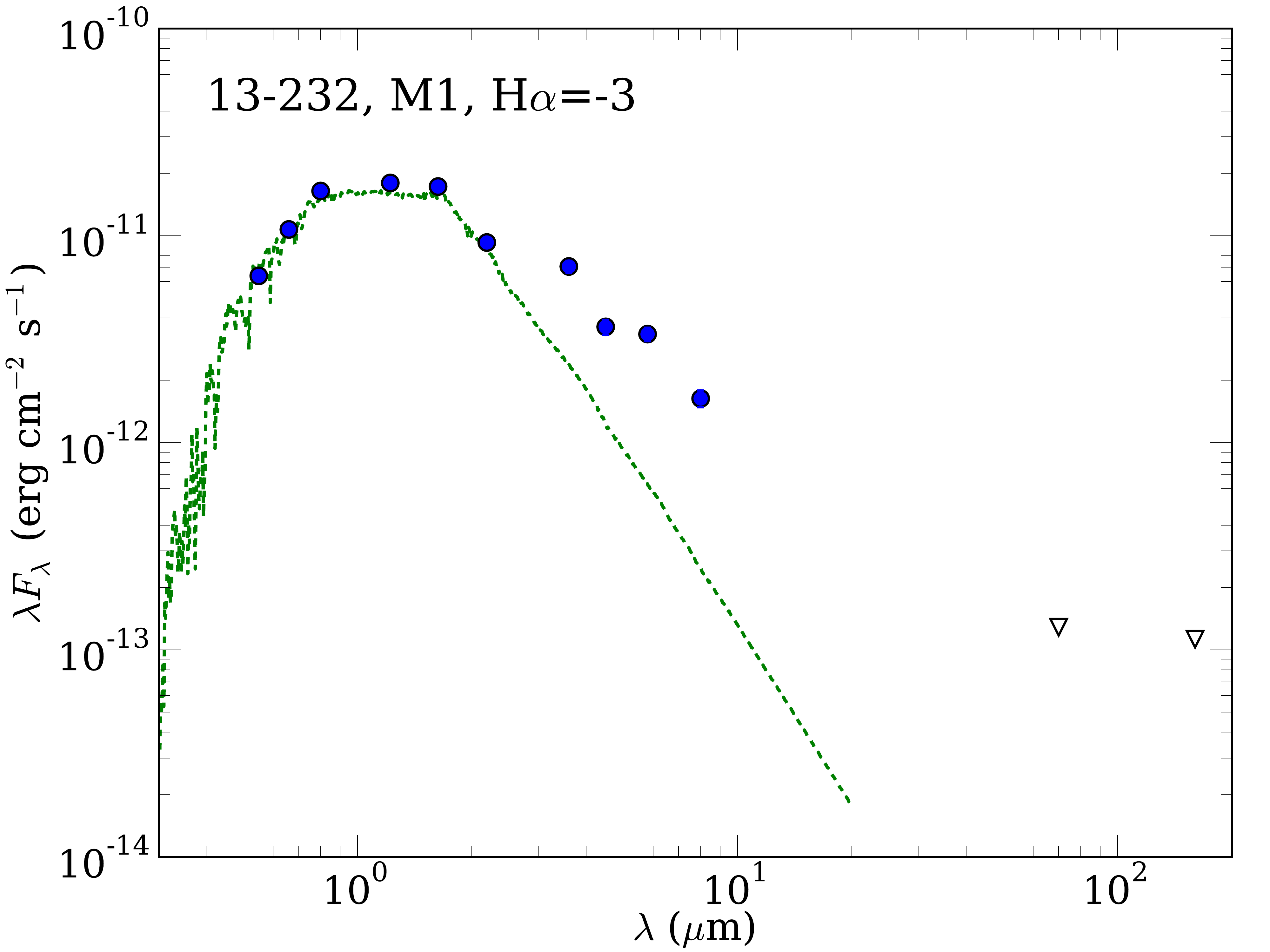} &
\includegraphics[width=0.24\linewidth]{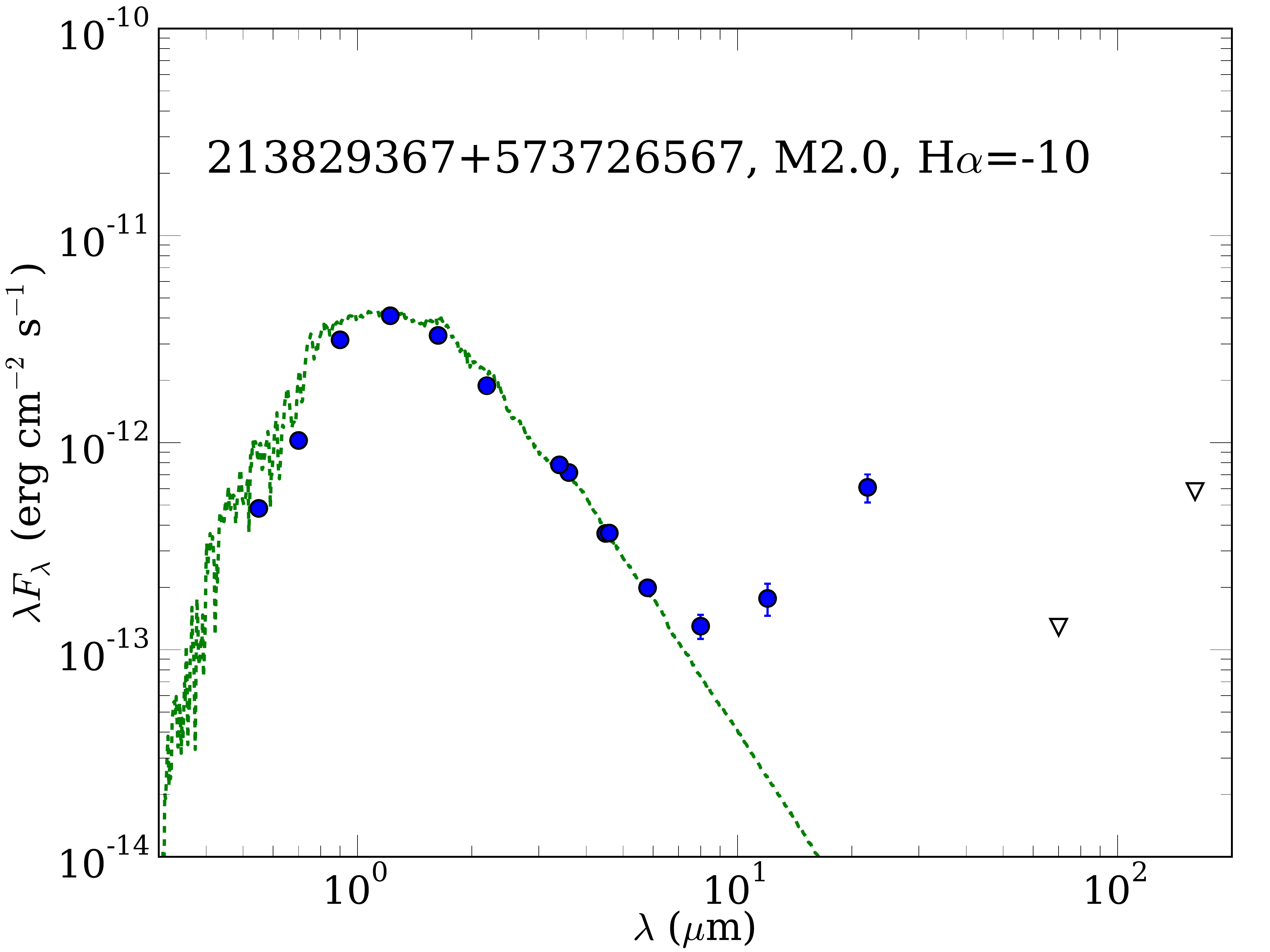} &
\includegraphics[width=0.24\linewidth]{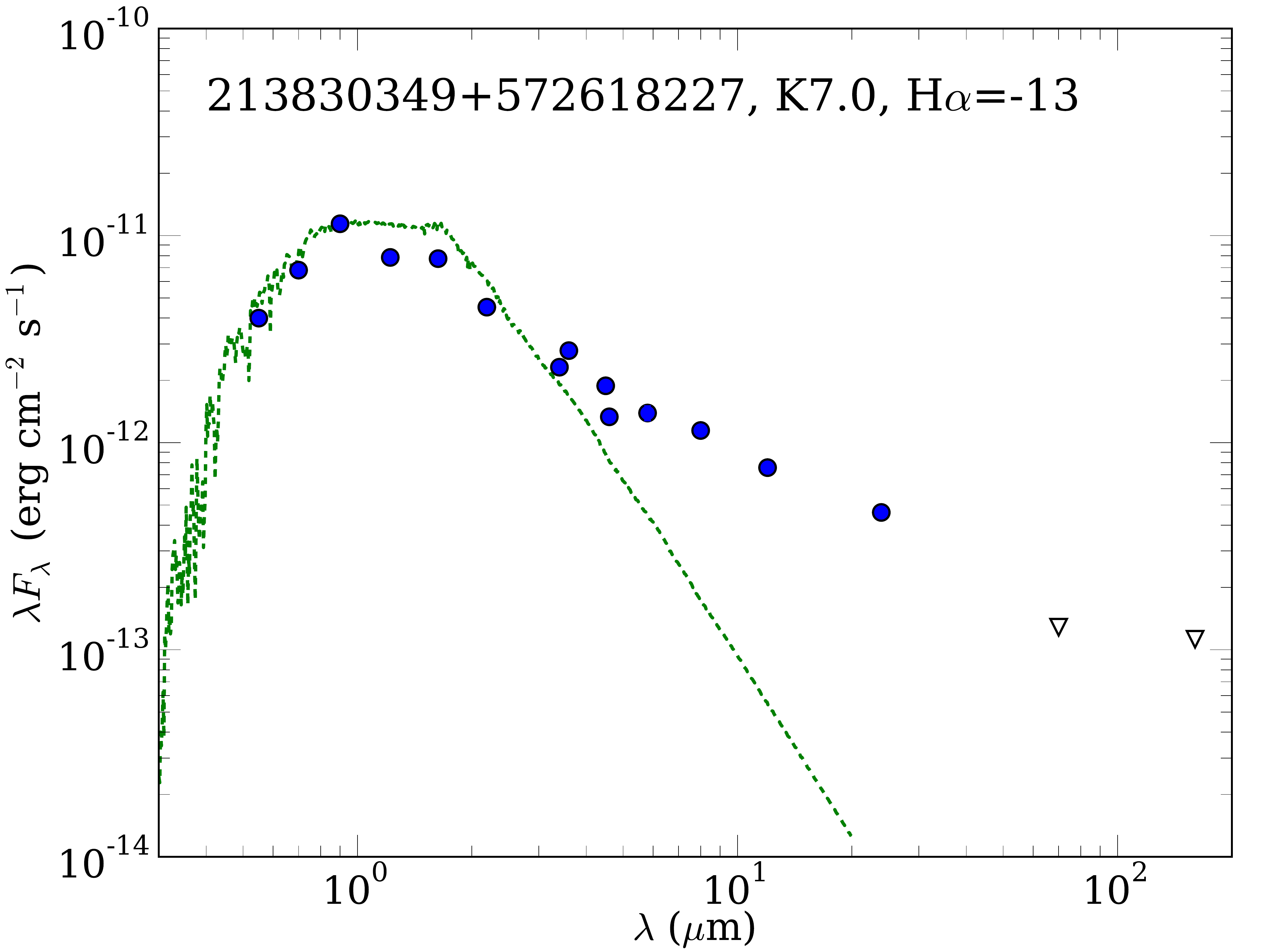} &
\includegraphics[width=0.24\linewidth]{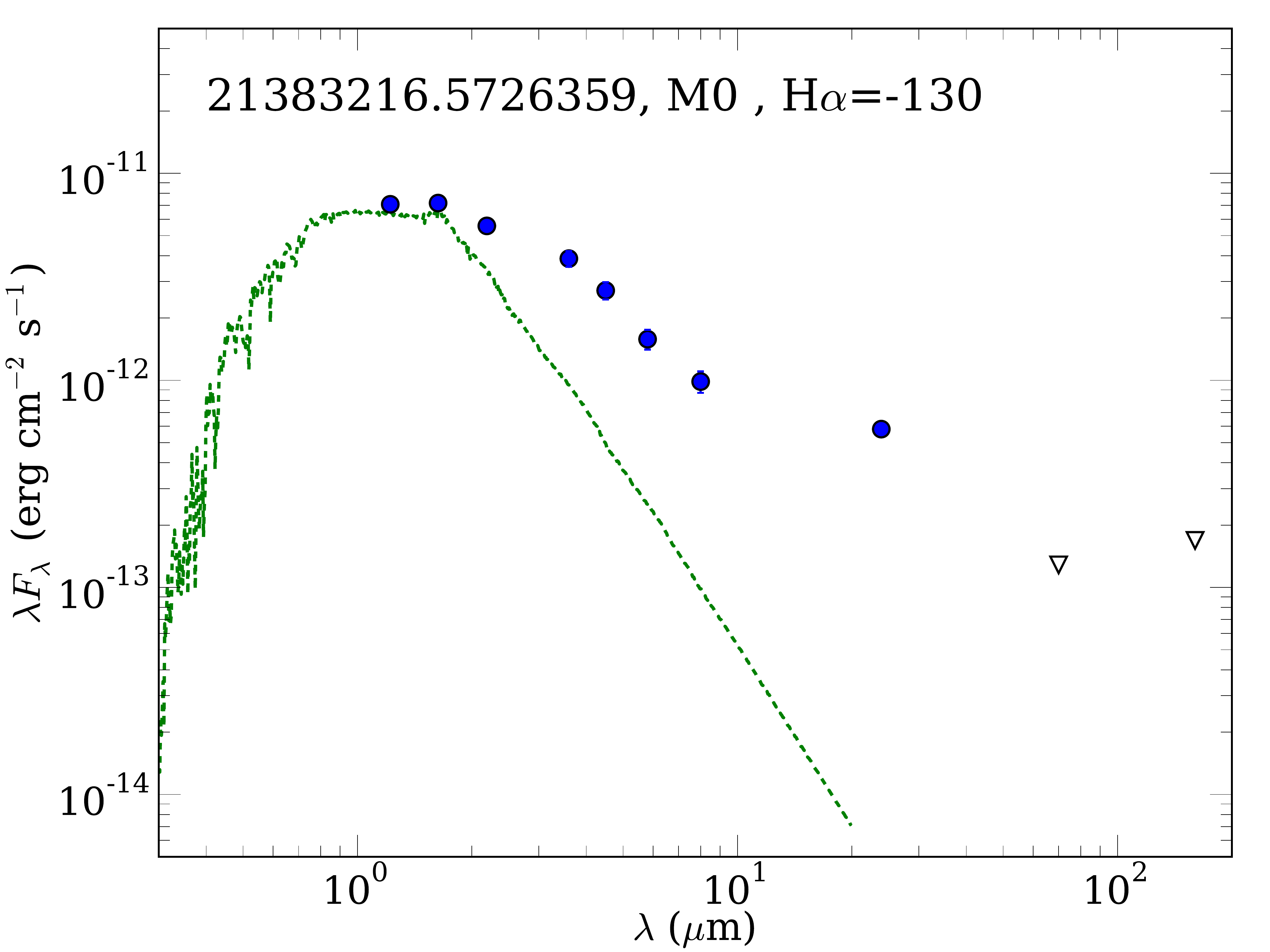} \\
\includegraphics[width=0.24\linewidth]{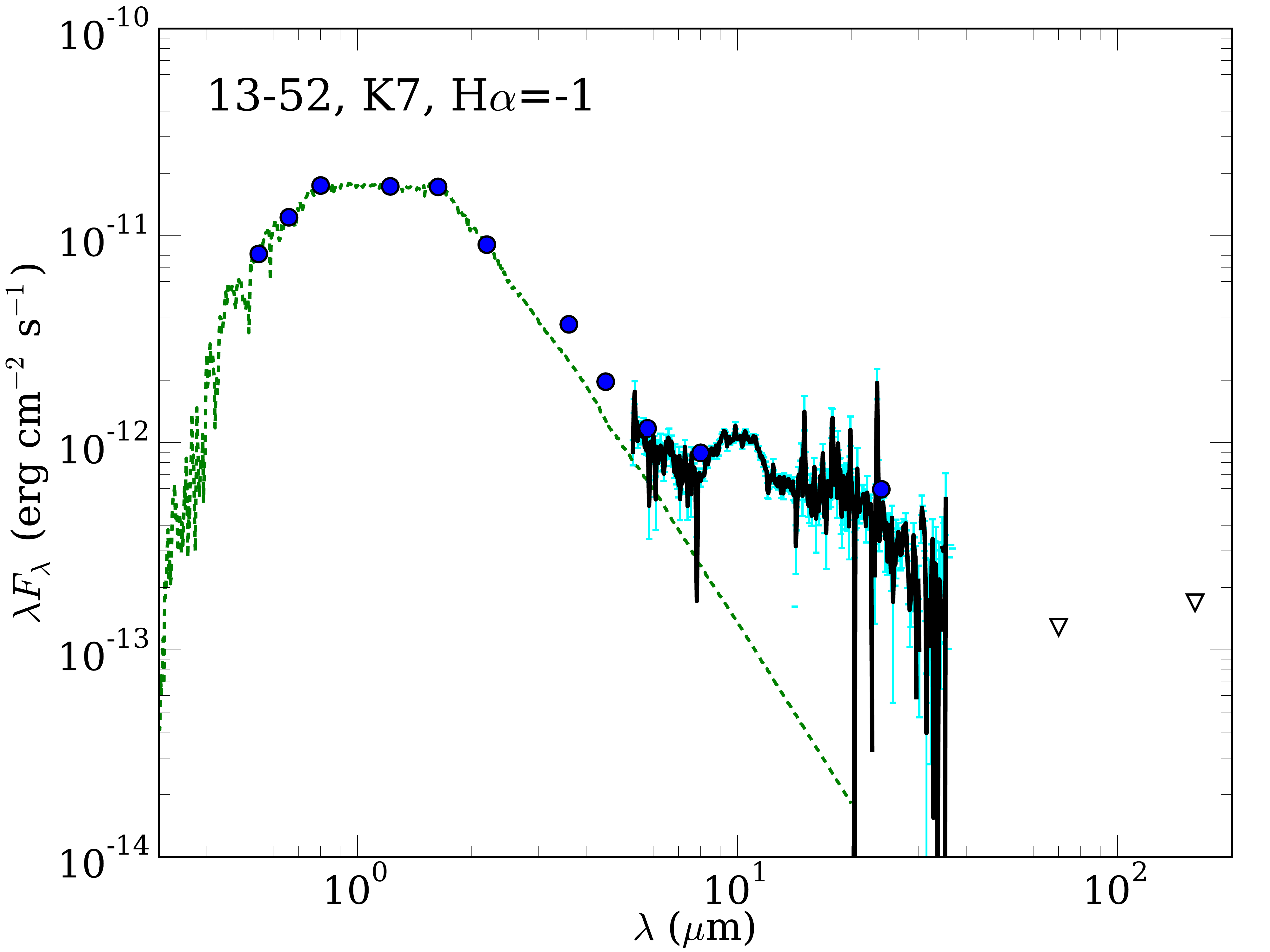} &
\includegraphics[width=0.24\linewidth]{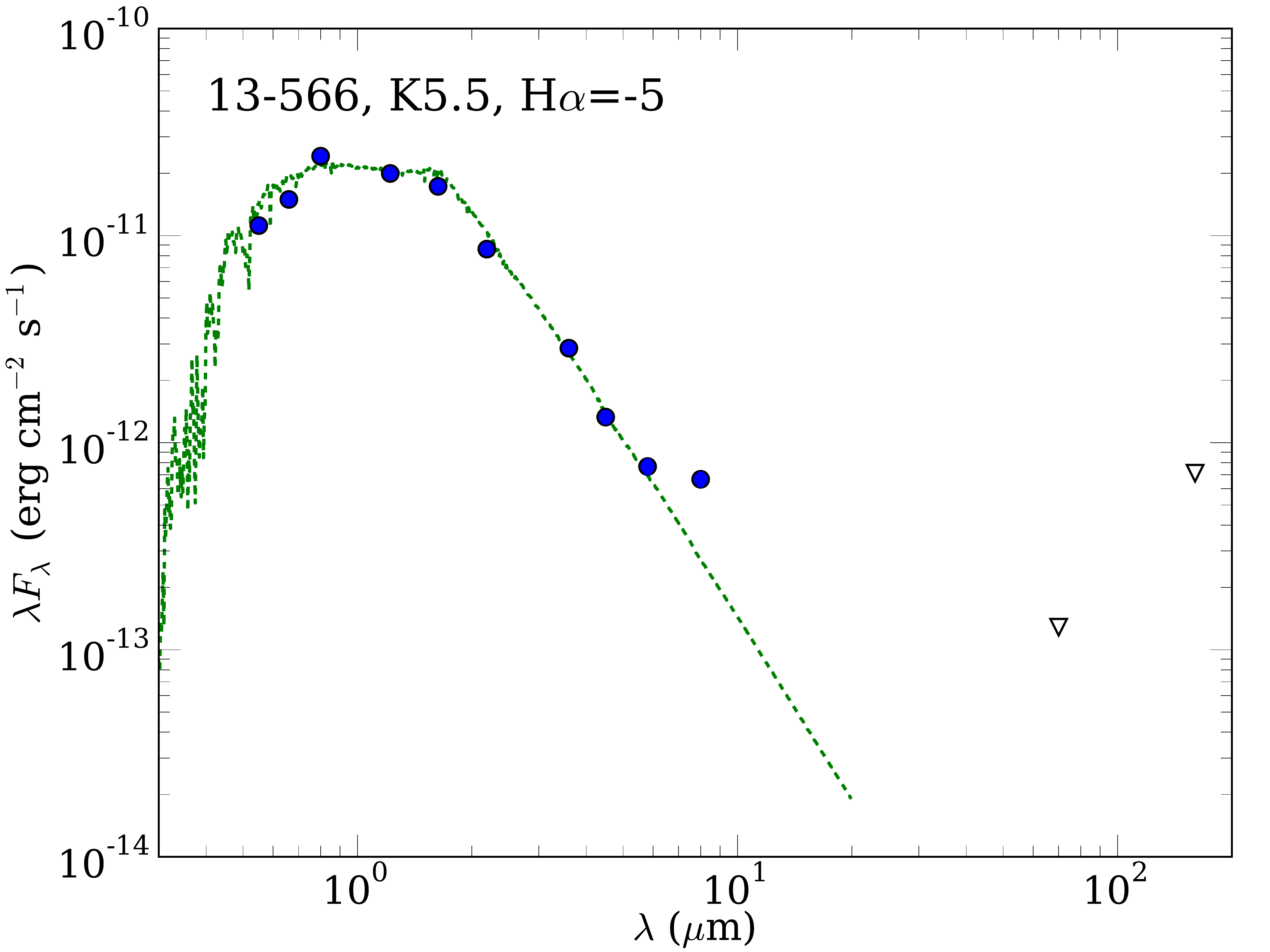} &
\includegraphics[width=0.24\linewidth]{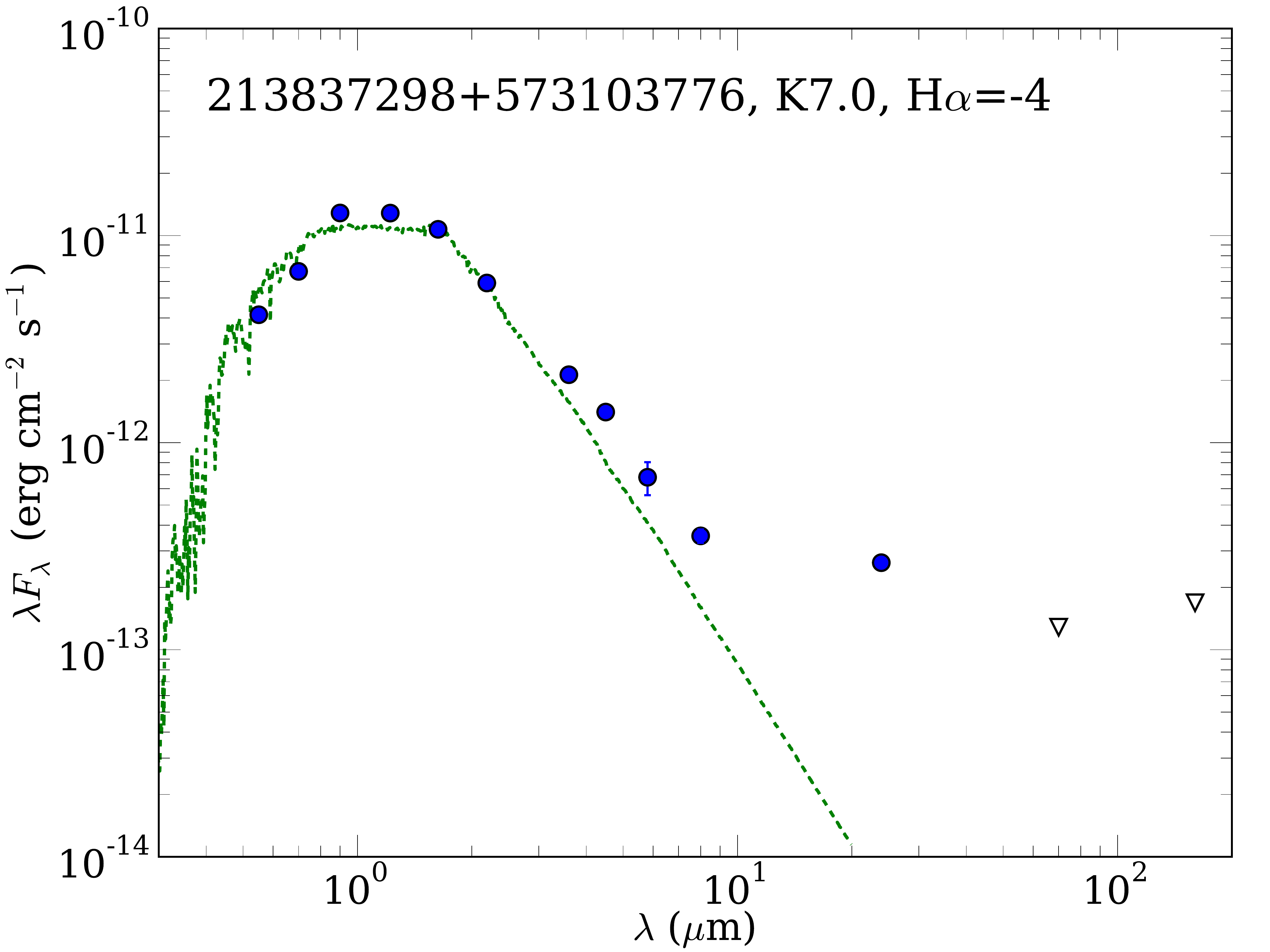} &
\includegraphics[width=0.24\linewidth]{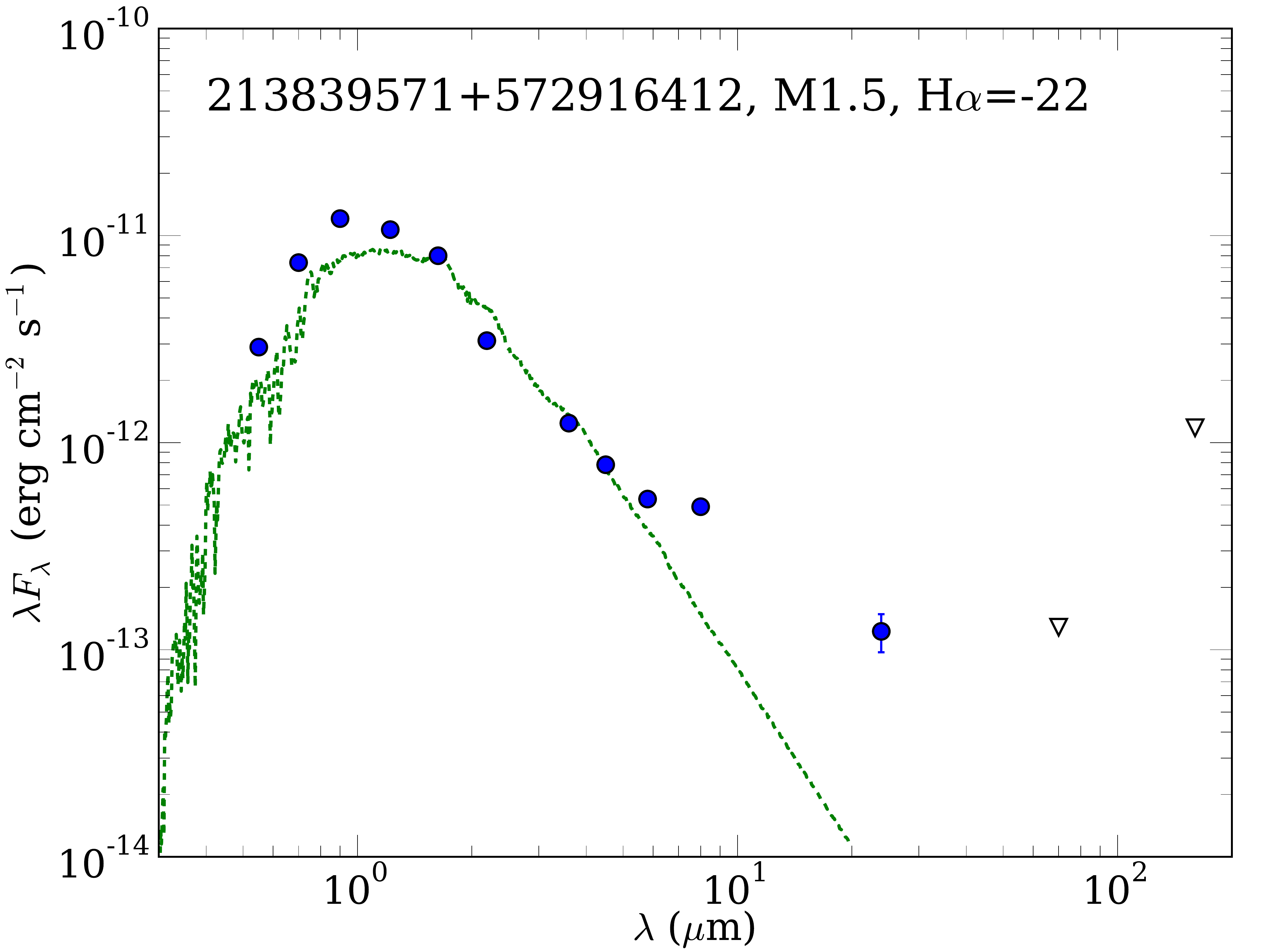} \\
\includegraphics[width=0.24\linewidth]{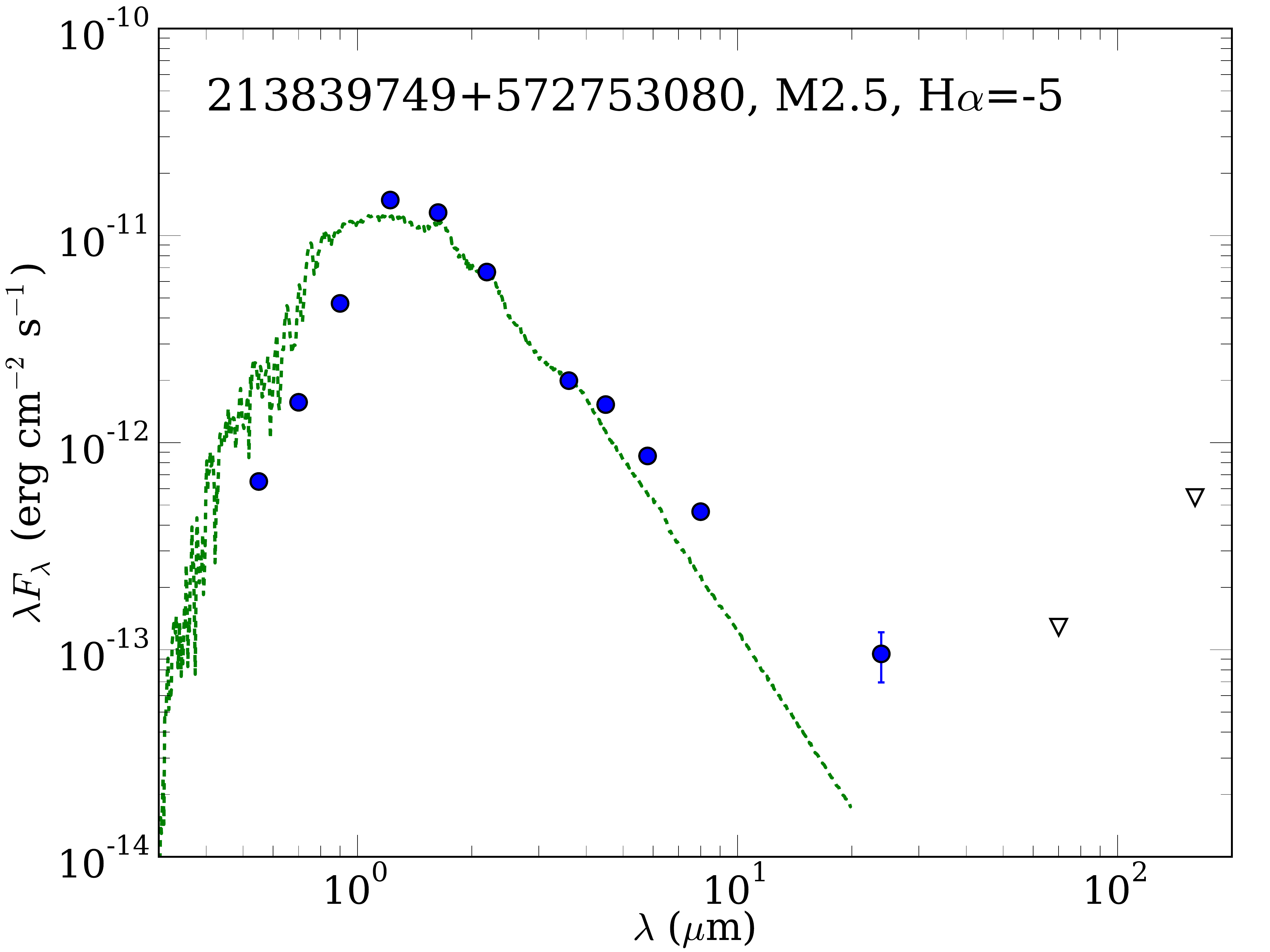} &
\includegraphics[width=0.24\linewidth]{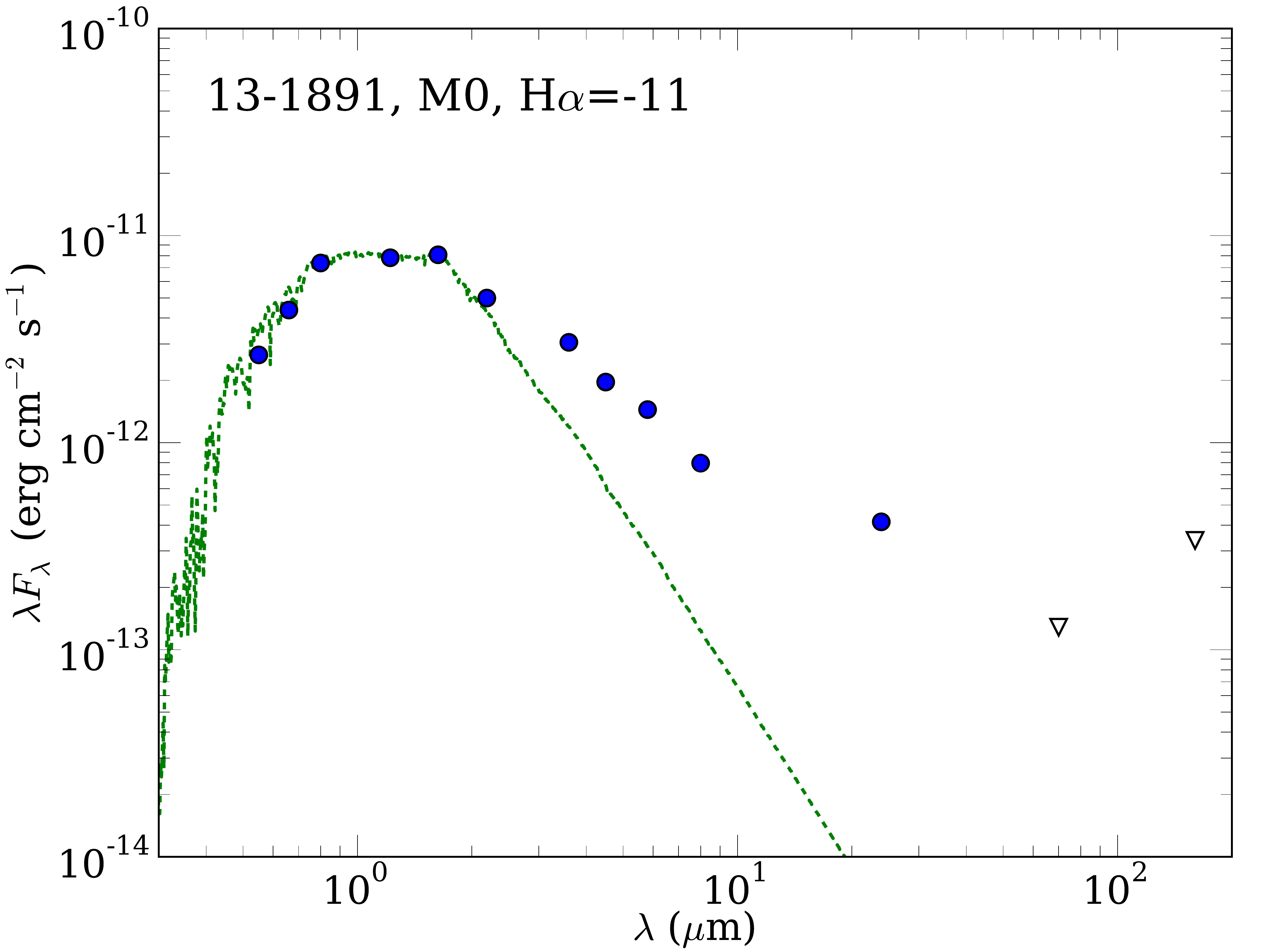} &
\includegraphics[width=0.24\linewidth]{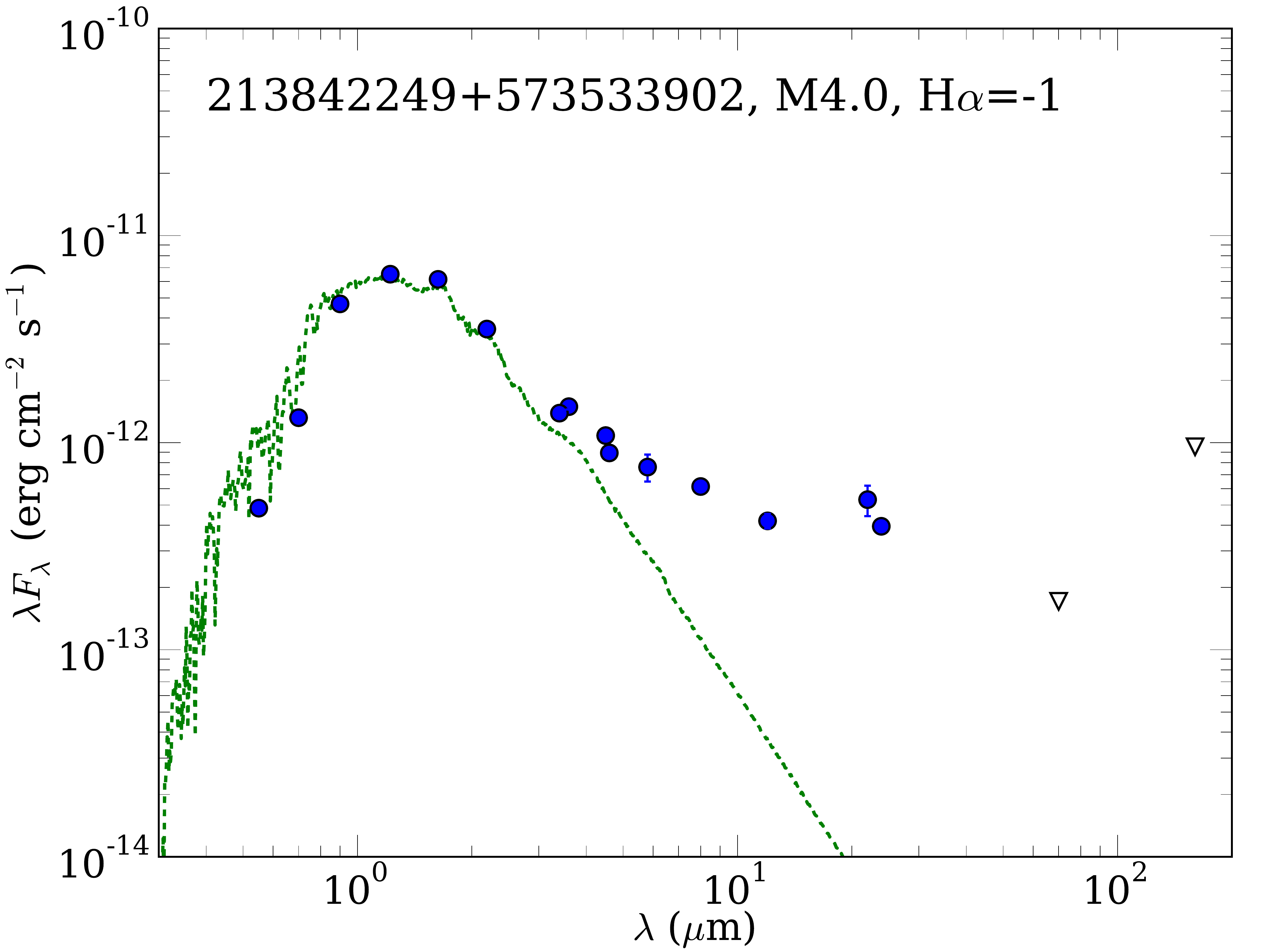} &
\includegraphics[width=0.24\linewidth]{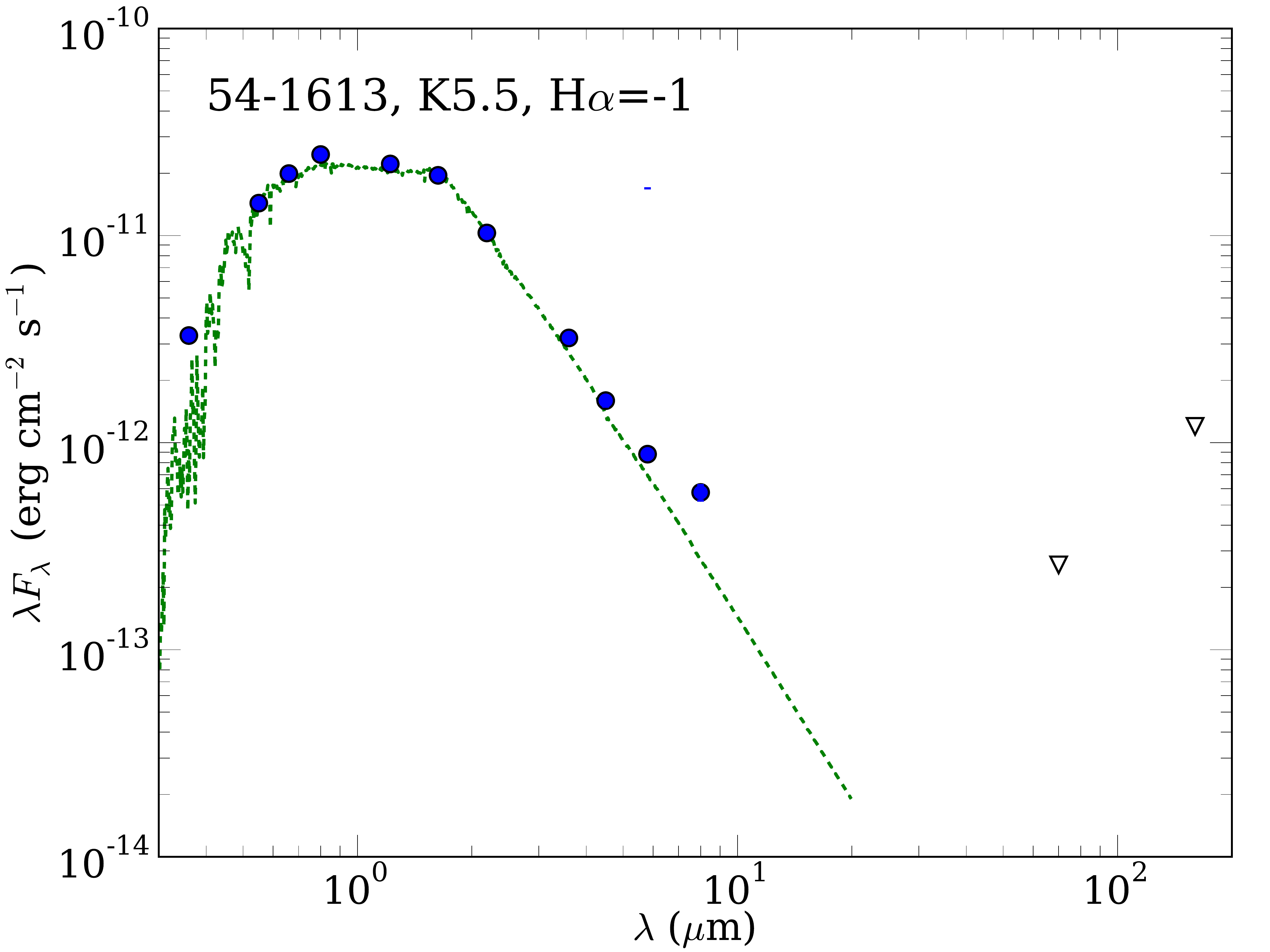} \\
\end{tabular}
\caption{SEDs of the objects with upper limits only. Symbols as in Figure \ref{uplims1-fig}.
\label{uplims2-fig}}
\end{figure*}

\begin{figure*}
\centering
\begin{tabular}{cccc}
\includegraphics[width=0.24\linewidth]{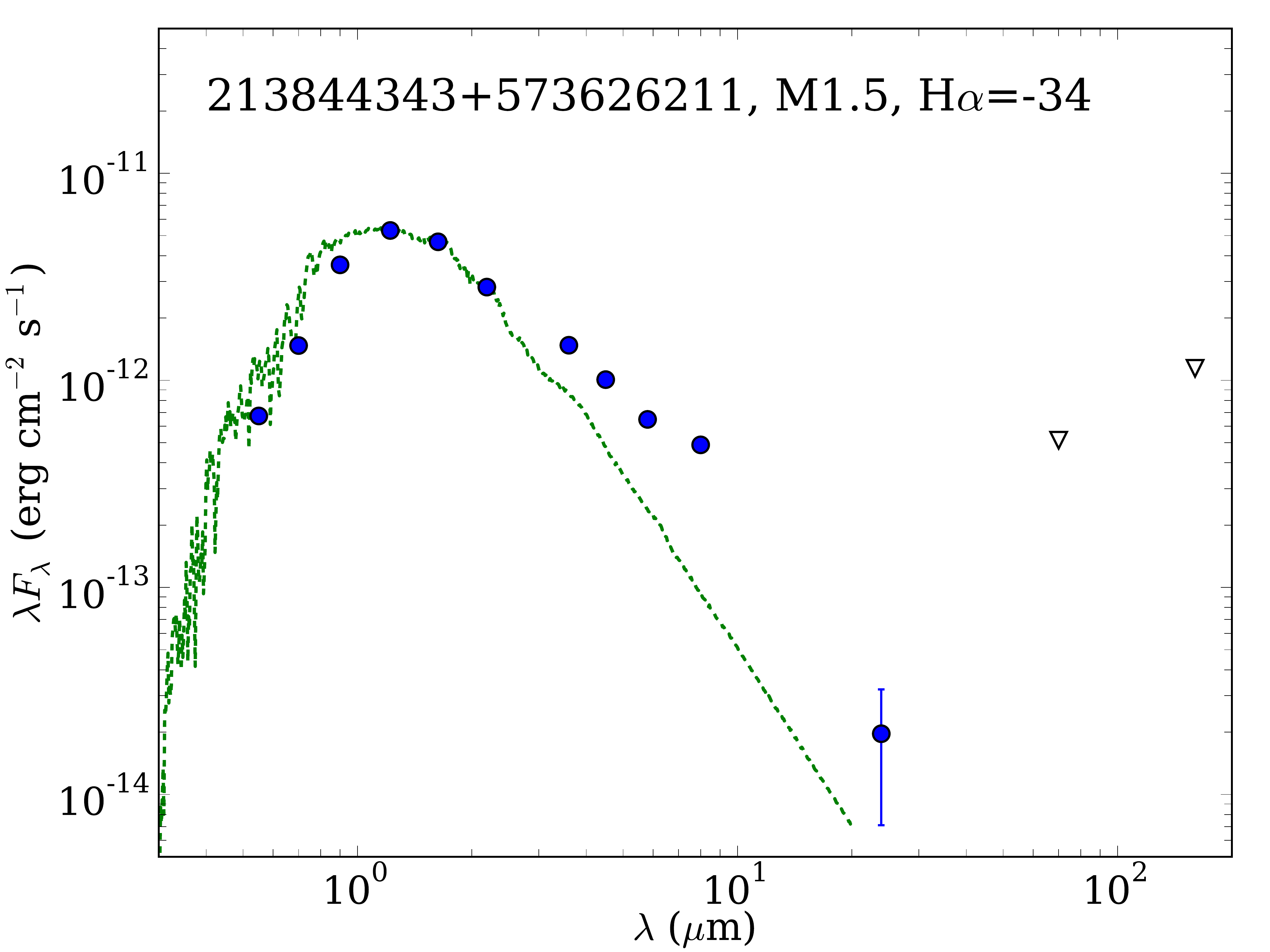} &
\includegraphics[width=0.24\linewidth]{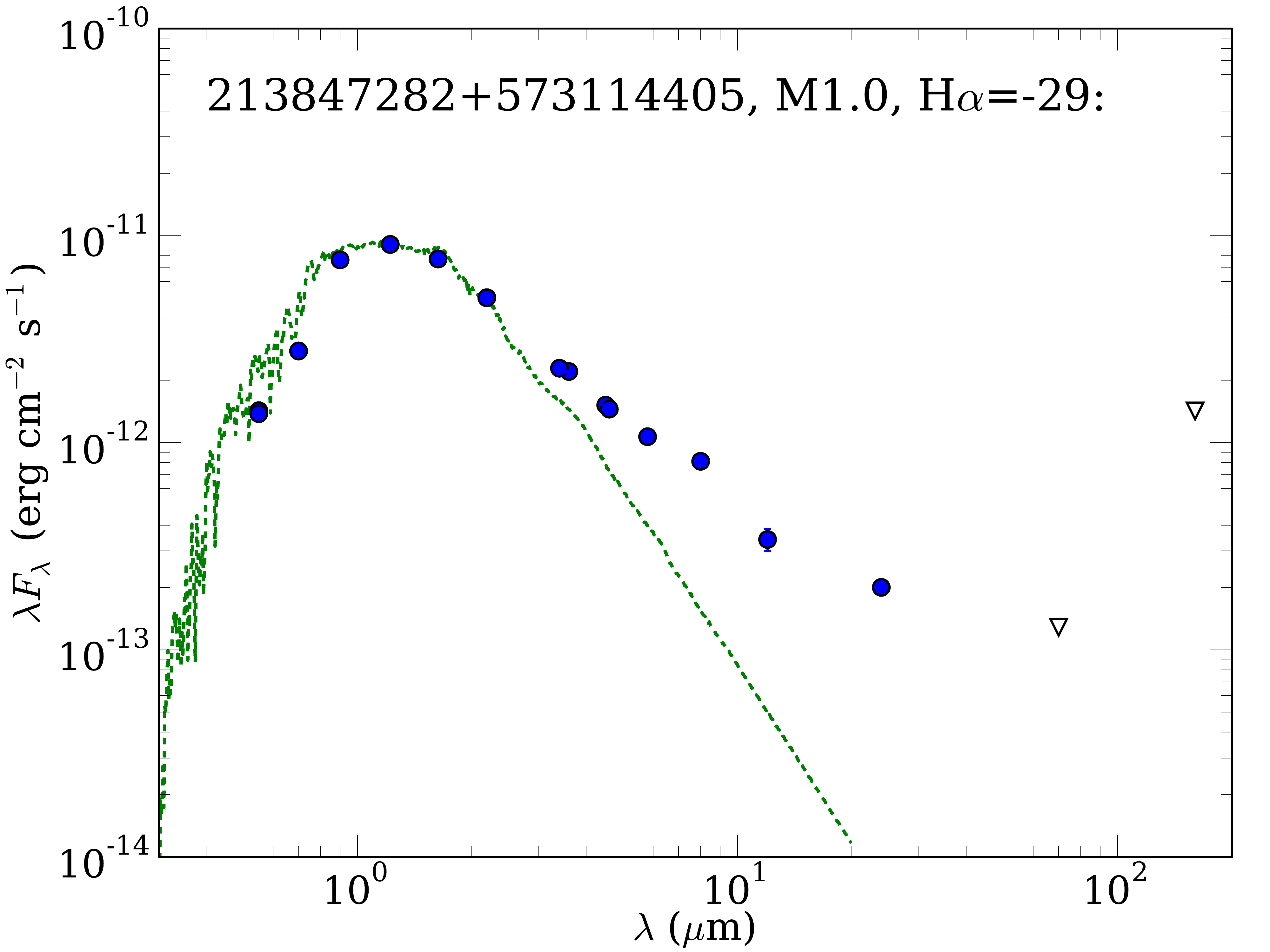} &
\includegraphics[width=0.24\linewidth]{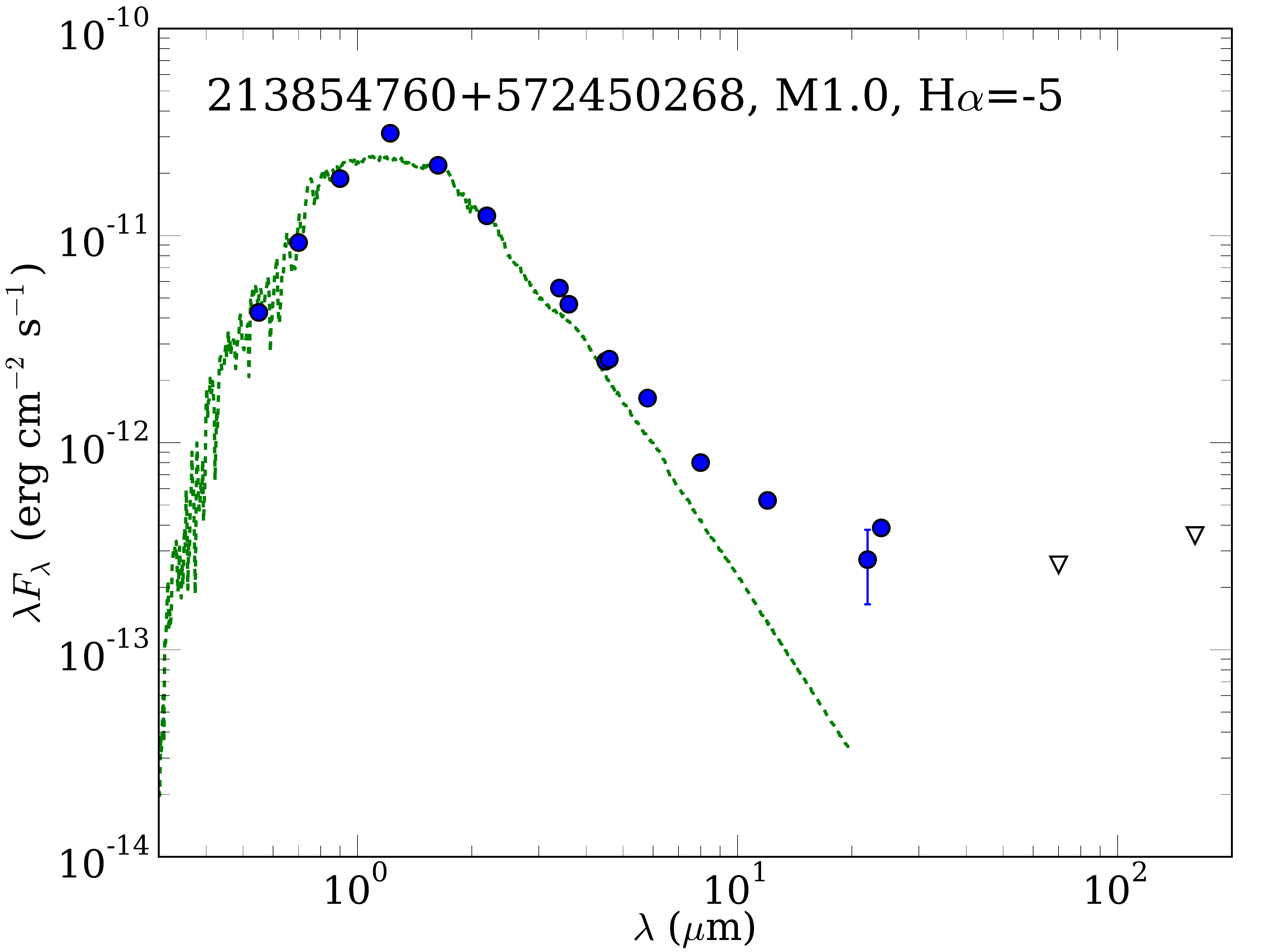} &
\includegraphics[width=0.24\linewidth]{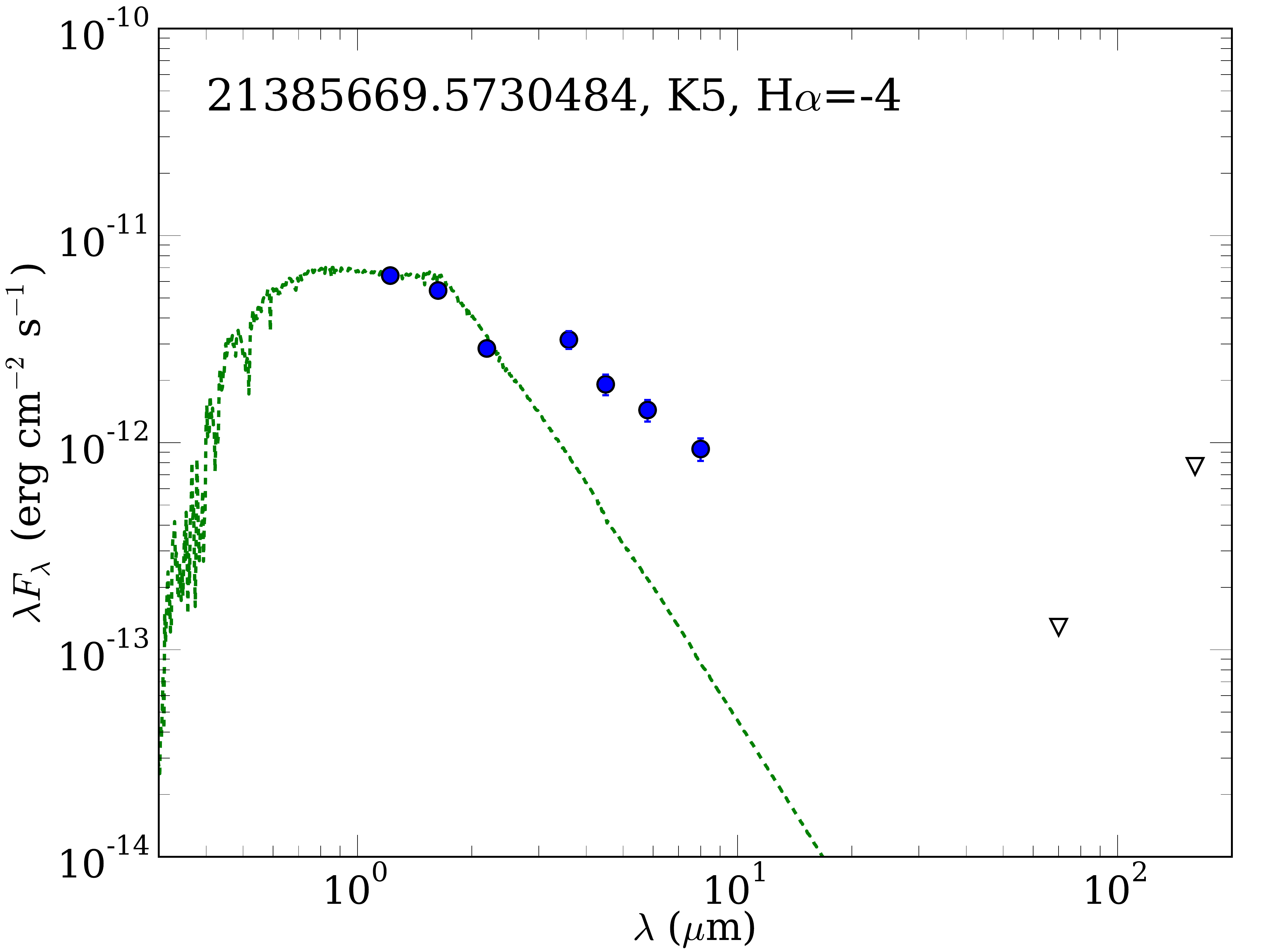} \\
\includegraphics[width=0.24\linewidth]{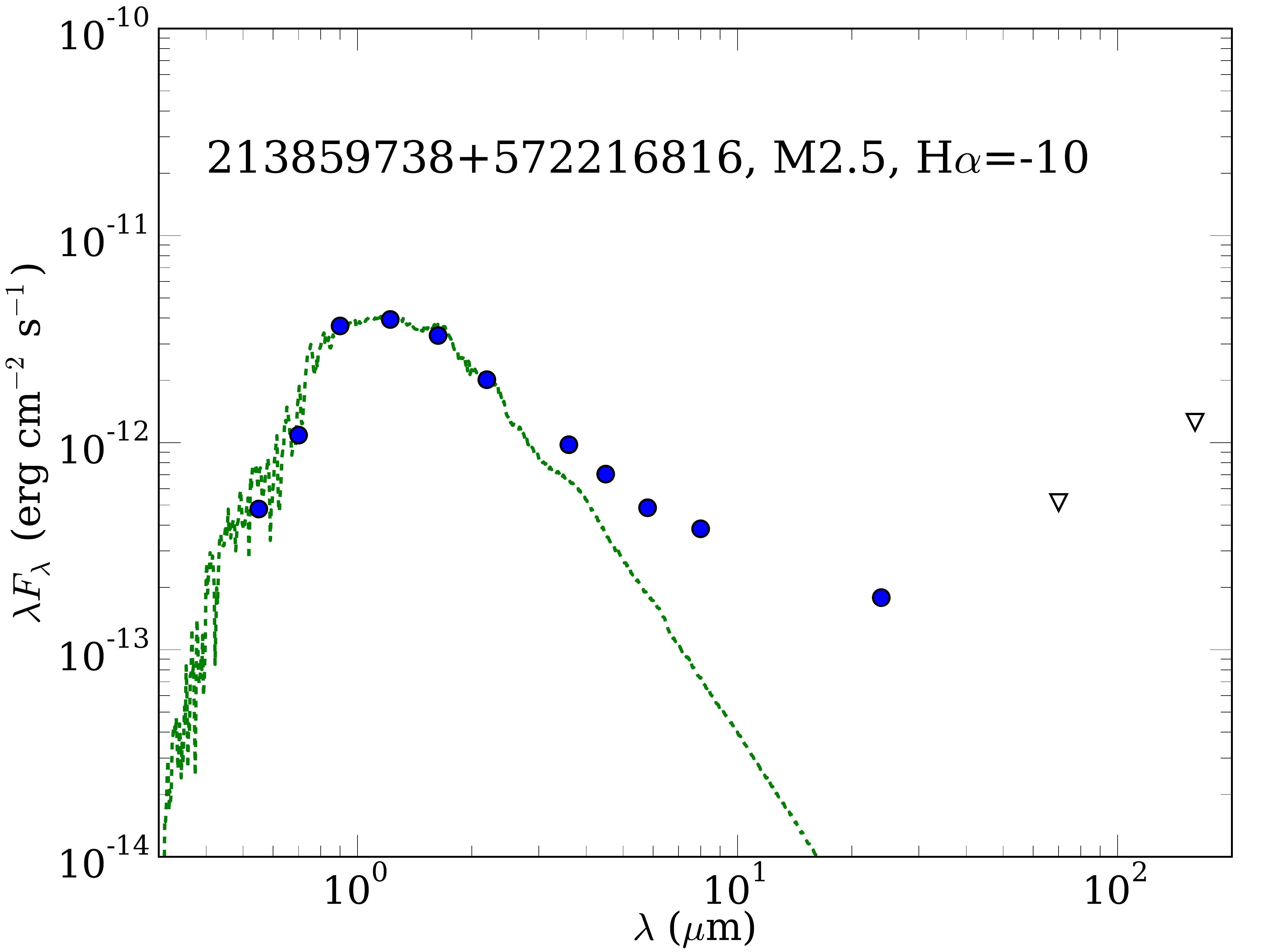} &
\includegraphics[width=0.24\linewidth]{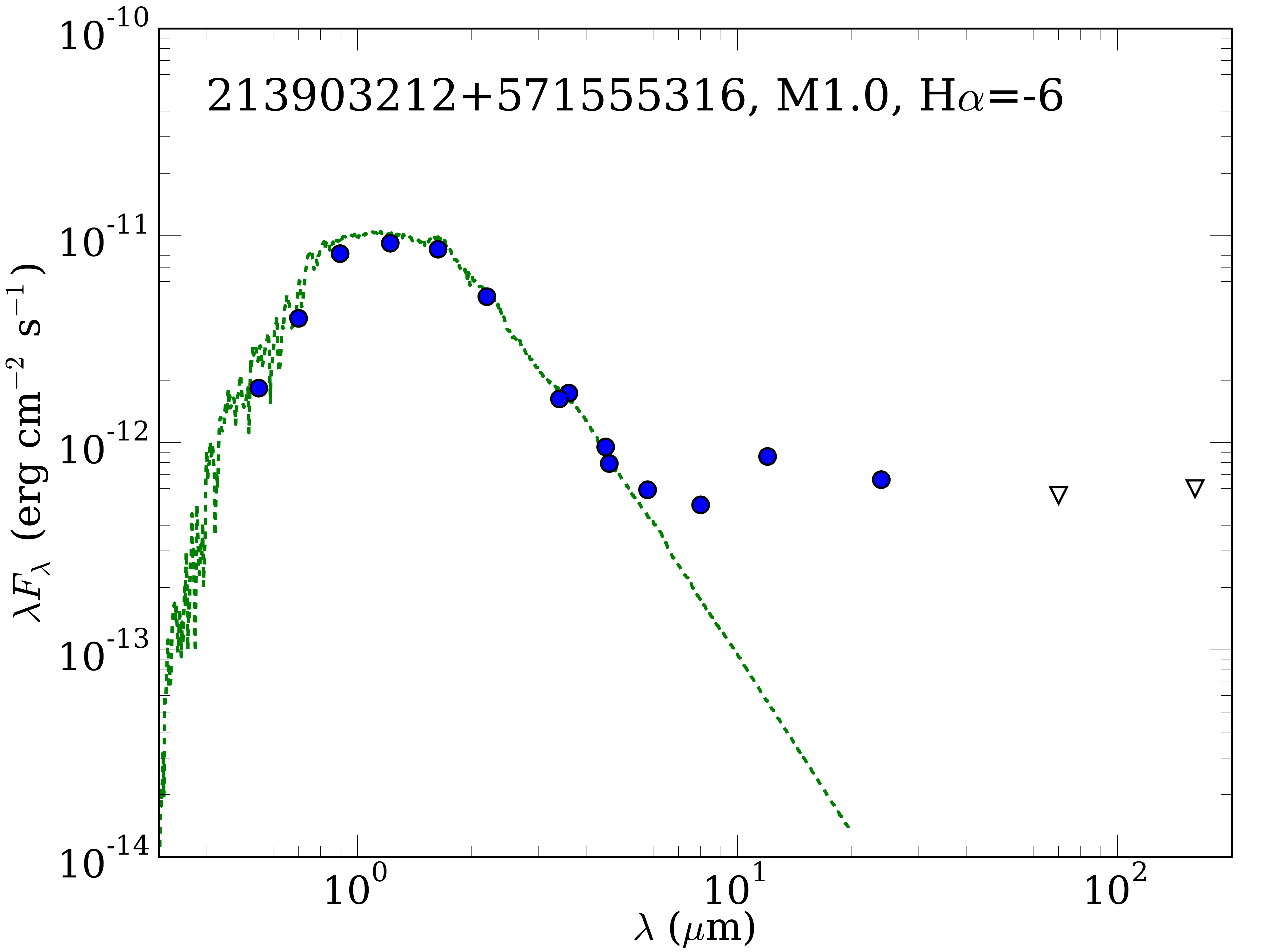} &
\includegraphics[width=0.24\linewidth]{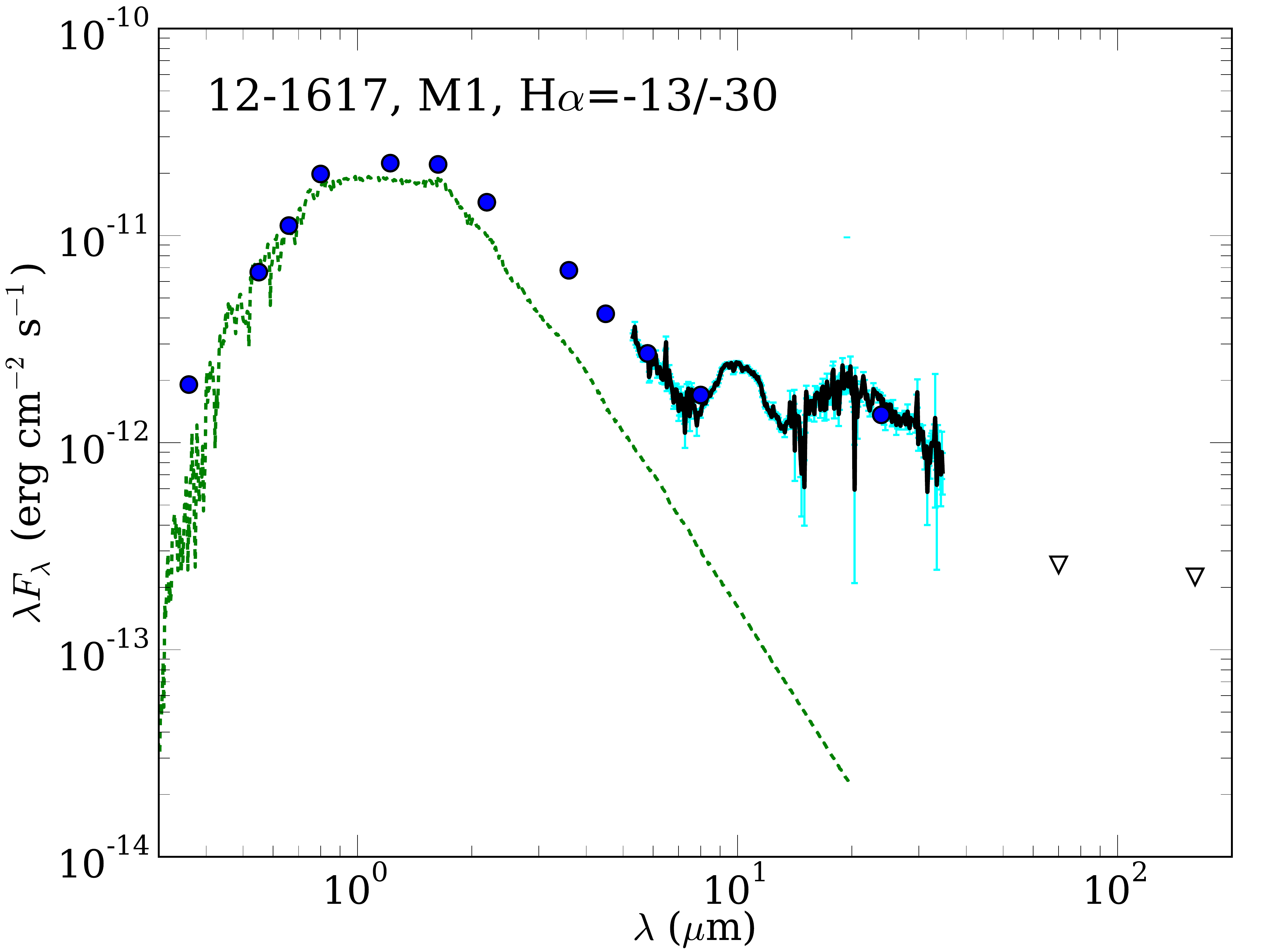} &
\includegraphics[width=0.24\linewidth]{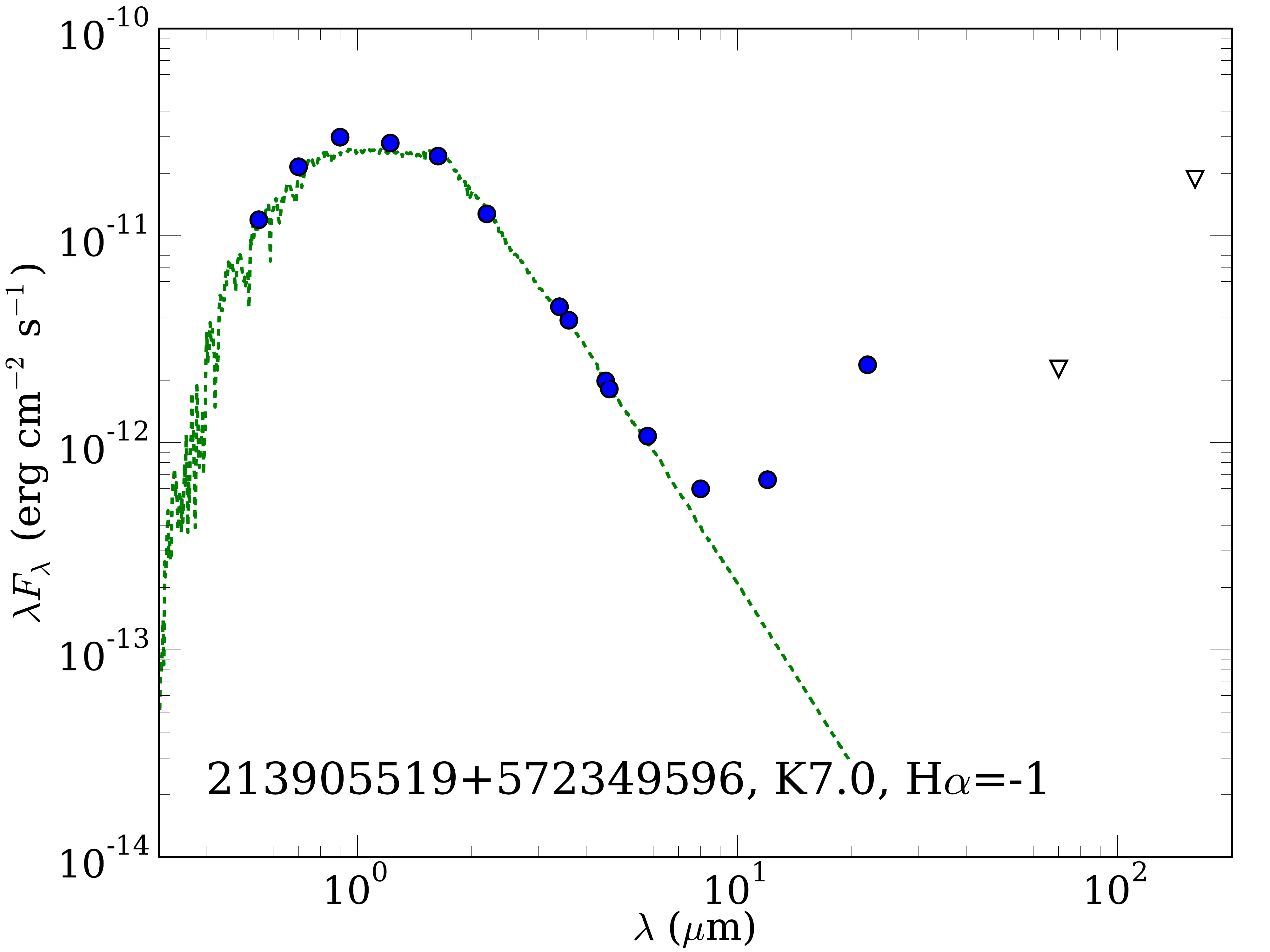} \\
\includegraphics[width=0.24\linewidth]{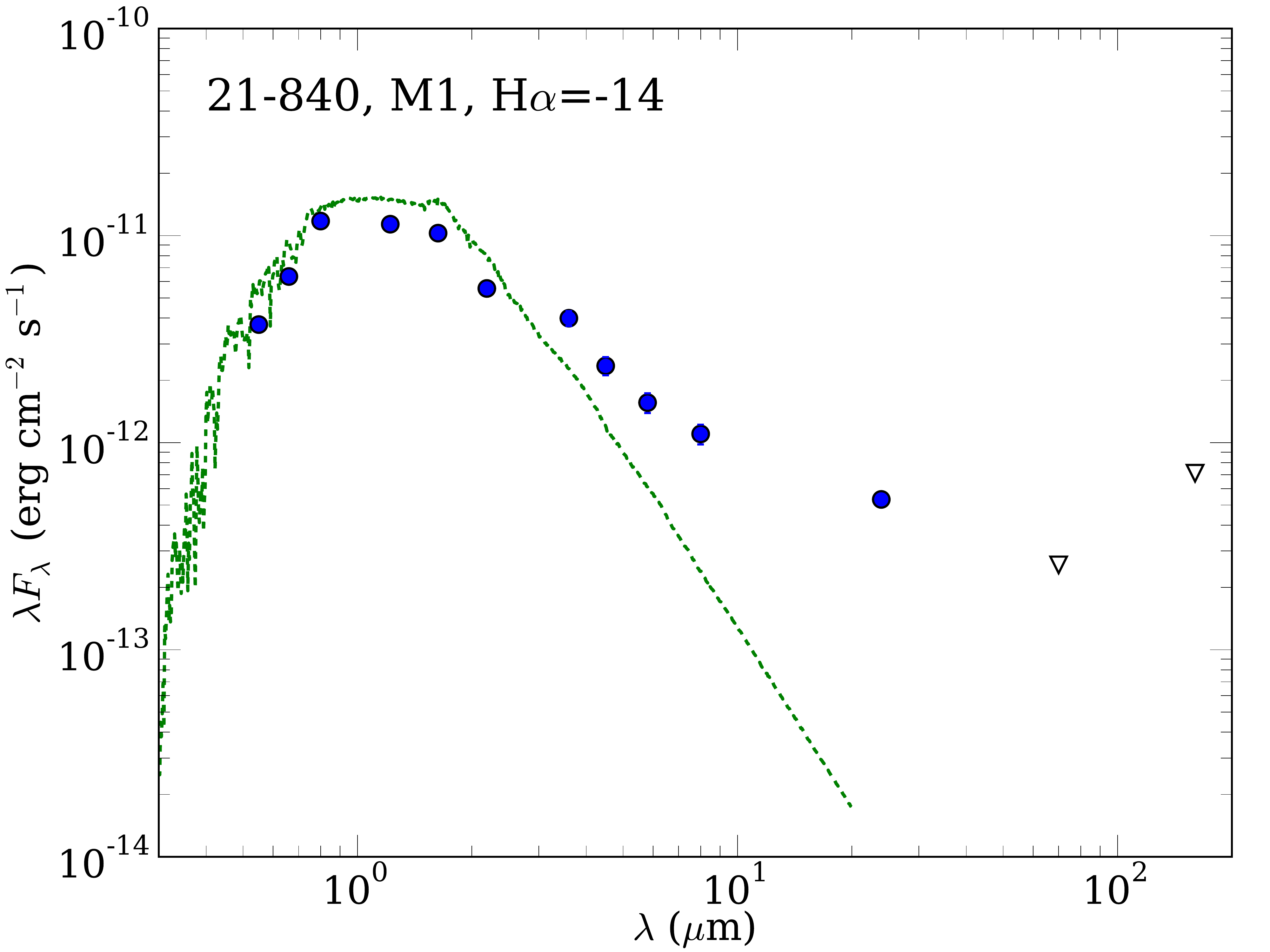} &
\includegraphics[width=0.24\linewidth]{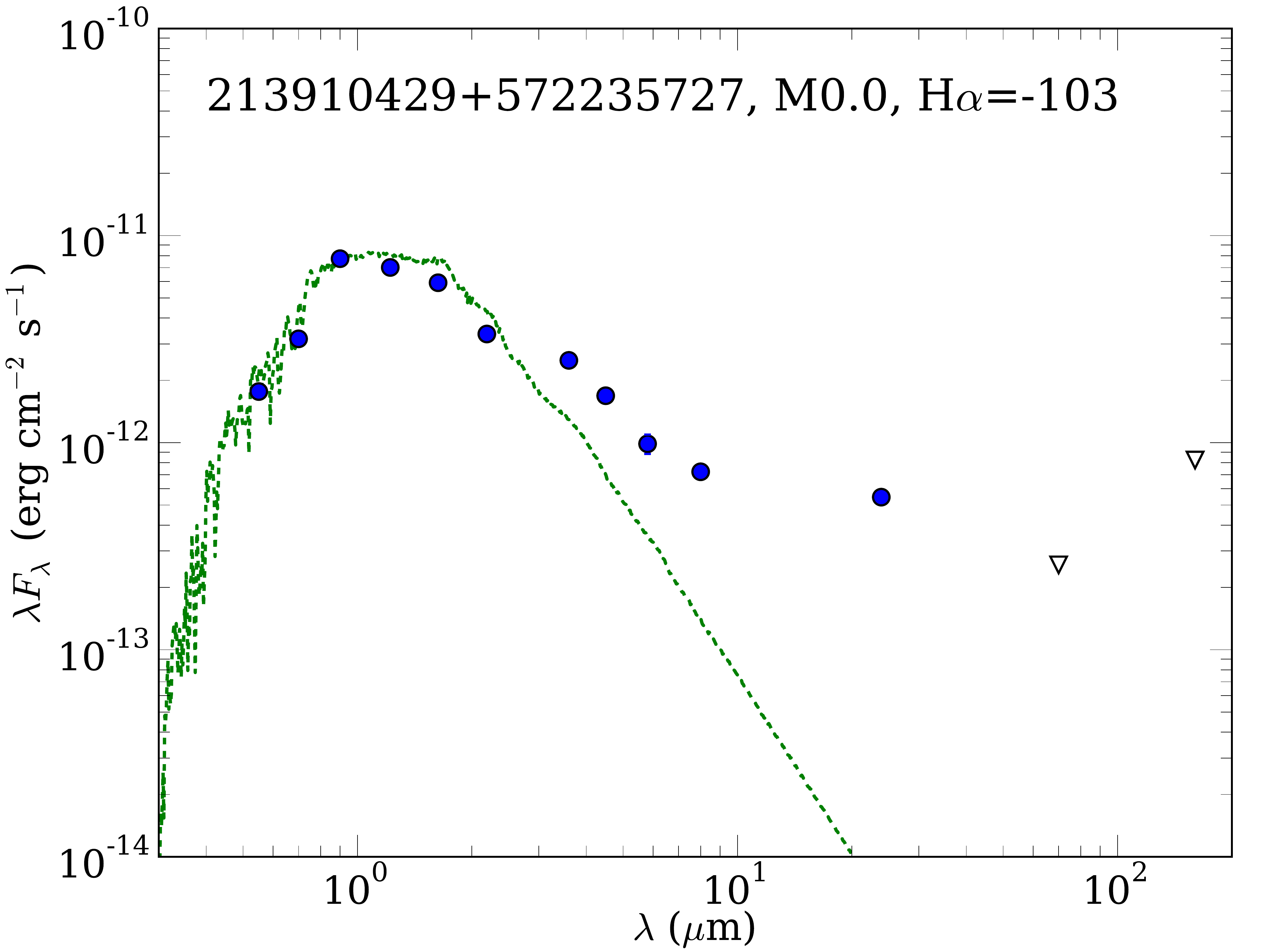} &
\includegraphics[width=0.24\linewidth]{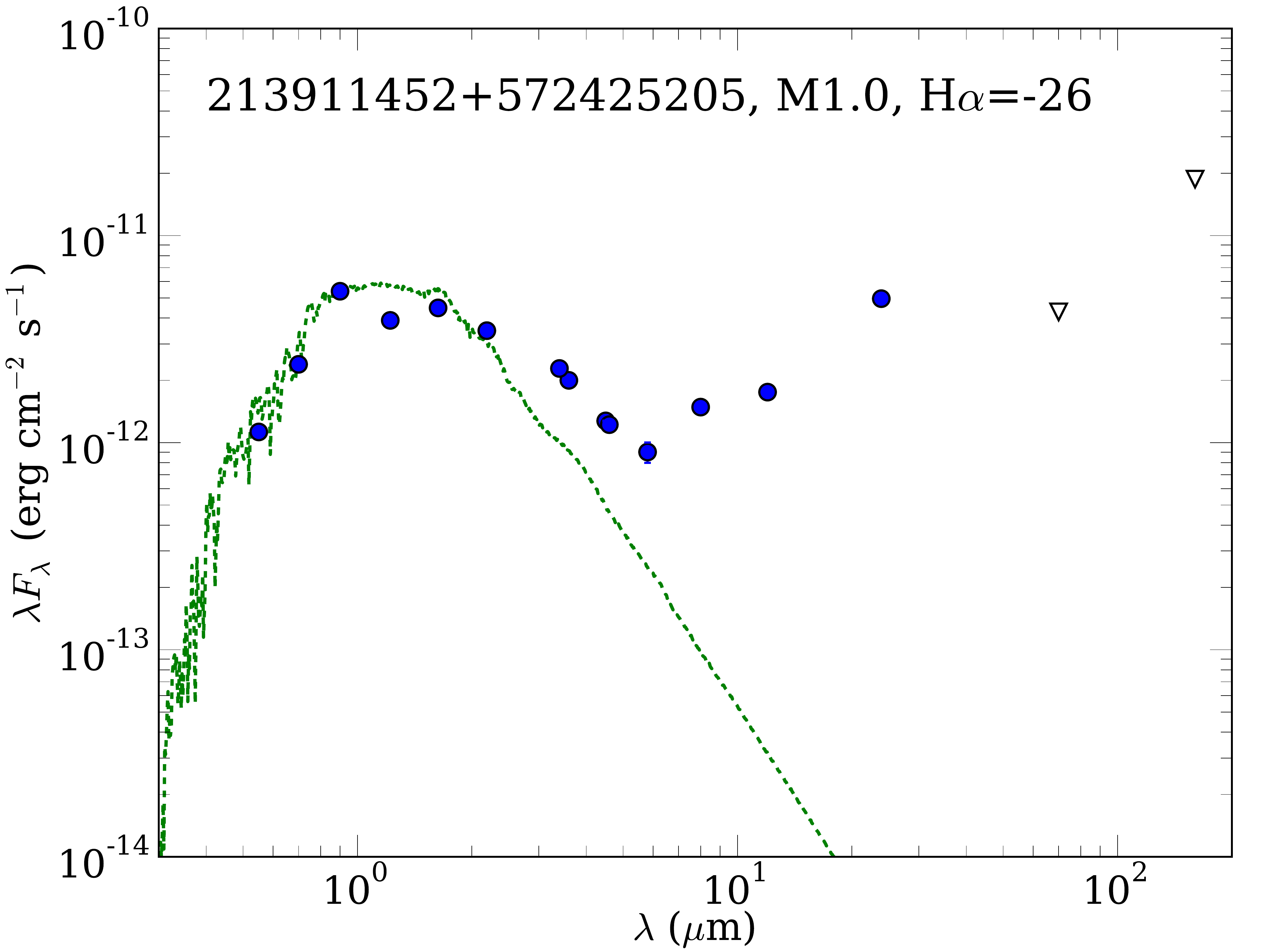} &
\includegraphics[width=0.24\linewidth]{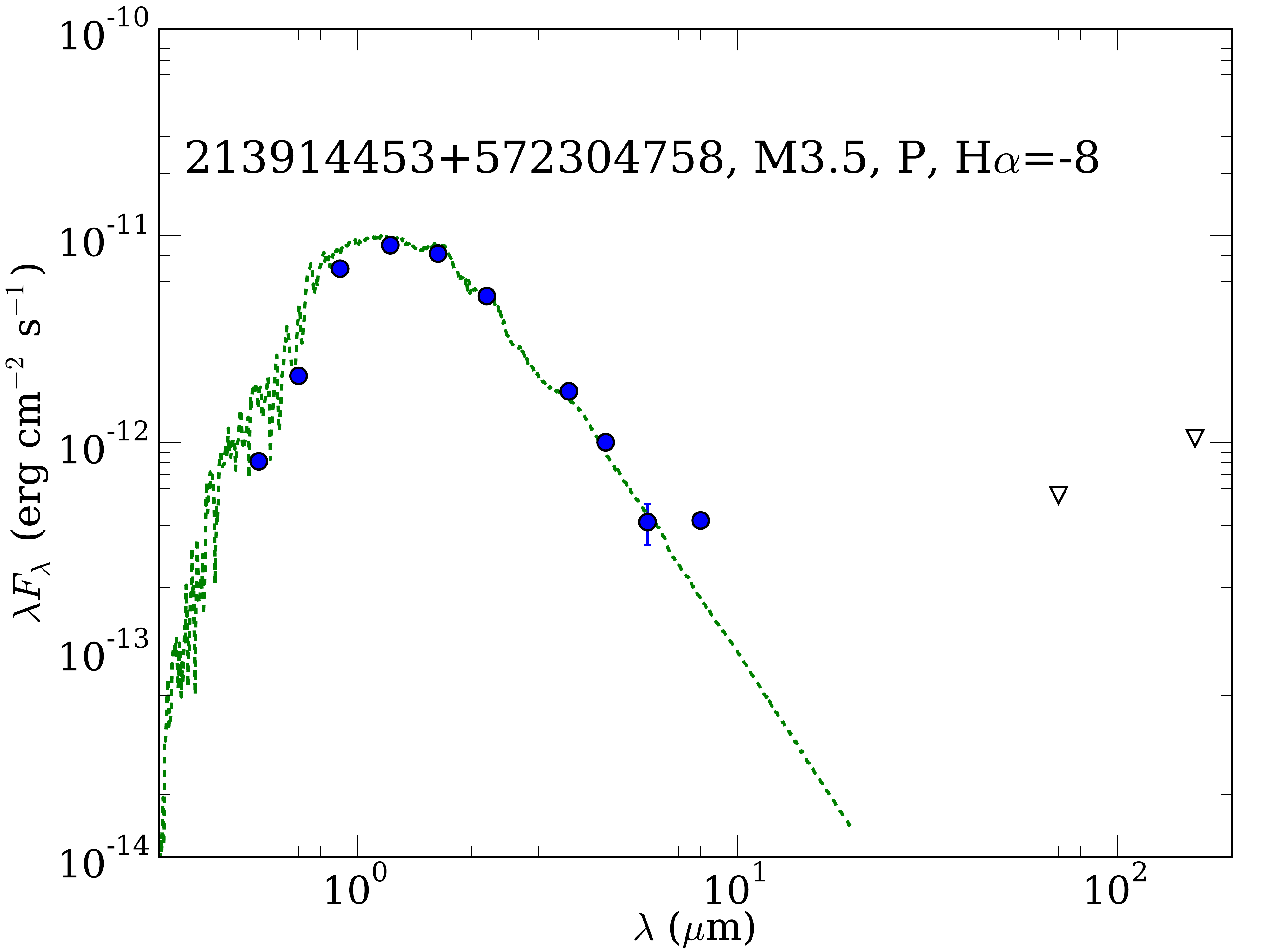} \\
\includegraphics[width=0.24\linewidth]{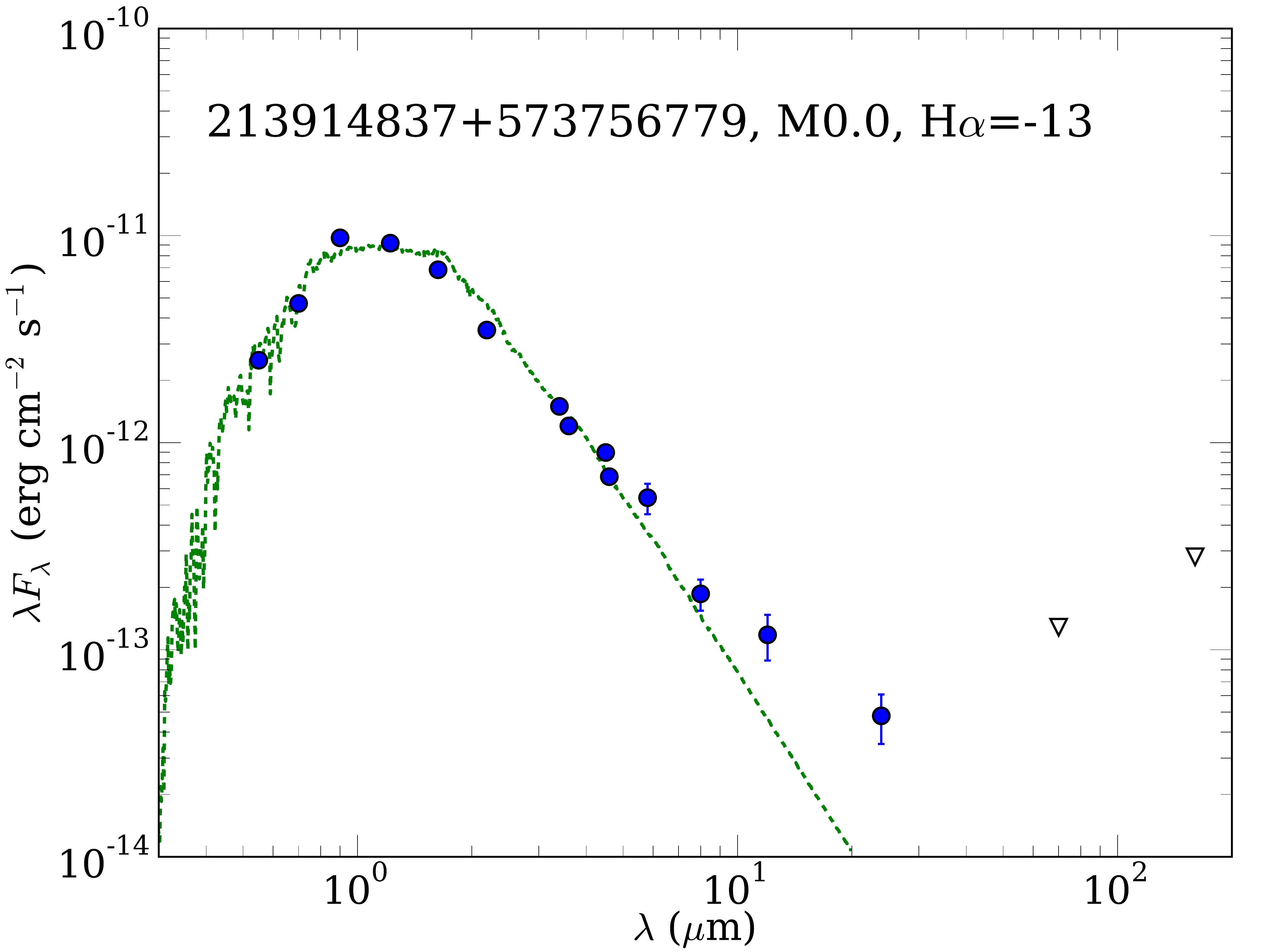} &
\includegraphics[width=0.24\linewidth]{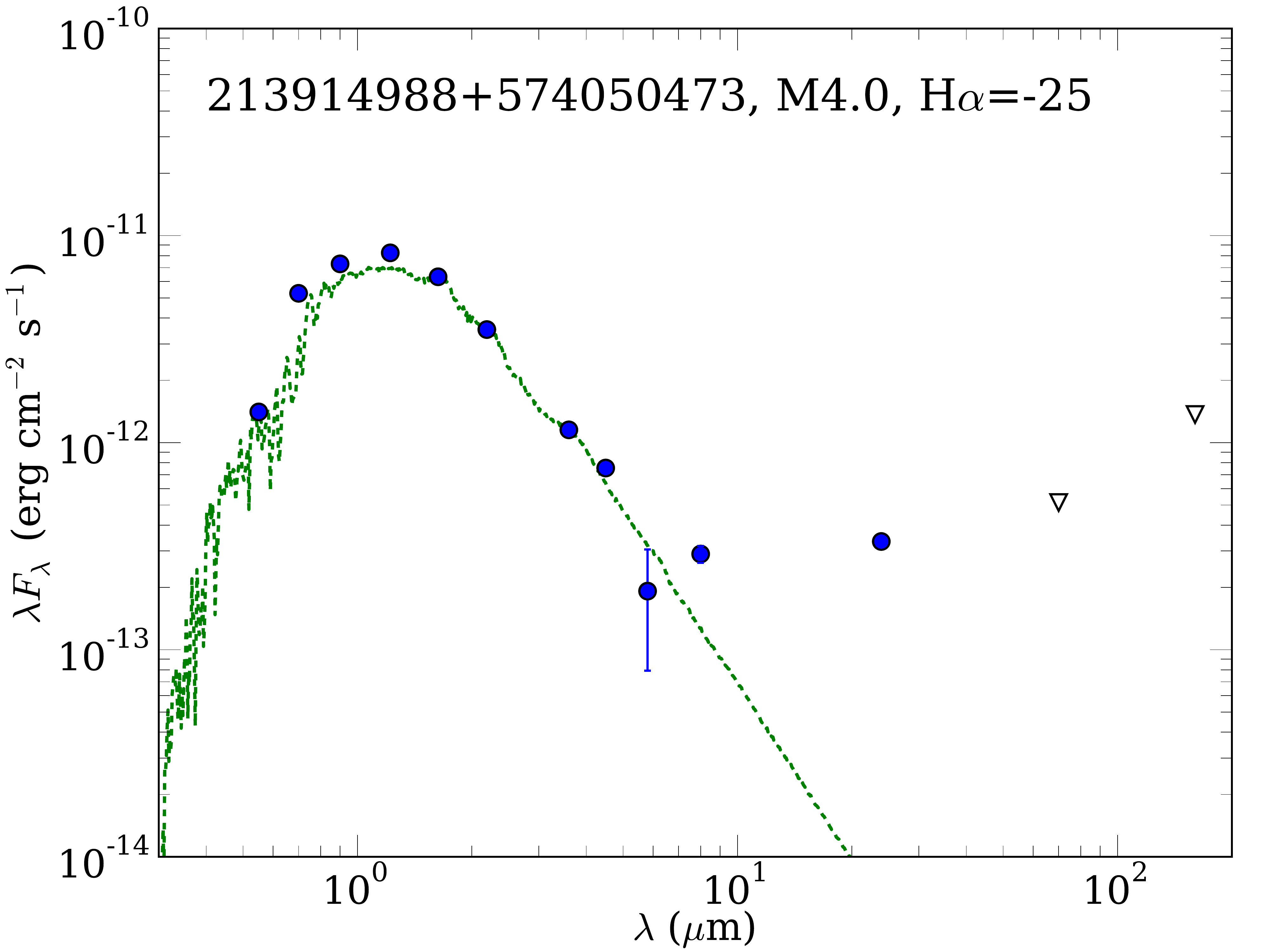} &
\includegraphics[width=0.24\linewidth]{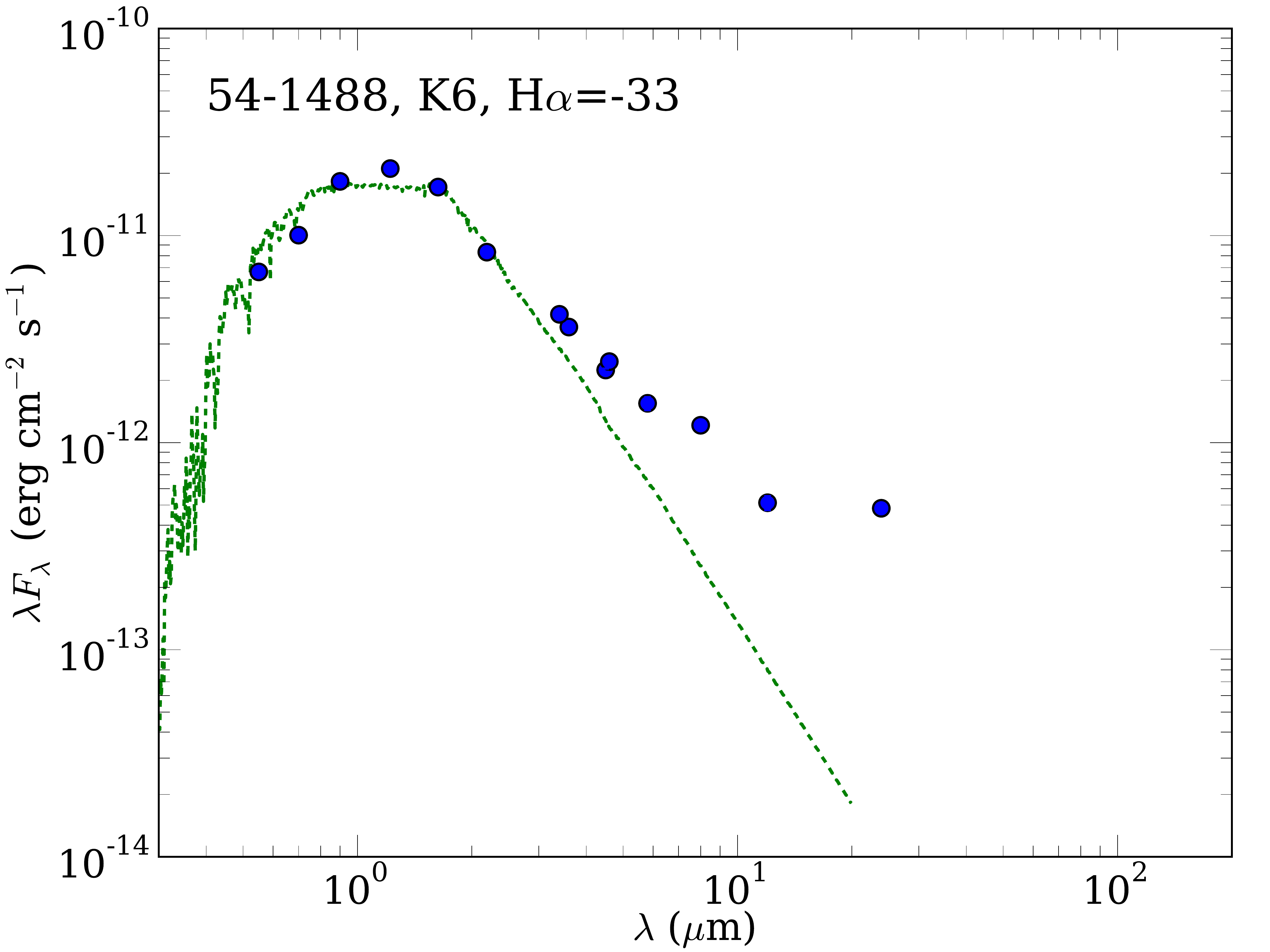} &
\includegraphics[width=0.24\linewidth]{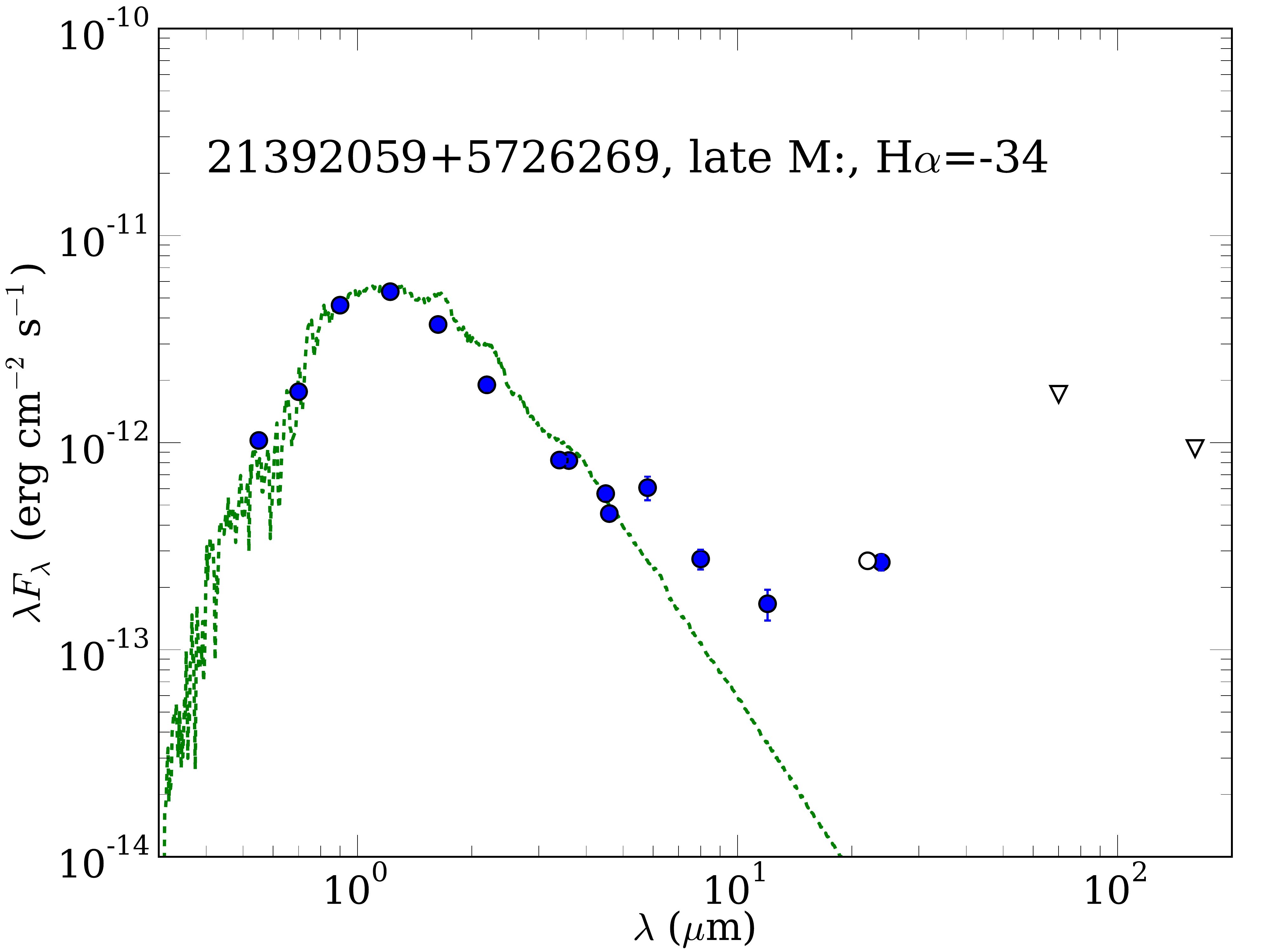} \\
\includegraphics[width=0.24\linewidth]{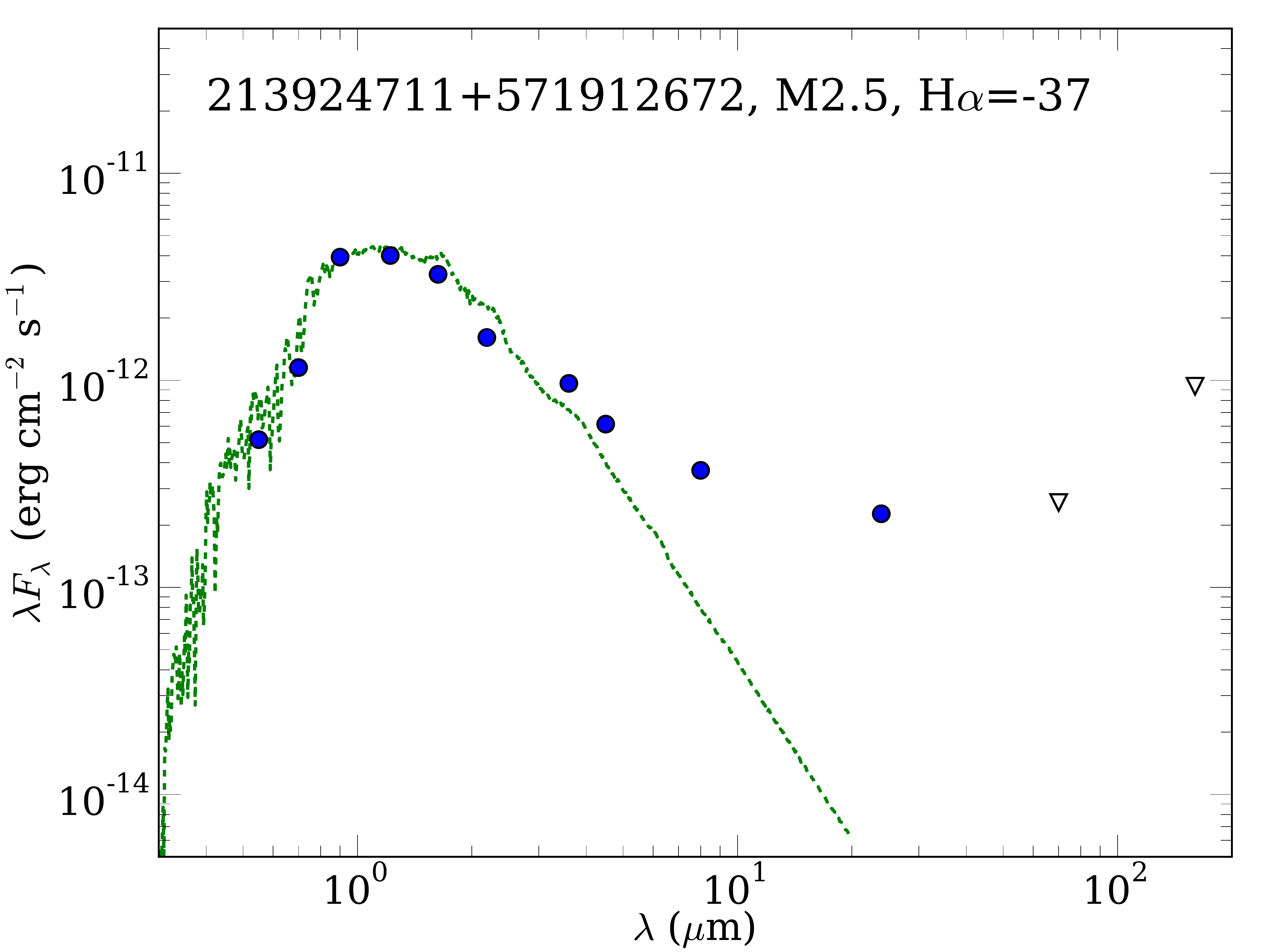} &
\includegraphics[width=0.24\linewidth]{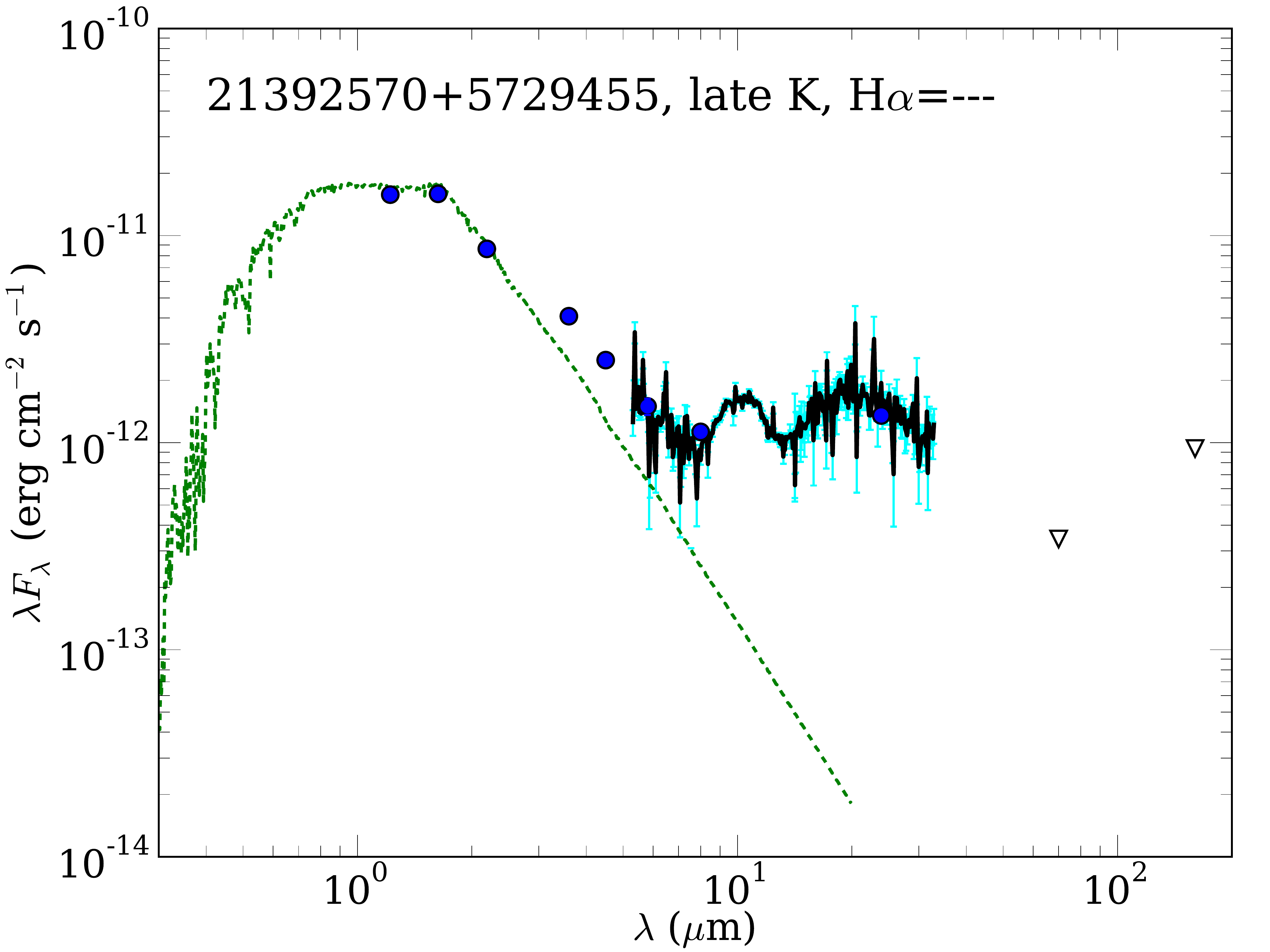} &
\includegraphics[width=0.24\linewidth]{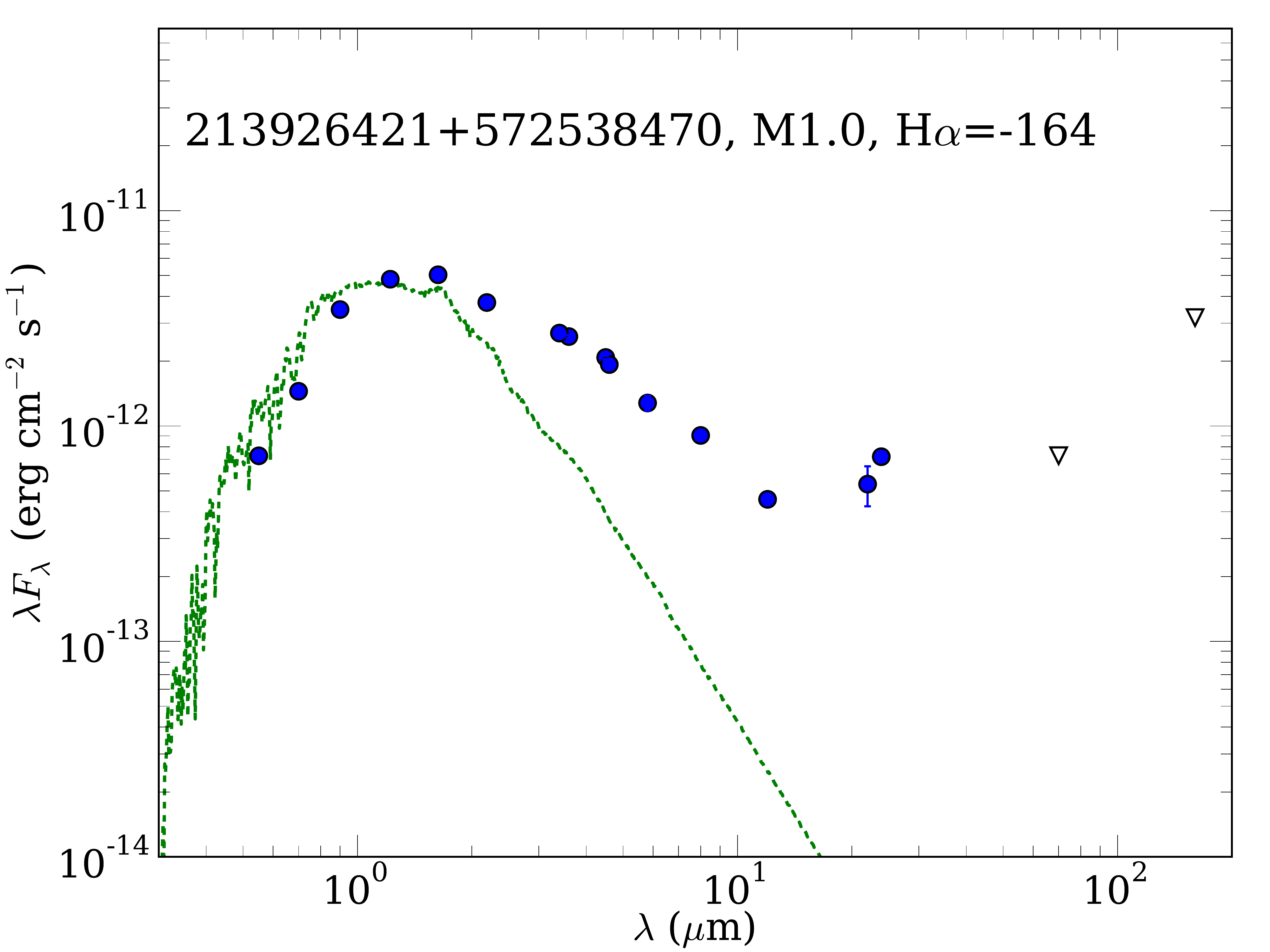} &
\includegraphics[width=0.24\linewidth]{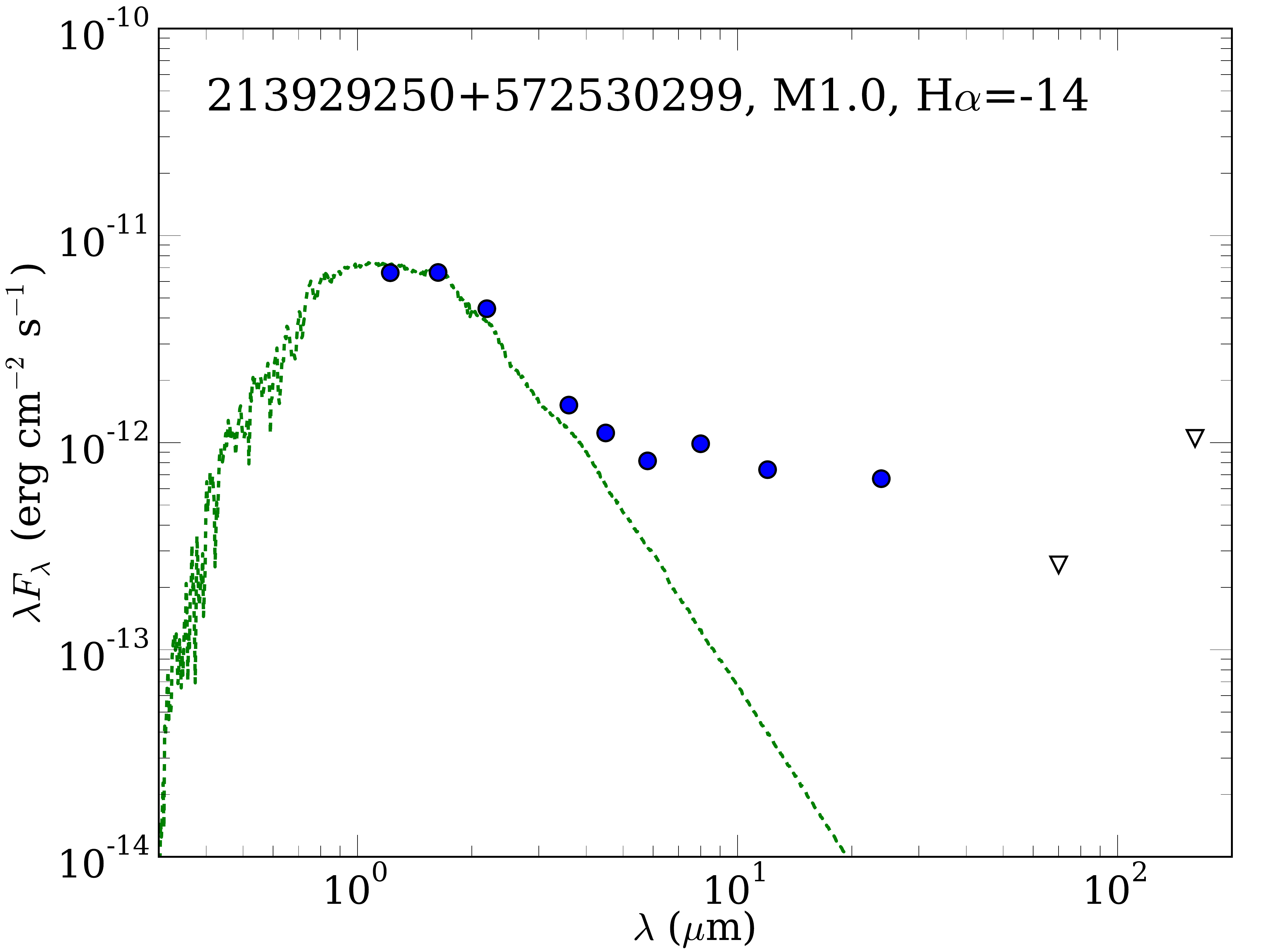} \\
\includegraphics[width=0.24\linewidth]{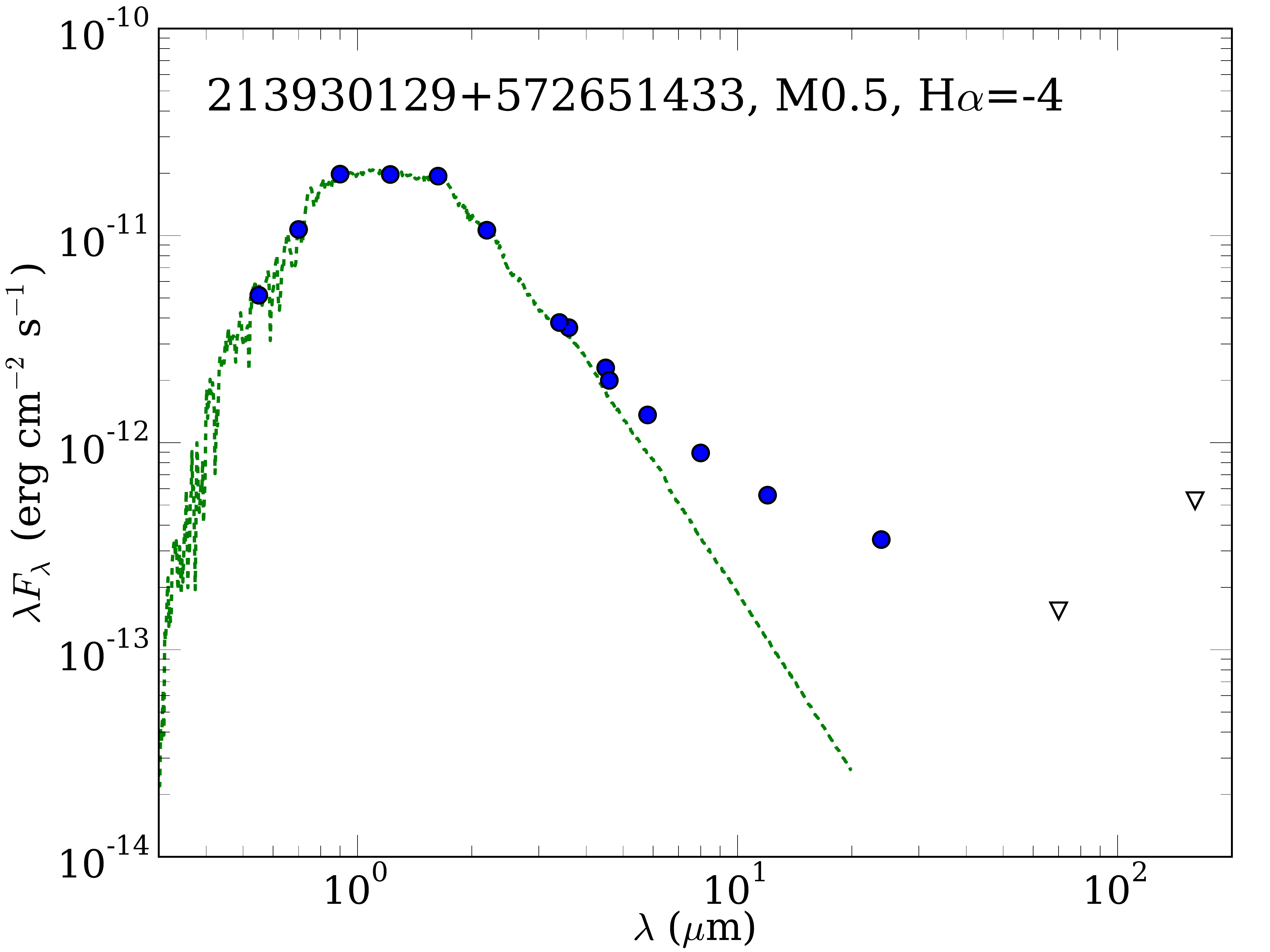} &
\includegraphics[width=0.24\linewidth]{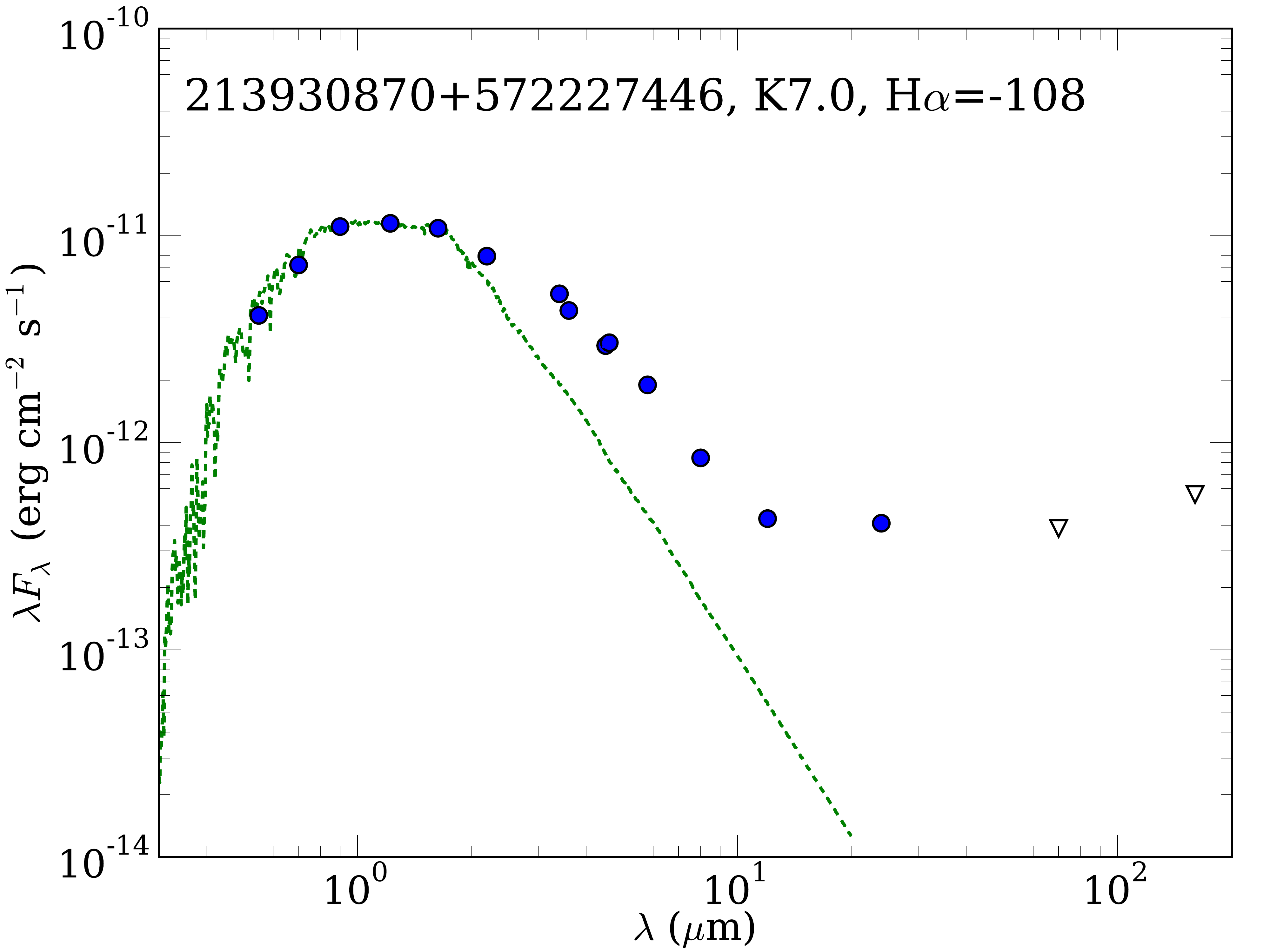} &
\includegraphics[width=0.24\linewidth]{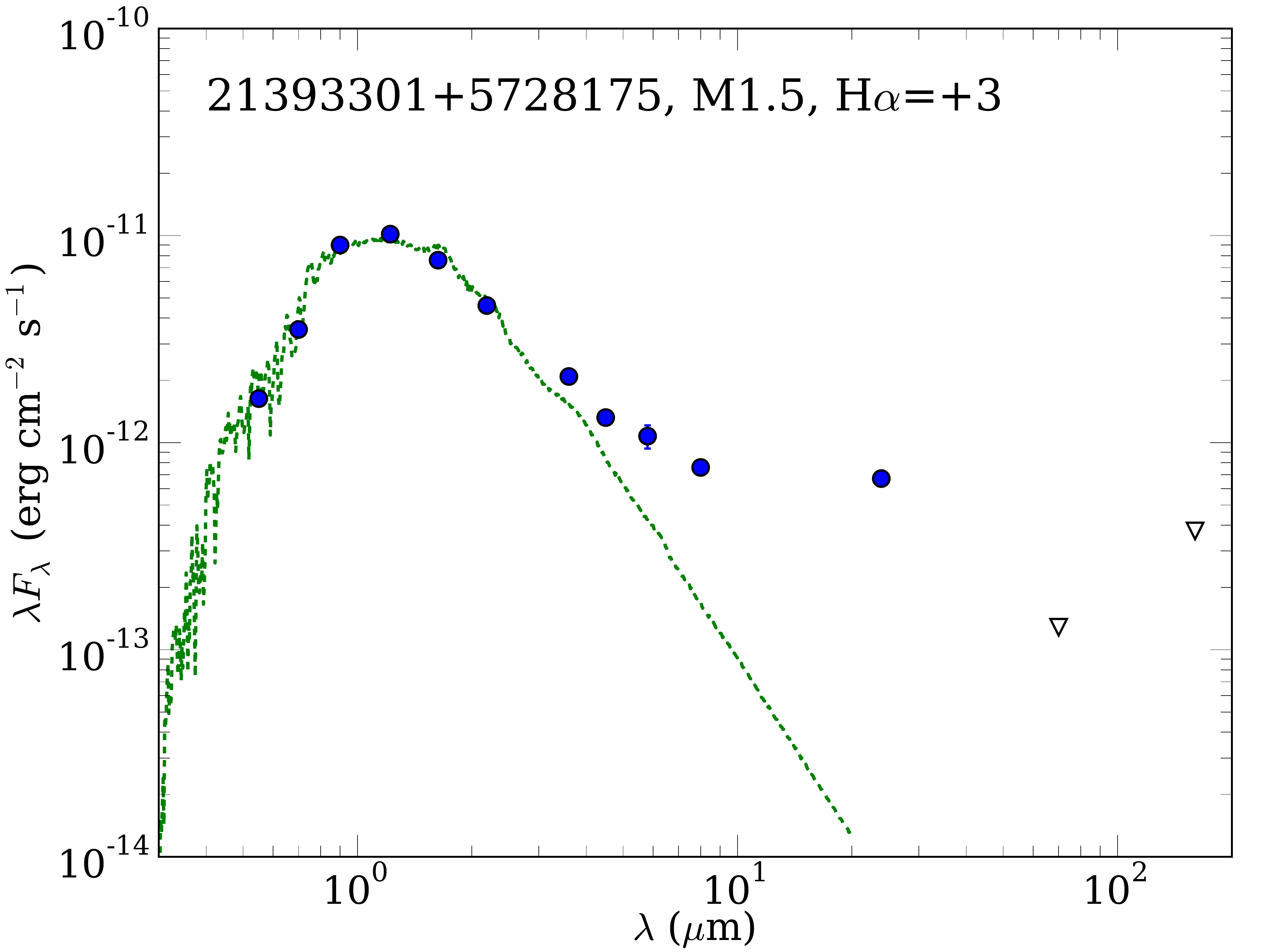} &
\includegraphics[width=0.24\linewidth]{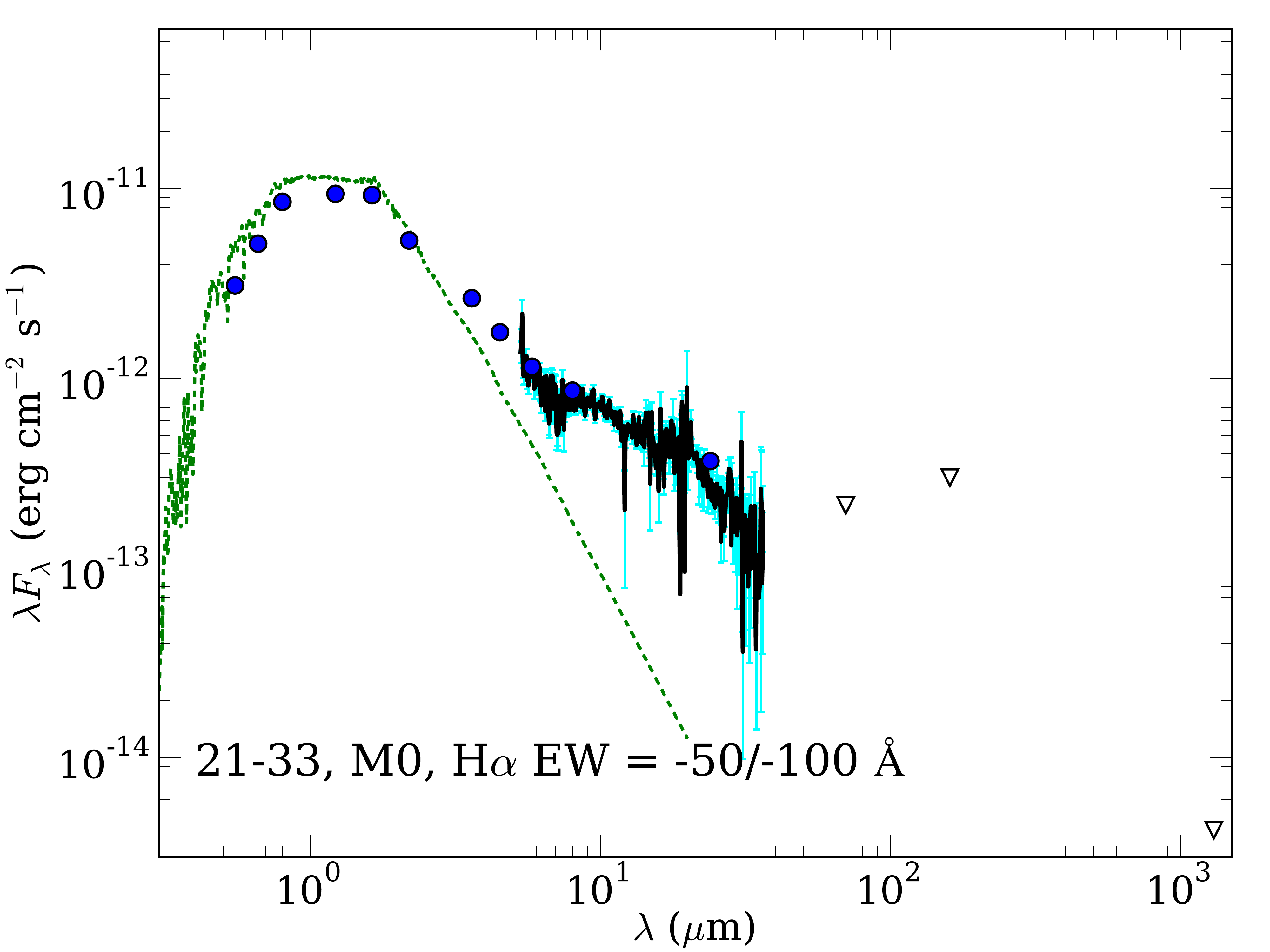} \\
\includegraphics[width=0.24\linewidth]{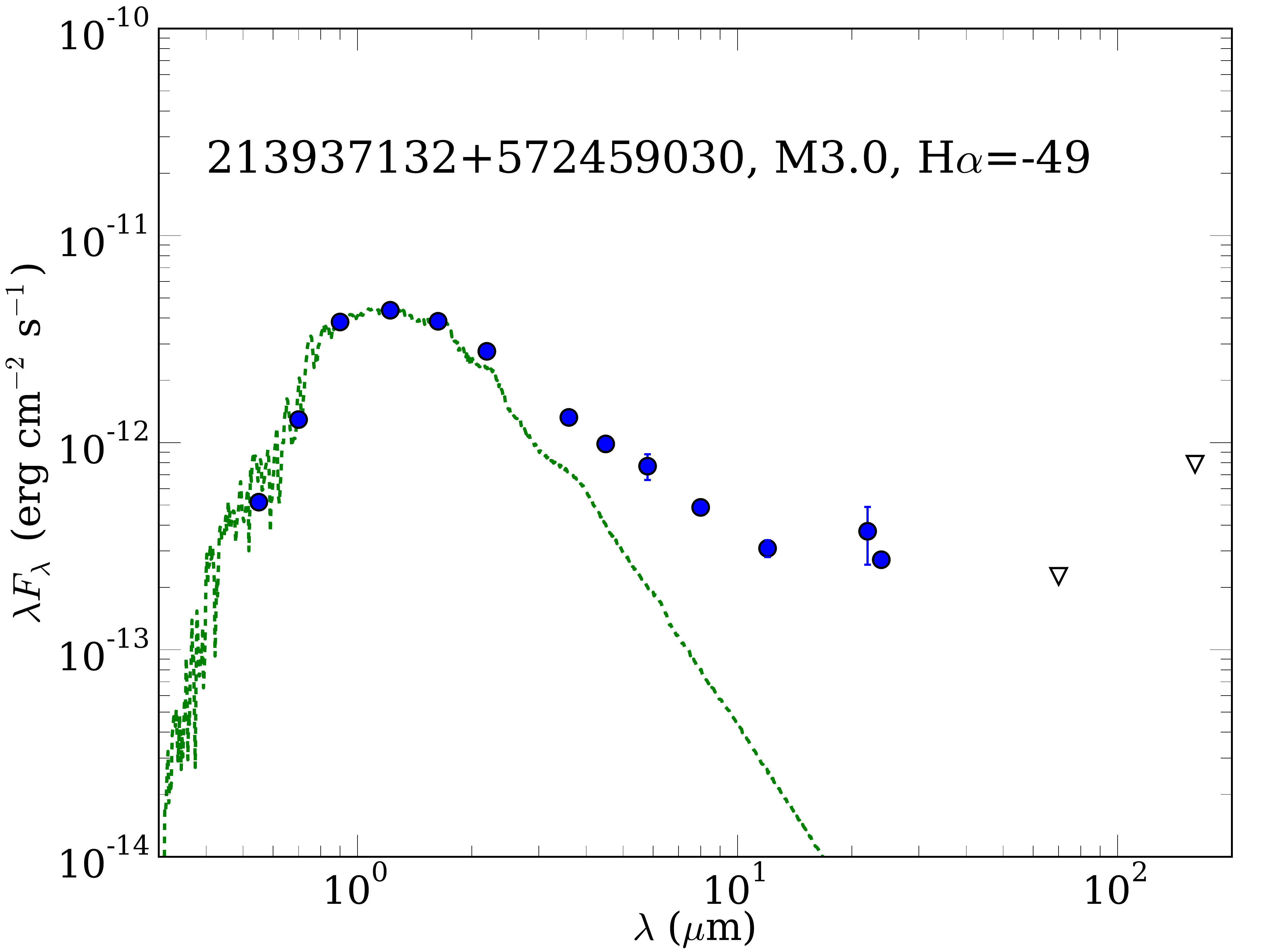} &
\includegraphics[width=0.24\linewidth]{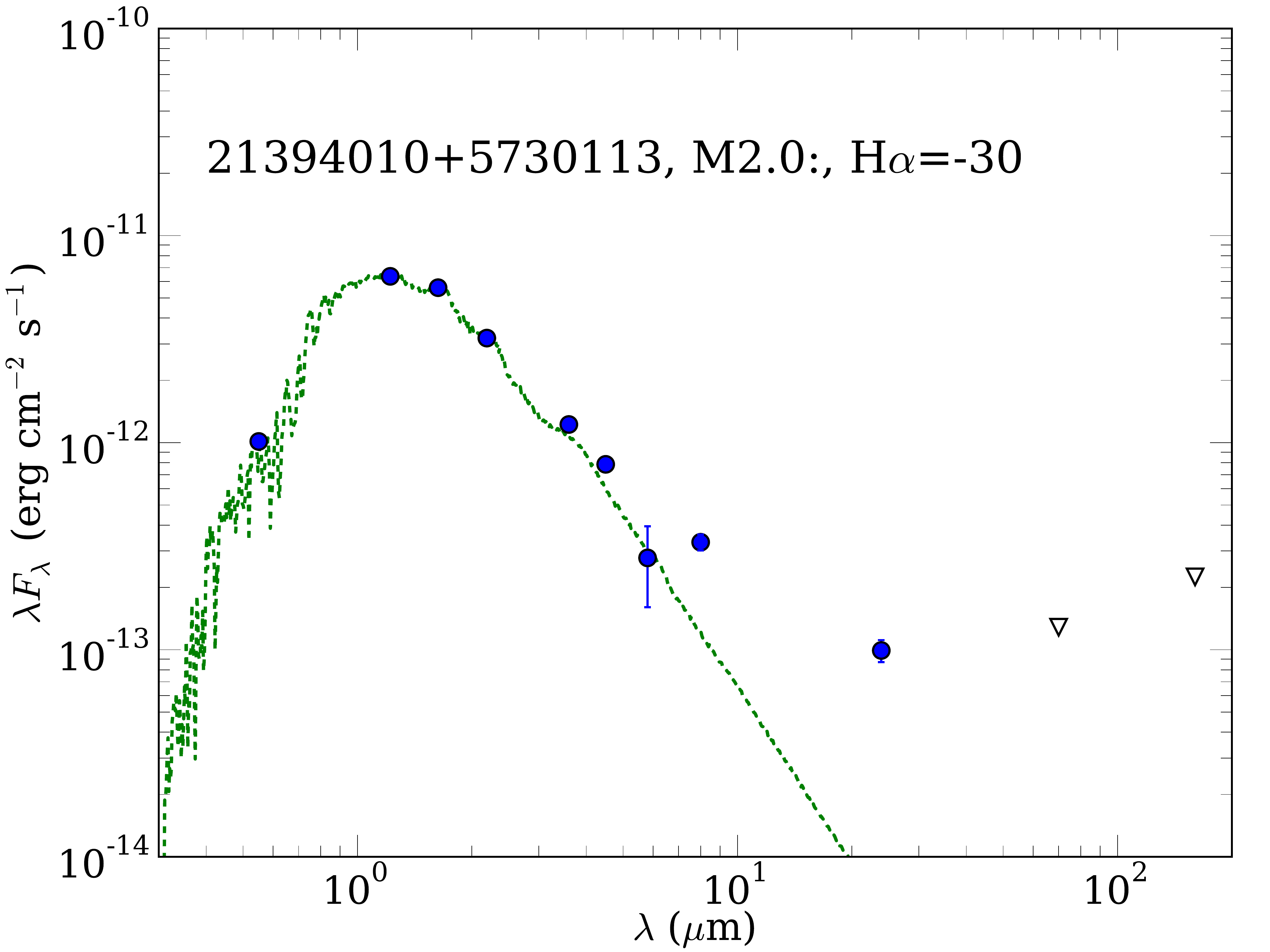} &
\includegraphics[width=0.24\linewidth]{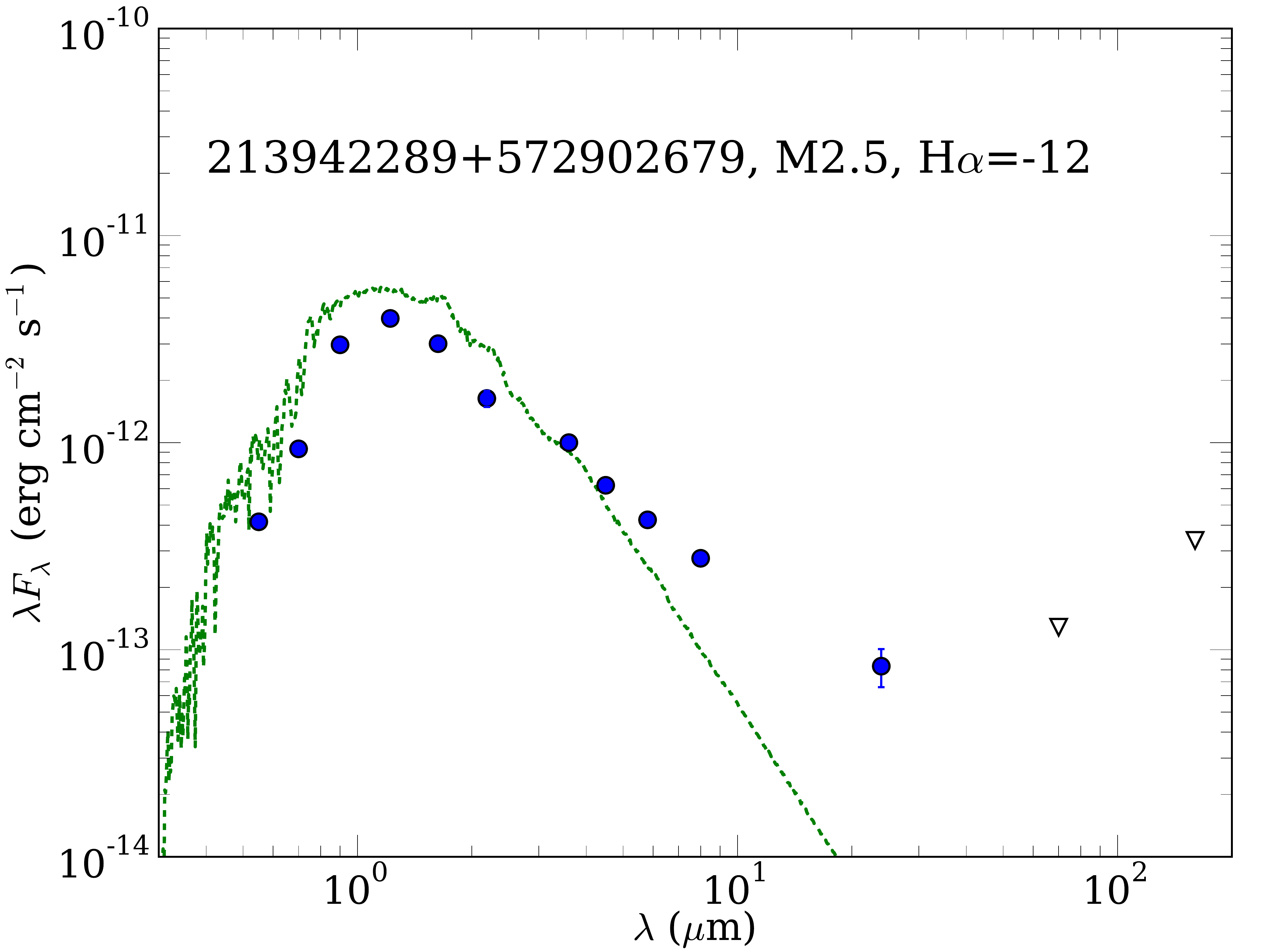} &
\includegraphics[width=0.24\linewidth]{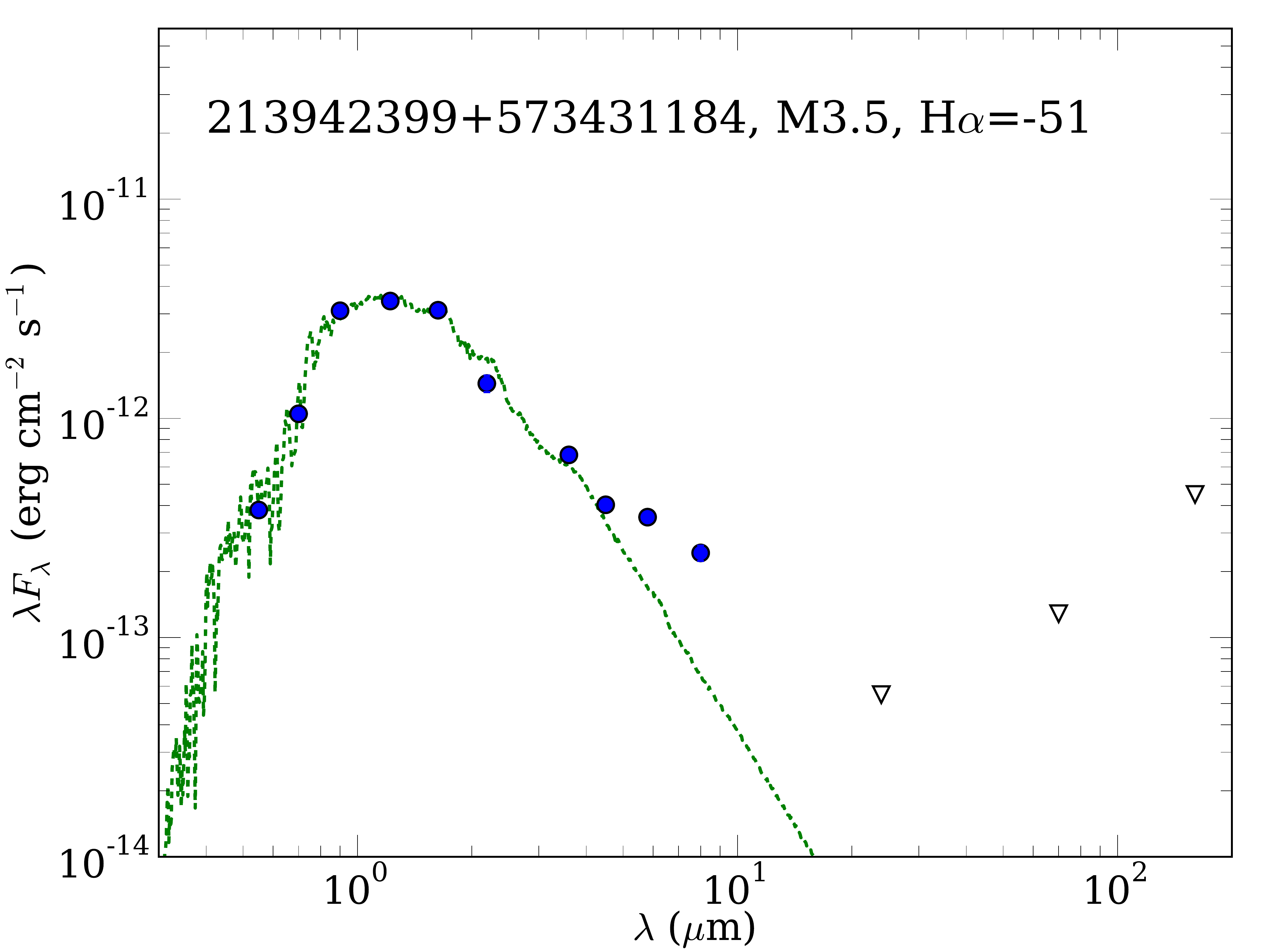} \\
\end{tabular}
\caption{SEDs of the objects with upper limits only. Symbols as in Figure \ref{uplims1-fig}.
\label{uplims3-fig}}
\end{figure*}

\begin{figure*}
\centering
\begin{tabular}{cccc}
\includegraphics[width=0.24\linewidth]{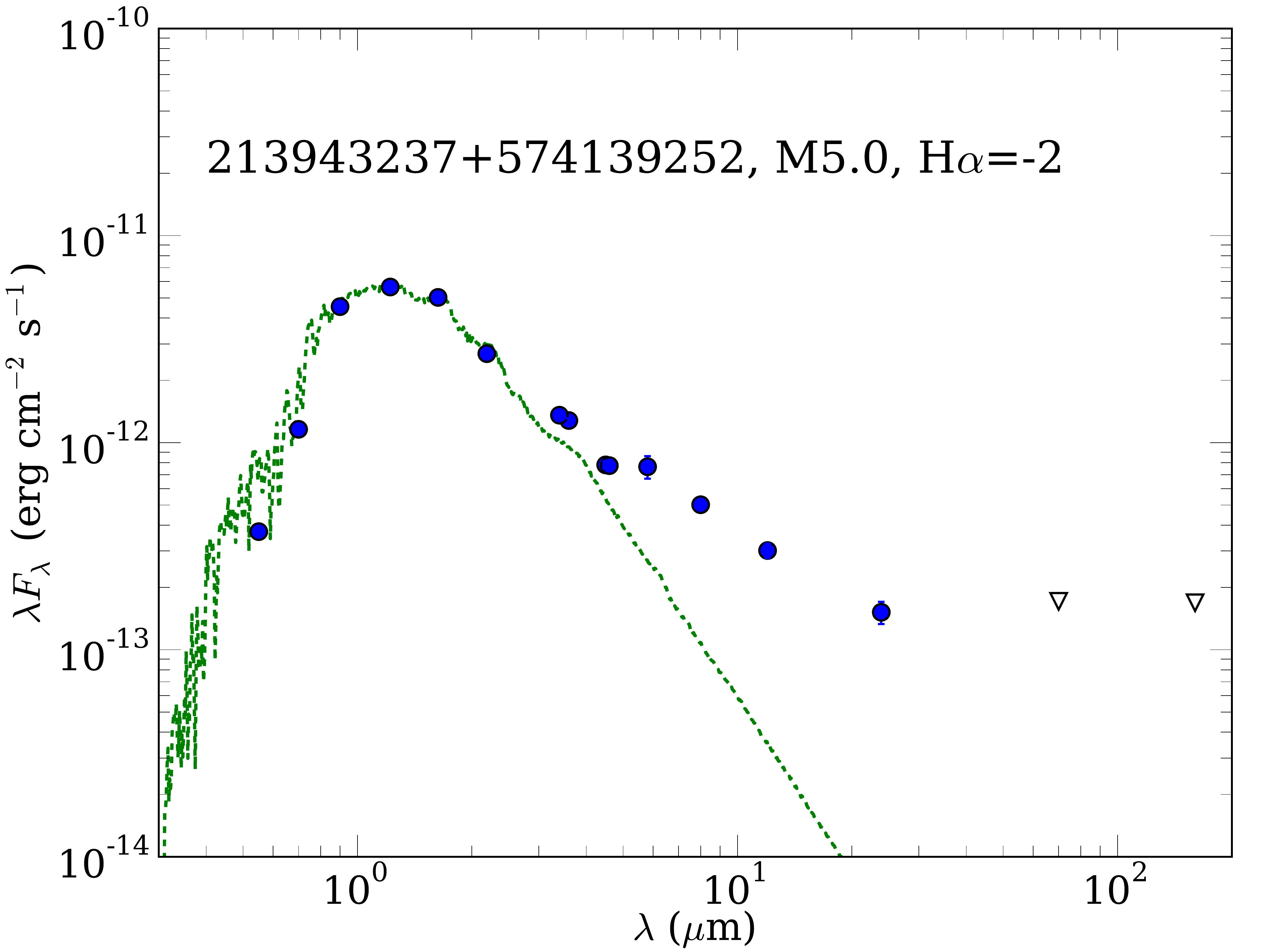} &
\includegraphics[width=0.24\linewidth]{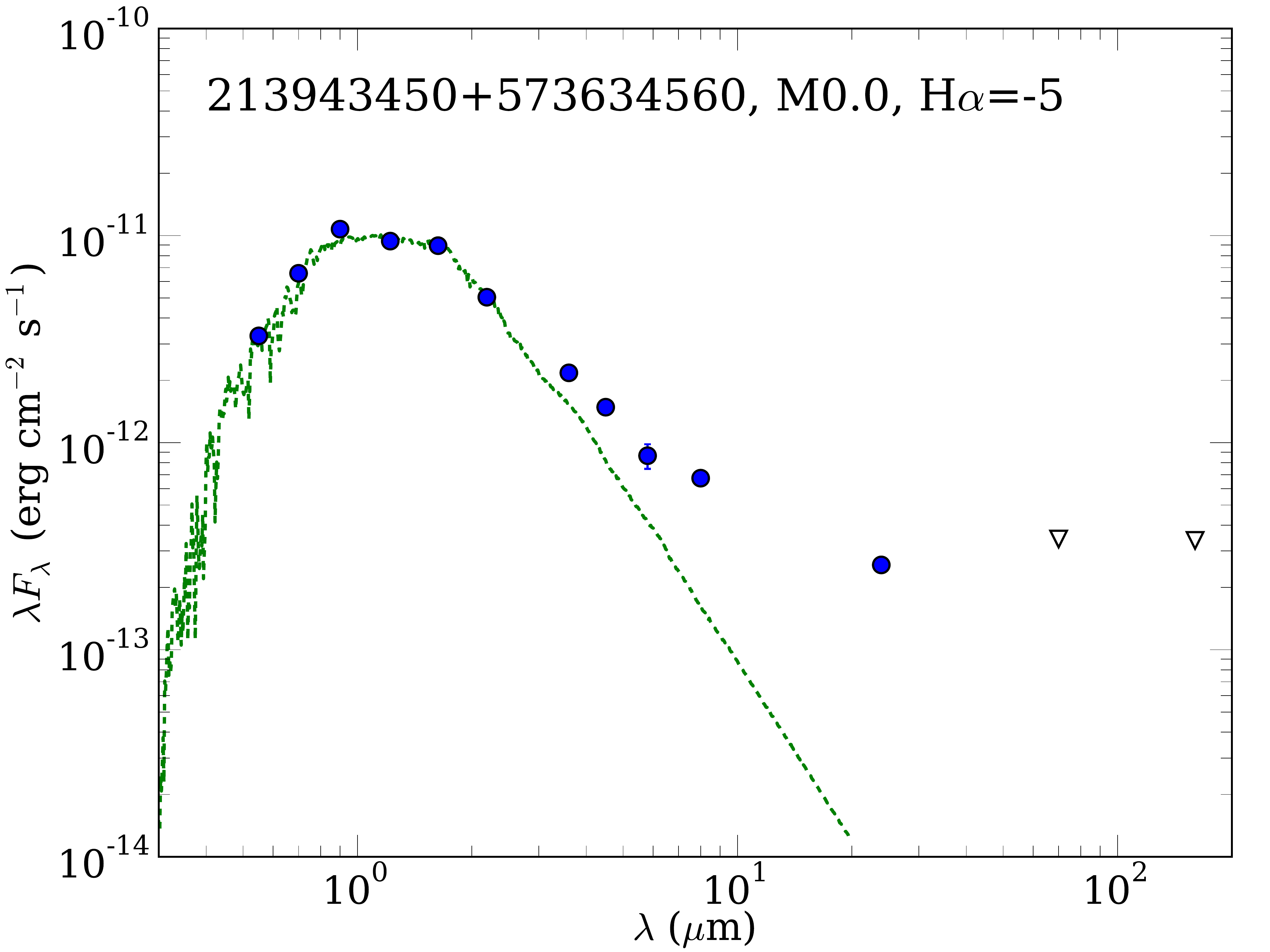} &
\includegraphics[width=0.24\linewidth]{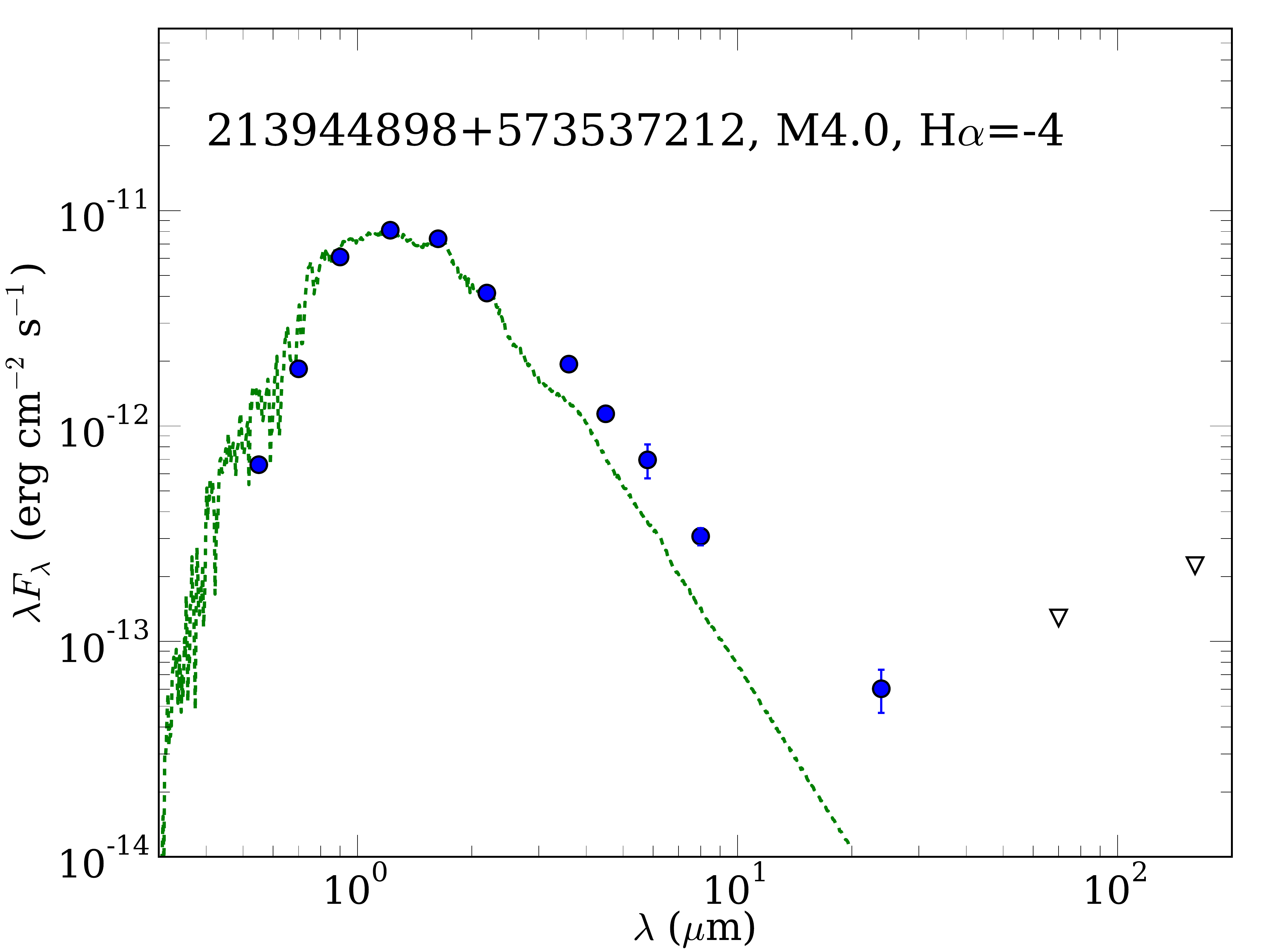} &
\includegraphics[width=0.24\linewidth]{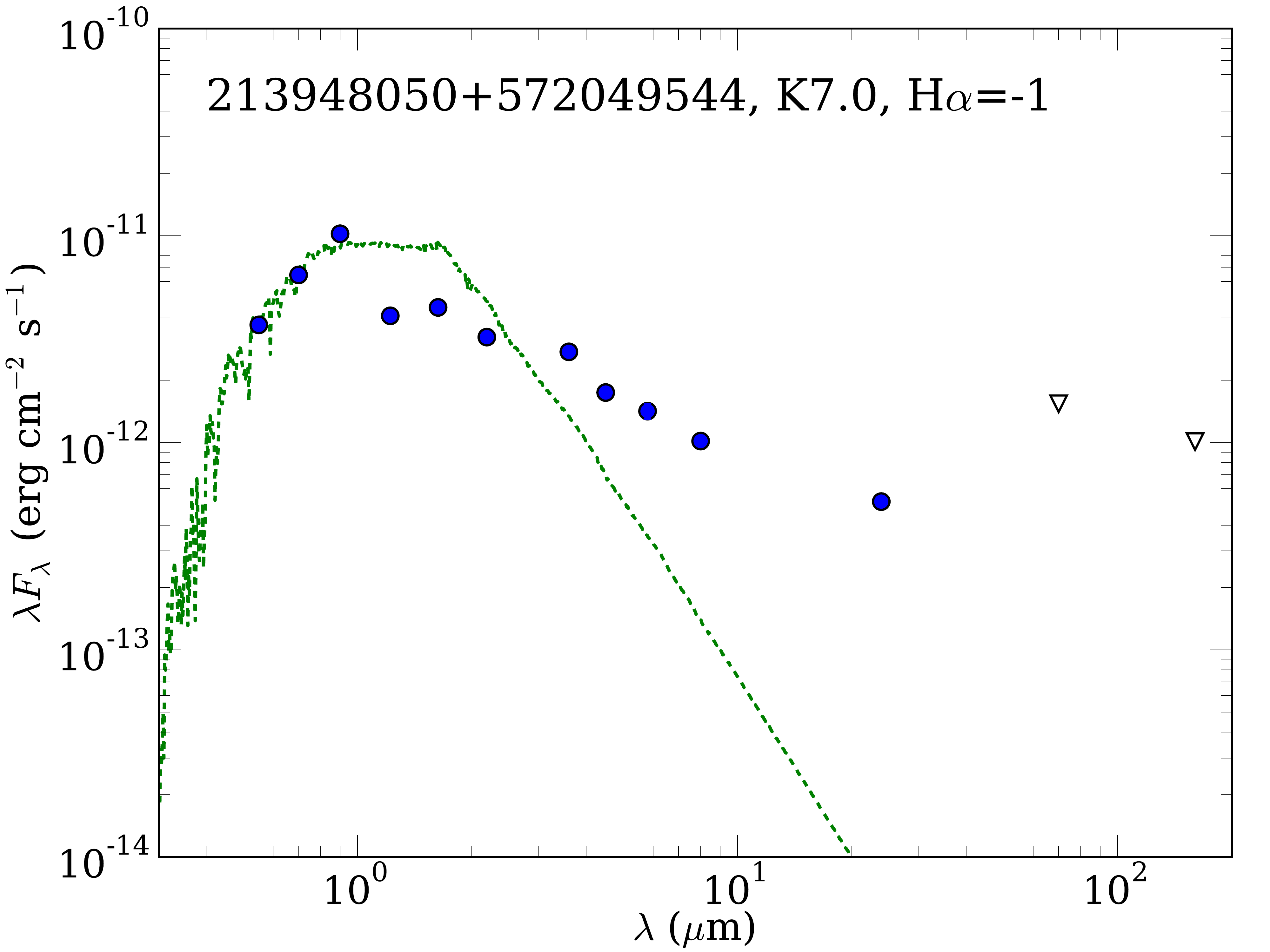} \\
\includegraphics[width=0.24\linewidth]{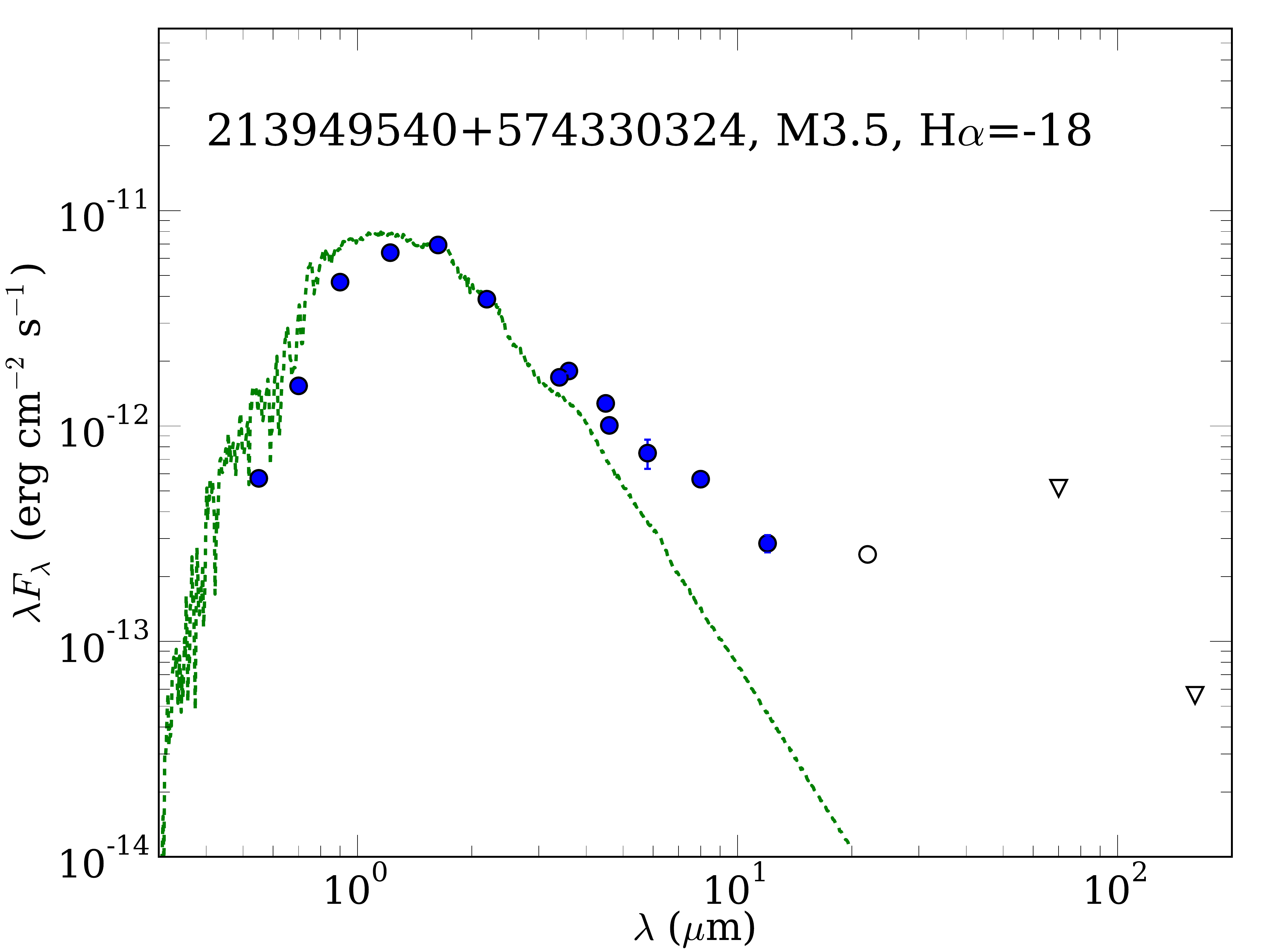} &
\includegraphics[width=0.24\linewidth]{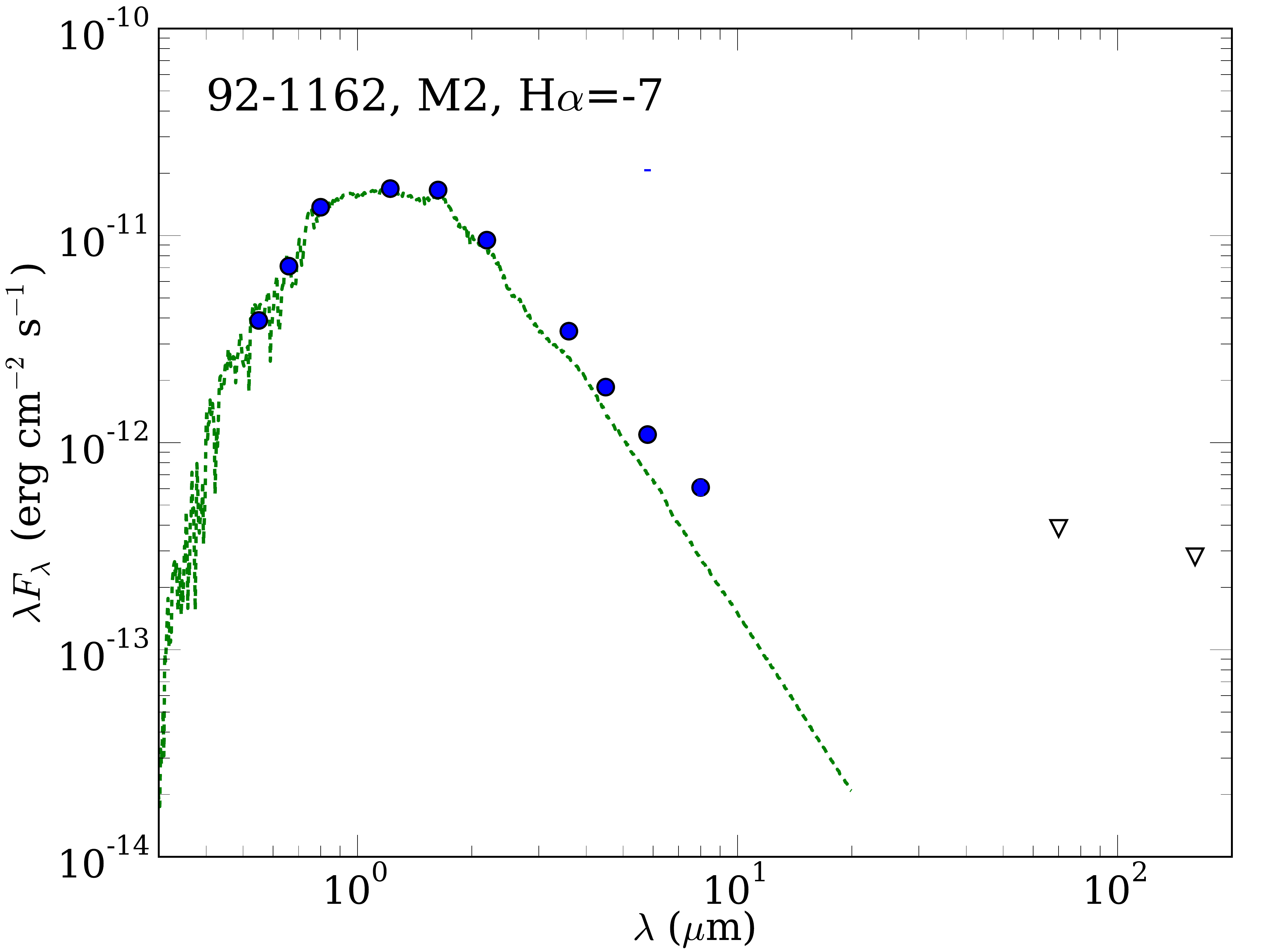} &
\includegraphics[width=0.24\linewidth]{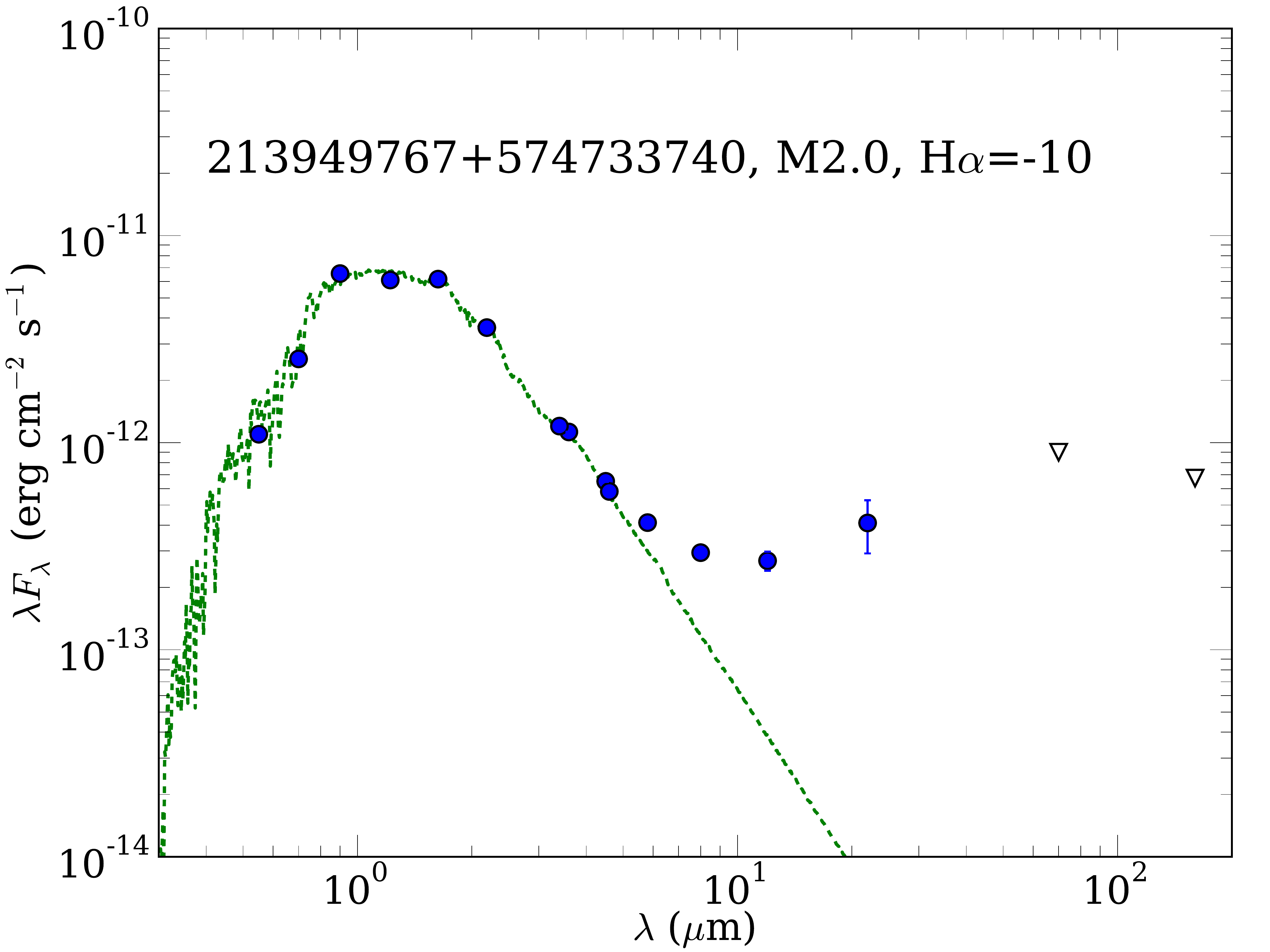} &
\includegraphics[width=0.24\linewidth]{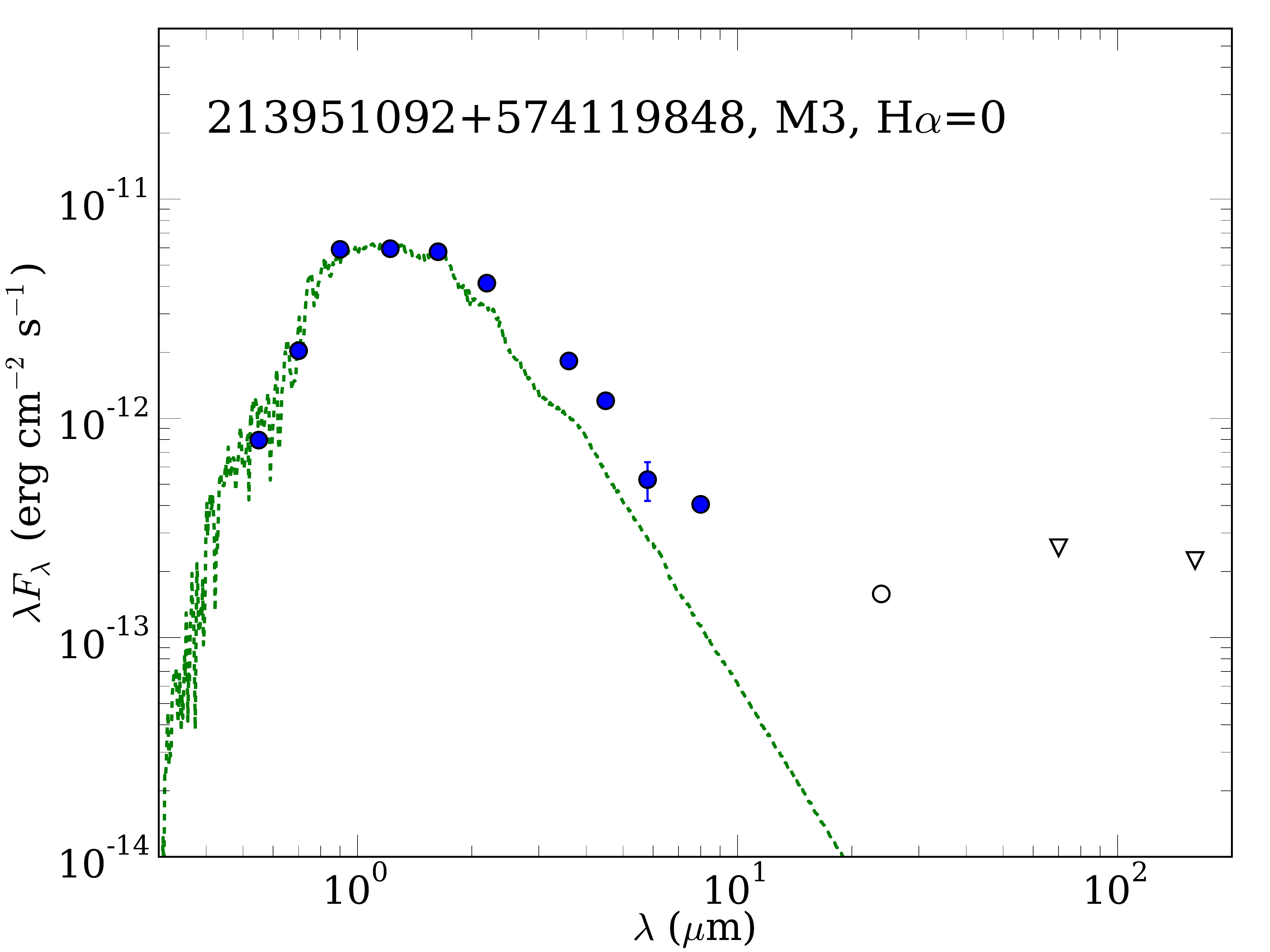} \\
\includegraphics[width=0.24\linewidth]{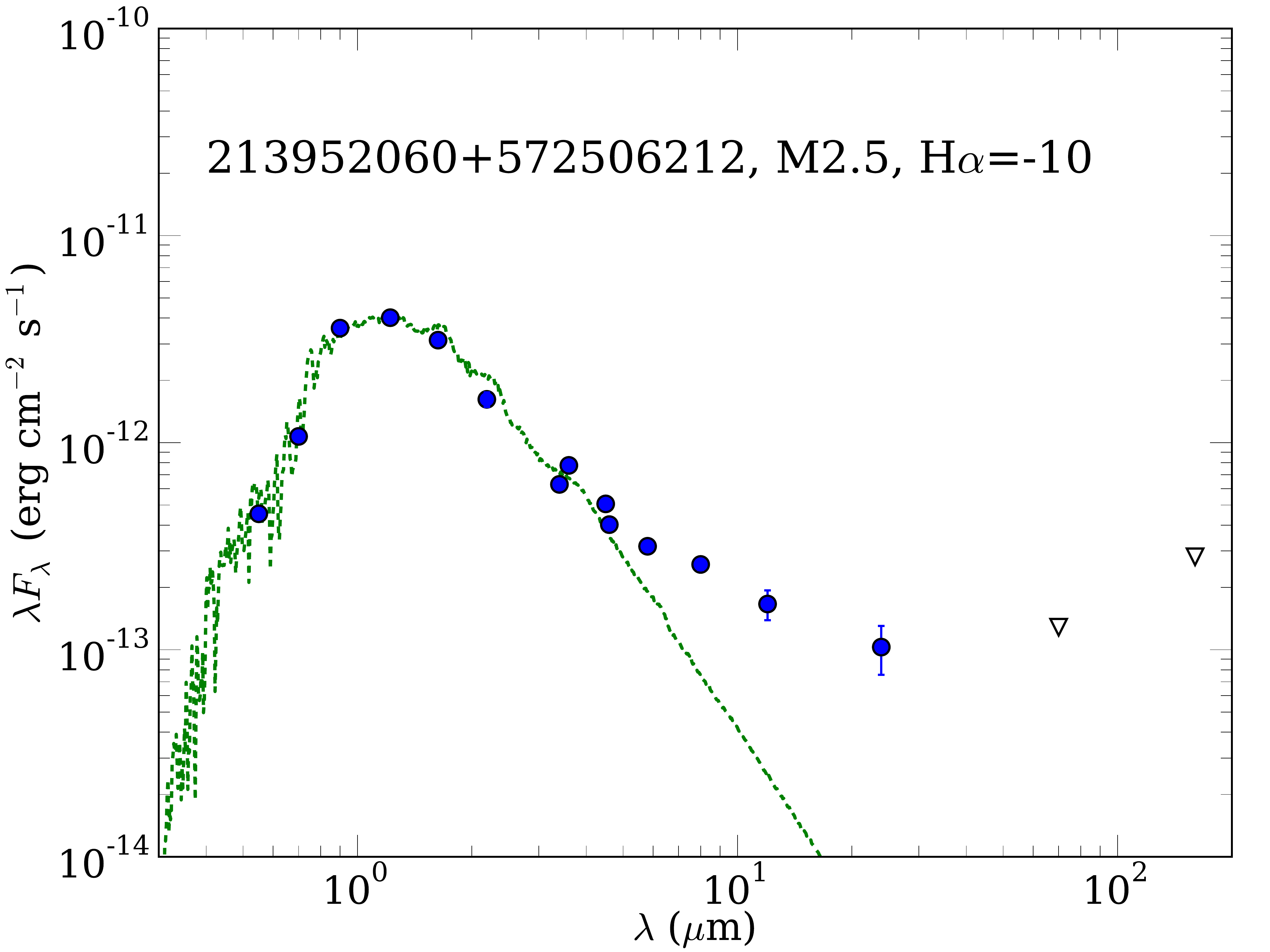} &
\includegraphics[width=0.24\linewidth]{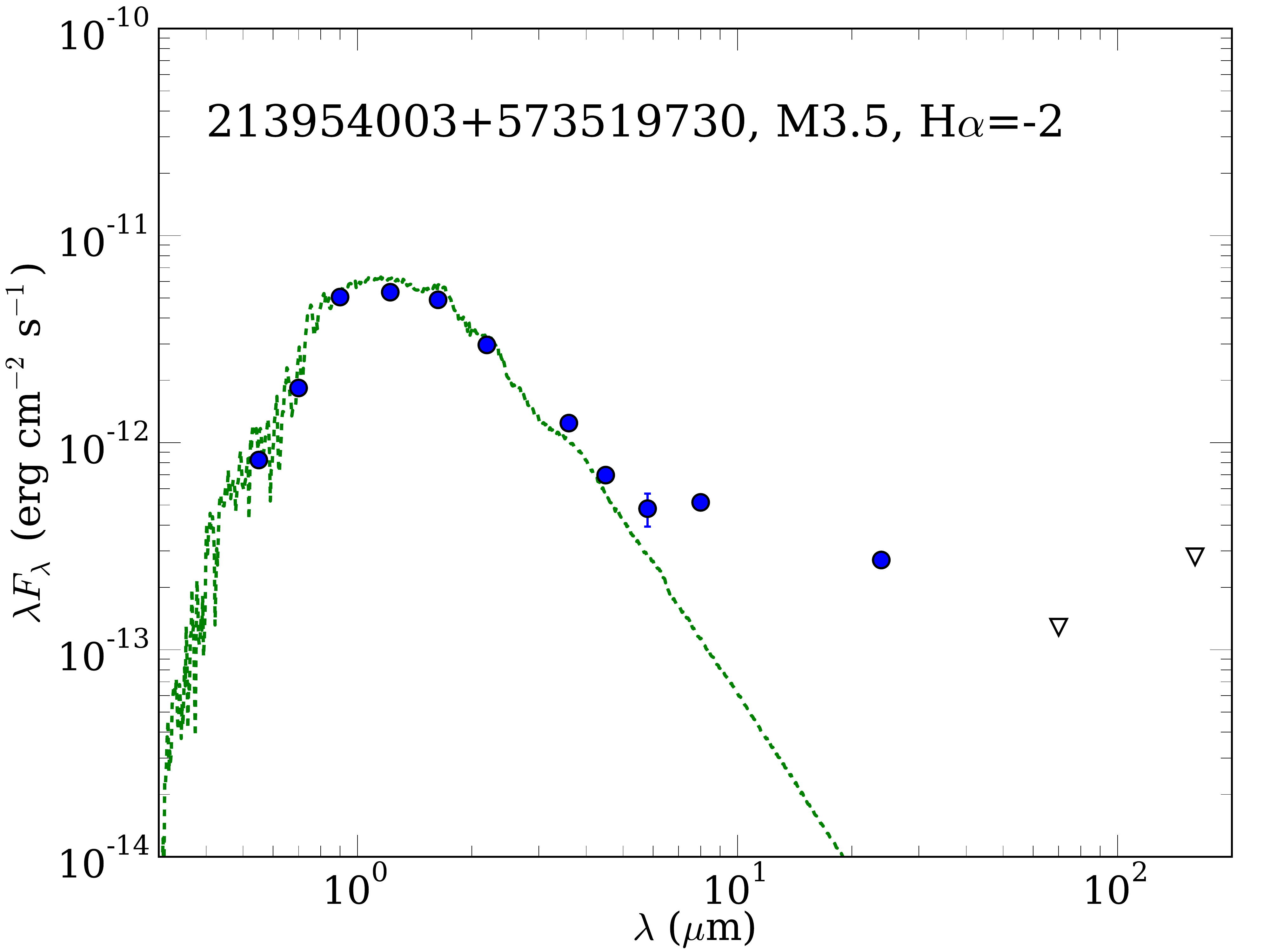} &
\includegraphics[width=0.24\linewidth]{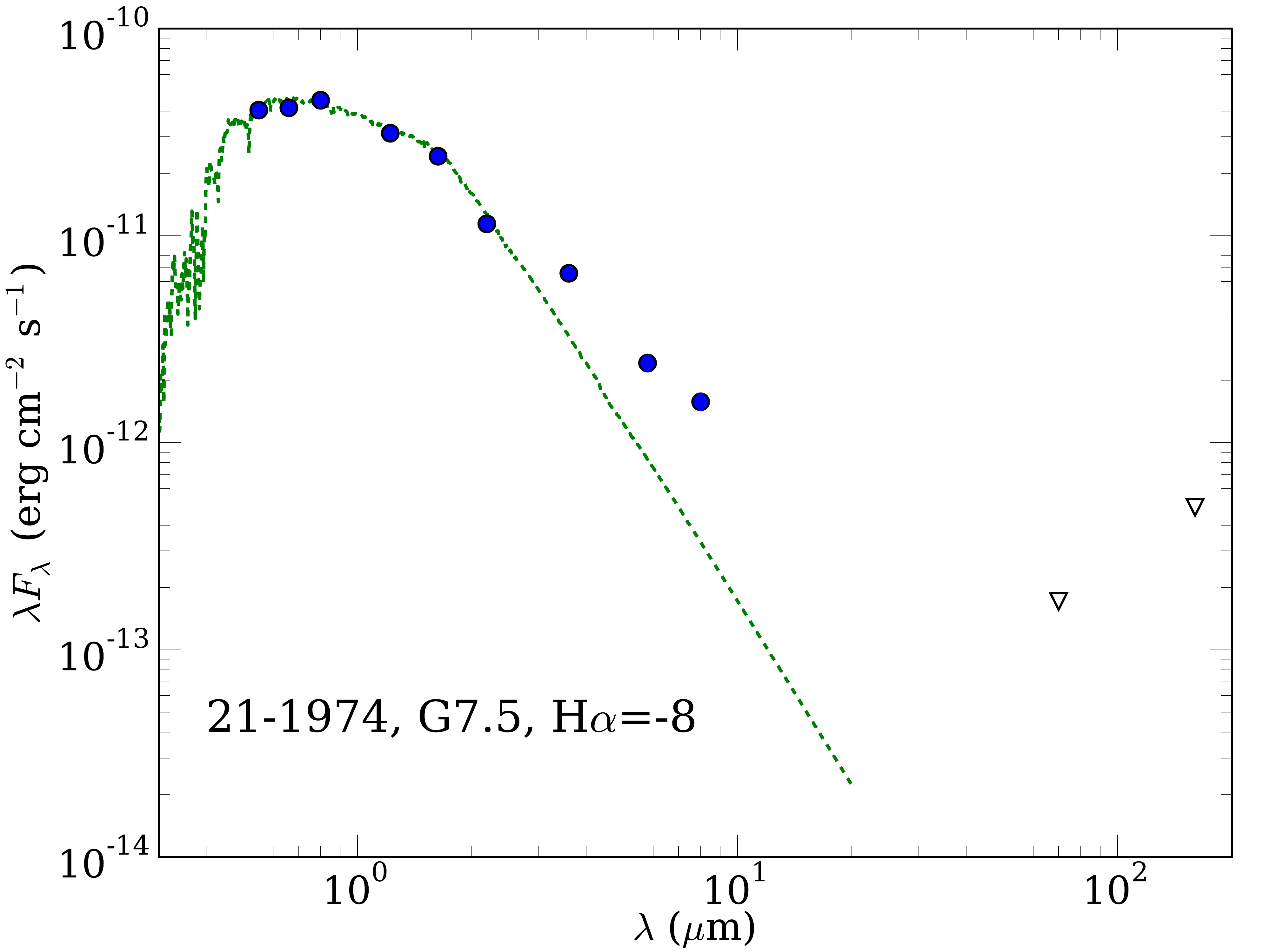} &
\includegraphics[width=0.24\linewidth]{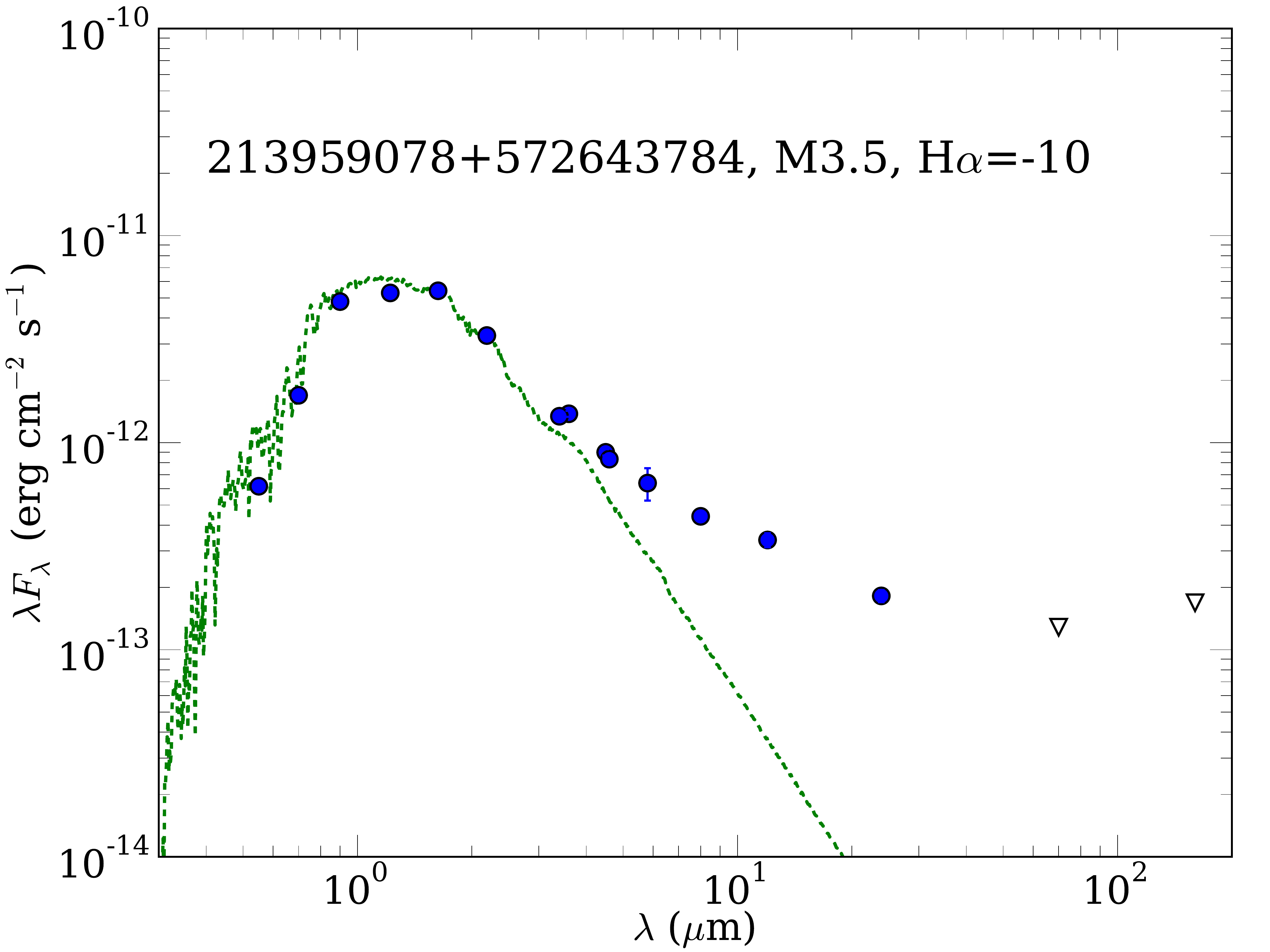} \\
\includegraphics[width=0.24\linewidth]{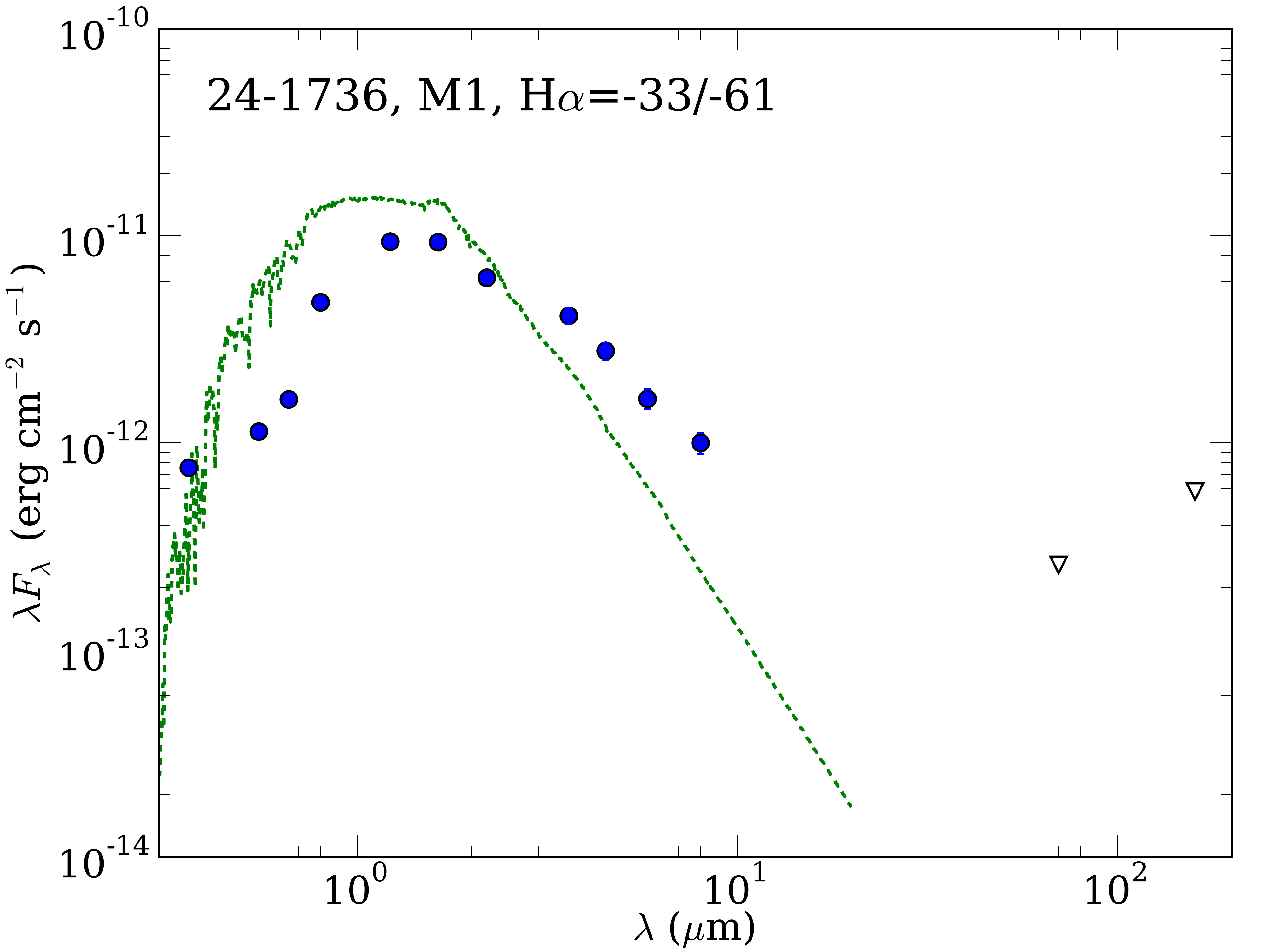} &
\includegraphics[width=0.24\linewidth]{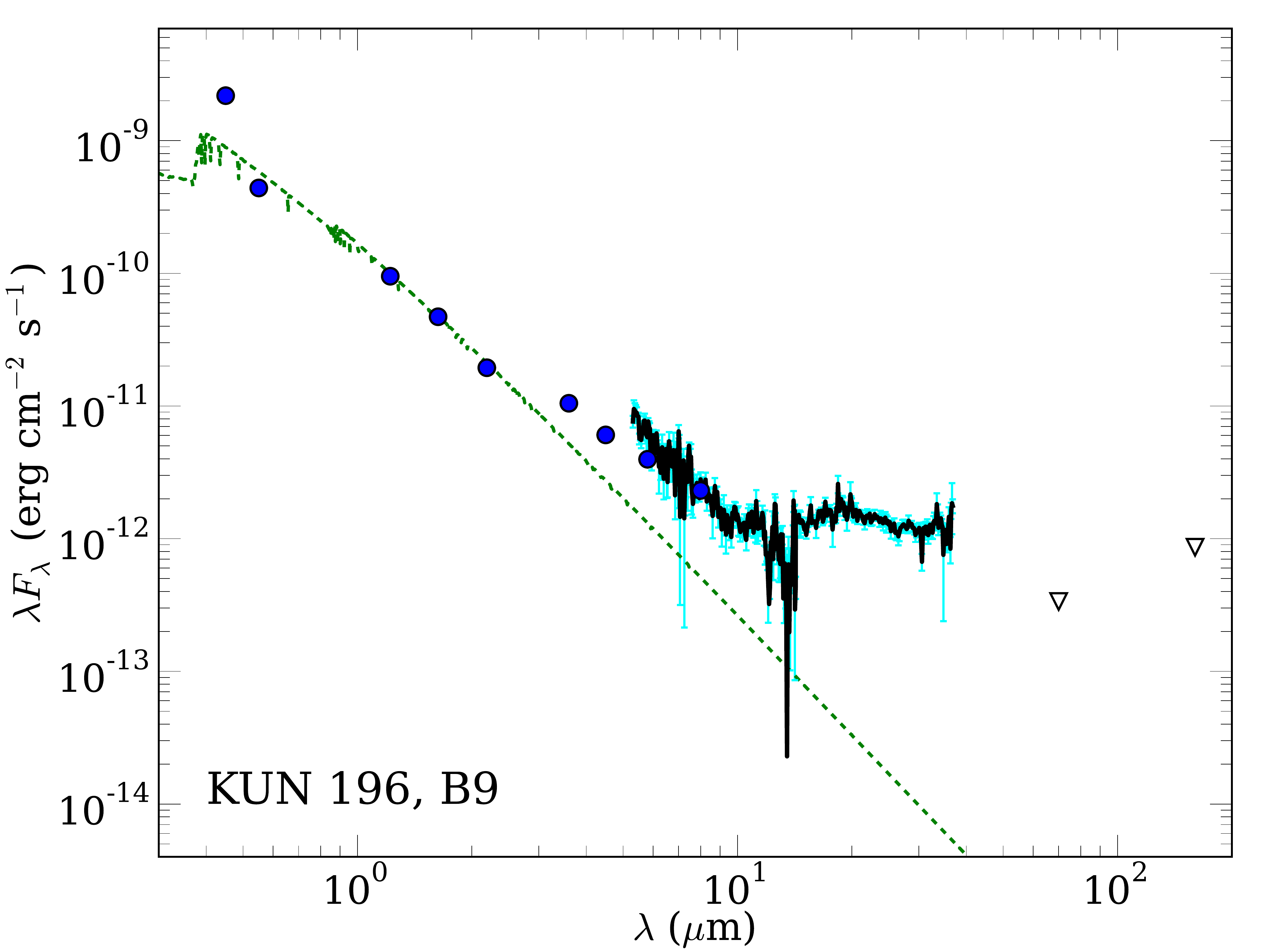} &
\includegraphics[width=0.24\linewidth]{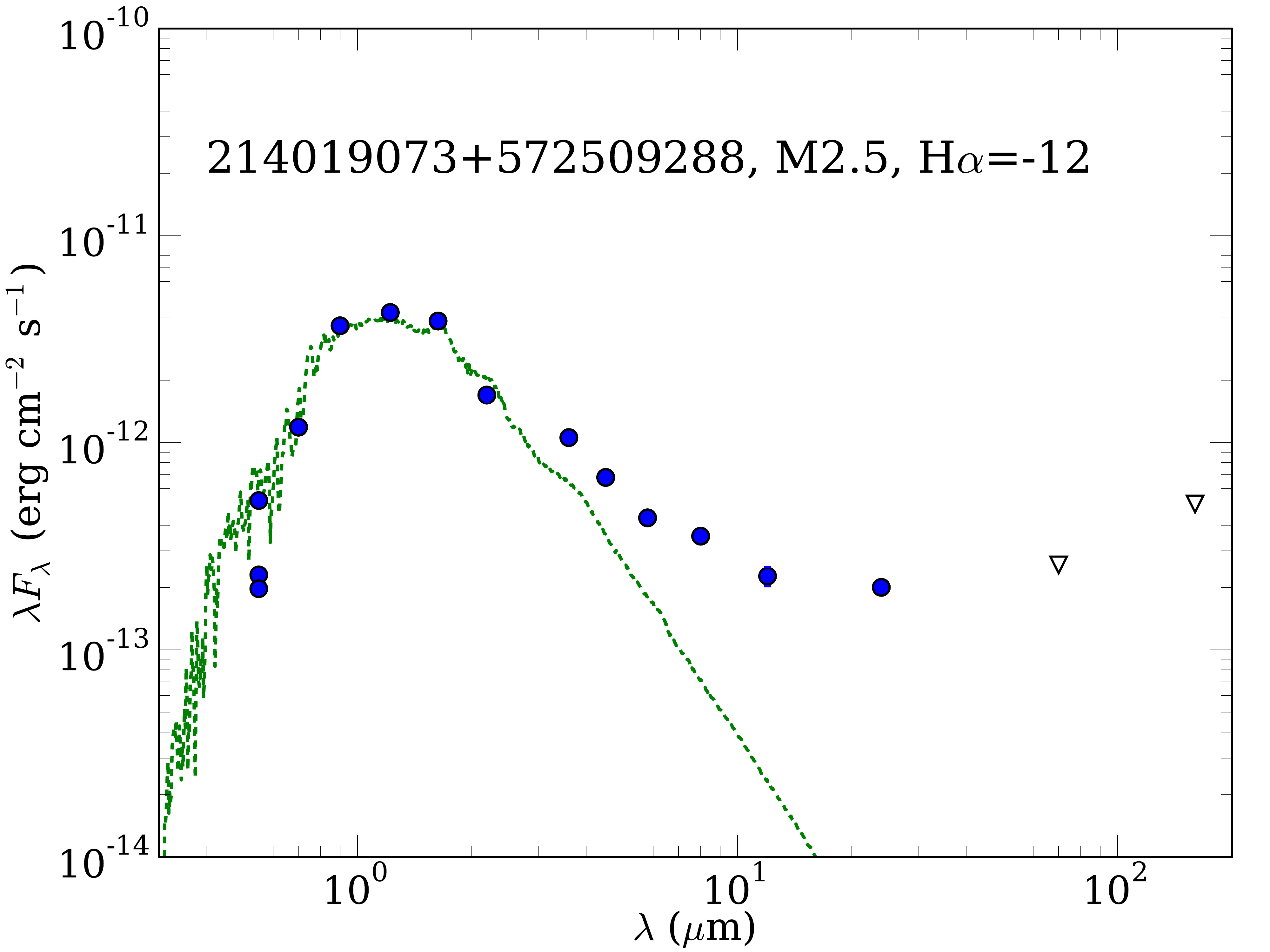} &
\includegraphics[width=0.24\linewidth]{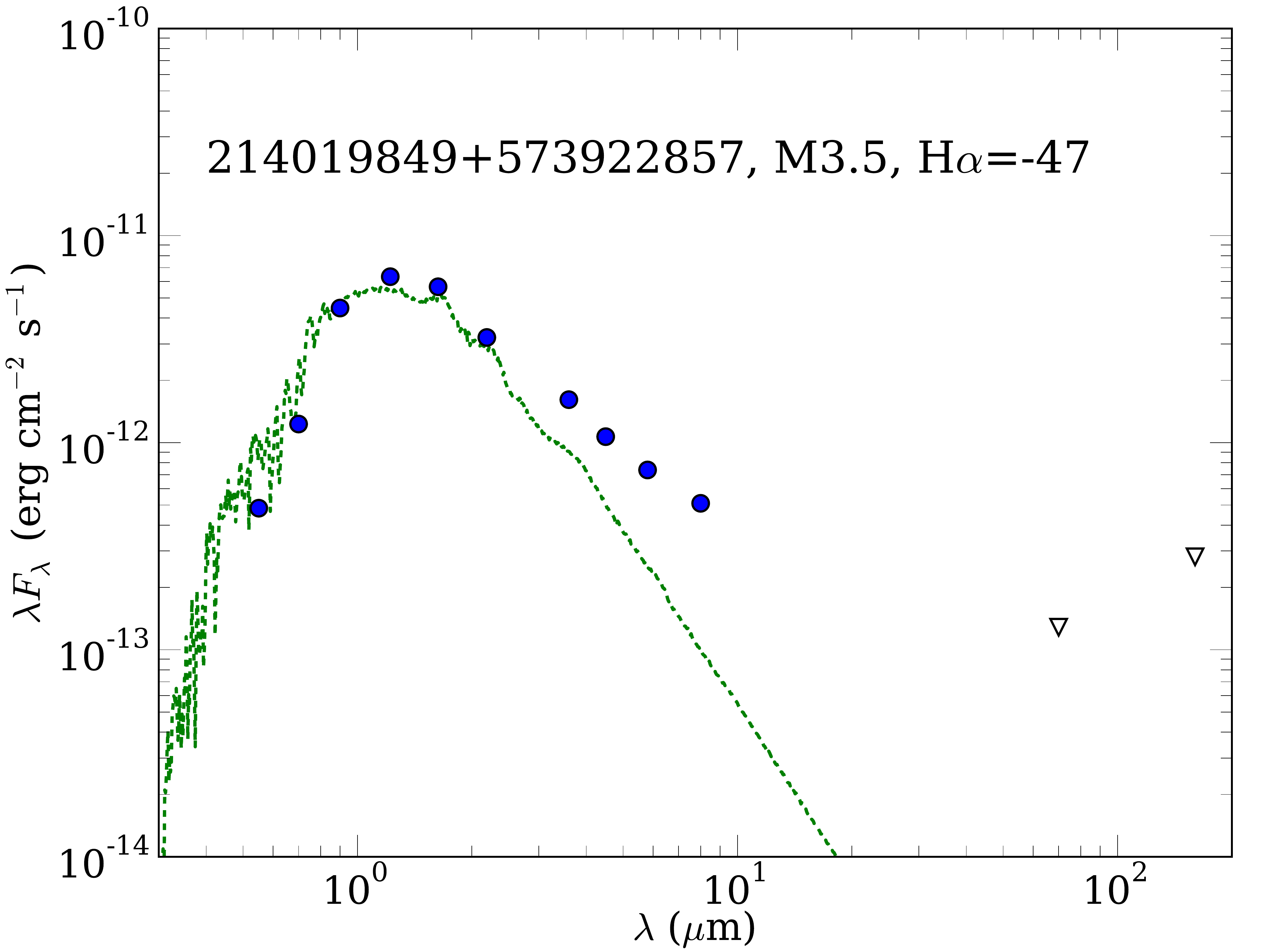} \\
\includegraphics[width=0.24\linewidth]{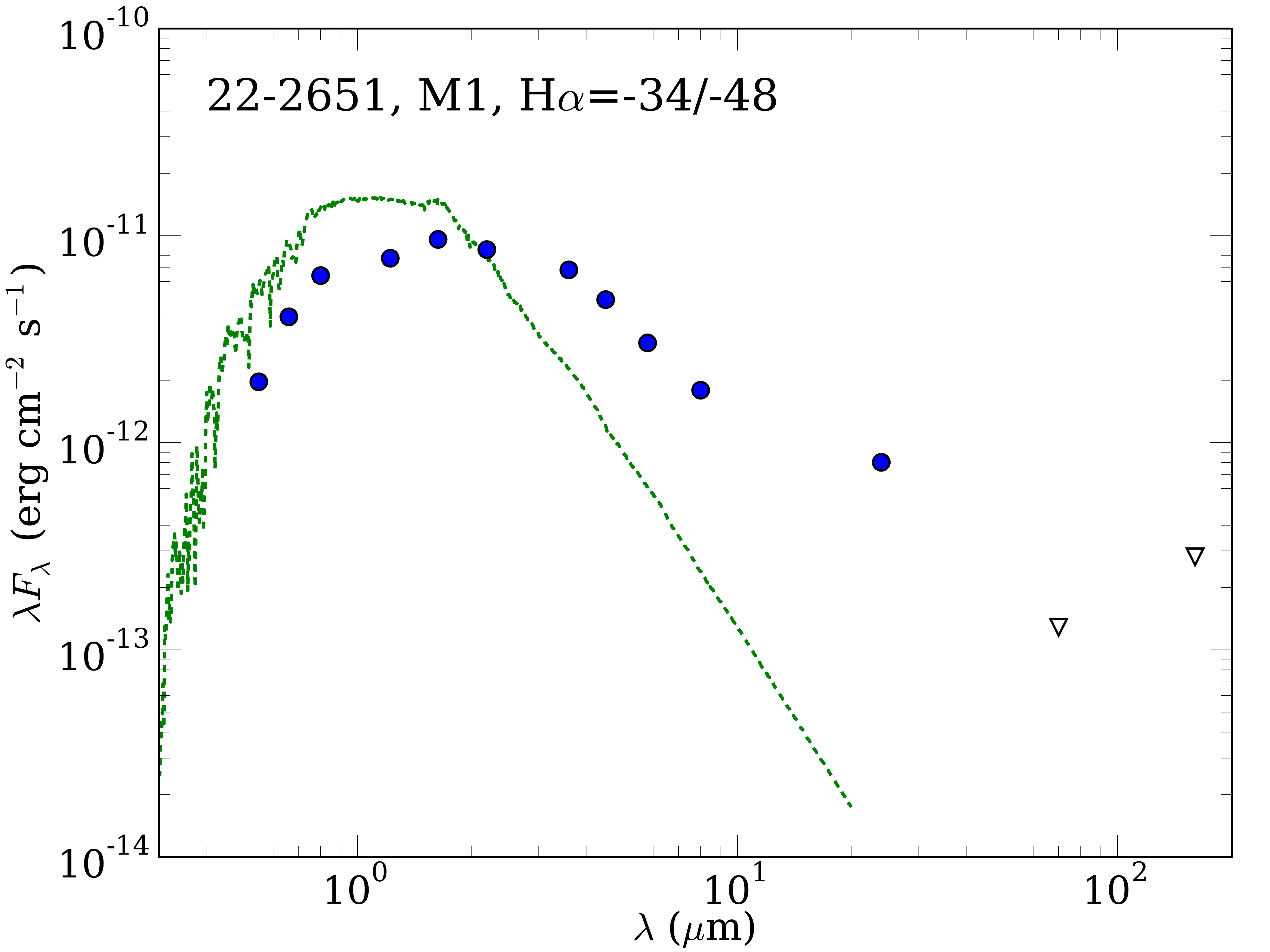} &
\includegraphics[width=0.24\linewidth]{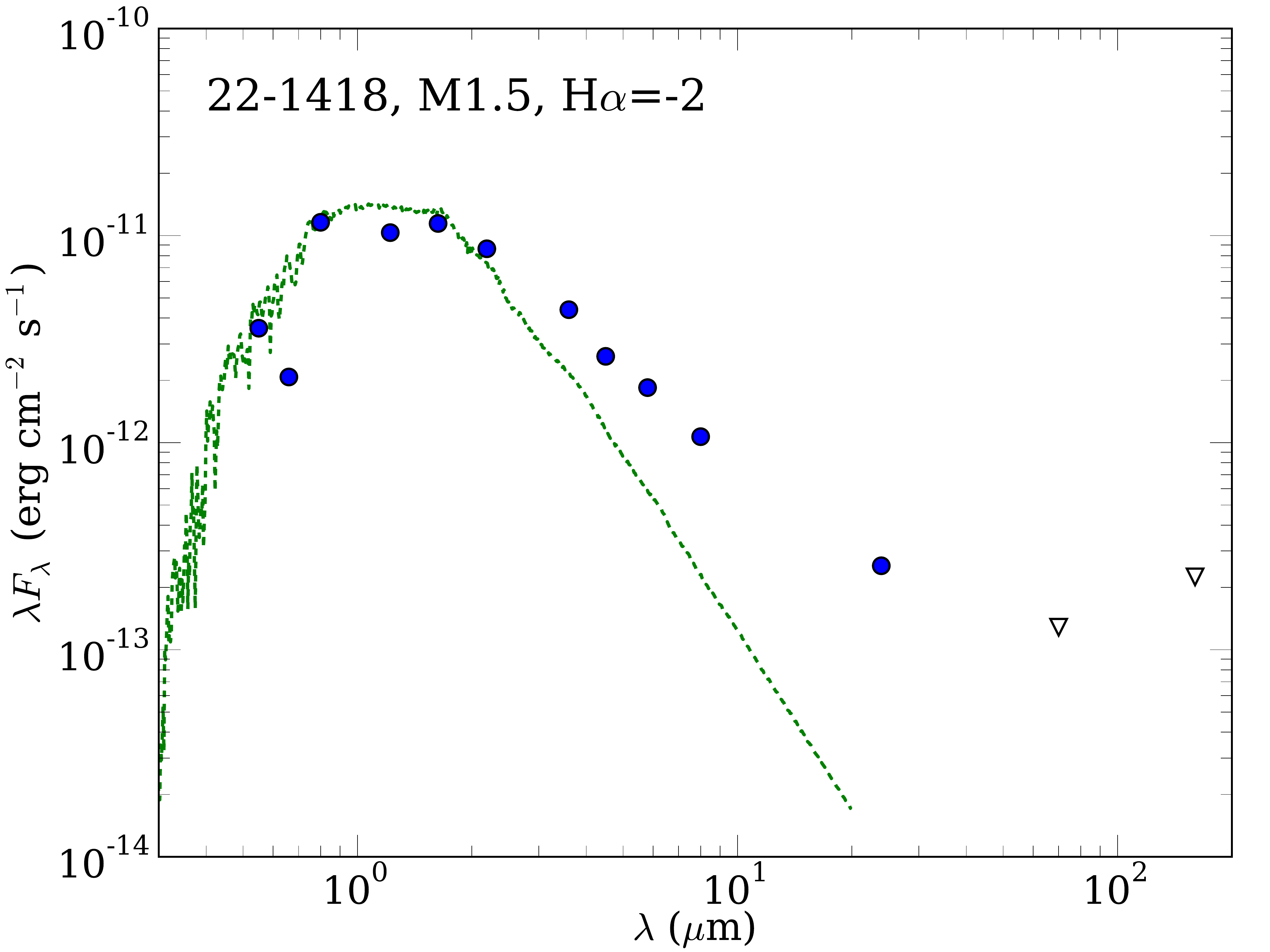} &
\includegraphics[width=0.24\linewidth]{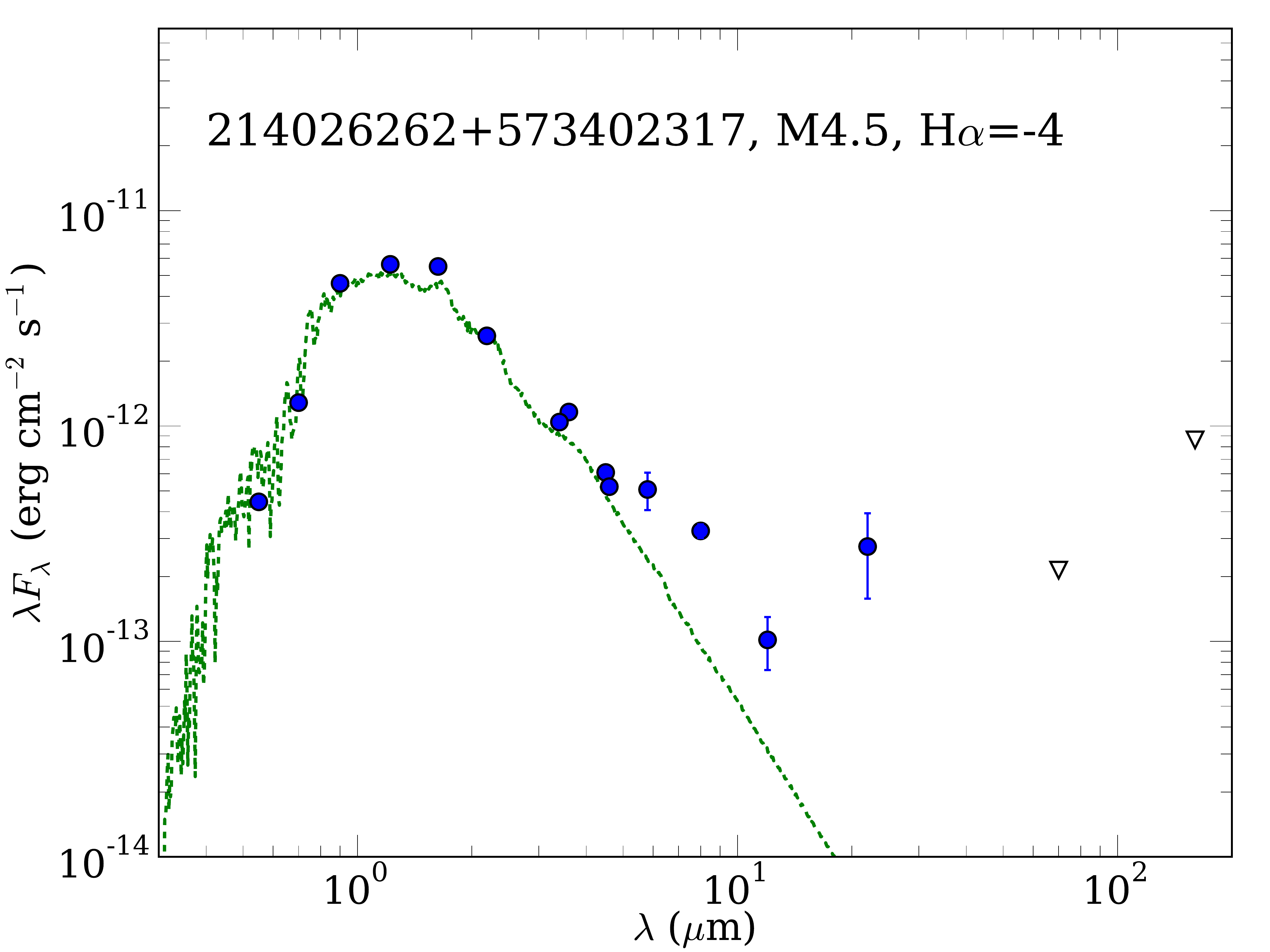} &
\includegraphics[width=0.24\linewidth]{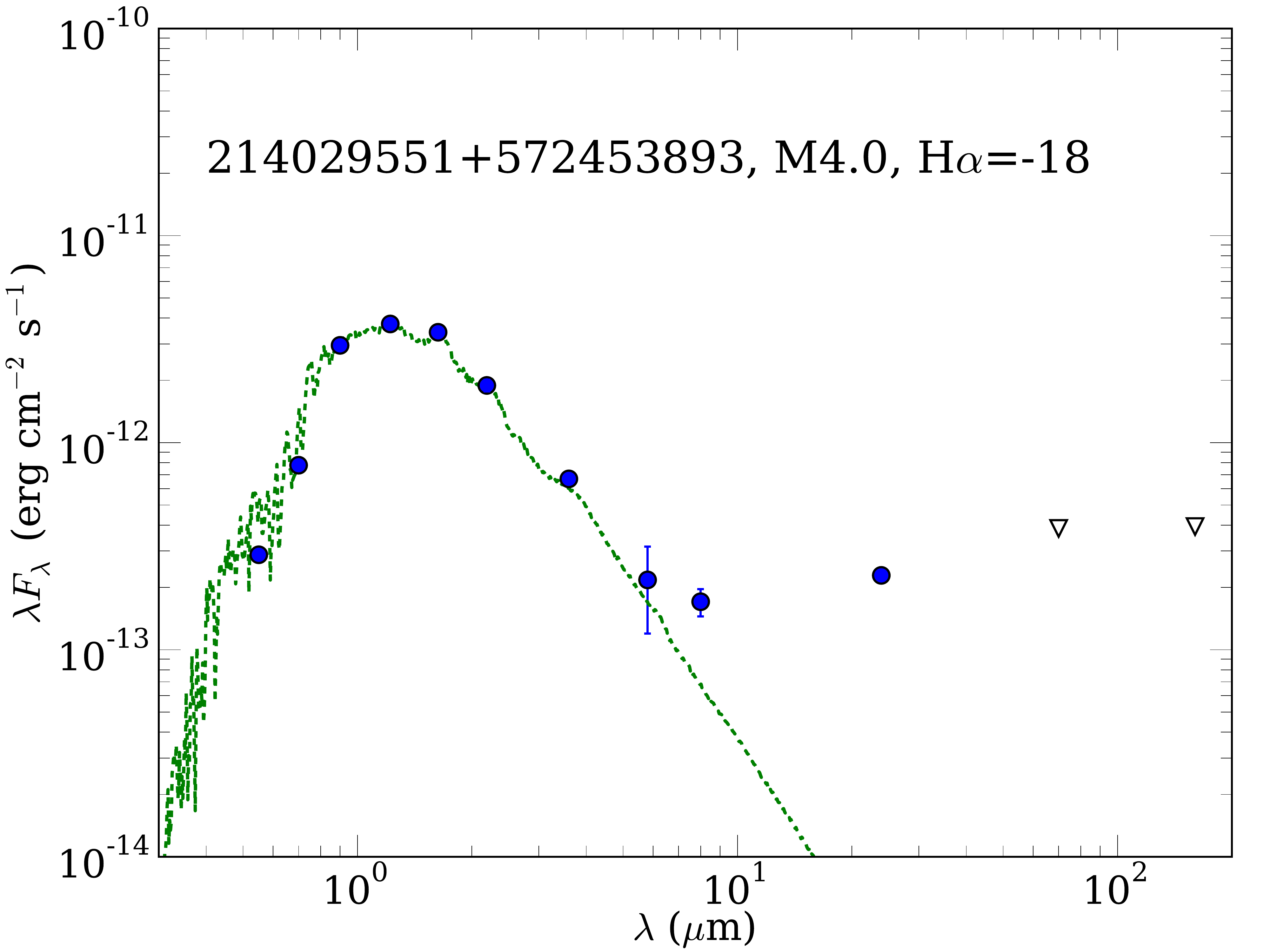} \\
\includegraphics[width=0.24\linewidth]{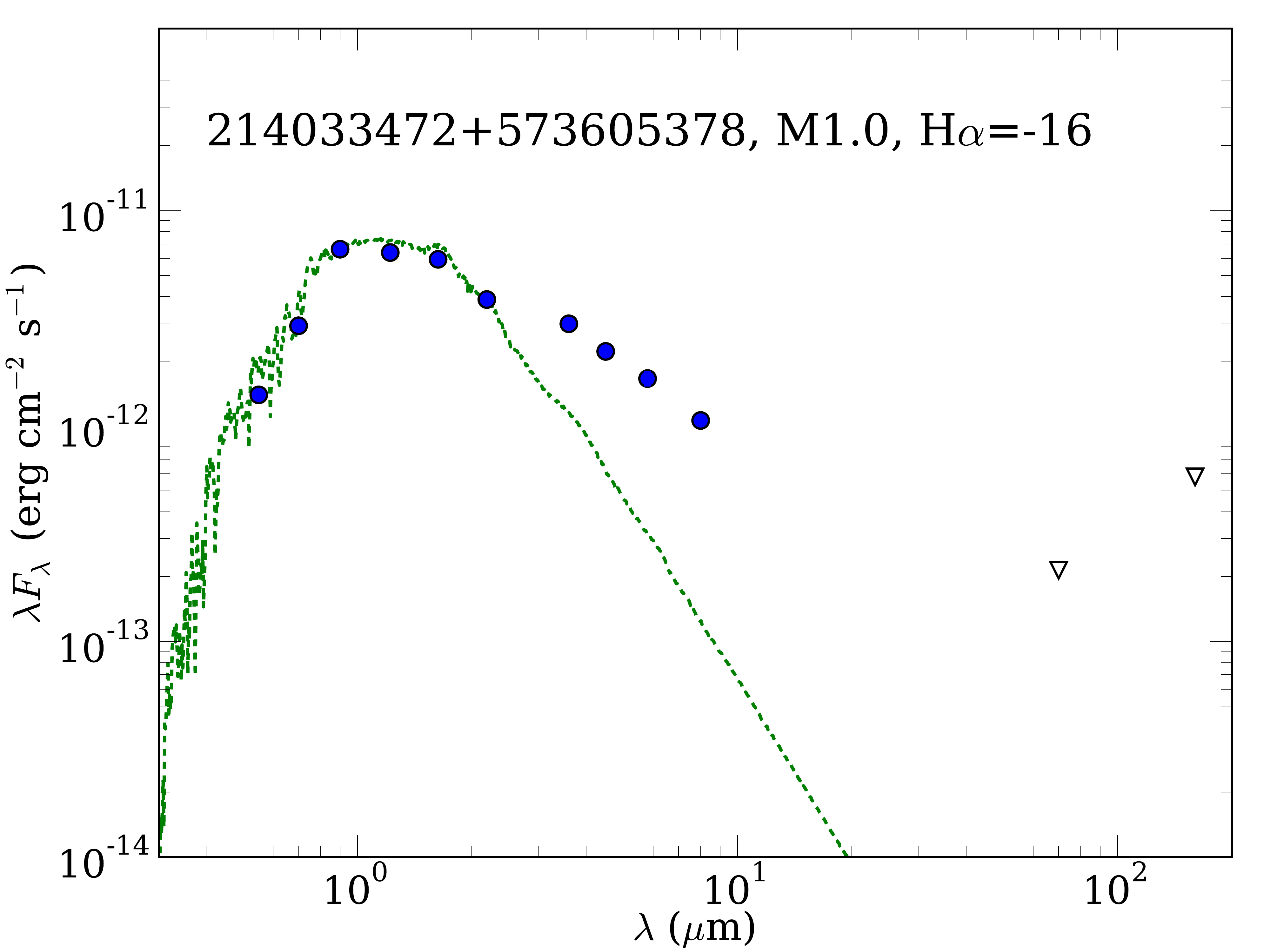} &
\includegraphics[width=0.24\linewidth]{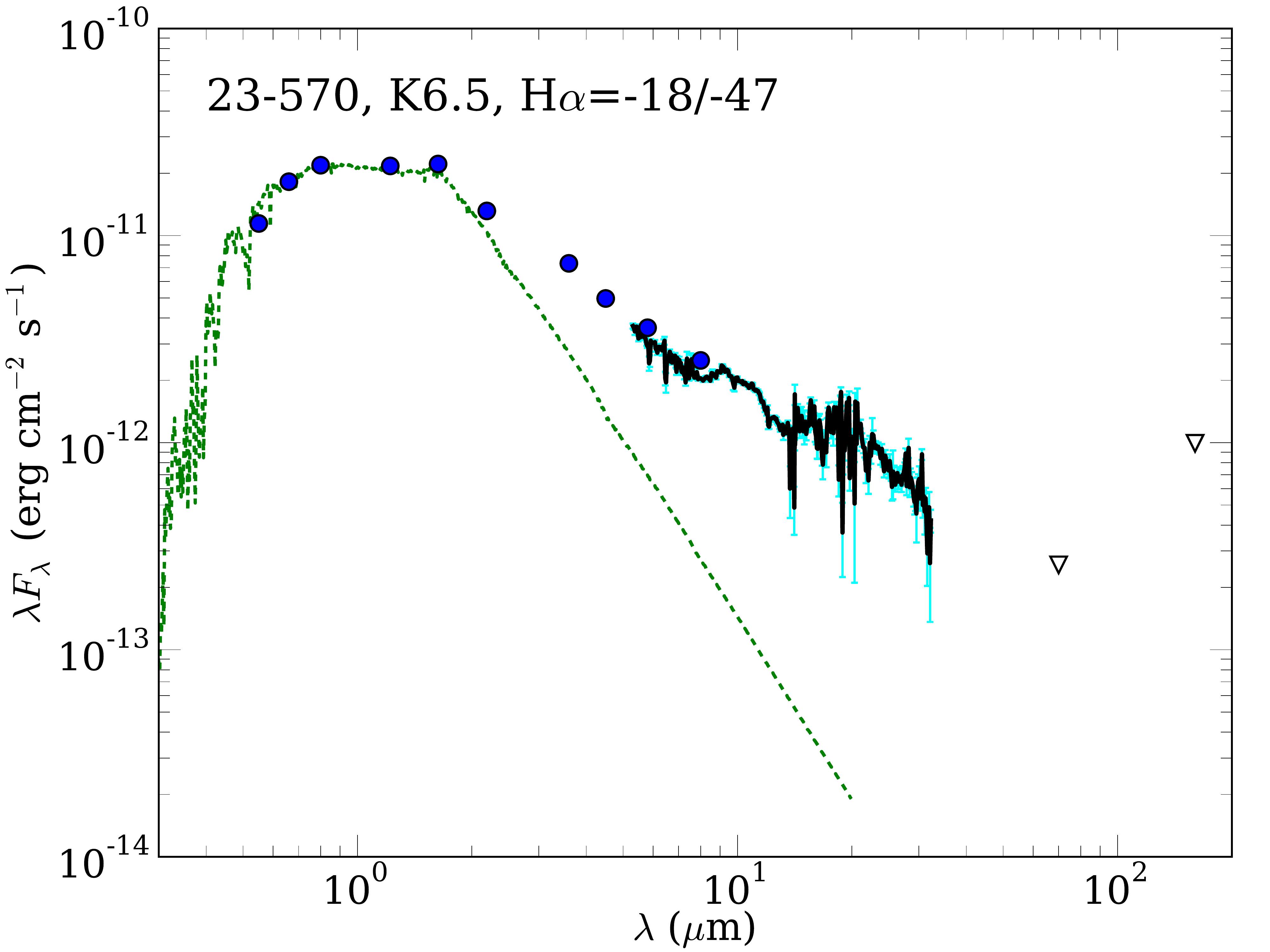} &
\includegraphics[width=0.24\linewidth]{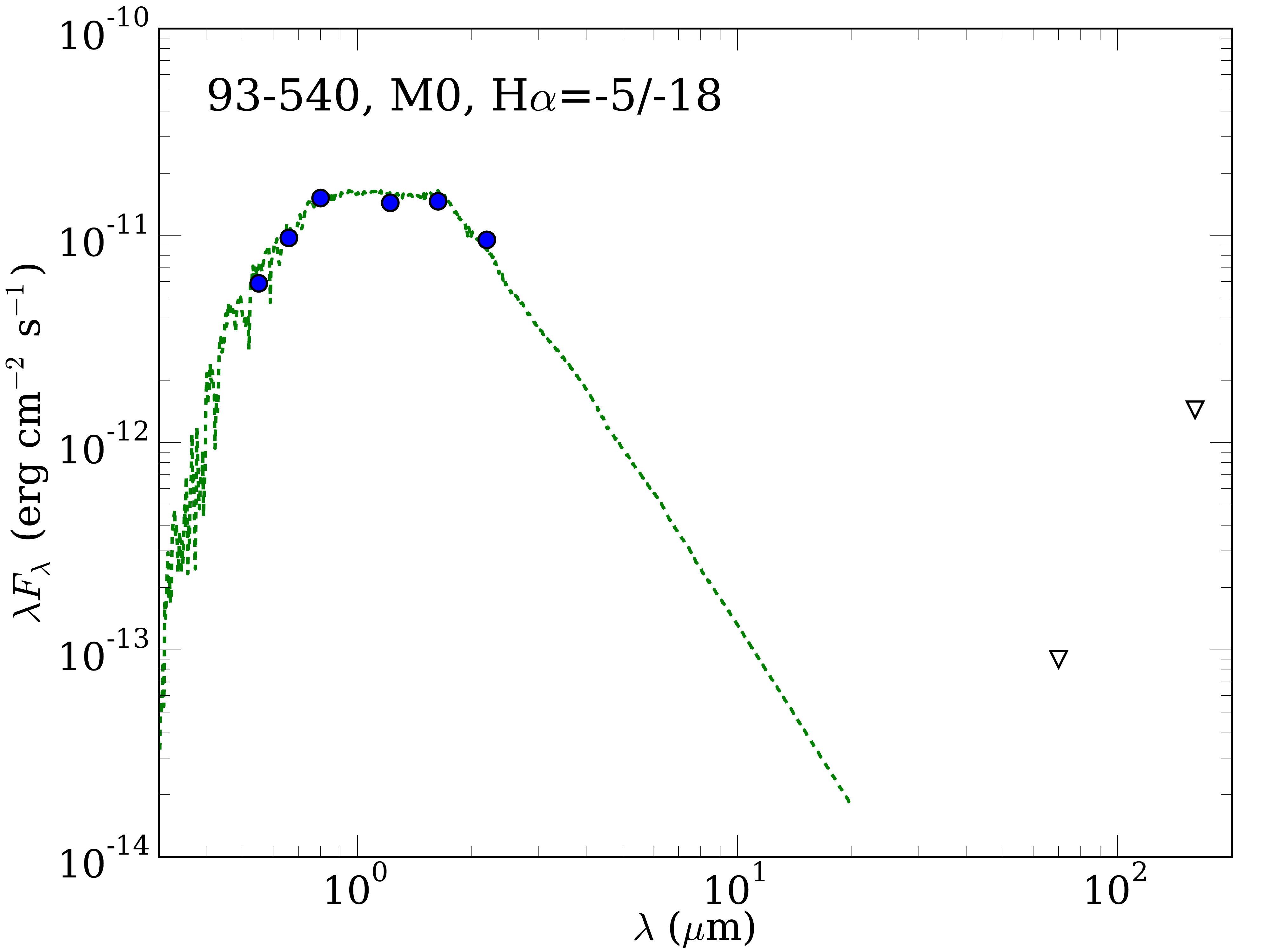} &
\includegraphics[width=0.24\linewidth]{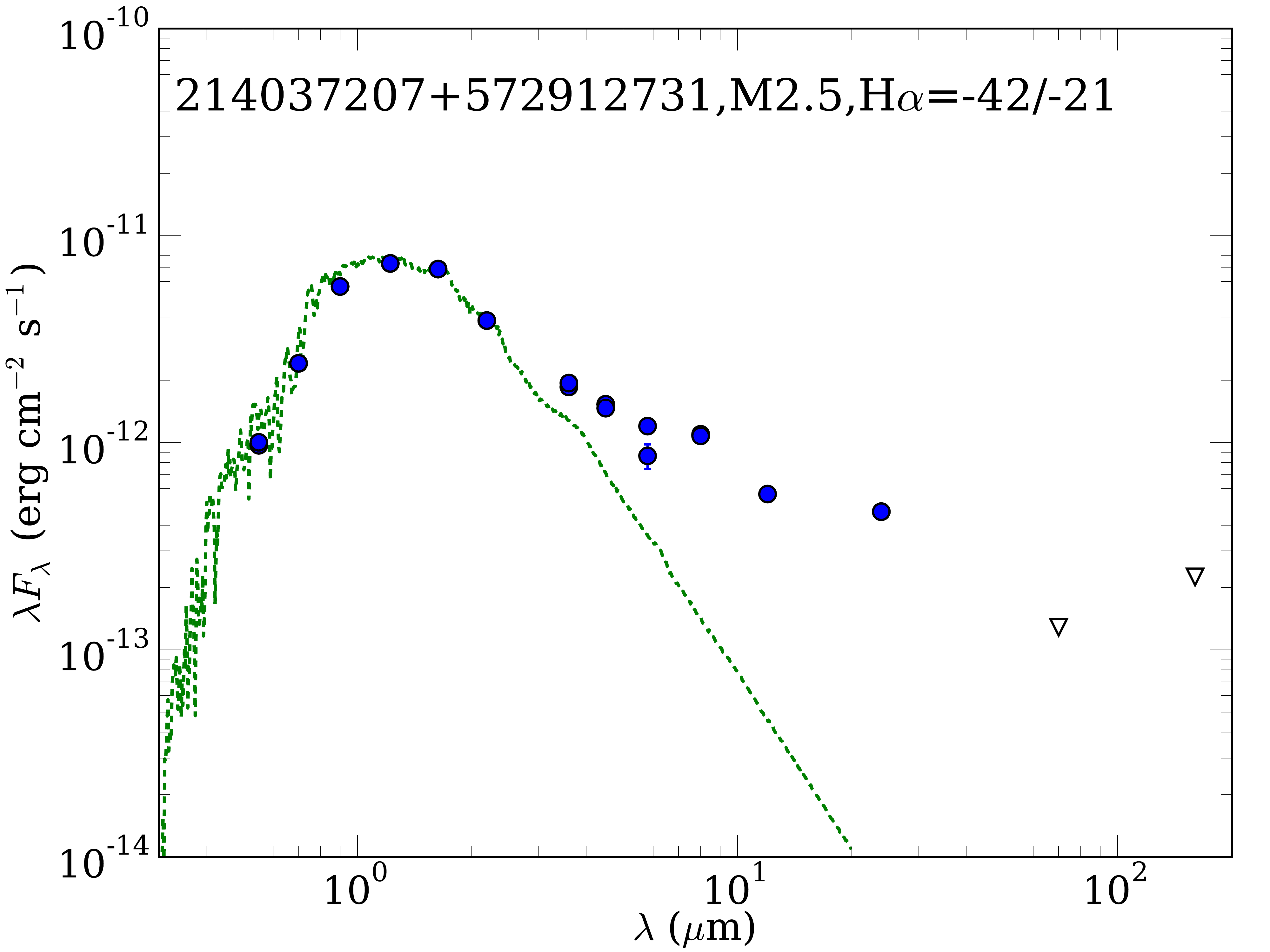} \\
\includegraphics[width=0.24\linewidth]{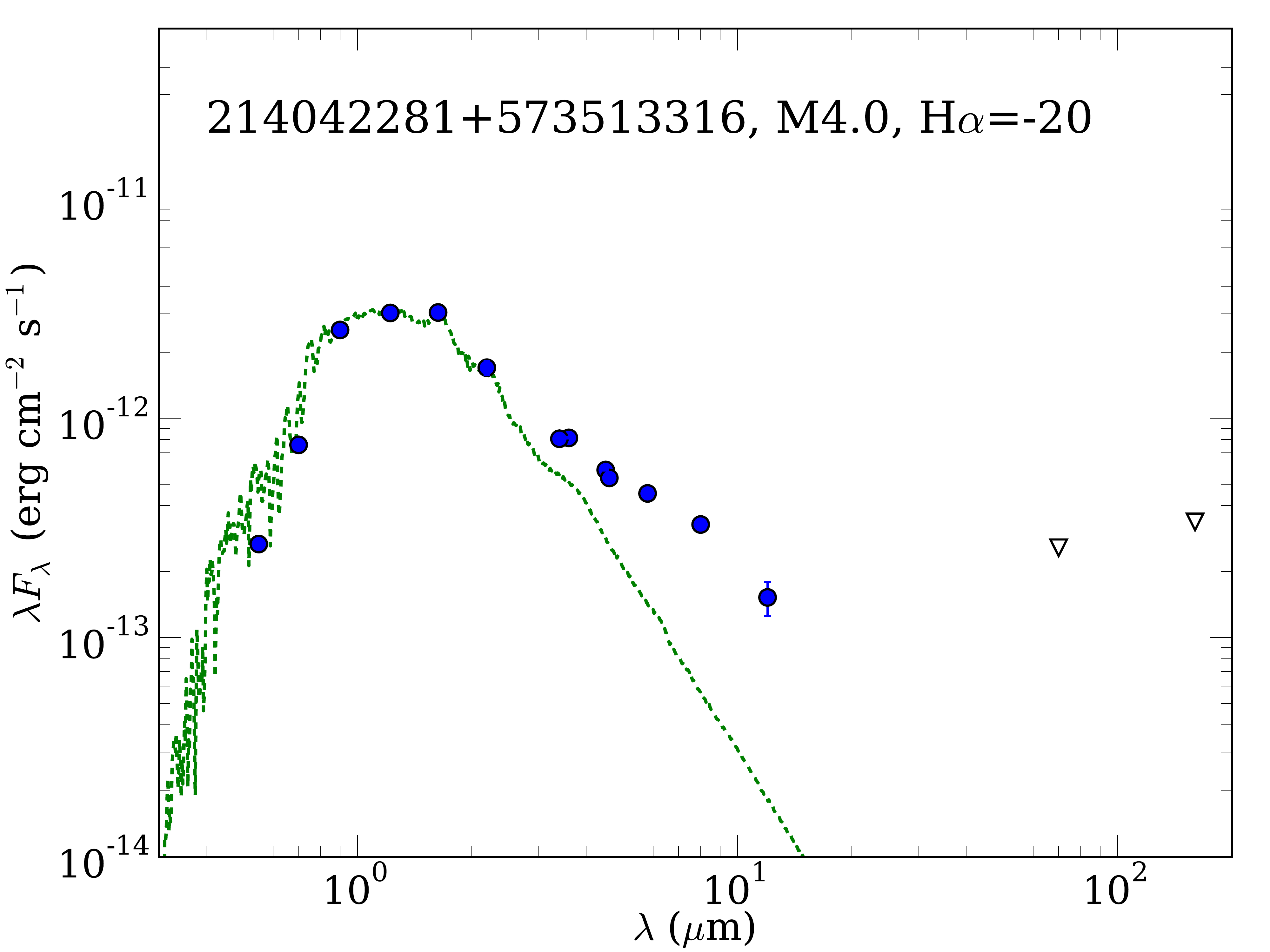} &
\includegraphics[width=0.24\linewidth]{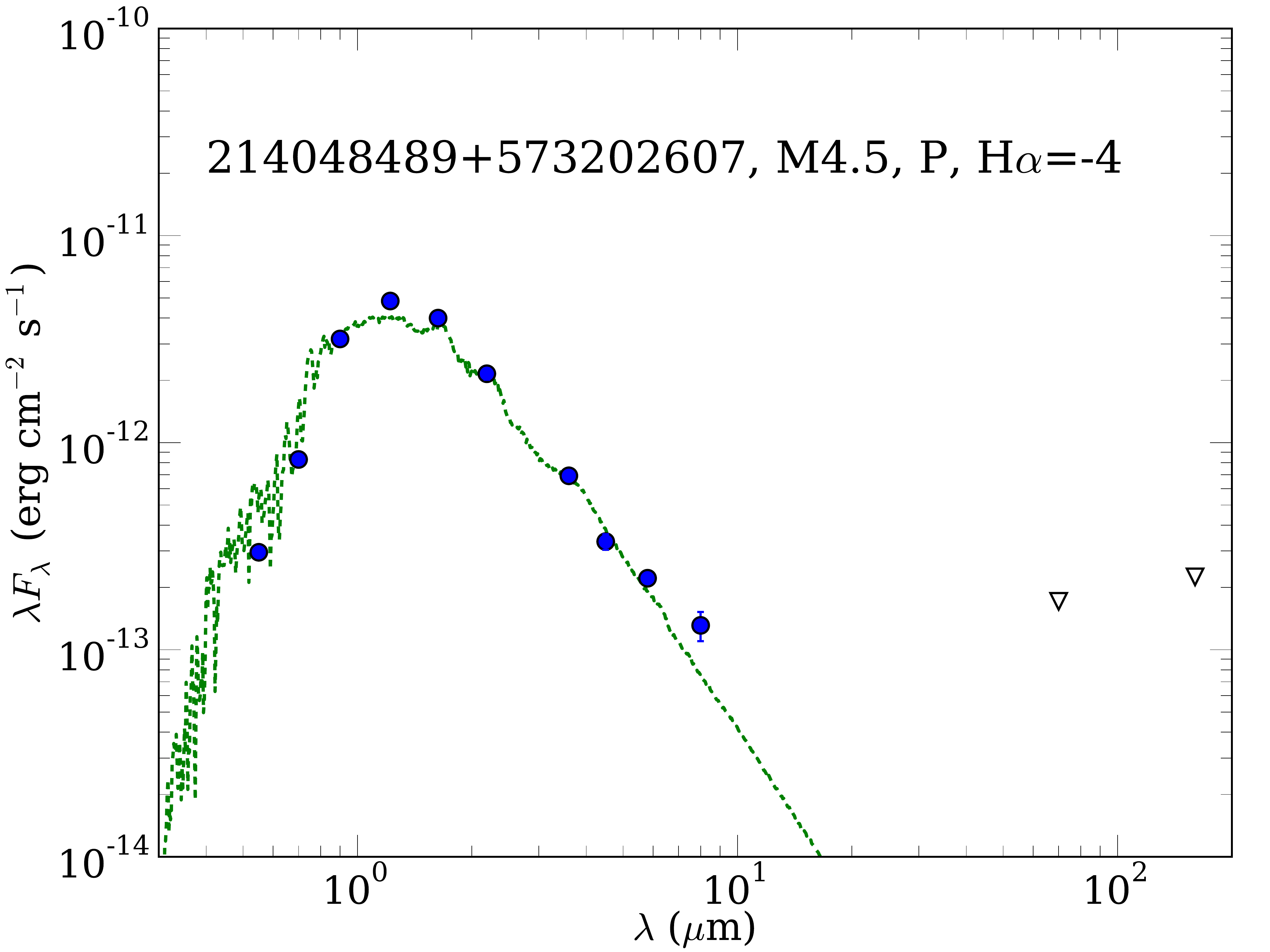} &
\includegraphics[width=0.24\linewidth]{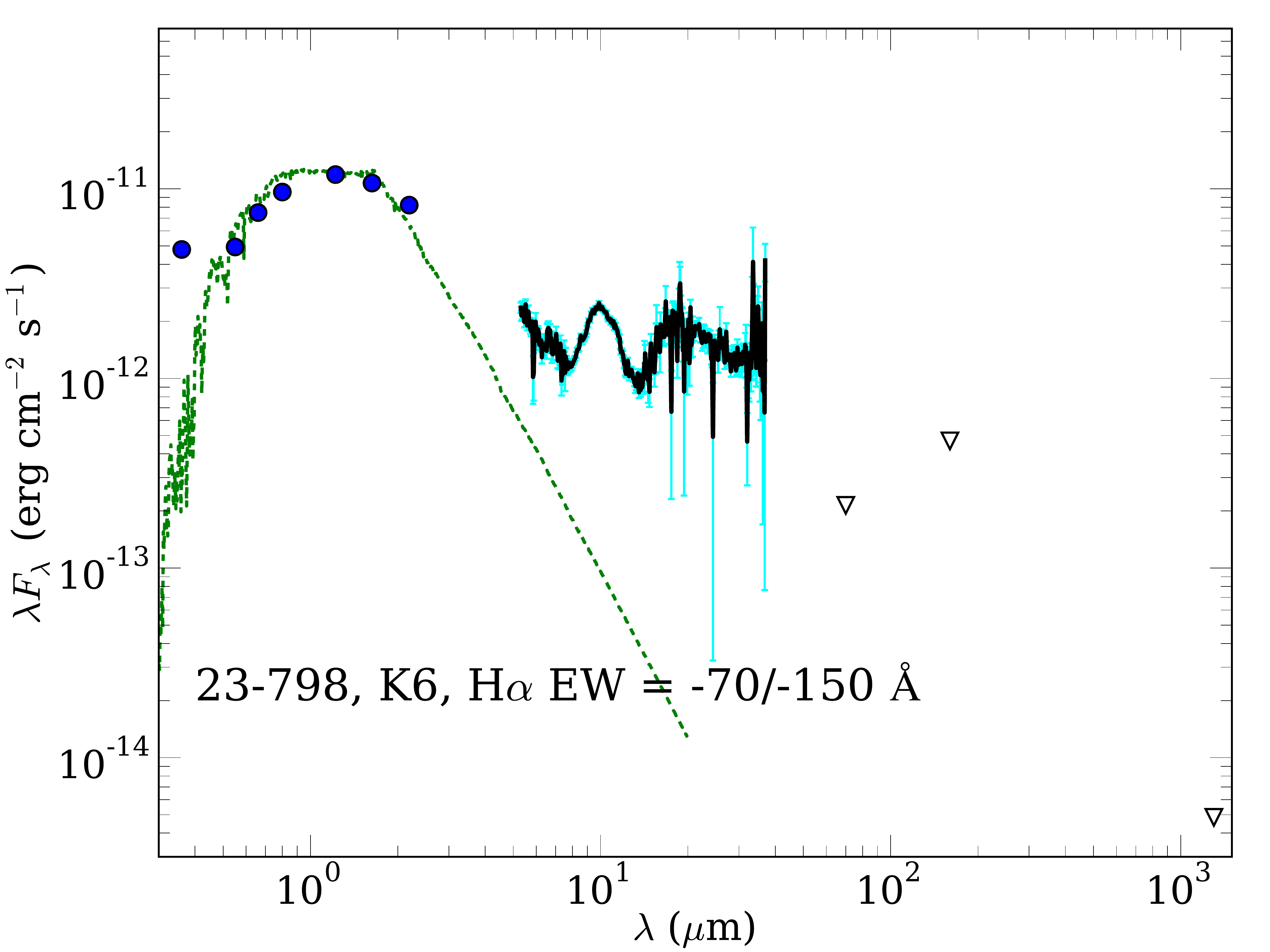} \\
\end{tabular}
\caption{SEDs of the objects with upper limits only. Symbols as in Figure \ref{uplims1-fig}.
\label{uplims4-fig}}
\end{figure*}

\begin{figure*}
\centering
\begin{tabular}{cccc}
\includegraphics[width=0.24\linewidth]{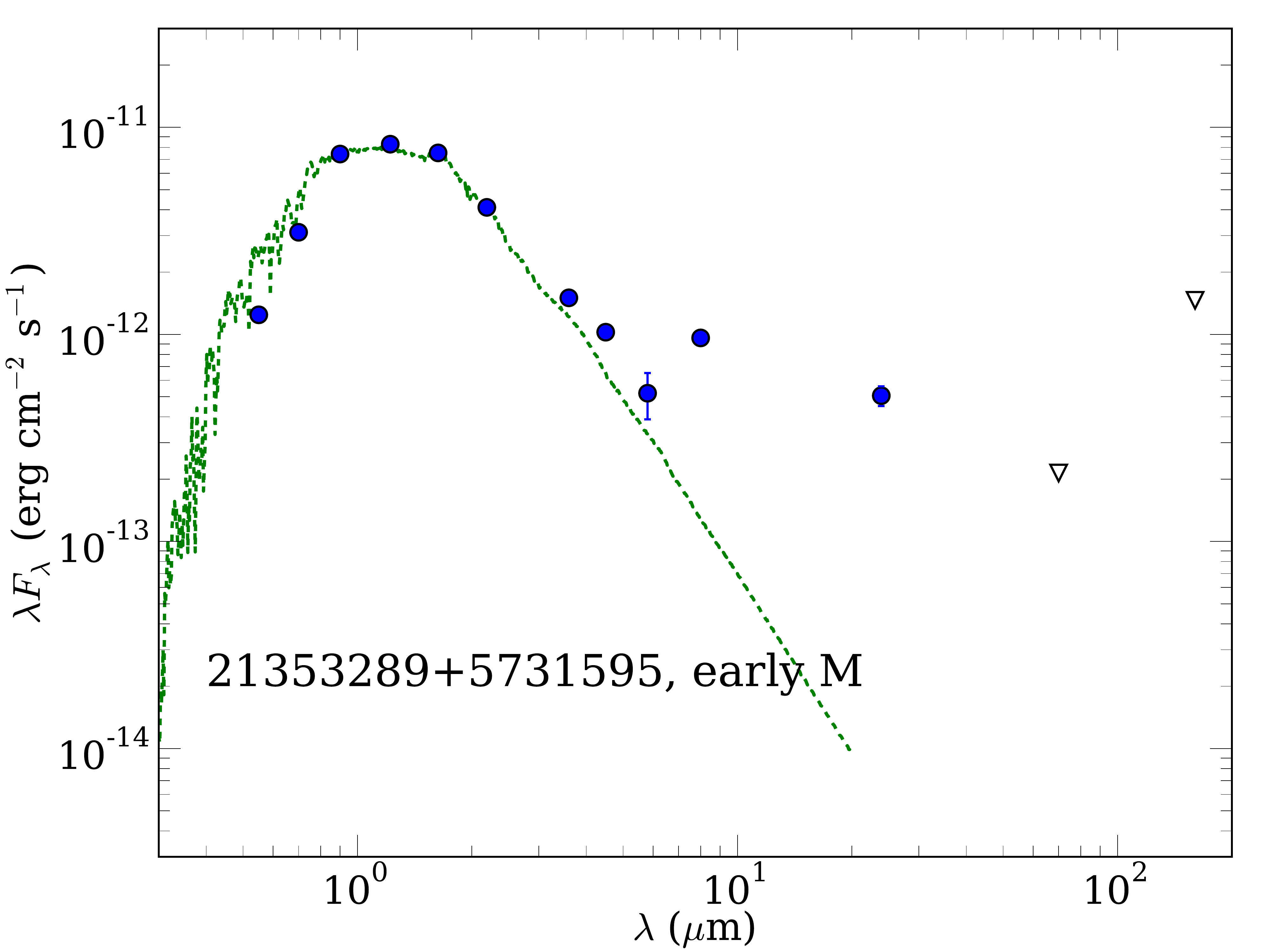} &
\includegraphics[width=0.24\linewidth]{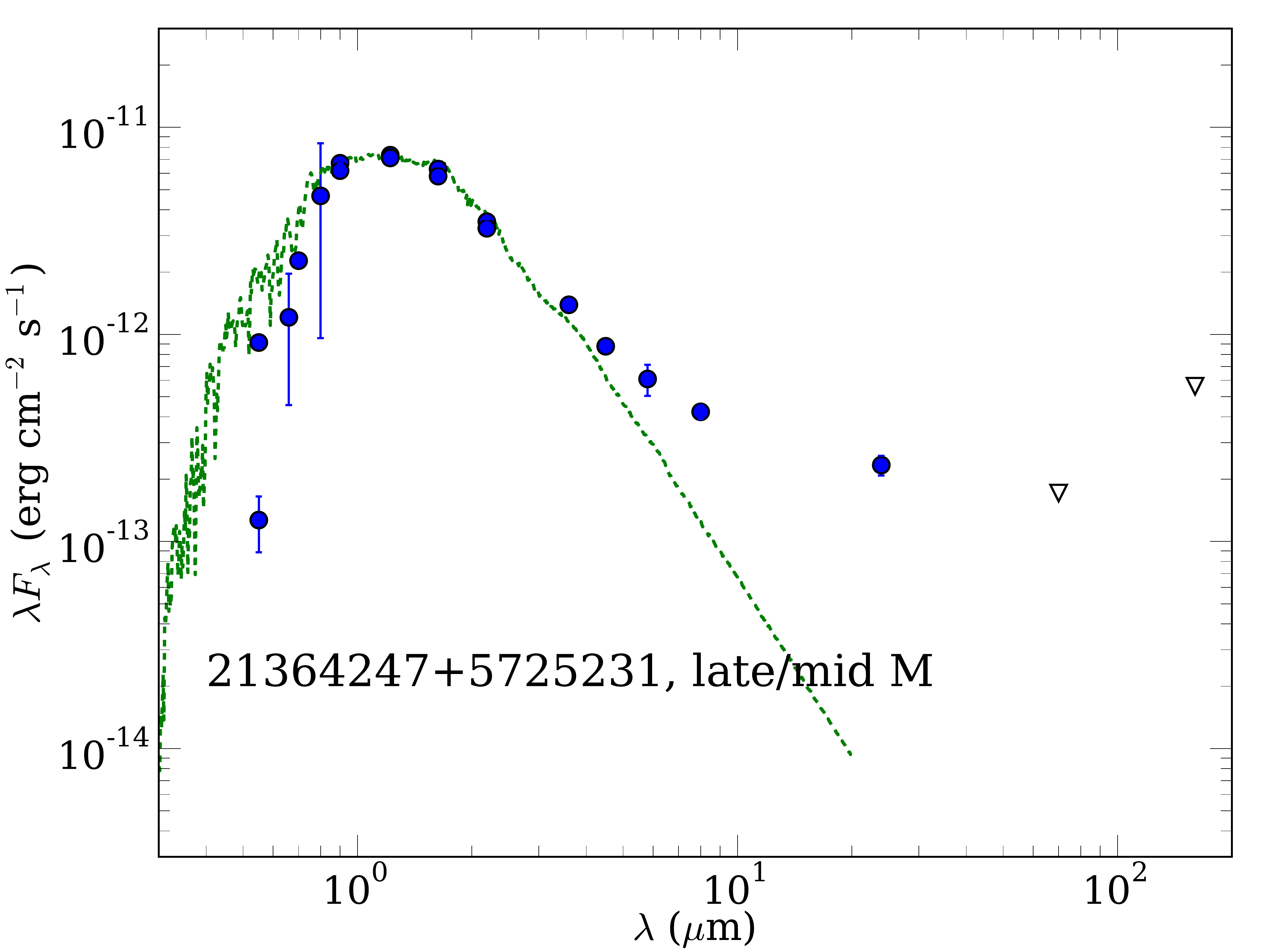} &
\includegraphics[width=0.24\linewidth]{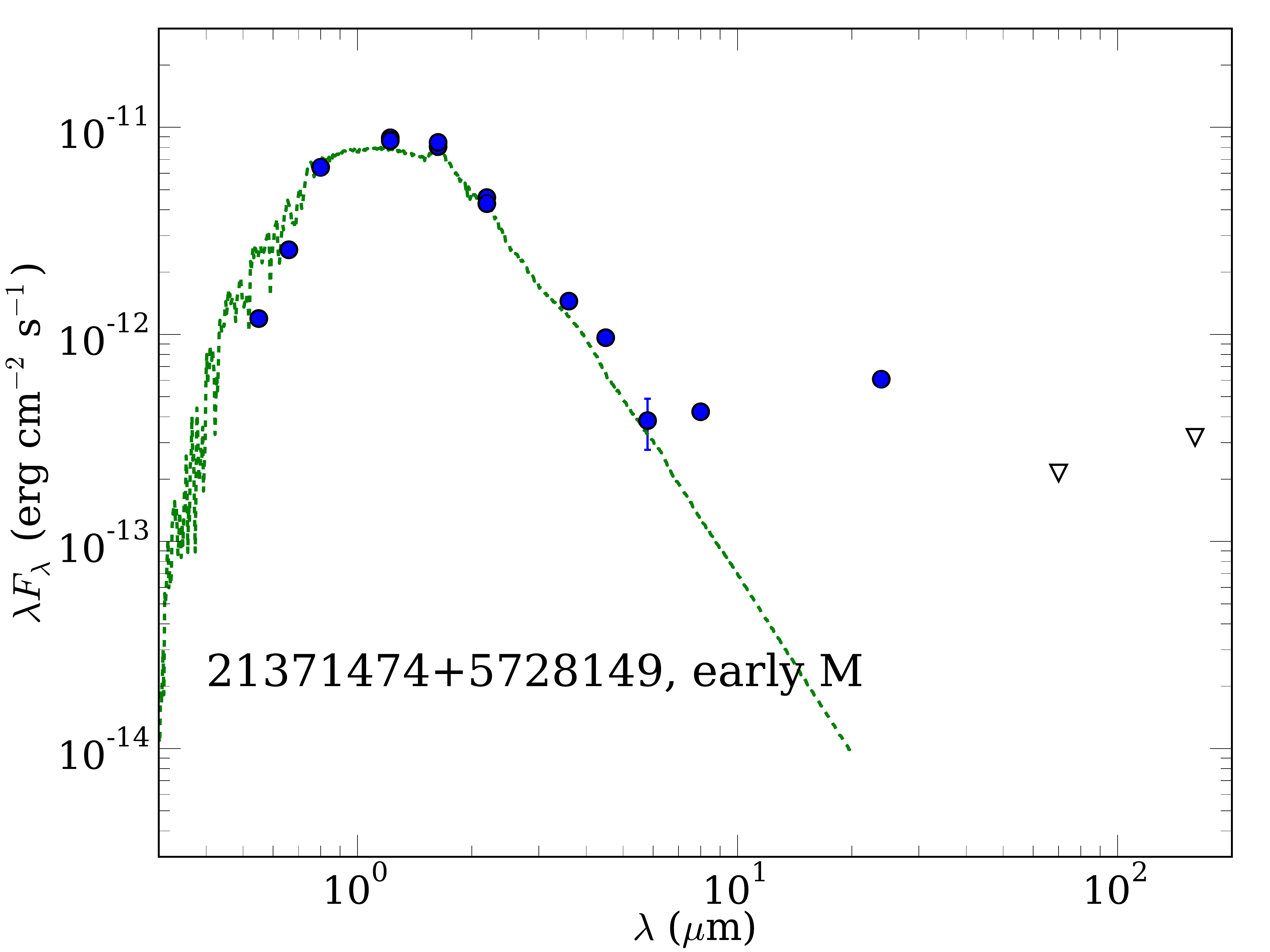} &
\includegraphics[width=0.24\linewidth]{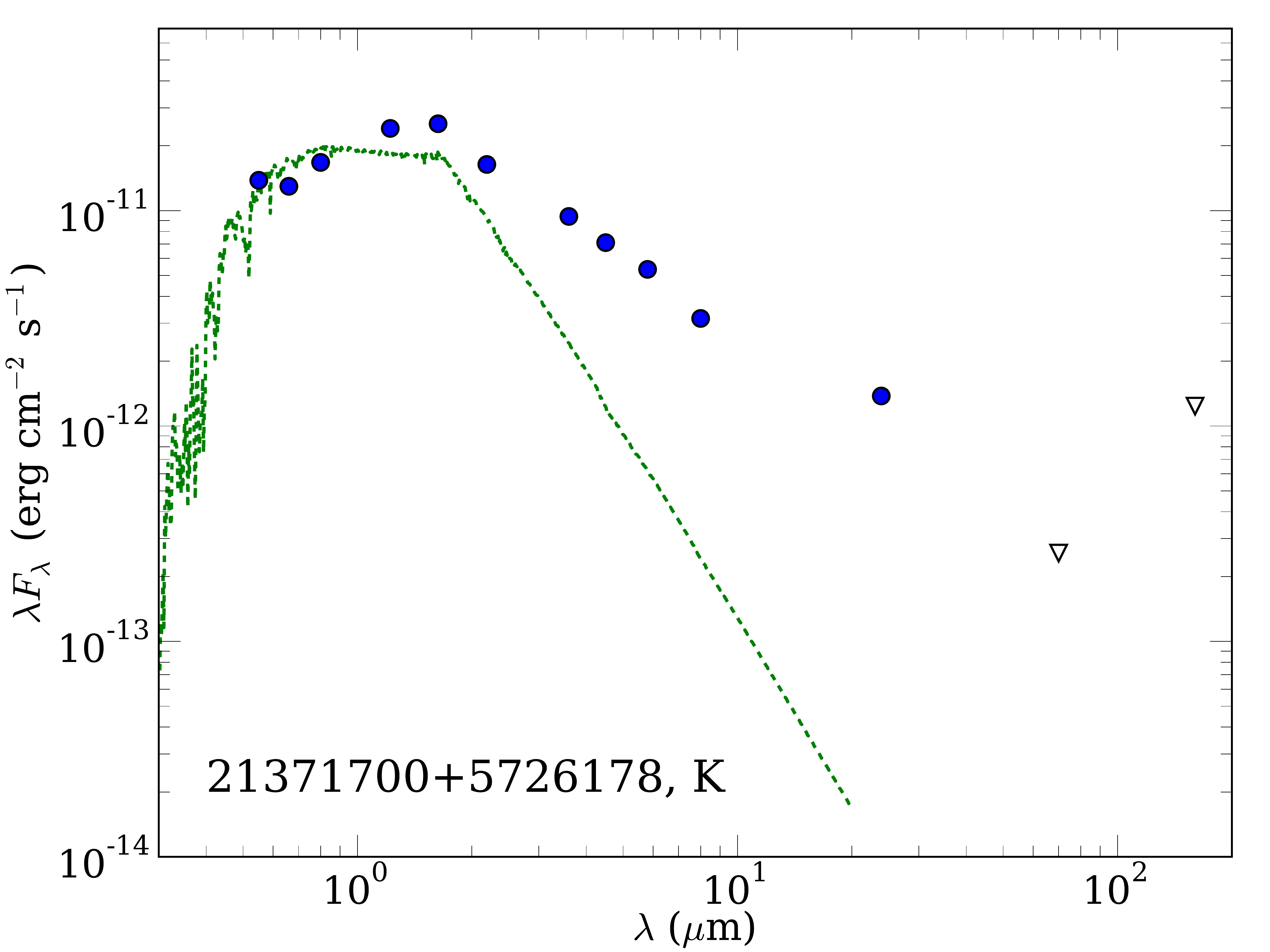} \\
\includegraphics[width=0.24\linewidth]{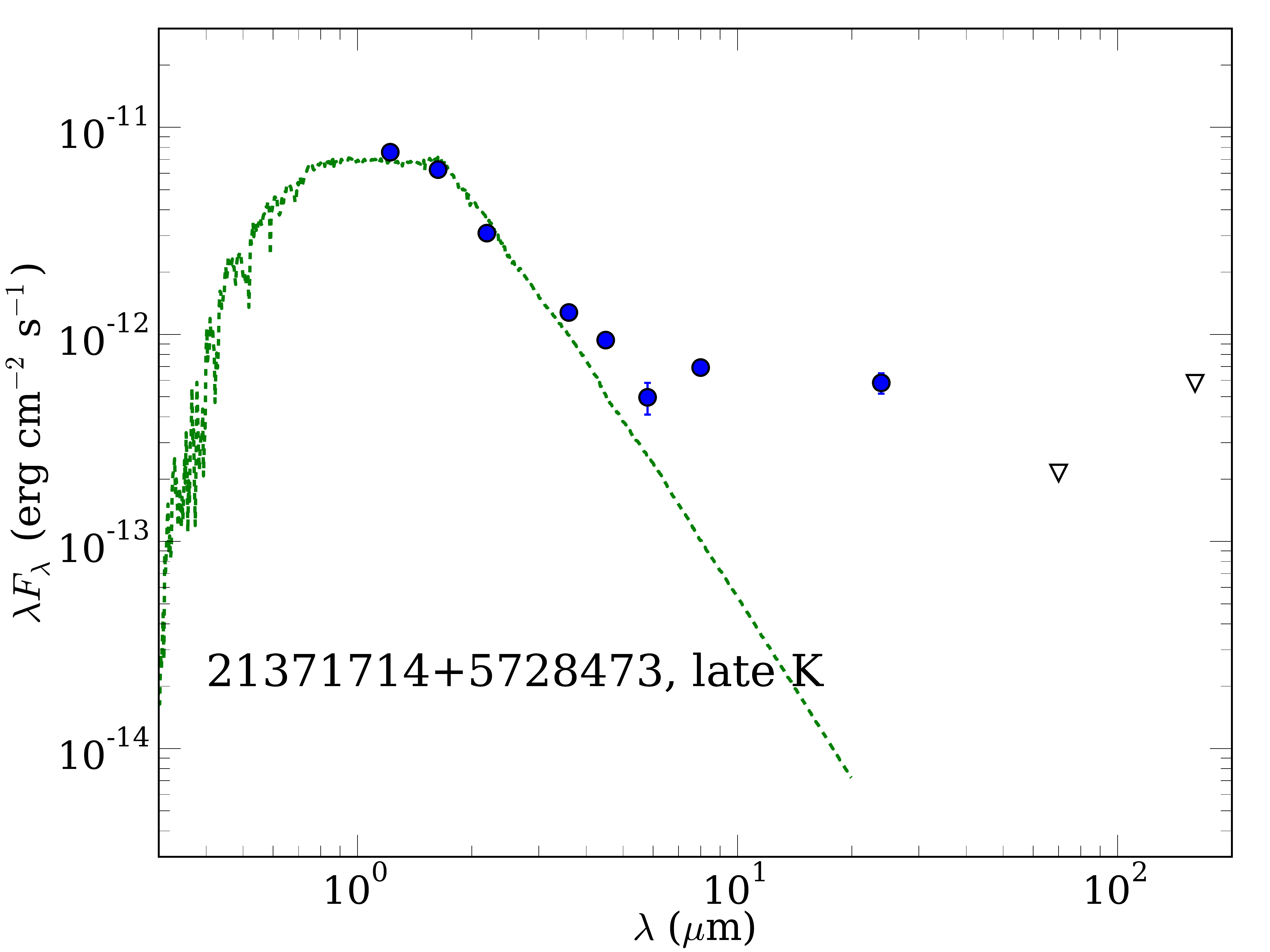} &
\includegraphics[width=0.24\linewidth]{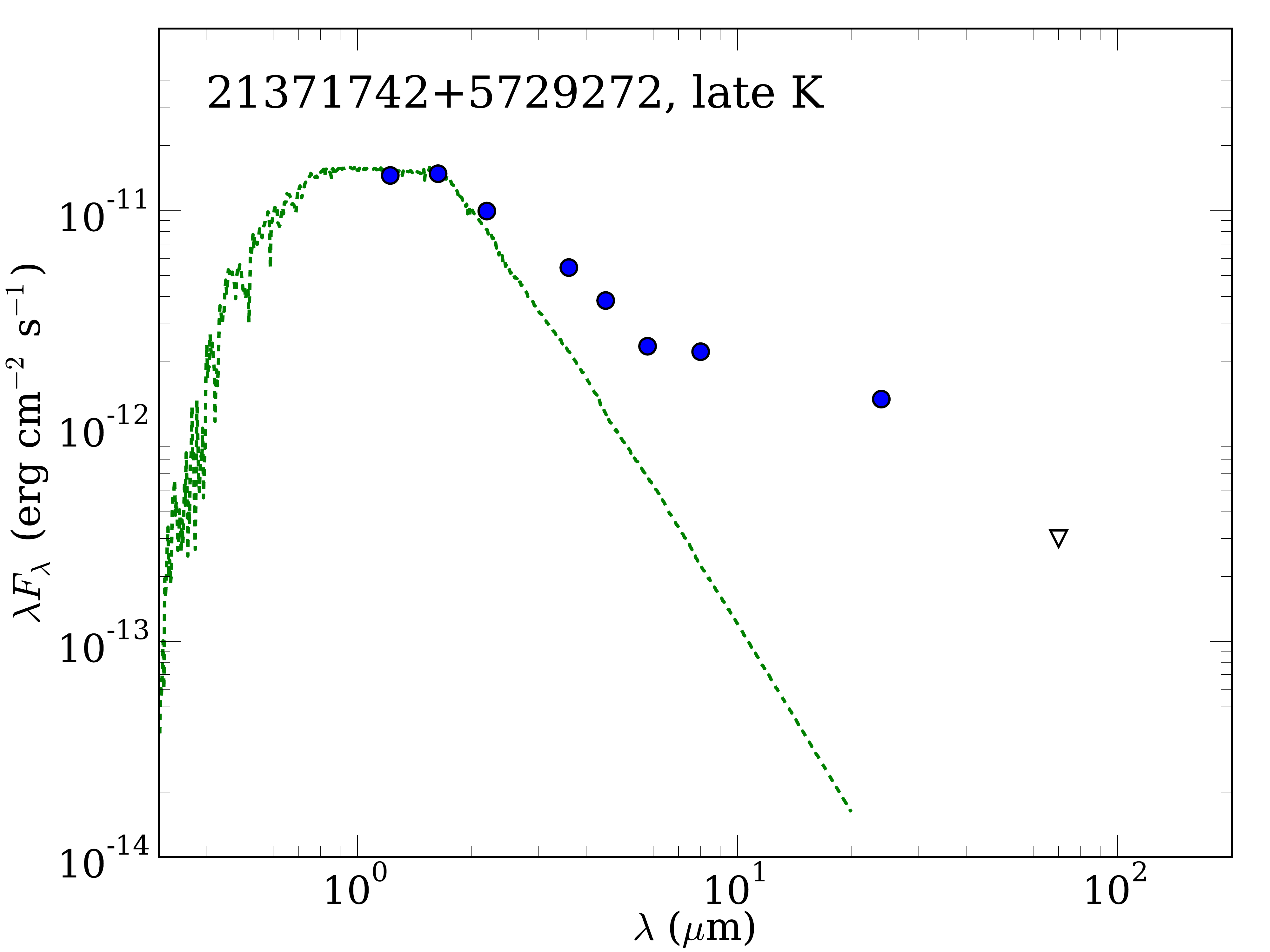} &
\includegraphics[width=0.24\linewidth]{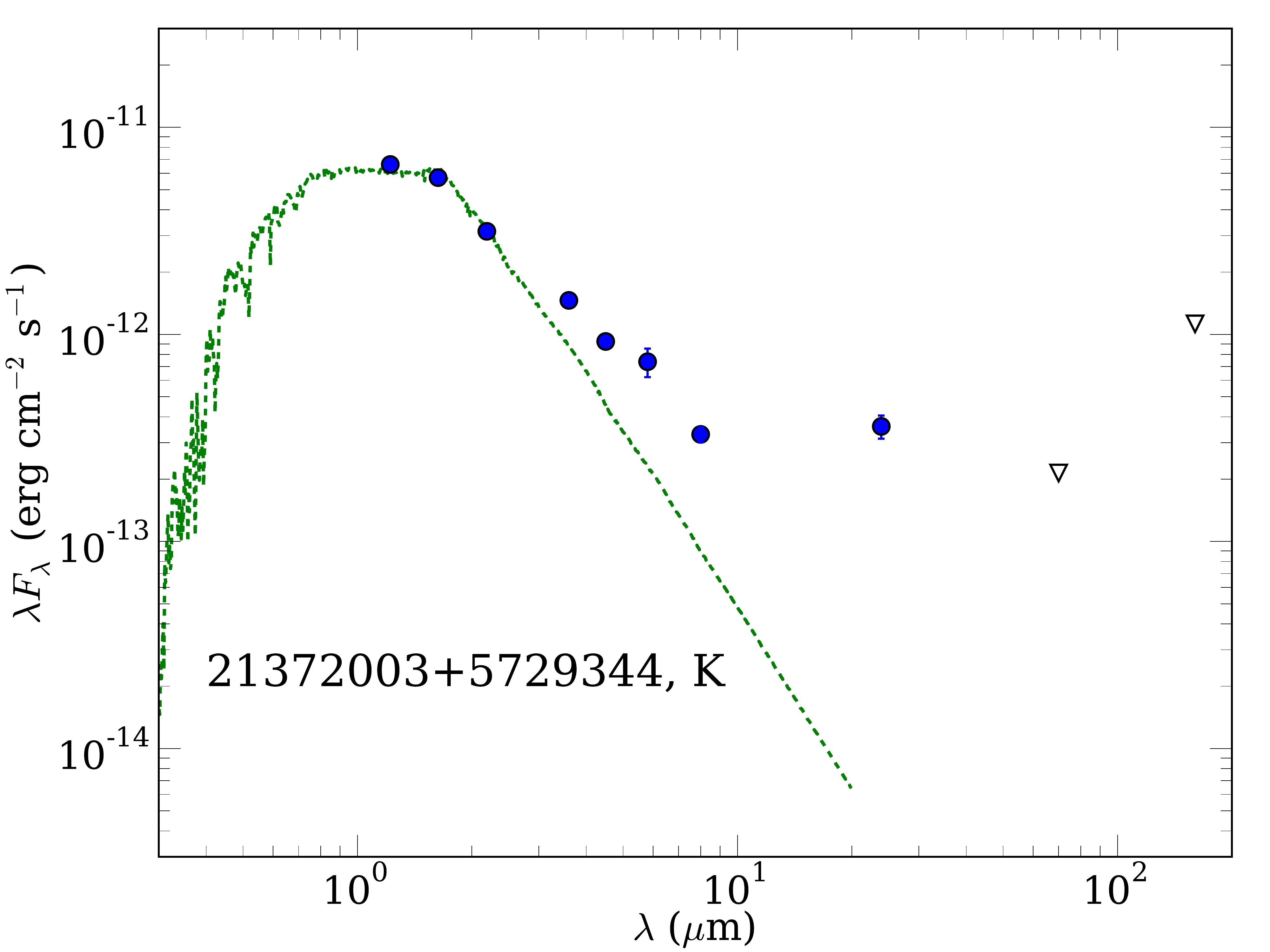} &
\includegraphics[width=0.24\linewidth]{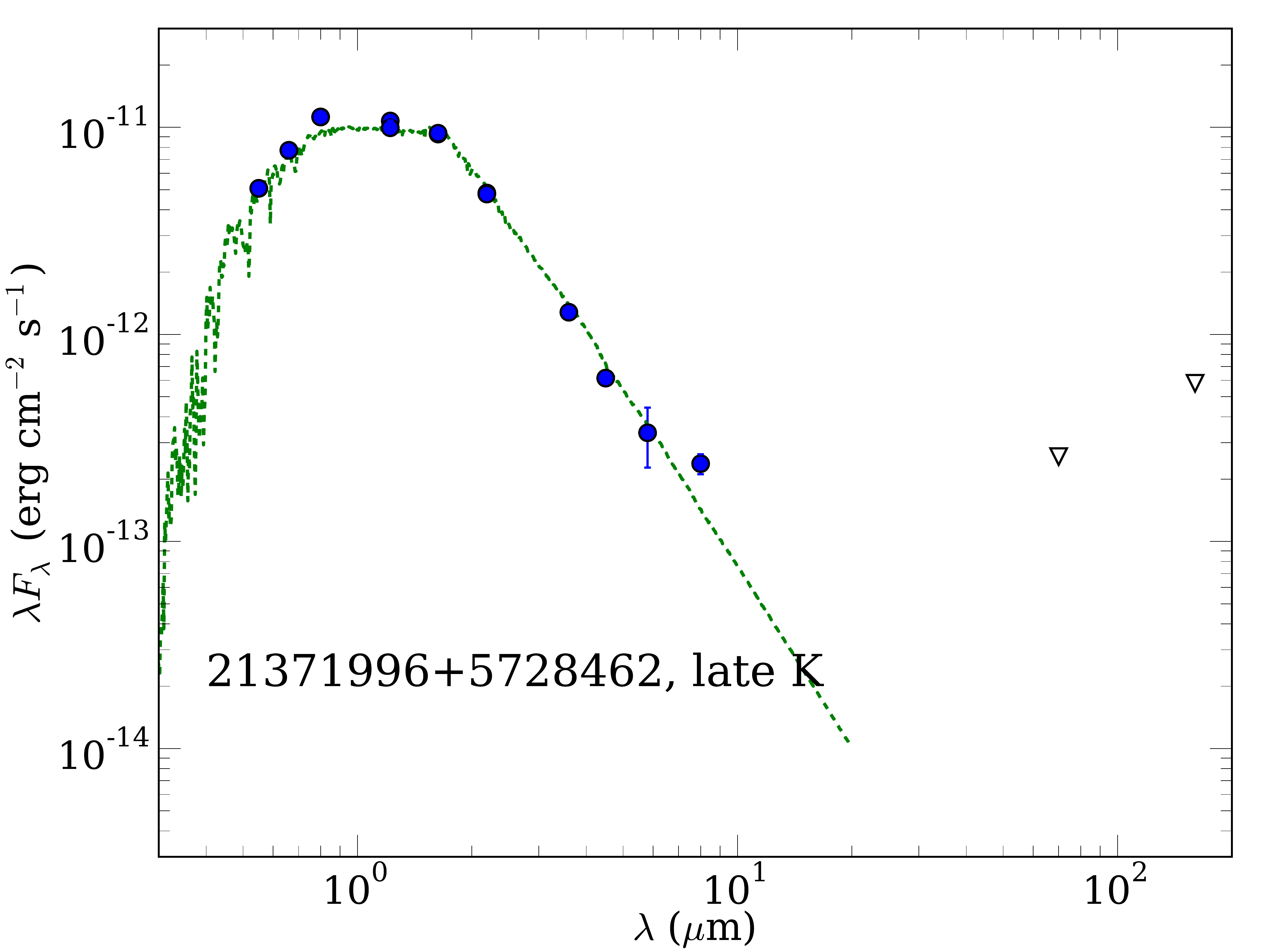} \\
\includegraphics[width=0.24\linewidth]{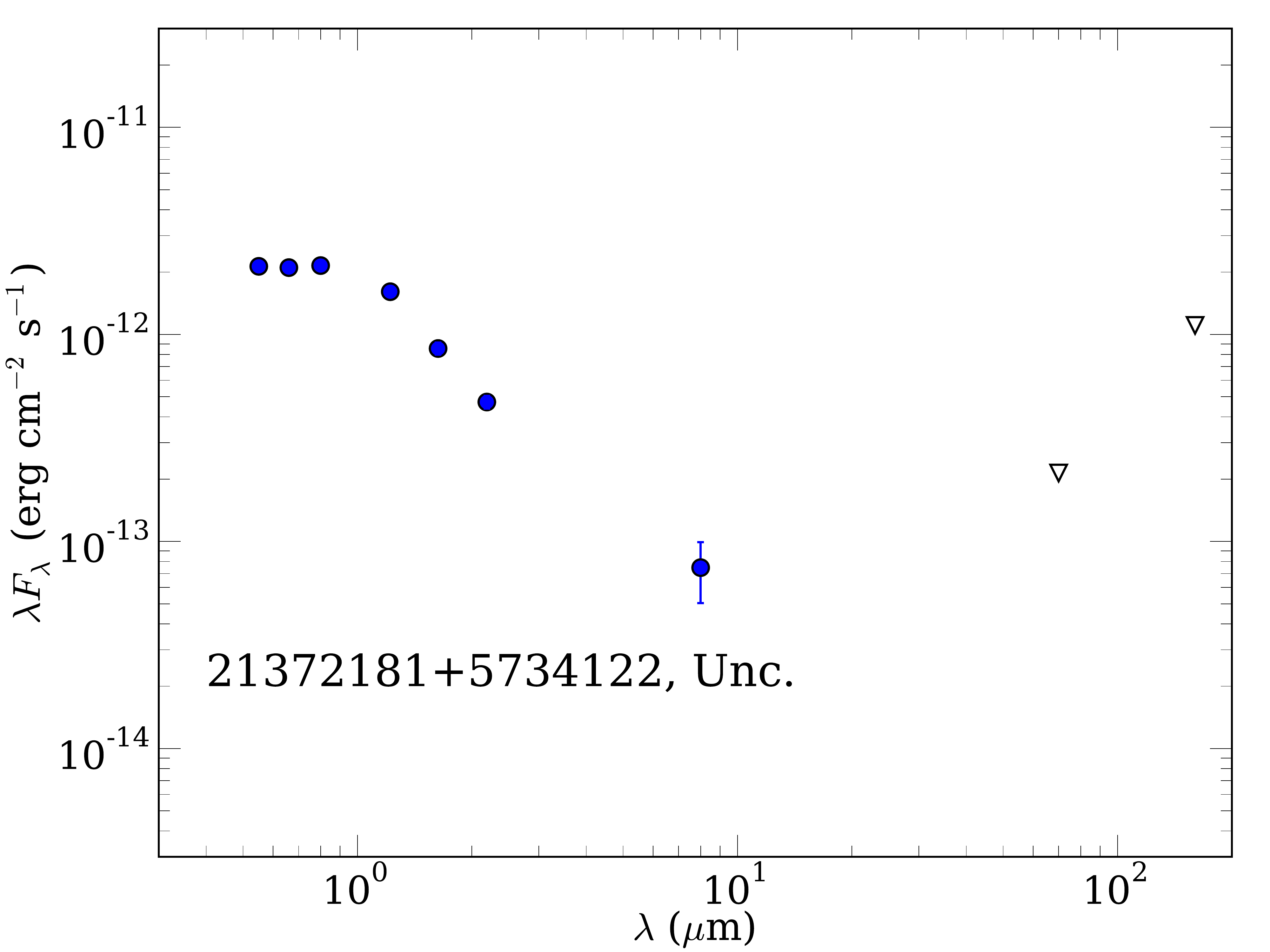} &
\includegraphics[width=0.24\linewidth]{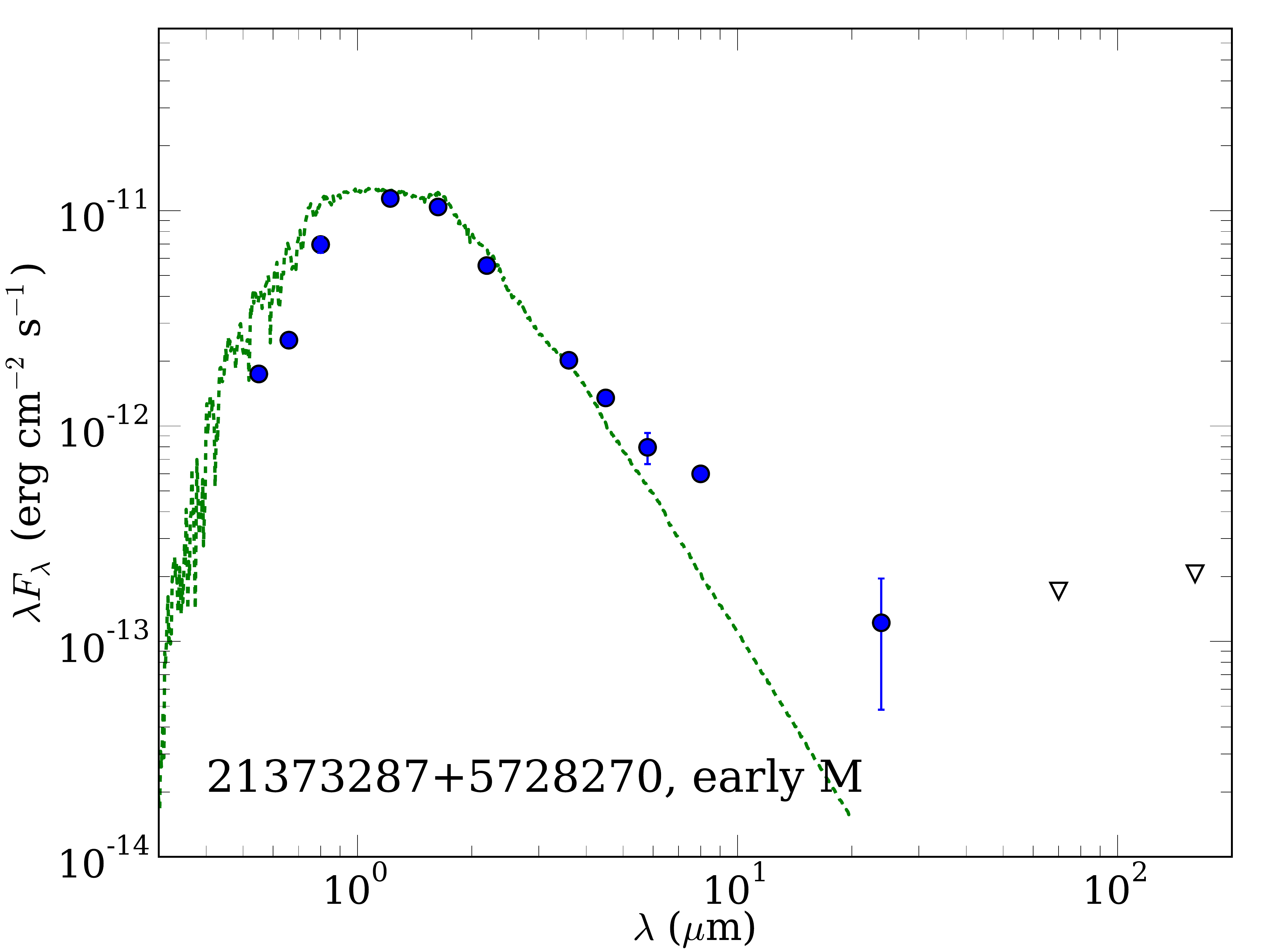} &
\includegraphics[width=0.24\linewidth]{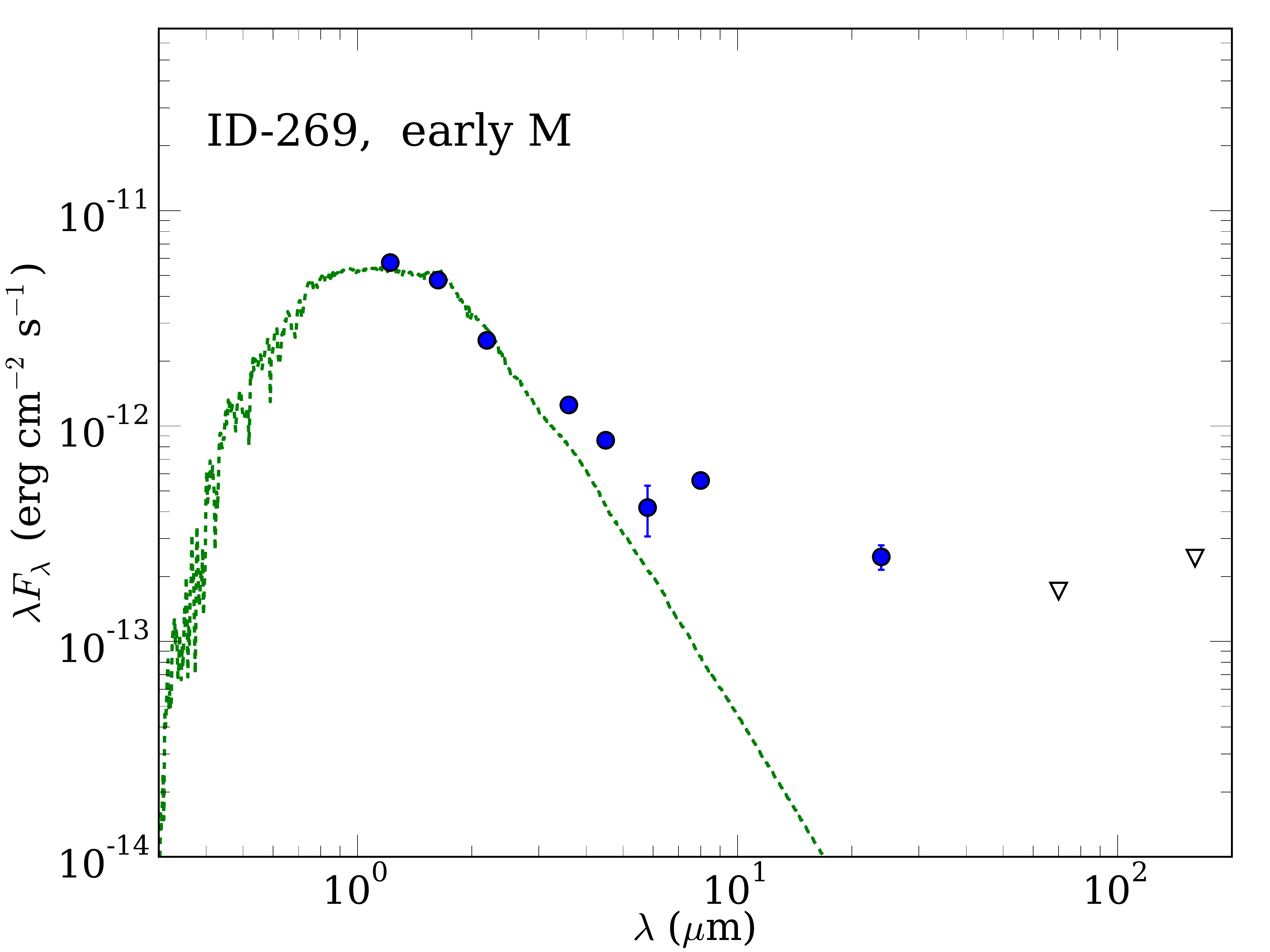} &
\includegraphics[width=0.24\linewidth]{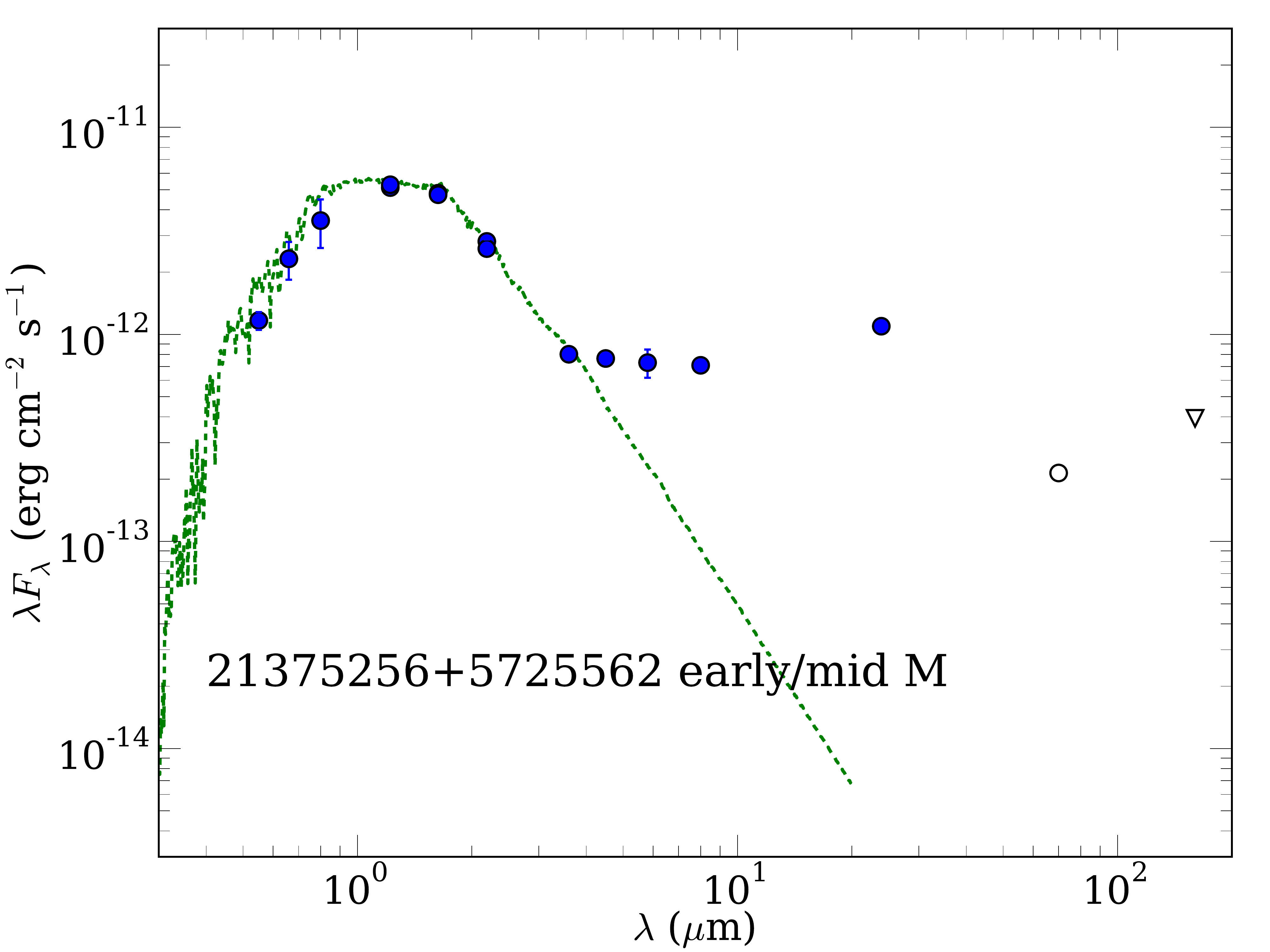} \\
\includegraphics[width=0.24\linewidth]{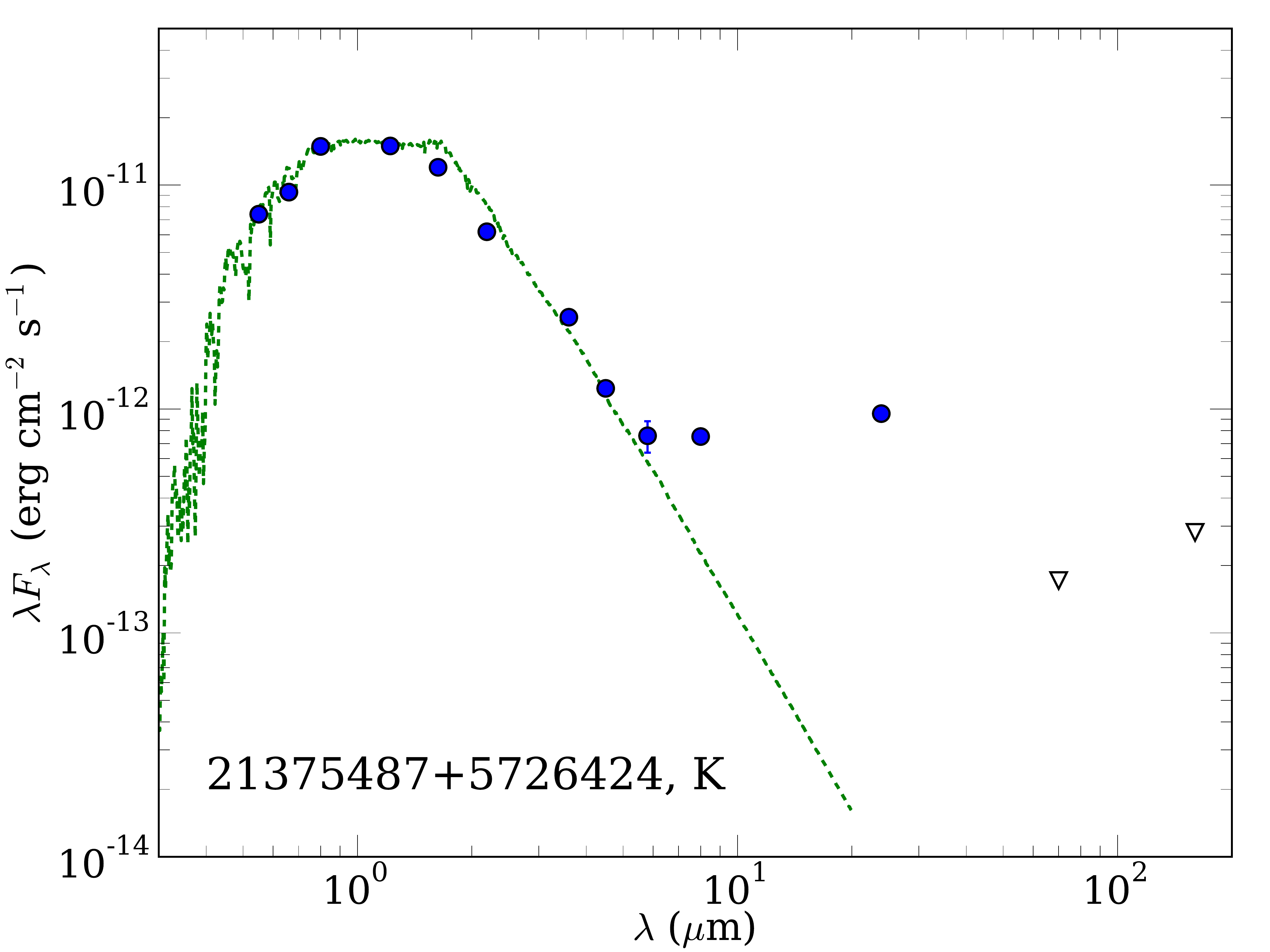} &
\includegraphics[width=0.24\linewidth]{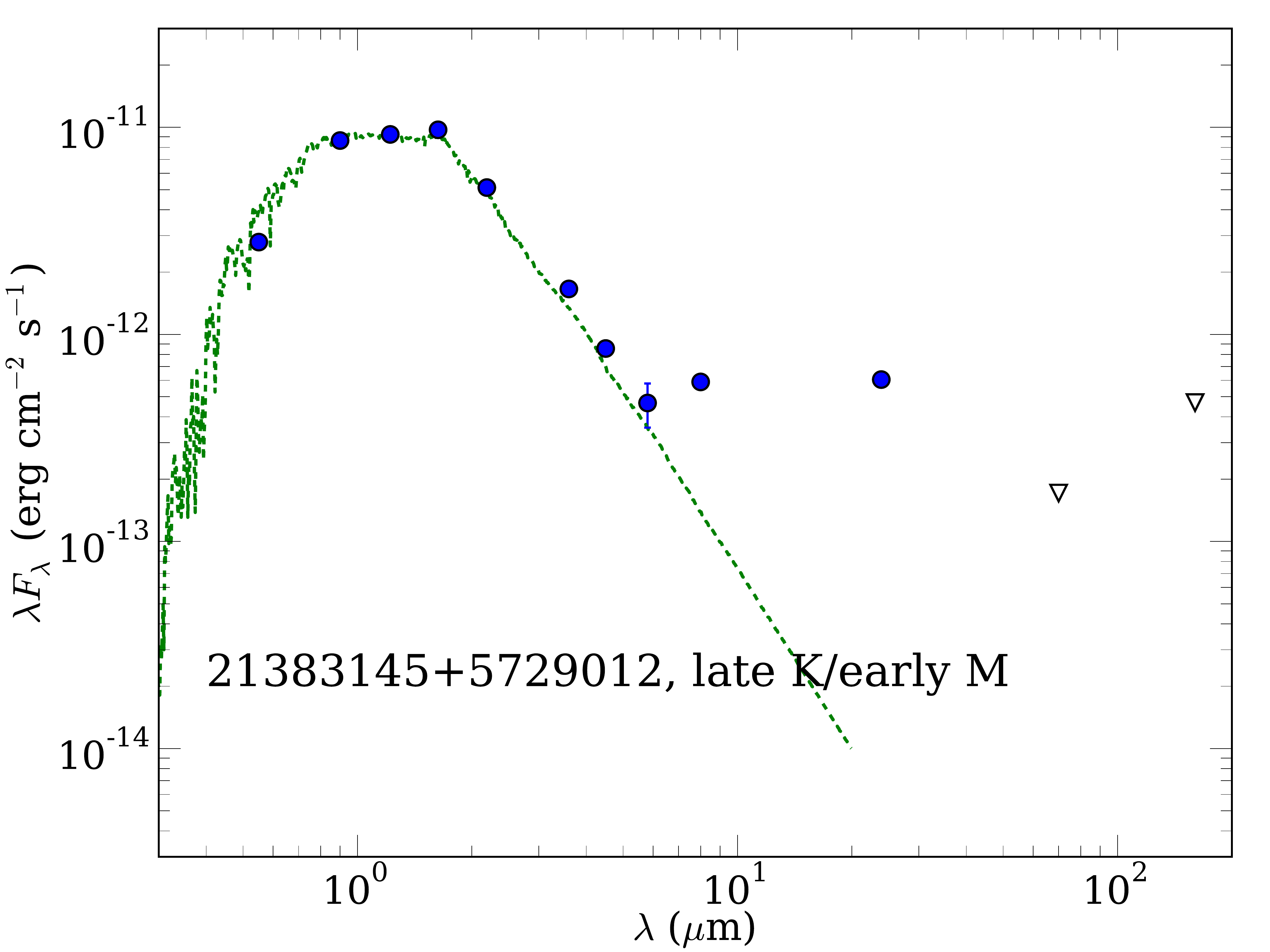} &
\includegraphics[width=0.24\linewidth]{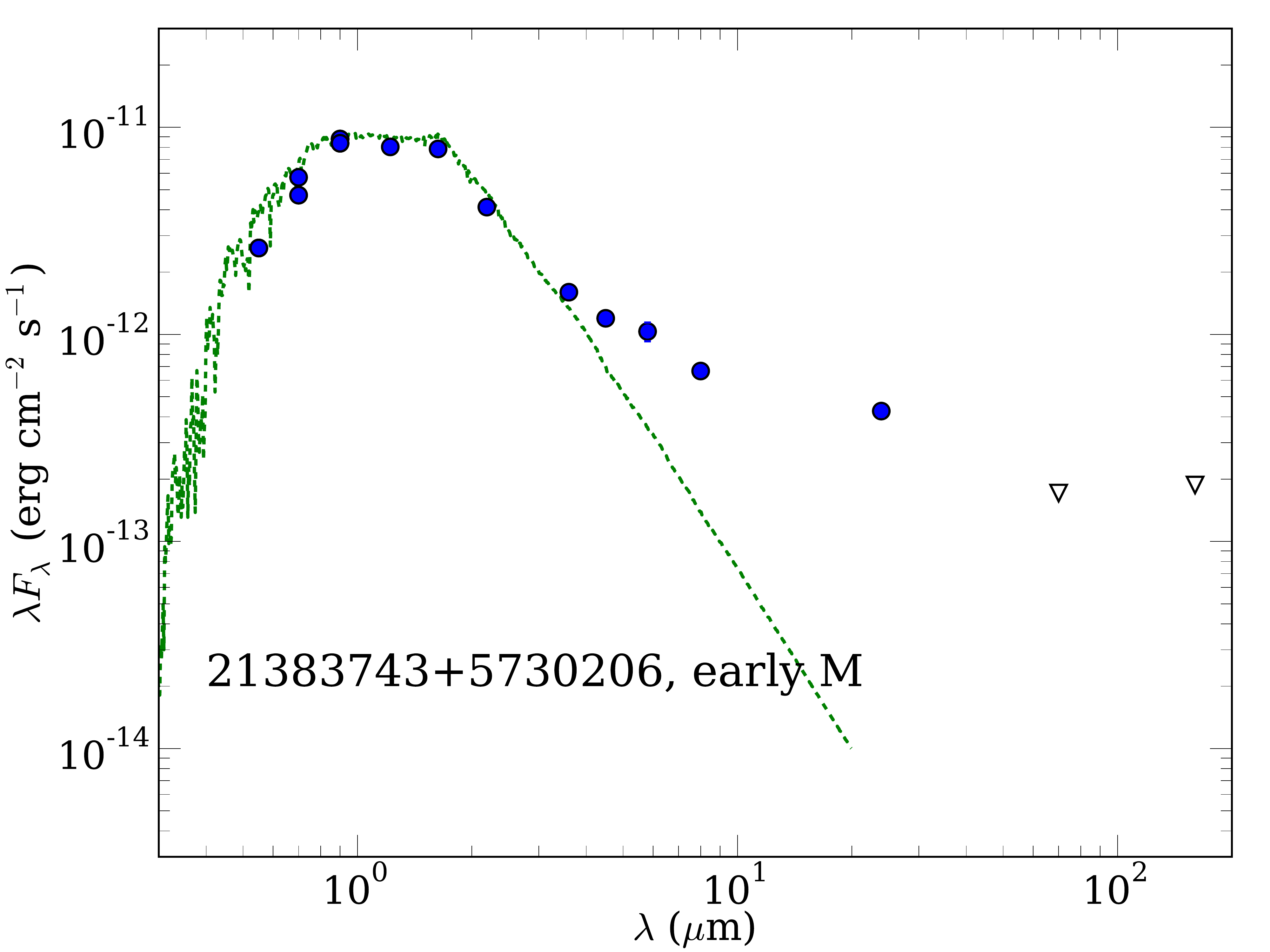} &
\includegraphics[width=0.24\linewidth]{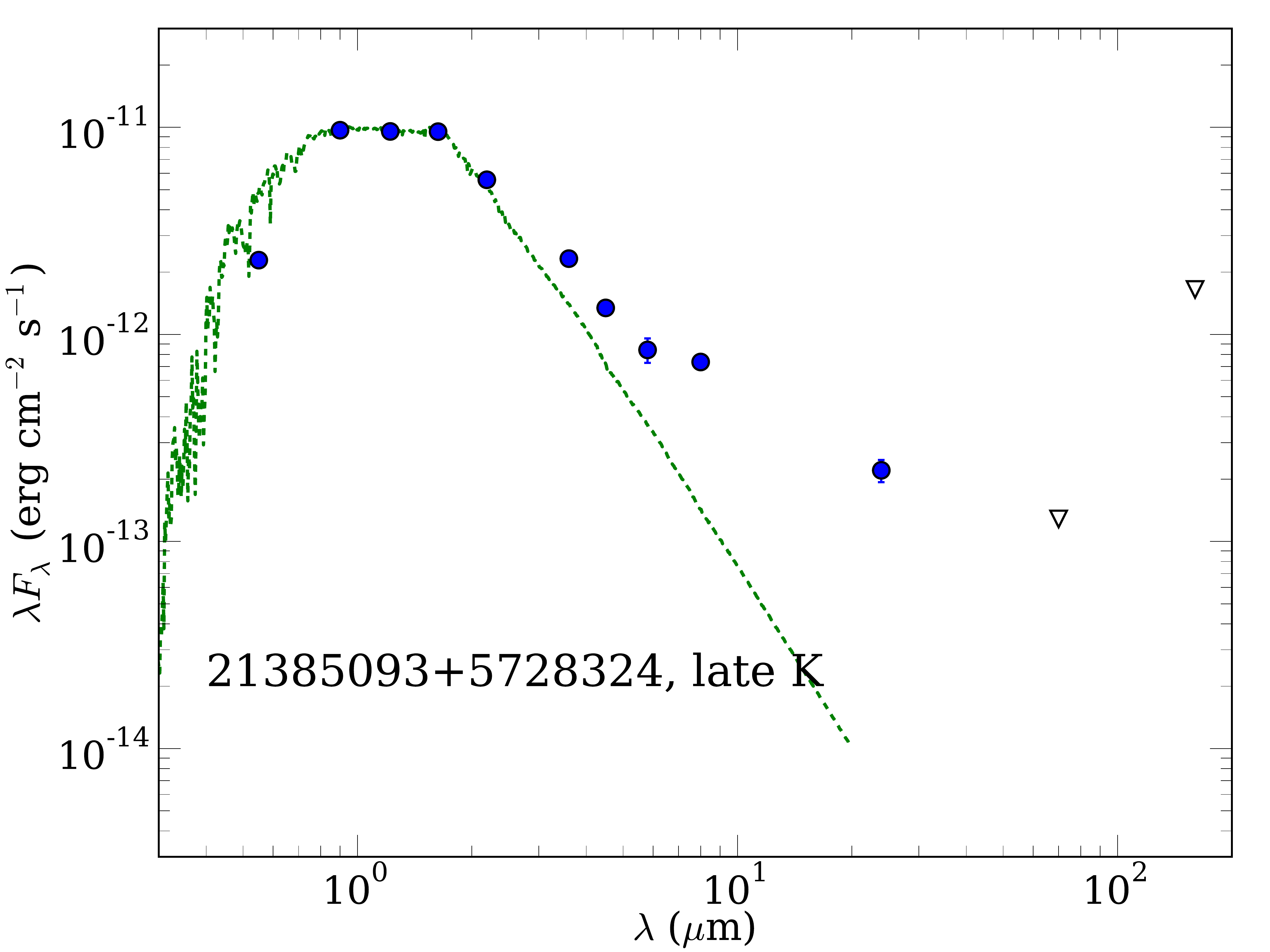} \\
\includegraphics[width=0.24\linewidth]{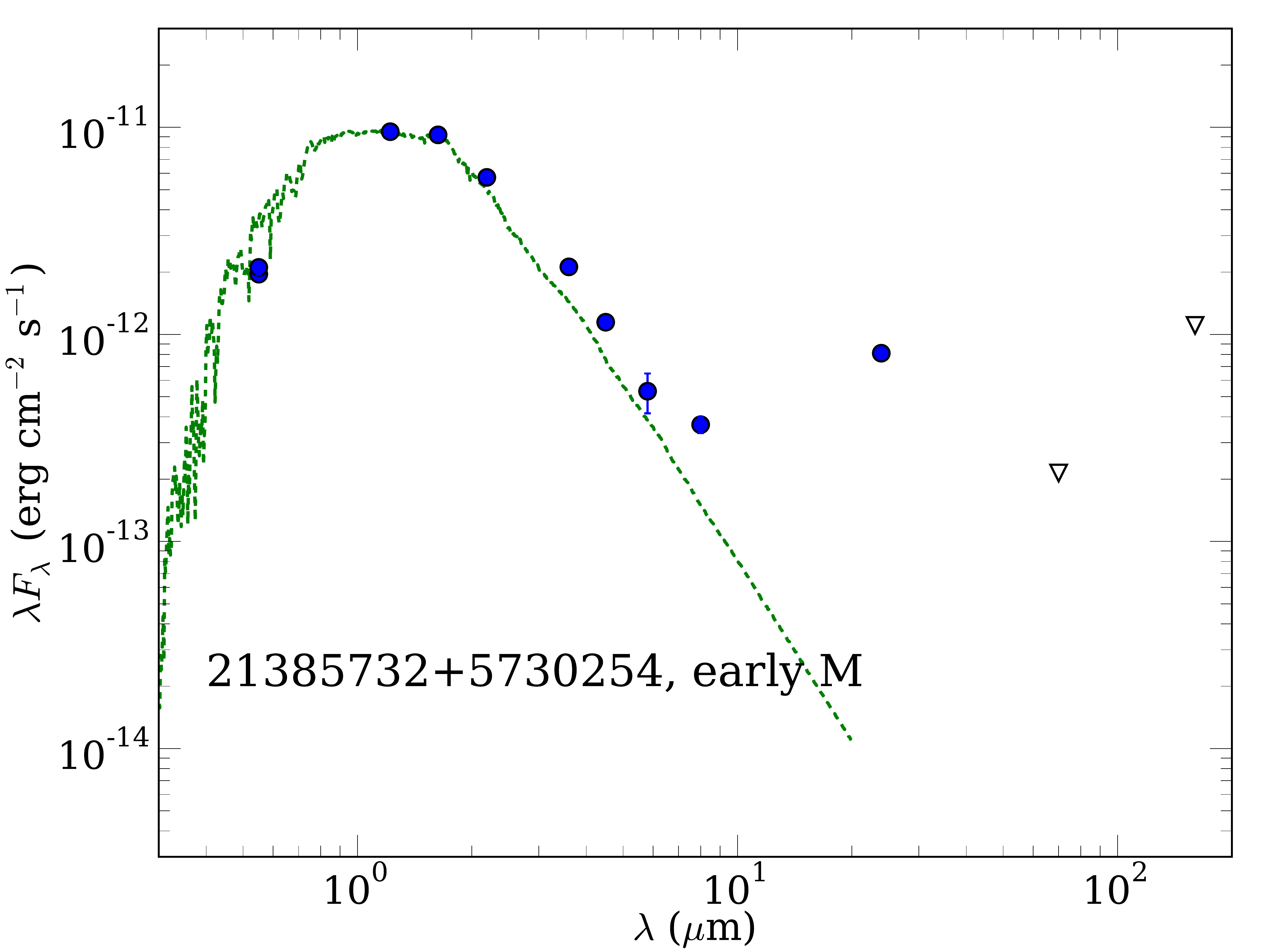} &
\includegraphics[width=0.24\linewidth]{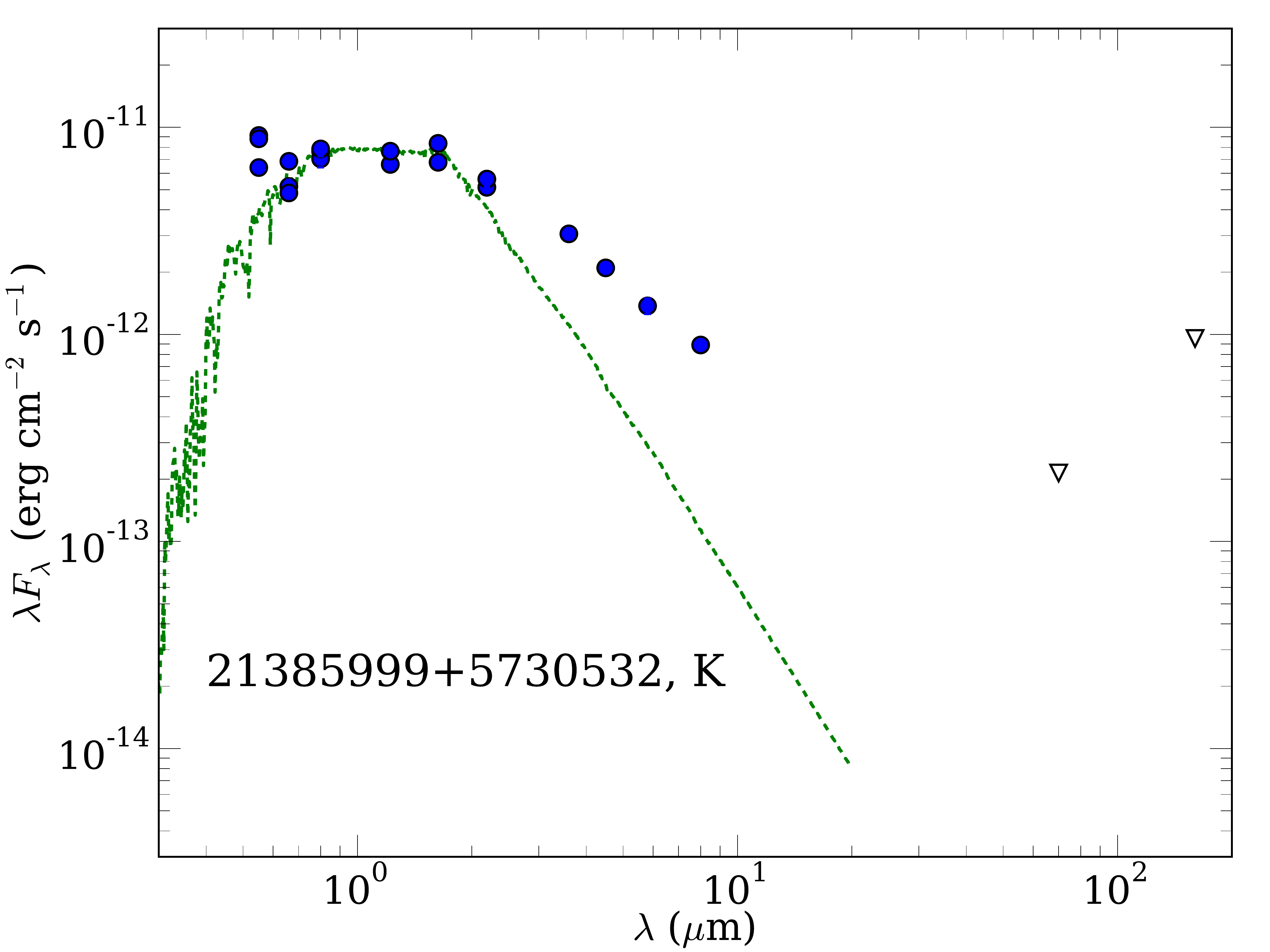} &
\includegraphics[width=0.24\linewidth]{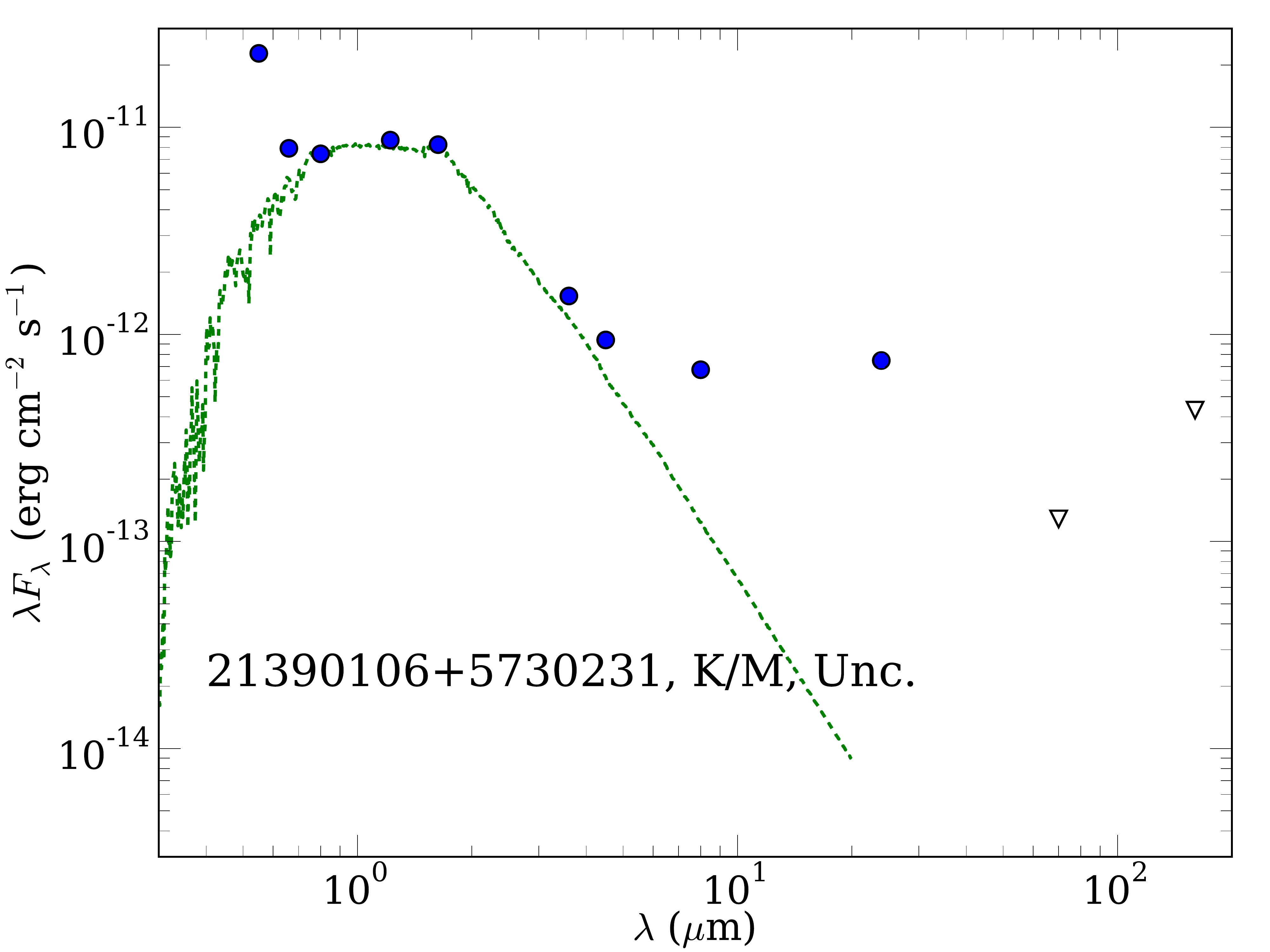} &
\includegraphics[width=0.24\linewidth]{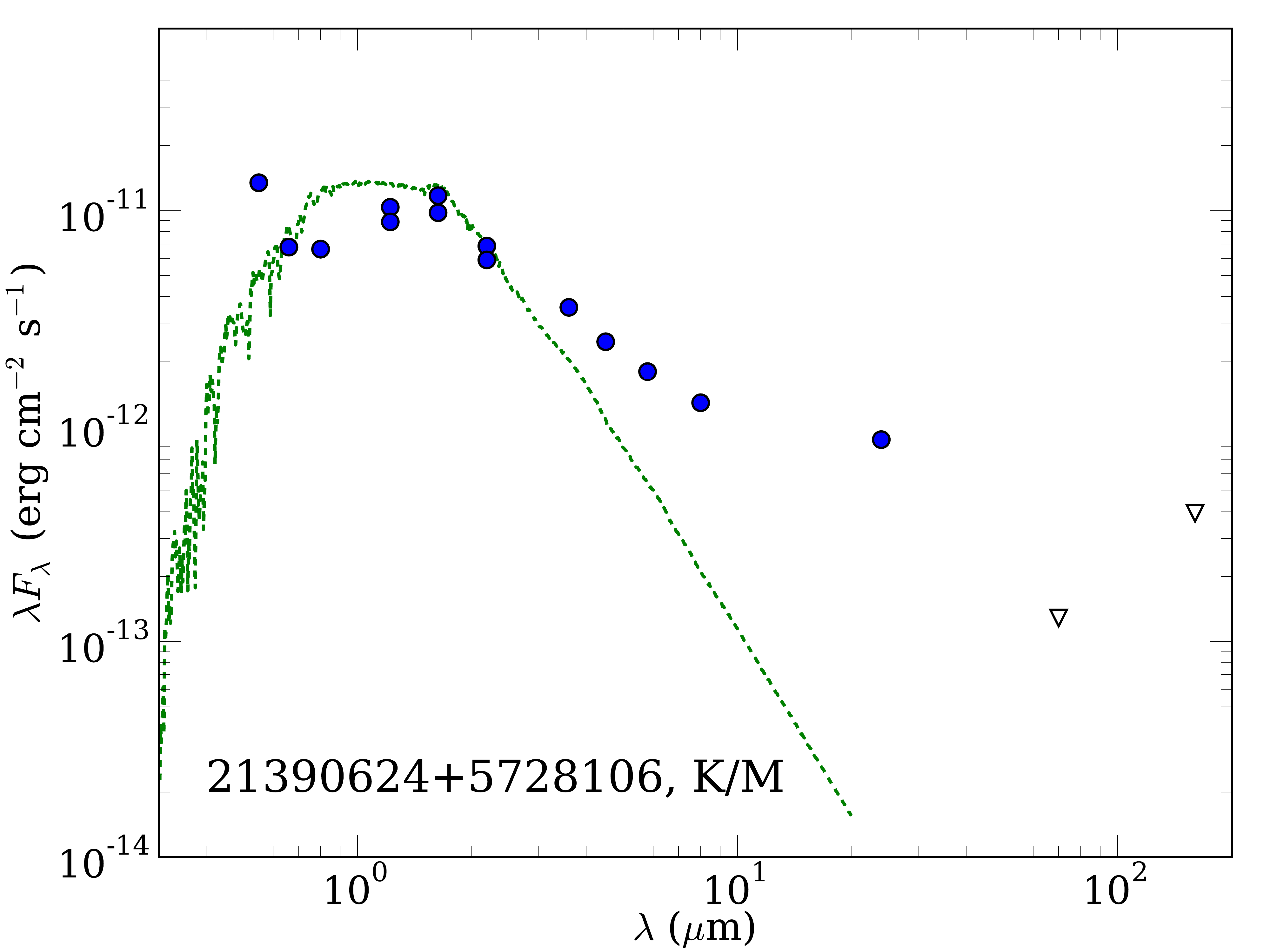} \\
\includegraphics[width=0.24\linewidth]{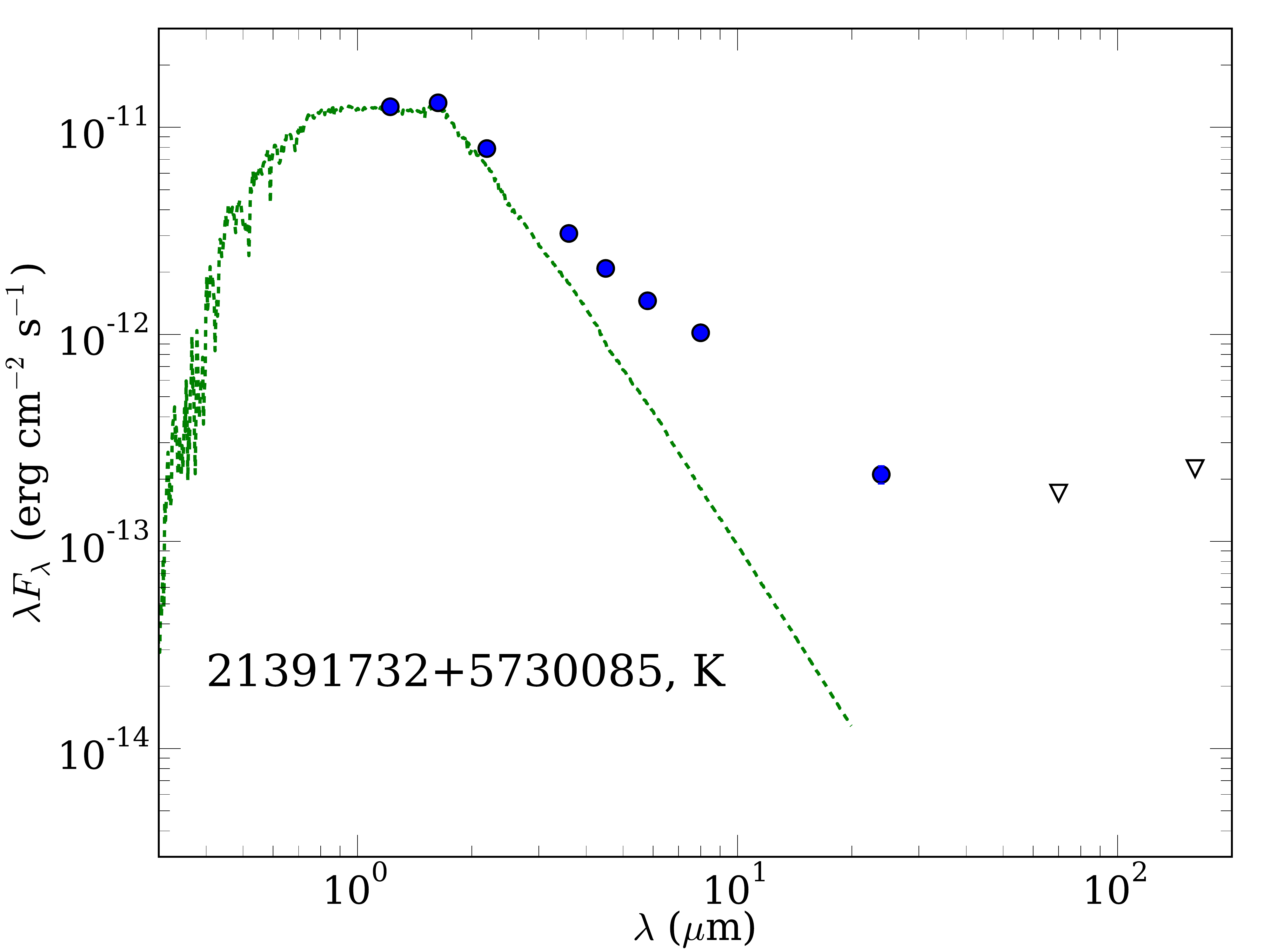} &
\includegraphics[width=0.24\linewidth]{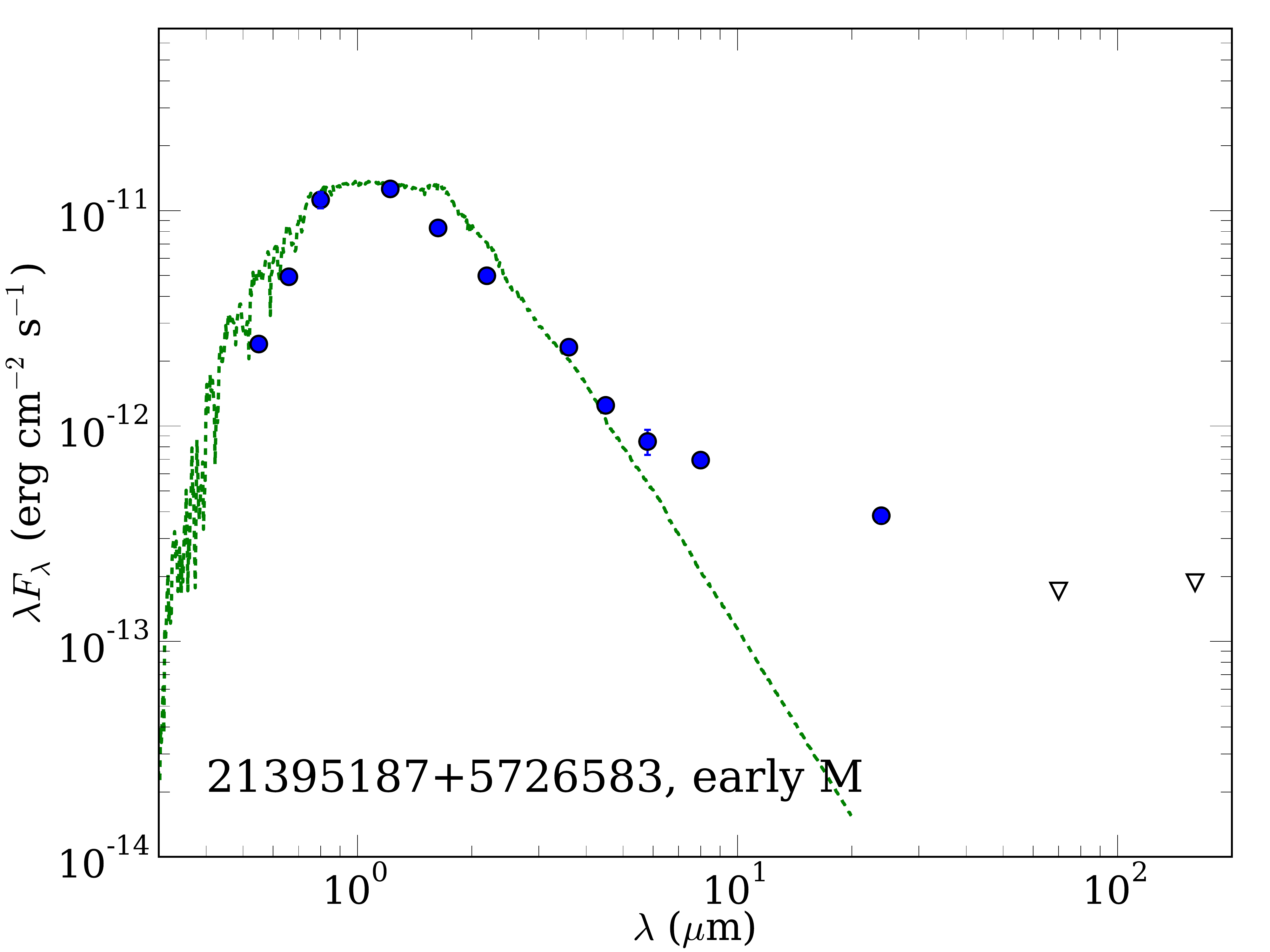} &
\includegraphics[width=0.24\linewidth]{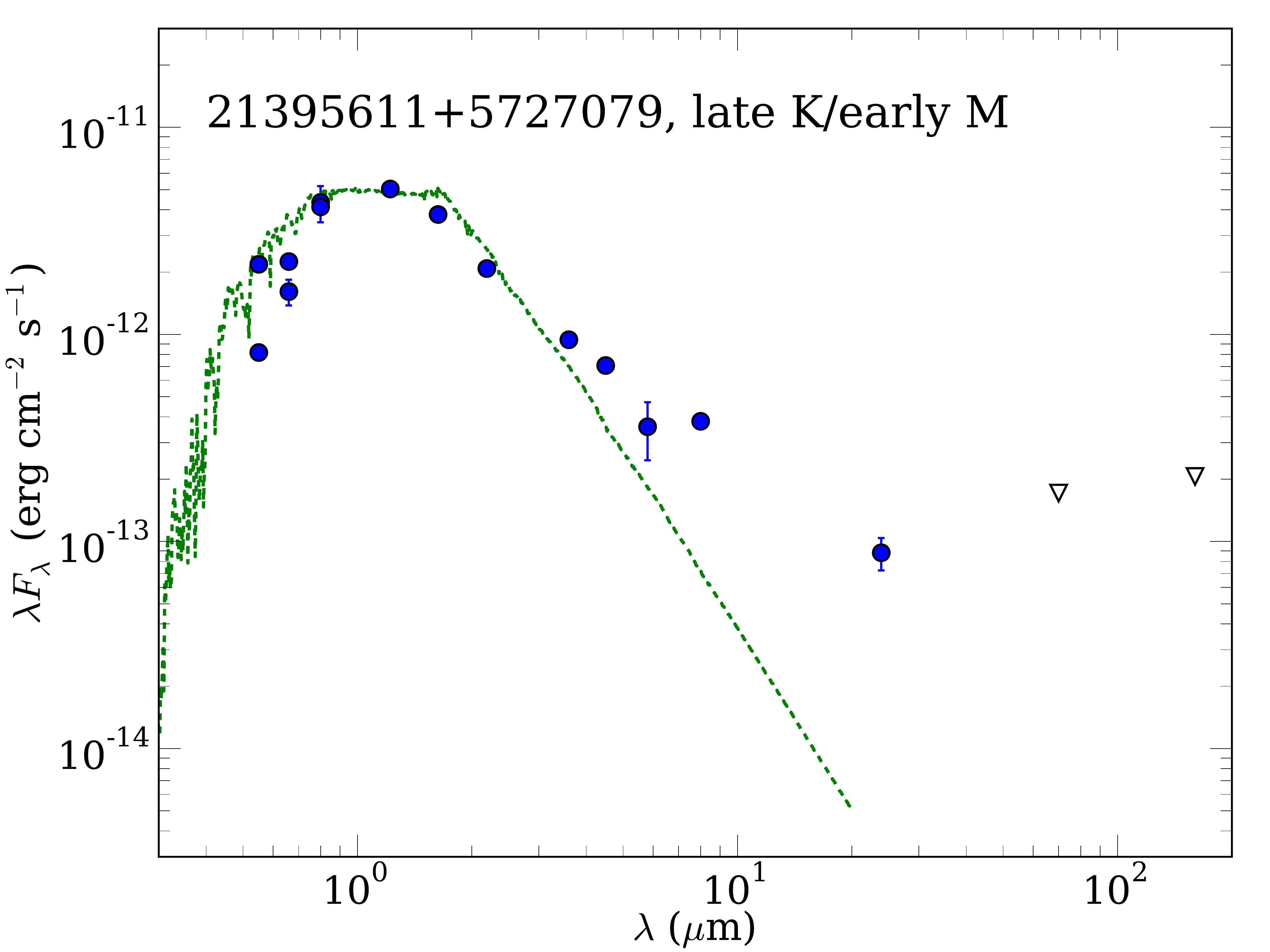} &
\includegraphics[width=0.24\linewidth]{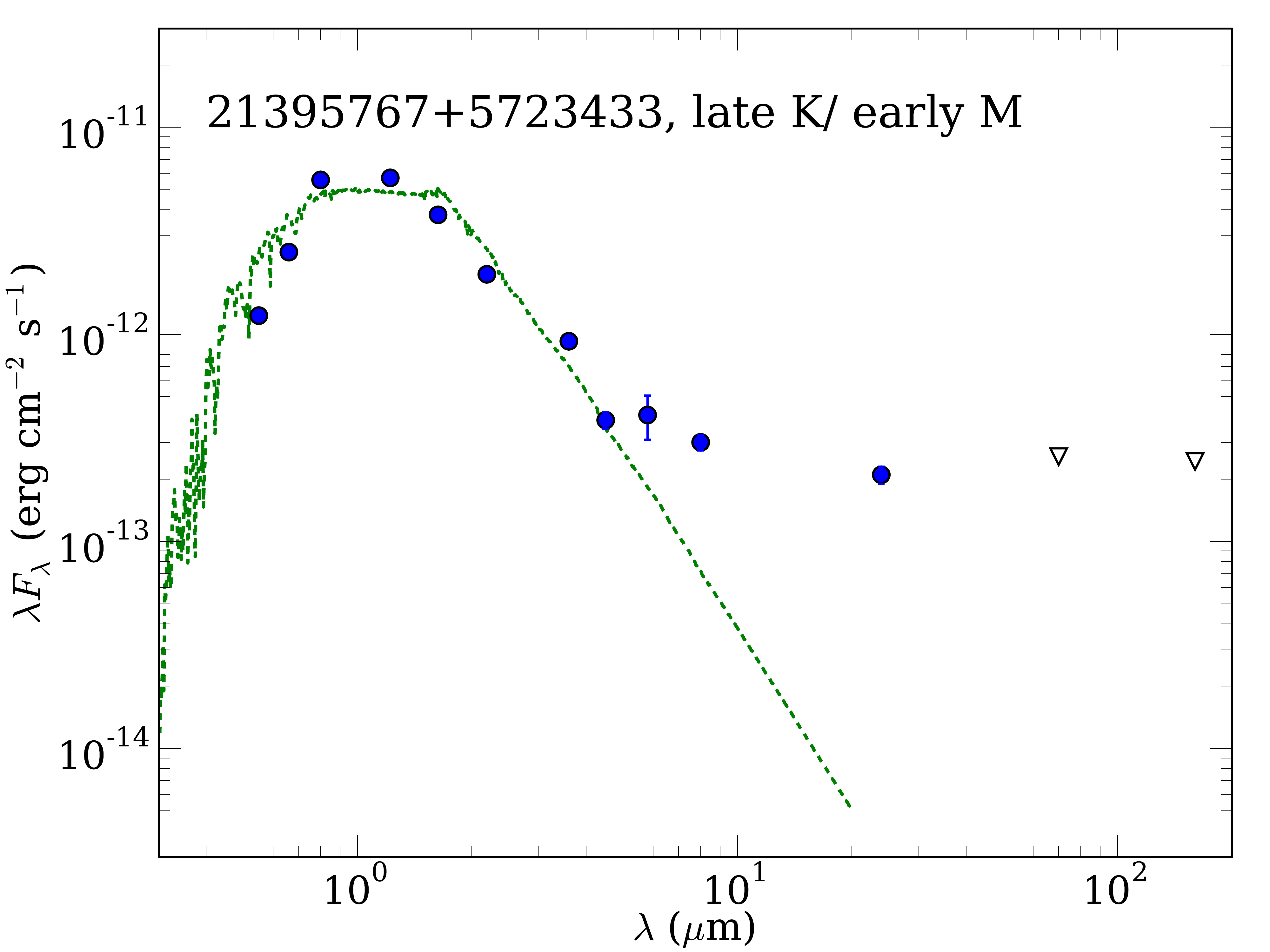} \\
\includegraphics[width=0.24\linewidth]{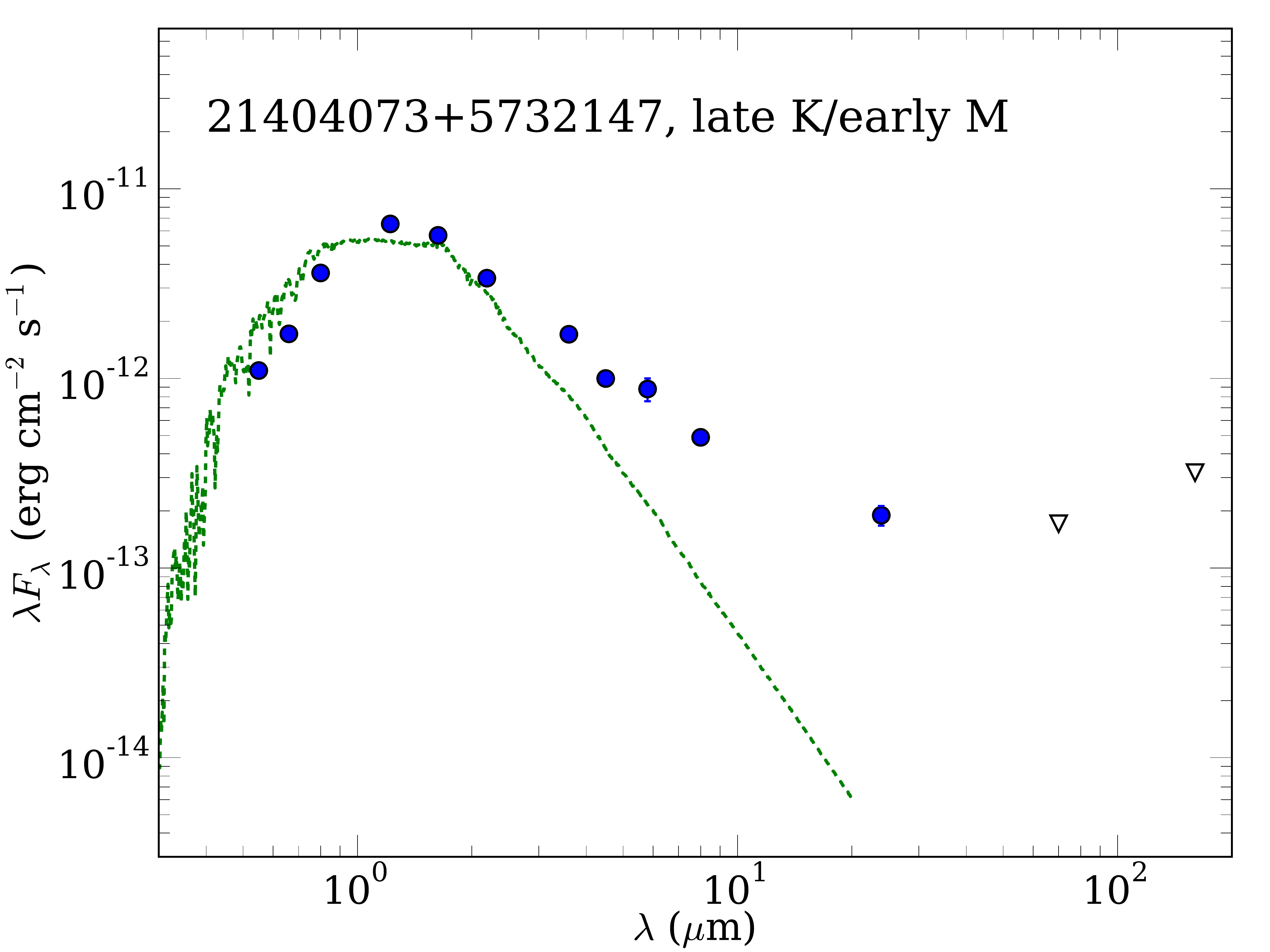} \\
\end{tabular}
\caption{SEDs of the objects with Herschel upper limits and
without optical spectroscopy confirmation (see Table \ref{otheruplims-table}). 
Only objects with evidence of a disk at other wavelengths and with significant
upper limits (i.e. not contaminated by nearby bright objects/cloud) are listed here.
Approximate spectral types (from SED fitting) are listed for the relevant objects.
Symbols as in Figure \ref{uplims1-fig}.\label{uplimsothers1-fig}}
\end{figure*}

This appendix contains full collection of SEDs of objects
observed with Herschel, together with
the complete Herschel photometry of the sources\footnote{Full photometry and spectroscopy, including
optical and IR data, is available upon request to the first author}. The first part (Table \ref{photometry-table},
Figures \ref{seds1-fig} to \ref{seds3-fig})
includes the sources previously known and confirmed from our spectroscopic surveys (Sicilia-Aguilar et al. 2005, 2006b; SA13).
The second part includes objects identified as YSO via Spitzer, H$\alpha$ photometry and/or X-ray observations
(Table \ref{other-table}, Figure \ref{otherseds-fig}). 
Table \ref{uplims-table} contains the upper limits for sources with disks 
identified via optical spectroscopy,
whose SEDs are displayed in Figures \ref{uplims1-fig}-\ref{uplims4-fig}. Table \ref{otheruplims-table} contains 
the upper limits for sources with disks from H$\alpha$ (B11), X-ray (M09; G12), 
and Spitzer (MC09) surveys, whose
SEDs are displayed in Figure \ref{uplimsothers1-fig}. Only sources with confirmed or potential IR excesses
and significan tupper limits (those that impose some contraints on the source SED)
are considered. Most of the sources located within strong nebular emission in and near the IC\,1396\,A globule are
excluded from the tables and figures, as their upper limits are typically several orders of magnitude
higher than any reasonable value expected from their Spitzer data. A few objects with uncertain mid-IR excesses are
also listed as upper limits. 

The optical data for the sources was compiled from the existing literature (Sicilia-Aguilar et al. 2005, 2010;
SA13; G12; B11; M09). Sources lacking this information in their original paper were searched for in our
optical photometry databases, which are presented in detail in Sicilia-Aguilar et al. (2005; including
VRI data from the Fred Lawrence Whipple Observatory) and Sicilia-Aguilar et al. (2010; with UVRI data
from LAICA/3.5m Calar Alto Telescope). For very bright sources such as HD\,206267 and 
CCDM J2137+5734, the optical data was obtained from SIMBAD. 
The sources were also matched to the 2MASS catalog (Cutri et al. 2003)
to obtain JHK photometry. The Spitzer data was compiled from the literature for the
optical spectroscopy members (SA13). For the remaining candidates, we
re-reduced the existing Spitzer data following the same procedures than in SA13, and also obtained the
WISE counterparts from the WISE All-Sky Catalog (Wright et al. 2010) for objects without Spitzer data.
All SEDs are plotted correcting for the measured extinction (when available), or using the cluster
average or best-fit extinction to the photospheric data of the star (if no measurement of A$_V$ is 
published).

The detection of photospheric fluxes at 70\,$\mu$m is limited to the O star HD\,206267, which makes all
upper limits on diskless intermediate- and low-mass stars irrelevant. Objects such as
cold TD or debris disks (with excesses only at 70\,$\mu$m and beyond) around low-mass stars
would typically remain undetected, so potential excesses around otherwise diskless objects
must be handled with care, as they can easily be the result of contamination by cloud emission
or background, unrelated objects.

Some of the disks are labelled as ``marginal detections". These are cases where we observe 
spatially variable, patchy, sky background in the proximities of the source and with similar
bightness than we would expect from source itself. This could be a sign of mismatch or contamination,
even if visual inspection reveals the presence of
a point-like source. For them, we consider that the flux of the object, even though more
uncertain than in clear detections as it could be contaminated, 
is probably close to the observed ``marginal detection".
This distinguishes them from plain upper limits, where there is no hint of point source emission at the
location of the source and thus the flux at the corresponding wavelength could in principle be much lower than the
actual upper limits. For all other analysis throughout the paper, marginal detections are considered as upper limits.

\begin{landscape}
\begin{longtable}{lccccccccl}
\caption{\label{photometry-table} PACS photometry of previously known CepOB2 members identified
by optical spectroscopy. The disk type is labeled as N (no disk), F (full disk),
TD (transition disk), PTD (pre-transitional disk), D (depleted disk), Db (debris disk),
 - (not enough information). Spectral types, extinction and H$\alpha$ EW from Sicilia-Aguilar et al. (2005, 2013).
Information on massive stars from Contreras et al.(2002).
Marginal detections (detected objects for which visual inspection suggests there could be some problems)
are marked as $\leq$ to distinguish from non-detections. 
The comments include 
"n" (nebular contamination in the region, although unless stated, the object is detected), "e" (near edge, 
the data are fine but the S/N is poorer than at the map center). The "Accretion" column lists either
the value of the accretion derived from U band photometry (Sicilia-Aguilar et al. 2005, 2010), or the presence (yes [Y], no [N],
uncertain [U], not measured [-]) of accretion as revealed by optical spectroscopy (Sicilia-Aguilar et al. 2005, 2006b). Only
high-resolution spectroscopy produces conclusive values in case of border-line objects with weak H$\alpha$ EW,
which are otherwise labelled as `uncertain'. 
The detectability limit for objects considered as non-accreting is 10$^{-11}$M$_\odot$/yr. Note that all objects
with accretion upper limits are confirmed to be accreting (via H$\alpha$ spectroscopy).
$^1$The 160\,$\mu$m point may be a background object. } \\
\hline\hline
ID  & Sp.T. & A$_V$  & H$\alpha$ EW & 2MASS ID  & F$_{70\mu\,m}$  &  F$_{160\mu\,m}$   &  Disk  & Accretion & Comments\\
    &       &  (mag) &  (\AA)        &           &  (Jy)           &   (Jy)             &  & (10$^{-8}$M$_\odot$/yr) & \\
\hline
\endfirsthead
\caption{Continued.}\\
\hline\hline
ID  & Sp.T. & A$_V$  & H$\alpha$ EW & 2MASS ID  & F$_{70\mu\,m}$  &  F$_{160\mu\,m}$   &  Disk  & Accretion & Comments\\
    &       &  (mag) &  (\AA)        &           &  (Jy)           &   (Jy)             &  & (10$^{-8}$M$_\odot$/yr) & \\
\hline
\endhead
\hline
01-580 & K4.5 & 1.6 & -96/--- &  21533707+6228469 & 0.010$\pm$0.002 & $<$0.005 & F  & 4$^{+3}_{-2}$ &  \\
\\
73-758 & K6.5 & 1.9 & -9/--- & 21350835+5736028 & $\leq$0.008$\pm$0.002 & $<$0.012 & TD & N:/U & \\        
72-1427 & M1.0 & 1.4 & -77/-16 & 21351627+5728222 & 0.027$\pm$0.004 & $\leq$0.067$\pm$0.017 & TD & 0.15$^{+0.28}_{-0.07}$ & n\\         
73-472 & K5.0 & 1.7 & -10/-23 & 21351861+5734092 & 0.084$\pm$0.012 & 0.078$\pm$0.016 & TD & Y &  \\         
MVA-1312 & B4 & 1.5 & --- & 21355654+5720529 & $\leq$0.044$\pm$0.008 & $<$0.4 & TD/Db & N & n(160\,$\mu$m) \\       
213639147+572953326 & G9.0 & 5.3 & -3/--- & 21363915+5729533 & $\leq$0.362$\pm$0.056 & $<$0.6 & F & Y: & near IC1396A \\      
213642470+572523186 & M3.5 & 1.4 & -27/--- & 21364247+5725231 & $\leq$0.037$\pm$0.007 & $<$0.03 & F & Y: & n \\       
213655201+573030103 & - & 1.6 & -45/--- & 21365520+5730301 & $\leq$0.178$\pm$0.036 & $<$8 & Class I & Y & \\        
213655283+572551668 & K2.5 & 2.6 & 1/--- & 21365527+5725516 & 0.028$\pm$0.004 & 0.031$\pm$0.009: & TD & N:/U &  \\         
11-2146 & K6.0 & 2.6 & -33/-28 & 21365767+5727331 & 0.064$\pm$0.010 & 0.075$\pm$0.019 & F & 0.99$^{+0.67}_{-0.37}$ & \\         
213701319+573418289 & M1/M3 & 1.4 & -2/-22/--- & 21365947+5731349 & $\leq$0.007$\pm$0.002 & --- & TD & U & n\\        
11-2322 & M1.0 & 0.9 & -23/-18 & 21370191+5728222 & 0.010$\pm$0.002 & --- & F & 1.6$^{+2.9}_{-0.8}$ & \\         
11-2037 & K4.5 & 1.6 & -50/-43 & 21370703+5727007 & 0.019$\pm$0.003 & $\leq$0.031$\pm$0.009 & F & $<$9 & \\         
213708137+573616213 & M4.0 & 1.7 & -45/--- & 21370814+5736162 & $\leq$0.020$\pm$0.004 & $<$0.025 & F & Y & e \\        
21370936.5729483 & M0.5 & 1.7 & -21/--- & 21370936+5729483 & 0.036$\pm$0.006 & --- & F & Y & \\        
14-125 & K5.0 & 1.7 & -13/-14 & 21371054+5731124 & $\leq$0.023$\pm$0.004 & --- & F &  0.34$^{+0.21}_{-0.12}$ & near IC1396A \\      
11-2131 & K6.5 & 2.3 & -10/--- & 21371215+5727262 & 0.019$\pm$0.003 & $\leq$0.044$\pm$0.010 & F & 0.25$^{+0.12}_{-0.12}$ & \\         
11-2397 & K7.0 & 1.2 & -81/--- & 21371450+5728409 & 0.026$\pm$0.004 & 0.030$\pm$0.009 & F & Y & \\         
11-2031 & K2 & 1.7 & -5/-5 & 21371591+5726591 & 0.017$\pm$0.003 & $\leq$0.027$\pm$0.011 & F & $<$1 &  \\      
213716349+572640200 & K7.0 & 6.0 & -17/--- & 21371635+5726402 & 0.075$\pm$0.011 & $\leq$0.065$\pm$0.014 & F & Y &  \\        
213724102+572411541 & K5.5 & 2.4 & -107/--- & 21372410+5724115 & 0.015$\pm$0.003 & $<$0.023 & F/PTD & Y &  \\        
21372447+5731359 & M1.0 & 3.1 & -62/--- & 21372447+5731359 & $\leq$0.009$\pm$0.002 & $<$0.08 & F & Y &   \\        
14-160 & K5.0 & 1.8 & -22/--- & 21372732+5731295 & 0.039$\pm$0.006 & $\leq$0.080$\pm$0.017 & F & Y & n(160\,$\mu$m)  \\    
14-1017 & M0 & 2.1 & -55/--- & 21372894+5736042 & 0.015$\pm$0.002 & $\leq$0.052$\pm$0.013 & F &  $<$1.5 & n(160\,$\mu$m)  \\        
14-335 & K6.5 & 1.5 & -20/--- & 21372915+5732534 & 0.020$\pm$0.003 & $<$0.012 & F & $<$0.14 &  \\         
213732341+572503204 & G9.0 & 1.8 & -47/--- & 21373234+5725032 & 0.015$\pm$0.003 & $<$0.009 & F & Y & \\         
213734113+573431198 & M0.0 & 2.1 & -56/--- & 21373411+5734311 & 0.011$\pm$0.002 & $<$0.016 & F & Y & \\         
213735713+573258349 & M2.0 & 1.5 & -62/--- & 21373571+5732583 & 0.008$\pm$0.001 & $<$0.015$\pm$0.004 & TD & Y & \\         
14-183 & K7/K5 & 2.0 & -65/-14 & 21373849+5731408 & 0.064$\pm$0.009 & 0.052$\pm$0.012 & PTD & 0.16$^{+0.12}_{-0.07}$ & \\         
213738830+572936901 & K3.5 & 5.1 & -7/--- & 21373883+5729369 & 0.050$\pm$0.007 & 0.035$\pm$0.007 & F & Y: & \\         
213740471+573433203 & M2.5 & 2.0 & -7/--- & 21374047+5734332 & 0.022$\pm$0.003 & $<$0.015 & TD & U & n \\        
CCDM J2137+5734w/e & B3/B5 & 1.6 & --- & 21374093+5733372 & 0.012$\pm$0.002 & $<$0.018 & TD/Db & N &  \\         
21374275.5733250 & F9.0 & 1.7 & -54/-32 & 21374275+5733250 & 0.042$\pm$0.006 & $<$0.017 & F & Y & \\         
213747963+573242323 & M1.0 & 1.4 & -61/--- & 21374795+5732423 & 0.026$\pm$0.004 & $\leq$0.036$\pm$0.008 & PTD & Y & n(160\,$\mu$m) \\        
12-1091 & G2.5 & 2.8 & -17/-2 & 21375762+5722476 & 0.017$\pm$0.004 & $\leq$0.058$\pm$0.009 & F &  $<$0.4 & n(160\,$\mu$m) \\     
13-1238 & M1.0 & 2.6 & -64/-31 & 21375926+5736162 & 0.024$\pm$0.004 & $<$0.018 & F & 0.6$^{+1.1}_{-0.3}$ & \\         
82-272 & G9.0 & 3.6 & -13/-15 & 21380350+5741349 & 0.044$\pm$0.007 & $\leq$0.058$\pm$0.013 & F & $<$0.9 & \\         
13-1426 & M0 & 3.2 & -109/-40 & 21380856+5737076 & 0.008$\pm$0.001 & $<$0.0012 & F &  0.6$^{+0.8}_{-0.3}$ & \\         
21380924+5720198 & -- & 1.7 & ---/-17 & 21380924+5720198 & 0.013$\pm$0.003 & 0.022$\pm$0.005 & F/D & Y & \\       
13-350 & M1 & 0.7 & -9/--- & 21381384+5731414 & $\leq$0.004$\pm$0.001 & $<$0.012 & TD & N &  \\        
MVA-426 & B7 & 1.4 & --- & 21380845+5726476 & 0.480$\pm$0.070 & 0.194$\pm$0.039 & F &  Y & \\       
213812641+572033696 & K7.0 & 2.3 & -159/--- & 21381264+5720336 & 0.010$\pm$0.003 & $<$0.009 & PTD & Y &  \\        
13-1877 & K7.0 & 1.9 & -68/-33 & 21381703+5739265 & 0.027$\pm$0.004 & 0.023$\pm$0.006 & F &  1.6$^{+1.2}_{-0.8}$ & \\         
13-277 & G8 & 2.0 & -10/-14 & 21381731+5731220 & 1.187$\pm$0.172 & 0.983$\pm$0.194 & F &  14$^{+6}_{-4}$ & \\         
213822810+574017294 & M0.0 & 2.5 & -18/--- & 21382281+5740172 & 0.012$\pm$0.002 & 0.023$\pm$0.008: & F & Y: & n(160\,$\mu$m)  \\        
213825831+574207487 & K2.0 & 4.6 & -2/--- & 21382583+5742074 & 0.021$\pm$0.003 & 0.014$\pm$0.005: & TD & U & \\         
13-236 & K2 & 1.8 & -47/-56 & 21382742+5731081 & 0.010$\pm$0.002 & $<$0.008 & F & 0.51$^{+0.26}_{-0.15}$ &  \\         
12-2113 & K6.0 & 2.5 & -28/-15 & 21382743+5727207 & 0.044$\pm$0.006 & 0.057$\pm$0.012 & F & 1.4$\pm$0.2 &   \\        
13-157 & K5.5 & 1.2 & -14/-20 & 21382804+5730464 & 0.007$\pm$0.001 & $<$0.007 & F &  4.2$^{+2.9}_{-1.5}$ &\\         
213830308+573255218 & M1.0 & 1.7 & -31/--- & 21383031+5732552 & 0.013$\pm$0.002 & $<$0.013 & TD & Y &  \\         
91-155 & M2.5 & 1.2 & -8/--- & 21383470+5741274 & 0.024$\pm$0.004 & $\leq$0.016$\pm$0.006 & F & U & nearby star  \\        
21384350.5727270 & M2.0 & 1.7 & -23/--- & 21384350+5727270 & $\leq$0.004$\pm$0.001 & $<$0.012 & TD & N:/U & n \\        
54-1547 & K5.5 & 1.1 & -34/-33 & 21384446+5718091 & 0.031$\pm$0.006 & 0.068$\pm$0.011 & PTD & 0.18$^{+0.13}_{-0.06}$ & \\         
91-506 & K6.5 & 1.4 & -47/-31 & 21385807+5743343 & 0.055$\pm$0.008 & 0.080$\pm$0.016 & PTD &  0.09$^{+0.07}_{-0.04}$ & \\         
13-1048 & M0.0 & 1.7 & -7/-8 & 21391088+5735181 & 0.015$\pm$0.002 & $<$0.016 & F & $<$0.3 &   \\       
13-1250 & K4.5 & 1.4 & -2/-4 & 21391213+5736164 & 0.016$\pm$0.002 & $<$0.019 & PTD & 0.33$^{+0.11}_{-0.10}$ &  \\         
21392541+5733202 & -- & 1.7 & --- & 21392541+5733202 & 0.011$\pm$0.002 & $<$0.06 & PTD & Y &  \\       
21393104+5747140 & -- & 1.7 & ---/-16 & 21393104+5747140 & 0.053$\pm$0.011 & 0.034$\pm$0.006 & F & Y & \\         
213933219+573300533 & M3.5 & 2.0 & -35/--- & 21393321+5733005 & $\leq$0.009$\pm$0.002 & $<$0.017 & TD & Y & n \\        
24-515 & M0.5 & 1.1 & -11/--- & 21393407+5733316 & $\leq$0.009$\pm$0.002 & $<$0.03 & TD &  0.10$^{+0.18}_{-0.05}$ & \\         
21-998 & K5.5 & 1.9 & -16/--- & 21393480+5723277 & 0.024$\pm$0.005 & $\leq$0.017$\pm$0.004 & F &  0.33$^{+0.19}_{-0.12}$ & \\         
92-393 & M2.0 & 1.9 & -34/-21 & 21394408+5742159 & 0.023$\pm$0.005 &  $<$0.008 & TD & Y & \\         
21-1536 & M0.0 & 1.8 & -26/--- & 21394570+5726242 & 0.016$\pm$0.003 &  $\leq$0.023$\pm$0.007 & PTD & 0.18$^{+0.29}_{-0.08}$ &  n(160\,$\mu$m) \\        
213945860+573051704 & K6.0 & 2.0 & -227/--- & 21394586+5730517 & 0.013$\pm$0.002 &  $\leq$0.016$\pm$0.005 & F & Y &  n(160\,$\mu$m) \\        
213954058+572933454 & K7.5 & 2.0 & -30/--- & 21395406+5729334 & 0.016$\pm$0.002 & $<$0.012 & PTD & Y &  \\         
213955699+571638273 & M0.5 & 1.4 & -31/--- & 21395569+5716382 &  $\leq$0.023$\pm$0.006 &  $\leq$0.031$\pm$0.008 & TD & Y &  \\         
21395813+5728335 & -- & 1.7 & ---/-15 & 21395813+5728335 & 0.017$\pm$0.003 &  $\leq$0.026$\pm$0.006:$^1$ & F & Y &  \\        
21400451.5728363 & K5.0 & 1.7 & -32/--- & 21400451+5728363 & 0.026$\pm$0.004 & 0.026$\pm$0.006 & F & 0.71$^{+0.36}_{-0.23}$ & \\         
93-720 & -- & 1.7 & ---/-16 & 21400999+5800036 & 0.027$\pm$0.006 & $<$0.033/0.048$\pm$0.008 & F & Y & \\         
24-1796 & K7.0 & 1.2 & -124/-73 & 21401182+5740121 & 0.027$\pm$0.005 & 0.029$\pm$0.006 & TD & 0.6$^{+0.4}_{-0.3}$ & n(160\,$\mu$m) \\        
21-2006 & K5.0 & 1.5 & -2/--- & 21401390+5728481 & $\leq$0.005$\pm$0.001 & $<$0.009 & F & 0.12$^{+0.09}_{-0.05}$ & \\         
21401438+5740507 & -- & 1.7 & ---/-46 & 21401438+5740507 & 0.012$\pm$0.002 & $<$0.05 & F & Y &  n \\       
214025369+573416284 & M2.0 & 1.4 & -72/--- & 21402536+5734162 & $\leq$0.007$\pm$0.002 & $<$0.05 & PTD & Y &  n \\       
23-405 & K5.0 & 1.1 & -14/-9 & 21403134+5733417 & 0.026$\pm$0.004 & 0.033$\pm$0.008 & F & 0.22$^{+0.14}_{-0.08}$ & \\         
214035490+572831395 & K3.0 & 1.7 & -2/--- & 21403549+5728314 & 0.016$\pm$0.003 & $<$0.012 & TD & U/Y: &   \\        
23-162 & K7.0 & 1.7 & -6/--- & 21404450+5731314 & 0.009$\pm$0.002 & $<$0.009 & F & Y & \\         
23-969 & K5.5 & 1.1 & -36/-26 & 21411497+5738149 & 0.021$\pm$0.005 & $\leq$0.021$\pm$0.006 & F & Y &  \\         
\hline
\end{longtable}
\end{landscape}

\clearpage

\begin{table*}
\centering
\caption{PACS photometry of CepOB2 members identified
by Spitzer/X-ray/H$\alpha$ photometry. All spectral types (except the one
from HD\,206267) are derived from SED fitting and thus subject to large uncertainties, 
assuming the extinction will be
within the usual cluster values A$_V$=1-3 mag.
 References: M09 (Mercer et al. 2009); MC09 (Morales-Calder\'{o}n et al. 2009);
 B11 (Barentsen et al. 2011); G12 (Getman et al. 2012). The disk type is labeled as N (no disk), F (full disk),
TD (transition disk), PTD (pre-transitional disk), D (depleted disk), - (not enough information).
The comments include "m" (marginal detection, marked as $\leq$ to distinguish from non-detection),
"n" (nebular contamination in the region resulting in very high upper limits), "e" (near edge, the data are 
fine but the S/N is poorer than at the map center). $^1$Given the presence of cloud and the absence of
mid-IR excess, the PACS excess is most likely nebular and the objects do not have evidence of disks.} 
\label{other-table}
\begin{tabular}{l c c c c c l}
\hline\hline
2MASS ID  &  Sp.T. &  F$_{70\mu\,m}$ (Jy)  &  F$_{160\mu\,m}$ (Jy) & Ref.  & Disk & Comments\\
\hline
21351021+5731475 & -- & $\leq$0.010$\pm$0.002 & $<$0.027 & B11 & - & No IRAC data\\
21351687+5732422 & late K/M & 0.009$\pm$0.002 & $<$0.045 & B11 & - & \\ 
21352723+5731301 & K/M & 0.016$\pm$0.003 & $<$0.036 & B11 &- & \\  
21353135+5731279 & early M & $\leq$0.010$\pm$0.002 & $<$0.015 & B11 & TD & \\ 
21354586+5736401 & late K/M & $\leq$0.009$\pm$0.002 & $<$0.015 & B11 & F & \\ 
21363802+5726579 & late M & $\leq$0.011$\pm$0.003 & $<$0.052 & MC09 & F & High extinction\\
21364964+5722270 & F/G & 0.033$\pm$0.007 & 0.039$\pm$0.010 & G12 & TD & \\ 
21370909+5725485 & late K/M & 0.023$\pm$0.004 & 0.054$\pm$0.012 & B11 & TD & n \\
21371389+5727270 & late K & 0.040$\pm$0.006 & 0.037$\pm$0.009: & B11 & F & \\ 
21371420+5736177 & K/M & 0.013$\pm$0.002 & $<$0.048 & G12 & F & \\ 
21371545+5727170 & K/M & 0.025$\pm$0.004 & 0.049$\pm$0.012 & G12 & F/PTD & n \\
21372152+5726123 & late K/M & 0.015$\pm$0.003 & $<$0.047 & G12 & F & \\ 
21372475+5729089 & early M & $\leq$0.007$\pm$0.002 & $<$0.022 & B11 & F & n \\
21373786+5728467 & mid K & 0.009$\pm$0.001:$^1$ & $<$0.019 & G12 & N:$^1$ & n\\ 
21373885+5732494 & late K & 0.008$\pm$0.002:$^1$ & $<$0.021 & G21 & N:$^1$ & n\\ 
21374292+5736314 & G/F & 0.405$\pm$0.059 & 0.338$\pm$0.067 & G12 & TD & \\ 
21374612+5734280 & early/mid M & $\leq$0.005$\pm$0.001 & $<$0.012 & B11 & F & \\ 
21385760+5729205 & O6.5 & 0.013$\pm$0.002 & $<$0.09 & M09 & N & HD\,206267 \\
21385963+5730080 & late K & 0.006$\pm$0.001: & $<$0.018 & M09 & F & \\ 
21390321+5730420 & late K/M & 0.004$\pm$0.001: & $<$0.12 & B11 & F & No optical counterpart\\
\hline
\end{tabular}
\end{table*}

\begin{landscape}
\begin{longtable}{lccccccccl}
\caption{\label{uplims-table} PACS upper limits for previously known CepOB2 members with disks.
The comments include
"n" (nebular contamination in the region resulting in very high upper limits), "e" (near edge, the data are 
fine but the S/N is poorer than at the map center).
Spectral types, extinction and H$\alpha$ EW from Sicilia-Aguilar et al. (2005, 2013).
Many of the objects in the IC1396A globule appear
without photometric data in the red since the upper limits that can be constrained are of a few Jy and thus not
informative. The disk type is labeled as N (no disk), F (full disk),
TD (transition disk), TD8 (transition disk with an excess at 8$\mu$m only, usually too faint to
be detected at 24$\mu$m), PTD (pre-transitional disk), D (depleted disk), U (uncertain, usually due to photometric
errors/contamination), - (not enough information). The "Accretion" column lists either
the value of the accretion derived from U band photometry (Sicilia-Aguilar et al. 2005, 2010), or the presence (yes [Y], no [N],
uncertain [U], not measured [-], added [:] indicates a border-line case) of accretion as revealed by optical 
spectroscopy (Sicilia-Aguilar et al. 2005, 2006b). Only
high-resolution spectroscopy produces conclusive values in case of border-line objects with weak H$\alpha$ EW. 
The detectability limit for objects considered as non-accreting is 10$^{-11}$M$_\odot$/yr. Note that all objects
with accretion upper limits are confirmed to be accreting (via H$\alpha$ spectroscopy). $^1$The object is clearly seen in the blue although due to
the surrounding nebular emission, the photometry is very uncertain. $^2$There is a clear point-source detection,
but given the small offset with respect to the optical counterparts and the lack of Spitzer excess,
it may correspond to a background source. } \\
\hline\hline
ID  & Sp.T. & A$_V$  & H$\alpha$ EW & 2MASS ID  & F$_{70\mu\,m}$  &  F$_{160\mu\,m}$   &  Disk  & Accretion & Comments\\
    &       &  (mag) &  (\AA)        &           &  (Jy)           &   (Jy)             &  & (10$^{-8}$M$_\odot$/yr) & \\
\hline
\endfirsthead
\caption{Continued.}\\
\hline\hline
ID  & Sp.T. & A$_V$  & H$\alpha$ EW & 2MASS ID  & F$_{70\mu\,m}$  &  F$_{160\mu\,m}$   &  Disk  & Accretion & Comments\\
    &       &  (mag) &  (\AA)        &           &  (Jy)           &   (Jy)             &  & (10$^{-8}$M$_\odot$/yr) & \\
\hline
\endhead
\hline
01-1152 & M1.5 & 0.9 & -5.5/---& 21531982+6234004 & $<$0.019 & $<$0.021 & TD/D & U &  \\  
DG-481 & A7.0 & 1.5 & --- & 21522113+6245034 & $<$0.003 & $<$0.010 & TD/Db & N & \\ 
\\                 
73-311 & M1.5 & 1.1 & -29/--- & 21352451+5733011 & $<$0.003 & $<$0.030 & D & Y & n(160\,$\mu$m) \\
73-71 & K6.0 & 2.1 & -12/-10 & 21353021+5731164 & $<$0.003 & $<$0.052 & D & Y & c\\ 
213542993+573337049 & M4.0 & 1.1 & -16/--- & 21354299+5733370 & $<$0.006 & $<$0.124 & D & U &  n \\
72-875 & M0.5 & 2.0 & -21/--- & 21354975+5724041 & $<$0.048$\pm$0.007: & $<$0.09 & TD: & 0.11$^{+0.23}_{-0.06}$ & n, extended \\        
73-194 & K6.5 & 1.5& -4/--- & 21355223+5732145 & $<$0.006 & $<$0.25 & N & N &  n\\ 
73-537 & G1.5 & 3.3 & -5/--- & 21360723+5734324 & $<$0.014 & $<$0.030 & N & N & \\ 
21360745.5734296 & -- & 1.7 & --- & 21360745+5734296 & $<$0.010 & $<$0.4 & F & -- &  n \\        
21362368+5732452 & --- & 1.6 & ---/--- & 21362368+5732452 & $<$0.12 & $<$0.68 & Unc. & U/N &  n \\
21362507.5727502 & M0.0 & 1.7 & -86/-78 & 21362507+5727502 & $<$0.007 & $<$0.05 & F & 0.09$^{+0.16}_{-0.05}$ & n, 160\,$\mu$m extended \\   
14-306 & K6.5 & 1.850 & -8/--- & 21362676+5732374 & $<$0.12 & $<$0.68 & N & N & n \\ 
213626897+573304351 & M0.0 & 3.6 & -21/--- & 21362690+5733043 & $<$0.12 & $<$0.68 & - & Y: &  n \\
213636909+573132683 & K7.0 & 3.3 & -20/--- & 21363691+5731326 & $<$0.24 & $<$4.6 & F & Y & n, truncated disk?\\
21364596.5729339 & --- & 1.6 & -47/--- & 21364596+5729339 & $<$3.7 & $<$3.7 & ClassI & Y & n \\
21364762.5729540 & K6.0 & 1.6 & -76/--- & 21364762+5729540 & $<$3.7 & $<$3.671 & F & Y &  n \\
11-1209 & K6 & 0.6 & -4/-6 & 21365850+5723257 & $<$0.006 & $<$0.012 & D & 0.56$^{+0.35}_{-0.23}$  &   \\
21365947.5731349 & M1/0.0 & 1.6/2.0 & -109/-30 & 21365947+5731349 & $<$1.3 & --- & F & Y & n$^1$ \\
11-1871 & M2.0 & 0.8 & -13/--- & 21370254+5726144 & $<$0.003 & $<$0.015 &  TD/D:/N & N: & 8$\mu$m excess only \\
14-287 & M0.0 & 2.2 & -35/-18 & 21370649+5732316 & $<$0.018 & --- & F & 0.21$^{+0.34}_{-0.11}$ & n \\
213708879+572107012 & M2.0 & 1.4 & -55/--- & 21370888+5721070 & $<$0.021 & $<$0.12 & TD/D: & Y & e\\ 
14-197 & K5.5 & 1.6 & -2/--- & 21372368+5731538 & 0.006$\pm$0.001$^2$ & 0.030$\pm$0.008$^2$ & N & N &  \\       
213726148+572330562 & M4.5 & 1.3 & -28/--- & 21372615+5723305 & $<$0.018 & $<$0.076 & TD & Y: & e\\ 
21372643+5731386 & K2.0 & 2.4 & -2/--- & 21372643+5731386 & $<$0.039 & $<$0.077 & TD:/N & U &  8$\mu$m excess only, nearby object\\
213727418+573123620 & M2.5 & 0.9 & -11/--- & 21372742+5731236 & $<$0.015 & $<$0.080 & F & U & \\ 
213733557+573550931 & M1.5 & 1.7 & -6/--- & 21373355+5735509 & $<$0.003 & $<$0.009 & D & N: & \\ 
11-1864 & G-K & 1.6 & -5/--- & 21373420+5726154 & $<$0.003 & $<$0.015 & TD:/N & N & \\
14-2148 & M1.5 & 1.5 & -2/--- & 21374184+5740400 & $<$0.006 & $<$0.030 & TD:/D:/N: &  N & 8$\mu$m excess only\\
213742167+573431486 & M2.0 & 1.4 & -60/-- & 21374216+5734314 & $<$0.015 &  $<$0.017 & TD  & n \\
213744131+573331130 & K7.5/M1 & 2.7/1.9 & -17/-19/--- & 21374412+5733311 & $<$0.043 & $<$0.016 & F & Y & \\        
213744543+572200213 & M5.0 & 1.1 & -51/--- & 21374453+5722002 & $<$0.015 & $<$0.13 & U & Y: & n, Spitzer data contaminated\\ 
213746871+573156208 & M1.5 & 2.9 & -42/--- & 21374687+5731562 & $<$0.006 & $<$0.021 & PTD & Y & m \\
213748237+572319411 & M3.5 & 0.7 & -8/-42 & 21374824+5723194 & $<$0.006 & $<$0.055 & TD & Y &  m \\
213748931+572320963 & K5.0 & 2.7 & -20/--- & 21374893+5723209 & $<$0.010 & $<$0.07 & F & Y & \\         
13-924 & K5.0 & 1.6 & -4/--- & 21375018+5733404 & $<$0.003 & $<$0.012 & N & N &  e \\      
12-2519 & K5.5 & 1.6 & -8/--- & 21375107+5727502 & $<$0.006 & $<$0.14 & D & 0.07$^{+0.03}_{-0.03}$ & 160\,$\mu$m extended \\  
213751210+572436151 & K7.5 & 3.3 & -15/--- & 21375121+5724361 & $<$0.003 & $<$0.018 & F/D & Y: & \\ 
12-1968 & K6.0 & 1.2 & -11/-8 & 21375487+5726424 & $<$0.003 & $<$0.008 & F & 0.34$^{+0.15}_{-0.13}$ & \\         
213756779+573448171 & M1.5 & 1.7 & -9/--- & 21375677+5734481 & $<$0.003 & $<$0.012 & TD & U & \\ 
13-669 & K1 & 2.183 & -22/-18 & 21380928+5733262 & $<$0.003 & $<$0.024 & F & 0.58$^{+0.20}_{-0.23}$ & n, Herschel flux too low\\
21380979+5729428 & --- & 1.6 & ---/--- & 21380979+5729428 & $<$0.005 & $<$0.009 & TD/D & N: & \\ 
213809997+572352782 & M2.0 & 2.2 & -50/--- & 21381000+5723527 & $<$0.005 & $<$0.057 & F & Y & \\ 
213810182+572708407 & M1.5 & 1.5 & -10/--- & 21381018+5727084 & $<$0.003 & $<$0.19 & PTD & U & Herschel flux too low\\ 
213812023+572500774 & M1.0 & 1.4 & -3/--- & 21381202+5725007 & $<$0.004 & $<$0.012 & F & N: & \\ 
54-1781 & M1.0 & 1.2 & -13/--- & 21381612+5719357 & $<$0.003 & $<$0.018 & TD & 0.3$^{+0.5}_{-0.1}$ & \\ 
213819411+572203907 & M2.0 & 2.7 & -14/--- & 21381941+5722039 & $<$0.012 & $<$0.018  & TD8 & U &  8$\mu$m excess only\\
213823950+572736175 & M1.0 & 1.4 & -18/--- & 21382395+5727361 & $<$0.004 & $<$0.029 & F & Y: & \\ 
213828028+574736432 & M2.5 & 1.0 & -4/--- & 21382803+5747364 & out & $<$0.042 & D & N: &  e \\
13-232 & M0.0 & 1.1 & -3/--- & 21382834+5731072 & $<$0.003 & $<$0.006 & F: & N & \\ 
213829367+573726567 & M2.0 & 2.0 & -10/--- & 21382936+5737265 & $<$0.003 & $<$0.031 & TD & U & lower than expected\\ 
213830349+572618227 & K7.0 & 2.1 & -13/--- & 21383035+5726182 & $<$0.003 & $<$0.006 & F & Y: & m\\ 
21383216.5726359 & M0.0 & 1.6 & -129/-43 & 21383216+5726359 & $<$0.003 & $<$0.009 & F & Y & m\\ 
13-52 & K7.0 & 1.3 & -1/--- & 21383255+5730161 & $<$0.003 & $<$0.009 & TD & N &  \\ 
13-566 & K5.5 & 2.4 & -5/--- & 21383481+5732500 & $<$0.003 & $<$0.038 &  TD8 & N & strong 8$\mu$m excess only\\
213837298+573103776 & K7.0 & 2.1 & -4/--- & 21383730+5731037 & $<$0.003 & $<$0.009 & TD & N: & \\ 
213839571+572916412 & M1.5 & 4.7 & -22/--- & 21383956+5729164 & $<$0.003 & $<$0.063 & TD/D & Y: &  \\ 
213839749+572753080 & M2.5 & 1.6 & -5/---& 21383956+5729164 & $<$0.003 &  $<$0.029 & D  & N: &  n \\
13-1891 & M0 & 1.0 & -11/--- & 21384001+5739303 & $<$0.003 & $<$0.018 & F & Y: & \\ 
213842249+573533902 & M4.0 & 0.8 & -1/--- & 21384224+5735339 & $<$0.004 & $<$0.021 & F & N: & m\\ 
54-1613 & K5.0 & 1.2 & -1/--- & 21384332+5718359 & $<$0.006 & $<$0.064 & TD & N &  \\ 
213844343+573626211 & M1.5 & 1.0 & -34/--- & 21384434+5736262 & $<$0.012 & $<$0.061 & D &  Y: &  anomalous SED\\        
213847282+573114405 & M1.0 & 1.6 & -29/--- & 21384727+5731144 & $<$0.003 & $<$0.076 & F & Y & \\ 
213854760+572450268 & M1.0 & 1.1 & -5/--- & 21385476+5724502 & $<$0.006 & $<$0.019 & D & U &  \\ 
21385669.5730484 & K5.0 & 1.6 & -4/--- & 21385669+5730484 & $<$0.003 & $<$0.04 & F &  U & \\ 
213859738+572216816 & M2.5 & 2.2 & -10/--- & 21385973+5722168 & $<$0.012 & $<$0.067 & F & U & n\\ 
12-1027 & M0 & 1.3 & -11/--- & 21390319+5722318 & $<$0.006 & $<$0.014 & N & N: &  \\    
213903212+571555316 & M1.0 & 0.8 & -6/--- & 21390320+5715553 & $<$0.013 & $<$0.032 & TD & U & n\\ 
12-1617 & M1.0 & 1.6 & -30/-13 & 21390468+5725128 & $<$0.006 & $<$0.012 & PTD & 0.16$^{+0.38}_{-0.09}$ & m \\
213905519+572349596 & K7.0/6.0 & 1.2/1.521 & -1/--- & 21390552+5723496 & $<$0.053 & $<$1.0 & TD8 & N: & n, 8$\mu$m excess only\\
21-840 & M1.0 & 2.2 & -14/--- & 21391012+5722323 & $<$0.006 & $<$0.038 & F & 0.06$^{+0.08}_{-0.03}$ & m \\
213910429+572235727 & M0.0 & 3.2 & -103/--- & 21391042+5722357 & $<$0.006 & $<$0.044 & F & Y & \\  
213911452+572425205 & M1.0 & 2.2 & -26/--- & 21391145+5724252 & $<$0.10 & $<$1.0 & F & Y &  n \\
213914453+572304758 & M3.5 & 1.4 & -8/--- & 21391444+5723047 & $<$0.013 & $<$0.056 &  TD8 & U &  8$\mu$m excess only\\
213914837+573756779 & M0.0 & 3.3 & -13/--- & 21391483+5737567 & $<$0.003 & $<$0.015 & TD & U & \\ 
213914988+574050473 & M4.0 & 3.1 & -25/--- & 21391498+5740504 & $<$0.012 & $<$0.073 & TD & U & \\ 
54-1488 & K7.0/6.0 & 2.2 & -13/-33/--- & 21391748+5717474 & $<$0.021 & $<$0.069 & PTD & Y: & e\\ 
21392059+5726269 & - & 1.6 & -34/--- & 21392059+5726269 & $<$0.004 & $<$0.052 & TD & Y: & n\\ 
213924711+571912672 & M2.5 & 2.8 & -37/--- & 21392471+5719126 & $<$0.006 & $<$0.050 & F & Y: & \\ 
21392570+5729455 & --- & 1.670 & ---/--- & 21392570+5729455 & $<$0.009 & $<$0.050 & TD & N &   m,n\\ 
213926421+572538470 & M1.0 & 3.2 & -164/--- & 21392641+5725384 & $<$0.004 & $<$0.017 & F & Y & \\ 
213929250+572530299 & M1.0 & 3.7 & -14/--- & 21392925+5725303 & $<$0.006 & $<$0.056 & PTD & U &  \\ 
213930129+572651433 & M0.5 & 2.0 & -4/--- & 21393012+5726514 & $<$0.004 & $<$0.028 & D & N:/U & \\   
213930870+572227446 & K7.0 & 2.3 & -108/--- & 21393086+5722274 & $<$0.009 & $<$0.015 & F & Y & \\
21393301+5728175 & M1.5 & 2.31 & 3/--- & 21393301+5728175 & $<$0.003 & $<$0.020 & F/TD & U & lower than expected \\ 
21-33 & M0 & 1.7 & -107/-51 & 21393561+5718220 & $<$0.005 & $<$0.016 & D &  0.07$^{+0.11}_{-0.04}$ & n(160\,$\mu$m)  \\      
213937132+572459030 & M3.0 & 1.0 & -49/--- & 21393713+5724590 & $<$0.005 & $<$0.030 & F & Y & m \\       
21394010+5730113 & M2.0 & 1.6 & -24/--- & 21394010+5730113 & $<$0.003 & $<$0.012 & TD/D & Y: & \\ 
213942289+572902679 & M2.5 & 2.1 & -12/--- & 21394229+5729026 & $<$0.003 & $<$0.018 & D & U & \\ 
213942399+573431184 & M3.5 & 2.0 & -51/--- & 21394240+5734311 & $<$0.003 & $<$0.024 & D & Y: & \\ 
213943237+574139252 & M5.0 & 1.8 & -2/--- & 21394324+5741392 & $<$0.004 & $<$0.009 & F & N: & \\ 
213943450+573634560 & M0.0 & 1.6 & -5/--- & 21394345+5736345 & $<$0.008 & $<$0.018 & F & N: & m\\ 
213944898+573537212 & M4.0 & 0.8 & -4/--- & 21394489+5735372 & $<$0.003 & $<$0.012 & D & N: & \\ 
24-820 & K6.5 & 1.2 & -2/--- & 21394746+5735059 & $<$0.003 & $<$0.012 & N & N &  \\ 
213948050+572049544 & K7.0 & 1.7 & -1/--- & 21394805+5720495 & $<$0.036 & $<$0.054 & F & U &  nearby object\\
24-78 & M2.0 & 1.3 & -6/--- & 21394936+5730546 & $<$0.003 & $<$0.020 & N & N & \\ 
213949540+574330324 & M3.5 & 0.7 & -18/--- & 21394954+5743303 & $<$0.012 & $<$0.088 & F & Y: & e\\ 
92-1162 & M2.0 & 1.4 & -7/--- & 21394974+5746468 & $<$0.009 & $<$0.015 & D:/N & N & weak IRAC excess only\\
213949767+574733740 & M2.0 & 1.9 & -10/--- & 21394977+5747337 & $<$0.021 & $<$0.036 & TD & U & \\ 
213951092+574119848 & M3.0 & 1.1 & 1/--- & 21395109+5741198 & $<$0.003 & $<$0.012 & D & N: & \\ 
213952060+572506212 & M2.5 & 2.3 & -10/--- & 21395205+5725062 & $<$0.003 & $<$0.015 & D & U & \\ 
213954003+573519730 & M3.5 & 0.8 & -2/--- & 21395400+5735197 & $<$0.003 & $<$0.015 & TD & N: & n\\ 
21-1974 & G7.5 & 2.5 & -8/--- & 21395861+5728404 & $<$0.004 & $<$0.026 & D & Y: & \\ 
213959078+572643784 & M3.5 & 0.5 & -10/--- & 21395908+5726437 & $<$0.003 & $<$0.009 & F & U &  \\        
21-895 & K5.0 & 1.4 & -1/--- & 21400321+5722505 & $<$0.015 & $<$0.024 & TD & 0.21$^{+0.09}_{-0.07}$ & \\ 
24-1736 & M1 & 1.0 & -61/-33 & 21401134+5739518 & $<$0.006 & $<$0.018 & F & 0.3$^{+0.5}_{-0.1}$ & \\ 
KUN-196 & B9 & 2.2 & --- & 21401508+5740513 & $<$0.008 & $<$0.046 & TD/Db & N &  n \\
214019073+572509288 & M2.5 & 2.0 & -12/--- & 21401907+5725092 & $<$0.006 & $<$0.027 & F & U & \\ 
214019849+573922857 & M3.5 & 1.3 & -47/--- & 21401984+5739228 & $<$0.003 & $<$0.015 & F & Y & \\ 
22-2651 & M1.5 & 0.9 & -34/-48 & 21402130+5726579 & $<$0.004 & $<$0.015 & F &  0.05$^{+0.08}_{-0.02}$ & lower than expected\\ 
22-1418 & M1.5 & 0.7 & -2/--- & 21402287+5727329 & $<$0.003 & $<$0.012 & D & $<$0.3 & \\ 
214026262+573402317 & M4.5 & 1.1 & -4/--- & 21402625+5734023 & $<$0.005 & $<$0.05 & D & N: &  \\        
214029551+572453893 & M4.0 & 1.0 & -18/--- & 21402955+5724538 & $<$0.009 & $<$0.021 & TD & U & \\ 
214033472+573605378 & M1.0 & 2.2 & -16/--- & 21403347+5736053 & $<$0.005 & $<$0.031 & F & Y: & n\\ 
23-570 & K6.0 & 1.3 & -47/-18 & 21403574+5734550 & $<$0.006 & $<$0.053 & D & 0.7$^{+0.5}_{-0.3}$ &  \\ 
93-540 & M0.0 & 2.2 & -18/-5 & 21403586+5758130 & $<$0.021 & $<$0.077 & - & Y &  m\\ 
21403721.5729127 & K7.5/M2 & 1.7/1.3 & -41/-42/-21 & 21403721+5729127 & $<$0.003 & $<$0.012 & F & Y &  \\ 
214042281+573513316 & M4.0 & 1.4 & -20/--- & 21404228+5735133 & $<$0.006 & $<$0.018 & F & U & \\ 
214048489+573202607 & M4.5 & 1.0 & -4/--- & 21404848+5732026 & $<$0.004 & $<$0.012 & TD: & N: & 8$\mu$m excess only\\
23-798 & K6 & 2.2 & -150/-70 & 21412864+5736432 & $<$0.005 & $<$0.025 & PTD & Y & n, lower than expected\\     
\hline
\end{longtable}
\end{landscape}

\begin{table*}
\caption{PACS upper limits of CepOB2 members identified
by Spitzer/X-ray/H$\alpha$ photometry.  All spectral types
are derived from SED fitting and thus subject to large uncertainties, 
assuming the extinction will be
within the usual cluster values A$_V$=1-3 mag.References: M09 (Mercer et al. 2009); MC09 (Morales-Calder\'{o}n et al. 2009);
 B11 (Barentsen et al. 2011); G12 (Getman et al. 2012). The disk type is labeled as N (no disk), F (full disk),
TD (transition disk), PTD (pre-transitional disk), D (depleted disk), - (not enough information).
The comments include "m" (marginal detection, marked as $\leq$ to distinguish from non-detection),
"n" (nebular contamination in the region resulting in very high upper limits), "e" (near edge, the data are 
fine but the S/N is poorer than at the map center). $^a$A nearby source 21371737+5729207 has bright excess,
original coordinates correspond to a point between both objects. } 
\label{otheruplims-table} 
\begin{tabular}{l  c c c c c l}
\hline\hline
2MASS ID  & Sp.T & F$_{70\mu\,m}$ (Jy)  &  F$_{160\mu\,m}$ (Jy) & Ref.  & Disk & Comments\\
\hline
21353289+5731595 & early M & $<$0.005 & $<$0.078 & B11 & F/PTD: & \\
21364033+5725455 & late M & $<$0.006 & $<$0.067 & MC09 & F & high extinction\\
21364247+5725231 & mid M & $<$0.004 & $<$0.030 & MC09 & F & \\
21365616+5726395 & --- & $<$0.004 & $<$0.032 & G12 & Unc./N & \\ 
21371040+5729034 & --- & $<$0.005 & --- & G12 & Unc. & n\\ 
21371474+5728149 & early M & $<$0.005 & $<$0.017 & G12 & TD & \\
21371689+5729199$^a$ & --- & $<$0.006 & --- & MC09 & F: & two nearby objects\\
21371695+5732209 & --- & $<$0.004 & $<$0.024 & G12 & Unc. & \\ 
21371700+5726178 & K & $<$0.006 & $<$0.066 & G12 & F & \\ 
21371714+5728473 & late K & $<$0.005 & $<$0.031 & B11 & F/PTD: & \\
21371742+5729272 & late K & $<$0.007 & --- & MC09 & F & nearby object \\
21371996+5728462 & late K & $<$0.006 & $<$0.031 & G12 & TD:/N & \\
21372003+5729344 & K & $<$0.005 & $<$0.060 & G12 & TD/PTD: & \\ 
21372009+5731351 & M & $<$0.006 & $<$0.018 & G12 & TD:/N &\\  
21372181+5734122 & --- & $<$0.005 & $<$0.059 & G12 & Unc. & \\ 
21373287+5728270 & early M & $<$0.004 & $<$0.011 & G12 & TD/D: & \\
21373380+5732166 & mid/late M & $<$0.004 & $<$0.10 & G12 & N: & \\
21374288+5734461 & M: & $<$0.005 & $<$0.012 & B11 & F & \\ 
21374851+5729587 & early M & $<$0.005 & $<$0.013 & G12 & TD:/D & \\
21374944+5733518 & --- & $<$0.003 & $<$0.012 & G12 & F & \\ 
21375256+5725562 & early/mid M & $<$0.005 & $<$0.021 & G12 & TD & \\
21375487+5726424 & K & $<$0.004 & $<$0.015 & G12 & TD & \\ 
21383145+5729012 & late K/early M & $<$0.004 & $<$0.025 & M09 & TD & \\
21383743+5730206 & early M & $<$0.004 & $<$0.010 & M09 & F & \\
21384370+5731032 & late K/early M & $<$0.003 & $<$0.076 & M09 & TD:/N & m \\
21385093+5728324 & late K & $<$0.003 & $<$0.088 & M09 & D & \\
21385732+5730254 & early M & $<$0.005 & $<$0.059 & M09 & TD & \\
21385999+5730532 & K & $<$0.005 & $<$0.051 & M09 & F & \\ 
21390106+5730231 & K/M & $<$0.003 & $<$0.023 & M09 & TD & \\ 
21390624+5728106 & K/M & $<$0.003 & $<$0.021 & M09 & F & \\ 
21391732+5730085 & K & $<$0.004 & $<$0.012 & M09 & F/D & \\ 
21395187+5726583 & early M & $<$0.004 & $<$0.010 & B11 & TD/D & \\
21395611+5727079 & late K/early M & $<$0.004 & $<$0.011 & B11 & TD/D: & \\
21395767+5723433 & late K/early M & $<$0.006 & $<$0.013 & B11 & TD & \\
21404073+5732147 & late K/early M & $<$0.004 & $<$0.017 & B11 & F & \\
\hline
\end{tabular}
\end{table*}

\clearpage

\section{RADMC models for disks with various disk structures \label{models-app}}

The following tables list the properties of the disks models used to interpret the
$\alpha$ spectral indices. The disk models do not reproduce any of the CepOB2
members in particular. Instead, we explore the parameter space that samples the behaviours observed
in different cluster members. The models are thus an extension of the parameter space explored in
Sicilia-Aguilar et al. (2011) for a subsample of representative disks with various SED morphologies.
All models were constructed using the 2D radiative-transfer MonteCarlo
code RADMC (Dullemond \& Dominik 2004) for a standard star with T$_{eff}$=4275 K, R$_*$=1.7 R$_\odot$,
and M$_*$=1 M$_\odot$ with a 200 AU disk. The disks are axisymmetric (2D) but the photon packages
are followed in 3D.
The RADMC code assumes a disk with well-mixed dust and gas and a standard gas-to-dust ratio
of 100. The dust temperatures of all species are the same. The inner rim of the disk is either fixed
at the dust destruction temperature (1500 K) or moved to cooler temperatures to simulate
inner holes. The models can be
either vertically iterated to reproduce hydrostatic equilibrium (which washes out any predetermined
structure parameters such as vertical pressure scale height H vs radius or disk thickness,
requiring that gas and dust are well-mixed), or to preproduce cases where gas and dust are not well-coupled
or there are additional variations in the vetical disk structure. In this latter case, we modify the
thickness of the disk  (H/R$\propto$R$^q$) at the
outer disk radius ((H/R)$_{out}$) to reproduce 
thicker/thinner disks, and modify the exponent q to obtain various degrees of flaring or global settling.
For the dust content we assume standard amorphous silicate grains with a grain size 
distribution between a$_{min}$-a$_{max}$ with a power-law dependency n(a)$\propto$a$^{-p}$,
which can be taken as the standard collisional distribution (p=-3.5) or changed. We also include
25\% of amorphous carbon grains with the same grain distribution as a source of continuum opacity. The grain properties
are maintained through the whole disk unless when radial variations are included (such as adding
small grain populations in the inner disk to reproduce pre-transitional and transitional disks).

\begin{landscape}
\begin{longtable}{l c c c c c c l}
\caption{\label{model-table} Summary of the RADMC models. All models are constructed for M$_*$=1.0\,M$_\odot$, R$_*$=1.7\,R$_\odot$ 
(typical of the low-mass stars in Cep\,OB2) and outer disk radius R$_{out}$=200\,AU.
Models with well-mixed dust and gas in hydrostatic equilibrium are marked in the disk structure parameters with "v".
Otherwise, we give the vertical scale height at the outer radius [(H/R)$_{out}$] and
the power-law exponent of the disk surface (H/R$\propto$R$^q$). The ID names of the models
include some information about the disk mass (massive [M], intermediate-mass [I], low-mass [L]),
the vertical scale height (vertical hydrostatic equilibrium [V] for models with well-mixed gas and dust, 
normal flaring with q=1/7 [F], settled/flat
disks with q=0 [S], very flared disks with q=1/4 [XF], together with the value of (H/R)$_{out}$ if no
vertical iteration is performed), indication of inner holes/gaps (inner hole in transitional disk [TD] with innermost
disk temperature, radial temperature at which radial properties change in pre-transitional disk [PTD]), followed by
the grain distribution (with p=2.5 or p=3.5) and a label in case we consider grains up to 100 $\mu$m only (sm).} \\
\hline\hline
Model & T$_{in}$ & M$_{disk}$   & a$_{min}$-a$_{max}$  & p                       & q                   & (H/R)$_{out}$ & Summary of disk properties   \\
      & (K) &  (M$_\odot$)      &  ($\mu$m)            & (n(a)$\propto$a$^{-p}$) & (H/R$\propto$R$^q$) &   &    \\
\hline
\endfirsthead
\caption{Continued.}\\
\hline\hline
Model & T$_{in}$ & M$_{disk}$   & a$_{min}$-a$_{max}$  & p                       & q                   & (H/R)$_{out}$ & Summary of disk properties   \\
      & (K) &  (M$_\odot$)      &  ($\mu$m)            & (n(a)$\propto$a$^{-p}$) & (H/R$\propto$R$^q$) &   &    \\
\hline
\endhead
\hline
I\_V\_p3.5    & 1500 & 1e-3 & 0.1-10000 & 3.5 & v & v   & Intermediate-mass, Well-mixed gas and dust \\
I\_S0.2\_p3.5 & 1500 & 1e-3 & 0.1-10000 & 3.5 & 0 & 0.2 & Intermediate-mass, Settled, Thick \\
I\_S0.3\_p3.5 & 1500 & 1e-3 & 0.1-10000 & 3.5 & 0 & 0.3 & Intermediate-mass, Settled,  Very Thick \\
I\_S0.1\_p3.5 & 1500 & 1e-3 & 0.1-10000 & 3.5 & 0 & 0.1 & Intermediate-mass, Settled, Thin   \\
M\_S0.1\_p3.5 & 1500 & 1e-2 & 0.1-10000 & 3.5 & 0 & 0.1 & Massive, Settled, Thin  \\
M\_V\_p3.5    & 1500 & 1e-2 & 0.1-10000 & 3.5 & v & v   & Massive, Well-mixed gas and dust  \\
M\_F0.1\_p3.5 & 1500 & 1e-2 & 0.1-10000 & 3.5 & 1/7 & 0.1 & Massive, Flared, Thin \\
M\_F0.05\_p3.5 & 1500 &  1e-2 & 0.1-10000 & 3.5 & 1/7 & 0.05 & Massive, Flared, Very Thin  \\
M\_F0.2\_p3.5 & 1500 &  1e-2 & 0.1-10000 & 3.5 & 1/7 & 0.2 & Massive, Flared, Thick  \\
M\_F0.3\_p3.5 & 1500 &  1e-2 & 0.1-10000 & 3.5 & 1/7 & 0.3 & Massive, Flared, Very Thick  \\
I\_F0.05\_p3.5 & 1500 & 1e-3 & 0.1-10000 & 3.5 & 1/7 & 0.05   & Intermediate-mass, Flared, Very Thin \\
I\_F0.2\_p3.5 & 1500 & 1e-3 & 0.1-10000 & 3.5 & 1/7 & 0.2   & Intermediate-mass, Flared, Thick \\
I\_F0.3\_p3.5 & 1500 & 1e-3 & 0.1-10000 & 3.5 & 1/7 & 0.3   & Intermediate-mass, Flared, Very Thick \\
I\_F0.1\_p3.5 & 1500 & 1e-3 & 0.1-10000 & 3.5 & 1/7 & 0.1   & Intermediate-mass, Flared, Thin \\
L\_V\_p3.5 & 1500 & 1e-4 & 0.1-10000 & 3.5 & v & v   & Low-mass, Well-mixed gas and dust \\
L\_F0.05\_p3.5 & 1500 & 1e-4 & 0.1-10000 & 3.5 & 1/7 & 0.05   & Low-mass, Flared, Very Thin \\
L\_F0.1\_p3.5 & 1500 & 1e-4 & 0.1-10000 & 3.5 & 1/7 & 0.1   & Low-mass, Flared, Thin \\
L\_F0.1\_p3.5 & 1500 & 1e-4 & 0.1-10000 & 3.5 & 1/7 & 0.2   & Low-mass, Flared, Thick \\
L\_F0.3\_p3.5 & 1500 & 1e-4 & 0.1-10000 & 3.5 & 1/7 & 0.3   & Low-mass, Flared, Very Thick \\
L\_S0.1\_p3.5 & 1500 & 1e-4 & 0.1-10000 & 3.5 & 0 & 0.1   & Low-mass, Settled, Very Thick \\
I\_XF0.1\_p3.5 & 1500 & 1e-3 & 0.1-10000 & 3.5 & 1/4 & 0.1   & Intermediate-mass, Very Flared, Thin \\
I\_XF0.1\_p3.5 & 1500 & 1e-3 & 0.1-10000 & 3.5 & 1/4 & 0.2   & Intermediate-mass, Very Flared, Thick \\
I\_F0.1\_p3.5\_sm & 1500 & 1e-3 & 0.1-100 & 3.5 & 1/7 & 0.1   & Intermediate-mass, Flared, Thin, Small Grains \\
I\_V\_p3.5\_sm & 1500 & 1e-3 & 0.1-100 & 3.5 & v & v   & Intermediate-mass, Well-mixed gas and dust, Small Grains \\
I\_F0.2\_p3.5\_sm & 1500 & 1e-3 & 0.1-100 & 3.5 & 1/7 & 0.2   & Intermediate-mass, Flared, Thick, Small Grains \\
I\_F0.1\_p2.5 & 1500 & 1e-3 & 0.1-10000 & 2.5 & 1/7 & 0.1   & Intermediate-mass, Flared, Thin, Distribution 2.5 \\
I\_F0.2\_p2.5 & 1500 & 1e-3 & 0.1-10000 & 2.5 & 1/7 & 0.2   & Intermediate-mass, Flared, Thick, Distribution 2.5 \\
I\_V\_p2.5 & 1500 & 1e-3 & 0.1-10000 & 2.5 & v & v   & Intermediate-mass, Flared, Well-mixed gas and dust, Distribution 2.5 \\
I\_V\_p2.5\_sm & 1500 & 1e-3 & 0.1-100 & 2.5 & v & v   & Intermediate-mass, Flared, Well-mixed gas and dust, Distribution 2.5, Small Grains \\
M\_F0.1\_p2.5 & 1500 & 1e-2 & 0.1-10000 & 2.5 & 1/7 & 0.1   & Massive, Flared, Thin, Distribution 2.5 \\
M\_F0.2\_p2.5 & 1500 & 1e-2 & 0.1-10000 & 2.5 & 1/7 & 0.2   & Massive, Flared, Thick, Distribution 2.5 \\
M\_V\_p2.5 & 1500 & 1e-2 & 0.1-10000 & 2.5 & v & v   & Massive, Flared, Well-mixed gas and dust, Distribution 2.5 \\
PTD200\_F0.1\_p3.5 & 1500/200 & 1e-3 & 0.1-2/0.1-10000 & 3.5 & 1/7 & 0.1   & Intermediate-mass, PTD (3.6 AU), Flared, Thin \\
PTD200\_F0.2\_p3.5 & 1500/200 & 1e-3 & 0.1-2/0.1-10000 & 3.5 & 1/7 & 0.2   & Intermediate-mass, PTD (3.6 AU, Flared, Thick \\
PTD200\_V\_p3.5 & 1500/200 & 1e-3 & 0.1-2/0.1-10000 & 3.5 & v & v   & Intermediate-mass, PTD (3.6 AU), Well-mixed gas and dust \\
PTD380\_F0.1\_p3.5 & 1500/380 & 1e-3 & 0.1-2/0.1-10000 & 3.5 & 1/7 & 0.1   & Intermediate-mass, PTD (1.0 AU), Flared, Thin \\
PTD380\_F0.2\_p3.5 & 1500/380 & 1e-3 & 0.1-2/0.1-10000 & 3.5 & 1/7 & 0.2   & Intermediate-mass, PTD (1.0 AU, Flared, Thick \\
PTD380\_V\_p3.5 & 1500/380 & 1e-3 & 0.1-2/0.1-10000 & 3.5 & v & v   & Intermediate-mass, PTD (1.0 AU), Well-mixed gas and dust \\
TD200\_F0.1\_p3.5 & 200 & 1e-3 & 0.1-10000 & 3.5 & 1/7 & 0.1   & Intermediate-mass, TD (3.6 AU), Flared, Thin \\
TD200\_F0.2\_p3.5 & 200 & 1e-3 & 0.1-10000 & 3.5 & 1/7 & 0.2   & Intermediate-mass, TD (3.6 AU, Flared, Thick \\
TD200\_V\_p3.5 & 200 & 1e-3 & 0.1-10000 & 3.5 & v & v   & Intermediate-mass, TD (3.6 AU),  Well-mixed gas and dust \\
TD380\_F0.1\_p3.5 & 380 & 1e-3 & 0.1-10000 & 3.5 & 1/7 & 0.1   & Intermediate-mass, TD (1.0 AU), Flared, Thin \\
TD380\_F0.2\_p3.5 & 380 & 1e-3 & 0.1-10000 & 3.5 & 1/7 & 0.2   & Intermediate-mass, TD (1.0 AU), Flared, Thick \\
TD380\_V\_p3.5 & 380 & 1e-3 & 0.1-10000 & 3.5 & v & v   & Intermediate-mass, TD (1.0 AU),  Well-mixed gas and dust \\
\hline
\end{longtable}
\end{landscape}

\begin{table*}[h!]
\centering
\caption{SED $\alpha$ spectral indices for the RADMC models at various wavelengths.} 
\label{alphamodel-table}
\begin{tabular}{l c c c c c}
\hline\hline
Model & $\alpha$(J-70$\mu$m) & $\alpha$(H-70$\mu$m) & $\alpha$(8-70$\mu$m) & $\alpha$(24-70$\mu$m)   & $\alpha$(70-160$\mu$m)   \\
\hline
1 / I\_V\_p3.5	& -0.84 & -0.90 & -0.33 & -0.65  & -1.77 \\
2 / I\_S0.2\_p3.5	& -0.91 & -1.01 & -1.07 & -1.30  & -2.07 \\
3 / I\_S0.3\_p3.5	& -0.80 & -0.92 & -1.12 & -1.39  & -2.17 \\
4 / I\_S0.1\_p3.5	& -1.07 & -1.17 & -1.04 & -1.20  & -1.89  \\
5 / M\_S0.1\_p3.5	& -0.86 & -0.95 & -1.10 & -1.35  & -1.28 \\
6 / M\_V\_p3.5	& -0.64 & -0.69 & -0.03 & -0.14  & -0.80 \\
7 / M\_F0.1\_p3.5	& -0.85 & -0.93 & -0.62 & -0.70  & -0.80 \\
8 / M\_F0.05\_p3.5	& -1.04 & -1.12 & -0.75 & -0.80  & -0.68 \\
9 / M\_F0.2\_p3.5	& -0.66 & -0.73 & -0.59 & -0.71  & -1.06 \\
10 / M\_F0.3\_p3.5	& -0.53 & -0.61 & -0.62 & -0.76  & -1.25 \\
11 / I\_F0.05\_p3.5	& -1.14 & -1.23 & -0.90 & -1.03  & -1.69 \\
12 / I\_F0.2\_p3.5	& -0.87 & -0.95 & -0.89 & -1.16  & -1.99 \\
13 / I\_F0.3\_p3.5	& -0.79 & -0.88 & -0.96 & -1.27  & -2.10 \\
14 / I\_F0.1\_p3.5	& -1.01 & -1.09 & -0.85 & -1.02  & -1.82 \\
15 / L\_V\_p3.5		& -1.17 & -1.26 & -0.91 & -1.58  & -2.48 \\
16 / L\_F0.05\_p3.5	& -1.34 & -1.44 & -1.24 & -1.64  & -2.67 \\
17 / L\_F0.1\_p3.5	& -1.25 & -1.36 & -1.27 & -1.75  & -2.71 \\
18 / L\_F0.2\_p3.5	& -1.16 & -1.27 & -1.36 & -1.96  & -2.76 \\
19 / L\_F0.3\_p3.5	& -1.11 & -1.22 & -1.45 & -2.10  & -2.80 \\
20 / L\_S0.1\_p3.5	& -1.29 & -1.39 & -1.39 & -1.82  & -2.83 \\
21 / I\_XF0.1\_p3.5	& -0.89 & -0.95 & -0.50 & -0.74  & -1.74 \\
22 / I\_XF0.1\_p3.5	& -0.76 & -0.83 & -0.52 & -0.91  & -1.93 \\
23 / I\_F0.1\_p3.5\_sm	& -0.83 & -0.90 & -0.51 & -0.41  & -0.77 \\
24 / I\_V\_p3.5\_sm	& -0.61 & -0.65 & 0.10 & 0.27  & -0.85 \\
25 / I\_F0.2\_p3.5\_sm	& -0.63 & -0.70 & -0.47 & -0.46  & -1.11 \\
26 / I\_F0.1\_p2.5	& -1.08 & -1.17 & -0.99 & -1.29  & -2.75 \\
27 / I\_F0.2\_p2.5	& -0.97 & -1.06 & -1.08 & -1.52  & -2.96 \\
28 / I\_V\_p2.5		& -1.01 & -1.09 & -0.65 & -1.15  & -2.85\\
44 / I\_V\_p2.5\_sm	& -0.89 & -0.96 & -0.43 & -0.63  & -0.77 \\
29 / M\_F0.1\_p2.5	& -0.91 & -0.99 & -0.73 & -0.92  & -1.95 \\
30 / M\_F0.2\_p2.5	& -0.76 & -0.84 & -0.77 & -1.05  & -2.28 \\
31 / M\_V\_p2.5		& -0.80 & -0.86 & -0.32 & -0.65  & -2.15 \\
32 / PTD200\_F0.1\_p3.5	& -0.91 & -0.98 & -0.54 & -0.73  & -2.13 \\
33 / PTD200\_F0.2\_p3.5	& -0.68 & -0.73 & -0.28 & -1.28  & -2.49 \\
34 / PTD200\_V\_p3.5	& -0.84 & -0.90 & -0.21 & -0.88  & -1.96 \\
35 / PTD380\_F0.1\_p3.5	& -1.01 & -1.08 & -0.71 & -1.09  & -1.89 \\
36 / PTD380\_F0.2\_p3.5	& -0.80 & -0.86 & -0.82 & -1.39  & -2.18 \\
37 / PTD380\_V\_p3.5	& -0.89 & -0.95 & -0.40 & -0.59  & -1.71 \\
38 / TD200\_F0.1\_p3.5	& -0.84 & -0.90 & -0.17 & -1.53  & -2.08 \\
39 / TD200\_F0.2\_p3.5	& -0.70 & -0.75 & -0.15 & -1.63  & -2.15 \\
40 / TD200\_V\_p3.5	& -0.88 & -0.94 & -0.19 & -1.43  & -1.76 \\
41 / TD380\_F0.1\_p3.5	& -0.93 & -0.99 & -0.88 & -1.36  & -1.96 \\
42 / TD380\_F0.2\_p3.5	& -0.77 & -0.82 & -0.91 & -1.47  & -2.13 \\
43 / TD380\_V\_p3.5	& -0.89 & -0.95 & -0.62 & -0.81  & -1.60 \\
\hline
\end{tabular}
\tablefoot{SED spectral indices ($\alpha$) for the disk models in Table \ref{model-table}.}
\end{table*}

\section{Notes on individual objects \label{individualobjects-app}}

Here we present a comparison of the fluxes predicted by our previous, Spitzer-based
disk models (Sicilia-Aguilar et al. 2011; SA13) and the actual Herschel observations.
Herschel/PACS fluxes and upper limits are highly consistent
with the Spitzer (MIPS/IRS) data. The measured far-IR fluxes are in good 
agreement with our
model predictions based on optical, Spitzer, and millimetre data, although some cases present
higher or lower fluxes than expected. 
Extended emission at 160\,$\mu$m 
likely affects the fluxes of objects like 72-1427 (Figure \ref{seds1-fig}), 21374763+573242323, 12-1091
(Figure \ref{seds2-fig}), and probably
others like 14-1017 (Figure \ref{seds1-fig}), 213945860+573051704, and 21400451+5728363 (Figure \ref{seds3-fig}), 
even though all of them are clearly
detected at this wavelength. 
The sharp turn-down/low mid-IR fluxes in objects like 13-1250, 
21392541+5733202, 21395813+5728335, and
21-2006 (Figure \ref{seds3-fig}) are also confirmed by Herschel/PACS. 
In particular, 13-1250 shows a remarkably steep SED
in the far-IR, a strong silicate feature and low
mid-IR fluxes, suggestive of a disk gap. A low far-IR flux could
indicate a low dust mass and/or strong flattening of the disk beyond what we had assumed in
Spitzer-based models. The star 21374275+5733250 (Figure \ref{seds2-fig}) 
also shows relatively lower-than-expected
fluxes in the far-IR, although since it is surrounded by nebulosity very close to the bright
binary CCDM J2137+5734, its 24\,$\mu$m flux could be contaminated.

\begin{figure*}
\centering
\begin{tabular}{cccc}
\includegraphics[width=0.24\linewidth]{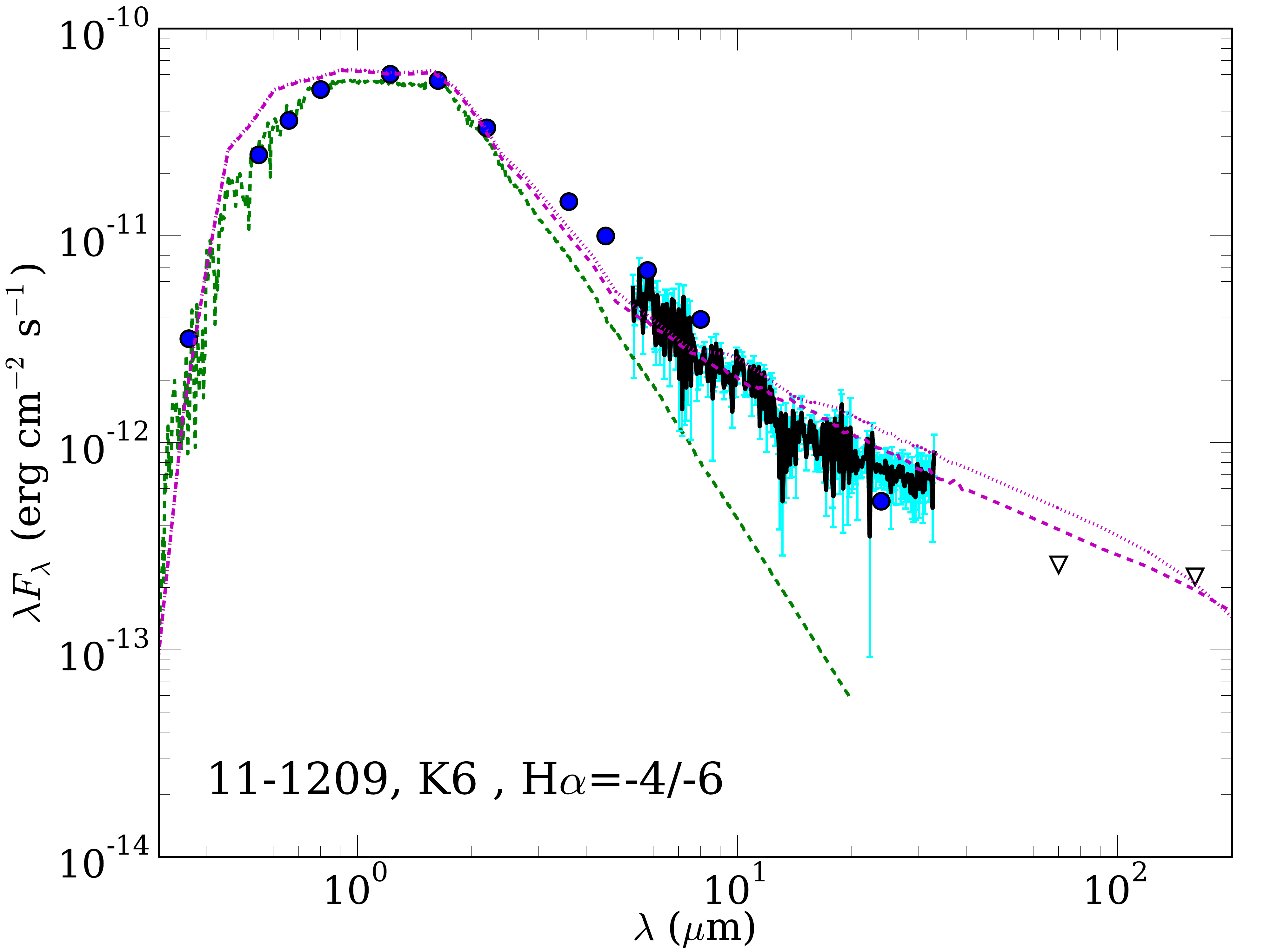} &
\includegraphics[width=0.24\linewidth]{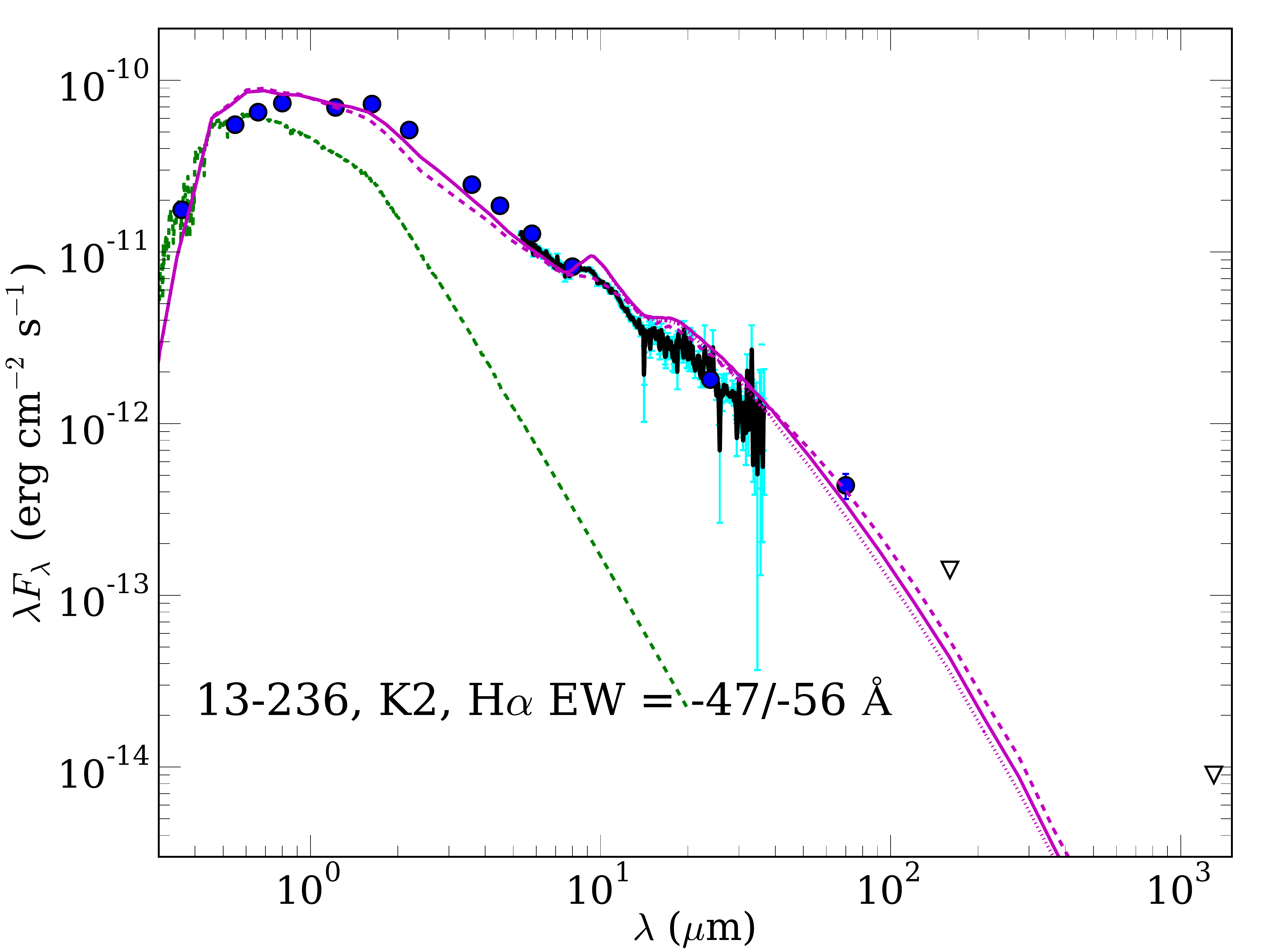} &
\includegraphics[width=0.24\linewidth]{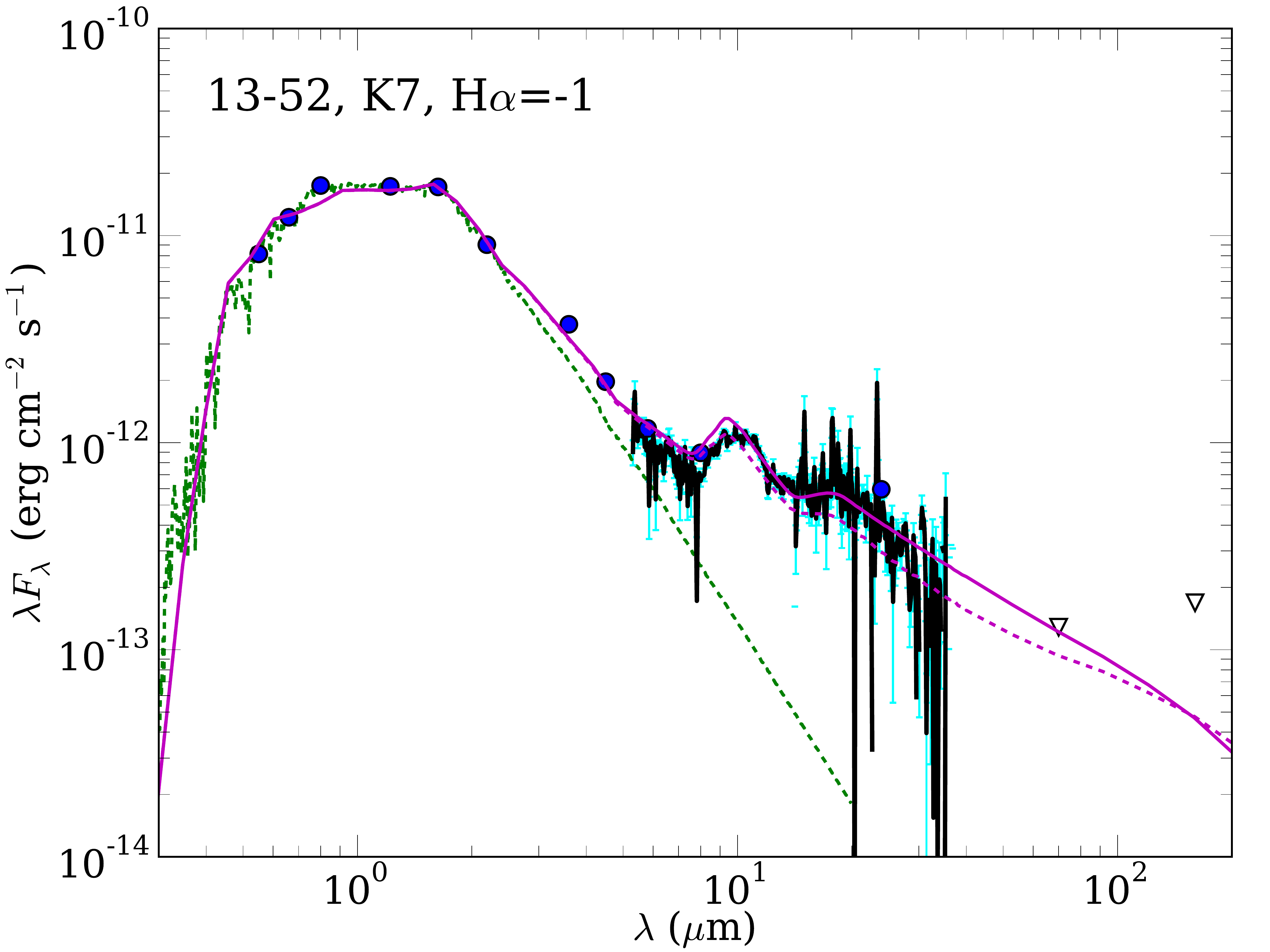} &
\includegraphics[width=0.24\linewidth]{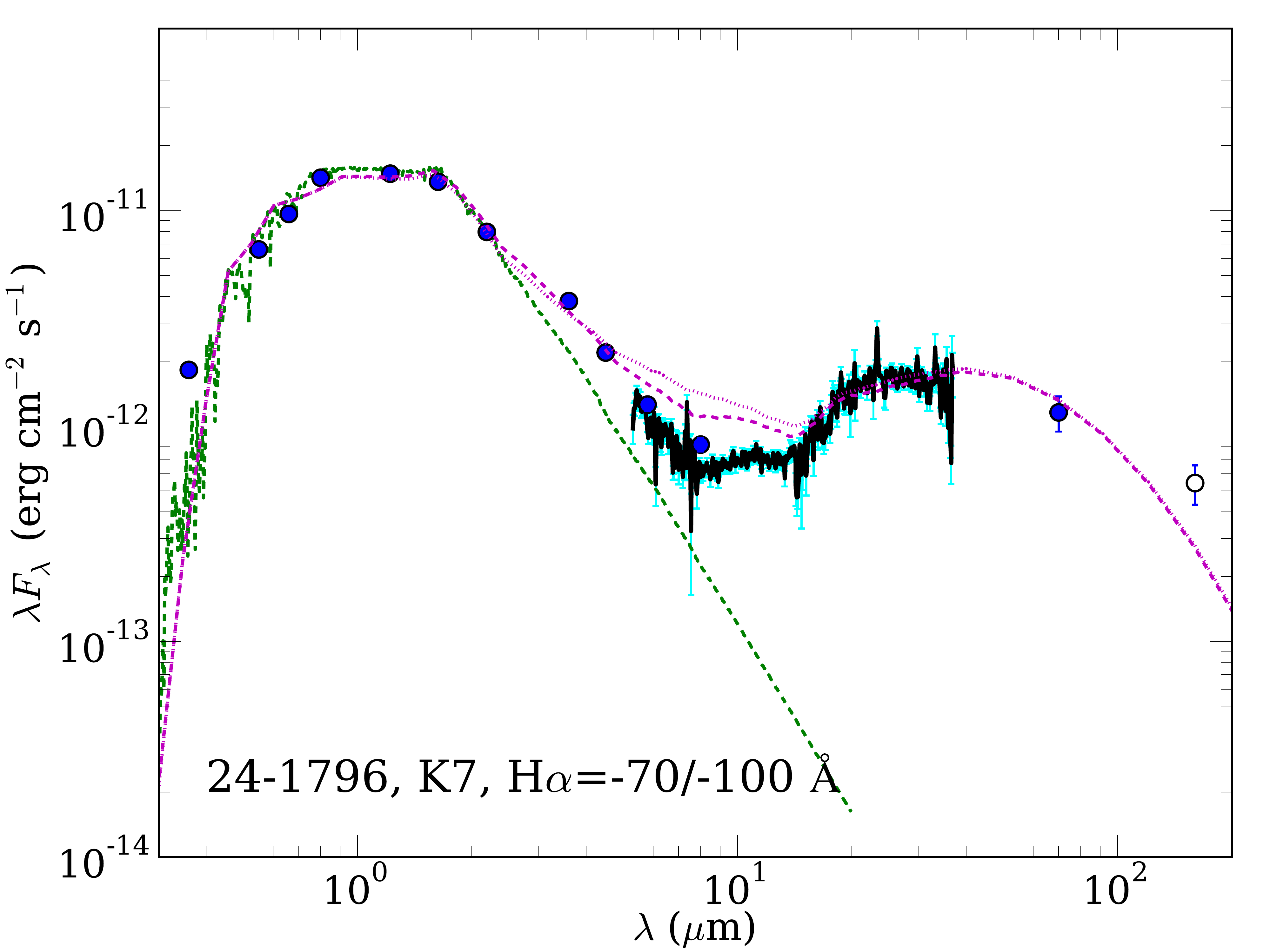} \\
\includegraphics[width=0.24\linewidth]{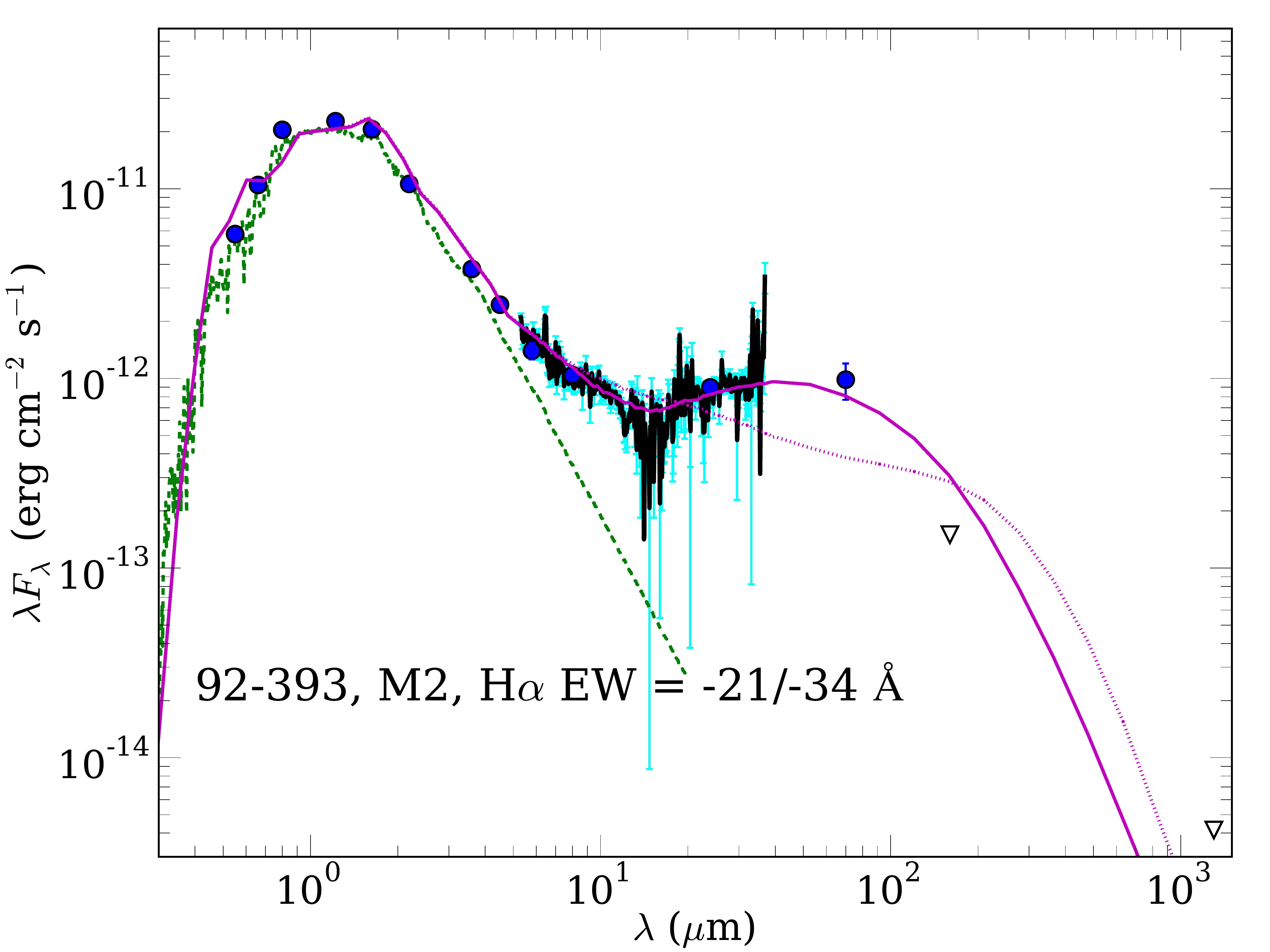} &
\includegraphics[width=0.24\linewidth]{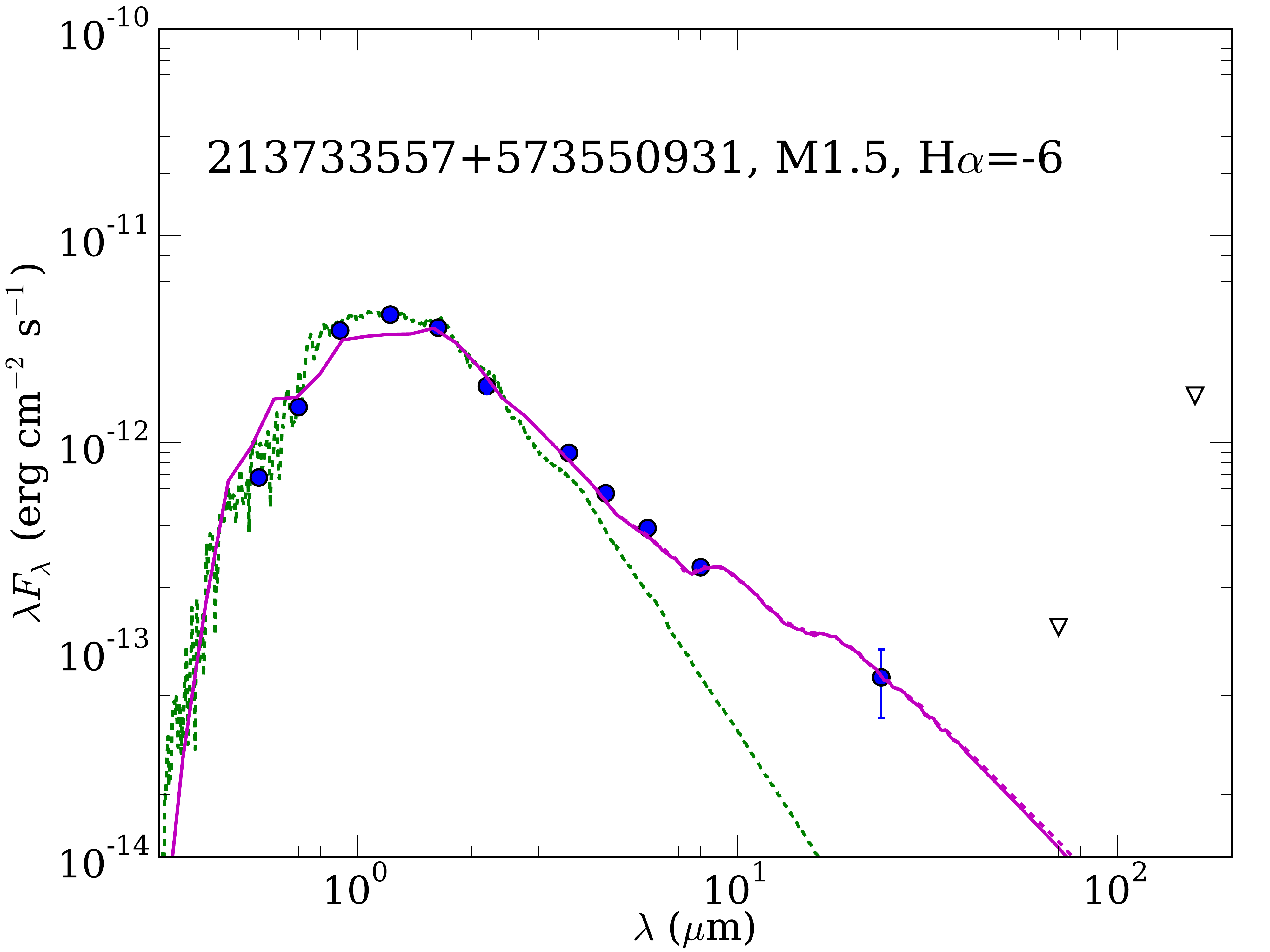} &
\includegraphics[width=0.24\linewidth]{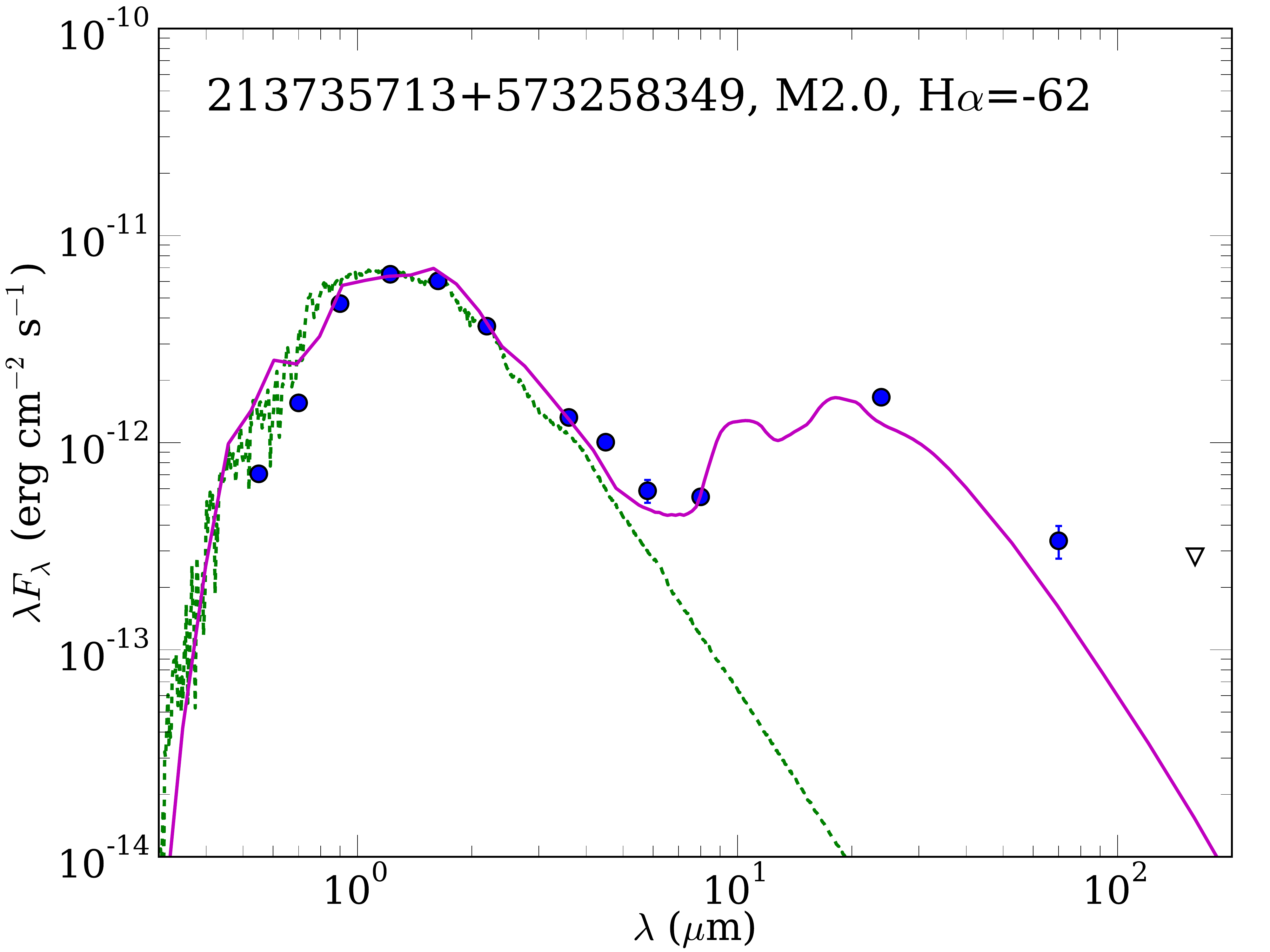} &
\includegraphics[width=0.24\linewidth]{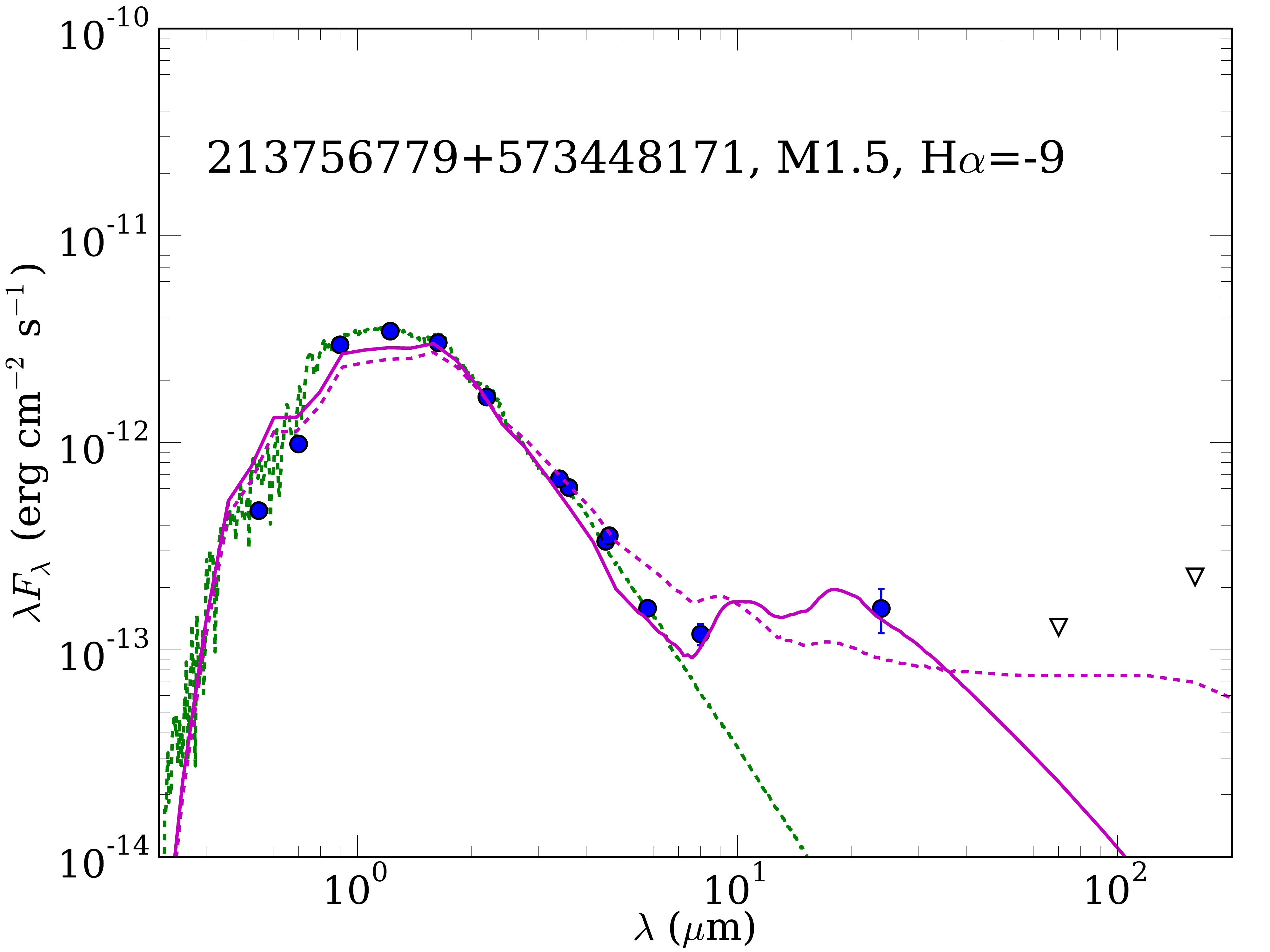} \\
\includegraphics[width=0.24\linewidth]{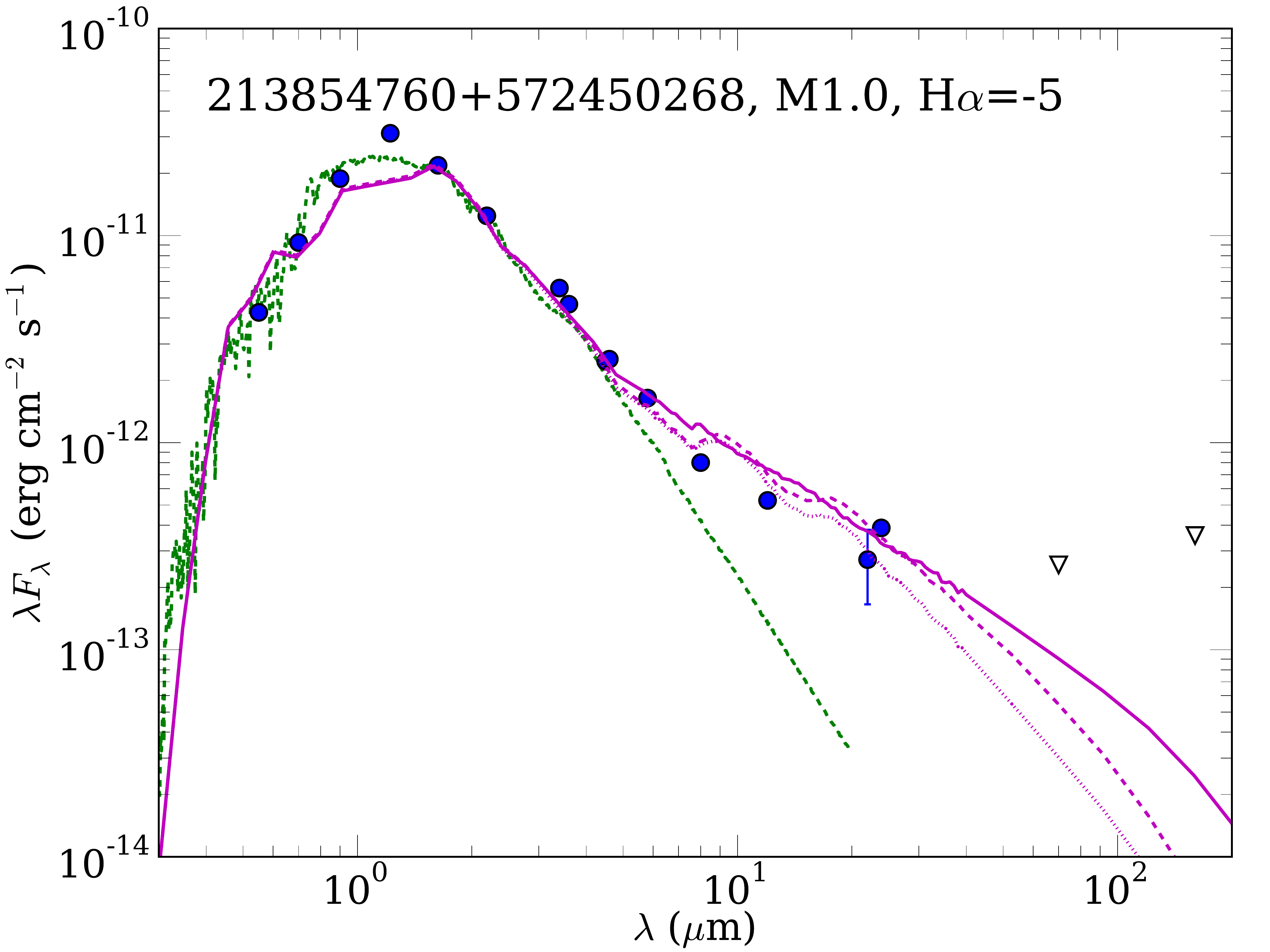} &
\includegraphics[width=0.24\linewidth]{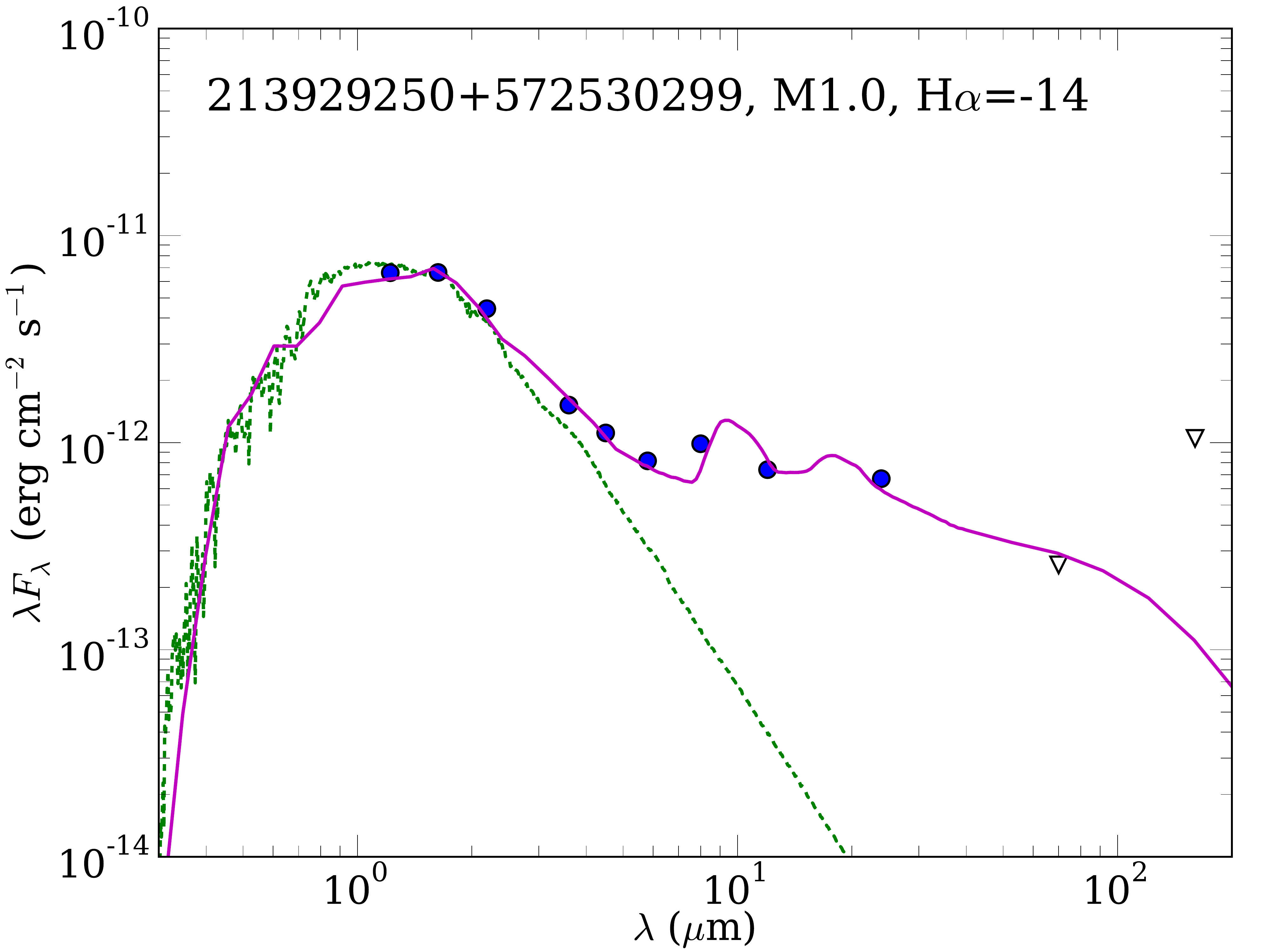} &
\includegraphics[width=0.24\linewidth]{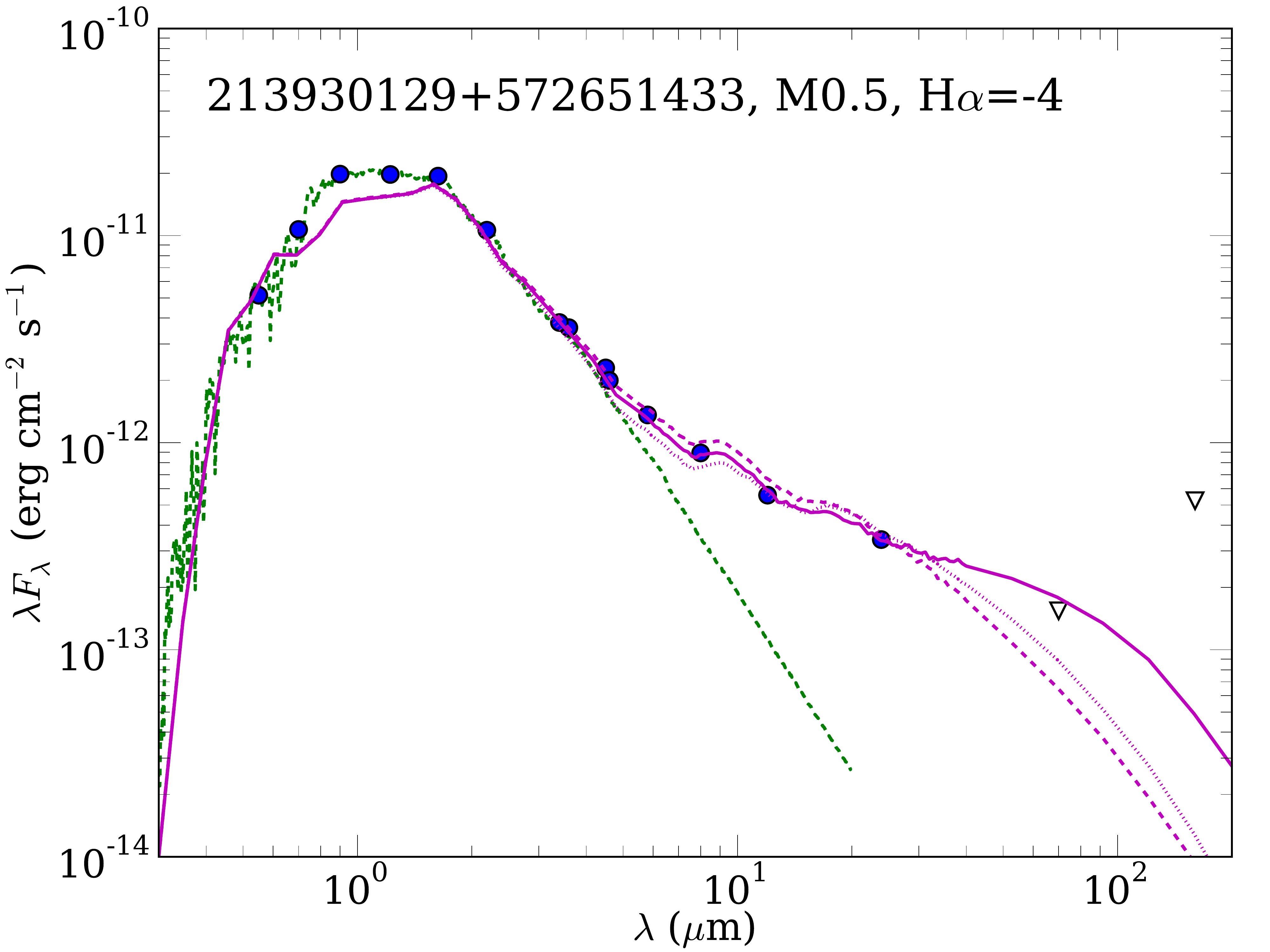} &
\includegraphics[width=0.24\linewidth]{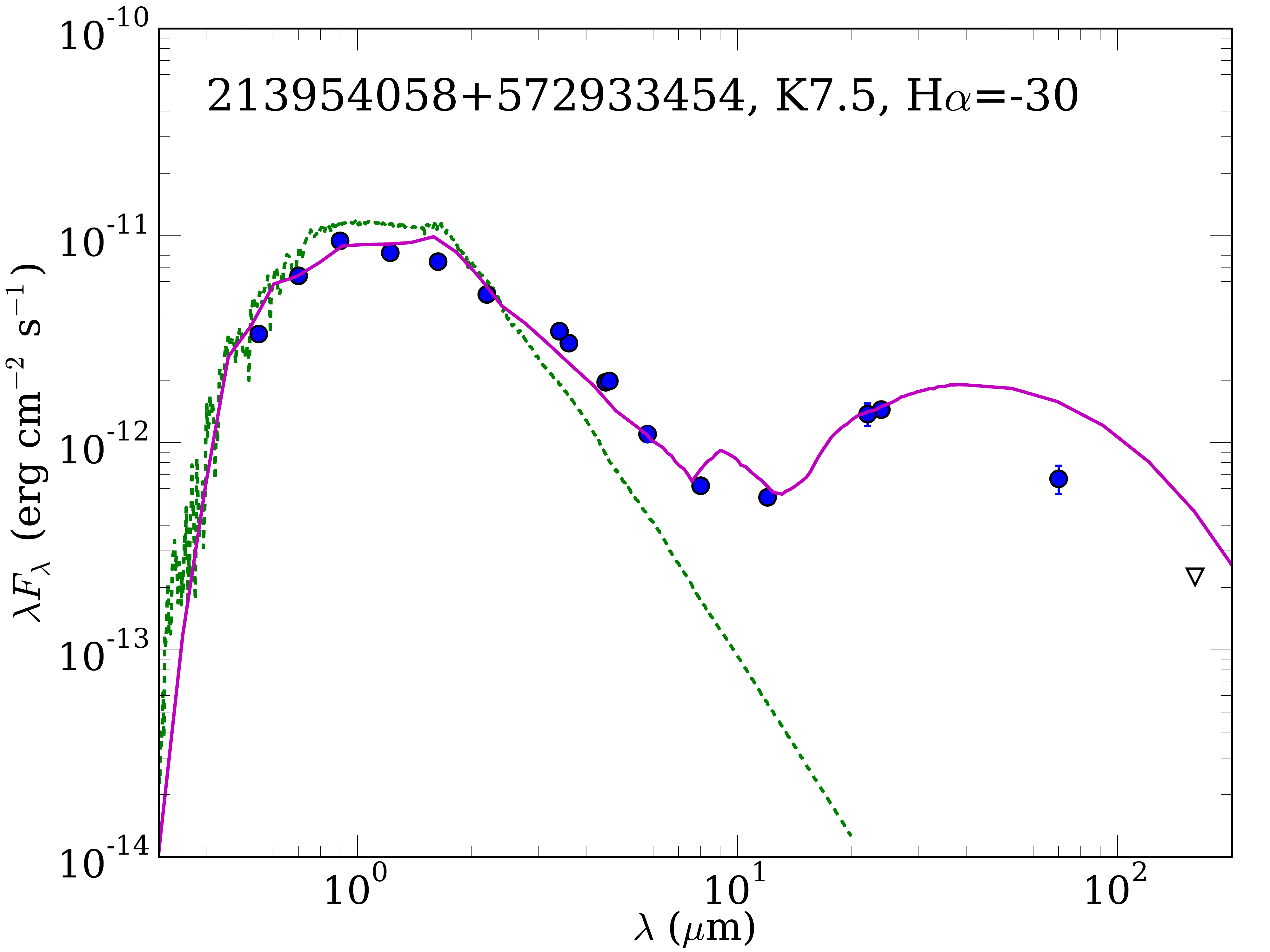} \\
\end{tabular}
\caption{Comparison of some of the Spitzer-based SED models (pink lines; Sicilia-Aguilar et al. 2011, 2013)
and predicted 70/160\,$\mu$m fluxes with the actual Herschel/PACS observations. Filled symbols mark detections at different wavelengths. Errorbars are shown in blue
for both photometry points (often smaller than the symbols) and the IRS spectra. Upper limits
are marked as inverted open triangles. Marginal detections (close to 3$\sigma$ or affected by nebulosity)
are marked as open circles. A photospheric MARCS model is displayed as a dotted line for comparison. \label{oldmodels-fig}}
\end{figure*}

Figure \ref{oldmodels-fig} presents the SEDs including the Herschel data together with 
our previous disk models derived from optical, Spitzer, and millimetre observations alone.
Individual disk modelling is beyond the scope of this paper. Further millimetre-wave
observations are essential to study the global evolution of these objects in more detail.
The 70 and 160\,$\mu$m model flux predictions for objects such as 13-236,  13-52, 24-1796 
are in excellent agreement with the observations. In cases such as 213930129+572651433, the
upper limits favour an interpretation of the disk in terms of a relatively low-mass/low-flaring, and not
simply as a massive but very settled disk. There is some disagreement (around 30\%)
between predicted and observed fluxes (or upper limits) in 92-393, 11-1209, 
213929250+572530299 (model predictions slightly higher than observed), and 213735713+573258349 (model
prediction sligthly lower than observed). This confirms that, although subject to strong
uncertainties in the disk mass and dust grain properties, the mid-IR based SED classification offers
a powerful prediction of the global evolutionary status of most protoplanetary disks.

\end{appendix}

\end{document}